\documentclass[seceq]{ptptex}
\usepackage{wrapft}
\usepackage{graphicx}
\usepackage{color}
\usepackage{dcolumn}
\usepackage{bm}
\usepackage{mathrsfs}
\usepackage{dsfont}
\usepackage[colorlinks]{hyperref}
\usepackage{hyperref}
\hypersetup{colorlinks=true,%
            citecolor=blue,%
            filecolor=blue,%
            linkcolor=blue,%
            urlcolor=blue}
\def\t#1{\tilde{#1}}
\def\wh#1{\widehat{#1}}
\def\h#1{\hat{#1}}
\def\b#1{\bar{#1}}
\def\ol#1{\overline{#1}}
\def\Sc#1{\textsc{#1}}

\notypesetlogo                       

\title{
On the Luttinger theorem concerning number of particles in the ground states of systems of interacting fermions%
}

\author{
Behnam \textsc{Farid}%
}

\inst{
Institute for Theoretical Physics, Department of Physics and Astronomy, University of Utrecht, Leuvenlaan 4, 3584 CE Utrecht, The Netherlands
}

\abst{
We analyze the original proof by Luttinger and Ward of the Luttinger theorem, according to which for uniform ground states of systems of (interacting) fermions, which may be metallic or insulating, the number of ${\bm k}$ points corresponding to $G_{\sigma}({\bm k};\mu) \ge 0$ is equal to the total number of particles with spin index $\sigma$ in these ground states. Here $G_{\sigma}({\bm k};\mu)$ is the single-particle Green function of particles with spin index $\sigma$ at the chemical potential $\mu$. For the cases where the two-body interaction potential is short-range (for which infrared divergences are ruled out), and in particular for lattice models (for which ultraviolet divergences are precluded), we explicitly demonstrate that this theorem is unconditionally valid, irrespective of the strength of the bare interaction potential (assumed to be repulsive). We arrive at this conclusion by amongst other things demonstrating that the perturbation series expansion for self-energy in terms of skeleton diagrams, as encountered in the proof of the Luttinger-Ward identity, is \emph{uniformly} convergent for almost all momenta and energies. We further investigate the mechanisms underlying some reported instances of failure of the Luttinger theorem. With one exception, for all the cases considered in this paper, we show that the apparent failures of the Luttinger theorem can be attributed either to shortcomings of the employed single-particle Green functions (and the associated self-energies) or to misapplication of this theorem. The one exceptional case brings to light the possibility of a genuine failure of the Luttinger theorem for \emph{insulating} ground states, which we show to be brought about by a \emph{false limit} that in principle can be reached on taking the zero-temperature limit without the value of $\mu$ coinciding with the zero-temperature limit of the chemical potential satisfying the equation of state at finite temperatures; no such ambiguity can arise for metallic states.
}

\begin{document}

\maketitle

\noindent {\scriptsize Preprint number: ITP-UU-2007/51}

{\scriptsize{\tableofcontents}}

\section{Introduction}
\label{s1}

The Luttinger theorem that we consider in this paper was first formulated and demonstrated by Luttinger and Ward \cite{LW60}; although cited by some as the original source, a subsequent paper by Luttinger \cite{JML60} merely quotes this theorem \cite{Note1}. Here we follow tradition and refer to this theorem as Luttinger theorem, although Luttinger-Ward theorem, to be distinguished from the Luttinger-Ward identity, which similarly originates in Ref.~\citen{LW60}, is a more appropriate appellation.

Over the course of the past fifteen years, the Luttinger theorem has been a subject of much attention and controversy, being in turns found to be both true and variously false. In general, the opponents of the theorem state, the proof having been based on the weak-coupling many-body perturbation theory, the theorem need not apply to strongly-correlated states. Curiously, this nostrum for all outstanding problems in condensed-matter physics disregards existence of exact non-perturbative proofs of the theorem for strongly correlated metallic states in one space dimension \cite{BB97,YOA97}, which at least implies that strong correlation need not be a bar to the validity of the Luttinger theorem. As we shall show in this paper, this reasoning of the opponents further only partially reflects the true nature of the proof provided by Luttinger and Ward \cite{LW60}.

In analyzing the original proof by Luttinger and Ward of the Luttinger theorem \cite{LW60} in the following sections, we complement this proof by four main elements. Firstly, we show that one principal instance of using weak-coupling perturbation series expansion by Luttinger and Ward \cite{LW60} is purely formal and serves merely as a stepping stone for arriving at a general result which can be deduced without recourse to perturbation theory. Secondly, we prove that the series expansion for self-energy as encountered in the proof of the Luttinger-Ward identity, in terms of linked skeleton diagrams, \emph{bare} two-body interaction potential and the \emph{exact} single-particle Green functions of both spin species, is not only convergent, but is \emph{uniformly} convergent for almost all momenta and energies. We note that, as in the treatment by Luttinger and Ward \cite{LW60}, our considerations in this paper explicitly relate to the cases where the two-body interaction potential is short range; more explicitly, we do not consider the cases where the contribution of any of the above-mentioned skeleton diagrams can be infrared divergent, necessitating reformulation of the above-mentioned series for self-energy in terms of the screened interaction potential \cite{JH57} prior to theoretical investigations. Thirdly, on the basis of the latter result and some general analytic properties of the single-particle Green function and self-energy in the complex energy plane, we demonstrate that the proof by Luttinger and Ward of the Luttinger theorem, based on the term-by-term verification of the Luttinger-Ward identity, is mathematically fully justified. And fourthly, we consider in some detail one aspect that in principle can lead to breakdown of the Luttinger theorem for \emph{metallic} ground states (GSs), related to the behaviour of self-energy as function of energy in the immediate neighbourhood of the underlying Fermi energy,\footnote{This aspect was considered by Luttinger and Ward \protect\cite{LW60}, however only for those metallic GSs which have since come to be known as \emph{conventional} Fermi liquids. Our consideration covers all possible metallic GSs.} and show that this breakdown cannot take place.

As regards the last statement, we show that although a specific equality, which is of vital significance to the validity of the Luttinger theorem, can be invalid for some ${\bm k}$ (as is indeed the case for the Luttinger-liquid metallic states in one space dimension), for stable GSs this invalidity \emph{cannot} persist over a set of ${\bm k}$ points of non-zero measure. On the basis of this observation and of some exact identity, we arrive at the conclusion that the Luttinger theorem is valid \emph{if and only if} the Luttinger-Ward identity is valid; deviation from this result signals a pathology in the underlying GS. We thus conclude that the Luttinger theorem is specifically valid for all stable uniform \emph{metallic} GSs.

In the light of the last conclusion, which is in conformity with that by Dzyaloshinski\v{\i} \cite{IED03} on the basis of a recent revaluation of the Luttinger theorem, in this paper we pay special attention to a number of prominent reports concerning breakdown of the Luttinger theorem. By carefully examining these reports, we clarify the mechanisms responsible for the observed failures of the Luttinger theorem. In all the cases concerning \emph{metallic} GSs, the problems turn out to be unequivocally external to the Luttinger theorem. One of the observations that we make in the process of examining two of the above-mentioned reports (Sections \ref{ss64} and \ref{ss65}) is of relevance to attempts towards experimental determination of the extent of the Fermi seas (or Luttinger seas) of the GSs of correlated systems with the aid of the angle-resolved photoemission spectroscopy. The experimentalist readers of this paper may consider to pay attention to this particular observation.

By analyzing a case, first detected by Rosch \cite{AR06}, concerning breakdown of the Luttinger theorem in the case of a Mott-insulating GS, we show that this apparent breakdown has its root in a false zero-temperature limit. In general, but not necessarily, for all insulating GSs a similar false limit, or limits, can be arrived at by evaluating a zero-temperature limit that is basic to the Luttinger theorem at a fixed chemical potential $\mu$ different from $\mu_{\infty}$, the zero-temperature \emph{limit} of the chemical potential satisfying the equation of state at finite temperatures. Although it is formally true that, for insulating GSs the $\mu$ in $G_{\sigma}({\bm k};\mu)$ is solely required to be located inside the gap region of the single-particle excitations of the underlying $N$-particle GS, for the specific case that we consider (Sec.~\ref{ss61}), we find that the expression of which the zero temperature limit has to be taken is a non-trivial function of $x\equiv \beta (\mu-\mu_{\infty})$, which, for sufficiently large $\beta \equiv 1/(k_{\Sc b} T)$, is to exponential accuracy equal to $\frac{1}{2}\tanh(\frac{1}{2} x)$; here $T$ is temperature. Whereas for $\mu=\mu_{\infty}$ this function is vanishing for \emph{all} $\beta$, on taking the limit $\beta\to\infty$ for $\mu\not=\mu_{\infty}$, it becomes impossible to recover the correct value corresponding to $\mu=\mu_{\infty}$ by subsequently effecting the limit $\mu \to\mu_{\infty}$. A problem such as encountered here amounts to a manifestation of the fact that in general not all \emph{repeated limits} of a multi-variable function need be equal (\S\S~302-306 in Ref.~\citen{EWH27}).

Although for insulating GSs the requirement $\mu=\mu_{\infty}$ may appear to be improvised at first glance, some reflection proves the contrary: for \emph{interacting} systems the conception of an insulating gap, i.e. a finite interval free from single-particle excitation energies, is principally a false one; in principle, such gap only truly exists at $\beta=\infty$ by virtue of the \emph{suppression} to zero of the amplitudes of the single-particle excitations whose corresponding energies are inside the gap region independently of $\beta$.\footnote{This aspect is best appreciated by considering the Lehmann representation of the thermal single-particle Green function, for which we refer the reader to appendix \protect\ref{sf}. } From this perspective, $\beta=\infty$ (or $T=0$) is a singular point so that assigning to the $\mu$ in $G_{\sigma}({\bm k};\mu)$ any other value than $\mu_{\infty}$, although in many instances harmless, is principally illegitimate; the requirement to identify $\mu$ with $\mu_{\infty}$ in the cases of insulating GSs, must be viewed in the same light as is done in the cases of metallic GSs. Consequently, the possibility of violation of the Luttinger theorem for insulating GSs, ensuing the choice $\mu\not=\mu_{\infty}$, cannot be rightfully viewed as signifying a shortcoming in the Luttinger theorem.

This paper contains a number of new exact results that have played a role in our present study and which may be of interest in investigations related to areas outside the restricted scope of this paper.

\section{Preliminaries}
\label{s2}

Considering the $N$-particle uniform GS of a single-band Hamiltonian $\wh{H}$, $N$ is obtained through
(appendix \ref{sc})
\begin{equation}
N = \sum_{\sigma} N_{\sigma},\;\;\; N_{\sigma} = \sum_{\bm k} {\sf
n}_{\sigma}({\bm k}), \label{e1}
\end{equation}
where ${\sf n}_{\sigma}({\bm k})$ is the GS momentum distribution
function corresponding to particles with spin $\sigma$, calculated
according to
\begin{equation}
{\sf n}_{\sigma}({\bm k}) = \frac{1}{\hbar}\int_{\mathscr{C}_+(\mu)}
\frac{{\rm d}z}{2\pi i}\; \t{G}_{\sigma}({\bm k};z). \label{e2}
\end{equation}
Here $\t{G}_{\sigma}({\bm k};z)$ is the single-particle Green function corresponding to the particles with spin $\sigma$ in the $N$-particle GS of $\wh{H}$, and $\mathscr{C}_+(\mu)$ a closed counter-clockwise oriented contour which begins and terminates at the point of infinity of the complex $z$ plane and intersects the real energy axis at the chemical potential $\mu$ corresponding to $N$ particles (more about this $\mu$ later, in particular in Sec.~\ref{ss23}); $\mathscr{C}_+(\mu)$ thus contains the interval $[-\infty,\mu]$ of the real energy axis. The summation $\sum_{\bm k}$ in Eq.~(\ref{e1}) covers the entire relevant wave-vector space; for instance, in the cases where $\wh{H}$ is defined on a Bravais lattice $\{ {\bm R}_j \}$, this sum covers the corresponding first Brillouin zone (1BZ).

Throughout this paper we assume that $[\wh{H}, \wh{N}_{\sigma}]_{-} =0$, $\forall\sigma$, where $\wh{N}_{\sigma}$ is the partial number operator corresponding to spin-$\sigma$ particles. Consequently, with some exceptions, in this paper we deal with $N_{\sigma}$, $\forall\sigma$, rather than $N \equiv \sum_{\sigma} N_{\sigma}$.

The physical Green function $G_{\sigma}({\bm k};\varepsilon)$, where
$\varepsilon \in \mathds{R}$, is deduced from $\t{G}_{\sigma}({\bm
k};z)$ according to
\begin{equation}
G_{\sigma}({\bm k};\varepsilon) = \t{G}_{\sigma}({\bm k};\varepsilon
\pm i 0^+) \equiv \lim_{\eta\downarrow 0} \t{G}_{\sigma}({\bm k};\varepsilon
\pm i\eta),\;\;\; \varepsilon \gtrless \mu. \label{e3}
\end{equation}
Conventionally, $\t{G}_{\sigma}({\bm k};\varepsilon + i 0^+)$ ($\t{G}_{\sigma}({\bm k};\varepsilon - i 0^+)$), $\varepsilon\in \mathds{R}$, is referred to as the retarded (advanced) Green function and is denoted by $G_{\sigma}^{\Sc r}({\bm k};\varepsilon)$ ($G_{\sigma}^{\Sc a}({\bm k};\varepsilon)$) \cite{FW03}.

With $\varepsilon_{\bm k}$ denoting the non-interacting single-particle energy dispersion underlying $\wh{H}$, and $\t{\Sigma}_{\sigma}({\bm k};z)$ self-energy (more precisely, the \emph{proper} self-energy \cite{FW03}), by the Dyson equation one has
\begin{equation}
\t{G}_{\sigma}({\bm k};z) = \frac{\hbar}{z-\varepsilon_{\bm
k}-\hbar\t{\Sigma}_{\sigma}({\bm k};z)}. \label{e4}
\end{equation}
The physical self-energy $\Sigma_{\sigma}({\bm k};\varepsilon)$ is deduced from $\t{\Sigma}_{\sigma}({\bm k};z)$ according to (cf. Eq.~(\ref{e3}))
\begin{equation}
\Sigma_{\sigma}({\bm k};\varepsilon) = \t{\Sigma}_{\sigma}({\bm
k};\varepsilon \pm i 0^+),\;\;\; \varepsilon \gtrless \mu.
\label{e5}
\end{equation}
In analogy with $\t{G}_{\sigma}({\bm k};\varepsilon \pm i 0^+)$, $\varepsilon\in \mathds{R}$, $\t{\Sigma}_{\sigma}({\bm k};\varepsilon + i 0^+)$ ($\t{\Sigma}_{\sigma}({\bm k};\varepsilon - i 0^+)$) is referred to as retarded (advanced) self-energy.

\subsection{Metals}
\label{ss21}

Since for metallic GSs
\begin{equation}
\varepsilon_{\bm k} + \hbar\Sigma_{\sigma}({\bm k};\varepsilon_{\Sc
f}) = \varepsilon_{\Sc f}\;\;\; \mbox{\rm for}\;\;\; {\bm k}\in
\mathcal{S}_{\Sc f;\sigma}, \label{e6}
\end{equation}
where $\varepsilon_{\Sc f}$ denotes the Fermi energy (which differs only infinitesimally from $\mu$ at zero temperature) and $\mathcal{S}_{\Sc f;\sigma}$ the Fermi surface, and since $\Sigma_{\sigma}({\bm k};\varepsilon_{\Sc f}) \in \mathds{R}$, $\forall {\bm k}$ (see later, specifically Sec.~\ref{ss21s2}), the number of ${\bm k}$ points in the \emph{interior} of the Fermi sea,\footnote{Our use here of the word `\emph{interior}' is imprecise; this use would be precise if $\Theta(0) = 0$, and the word `\emph{closure}' would be appropriate if $\Theta(0)=1$. Depending on the circumstances, it is likely that $\Theta(0) = \frac{1}{2}$ may be the most appropriate convention (see Sec.~\protect\ref{ss25}).} which we denote by $N_{\Sc f;\sigma}$, is equal to
\begin{equation}
N_{\Sc f;\sigma} {:=} \sum_{\bm k} \Theta\big(\varepsilon_{\Sc f} -\varepsilon_{\bm k}-\hbar\Sigma_{\sigma}({\bm k};\varepsilon_{\Sc f})\big).
\label{e7}
\end{equation}

With $\mathcal{S}_{\Sc f;\sigma}$ considered to be the \emph{boundary} of the underlying Fermi sea, it is important to realise that Fermi sea need not be a \emph{closed} set so that knowledge of $\mathcal{S}_{\Sc f;\sigma}$ is not in general sufficient to specify the corresponding Fermi sea (here and later we use the terms \emph{interior}, \emph{boundary}, \emph{frontier}, \emph{open}, \emph{closed} and \emph{closure} in their technical sense, as specified in, e.g., Ref.~\citen{EWH27}, \S~55, or in Ref.~\citen{HS91}, Ch.~4, \S~3).\footnote{The notion of \emph{frontier} has not been defined in Ref.~\protect\citen{HS91}. Here we employ this notion as defined in Ref.~\protect\citen{EWH27}, pp. 77 and 78: ``\emph{Those points of a set $G$ which are not interior points of $G$, together with those points of $C(G)$ [the \emph{complement} of $G$ with respect to the embedding set] which are not interior points of $C(G)$, form a set which is called the frontier of $G$ or of $C(G)$}''. Note that a point belonging to the \emph{boundary} of $G$ is a point of the \emph{frontier} of $G$, however a point of the \emph{frontier} of $G$ need not be a point of the \emph{boundary} of $G$, the latter set being empty for $G$ an \emph{open} set. For completeness, for both mathematical and physical consistency, the embedding ${\bm k}$ space, such as the $\mathrm{1BZ}$ in the case of a system defined on a Bravais lattice, should be considered as \emph{open}.} Nonetheless, the quantity $N_{\Sc f;\sigma}$ as defined in Eq.~(\ref{e7}) is determinate for metallic GSs, aside from the mathematical ambiguity corresponding to the value of $\Theta(x)$ at $x=0$, which is unimportant for macroscopic states. Evidently, for macroscopic states both sides of Eq.~(\ref{e7}) should be divided by a macroscopic quantity, such as the `volume' of the system, in order for the result in Eq.~(\ref{e7}) to be meaningful; only in this sense is the ambiguity of $\Theta(x)$ at $x=0$ unimportant. The same statement applies to similar expressions to be met below.

The fact that Fermi sea may not be \emph{closed} is however of experimental consequence, as has been pointed out earlier \cite{ET05,KRT06}; the extent of the Fermi sea as deduced from the measured Fermi surface need not be reliable. As regards metallic states, it is only here that the notion of the Luttinger surface $\mathcal{S}_{\Sc l;\sigma}$ has any significance; this notion plays no role insofar as the theoretical calculation of $N_{\Sc f;\sigma}$, according to Eq.~(\ref{e7}), is concerned. We shall formally define $\mathcal{S}_{\Sc l;\sigma}$ later in Sec.~\ref{ss24} while discussing non-metallic GSs. For now, however, we mention that for metallic states $\mathcal{S}_{\Sc f;\sigma} \cup \mathcal{S}_{\Sc l;\sigma}$ constitutes the \emph{frontier} of the underlying Fermi sea.

An example should be clarifying. Let $E_{\sigma}({\bm k}) \equiv \varepsilon_{\bm k} + \hbar \Sigma_{\sigma}({\bm k};\varepsilon_{\Sc f})$ so that Fermi sea is formally defined as $\{ {\bm k}\,\|\, E_{\sigma}({\bm k}) \le\varepsilon_{\Sc f}\}$. Consider ${\bm k} \equiv k \h{\bm n}$, where $\h{\bm n}$ is a unit vector centred at ${\bm k}={\bm 0}$ and pointing in some predetermined direction in the ${\bm k}$ space. Suppose that $E_{\sigma}({\bm k})$ continuously and strictly monotonically increases from some value less than $\varepsilon_{\Sc f}$ at $k=0$ to $+\infty$ at $k=k_{\star}$ and that it continuously and strictly monotonically increases from $-\infty$ at $k=k_{\star}$ to $+\infty$ at $k=+\infty$. By continuity one must have $E_{\sigma}({\bm k}) = \varepsilon_{\Sc f}$ at $k=k_1$ and $k=k_2$, where $0 < k_1 < k_{\star}$ and $k_{\star} < k_2 <\infty$. Thus by definition one has ${\bm k}_1, {\bm k}_2 \in \mathcal{S}_{\Sc f;\sigma}$, where ${\bm k}_j \equiv k_j \h{\bm n}$, $j=1,2$; since at ${\bm k} = {\bm k}_{\star} \equiv k_{\star} \h{\bm n}$ the equation $E_{\sigma}({\bm k}) = \varepsilon_{\Sc f}$ is not satisfied, ${\bm k}_{\star} \not\in \mathcal{S}_{\Sc f;\sigma}$. Yet, without the knowledge of $k_{\star}$ it is not possible to determine the region belonging to the underlying Fermi sea in the direction of $\h{\bm n}$ solely from the knowledge of the Fermi points ${\bm k}_1$ and ${\bm k}_2$; the additional knowledge concerning ${\bm k}_{\star}$ is indispensable for this task. In this example, ${\bm k}_{\star}$ belongs to the \emph{frontier} of the Fermi sea and its complementary space $\{{\bm k}\,\|\, E_{\sigma}({\bm k}) >\varepsilon_{\Sc f}\}$ in the underlying ${\bm k}$ space, and not to the \emph{boundary} of the underlying Fermi sea, that is $\mathcal{S}_{\Sc f;\sigma}$.\footnote{With reference to the previous footnote, in this example $G \equiv [0,k_1]\cup (k_{\star},k_2]$ and $C(G) \equiv (k_1,k_{\star}] \cup (k_2,\infty)$. Evidently, $k_{*}$ is neither an \emph{interior} point of $G$ nor that of $C(G)$.}

\subsubsection{Remarks}
\label{ss21s1}

It should be noted that the \emph{definition} of $N_{\Sc f;\sigma}$ depends solely on the existence of $\mathcal{S}_{\Sc f;\sigma}$ (that is that Eq.~(\ref{e6}) is satisfied for at least a single ${\bm k}$ for the given value of $\varepsilon_{\Sc f}$) and \emph{not} on the way in which ${\bm\nabla}_{\!\bm k}\Sigma_{\sigma}({\bm k};\varepsilon_{\Sc f})$ and $\partial\t{\Sigma}_{\sigma}({\bm k};z)/\partial z$, ${\bm k}\in \mathcal{S}_{\Sc f;\sigma}$, behave for respectively ${\bm k}$ in a neighbourhood of $\mathcal{S}_{\Sc f;\sigma}$ and $z$ in a neighbourhood of $\varepsilon_{\Sc f}$. Consequently, $N_{\Sc f;\sigma}$ is well-defined irrespective of whether the underlying metallic GS is a Fermi liquid or otherwise. Notably, $N_{\Sc f;\sigma}$ is well-defined for the cases where $\mathcal{S}_{\Sc f;\sigma} \not=\varnothing$ but $\mathcal{S}_{\Sc f;\sigma}$ is interspersed by finite pseudogap regions. With ${\bm k}_{\rm pg}$ denoting a point in the pseudogap region of the putative Fermi surface, $\mathrm{Re}[\Sigma_{\sigma}({\bm k};\varepsilon_{\Sc f})]$ should undergo a \emph{finite} positive discontinuity on transposing ${\bm k}$ from infinitesimally inside the Fermi sea through ${\bm k}_{\rm pg}$ to infinitesimally outside the Fermi sea \cite{BF03}. In the event that $\mathrm{Re}[\Sigma_{\sigma}({\bm k};\varepsilon_{\Sc f})]$ is \emph{lower semi-continuous} at ${\bm k}={\bm k}_{\rm pg}$ (\S~230 in Ref.~\citen{EWH27}, Ch.~9, \S~2 in Ref.~\citen{HS91}), ${\bm k}_{\rm pg}$ is \emph{formally} a point of $\mathcal{S}_{\Sc f;\sigma}$. Otherwise, by a slight modification of the conventional definition of $\mathcal{S}_{\Sc l;\sigma}$, which we shall introduce later in Sec.~\ref{ss24}, ${\bm k}_{\rm pg}$ can be made to count as a point of $\mathcal{S}_{\Sc l;\sigma}$. For physical reason, we shall consider ${\bm k}_{\rm pg} \in \mathcal{S}_{\Sc l;\sigma}$ even in the cases where $\mathrm{Re}[\Sigma_{\sigma}({\bm k};\varepsilon_{\Sc f})]$ is \emph{lower semi-continuous} at ${\bm k}_{\rm pg}$.

\subsubsection{Remarks}
\label{ss21s2}

In the extant literature one often encounters the expression for $N_{\Sc f;\sigma}$ in terms of $\mathrm{Re}[\Sigma_{\sigma}({\bm k};\varepsilon_{\Sc f})]$, or $\mathrm{Re}[\Sigma_{\sigma}({\bm k};\mu)]$, which is suggestive of the possibility that  $\Sigma_{\sigma}({\bm k};\varepsilon_{\Sc f})$, or $\Sigma_{\sigma}({\bm k};\mu)$, may be complex-valued. In fact, the computational results reported in Refs.~\citen{SLGB96a,SLGB96b,LSGB95,LSGB96} (to be examined in some detail in Sec.~\ref{ss63}) positively indicate violation of the result
\begin{equation}
\mathrm{Im}[\Sigma_{\sigma}({\bm k};\mu)] = 0,\;\;\;\forall {\bm k}.
\label{e8}
\end{equation}
In appendix \ref{sc} we shall deduce this result both on the basis of the Lehmann representation of $\t{G}_{\sigma}({\bm k};z)$ and of the spectral representation of this function in terms of the single-particle spectral function $A_{\sigma}({\bm k};\varepsilon)$; both approaches lead us directly to the fundamental property $\mathrm{Im}[G_{\sigma}({\bm k};\mu)]=0$, $\forall {\bm k}$. For now the following two remarks are in order.

Firstly, since $\varepsilon_{\bm k}, \varepsilon_{\Sc f} \in \mathds{R}$, without
\begin{equation}
\mathrm{Im}[\Sigma_{\sigma}({\bm k};\varepsilon_{\Sc f})] = 0
\label{e9}
\end{equation}
for \emph{some} ${\bm k}$, Eq.~(\ref{e6}) would not be satisfied for any ${\bm k}$, contradicting the assumption that the GS under consideration is metallic, with $\varepsilon_{\Sc f}$ its underlying Fermi energy. In other words, Eq.~(\ref{e8}) must at least be satisfied for \emph{all} ${\bm k}\in \mathcal{S}_{\Sc f;\sigma}$. Remarkably, even by making allowance for finite-temperature effects, the results reported in Refs.~\citen{SLGB96a,SLGB96b,LSGB95,LSGB96} are in clear violation of this fundamental fact.

Secondly, since $\t{\Sigma}_{\sigma}({\bm k};z)$ is analytic everywhere in the complex $z$ plane away from the real axis (appendix \ref{sc}),\cite{JML61} it follows that the approach
\begin{equation}
\t{\Sigma}_{\sigma}({\bm k}; \mu + \eta\,\mathrm{e}^{\pm i\vartheta}) \to \Sigma_{\sigma}({\bm k};\mu)\;\;\;\mbox{\rm for}\;\;\; \eta\downarrow 0,\;\; 0<\vartheta<\pi, \label{e10}
\end{equation}
is continuous by the Taylor theorem (see \S\S~3.22 and 5.4 in Ref.~\citen{WW62}); in fact, this continuity is uniform (see \S~3.61, \emph{Corollary} (i), in Ref.~\citen{WW62}, \S~217 in Ref.~\citen{EWH27}). The exclusion of $\vartheta=0, \pi$ from the range of variation of $\vartheta$ is in keeping with the prescription in Eq.~(\ref{e5}). Since $\Sigma_{\sigma}({\bm k};\mu)$ can be unbounded for some ${\bm k}$ (see Sec.~\ref{ss24} where we discuss the Luttinger surface), Eq.~(\ref{e10}) as well as the following related expressions are only meaningful under the assumption that $\vert\Sigma_{\sigma}({\bm k};\mu)\vert<\infty$. In this connection, one should bear in mind that for the ${\bm k}$ points at which $\vert\Sigma_{\sigma}({\bm k};\mu)\vert=\infty$, whereby $G_{\sigma}({\bm k};\mu)=0$, the very question with regard to the `value' of $\mathrm{Im}[\Sigma_{\sigma}({\bm k};\mu)]$ is utterly meaningless.

By writing
\begin{equation}
\t{\Sigma}_{\sigma}({\bm k};z) = \Sigma_{\sigma}({\bm k};\mu) + \delta\t{\Sigma}_{\sigma}^{\pm}({\bm k};z),\;\;\; \mathrm{Im}(z) \gtrless 0, \label{e11}
\end{equation}
on account of the result in Eq.~(\ref{e10}) it follows that the approach
\begin{equation}
\delta\t{\Sigma}_{\sigma}^{\pm}({\bm k}; \mu + \eta\,\mathrm{e}^{\pm i\vartheta}) \to 0\;\;\;\mbox{\rm for}\;\;\; \eta\downarrow 0,\;\;
0<\vartheta<\pi, \label{e12}
\end{equation}
is uniformly continuous. By virtue of this property, for a given $\epsilon>0$ there exists an $\eta_{\epsilon}>0$ for which
\begin{equation}
\left| \delta\t{\Sigma}_{\sigma}^{\pm}({\bm k};\mu + \eta\,\mathrm{e}^{\pm i\vartheta})\right| <\epsilon\;\;\;\mbox{\rm for any}\;\;\; 0\le\eta <\eta_{\epsilon},\;\; \forall\vartheta \in (0,\pi). \label{e13}
\end{equation}

In appendix \ref{sc} we show that stability of the GS under consideration implies that \cite{JML61}
\begin{equation}
\mathrm{Im}[\t{\Sigma}_{\sigma}({\bm k};z)] \gtrless 0 \;\;\;\mbox{\rm for}\;\;\; \mathrm{Im}(z) \lessgtr 0,\;\;\forall {\bm k}. \label{e14}
\end{equation}
Let us now suppose that $\Sigma_{\sigma}({\bm k};\mu) \not\in \mathds{R}$ for some ${\bm k}$, say ${\bm k}_0$. With
\begin{equation}
\epsilon = \left|\mathrm{Im}[\Sigma_{\sigma}({\bm k}_0;\mu)]\right|, \label{e15}
\end{equation}
the above considerations show that irrespective of the magnitude of $\epsilon\not=0$, there exists a finite closed neighbourhood of $\mu$ for which one of the inequalities in Eq.~(\ref{e14}) will be violated; this neighbourhood is in the upper-half of the $z$ plane for $\mathrm{Im}[\Sigma_{\sigma}({\bm k}_0;\mu)] >0$ and in the lower-half part for $\mathrm{Im}[\Sigma_{\sigma}({\bm k}_0;\mu)] <0$. \emph{It follows that for a stable GS the result in Eq.~(\ref{e8}) cannot be violated at any ${\bm k}$ for which $\Sigma_{\sigma}({\bm k};\mu)$ is bounded.} In Sec.~\ref{ss63} we shall demonstrate that the apparent violation of Eq.~(\ref{e8}) by the results reported in Refs.~\citen{SLGB96a,SLGB96b,LSGB95,LSGB96} arises from the use in the underlying calculations of an incomplete spectral representation for the `retarded' self-energy.

Following the above observation, and in view of Eq.~(\ref{e5}), for the physical $\Sigma_{\sigma}({\bm k};\varepsilon)$ the inequalities in Eq.~(\ref{e14}) imply that
\begin{equation}
\mathrm{Im}[\Sigma_{\sigma}({\bm k};\varepsilon)] \ge
0,\;\;\varepsilon \le\mu\;\;\; \mbox{\rm and}\;\;\; \mathrm{Im}[\Sigma_{\sigma}({\bm
k};\varepsilon)] \le 0,\;\,\varepsilon \ge\mu,\;\;\forall {\bm k}. \label{e16}
\end{equation}
Although these results are in conformity with the property expressed in Eq.~(\ref{e8}), they do not necessarily imply the latter property.

\subsubsection{Historical background}
\label{ss21s3}

The combination of the results in Eqs.~(\ref{e8}) and (\ref{e16}), expressed in the form (cf. Eq.~(\ref{e12}))
\begin{equation}
\mathrm{Im}[\Sigma_{\sigma}({\bm k};\varepsilon)] \to
0\;\;\;\mbox{\rm for}\;\;\; \varepsilon\to \mu,\;\;\;
\forall {\bm k},\label{e17}
\end{equation}
was demonstrated by Luttinger \cite{JML61} to all orders of perturbation theory; that according to Luttinger $\mathrm{Im}[\Sigma_{\sigma}({\bm k};\varepsilon)]$ should vanish quadratically for $\varepsilon\to\varepsilon_{\Sc f}$ (a hallmark of the conventional Fermi liquids), is a direct consequence of an implicit assumption in the analysis by Luttinger, rather than of the use of the
perturbation theory \emph{per se} \cite{BF99}. Earlier, by considering ${\bm k}$ to be \emph{in the vicinity of the Fermi surface} of an isotropic interacting Fermi gas, the result in Eq.~(\ref{e17}) was stated by Hugenholtz \cite{NMH57b} (see p.~544 in Ref.~\citen{NMH57b}) as following from some pertinent equation, first deduced by Hugenholtz himself in Ref.~\citen{NMH57a}, concerning the self-energy in macroscopic systems (see Ref.~\citen{Note2}); explicitly, Hugenholtz stated that $\mathrm{Im}[\Sigma_{\sigma}({\bm k};\varepsilon)]$ would vanish quadratically for $\varepsilon \to\varepsilon_{\Sc f}$. A similar result, but one that in the contemporary terminology can be said to correspond to both conventional and unconventional Fermi liquids, was presented by Hugenholtz in Ref.~\citen{NMH57a} as a working assumption (see Eq.~(10.7) herein); had Hugenholtz specified this working assumption more sharply, the corresponding expression would have also taken account of marginal Fermi liquids \cite{VLS-RAR89}. For ${\bm k}$ close to the Fermi surface of an isotropic Fermi gas, DuBois \cite{DB59} demonstrated Eq.~(\ref{e17}) to second order in perturbation theory (see Eq.~(3.22) in Ref.~\citen{DB59}). For a dilute gas of fermions interacting through a hard-core potential, Galitski\v{\i} \cite{VMG58} (see also Ch. 4, \S~11 in Ref.~\citen{FW03}) showed that the on-the-mass-shell value of $\mathrm{Im}[\Sigma_{\sigma}({\bm k};\varepsilon)]$ is vanishing for ${\bm k}$ approaching the underlying Fermi surface.

\subsubsection{Statement of the Luttinger theorem for metallic GSs}
\label{ss21s4}

According to the Luttinger theorem \cite{LW60} under consideration
\begin{equation}
N_{\sigma} = N_{\Sc f;\sigma},\;\;\forall\sigma, \label{e18}
\end{equation}
or, following Eqs.~(\ref{e1}) and (\ref{e7}),
\begin{equation}
\sum_{\bm k} \Big\{ {\sf n}_{\sigma}({\bm k}) - \Theta\big(\mu
-\varepsilon_{\bm k}-\hbar\Sigma_{\sigma}({\bm k};\varepsilon_{\Sc
f})\big) \Big\} = 0,\;\;\forall\sigma. \label{e19}
\end{equation}
That the Luttinger theorem amounts to a very remarkable result is
best appreciated by realising that according to this theorem
it is as though on and in the interior of $\mathscr{C}_+(\mu)$ one
had (cf. Eq.~(\ref{e4}))
\begin{equation}
\t{G}_{\sigma}({\bm k};z) \equiv \frac{\hbar}{z-\varepsilon_{\bm
k}-\hbar\Sigma_{\sigma}({\bm k};\varepsilon_{\Sc f})},\;\;\;
\forall{\bm k}, \label{e20}
\end{equation}
which is a manifestly erroneous expression for interacting GSs for which
the self-energy $\t{\Sigma}_{\sigma}({\bm k};z)$ is a non-trivial
function of $z$. This aspect is reflected in the fact that for interacting GSs, ${\sf n}_{\sigma}({\bm k})$ takes values different from solely $0$ and $1$; consequently, for these GSs the summand of the sum on the LHS of Eq.~(\ref{e19}) is not an identically vanishing function of ${\bm k}$.\footnote{See Eqs.~(\ref{e46}) and (\ref{e65}) and contrast the ${\sf n}_{\sigma}({\bm k})$ and $\lim_{\beta\to\infty} \b{\nu}_{\sigma}^{(1)}({\bm k})$ in Figs.~\ref{f2} and \ref{f3}.}

\subsection{Non-metals and metals}
\label{ss22}

Non-metals are distinguished from metals by the fact that for the former $\mathcal{S}_{\Sc f;\sigma}$ is an empty set. Consequently, $N_{\Sc f;\sigma}$, as defined in Eq.~(\ref{e7}), is meaningless for non-metals. In contrast, the `Luttinger number'
\begin{equation}
N_{\Sc l;\sigma} {:=} \sum_{\bm k} \Theta\big(G_{\sigma}^{-1}({\bm
k};\mu)\big) \equiv \sum_{\bm k} \Theta\big(G_{\sigma}({\bm
k};\mu)\big) \label{e21}
\end{equation}
is well-defined, irrespective of whether the underlying GS is metallic or otherwise, as in this definition no explicit reference is made to $\varepsilon_{\Sc f}$ and thus to $\mathcal{S}_{\Sc f;\sigma}$. Since for metallic states, and zero temperature, the appropriate $\mu$ only infinitesimally differs from $\varepsilon_{\Sc f}$, \emph{for these states} $N_{\Sc f;\sigma} \equiv N_{\Sc l;\sigma}$. We note that for both metallic and insulating GSs, $G_{\sigma}({\bm k};\mu) \in \mathds{R}$, $\forall {\bm k}$ (see appendix \ref{sc} as well as Sec.~\ref{ss21s2} above). In analogy with the notion of Fermi sea, we refer to the set of points contributing to $N_{\Sc l;\sigma}$ as the `Luttinger sea'.

\subsubsection{Statement of the Luttinger theorem for metallic and non-metallic GSs; the \emph{generalised} Luttinger theorem}
\label{ss22s1}

The generalised Luttinger theorem that we consider in this paper
states that \cite{IED03,AMT03}
\begin{equation}
N_{\sigma} = N_{\Sc l;\sigma},\;\; \forall\sigma, \label{e22}
\end{equation}
irrespective of whether the underlying GS is metallic or otherwise.
For (macroscopic) metallic GSs, the set of ${\bm k}$ points for which $G_{\sigma}^{-1}({\bm k};\mu) = 0$, or $G_{\sigma}({\bm k};\mu) = 0$, is of measure zero in the embedding ${\bm k}$ space, however for insulating GSs this rule may be violated in idealised cases; in Sec.~\ref{ss61} we shall encounter one such instance. In the cases where this set is a subset of finite measure, the $\Theta$ function in Eq.~(\ref{e21}) should be replaced by the function of which it is the zero-temperature limiting function (Sec.~\ref{ss43} and Sec.~\ref{ss61s2}).

\subsection{Two chemical potentials}
\label{ss23}

The Luttinger theorem under consideration has its root in the finite-temperature formalism of interacting fermion systems, so that the functions and quantities that one encounters in the context of this theorem are in principle all zero-temperature \emph{limits} of their finite-temperature counterparts.\footnote{See Sec.~\protect\ref{s3} as well as appendix \protect\ref{sf} for some pertinent technical details.} Further, this theorem is based on considerations specific to grand-canonical ensembles, with the thermodynamic grand potential $\Omega(\beta,\mu,V)$ playing a key role in its formulation. Here $V$ is the volume of the systems in the ensemble. Consequently, the chemical potential $\mu$ in the context of the Luttinger theorem is originally a thermodynamic variable: it is in principle an arbitrary constant, to be varied at will; experimentally, or even axiomatically (\S~7.3 in Ref.~\citen{KH87}), it is an external parameter determined by the larger system with which the system under consideration is in contact. For the purpose of dealing with $N$-particle GSs, however, it is to be determined from the requirement
\begin{equation}
\b{N} = N, \label{e23}
\end{equation}
where $\b{N}$ is the mean-value of the number of particles in the ensemble, obtained through the relationship
\begin{equation}
\b{N} \equiv -\frac{\partial\Omega(\beta,\mu,V)}{\partial\mu}. \label{e24}
\end{equation}
In this paper we shall denote the \emph{solution} of Eq.~(\ref{e23}), which is a function of $\beta$, $N$ and $V$, by $\mu(\beta,N,V)$, as well as $\mu_{\beta}$ for conciseness. In the abstract as well as in Sec.~\ref{s1} we have referred to $\mu(\beta,N,V)$ as the chemical potential satisfying the equation of state.

In view of the above statements, in considering the Luttinger theorem it is relevant to recognise that $\t{G}_{\sigma}({\bm k};z)$ and $\t{\Sigma}_{\sigma}({\bm k};z)$ are \emph{implicit} functions of the constant parameter $\mu$ which in a broader context would be arbitrary and in principle unrelated to $\mu(\beta,N,V)$. Such arbitrariness, as regards the \emph{implicit} dependence on $\mu$ of $\t{G}_{\sigma}({\bm k};z)$ and $\t{\Sigma}_{\sigma}({\bm k};z)$, is evidently out of question in the context of the Luttinger theorem: the relationships in Eqs.~(\ref{e2}) and (\ref{e21}) have direct bearing on $N$-particle GSs; stated differently, the Green functions in these expressions are necessarily implicit functions of the $\mu(\beta,N,V)$ corresponding to the \emph{limit} $\beta\to\infty$, that is of
\begin{equation}
\mu_{\infty} \equiv \lim_{\beta\to\infty} \mu(\beta,N,V). \label{e25}
\end{equation}
As we shall indicate in the next paragraph, this strict rule may be deviated from somewhat in the cases of \emph{insulating} $N$-particle GSs. This deviation should be avoided however, as without identifying $\mu$ with $\mu_{\infty}$ at finite values of $\beta$ (or with $\mu_{\beta}$ in the event that $\mu_{\beta}$ is certain to approach $\mu_{\infty}$ faster than $1/\beta$ for $\beta\to\infty$), an essential limiting process corresponding to $\beta\to\infty$, referred to in Sec.~\ref{s1}, may in some cases be ill-defined. This has its root in the possibility of some relevant and non-trivial functions becoming dependent on the dimensionless combination $\beta (\mu -\mu_{\infty})$, whereby for $\beta\to\infty$ one may arrive at different limits, depending on whether $\mu$ is held constant and different from $\mu_{\infty}$ on taking the limit $\beta\to\infty$, or identified with $\mu_{\infty}$ (or $\mu_{\beta}$) prior to taking this limit.

For the $N$-particle GS of the system under consideration, one naturally has $\mu_{\infty} \in (\mu_N^-,\mu_N^+)$ (appendix \ref{sc}).\footnote{In appendix \ref{sc} one encounters both $\mu_{N}^{\pm}$ and $\mu_{N;\sigma}^{\pm}$. For the present discussions it is immaterial whether one uses $\mu_{N}^{\pm}$ or $\mu_{N;\sigma}^{\pm}$. We note however that $(\mu_{N}^-,\mu_{N}^+) \subseteq (\mu_{N;\sigma}^-,\mu_{N;\sigma}^+)$, $\forall\sigma$. } For \emph{metallic} $N$-particle GSs, $\mu_{N}^+ - \mu_{N}^- = O(1/N)$ so that the condition $\mu_{\infty} \in (\mu_N^-,\mu_N^+)$ and the requirement $\mu \in (\mu_N^-,\mu_N^+)$ imply that up to an infinitesimal correction $\mu=\mu_{\infty}$. In contrast, since for \emph{insulating} $N$-particle GSs the width of the interval $(\mu_N^-,\mu_N^+)$ is finite, the condition $\mu_{\infty} \in (\mu_N^-,\mu_N^+)$ and the requirement $\mu \in (\mu_N^-,\mu_N^+)$ do not necessitate equality of $\mu$ and $\mu_{\infty}$. In fact, one can show that for these GSs $\t{G}_{\sigma}({\bm k};z)$ and $\t{\Sigma}_{\sigma}({\bm k};z)$ do not vary for variation of $\mu$ inside the interval $(\mu_N^-,\mu_N^+)$, in contrast to the finite-temperature counterparts of these functions which manifestly depend on $\mu$ and vary for variation of $\mu$ inside $(\mu_N^-,\mu_N^+)$ (appendix \ref{sf}). In Sections \ref{ss41} and \ref{ss51} we shall expose the way in which the validity at finite temperatures of two main contributory elements to the Luttinger theorem crucially depends on the equality of the value of the implicit $\mu$ with that of the explicit $\mu$ in the problem. In view of this and of the above-mentioned fact that $\t{G}_{\sigma}({\bm k};z)$ does not vary for variations of $\mu$ inside $(\mu_{N}^-,\mu_{N}^+)$, it follows that in principle the Luttinger theorem, Eq.~(\ref{e22}), should apply for any value of the explicit $\mu$ in Eq.~(\ref{e21}) inside the interval $(\mu_{N}^-,\mu_{N}^+)$; this value may thus be different from $\mu_{\infty}$. With reference to our remark in the previous paragraph, this formal argument, with regard to the freedom of assigning to $\mu$ an arbitrary value from inside $(\mu_{N}^-,\mu_{N}^+)$, is not fully warranted by virtue of the possibility that the choice $\mu\not= \mu_{\infty}$ (or $\mu\not=\mu_{\beta}$) may give rise to a false limit for $\beta\to\infty$, thereby undermining the Luttinger theorem.

\subsection{Luttinger surface and Fermi surface}
\label{ss24}

The set of ${\bm k}$ points at which
\begin{equation}
G_{\sigma}({\bm k};\mu) = 0 \label{e26}
\end{equation}
has come to be known as the `Luttinger surface', which in this paper we denote by $\mathcal{S}_{\Sc l;\sigma}$. For the reason that we shall specify below, we propose that $\mathcal{S}_{\Sc l;\sigma}$ be more generally defined as the set of ${\bm k}$-points in the infinitesimal neighbourhoods of which $G_{\sigma}({\bm k};\mu)$ is \emph{bounded} and changes sign; by so doing one bypasses the problem arising from $G_{\sigma}({\bm k};\mu)$ undergoing a finite discontinuity accompanied by a change of sign in $G_{\sigma}({\bm k};\mu)$, whereby the equation $G_{\sigma}({\bm k};\mu) = 0$ may have no solution, despite the fact that this point of discontinuity shares all characteristic aspects common to solutions of Eq.~(\ref{e26}). It should be further noted that, in particular for macroscopic systems, a point at which $G_{\sigma}({\bm k};\mu)$ merely vanishes but retains the same sign in its infinitesimal neighbourhood, is of no consequence to the sums on the RHS of Eq.~(\ref{e21}) and therefore need not be counted as a point of $\mathcal{S}_{\Sc l;\sigma}$.

The above-mentioned adjective `\emph{bounded}' is essential in order to
differentiate $\mathcal{S}_{\Sc l;\sigma}$ from $\mathcal{S}_{\Sc
f;\sigma}$ in the case of metallic GSs; in this connection, although
functions need not be discontinuous at the points where they are
unbounded (\S~219 in Ref.~\citen{EWH27}), such points are necessarily
points of discontinuity if divergence is accompanied by a change of
sign in the function under consideration.

With the above extension of the definition of $\mathcal{S}_{\Sc l;\sigma}$, one observes that indeed the \emph{frontier} of the Fermi sea of an arbitrary metallic GS coincides with $\mathcal{S}_{\Sc f;\sigma}\cup \mathcal{S}_{\Sc l;\sigma}$ (Sec.~\ref{ss21s1}) so that knowledge of both $\mathcal{S}_{\Sc f;\sigma}$ \emph{and} $\mathcal{S}_{\Sc l;\sigma}$ suffices to determine the extent of the Fermi sea corresponding to an arbitrary metallic GS.

\subsubsection{Remarks}
\label{ss24s1}

Recent numerical calculations \cite{BGBG06}, based on a generalised dynamical mean-field theory (referred to as the ch-DMFT in Ref.~\citen{BGBG06}), concerning coupled one-dimensional fermionic chains, show that transition of the one-dimensional Mott insulating state to a two-dimensional metal is signalled by a discontinuity of\footnote{The function $\mathscr{S}_{\!\sigma}({\bm k};\zeta_m)$ is defined in Sec.~\protect\ref{s3}.} $\mathrm{Re}[\mathscr{S}_{\!\sigma}({\bm k};\zeta_0)]$ (corresponding to $T=0.1$ in units of the nearest-neighbour hopping parameter $t$) and divergence of $\mathrm{Re}[\t{\Sigma}_{\sigma}({\bm k};\mu+i 0^+)] \equiv \Sigma_{\sigma}({\bm k};\mu)$ at $k_{\parallel} = \pi/2$ (in units of the inverse of the lattice constant in the chain direction). Although in these calculations $\t{\Sigma}_{\sigma}({\bm k};\mu+i 0^+)$ has been deduced with the aid of an empirical analytic-continuation procedure, relying on the values $\{\mathscr{S}_{\!\sigma}({\bm k};\zeta_m)\,\|\, m\in \mathds{Z}\}$ corresponding to low temperatures, the computational results provide a concrete example of a case where in a region of the ${\bm k}$ space $\Sigma_{\sigma}({\bm k};\mu)$ is discontinuously divergent and $G_{\sigma}({\bm k};\mu)$ continuously passes through zero.

\subsection{Finite systems}
\label{ss25}

In the cases where the ${\bm k}$ space consists of a finite number of points,\footnote{Since we are dealing with \emph{uniform} GSs, these finite systems must be defined on finite lattices without boundary.} it is relevant that $\Theta(x)$ be appropriately defined at $x=0$. With reference to the fact that the $\Theta\big(G_{\sigma}^{-1}({\bm k};\mu)\big)$ in Eq.~(\ref{e21}) is the limit for $\eta\downarrow 0$ of $-\frac{1}{\pi}\mathrm{Arctan}(y/x)$, where $\mathrm{Arctan}(y/x)$ is defined in Eq.~(\ref{e49}) below, in which $y$ stands for  $\mathrm{Im}[\t{G}_{\sigma}^{-1}({\bm k};\mu+i\eta)]$ and $x$ for $\mathrm{Re}[\t{G}_{\sigma}^{-1}({\bm k};\mu+i\eta)]$ (Eq.~(\ref{e53}) below), we propose that for the specific ${\bm k}$ at which $G_{\sigma}^{-1}({\bm k};\mu)$ turns out to be vanishing, one employ the expression $-\frac{1}{\pi}\mathrm{Arctan}(y/x)$ and take the limit $\eta\downarrow 0$ explicitly. These considerations are also relevant in practical calculations concerning macroscopic systems, where $G_{\sigma}^{-1}({\bm k};\mu)$ is by necessity explicitly calculated at a finite number of ${\bm k}$ points.

\section{Generalities}
\label{s3}

Let $\mathscr{G}_{\sigma}({\bm k};\zeta_m)$ and
$\mathscr{S}_{\!\sigma}({\bm k};\zeta_m)$ denote the
finite-temperature Green function and self-energy in the grand-canonical ensemble, where
\begin{equation}
\zeta_{m} \equiv i\hbar\omega_m + \mu, \label{e27}
\end{equation}
in which
\begin{equation}
\omega_{m} = \frac{(2 m+1) \pi}{\beta\hbar},\;\; m \in \mathds{Z},
\label{e28}
\end{equation}
is the $m$th fermionic Matsubara frequency with $1/\beta \equiv k_{\Sc b} T$, where $T$ denotes temperature \cite{TM55,FW03}. Provided that the value of the thermodynamic variable $\mu$ satisfies $\mu \in (\mu_{N}^-,\mu_{N}^+)$, or that it is identified with $\mu_{\beta} \equiv \mu(\beta,N,V)$, one has (Sec.~\ref{ss23}, appendices \ref{sc} and \ref{sf})
\begin{equation}
\lim_{\beta\to\infty}\mathscr{G}_{\sigma}({\bm k};z) = \t{G}_{\sigma}({\bm k};z),
\label{e29}
\end{equation}
and
\begin{equation}
\lim_{\beta\to\infty}\mathscr{S}_{\!\sigma}({\bm k};z) = \t{\Sigma}_{\sigma}({\bm k};z),
\label{e30}
\end{equation}
where $\t{G}_{\sigma}({\bm k};z)$ is the zero-temperature Green function corresponding to the $N$-particle GS of the $\wh{H}$ under consideration, and $\t{\Sigma}_{\sigma}({\bm k};z)$ its associated zero-temperature self-energy.

In this paper we shall have occasion to calculate the finite-temperature \emph{mean value} of the number of spin-$\sigma$ particles in the grand canonical ensemble, $\b{N}_{\sigma}$, on the basis of the following expression, which is an alternative to that in Eq.~(\ref{e24}):
\begin{equation}
\b{N}_{\sigma} = \sum_{\bm k} \int_{-\infty}^{\infty} \frac{{\rm d}\varepsilon}{\hbar}\; \frac{\mathscr{A}_{\sigma}({\bm k};\varepsilon)}{\mathrm{e}^{\beta (\varepsilon-\mu)}+1}, \label{e31}
\end{equation}
where $\mathscr{A}_{\sigma}({\bm k};\varepsilon)$ is the finite-temperature single-particle spectral function, defined according to (cf. Eq.~(\ref{ec40}); see also appendix \ref{sf})
\begin{equation}
\mathscr{A}_{\sigma}({\bm k};\varepsilon) \equiv \pm\frac{1}{\pi}\, \mathrm{Im}[\mathscr{G}_{\sigma}({\bm k};\varepsilon\mp i 0^+)]. \label{e32}
\end{equation}
Following Eq.~(\ref{e29}), one has (appendix \ref{sf})
\begin{equation}
\lim_{\beta\to\infty} \mathscr{A}_{\sigma}({\bm k};\varepsilon) = A_{\sigma}({\bm k};\varepsilon). \label{e33}
\end{equation}
The expression for $N_{\sigma}$ in Eq.~(\ref{e1}), allied with that in Eq.~(\ref{e2}), follows directly from that in Eq.~(\ref{e31}); clearly, the $\mu$ in $\mathscr{C}_+(\mu)$ has its origin in the \emph{explicit} $\mu$ on the RHS of Eq.~(\ref{e31}). In this connection, assuming that the isothermal compressibility $\kappa_{\beta}$ of the system under consideration is non-vanishing and finite for $\beta\to\infty$, on the basis of the fluctuation-dissipation theorem one deduces that for $\beta<\infty$ fluctuation in the particle numbers contributing to the ensemble average $\b{N}$ scales like $\b{N}^{1/2}$ as $\b{N}\to\infty$ (\S~7.4 in Ref.~\citen{KH87}). This fluctuation is however identically vanishing for $\beta=\infty$ and any finite value of $\b{N}$, signifying $\beta=\infty$ as a singular point (Sec.~\ref{s1}). This statement applies irrespective of whether the $N$-particle GS of the system under consideration is metallic or insulating; although in the latter case $\kappa_{\beta}$ decays exponentially towards zero for $\beta\to\infty$, nonetheless one has $\kappa_{\beta} >0$ for $\beta\to\infty$ but $\beta<\infty$. From this perspective, one observes that, as regards $\mu$, \emph{in applying the Luttinger theorem no distinction should be made between $N$-particle metallic states and $N$-particle insulating states}; in both cases $\mu_{\infty}$ is the only legitimate value to be assigned to $\mu$ (Sec.~\ref{s1}), even though for a host of insulating $N$-particle states a $\mu$ satisfying $\mu \in (\mu_{N}^-,\mu_{N}^+)$ and $\mu\not=\mu_{\infty}$ may prove appropriate.

The result in Eq.~(\ref{e31}) is obtained from an expression for $\b{N}_{\sigma}$, originally deduced by Luttinger and Ward for interacting ensembles (appendix B in Ref.~\citen{LW60}), through first replacing in this expression a pertinent summation with respect to Matsubara frequencies by its equivalent Mellin-Barnes-type integral representation (\S\S~14.5 and 16.4 in Ref.~\citen{WW62}) and subsequently effecting an appropriate contour deformation.

\subsubsection{Remarks}
\label{ss30s1}

It is common practice to use the notations $\mathscr{G}_{\sigma}({\bm k};i\hbar\omega_m)$ and $\mathscr{S}_{\!\sigma}({\bm k};i\hbar\omega_m)$,\footnote{In Ref.~\protect\citen{FW03} one even encounters $\mathscr{G}({\bm k};\omega_m)$ and $\Sigma^{\star}({\bm k};\omega_m)$.} instead of respectively $\mathscr{G}_{\sigma}({\bm k};\zeta_m)$ and $\mathscr{S}_{\!\sigma}({\bm k};\zeta_m)$ that we employ in this paper (following Luttinger and Ward \cite{LW60}). Our adopted notation enables us to bypass the need to introduce the so-called ``real-time'' Green function, which is distinct from the Matsubara Green function and which is conventionally employed for calculating ensemble averages at finite temperatures and in particular in the zero-temperature \emph{limit} (see, for instance, Ch. 9, \S~31 in Ref.~\citen{FW03} and Ch. 5, \S~2 in Ref.~\citen{NO98}). If we had adopted the commonly-used notation, we would have $\t{G}_{\sigma}({\bm k};z)$ and $\t{\Sigma}_{\sigma}({\bm k};z)$ as the zero-temperature limits of $\mathscr{G}_{\sigma}({\bm k};z-\mu)$ and $\mathscr{S}_{\!\sigma}({\bm k};z-\mu)$ respectively. It is for this reason that in many texts the equivalents of our functions $\t{G}_{\sigma}({\bm k};z)$ and $\t{\Sigma}_{\sigma}({\bm k};z)$ are explicit functions of $z+\mu$.

\section{Elements relevant to the Luttinger theorem}
\label{s4}

Here we present an overview of the details underlying the proof of the Luttinger theorem under consideration \cite{LW60}. In the subsequent sections we shall subject these details to rigorous examination. To facilitate comparison, in Table~\ref{t1} we present the symbols of the main functions to be encountered below along with those of their equivalents in Ref.~\citen{LW60}.

\begin{table}[t!]
\caption{Some functions encountered in the paper by Luttinger and Ward\protect\cite{LW60} (LW) and their equivalents in the present paper (Present). } \vspace{2pt}
\label{t1}
\begin{center}
\begin{tabular}{ccc}
\hline\hline
LW & Present  & Specification\\
\hline
$G_r(\zeta_l)$ & $\mathscr{S}_{\!\sigma}({\bm k};\zeta_m)$ & $r \equiv ({\bm k},\sigma)$, $l\equiv m$ \\
$G_{r n}'(\zeta_l)$ & $\mathscr{S}_{\sigma}^{\sharp (\nu)}({\bm k};\zeta_m)$ & $n \equiv \nu$ \\
$G_{r n}''(\zeta_l)$ & $\mathscr{S}_{\sigma}^{(\nu)}({\bm k};\zeta_m)$ & --- \\
$S_r(\zeta_l)$ & $\mathscr{G}_{\sigma;0}({\bm k};\zeta_m)$ & ---\\
$S_r'(\zeta_l)$ & $\mathscr{G}_{\sigma}({\bm k};\zeta_m)$ & ---\\
\hline
\end{tabular}
\end{center}
\end{table}

\subsection{An overview of details}
\label{ss41}

For the average number of particles $\b{N}$ in the grand-canonical
ensemble, Luttinger and Ward \cite{LW60} obtained
\begin{equation}
\b{N} = \sum_{\sigma} \big(\b{N}_{\sigma}^{(1)} +
\b{N}_{\sigma}^{(2)}\big) \equiv \sum_{{\bm k},\sigma} \big(
\b{\nu}_{\sigma}^{(1)}({\bm k}) + \b{\nu}_{\sigma}^{(2)}({\bm k})
\big), \label{e34}
\end{equation}
where
\begin{equation}
\b{\nu}_{\sigma}^{(1)}({\bm k}) \equiv \frac{1}{\beta} \sum_m
\mathrm{e}^{\zeta_m 0^+}\, \frac{\partial}{\partial\zeta_m}
\ln\big(-\beta\hbar\, \mathscr{G}_{\sigma}^{-1}({\bm
k};\zeta_{m})\big), \label{e35}
\end{equation}
\begin{equation}
\b{\nu}_{\sigma}^{(2)}({\bm k}) \equiv \frac{1}{\beta} \sum_m
\mathrm{e}^{\zeta_m 0^+}\, \mathscr{G}_{\sigma}({\bm k};\zeta_m)
\frac{\partial}{\partial\zeta_m}\,\mathscr{S}_{\!\sigma}({\bm
k};\zeta_m). \label{e36}
\end{equation}
The considerations in Sec.~\ref{s5} will shed light on the degree to which the validity of the expression in Eq.~(\ref{e34}) may be dependent on the validity of the weak-coupling many-body perturbation theory. We draw attention to the fact that $\b{\nu}_{\sigma}^{(1)}({\bm k})$ and $\b{\nu}_{\sigma}^{(2)}({\bm k})$ depend on $\mu$ both explicitly (through $\zeta_m$, Eq.~(\ref{e27})) and implicitly. Since Eq.~(\ref{e34}) has its root in Eq.~(\ref{e24}) (see Sec.~\ref{ss51}), it follows that the explicit $\mu$ in Eqs.~(\ref{e35}) and (\ref{e36}) must be assigned the same value as the $\mu$ on which $\b{\nu}_{\sigma}^{(1)}({\bm k})$ and $\b{\nu}_{\sigma}^{(2)}({\bm k})$ implicitly depend; this common value amounts to the value of the thermodynamic variable $\mu$ in $\Omega(\beta,\mu,V)$. To be more specific, this $\mu$ need \emph{not} be equal to $\mu(\beta,N,V)$.

In calculating the zero-temperature limit (that is $\beta\to\infty$) of $\b{N}$, Luttinger and Ward \cite{LW60} treated the two sums with respect to Matsubara frequencies in Eqs.~(\ref{e35}) and (\ref{e36}) differently.

Concerning $\b{\nu}_{\sigma}^{(1)}({\bm k})$, Luttinger and Ward \cite{LW60} employed the exact Mellin-Barnes-type integral representation of the sum with respect to $m$ (\S\S~14.5 and 16.4 in Ref.~\citen{WW62}), deducing thus
\begin{equation}
\b{\nu}_{\sigma}^{(1)}({\bm k}) = \int_{\Gamma^-} \frac{{\rm
d}\zeta}{2\pi i}\; \frac{\mathrm{e}^{\zeta\, 0^+}}{\mathrm{e}^{\beta
(\zeta-\mu)} +1}\; \frac{\partial}{\partial\zeta}\,
\ln\big(-\beta\hbar\, \mathscr{G}_{\sigma}^{-1}({\bm k};\zeta)\big),
\label{e37}
\end{equation}
where $\Gamma^-$ consists of two clock-wise oriented closed contours $\Gamma_1^-$ and $\Gamma_2^-$ (we thus write $\Gamma^- = \Gamma_1^- \oplus \Gamma_2^-$), of which $\Gamma_1^-$ ($\Gamma_2^-$) encloses the line $\mu+i y$ in the $\zeta$ plane corresponding to $y> 0$ ($y< 0$), the finite intercept of $\Gamma_1^-$ ($\Gamma_2^-$) with the latter line being at $y_0 \in (0,\hbar\omega_0)$ ($y_0 \in (-\hbar\omega_0,0)$) (cf. Fig.~5 in Ref.~\citen{LW60}). By deforming the contour $\Gamma^-$ into the counter-clock-wise oriented contour $\Gamma_0^+$, which encloses the entire real axis of the $\zeta$ plane and none of the discrete energies $\zeta_m$, Luttinger and Ward \cite{LW60} obtained that
\begin{equation}
\lim_{\beta\to\infty} \b{N}_{\sigma}^{(1)} = N_{\Sc
f;\sigma}, \label{e38}
\end{equation}
where $N_{\Sc f;\sigma}$ is defined in Eq.~(\ref{e7}). Since the treatment by Luttinger and Ward equally applies to insulating GSs, by analogy one has
\begin{equation}
\lim_{\beta\to\infty} \b{N}_{\sigma}^{(1)} = N_{\Sc l;\sigma}, \label{e39}
\end{equation}
where $N_{\Sc l;\sigma}$ is defined in Eq.~(\ref{e21}).

Concerning $\b{\nu}_{\sigma}^{(2)}({\bm k})$, making use of (cf. Eqs.~(\ref{e27}) and (\ref{e28}))
\begin{equation}
{\rm d}\zeta_m \equiv \zeta_{m+1} - \zeta_m = \frac{2\pi
i}{\beta},\;\;\;\forall m, \label{e40}
\end{equation}
Luttinger and Ward \cite{LW60} employed
\begin{equation}
\lim_{\beta\to\infty} \frac{1}{\beta} \sum_m f_{\beta}(\zeta_m) =
\int_{\mathscr{C}(\mu)} \frac{{\rm d}\zeta}{2\pi i}\;
\lim_{\beta\to\infty} f_{\beta}(\zeta), \label{e41}
\end{equation}
where $\mathscr{C}(\mu)$ is in principle parameterised according to
\begin{equation}
\zeta=\mu+iy,\;\;\;\mbox{\rm where}\;\;\;
y\uparrow_{-\infty}^{+\infty}. \label{e42}
\end{equation}
Owing to the fact that the function $\lim_{\beta\to\infty} f_{\beta}(\zeta)$ for which the result in Eq.~(\ref{e41}) is applied decays sufficiently fast for $\vert \zeta\vert \to\infty$, one may identify $\mathscr{C}(\mu)$ with both $\mathscr{C}_+(\mu)$, Eq.~(\ref{e1}), and $\mathscr{C}_-(\mu)$, which is similar to $\mathscr{C}_+(\mu)$ however is clock-wise oriented and contains the interval $[\mu,\infty]$ of the real energy axis. Thus Luttinger and Ward \cite{LW60} obtained that (cf. Eqs.~(\ref{e29}) and (\ref{e30}))
\begin{equation}
\lim_{\beta\to\infty} \b{\nu}_{\sigma}^{(2)}({\bm k}) =
\int_{\mathscr{C}(\mu)} \frac{{\rm d}z}{2\pi i}\;
\t{G}_{\sigma}({\bm k};z) \frac{\partial}{\partial z}\,
\t{\Sigma}_{\sigma}({\bm k};z),\;\;\forall\sigma. \label{e43}
\end{equation}
Luttinger and Ward \cite{LW60} subsequently demonstrated that to all
orders of perturbation theory
\begin{equation}
\lim_{\beta\to\infty} \b{N}_{\sigma}^{(2)} \equiv \sum_{\bm
k} \lim_{\beta\to\infty}\b{\nu}_{\sigma}^{(2)}({\bm k}) =
0,\;\;\forall\sigma. \label{e44}
\end{equation}
This result is referred to as the Luttinger-Ward \emph{identity}. In Sec.~\ref{s5} we shall investigate the extent to which validity of this identity is dependent on the viability of the weak-coupling many-body perturbation theory. Combining the results in Eqs.~(\ref{e39}) and (\ref{e44}), one arrives at the Luttinger theorem, Eqs.~(\ref{e18}) and (\ref{e22}).

Although Eq.~(\ref{e41}) applies for an arbitrary $\mu$, since in arriving at Eq.~(\ref{e43}) we have replaced $\mathscr{G}_{\sigma}({\bm k};z)$ and $\mathscr{S}_{\!\sigma}({\bm k};z)$ by their zero-temperature limits corresponding to the $N$-particle GS of $\wh{H}$, it follows that the explicit $\mu$ in Eq.~(\ref{e43}) is required to satisfy $\mu \in (\mu_{N}^-, \mu_{N}^+)$; the choice $\mu=\mu_{\infty}$ appears not to be necessary. As we have indicated in Sections \ref{s1} and \ref{ss23}, and as we shall explicitly show in Sec.~\ref{ss61s2}, even though the choice $\mu \in (\mu_{N}^-, \mu_{N}^+)$, with $\mu\not=\mu_{\infty}$ in the cases of insulating $N$-particle GSs for which $(\mu_{N}^-, \mu_{N}^+)$ is finite, is formally permitted, this choice is likely to lead to failure. The appropriate choice is to identify the $\mu$ in $\mathscr{C}(\mu)$ with $\mu_{\infty}$.

With reference to the last statement, we draw attention to the fact that, the prescription to identify $\mu$ with $\mu_{\infty}$ amounts to a treatment of insulating $N$-particle GSs on the same footing as metallic $N$-particle GSs. In this connection, it is important to realise that according to the Lehmann representation for $\mathscr{G}_{\sigma}({\bm k};z)$ (appendix \ref{sf}), the function $\mathscr{G}_{\sigma}({\bm k};z)$ corresponding to an $\wh{H}$ whose $N$-particle GS is insulating, has, for in particular $\mu \in (\mu_{N}^-, \mu_{N}^+)$, a qualitatively similar analytic structure as the $\t{G}_{\sigma}({\bm k};z)$ pertaining to an $N$-particle metallic GS; the principal difference between the two functions, for $\mu \in (\mu_{N}^-, \mu_{N}^+)$ and sufficiently large $\beta$, corresponds to the exponential \emph{suppression} (to be contrasted with the \emph{full elimination}) of the spectral weights of the excitations described by $\mathscr{G}_{\sigma}({\bm k};z)$, for $\beta<\infty$, inside the finite interval $(\mu_{N}^-,\mu_{N}^+)$; these excitations relate to the $N$- and $N\pm 1$-particle eigenstates of $\wh{H}$ and are therefore independent of $\beta$.\footnote{Such in-gap excitations are specific to \emph{interacting} systems. Recall that the Lehmann representation of $\mathscr{G}_{\sigma}({\bm k};z)$ involves summations over the compound indices $s$ and $s'$; the apparent `gap' is brought about through a quenching of $s$ to $0$, the compound index associated with the $N$-particle GS of $\wh{H}$, in the zero-temperature limit (see Eqs.~(\protect\ref{ef12}), (\protect\ref{ef20}) and (\protect\ref{ef21})).} The fact that $\t{G}_{\sigma}({\bm k};z)$ is free from singularities inside the real interval $(\mu_{N}^-, \mu_{N}^+)$ and that $\t{G}_{\sigma}({\bm k};z)$ does not vary for variations of $\mu$ when $\mu \in (\mu_{N}^-, \mu_{N}^+)$, signify that $\beta=\infty$ is a singular point of the theory along the $\beta$ axis (Sections \ref{s1}, \ref{ss23} and \ref{s3}). Viewed from this perspective, one observes that the choice $\mu=\mu_{\infty}$ for the cases where the underlying $N$-particle GSs are insulating, is in fact the most natural one.

We note that Luttinger and Ward \cite{LW60} explicitly emphasized robustness of the transformation in Eq.~(\ref{e41}) in the context of deducing Eq.~(\ref{e44}) from Eq.~(\ref{e36}).\footnote{See footnote 9 on page 1424 in Ref.~\protect\citen{LW60} and the text to which this footnote corresponds.} Our considerations in this paper will confirm that this is indeed the case (for some technical details see appendix \ref{sd}).

\subsubsection{Remarks}
\label{ss41s1}

In this paper the \emph{superscripts} `$\mp$' of the basic symbols denoting contours signify the orientations of these: `$-$' signifies clockwise and `$+$' counter-clockwise orientations. Accordingly, we shall denote the contour obtained through reversing the orientation of a given contour by the same basic symbol but with the complementary superscript; thus, we denote the contour that has the reverse orientation of, say, $\Gamma_1^{-}$ by $\Gamma_1^{+}$. Further, the \emph{subscripts} $1$ and $2$ attached to the same symbol (such as in $\Gamma_1^-$ and $\Gamma_2^-$, the symbol being $\Gamma^-$) indicate the respective contours to be respectively in the upper and the lower half of the pertinent complex plane.

\subsection{Two observations}
\label{ss42}

Starting from the exact expression for $\b{\nu}_{\sigma}^{(1)}({\bm k})$ in Eq.~(\ref{e37}) and using the Dyson equation, one trivially obtains that
\begin{equation}
\b{\nu}_{\sigma}^{(1)}({\bm k}) = \int_{\Gamma^-}
\frac{{\rm d}\zeta}{2\pi i\hbar}\; \frac{\mathrm{e}^{\zeta\,
0^+}}{\mathrm{e}^{\beta (\zeta-\mu)} +1}\; \mathscr{G}_{\sigma}({\bm
k};\zeta) -\int_{\Gamma^-} \frac{{\rm d}\zeta}{2\pi i}\;
\frac{\mathrm{e}^{\zeta\, 0^+}}{\mathrm{e}^{\beta (\zeta-\mu)} +1}\;
\mathscr{G}_{\sigma}({\bm k};\zeta) \frac{\partial}{\partial\zeta}\,
\mathscr{S}_{\!\sigma}({\bm k};\zeta), \label{e45}
\end{equation}
leading to the identity
\begin{equation}
{\sf n}_{\sigma}({\bm k}) = \lim_{\beta\to\infty} \b{\nu}_{\sigma}^{(1)}({\bm k}) + \lim_{\beta\to\infty} \b{\nu}_{\sigma}^{(2)}({\bm
k}), \label{e46}
\end{equation}
where ${\sf n}_{\sigma}({\bm k})$ is the GS momentum distribution function, Eq.~(\ref{e2}), and $\lim_{\beta\to\infty} \b{\nu}_{\sigma}^{(2)}({\bm k})$ is \emph{the short-hand notation}\footnote{That is, the question whether the expression in Eq.~(\ref{e43}) is invariably deducible from the defining expression in Eq.~(\ref{e36}) is of no relevance here.} for the expression on the RHS of Eq.~(\ref{e43}). We shall employ the result in Eq.~(\ref{e46}) in Sec.~\ref{ss62s3}.

The result in Eq.~(\ref{e46}) leads one to the following two observations. Firstly, in view of Eq.~(\ref{e1}) the exactness of the result in Eq.~(\ref{e46}) suggests the possibility that, insofar as the Luttinger theorem is concerned, it should be immaterial on which grounds Luttinger and Ward \cite{LW60} may have deduced the expression in Eq.~(\ref{e34}), i.e. their possible reliance on the weak-coupling many-body perturbation theory in arriving at this result would have no significance. Secondly, since in arriving at the expression in Eq.~(\ref{e46}) we have made no use of the expression in Eq.~(\ref{e41}), it appears that this expression would be redundant.

As we shall see later, both of these observations turn out to be premature: both the expression in Eq.~(\ref{e34}) and the transformation in Eq.~(\ref{e41}) turn out to be relevant in the proof of the Luttinger-Ward identity, in a way that is not apparent from the expressions in Eqs.~(\ref{e34}) and (\ref{e41}). For now we mention that Eq.~(\ref{e34}) is related to a function, denoted by $Y'$, Eq.~(\ref{e72}) below, which is defined in terms of the perturbative contributions to the self-energy $\mathscr{S}_{\!\sigma}({\bm k};z)$, in terms of $\{\mathscr{G}_{\sigma'}({\bm k};z)\}$, and which plays a vital role in the proof of the Luttinger-Ward identity. We remark that although the proof of the Luttinger theorem as presented by Abrikosov, Gor'kov, and Dzyaloshinski\u{\i} \cite{AGD75} (pp. 166-168 in Ref.~\citen{AGD75}) appears to bypass the expressions in Eqs.~(\ref{e34}) and (\ref{e41}), a careful examination of this proof reveals that both of these expressions are implicit in the considerations by the latter authors.

\subsection{Equality of $\sum_{{\bf k}} \lim_{\beta\to\infty} \b{\nu}_{\sigma}^{(1)}({\bf k})$ with the Luttinger number $N_{\Sc l;\sigma}$}
\label{ss43}

Following Eq.~(\ref{e37}), the contour integral over $\Gamma^-$ can be expressed as one over $\Gamma_0^+$ (see Fig.~5 in Ref.~\citen{LW60}). This contour can be considered as consisting of two straight lines parallel to the real energy axis, along one of which one has $\zeta = \varepsilon + i 0^+$, where $\varepsilon \downarrow_{-\infty}^{+\infty}$, and along the other $\zeta = \varepsilon - i 0^+$, where $\varepsilon \uparrow_{-\infty}^{+\infty}$ (recall that $\Gamma_0^+$ is a counter-clockwise oriented closed contour). On applying integration by parts, taking into account that owing to $\mathrm{e}^{\zeta 0^+}$ and $1/[\mathrm{e}^{\beta (\zeta-\mu)}+1]$ there are no contributions arising from $\varepsilon=-\infty$ and $+\infty$ respectively, from the expression in Eq.~(\ref{e37}) one deduces that
\begin{equation}
\b{\nu}_{\sigma}^{(1)}({\bm k}) = \int_{-\infty}^{\infty} \frac{{\rm
d}\varepsilon}{\pi}\; \mathrm{Im}\big[ \ln\big(-\beta\hbar\,
\mathscr{G}_{\sigma}^{-1}({\bm k};\varepsilon+i 0^+)\big)\big]\,
\frac{\partial}{\partial\varepsilon}\, \frac{1}{\mathrm{e}^{\beta
(\varepsilon - \mu)}+1}, \label{e47}
\end{equation}
where we have used the fact that $\mathscr{G}_{\sigma}^{-1}({\bm
k};\varepsilon -i 0^+) = (\mathscr{G}_{\sigma}^{-1}({\bm
k};\varepsilon +i 0^+))^*$ for $\varepsilon\in \mathds{R}$.
Following the same procedure as employed in obtaining the Sommerfeld
expansion (appendix C in Ref.~\citen{AM76}; see Ref.~\citen{Note3}), from Eq.~(\ref{e47}) one readily deduces that \cite{JML60}
\begin{eqnarray}
&&\hspace{-1.2cm}\b{\nu}_{\sigma}^{(1)}({\bm k}) = -\frac{1}{\pi}\,
\mathrm{Arctan}\Big(\frac{\mathrm{Im}[\mathscr{G}_{\sigma}^{-1}({\bm
k};\mu+i 0^+)]}{\mathrm{Re}[\mathscr{G}_{\sigma}^{-1}({\bm k};\mu+i
0^+)]}\Big)\nonumber\\
&&\hspace{0.45cm} -\frac{\pi}{6}\,\frac{1}{\beta^2} \, \mathrm{Im}\Big[
\left.\frac{\partial}{\partial\varepsilon}\,\Big(\mathscr{G}_{\sigma}({\bm
k};\varepsilon+i 0^+)\, \frac{\partial}{\partial\varepsilon}\,
\mathscr{G}_{\sigma}^{-1}({\bm k};\varepsilon+i
0^+)\Big)\right|_{\varepsilon=\mu}\Big] +\dots~, \label{e48}
\end{eqnarray}
where
\begin{equation}
\mathrm{Arctan}\Big(\frac{y}{x}\Big) \equiv
\arctan\Big(\frac{y}{x}\Big) - \Theta(x)\, \mathrm{sgn}(y)\,\pi,\;\;
x,y\in \mathds{R}, \label{e49}
\end{equation}
in which $\arctan(x)$ is the principal branch of the many-valued
function $\tan^{-1}(x)$ (\S~4.4 in Ref.~\citen{AS72}), satisfying
\begin{equation}
-\frac{\pi}{2} \le \arctan(x) \le \frac{\pi}{2},\;\;\; x\in
\mathds{R}. \label{e50}
\end{equation}
Note that the first term on the RHS of Eq.~(\ref{e48}) only implicity depends on $\beta$.

In arriving at the first term on the RHS of Eq.~(\ref{e48}) we have
used
\begin{equation}
\mathrm{Im}[\ln(-z)] = \mathrm{Im}[\ln(z)] -
\mathrm{sgn}(\mathrm{Im}[z])\, \pi \label{e51}
\end{equation}
and, with $z=x+i y$, $x,y\in\mathds{R}$,
\begin{equation}
\mathrm{Im}[\ln(z)] = \arctan\Big(\frac{y}{x}\Big) + \Theta(-x)\,
\mathrm{sgn}(y)\, \pi. \label{e52}
\end{equation}

From Eq.~(\ref{e47}) one deduces that (cf. Eq.~(\ref{e29}))
\begin{equation}
\lim_{\beta\to\infty} \b{\nu}_{\sigma}^{(1)}({\bm k}) =
-\frac{1}{\pi}\,
\mathrm{Arctan}\Big(\frac{\mathrm{Im}[\t{G}_{\sigma}^{-1}({\bm
k};\mu+i 0^+)]}{\mathrm{Re}[\t{G}_{\sigma}^{-1}({\bm k};\mu+i
0^+)]}\Big), \label{e53}
\end{equation}
or, equivalently (using the expression in Eq.~(\ref{e49})),
\begin{eqnarray}
&&\hspace{-1.0cm}\lim_{\beta\to\infty} \b{\nu}_{\sigma}^{(1)}({\bm k}) =
\mathrm{sgn}\big(\mathrm{Im}[\t{G}_{\sigma}^{-1}({\bm k};\mu+i
0^+)]\big)\,\Theta\big(\mathrm{Re}[\t{G}_{\sigma}^{-1}({\bm k};\mu+i
0^+)]\big) \nonumber\\
&&\hspace{1.5cm} -\frac{1}{\pi}\,
\arctan\Big(\frac{\mathrm{Im}[\t{G}_{\sigma}^{-1}({\bm k};\mu+i
0^+)]}{\mathrm{Re}[\t{G}_{\sigma}^{-1}({\bm k};\mu+i 0^+)]}\Big).
\label{e54}
\end{eqnarray}
Although $\mu$ can, as a thermodynamic variable corresponding to the grand-canonical ensemble under investigation, take any arbitrary value, since by definition $\t{G}_{\sigma}({\bm k};z)$ corresponds to the $N$-particle GS of $\wh{H}$, the explicit $\mu$ in Eqs.~(\ref{e53}) and (\ref{e54}) can no longer be chosen to be outside the interval $(\mu_{N;\sigma}^-,\mu_{N;\sigma}^+)$, $\forall\sigma$ (appendix \ref{sc}). \emph{This restriction will be in force for the remaining part of this section.} With reference to the remarks following Eq.~(\ref{e36}), here and in the following $\mu$ cannot further deviate from the value on which $\t{G}_{\sigma}({\bm k};z)$, as the zero-temperature \emph{limit} of $\mathscr{G}_{\sigma}({\bm k};z)$, implicitly depends.

From the Dyson equation, Eq.~(\ref{e4}), one has
\begin{equation}
\mathrm{Re}[\t{G}_{\sigma}^{-1}({\bm k};\mu+i 0^+)] = \mu
-\varepsilon_{\bm k} -\hbar\,\mathrm{Re}[\t{\Sigma}_{\sigma}({\bm
k};\mu+i 0^+)], \label{e55}
\end{equation}
\begin{equation}
\mathrm{Im}[\t{G}_{\sigma}^{-1}({\bm k};\mu+i 0^+)] = 0^+
-\hbar\,\mathrm{Im}[\t{\Sigma}_{\sigma}({\bm k};\mu+i 0^+)].
\label{e56}
\end{equation}
The assumed stability of the underlying GS implies the inequalities in Eq.~(\ref{e14}) (appendix \ref{sc}), and thus those in Eq.~(\ref{e16}), whereby the RHS of Eq.~(\ref{e56}) is \emph{positive} for all ${\bm k}$; the $0^+$ on the RHS of Eq.~(\ref{e56}) prevents ambiguity which would arise on identifying $\mathrm{Im}[\t{\Sigma}_{\sigma}({\bm k};\mu + i 0^+)]\le 0$ with zero. It follows that the $\mathrm{sgn}\big(\mathrm{Im}[\t{G}_{\sigma}^{-1}({\bm k};\mu+i 0^+)]\big)$ on the RHS of Eq.~(\ref{e54}) can be identified with unity for all ${\bm k}$. Thus, since $\mu \in (\mu_{N;\sigma}^-,\mu_{N;\sigma}^+)$, on the basis of Eq.~(\ref{e8}) the result in Eq.~(\ref{e54}) can be expressed as
\begin{eqnarray}
&&\hspace{-1.0cm}\lim_{\beta\to\infty} \b{\nu}_{\sigma}^{(1)}({\bm
k}) = \Theta\big(\mu-\varepsilon_{\bm k}
-\hbar\, \Sigma_{\sigma}({\bm k};\mu)\big) \nonumber\\
&&\hspace{1.5cm} -\frac{1}{\pi}\, \arctan\Big(\left.\frac{0^+
-\hbar\,\mathrm{Im}[{\Sigma}_{\sigma}({\bm k};\varepsilon)]}{\mu
-\varepsilon_{\bm k} -\hbar\,\mathrm{Re}[{\Sigma}_{\sigma}({\bm
k};\varepsilon)]}\right|_{\varepsilon\downarrow \mu}\Big),
\label{e57}
\end{eqnarray}
where
\begin{equation}
\Theta\big(\mu-\varepsilon_{\bm k} -\hbar\, \Sigma_{\sigma}({\bm
k};\mu)\big) \equiv \Theta\big(G_{\sigma}^{-1}({\bm k};\mu)\big)
\equiv \Theta\big(G_{\sigma}({\bm k};\mu)\big). \label{e58}
\end{equation}
In Eq.~(\ref{e57}), $\varepsilon\downarrow \mu$ signifies the approach of $\varepsilon$ towards $\mu$ from above (cf. Eq.~(\ref{e3})).

For \emph{insulating} $N$-particle GSs one has (cf. Eq.~(\ref{e8}))
\begin{equation}
\mathrm{Im}[\Sigma_{\sigma}({\bm k};\varepsilon)] \equiv
0\;\;\;\mbox{\rm for}\;\;\;\varepsilon \in (\mu_{N;\sigma}^-,
\mu_{N;\sigma}^+),\;\;\; \forall {\bm k},  \label{e59}
\end{equation}
where $\mu_{N;\sigma}^+ - \mu_{N;\sigma}^-$ is positive and finite and where the chemical potential corresponding to $N$ particles satisfies $\mu_{N;\sigma}^- < \mu < \mu_{N;\sigma}^+$, $\forall\sigma$ (appendix \ref{sc}). For these GSs, and $\mu$ satisfying $\mu_{N;\sigma}^- < \mu < \mu_{N;\sigma}^+$, one can therefore unequivocally identify the second term on the RHS of Eq.~(\ref{e57}) with zero for all ${\bm k}$.

Although for \emph{metallic} $N$-particle GSs the second term on the RHS of Eq.~(\ref{e57}) cannot be \emph{a priori} identified with zero for all ${\bm k}$, it can be shown that for these GSs the deviation from zero of this term can at most extent over a zero-measure subset of the available ${\bm k}$ space in the neighbourhood of the underlying $\mathcal{S}_{\Sc f;\sigma}$. The proof of this statement is as follows: since Eq.~(\ref{e8}) applies for all GSs and \emph{all} ${\bm k}$, in order for the second term on the RHS of Eq.~(\ref{e57}) to be non-vanishing over a finite subset of the underlying ${\bm k}$ space, it is necessary that over this subset one has $\varepsilon_{\bm k} +\hbar\Sigma_{\sigma}({\bm k};\mu) = \mu$. This condition is not feasible for a stable GS, since it implies an extended $d$-dimensional Fermi ``surface'' in a $d$-dimensional ${\bm k}$ space (cf. Eq.~(\ref{e6}) and recall that for metallic GSs $\mu=\varepsilon_{\Sc f}$).

For illustration, for Fermi liquids \cite{JML61} and marginal Fermi liquids \cite{VLS-RAR89}, where one has (see, e.g., Ref.~\citen{BF99})
\begin{equation}
\frac{\mathrm{Im}[\Sigma_{\sigma}({\bm
k};\varepsilon)]}{\mathrm{Re}[\Sigma_{\sigma}({\bm k};\varepsilon)]
-\Sigma_{\sigma}({\bm k};\mu)} \to 0\;\;\; \mbox{\rm as}\;\;\;
\varepsilon \to\mu,\;\;\; \forall {\bm k}, \label{e63}
\end{equation}
the second term on the RHS of Eq.~(\ref{e57}) can be unequivocally equated with zero for all ${\bm k}$. For the metallic states of the one-dimensional Luttinger model for spin-less fermions
\cite{JV94,JV93}, however, where (see appendix \ref{sd})
\begin{equation}
\frac{\mathrm{Im}[\Sigma_{\sigma}({\bm
k};\varepsilon)]}{\mathrm{Re}[\Sigma_{\sigma}({\bm k};\varepsilon)]
-\Sigma_{\sigma}({\bm k};\mu)} \to \tan(\pi\gamma_0({\bm k}))\;\;\;
\mbox{\rm as}\;\;\; \varepsilon \to\mu, \label{e64}
\end{equation}
the second term on the RHS of Eq.~(\ref{e57}) is not vanishing for all ${\bm k}$. However, one can readily verify that this is only the case for ${\bm k}$ in the \emph{infinitesimal} neighbourhoods of the underlying Fermi points; away from these neighbourhoods, the change in $\varepsilon_{\bm k}+\hbar\Sigma_{\sigma}({\bm k};\mu)$ causes this term rapidly to approach zero. Consequently, with the exception of the points in the last-mentioned neighbourhoods, forming a subset of measure zero of the real ${\bm k}$ axis (see our statement in the previous paragraph), the second term on the RHS of Eq.~(\ref{e57}) can be equated with zero also in the case of the metallic GSs of the one-dimensional Luttinger model for spin-less fermions.\footnote{In appendix \protect\ref{sd} we consider the consequence for $\lim_{\beta\to\infty}\b{\nu}_{\sigma}^{(2)}({\bm k})$, Eq.~(\protect\ref{e43}), of the peculiar behaviour of self-energy as reflected in Eq.~(\protect\ref{e64}), and show that this function is perfectly well-defined.} We recall that the validity of the Luttinger theorem at hand has been rigorously demonstrated for the metallic GSs of this model \cite{BB97,YOA97}.

We conclude that barring possible subsets of measure zero in the neighbourhoods of the Fermi surfaces of metallic GSs, for \emph{all} uniform GSs one has
\begin{equation}
\lim_{\beta\to\infty}\b{\nu}_{\sigma}^{(1)}({\bm k})=
\Theta\big(G_{\sigma}^{-1}({\bm k};\mu)\big) \equiv
\Theta\big(G_{\sigma}({\bm k};\mu)\big). \label{e65}
\end{equation}
Hereby is the validity of the results in Eqs.~(\ref{e38}) and (\ref{e39}) established. It remains therefore to investigate the domain of validity of the Luttinger-Ward identity, Eq.~(\ref{e44}). We note that in view of the result in Eq.~(\ref{e65}) and of the identity in Eq.~(\ref{e46}), \emph{the Luttinger theorem is valid if and only if the Luttinger-Ward identity is valid}.

\subsubsection{Remarks}
\label{ss43s1}

Comparing the expressions in Eqs.~(\ref{e57}) and (\ref{e65}), one observes that breakdown of the result Eq.~(\ref{e17}) would render the result in Eq.~(\ref{e65}) invalid, however since the sum with respect to ${\bm k}$ of the second term on the RHS of Eq.~(\ref{e57}) may vanish, failure of the result Eq.~(\ref{e17}) can only potentially, but not necessarily, lead to violation of the results in Eqs.~(\ref{e38}) and (\ref{e39}).

Numerical results by Schmalian \emph{et al.} \cite{SLGB96a,SLGB96b} and Langer \emph{et al.} \cite{LSGB95,LSGB96} for the metallic states of the Hubbard Hamiltonian in two space dimensions and close to half-filling exhibit a failure of the result in Eq.~(\ref{e17}) which in addition gives rise to violation of the equality in Eq.~(\ref{e38}). In Sec.~\ref{ss63} we shall demonstrate that these numerical results are adversely affected by a computational artefact.

\section{Role of perturbation theory in the proof of the Luttinger theorem}
\label{s5}

As we have indicated in Sec.~\ref{s1}, a reason that often is put forward for rationalising the observed breakdowns of the Luttinger theorem, is the supposed failure of the weak-coupling many-body perturbation theory in correctly describing strongly-correlated GSs. It is therefore paramount to establish the true function of the perturbation theory in the original proof by Luttinger and Ward \cite{LW60} of the Luttinger theorem. Below we examine all instances where Luttinger and Ward \cite{LW60} employed this theory in their proof and demonstrate that the perturbation theory as utilised by Luttinger and Ward is either a formal instrument for obtaining a non-perturbative result, or, where this is not the case, it does \emph{not} break down.

The following details provide an overview of the contents of this section.

Luttinger and Ward \cite{LW60} employed the weak-coupling many-body perturbation theory in three main instances. In the first instance, they deduced an expression, Eq.~(\ref{e66}) below, for the grand potential $\Omega$ as a function of the coupling-constant of interaction, $\lambda$; the role of the grand potential in the context of the Luttinger theorem consists of providing an expression for the mean-value of particles, $\b{N}$, in the grand-canonical ensemble, Eq.~(\ref{e24}). We shall show that this expression for $\Omega$ can be deduced without recourse to perturbation theory.

The second instance, where perturbation theory features in the work by Luttinger and Ward, concerns the definition of a functional $Y(\lambda)$, Eq.~(\ref{e68}) below, in which a functional $Y'(\lambda)$, Eq.~(\ref{e72}) below, is defined in terms of perturbative contributions to the proper self-energy $\mathscr{S}_{\!\sigma}({\bm k};z)$; through establishing equality of $\Omega(\lambda)$ with $Y(\lambda)$, Eq.~(\ref{e71}) below, Luttinger and Ward \cite{LW60} arrived at the expression for $\b{N}$ presented in Eq.~(\ref{e34}). As we have indicated Sec.~\ref{ss42}, were it not for the fact that $Y'$ plays a vital role in the proof of the Luttinger-Ward identity, even failure of Eq.~(\ref{e34}) would not have a direct consequence for the validity of the Luttinger theorem. We shall obtain the key element for establishing the exactness of the equality $\Omega(\lambda) =Y(\lambda)$ in the process of investigating the validity of the Luttinger-Ward identity, Eq.~(\ref{e44}), whose proof relies on the same perturbation series expansion for $\mathscr{S}_{\!\sigma}({\bm k};z)$ as is encountered in the expression for $Y'(\lambda)$.

The third instance of using perturbation theory concerns, as we have just indicated, the proof of the Luttinger-Ward identity, Eq.~(\ref{e44}). We shall demonstrate that the perturbation series expansion for the self-energy $\t{\Sigma}_{\sigma}({\bm k};z)$ that features in the original proof by Luttinger and Ward of this identity is uniformly convergent for almost all ${\bm k}$ and $z$. On the basis of this property we shall rigorously demonstrate the exactness of the Luttinger-Ward identity, with the proviso that $\mu=\mu_{\infty}$ (Sections \ref{ss23} and \ref{ss41}). The same property, concerning the adopted perturbation series for $\t{\Sigma}_{\sigma}({\bm k};z)$, provides an \emph{a posteriori} proof for the exactness of the equality $\Omega(\lambda) = Y(\lambda)$, Eq.~(\ref{e71}) below, through establishing that indeed $Y'$ is a well-defined functional.

\subsection{The first two instances of the use of perturbation theory}
\label{ss51}

The starting point of the work by Luttinger and Ward \cite{LW60} is
the finite-temperature perturbation series expansion due to Bloch and de Dominicis \cite{BdD58} of the grand potential $\Omega$,
\begin{equation}
\Omega = \sum_{\nu=0}^{\infty} \Omega_{\nu}, \label{e9a}
\end{equation}
where $\Omega_{\nu}$ denotes the total contribution of the $\nu$th order closed linked diagrams to $\Omega$. \emph{Use of Eq.~(\ref{e9a}) marks the first instance where perturbation theory plays at least a formal role in the proof of the Luttinger theorem.}

A notable characteristic of the Bloch-de Dominicis formalism is its reliance on a \emph{generalized} Wick theorem \cite{BdD58,TM55,MG60,FW03} which establishes an exact relationship between the interacting ensemble-average of a time-ordered product consisting of equal numbers of creation and annihilation operators, with a finite series of non-interacting ensemble-averages of fully contracted operators; the only condition on which the validity of this theorem rests is that the non-interacting Hamiltonian $\wh{H}_0$, defining the non-interacting grand-canonical statistical operator $\h{\rho}_0 \equiv \mathrm{e}^{-\beta (\wh{H}_0-\mu\wh{N})}$, commute with the total-number operator $\wh{N}$. It follows that so long as $[\wh{H}_0,\wh{N}]_{-} = 0$ and so long as the Fock space spanned by the eigenstates of $\wh{H}_0$ coincides with that spanned by the eigenstates of the interacting Hamiltonian $\wh{H}$, the series in Eq.~(\ref{e9a}) is \emph{formally} exact. As will become evident below, for our present considerations it will not be necessary to investigate the convergence of the series in Eq.~(\ref{e9a}); the mere fact that this series amounts to a complete representation of $\Omega$ will suffice.

The above statement, that in the context of the present work it is not necessary to investigate the convergence property of the series in Eq.~(\ref{e9a}), is based on the fact that Luttinger and Ward \cite{LW60} did \emph{not} use the series on the RHS of Eq.~(\ref{e9a}) in its explicit form. Instead, they used this series as a stepping-stone for deducing the general relationship (see Eqs.~(41), (42) and (43) in Ref.~\citen{LW60}; use of Table \ref{t1} should prove helpful)
\begin{eqnarray}
\Omega(\lambda) &=& \Omega_0 + \sum_{{\bm k},\sigma} \frac{1}{2\beta}
\sum_{m} \mathrm{e}^{\zeta_m 0^+}\mathscr{G}_{\sigma;0}({\bm
k};\zeta_m)\! \int_0^{\lambda} \frac{{\rm d}\lambda'}{\lambda'}\;
\mathscr{S}_{\!\sigma;\lambda'}^{\sharp}({\bm k};\zeta_m) \nonumber\\
&\equiv& \Omega_0 + \sum_{{\bm k},\sigma} \frac{1}{2\beta}
\sum_{m}\mathrm{e}^{\zeta_m 0^+}\!\! \int_0^{\lambda} \frac{{\rm
d}\lambda'}{\lambda'}\; \mathscr{G}_{\sigma;\lambda'}({\bm
k};\zeta_m)\, \mathscr{S}_{\!\sigma;\lambda'}({\bm k};\zeta_m),
\label{e66}
\end{eqnarray}
where $\lambda$ denotes the dimensionless coupling constant of interaction; the function $\mathscr{S}_{\!\sigma;\lambda'}^{\sharp}({\bm k};\zeta_m)$ is the \emph{total} thermal self-energy, including both the \emph{proper} $\mathscr{S}_{\!\sigma;\lambda'}({\bm k};\zeta_m)$ and \emph{improper} self-energies \cite{FW03}, corresponding to the case where the coupling constant of interaction is equal to $\lambda'$; similarly for the interacting thermal Green function $\mathscr{G}_{\sigma;\lambda'}({\bm k};\zeta_m)$. With the actual value of the coupling constant of interaction being equal to unity, $\mathscr{S}_{\!\sigma}^{\sharp}({\bm k};\zeta_m)$, $\mathscr{S}_{\!\sigma}({\bm k};\zeta_m)$ and $\mathscr{G}_{\sigma}({\bm k};\zeta_m)$ are the short-hand notations for $\mathscr{S}_{\!\sigma;1}^{\sharp}({\bm k};\zeta_m)$, $\mathscr{S}_{\!\sigma;1}({\bm k};\zeta_m)$ and $\mathscr{G}_{\sigma;1}({\bm k};\zeta_m)$ respectively. We note that from the Dyson equations corresponding to the total self-energy $\mathscr{S}_{\!\sigma;\lambda}^{\sharp}({\bm k};\zeta_m)$ and the proper self-energy $\mathscr{S}_{\!\sigma;\lambda}({\bm k};z)$, one obtains that \begin{equation}
\mathscr{G}_{\sigma;0}({\bm k};\zeta_m)\,
\mathscr{S}_{\!\sigma;\lambda}^{\sharp}({\bm k};\zeta_m) \equiv \mathscr{G}_{\sigma;\lambda}({\bm k};\zeta_m)\,
\mathscr{S}_{\!\sigma;\lambda}({\bm k};\zeta_m), \label{e67}
\end{equation}
on the strength of which the last expression on the RHS of Eq.~(\ref{e66}) is obtained from the preceding expression.

On the basis of the expression in Eq.~(\ref{e66}) Luttinger and Ward \cite{LW60} demonstrated that their $Y(\lambda)$ functional, defined as
\begin{equation}
Y(\lambda) {:=} -\sum_{{\bm k},\sigma} \frac{1}{\beta} \sum_m
\mathrm{e}^{\zeta_m 0^+}\, \Big\{\ln\big(-\beta\hbar\,
\mathscr{G}_{\sigma;\lambda}^{-1}({\bm k};\zeta_m)\big)+
\mathscr{G}_{\sigma;\lambda}({\bm k};\zeta_m)\,
\mathscr{S}_{\!\sigma;\lambda}({\bm k};\zeta_m)\Big\} + Y'(\lambda),
\label{e68}
\end{equation}
satisfies both
\begin{equation}
Y(0) =\Omega_0 \label{e69}
\end{equation}
and
\begin{equation}
\lambda\, \frac{\partial\Omega(\lambda)}{\partial\lambda} =
\lambda\, \frac{\partial
Y(\lambda)}{\partial\lambda},\;\;\;\forall\lambda. \label{e70}
\end{equation}
Through integrating both sides of this equation, subject to the
initial condition in Eq.~(\ref{e69}), Luttinger and Ward
\cite{LW60} arrived at the conclusion that\footnote{For a fundamental aspect associated with calculating, for instance, $\Omega(\lambda)$ through integrating $\partial\Omega(\lambda)/\partial\lambda$, we refer the reader to Sec.~\protect\ref{ssf3s3}.}
\begin{equation}
\Omega(\lambda) = Y(\lambda),\;\;\; \forall\lambda. \label{e71}
\end{equation}
The functional $Y'$ on the RHS of Eq.~(\ref{e68}) is defined as
\begin{equation}
Y' {:=} \sum_{\nu=1}^{\infty} Y^{\prime(\nu)} \equiv \sum_{\sigma}\sum_{\nu=1}^{\infty} Y_{\sigma}^{\prime(\nu)} \equiv \sum_{\sigma}\sum_{\nu=1}^{\infty}\frac{1}{2\nu}
\sum_{{\bm k}} \frac{1}{\beta} \sum_m \mathrm{e}^{\zeta_m
0^+}\mathscr{G}_{\sigma}({\bm
k};\zeta_m)\,\mathscr{S}_{\!\sigma}^{(\nu)}({\bm k};\zeta_m),
\label{e72}
\end{equation}
where $\mathscr{S}_{\!\sigma}^{(\nu)}({\bm k};\zeta_m)$ denotes the total contribution of all skeleton self-energy diagrams \cite{LW60} of order $\nu$ in terms of the \emph{bare} two-particle interaction potential and the \emph{interacting} Green functions $\{\mathscr{G}_{\sigma'}({\bm k};\zeta_m)\,\|\, \sigma' = \uparrow,\downarrow\}$; in other words, the $\nu$ in $\mathscr{S}_{\!\sigma}^{(\nu)}({\bm k};\zeta_m)$ counts the order of self-energy contributions in accordance with their \emph{explicit} dependence on the bare interaction potential and \emph{not} their total dependence on this potential, which is of infinite order. We point out that since skeleton self-energy diagrams by definition do not contain self-energy insertions \cite{LW60}, it follows that these diagrams constitute a subset of the \emph{proper} self-energy diagrams.

Following (cf. Eqs.~(\ref{e24}) and (\ref{e71}))
\begin{equation}
\b{N} = -\frac{\partial \Omega}{\partial\mu} \equiv  -\frac{\partial
Y}{\partial\mu},\label{e73}
\end{equation}
and on the basis of Eq.~(\ref{e68}), Luttinger and Ward \cite{LW60} arrived at Eq.~(\ref{e34}). The contribution $\b{N}_{\sigma}^{(2)}$ to $\b{N}_{\sigma}$, with $\b{\nu}_{\sigma}^{(2)}({\bm k})$ as defined in Eq.~(\ref{e36}), arises from the combination of $Y'$ and the second term inside the curly braces on the RHS of Eq.~(\ref{e68}).

The contribution $-\partial Y'/\partial\mu$ is calculated on the basis of the property (Eq.~(\ref{e27}))
\begin{equation}
\frac{\partial}{\partial\mu} = \frac{\partial}{\partial\zeta_m},
\label{e74}
\end{equation}
from which one has
\begin{equation}
\frac{\partial}{\partial\mu} \sum_{{\bm k},m} \mathscr{G}_{\sigma}({\bm
k};\zeta_m)\, \mathscr{S}_{\!\sigma}^{(\nu)}({\bm k};\zeta_m) = 2\nu
\sum_{{\bm k},m} \mathscr{G}_{\sigma}({\bm k};\zeta_m)\, \frac{\partial
\mathscr{S}_{\!\sigma}^{(\nu)}({\bm k};\zeta_m)}{\partial\zeta_m}.
\label{e75}
\end{equation}
The validity of this expression can be explicitly established for any finite value of $\nu$. One can further readily verify that without all $\mathscr{G}_{\sigma}$, $\forall\sigma$, on which $\mathscr{S}_{\!\sigma}^{(\nu)}$ depends\footnote{The pertinent functional form of $\mathscr{S}_{\!\sigma}^{(\nu)}({\bm k};z)$ is fully specified by the collection of the $\nu$th-order skeleton diagrams.} being identical to the explicit $\mathscr{G}_{\sigma}$ that one encounters in Eq.~(\ref{e75}), the LHS of Eq.~(\ref{e75}) would not be identical to $2\nu$ times the same quantity for all $\sigma$ (exceptions cannot be excluded); the interchangeability of \emph{all}, explicit and implicit, Green functions gives rise to a cyclic property, the consequence of which is clearly manifested by Eq.~(\ref{e75}). In this light, it should not come as a surprise that Eq.~(\ref{e75}) remains valid on replacing \emph{all} the underlying interacting Green functions by, for instance, their non-interacting counterparts.

We remark that, through $\zeta_m$ (cf. Eq.~(\ref{e27}) both sides of Eq.~(\ref{e75}) \emph{explicitly} depend on the thermodynamic variable $\mu$ of the underlying grand-canonical ensemble. In addition, both sides depend \emph{implicitly} on $\mu$ through the implicit dependence of $\mathscr{G}_{\sigma}({\bm k};z)$ and $\mathscr{S}_{\!\sigma}^{(\nu)}({\bm k};z)$ on $\mu$. For exactly the same reason as we presented in the previous paragraph, the general validity of Eq.~(\ref{e75}) is also dependent on the equality of the values of the two chemical potentials; the common value for the two is however entirely arbitrary and need not be equal to, for instance, $\mu(\beta,N,V)$.

It is relevant to point out that the above-mentioned `cyclic property', which is responsible for the validity of Eq.~(\ref{e75}), is dependent on the admissability of exchanging orders of a multiplicity of infinite sums. In this paper we shall not present the mathematical justification for such exchanges. Suffice it to mention however that this justification can be presented and that the underlying details are very akin to those that we shall present in Sections \ref{ss53s12}, \ref{ss53s13} and \ref{ss53s14}.

Involvement of $Y'$ in the proof of the Luttinger theorem implies reliance of this proof on the perturbation series expansion for the self-energy $\mathscr{S}_{\!\sigma}({\bm k};z)$. \emph{This constitutes the second instance where perturbation theory plays a role in the proof by Luttinger and Ward \cite{LW60} of the Luttinger theorem.} Note that since the expression for $\b{N}_{\sigma}^{(2)}$, or $\b{\nu}_{\sigma}^{(2)}({\bm k})$, Eq.~(\ref{e36}), does not involve any summation with respect to $\nu$, it amounts to a non-perturbative result; perturbation theory plays a role only in the proof of the Luttinger-Ward identity, Eq.~(\ref{e44}).

\subsubsection{Remarks}
\label{ss51s1}

The above-mentioned property concerning the formal exactness of the finite-temperature series in Eq.~(\ref{e9a}) is not necessarily shared by the strictly \emph{zero-temperature} many-body series expansions specific to GS properties. These rely on the use of the Wick theorem \cite{GCW50,FW03} (which amounts to an exact operator identity) and the fact that expectation values of the normal-ordered products of creation and annihilation operators with respect to the GS of $\wh{H}_0$, defining the vacuum state, are vanishing. Validity of these series, therefore, crucially depends on the $N$-particle GS of $\wh{H}_0$ being adiabatically connected with that of $\wh{H}$.\footnote{See in particular the last part of the discussions, presented in Ref.~\citen{FW03}, pp. 61-64, concerning the Gell-Mann and Low theorem.} For instance, if interaction gives rise to a symmetry breaking, the above-mentioned two GSs cannot be adiabatically connected. Evidently, no such centrality is given to the $N$-particle GS of $\wh{H}_0$ in finite-temperature series expansions.

For macroscopic systems, the superiority of the finite-temperature series expansion of the total energy in the zero-temperature \emph{limit} in comparison with the zero-temperature series expansion of the GS total energy, due to Brueckner and Goldstone, was explicitly demonstrated by Kohn and Luttinger \cite{KL60} who established the possibility of the existence of so-called anomalous contributions to the GS total energy of macroscopic systems which are missing in the strictly zero-temperature Brueckner-Goldstone series.

\subsubsection{Technicalities}
\label{ss51s2}

Underlying the result in Eq.~(\ref{e70}) lies the observation that for a fixed $\mu$, which is a thermodynamic variable (to be distinguished from the chemical potential $\mu(\beta,\b{N},V)$ corresponding to a predetermined value for the mean number of particles $\b{N}$ in the ensemble), the first derivative with respect to $\lambda$ of the first term on the RHS of Eq.~(\ref{e68}) is vanishing, so that
\begin{equation}
\frac{\partial Y(\lambda)}{\partial\lambda} \equiv \frac{\partial Y'(\lambda)}{\partial\lambda}. \label{e76}
\end{equation}
The constancy to linear order in $\lambda$ of the first term on the RHS of Eq.~(\ref{e68}) for a fixed $\mu$, is a direct consequence of two facts: firstly, this term is an explicit function, or a \emph{local} functional, of $\mathscr{S}_{\!\sigma;\lambda}({\bm k};\zeta_m)$ (whence originates the dependence on $\lambda$ of the first term on the RHS of Eq.~(\ref{e68})), and, secondly,
\begin{equation}
\frac{\delta Y}{\delta \mathscr{S}_{\!\sigma;\lambda}({\bm k};\zeta_m)} = 0,\;\;\; \forall {\bm k}, m, \sigma. \label{e77}
\end{equation}
This variational relationship, whose validity can be readily verified, was first demonstrated by Luttinger and Ward \cite{LW60}.

We note that $Y'$, and thus $Y$, does \emph{not} explicitly depend on the non-interacting Green function $\mathscr{G}_{\sigma;0}({\bm k};\zeta_m)$; an implicit dependence remains through the initial condition in Eq.~(\ref{e69}), requiring that
\begin{equation}
\lim_{\lambda\downarrow 0}\mathscr{G}_{\sigma;\lambda}({\bm
k};\zeta_m) = \mathscr{G}_{\sigma;0}({\bm k};\zeta_m).\label{e78}
\end{equation}
Since $\{ \mathscr{S}_{\sigma;\lambda}^{(\nu)}({\bm k};\zeta_m) \}$ are continuous functions of $\lambda$ and since $\sum_{\nu=1}^{\infty} \mathscr{S}_{\sigma;\lambda}^{(\nu)}({\bm k};\zeta_m)$ converges \emph{uniformly} for all ${\bm k}$, $m$ and $\lambda \ge 0$, the possibility of violation of the equality in Eq.~(\ref{e78}) is ruled out (\S~45 in Ref.~\citen{TB65}).\footnote{A typical example of a non-uniformly convergent series is $S(x)\equiv \sum_{j=0}^{\infty} x^2/(1+x^2)^j$, which is absolutely convergent for all $x\in \mathds{R}$ (\S~3.3 in Ref.~\citen{WW62}, \S~44 in Ref.~\citen{TB65}). For $x\in \mathds{R}$ and $x\not= 0$ one obtains that $S(x) = 1+x^2$, which leads to $\lim_{x\to 0} S(x) = 1$, to be contrasted with $S(0) = 0$. } Further, since the lower boundary of the sum with respect to $\nu$ on the RHS of Eq.~(\ref{e72}) is equal to $1$, indeed $Y'(0) = 0$, which is essential for the validity of Eq.~(\ref{e69}).\footnote{A similar argument as in the previous footnote applies here, so that $\lim_{\lambda\downarrow 0} Y'(\lambda) = Y'(0)$. See Sec.~\protect\ref{ss51s4}.}

The expression in Eq.~(\ref{e66}) is obtained from that in Eq.~(\ref{e9a}) on the basis of the \emph{identity}
\begin{equation}
\lambda\, \frac{\rm d}{{\rm d}\lambda}
\mathscr{S}_{\!\sigma;\lambda}^{\sharp(\nu)}({\bm k};\zeta_m) =
\nu\,\mathscr{S}_{\!\sigma;\lambda}^{\sharp(\nu)}({\bm k};\zeta_m),
\label{e79}
\end{equation}
where $\mathscr{S}_{\!\sigma;\lambda}^{\sharp(\nu)}({\bm
k};\zeta_m)$ is the sum of all self-energy contributions (including
both proper and improper parts \cite{FW03}) of order $\nu$ expressed
in terms of the \emph{non-interacting} single-particle Green
functions $\{\mathscr{G}_{\sigma';0}({\bm k};z)\,\|\, \sigma'\}$.
Making use of the identity in Eq.~(\ref{e79}) and of the fact that $\mathscr{S}_{\!\sigma;\lambda}^{\sharp(\nu)}({\bm k};\zeta_m) \equiv 0$ for $\lambda=0$, one readily obtains that (cf.
Eq.~(\ref{e66}))
\begin{equation}
\sum_{\nu=1}^{\infty} \frac{1}{\nu}\,
\mathscr{S}_{\!\sigma;\lambda}^{\sharp(\nu)}({\bm k};\zeta_m) \equiv
\sum_{\nu=1}^{\infty} \int_{0}^{\lambda} \frac{{\rm
d}\lambda'}{\lambda'}\;\mathscr{S}_{\!\sigma;\lambda'}^{\sharp(\nu)}({\bm
k};\zeta_m) \equiv \int_{0}^{\lambda} \frac{{\rm
d}\lambda'}{\lambda'}\;\mathscr{S}_{\!\sigma;\lambda'}^{\sharp}({\bm
k};\zeta_m). \label{e80}
\end{equation}

\subsubsection{Non-perturbative derivation of Eq.~(\protect\ref{e66})}
\label{ss51s3}

The last expression in Eq.~(\ref{e66}) can be obtained without recourse to the use of the weak-coupling many-body perturbation theory for $\Omega(\lambda)$. A comprehensive exposition of this alternative derivation of the last expression in Eq.~(\ref{e66}) can be found in Ref.~\citen{FW03} (see the text between Eqs.~(23.16) and (23.22) in Ref.~\citen{FW03}); this derivation relies solely on the cyclic property of the trace operator, leading to the expression
\begin{equation}
\frac{\partial\Omega(\lambda)}{\partial\lambda} = \langle
\wh{H}_1 \rangle_{\lambda} \equiv
\mathrm{Tr}[\mathrm{e}^{-\beta (\wh{H}_{\lambda} - \mu
\wh{N})}\, \wh{H}_1], \label{e81}
\end{equation}
where $\wh{H}_1 \equiv (\wh{H}_{\lambda} -\wh{H}_0)/\lambda$, in which $\wh{H}_{\lambda} = \wh{H}_0 + \lambda\wh{H}_1$ is the total Hamiltonian. The last expression on the RHS of Eq.~(\ref{e66}) is obtained from the
expression in Eq.~(23.22) of Ref.~\citen{FW03} on employing
\begin{equation}
\mathscr{G}_{\sigma;0}^{-1}({\bm
k};\zeta_m)\,\mathscr{G}_{\sigma;\lambda'}({\bm k};\zeta_m) =
\mathscr{G}_{\sigma;\lambda'}({\bm k};\zeta_m)
\mathscr{S}_{\!\sigma;\lambda'}({\bm k};\zeta_m) + 1,
\label{e82}
\end{equation}
which follows from the Dyson equation, and making use of the fact that
(Eq.~(26.9) in Ref.~\citen{FW03})
\begin{equation}
\sum_{m} \mathrm{e}^{\zeta_m 0^+} = 0. \label{e83}
\end{equation}

The counterpart of the expression in Eq.~(\ref{e81}) for $\beta=\infty$, and specific to the GS total energy $E_0(\lambda)$, is known as the Hellmann-Feynman theorem; the expression for $\partial E_0(\lambda)/\partial\lambda$ can be described in terms of $\{ \t{G}_{\sigma';\lambda}({\bm k};z)\}$ through reliance solely on the canonical anti-commutation relations of the creation and annihilation operators (see the text between Eqs.~(7.12) and (7.32) in Ref.~\citen{FW03}).

\subsubsection{$Y'$ is well-defined irrespective of the strength of the interaction potential}
\label{ss51s4}

The dependence of $Y'$, Eq.~(\ref{e72}), on the perturbative contributions $\{ \mathscr{S}_{\!\sigma}^{(\nu)}({\bm k};z)\}$ \emph{cannot} be of consequence to the validity or otherwise of the Luttinger theorem. This follows from the fact that the series
\begin{equation}
\mathscr{S}_{\!\sigma}({\bm k};\zeta_m) = \sum_{\nu=1}^{\infty}
\mathscr{S}_{\!\sigma}^{(\nu)}({\bm k};\zeta_m) \label{e84}
\end{equation}
is uniformly convergent for all ${\bm k}$ and $m$, so that on account of the generalised Abel theorem, or Hardy's theorem (\S~3.35 in Ref.~\citen{WW62}), the series
\begin{equation}
\sum_{\nu=1}^{\infty} \frac{1}{2\nu}
\mathscr{S}_{\!\sigma}^{(\nu)}({\bm k};\zeta_m) \nonumber
\end{equation}
is similarly uniformly convergent for all ${\bm k}$ and $m$. As a general rule, infinite series of functions that converge uniformly in some domain, behave like finite series in that domain (\S~3.33 in Ref.~\citen{WW62}). We point out that in this paper we shall not explicitly deal with the series in Eq.~(\ref{e84}), instead with the series
\begin{equation}
\t{\Sigma}_{\sigma}({\bm k};z) = \sum_{\nu=1}^{\infty}
\t{\Sigma}_{\sigma}^{(\nu)}({\bm k};z), \label{e85}
\end{equation}
which in Sec.~\ref{ss53} we shall rigorously prove to be uniformly convergent for all ${\bm k}$ and all $z$ over the entire complex plane outside the real axis. The proof of the uniformity of convergence of the series in Eq.~(\ref{e84}) for all ${\bm k}$ and $m$ is in almost all respects identical to that of the uniformity of convergence of the series in Eq.~(\ref{e85}) for all ${\bm k}$ and all $z$, with $\mathrm{Im}(z)\not=0$.

We have thus demonstrated that $Y'$ is well-defined, irrespective of the strength of the bare two-body interaction potential, which is assumed to be short-ranged.

\subsubsection{Summary}
\label{ss51s5}

In the light of the above considerations, and pending the proof of the uniformity of convergence of the series in Eq.~(\ref{e85}) for almost all $({\bm k},z,\sigma)$, we are in a position to state that Eq.~(\ref{e71}) is exact and that its derivation by Luttinger and Ward on the basis of perturbation theory only insofar depends on this theory as $Y'$ is defined in terms of the perturbative contributions $\{\mathscr{S}_{\!\sigma}^{(\nu)}({\bm k};z)\}$ to $\mathscr{S}_{\!\sigma}({\bm k};z)$; anticipating what follows (see in particular Sections \ref{ss53s2} and \ref{ss53s4}), we have shown that $Y'$ is well-defined irrespective of the strength of the bare two-body interaction potential. As regards use of perturbation series in arriving at the last expression in Eq.~(\ref{e66}), this use has been non-essential, since the latter expression can be obtained without recourse to perturbation theory.

\subsection{The third instance of the use of perturbation theory; points (i), (ii) and (iii)}
\label{ss52}

Perturbation theory plays an essential role in the proof of the Luttinger-Ward identity, Eq.~(\ref{e44}). On account of Eq.~(\ref{e85}), Luttinger and Ward \cite{LW60} employed the expression
\begin{equation}
\lim_{\beta\to\infty} \b{N}_{\sigma}^{(2)} =
\sum_{\nu=1}^{\infty} \sum_{\bm k} \int_{\mathscr{C}(\mu)}
\frac{{\rm d}z}{2\pi i}\; \t{G}_{\sigma}({\bm k};z)
\frac{\partial}{\partial z}\, \t{\Sigma}_{\sigma}^{(\nu)}({\bm
k};z).\label{e86}
\end{equation}
Note that the lower bound of the sum with respect to $\nu$ on the RHS of Eq.~(\ref{e86}) can be identified with $2$ owing to the fact that $\t{\Sigma}_{\sigma}^{(1)}({\bm k};z)$ is independent of $z$, whereby $\partial\t{\Sigma}_{\sigma}^{(1)}({\bm k};z)/\partial z\equiv 0$, $\forall {\bm k},z,\sigma$.

On arriving at the expression in Eq.~(\ref{e86}), Luttinger and Ward \cite{LW60} subsequently demonstrated that
\begin{equation}
\sum_{\bm k} \int_{\mathscr{C}(\mu)} \frac{{\rm d}z}{2\pi i}\;
\t{G}_{\sigma}({\bm k};z) \frac{\partial}{\partial z}\,
\t{\Sigma}_{\sigma}^{(\nu)}({\bm k};z) = 0,\;\;\;\forall\nu\in \mathds{N}.
\label{e87}
\end{equation}
The proof of this expression relies on the observation that, following the expression in Eq.~(\ref{e75}) and in view of the result in Eq.~(\ref{e41}),
\begin{equation}
\sum_{\bm k} \int_{\mathscr{C}(\mu)} \frac{{\rm d}z}{2\pi i}\;
\t{G}_{\sigma}({\bm k};z) \frac{\partial}{\partial z}\,
\t{\Sigma}_{\sigma}^{(\nu)}({\bm k};z) = \lim_{\beta\to\infty} \frac{\partial Y_{\sigma}^{\prime(\nu)}}{\partial\mu},
\label{e88}
\end{equation}
where $Y_{\sigma}^{\prime(\nu)}$ is defined in Eq.~(\ref{e72}). On employing the expression in Eq.~(\ref{e74}), the function $\partial Y_{\sigma}^{\prime(\nu)}/\partial\mu$ can be expressed in terms of a superposition of contributions involving derivatives with respect to the internal energy variables of the Green functions that comprise the expression for $Y_{\sigma}^{\prime(\nu)}$; since $Y_{\sigma}^{\prime(\nu)}$ consists of the contributions corresponding to \emph{closed} Feynman diagrams, through repeated application of integration by parts one readily verifies that all contributions to the expression on the RHS of Eq.~(\ref{e88}) pair-wise cancel, this on account of conservation of energy at the vertices of these diagrams \cite{LW60}. Were it not for the conservation of energy at vertices of Feynman diagrams, each of the last-mentioned contributions would be identically vanishing \cite{LW60}.

\subsubsection{Remark on a significant aspect of the zero-temperature limit}
\label{ss52s1}

The above-mentioned pair-wise cancelations of contributions takes place on account of the zero-temperature limit $\beta\to\infty$, whereby \emph{sums} with respect to Matsubara frequencies are transformed into \emph{integrals}, according to Eq.~(\ref{e41}), leading to the vanishing of the Matsubara sums of total-derivative functions in this limit (consult point (1) in the following section).

\subsubsection{Remarks on two implicit assumptions}
\label{ss52s2}

The above-mentioned pair-wise cancelations, leading to Eq.~(\ref{e87}), is conditional on the following two properties which are implicit in the considerations by Luttinger and Ward \cite{LW60}:
\begin{itemize}
\item[(1)] the contour $\mathscr{C}(\mu)$ in Eq.~(\ref{e87}) is \emph{closed}, and
\item[(2)] $\t{G}_{\sigma}({\bm k};z)$ times any term contributing to $\t{\Sigma}_{\sigma}^{(\nu)}({\bm k};z)$, $\forall\nu$, is continuously differentiable with respect to $z$ along $\mathscr{C}(\mu)\backslash \mu$.
\end{itemize}

Concerning (1), if $\mathscr{C}(\mu)$ were not closed, each application of integration by parts, referred to above, would leave boundary contributions which in general would add up to a non-vanishing value, in violation of Eq.~(\ref{e87}). Explicitly, let $\mathscr{C}(\mu)$ be parameterised according to $z=\mu+i y$, where $y \uparrow_{-\infty}^{+\infty}$. Now consider the case where one restricts the interval of integration with respect to $y$ to $[-E,E]$, where $E>0$ is a \emph{finite} energy which we assume to be sufficiently large so that for $y\approx \pm E$ the leading-order terms of the asymptotic series expansions of $\t{G}_{\sigma}({\bm k};z)$ and $\t{\Sigma}_{\sigma}^{(\nu)}({\bm k};z)$, $\forall\nu$, corresponding to $\vert z\vert\to\infty$, are sufficiently accurate; one has: $\t{G}_{\sigma}({\bm k};z) \sim \hbar/z$ and $\t{\Sigma}_{\sigma}^{(\nu)}({\bm k};z) \sim {\sf S}_{\sigma}^{(\nu)}({\bm k})/z^{\nu-1}$, where ${\sf S}_{\sigma}^{(\nu)}({\bm k})\not\equiv 0$, as $\vert z\vert\to\infty$ (appendix \ref{sc}, Sec.~\ref{ssc7}). In view of these explicit expressions, it is evident that the above-mentioned boundary contributions are not identically vanishing for $E<\infty$. These expressions further show that for sufficiently large $E$ the significance of these boundary contributions decreases for increasing values of $\nu$.

Concerning (2), two remarks are in order: firstly, the property indicated here is in principle, although not necessarily in practice (see later), fully satisfied on account of the analyticity of $\t{G}_{\sigma}({\bm k};z)$, and of all contributions of which $\t{\Sigma}_{\sigma}^{(\nu)}({\bm k};z)$ consists, everywhere away from the real axis of the $z$ plane, and, secondly, this property is a prerequisite for the permissibility of applying integration by parts to contour integrals over the entire $\mathscr{C}(\mu)\backslash \mu$ (\S~351 in Ref.~\citen{EWH27}; see also Ref.~\citen{Note4}): if $\t{G}_{\sigma}({\bm k};z)$ times at least one of the contributions corresponding to $\t{\Sigma}_{\sigma}^{(\nu)}({\bm k};z)$ were not \emph{continuously differentiable} with respect to $z$ for some $z$ along $\mathscr{C}(\mu)\backslash \mu$, say for $z_j$, $j=1,\dots,m$, then prior to applying integration by parts one would have to express the pertinent integral over $\mathscr{C}(\mu)$ in terms of a superposition of integrals over subintervals of which $z_j$, $j=1,\dots,m$, are boundary points; application of integration by parts to the latter integrals would result in a number of non-vanishing boundary contributions which in general do not add up to zero.

By the same reasoning, in any scheme where the derivatives with respect to $z$ of $\t{G}_{\sigma}({\bm k};z)$ and $\t{\Sigma}_{\sigma}({\bm k};z)$ along $\mathscr{C}(\mu)$ are discontinuous, one should expect violation of the Luttinger-Ward identity (see point (2) above). In Sec.~\ref{ss62} we shall deal with one concrete example, corresponding to a model due to Chubukov and collaborators \cite{CMS96,CM97}, where the observed breakdown of the Luttinger theorem is caused by this very mechanism.

\subsubsection{Observations and remarks}
\label{ss52s3}

Let $\t{G}_{\sigma;0}({\bm k};z)$ denote the Green function corresponding to the $N$-particle GS of a mean-field Hamiltonian $\wh{H}_0$; $\wh{H}_0$ need not be the non-interacting part of the $\wh{H}$ under consideration. One has
\begin{equation}
\t{G}_{\sigma;0}({\bm k};z) = \frac{\hbar}{z - \t{\varepsilon}_{{\bm k};\sigma}}, \label{e89}
\end{equation}
where
\begin{equation}
v_{\sigma}({\bm k}) \equiv
\t{\varepsilon}_{{\bm k};\sigma} -\varepsilon_{\bm k} \label{e90}
\end{equation}
amounts to the mean-field potential. Evidently, $v_{\sigma}({\bm k})\equiv 0$ for $\wh{H}_0$ coinciding with the non-interacting part of $\wh{H}$.

Introducing the short-hand notation
\begin{equation}
\t{\Sigma}_{\sigma;0}^{(\nu)}({\bm k};z) {:=} \t{\Sigma}_{\sigma}^{(\nu)}({\bm k};z;[\{\t{G}_{\sigma';0}\}]), \label{e91}
\end{equation}
where the expression on the RHS indicates that the underlying $\nu$th-order skeleton self-energy diagrams are evaluated in terms of $\{\t{G}_{\sigma';0}({\bm k};z)\,\|\,\sigma'\}$, and
\begin{equation}
\t{\Sigma}_{\sigma;0}({\bm k};z) {:=} \sum_{\nu=1}^{\infty} \t{\Sigma}_{\sigma;0}^{(\nu)}({\bm k};z) -\frac{1}{\hbar} v_{\sigma}({\bm k}), \label{e92}
\end{equation}
one can demonstrate that, for $\mu^{(0)}$ denoting the chemical potential corresponding to the $N$-particle GS of $\wh{H}_0$, one has (cf. Eqs.~(\ref{e87}) and (\ref{e44}))
\begin{equation}
\sum_{\bm k} \int_{\mathscr{C}(\mu^{(0)})} \frac{{\rm d}z}{2\pi i}\;
\t{G}_{\sigma;0}({\bm k};z) \frac{\partial}{\partial z} \t{\Sigma}_{\sigma;0}^{(\nu)}({\bm k};z) = 0,\;\; \forall\nu\ge 1, \label{e93}
\end{equation}
\begin{equation}
\sum_{\bm k} \int_{\mathscr{C}(\mu^{(0)})} \frac{{\rm d}z}{2\pi i}\;
\t{G}_{\sigma;0}({\bm k};z) \frac{\partial}{\partial z} \t{\Sigma}_{\sigma;0}({\bm k};z) = 0. \label{e94}
\end{equation}
We should emphasise that the expression in Eq.~(\ref{e92}) is formal, in that the series on the RHS of this equation may not converge; in other words, $\t{\Sigma}_{\sigma;0}({\bm k};z)$ may not exist. The reason for this fact will be clarified in Sec.~\ref{ss53s2} (see also the simple example in Sec.~\ref{s7}). Since, however, the functions $\{\t{\Sigma}_{\sigma;0}^{(\nu)}({\bm k};z)\}$ correspond to skeleton diagrams (see Sec.~\ref{ss53s1}), truncation of the sum on the RHS of Eq.~(\ref{e92}) to a finite sum leads to a function which is always bounded almost everywhere.

The proof of Eq.~(\ref{e93}) consists of the observation, discussed earlier, that the expression in Eq.~(\ref{e75}) remains valid on replacing \emph{all} $\{\mathscr{G}_{\sigma'}({\bm k};z)\}$ (i.e. both the explicit and the implicit $\{\mathscr{G}_{\sigma'}({\bm k};z)\}$) in this expression by a different suitable function. The proof of Eq.~(\ref{e94}) coincides with that of the Luttinger-Ward identity, Eq.~(\ref{e44}), on the basis of the result in Eq.~(\ref{e87}). The identity in Eq.~(\ref{e94}) should not be confused with the Luttinger-Ward identity corresponding to $\t{G}_{\sigma;0}({\bm k};z)$ whose associated self-energy is equal to $\hbar^{-1} v_{\sigma}({\bm k})$; this self-energy being independent of $z$, the Luttinger-Ward identity corresponding to $\t{G}_{\sigma;0}({\bm k};z)$ is trivially satisfied.

Note that on account of Eqs.~(\ref{e1}) and (\ref{e2}) and by the definition of $\mu^{(0)}$, one has
\begin{equation}
\sum_{\bm k} \Theta\big(G_{\sigma;0}^{-1}({\bm k};\mu^{(0)})\big) \equiv \sum_{\bm k} \Theta\big(G_{\sigma;0}({\bm k};\mu^{(0)})\big) = N_{\sigma}. \label{e95}
\end{equation}
With reference to Eq.~(\ref{e21}), one observes that the Luttinger number corresponding to $\t{G}_{\sigma;0}({\bm k};z)$ is equal to the number of spin-$\sigma$ articles in the $N$-particle GS of $\wh{H}_0$. This result, which is in agreement with the Luttinger theorem, Eq.~(\ref{e22}), also conforms with the validity of the Luttinger-Ward identity corresponding to $\t{G}_{\sigma;0}({\bm k};z)$, referred to above.

With
\begin{equation}
\t{G}_{\sigma}'({\bm k};z) {:=} \big(1-\t{G}_{\sigma;0}({\bm k};z) \t{\Sigma}_{\sigma;0}({\bm k};z)\big)^{-1} \t{G}_{\sigma;0}({\bm k};z), \label{e96}
\end{equation}
which amounts to the \emph{non-selfconsistent} Green function corresponding to $\t{\Sigma}_{\sigma;0}({\bm k};z)$ (assuming that this function exists) through the Dyson equation, in general
\begin{equation}
\sum_{\bm k} \int_{\mathscr{C}(\mu^{(0)})} \frac{{\rm d}z}{2\pi i}\;
\t{G}_{\sigma}'({\bm k};z) \frac{\partial}{\partial z} \t{\Sigma}_{\sigma;0}^{(\nu)}({\bm k};z) \not= 0,\;\; \nu\ge 2, \label{e97}
\end{equation}
\begin{equation}
\sum_{\bm k} \int_{\mathscr{C}(\mu^{(0)})} \frac{{\rm d}z}{2\pi i}\;
\t{G}_{\sigma}'({\bm k};z) \frac{\partial}{\partial z} \t{\Sigma}_{\sigma;0}({\bm k};z) \not= 0. \label{e98}
\end{equation}
The reason underlying these results lies in the fact that in general the expression in Eq.~(\ref{e75}) fails to hold by replacing the \emph{explicit} $\mathscr{G}_{\sigma}({\bm k};\zeta_m)$ in this expression by a function which is different from that on which $\mathscr{S}_{\!\sigma}^{(\nu)}({\bm k};\zeta_m)$ implicitly depends. In the present case, the function $\t{G}_{\sigma}'({\bm k};z)$ is different from the function $\t{G}_{\sigma;0}({\bm k};z)$, $\forall\sigma$, in terms of which $\t{\Sigma}_{\sigma;0}^{(\nu)}({\bm k};z)$ and $\t{\Sigma}_{\sigma;0}({\bm k};z)$ are evaluated.

\subsubsection{Summary}
\label{ss52s4}

The third instance of using perturbation theory in the proof of the Luttinger theorem has direct bearing on the proof of the Luttinger-Ward identity, Eq.~(\ref{e44}). Assuming that the $\sum_{\nu=1}^{\infty}$ on the RHS of Eq.~(\ref{e86}) can be rightfully transposed to the position between $\partial/\partial z$ and $\t{\Sigma}_{\sigma}^{(\nu)}({\bm k};z)$, under the conditions (1) and (2), specified in Sec.~\ref{ss52s2}, the Luttinger-Ward identity immediately follows. It is therefore necessary to investigate the admissability in Eq.~(\ref{e86}) of
\begin{itemize}
\item[(i)] exchanging the order of $\sum_{\nu=1}^{\infty}$ with that of $\frac{\partial}{\partial z}$,
\item[(ii)] exchanging the order of $\sum_{\nu=1}^{\infty}$ with that of $\int_{\mathscr{C}(\mu)} {\rm d}z\;$, and
\item[(iii)] exchanging the order of $\sum_{\nu=1}^{\infty}$ with that of $\sum_{\bm k}$,
\end{itemize}
where the latter sum is transformed into an integral in the thermodynamic limit. Below we shall carry out this investigation. \emph{For the cases where these re-orderings are legitimate, and provided that $\mu=\mu_{\infty}$, the Luttinger-Ward identity, Eq.~(\ref{e44}), is necessarily exact.}

\subsection{Detailed considerations}
\label{ss53}

In this section we investigate wether the three operations (i), (ii) and (iii), indicated in the closing part of the previous section, are permissable. To this end we first establish that the series in Eq.~(\ref{e85}) is convergent. Subsequently, we show that this convergence is in fact a \emph{uniform} one (Ch. III in Ref.~\citen{WW62}, Ch. VII in Ref.~\citen{TB65}) for almost all ${\bm k}$ and $z$. Equipped with this knowledge, we proceed with investigating the admissibility of the above-mentioned steps (i), (ii) and (iii). Both here and later, ${\bm k}$ can be anywhere in the relevant ${\bm k}$ space. In what follows we assume that (a) the bare two-body interaction potential in the system under consideration is short range, and that (b) either the underlying ${\bm k}$ space is bounded or $\wh{H}$ is defined on a lattice.

\subsubsection{General remarks concerning skeleton self-energy diagrams}
\label{ss53s1}

Skeleton self-energy diagrams have at least two special properties which have not been widely discussed in the literature, although Luttinger \cite{JML61} has called attention to these and fruitfully made use of them in his considerations regarding analytic properties of the self-energy $\t{\Sigma}_{\sigma}({\bm k};z)$. These properties are readily uncovered \cite{JML61} by expressing each conventional skeleton self-energy diagram in terms of \emph{time-ordered} skeleton diagrams (\S~3.2 in Ref.~\citen{NO98}), such as encountered in the framework of the time-dependent many-body perturbation theory due to Goldstone \cite{JG57}. Here we review one of the above-mentioned two properties which is of direct relevance to our considerations in this paper.

A (skeleton) self-energy diagram can be expressed in terms of a finite number of time-ordered (skeleton) diagrams \cite{JML61}. On replacing $\{ \t{G}_{\sigma'}({\bm k};z)\,\|\, \sigma'\}$, represented by solid lines in these diagrams, by the single-particle Green functions $\{ \t{G}_{\sigma';0}({\bm k};z)\,\|\, \sigma'\}$ corresponding to a \emph{mean-field} Hamiltonian (not necessarily the non-interacting part $\wh{H}_0$ of the full Hamiltonian $\wh{H}$; see Sec.~\ref{ss52s3}), the integrals with respect to the internal energy variables encountered in the analytic expressions corresponding to the latter diagrams can be evaluated explicitly \cite{JML61}. The resulting expressions reveal the analytic properties of the corresponding $\t{\Sigma}_{\sigma}^{(\nu)}({\bm k};z)$, $\forall\nu$, under the hypothetical condition that $\t{G}_{\sigma}({\bm k};z) \equiv \t{G}_{\sigma;0}({\bm k};z)$, $\forall\sigma$. In Sec.~\ref{ss52s3} we denoted this $\t{\Sigma}_{\sigma}^{(\nu)}({\bm k};z)$ by $\t{\Sigma}_{\sigma;0}^{(\nu)}({\bm k};z)$.

Although the condition $\t{G}_{\sigma}({\bm k};z) \equiv \t{G}_{\sigma;0}({\bm k};z)$, $\forall\sigma$, is admittedly restrictive, it is significant that on employing the spectral representation of $\t{G}_{\sigma}({\bm k};z)$ one obtains analytic expressions for time-ordered (skeleton) self-energy diagrams that closely resemble the latter approximate expressions specific to $\t{G}_{\sigma}({\bm k};z) \equiv \t{G}_{\sigma;0}({\bm k};z)$, $\forall\sigma$ \cite{JML61}. In particular, in the cases of Fermi-liquid metallic states these expressions can be considerably simplified, enabling one to deduce the asymptotic series expansion of $\t{\Sigma}_{\sigma}^{(\nu)}({\bm k};z)$, $\forall\nu$, for $z\to \mu$ \cite{JML61}. The same can be achieved, albeit more elaborately, by assuming the underlying GSs to be a marginal Fermi-liquids or Luttinger liquids.

The specific property of skeleton diagrams that is of direct interest to our investigations in this paper, consists of the fact that the \emph{inner kernel} (see later) of the contribution corresponding to an arbitrary skeleton self-energy diagram has \emph{simple poles} (\S~5.61 in Ref.~\citen{WW62}) along the real axis of the $z$ plane when contributions of skeleton diagrams are determined in terms of $\{ \t{G}_{\sigma';0}({\bm k};z)\,\|\, \sigma'\}$ (see above); in the terminology of Luttinger, one has ``\emph{no repeated denominators}'' \cite{JML61}. This property follows from the fact that by definition a skeleton self-energy diagram does not contain a self-energy part capable of being excised through cutting two internal solid lines representing Green functions \cite{JML61}.

As for what we have called `\emph{inner kernel}', on evaluating the integrals with respect to the internal energy variables encountered in the analytic expression corresponding to a time-ordered skeleton diagram associated with $\t{\Sigma}_{\sigma}^{(\nu)}({\bm k};z)$, one can frame the resulting expression in a canonical form, consisting of a repeated sum $\sum_{{\bm k}_1} \sum_{{\bm k}_2}\dots$, over the relevant internal wavevectors, of an `inner kernel'. Thus the `inner kernel' corresponding to $\t{\Sigma}_{\sigma}^{(\nu)}({\bm k};z)$ consists of a superposition of a finite number of functions, each of which depends on both the external variables ${\bm k}$, $z$, and the internal wavevectors ${\bm k}_1$, ${\bm k}_2$, \dots~. For the reason indicated above, each of these constituent functions has simples poles along the real axis of the $z$ plane.

For finite systems,\footnote{As we have indicated earlier, since we are dealing with \emph{uniform} GSs, these finite systems must be defined on finite lattices without boundary.} where a wave-vector sum $\sum_{{\bm k}_i}$ accounts for a finite number of terms, there is no conceptual advantage to be had in making a distinction between the analytic properties of a contribution corresponding to a skeleton diagram and those of its corresponding `inner kernel'; thus for exactly the same reason as given above, the contribution of a skeleton self-energy diagram, evaluated in terms of $\{\t{G}_{\sigma';0}({\bm k};z)\}$, corresponding to a finite system is a function which possesses simple poles along the real axis of the $z$ plane.

For macroscopic systems, on the other hand, where wave-vector sums are transformed into wave-vector integrals (see Eq.~(\ref{e147}) below), the simple poles of an `inner kernel' corresponding to $\t{\Sigma}_{\sigma}^{(\nu)}({\bm k};z)$ do not result in contributions to $\t{\Sigma}_{\sigma}^{(\nu)}({\bm k};z)$ expressible in terms of functions with simple poles. To appreciate this aspect, consider the function $f(k;z) \equiv 1/(k-z)$ which has a simple pole at $z=k$. On the other hand
\begin{equation}
g(z) \equiv \int_{k_0}^{k_1} {\rm d}k\; f(k;z) = \ln(k_1-z) -
\ln(k_0-z) \label{e99}
\end{equation}
is logarithmically divergent at $z=k_1$ and $z=k_0$; at these points, $g(z)$ is \emph{essentially singular }(\S\S~5.61 - 5.7 in Ref.~\citen{WW62}) and thus not expressible in terms of a function
that has poles, of arbitrary multiplicity, at $k=k_0, k_1$.

The above discussions reveal that on evaluating skeleton self-energy diagrams in terms of $\{ \t{G}_{\sigma';0}({\bm k};z)\,\|\, \sigma'\}$, $\t{\Sigma}_{\sigma}^{(\nu)}({\bm k};z)$, $\forall\nu$, can be undefined only on account of two possibilities, namely
\begin{itemize}
\item[(1)] of undefined matrix elements of the two-body
interaction potential, and/or
\item[(2)] of undefined repeated sums of the form $\sum_{{\bm k}_1}
\sum_{{\bm k}_2}\dots$ over internal wavevectors.
\end{itemize}
The possibility (1) is ruled out for short-range two-body potentials
and the ${\bm k}$ spaces corresponding to discrete lattices for which the inverse of the shortest distance between lattice points is finite. As for the possibility (2), this cannot be realised for finite systems, where $\sum_{{\bm k}_i}$ amounts to a sum over a finite number of terms; regarding macroscopic systems, this prospect is fundamentally excluded for lattice models.

Although the above observations correspond to the cases where skeleton self-energy diagrams are evaluated in terms of $\{\t{G}_{\sigma';0}({\bm k};z)\,\|\, \sigma'\}$, use of $\{ \t{G}_{\sigma'}({\bm k};z)\,\|\, \sigma'\}$ cannot place existence of $\t{\Sigma}_{\sigma}^{(\nu)}({\bm k};z)$, $\forall\nu$, in jeopardy. This follows from the fact that singularities of $\t{G}_{\sigma}({\bm k};z)$, $\forall\sigma$, along the real axis of the $z$ plane, cannot be stronger than simple poles, for interaction gives rise to transfer to a continuum of energies of the spectral weight associated with the simple pole of $\t{G}_{\sigma;0}({\bm k};z)$ at $z=\t{\varepsilon}_{{\bm k};\sigma}$, Eq.~(\ref{e89}). In this connection, one should recall the sum rule $\hbar^{-1} \int_{-\infty}^{\infty} {\rm d}\varepsilon\; A_{\sigma;\lambda}({\bm k};\varepsilon) = 1$ (appendix \ref{sc}), which applies for all $\lambda \in [0,1]$.

\subsubsection{On the convergence of $\sum_{\nu=1}^{\infty}\t{\Sigma}_{\sigma}^{(\nu)}({\bf k};z)$}
\label{ss53s2}

In order to establish the convergence of the series in Eq.~(\ref{e85}), we first express this series in a more explicit form; this form should reflect the fact that, in contrast to ordinary series of the form $\sum_{\nu=1}^{\infty} a_{\nu}$ in terms of a \emph{predetermined} sequence $\{ a_{\nu}\}$, the series in Eq.~(\ref{e85}) is expressed in terms of $\{\t{\Sigma}_{\sigma}^{(\nu)}({\bm k};z)\,\|\, \nu\}$, where $\t{\Sigma}_{\sigma}^{(\nu)}({\bm k};z)$ is a functional of $\{\t{\Sigma}_{\sigma'}({\bm k};z)\,\|\,\sigma'\}$. The functional dependence of $\t{\Sigma}_{\sigma}^{(\nu)}({\bm k};z)$, $\nu=1,2,\dots$, on $\{\t{\Sigma}_{\sigma'}({\bm k};z)\,\|\,\sigma'\}$ follows from the fact that $\t{\Sigma}_{\sigma}^{(\nu)}({\bm k};z)$ is determined in terms of the exact single-particle Green functions $\{\t{G}_{\sigma'}({\bm k};z)\,\|\, \sigma'\}$ where $\t{G}_{\sigma'}({\bm k};z)$ is, through the Dyson equation, an explicit function of $\t{\Sigma}_{\sigma'}({\bm k};z)$. We make these facts explicit by the notation
\begin{equation}
\t{\Sigma}_{\sigma}^{(\nu)}({\bm k};z) \equiv
\t{\Sigma}_{\sigma}^{(\nu)}({\bm k};z;[\{\t{\Sigma}_{\sigma'}\}]).
\label{e100}
\end{equation}
Since $\t{\Sigma}_{\sigma}^{(\nu)}({\bm k};z;[\{\t{\Sigma}_{\sigma'}\}])$ corresponds to the combined contributions of all $\nu$th-order skeleton diagrams \cite{LW60}, the functional form of $\t{\Sigma}_{\sigma}^{(\nu)}({\bm k};z;[\{\t{\Sigma}_{\sigma'}\}])$ is fully specified.

Following Eq.~(\ref{e100}), for a given integer $m$ we define
$\t{\Sigma}_{\sigma}^{[m]}({\bm k};z)$ as the \emph{solution} of the
equation
\begin{equation}
\t{\Sigma}_{\sigma}^{[m]}({\bm k};z) = \sum_{\nu=1}^{I(m)}
\t{\Sigma}_{\sigma}^{(\nu)}({\bm
k};z;[\{\t{\Sigma}_{\sigma'}^{[m]}\}]),\;\;\; \forall\sigma,
\label{e101}
\end{equation}
where $I(x)$ is a monotonically increasing functions of $x$ which takes integer values for integer values of $x$; one may, for instance, have $I(x) \equiv \kappa\,x$, where $\kappa\in \mathds{N}$. Although it would seem natural to identify $I(x)$ with $x$, we have avoided this choice so as to circumvent the possibility that with $I(x)\equiv x$ Eq.~(\ref{e101}) may have no solution for some specific values of $m$; by leaving $I(x)$ only minimally specified, we achieve that formally all elements of the infinite sequence of the \emph{self-consistent} partial sums $\{\t{\Sigma}_{\sigma}^{[m]}({\bm k};z) \,\|\, m\in \mathds{N}\}$ exist. We should emphasise that $\t{\Sigma}_{\sigma}^{[m]}({\bm k};z)$, $\forall m$, is supposed to be analytic everywhere on the complex $z$ plane away from the real axis and further that for $\vert z\vert\to\infty$ it decays in the same way as expected from the exact $\t{\Sigma}_{\sigma}({\bm k};z)$ (appendix \ref{sc}). These two conditions are prerequisite of the function space in which Eq.~(\ref{e101}) is solved.

With the sequence $\{ \t{\Sigma}_{\sigma}^{[m]}({\bm k};z) \,\|\, m\}$ determined according to the above prescription, from the expression in Eq.~(\ref{e85}) one deduces that for the exact self-energy $\t{\Sigma}_{\sigma}({\bm k};z)$ one must have
\begin{equation}
\t{\Sigma}_{\sigma}({\bm k};z) = \lim_{m\to\infty}
\t{\Sigma}_{\sigma}^{[m]}({\bm k};z). \label{e102}
\end{equation}
Equations (\ref{e101}) and (\ref{e102}) furnish the abstract notation in Eq.~(\ref{e85}) with an operational meaning; in contrast to $\t{\Sigma}_{\sigma}^{(\nu)}({\bm k};z;[\{\t{\Sigma}_{\sigma'}\}])$, which, unless one already knows the exact $\t{\Sigma}_{\sigma}({\bm k};z)$, cannot be calculated, not even for $\nu=1$, $\t{\Sigma}_{\sigma}^{(\nu)}({\bm k};z;[\{\t{\Sigma}_{\sigma'}^{[m]}\}])$ can in principle be calculated for an arbitrary $\nu$. It should be noted that all elements of the sequence $\{ \t{\Sigma}_{\sigma}^{[m]}({\bm k};z)\,\|\, m\}$ correspond to the same (and the actual) two-body interaction potential; in generating this sequence, no explicit reference is made to some non-interacting problem.

Concerning the expression in Eq.~(\ref{e102}), according to which
$\t{\Sigma}_{\sigma}({\bm k};z)$ is the limit of the infinite
sequence $\{\t{\Sigma}_{\sigma}^{[1]}({\bm k};z),
\t{\Sigma}_{\sigma}^{[2]}({\bm k};z),\dots\}$, the following remarks
are in order:
\begin{itemize}
\item[(a)] For any set of parameters $({\bm k},z,\sigma)$ for which $\t{\Sigma}_{\sigma}({\bm k};z)$ is a bounded function, $\t{\Sigma}_{\sigma}^{[m]}({\bm k};z)$ \emph{cannot} diverge as $m\to\infty$. This is a direct consequence of the fact that, for any value of $m$, $\t{\Sigma}_{\sigma}^{[m]}({\bm k};z)$ is presupposed to be a well-defined \emph{solution} of Eq.~(\ref{e101}); dividing both sides of Eq.~(\ref{e101}) by this solution, the resulting sum is identically equal to unity for any arbitrary value of $m$, no matter how large $m$ may be (see Sec.~\ref{s7} for a simple example illustrating this case).
\item[(b)] From the purely mathematical point of view, it is \emph{a priori} not evident that boundedness of $\t{\Sigma}_{\sigma}^{[m]}({\bm k};z)$ for $m\to\infty$ should necessarily preclude the possibility of $\t{\Sigma}_{\sigma}^{[m]}({\bm k};z)$ \emph{oscillating} between two finite \emph{limits of indeterminacy} for increasing values of $m$ (\S~331 in Ref.~\citen{EWH07}; see Sec.~\ref{ss53s3} for a brief introduction of the relevant notions). Since in Sec.~\ref{ss53s5} we shall consider the problem of \emph{oscillation} of series in some detail, we suffice here to mention that in the case at hand this possibility is ruled out by the physical fact that $\mathrm{Im}[\t{\Sigma}_{\sigma}^{(\nu)}({\bm k};\varepsilon-i 0^+;[\{\t{\Sigma}_{\sigma'}^{[m]}\}])] \ge 0$, $\forall\varepsilon \in \mathds{R}$, $\forall\nu$. It follows that the infinite sequence $\{ \t{\Sigma}_{\sigma}^{[m]}({\bm k};z)\,\|\, m\}$ indeed \emph{converges} to a finite limit.
\item[(c)] In spite of the convergence of $\{\t{\Sigma}_{\sigma}^{[m]}({\bm k};z) \,\|\, m\}$ to a bounded limit, which limit we denote by $\t{\Sigma}_{\sigma}({\bm k};z)$, the limit function $\t{\Sigma}_{\sigma}({\bm k};z)$ need not be unique: the nonlinearity of Eq.~(\ref{e101}) implies that the limit function $\t{\Sigma}_{\sigma}({\bm k};z)$ may be one of a multiplicity of solutions, corresponding to different uniform states (not necessarily GSs) of the system that are compatible with the symmetry of the underlying ${\bm k}$ space (such as the $\mathrm{1BZ}$ for the cases where the model under consideration is defined on a Bravais lattice).\footnote{Recall that lattice coordinates are not dynamical variables in our considerations so that lattice distortions, and thus distortions of the corresponding $\mathrm{1BZ}$, are \emph{a priori} ruled out.} This potentiality should not be misconstrued as an indication that $\t{\Sigma}_{\sigma}^{(1)} + \t{\Sigma}_{\sigma}^{(2)} + \dots$ were an \emph{oscillating} series (in the sense referred to above); the requirement of self-consistency, enforced by Eq.~(\ref{e101}), implies that two distinct limit functions in fact are limits of two distinct infinite series corresponding to two different sequences. Assuming that the true GS of the system under consideration (conform the prescribed symmetry of the underlying ${\bm k}$ space) is non-degenerate, only one of the possible several limit functions can be the sought-after self-energy corresponding to the GS.
\end{itemize}
Hereby is the proof of convergence of the series in Eq.~(\ref{e85}) completed. We point out that for $\mu=\mu_{\infty}^{[m]}$ the Luttinger-Ward identity can be shown to be exactly satisfied by $\t{\Sigma}_{\sigma}^{[m]}({\bm k};z)$ and the associated single-particle Green function $\t{G}_{\sigma}^{[m]}({\bm k};z)$. The details underlying the proof of this statement are similar to those discussed in Sections \ref{ss52} and \ref{ss52s3}; here $\mu_{\infty}^{[m]}$ denotes the counterpart of $\mu_{\infty}$ corresponding to $m$. The validity of the Luttinger-Ward identity for $\t{\Sigma}_{\sigma}^{[m]}({\bm k};z)$ and $\t{G}_{\sigma}^{[m]}({\bm k};z)$ is relevant in that it shows that from the perspective of this identity, any of the possible multiplicity of solutions referred to above (item (c)) is equally acceptable; validity of the Luttinger-Ward identity is only a necessary condition to be satisfied by the exact $\t{G}_{\sigma}({\bm k};z)$ and the corresponding $\t{\Sigma}_{\sigma}({\bm k};z)$.

\subsubsection{Some mathematical notions concerning oscillating series}
\label{ss53s3}

Let $u_1 + u_2 + u_3+\dots$ be an \emph{oscillating} series of the real sequence $\{ u_1, u_2,\dots\}$; for $\{ u_1, u_2,\dots\}$ a complex sequence, one proceeds analogously, thorough separately considering the two real sequences $\{ u_1', u_2', \dots\}$ and $\{u_1'', u_2'',\dots\}$, where $u_j' + i u_j'' \equiv u_j$. According to the general theory of infinite series (see in particular \S~331 in Ref.~\citen{EWH07} and \S\S~52-54 in Ref.~\citen{EWH27}), the partial sums $s_1=u_1 $, $s_2=u_1+u_2$, $\dots$ of the oscillating series $u_1 + u_2 + u_3 + \dots$ constitute a set $S$ whose subset of limiting points, called \emph{derived} (or \emph{derivative}) set and denoted by $S'$, may consist of either a finite (at least two) or an infinite number of elements. This set being closed, it contains upper and lower boundaries, $U$ and $L$ respectively, referred to as \emph{limits of indeterminacy}; the set $S'$ may consist of the entire closed interval $[L,U]$, or it may be a non-dense subset of $[L,U]$. Different infinite subsets of the infinite set $\{s_1,s_2,\dots\}$, of the form $\{ s_{n_1}, s_{n_2},\dots\}$, where $n_1 <n_2 <\dots$, converge to different points of $S'$. Evidently, the \emph{derived set} corresponding to the set of partial sums of a non-oscillating convergent series consists solely of a single point, which is formally equivalent to the case where the two points $L$ and $U$ of $S'$ coincide.

\subsubsection{On the uniformity of convergence of
$\sum_{\nu=1}^{\infty}\t{\Sigma}_{\sigma}^{(\nu)}({\bf k};z)$ for
almost all ${\bf k}$ and $z$}
\label{ss53s4}

Having established convergence of the series in Eq.~(\ref{e85}), in this section we demonstrate that this convergence is \emph{uniform} for almost all ${\bm k}$ and $z$ (Ch. III in Ref.~\citen{WW62}, Ch. VII in\footnote{See in particular \S~49.1 in Ref.~\citen{TB65}, \emph{Historical Note on Uniform Convergence}.} Ref.~\citen{TB65}, \S\S~67-76 in Ref.~\citen{EWH26}). By `almost all' we mean what is conventionally meant, namely that the series in Eq.~(\ref{e85}) may not converge uniformly for sets of ${\bm k}$ and $z$ that are of measure zero in their respective embedding sets. We subdivide our demonstration into \emph{four} elementary steps.

\subsubsection{First step}
\label{ss53s5}

For reasons that will become evident, we first consider the convergence property of the series in Eq.~(\ref{e85}) for the specific case where $z=\varepsilon-i 0^+$ in which $\varepsilon\in \mathds{R}$; the choice $z=\varepsilon + i 0^+$ would be equally appropriate. We thus consider the series
\begin{equation}
\t{\Sigma}_{\sigma}({\bm k};\varepsilon-i 0^+) =
\sum_{\nu=1}^{\infty} \t{\Sigma}_{\sigma}^{(\nu)}({\bm
k};\varepsilon-i 0^+),\;\;\;\varepsilon\in \mathds{R}. \label{e103}
\end{equation}
Following the considerations in the previous section, this series is convergent for all ${\bm k}$ and $\varepsilon$ at which $\t{\Sigma}_{\sigma}({\bm k};\varepsilon-i 0^+)$ is bounded. Since $\t{\Sigma}_{\sigma}({\bm k};z)$ is analytic for all $z$ away from the real axis (appendix \ref{sc}), it follows that $\t{\Sigma}_{\sigma}({\bm k};\varepsilon-i \eta)$ is bounded for all $\varepsilon, \eta \in \mathds{R}$ and $\eta\not=0$; consequently, over the set of $\varepsilon$ for which $\t{\Sigma}_{\sigma}({\bm k};\varepsilon-i 0^+)$ (which denotes the \emph{limit} of $\t{\Sigma}_{\sigma}({\bm k};\varepsilon-i \eta)$ for $\eta\downarrow 0$) is unbounded, this unboundedness is removed on displacing $\varepsilon$ by a finite amount (no matter how small) into the lower half of the complex plane.

For completeness, although $\t{\Sigma}_{\sigma}({\bm k};\varepsilon-i 0^+)$ need not be unbounded for any ${\bm k}$ and $\varepsilon$, for the subset of $\mathcal{S}_{\Sc l;\sigma}$ over which $G_{\sigma}({\bm k};\mu) = 0$ (Sec.~\ref{ss24}), $\t{\Sigma}_{\sigma}({\bm k};z)$ is necessarily unbounded at $z=\mu$; evidently, even for $\mathcal{S}_{\Sc l;\sigma} \not= \emptyset$, the last-mentioned subset may be empty (Sec.~\ref{ss24}). Be it as it may, since in $d$ dimensions $\mathcal{S}_{\Sc l;\sigma}$ is a $(d-1)$-dimensional subset of the underlying ${\bm k}$ space, it follows that if the series in Eq.~(\ref{e103}) diverges, it does so over a subset of measure zero of the relevant ${\bm k}$ space, whatever the value of $\varepsilon$ may be.

We investigate the convergence property of the series in Eq.~(\ref{e103}) by separately considering the related series (note that $\mathrm{Im}[\t{\Sigma}_{\sigma}^{(1)}({\bm k};z)]\equiv 0$, $\forall {\bm k}, z$)
\begin{equation}
\mathrm{Im}[\t{\Sigma}_{\sigma}({\bm k};\varepsilon-i 0^+)] =
\sum_{\nu=2}^{\infty} \mathrm{Im}[\t{\Sigma}_{\sigma}^{(\nu)}({\bm
k};\varepsilon-i 0^+)],\;\;\; \varepsilon\in \mathds{R}. \label{e104}
\end{equation}
and
\begin{equation}
\mathrm{Re}[\t{\Sigma}_{\sigma}({\bm k};\varepsilon-i 0^+)]
=\sum_{\nu=1}^{\infty} \mathrm{Re}[\t{\Sigma}_{\sigma}^{(\nu)}({\bm
k};\varepsilon-i 0^+)]. \label{e105}
\end{equation}
It is relevant to point out that in general it is not possible to express a convergent \emph{infinite} series, such as $\sum_{\nu=1}^{\infty} (a_{\nu} + b_{\nu})$, as $\sum_{\nu=1}^{\infty} a_{\nu} + \sum_{\nu=1}^{\infty} b_{\nu}$, for in spite of the convergence of the first series, the latter two series may not be convergent. This is however not the case if $a_{\nu} \equiv \mathrm{Re}[f_{\nu}]$ and $b_{\nu} \equiv i\mathrm{Im}[f_{\nu}]$, where $\{ f_{\nu} \}$ is a complex sequence; evidently, $\sum_{\nu=1}^{\infty} f_{\nu}$ cannot be convergent if at least one of $\sum_{\nu=1}^{\infty} \mathrm{Re}[f_{\nu}]$ and $\sum_{\nu=1}^{\infty} \mathrm{Im}[f_{\nu}]$ is not convergent (\S~75 in Ref.~\citen{TB65}). It follows that the series in Eqs.~(\ref{e104}) and (\ref{e105}) are both convergent for almost all ${\bm k}$ and $\varepsilon$.

The assumed stability of the GS under consideration implies not only
that (see Eqs.~(\ref{e16}) and (\ref{e5}), as well as
Eqs.~(\ref{ec54}) and (\ref{ec58}))
\begin{equation}
\mathrm{Im}[\t{\Sigma}_{\sigma}({\bm k};\varepsilon-i 0^+)] \ge
0,\;\;\; \forall\varepsilon\in \mathds{R}, \label{e106}
\end{equation}
but also that
\begin{equation}
\mathrm{Im}[\t{\Sigma}_{\sigma}^{(\nu)}({\bm k};\varepsilon-i 0^+)]
\ge 0,\;\;\; \forall\varepsilon\in \mathds{R},\; \forall\nu.
\label{e107}
\end{equation}
The latter result can be understood on physical grounds:\cite{Note5} $\mathrm{Im}[\t{\Sigma}_{\sigma}({\bm k};\varepsilon-i0^+)]$, which amounts to a measure for decay of elementary excitations characterised by ${\bm k}$, $\varepsilon$ and $\sigma$, consists of the cumulative contributions of decay processes brought about by all possible modes of particle-particle interaction (`scattering events'); $\mathrm{Im}[\t{\Sigma}_{\sigma}^{(\nu)}({\bm k};\varepsilon-i 0^+)]<0$, for a specific $\nu$, would imply that $\nu$th-order scattering events collectively countered the above-mentioned decay, a regeneration phenomenon whose emergence would signal instability of the underlying GS. By the same reasoning, one concludes that the inequality in Eq.~(\ref{e107}) applies more generally to all $\mathrm{Im}[\t{\Sigma}_{\sigma}^{(\nu;j)}({\bm k};\varepsilon-i 0^+)]$, $j=1,\dots, N^{(\nu)}$, where $\t{\Sigma}_{\sigma}^{(\nu;j)}({\bm k};z)$ is the contribution of the $j$th self-energy diagram in the set of connected skeleton diagrams representing $\t{\Sigma}_{\sigma}^{(\nu)}({\bm k};z)$. We shall have no use for this more strong result in this paper.

The result in Eq.~(\ref{e107}) is of a far-reaching mathematical consequence, as it establishes the series on the RHS of Eq.~(\ref{e104}) as one consisting of non-negative terms. In consequence of this and of the \emph{continuity} of both $\mathrm{Im}[\t{\Sigma}_{\sigma}({\bm k};\varepsilon-i 0^+)]$ and $\mathrm{Im}[\t{\Sigma}_{\sigma}^{(\nu)}({\bm k};\varepsilon-i 0^+)]$ for almost all ${\bm k}$ and $\varepsilon$ (see Sec.~\ref{ss53s6}), convergence of this series, established in Sec.~\ref{ss53s2}, implies, by Dini's theorem, its \emph{uniform} convergence for almost all ${\bm k}$ and $\varepsilon$ (\S~347 in Ref.~\citen{EWH07}, \S~79 in Ref.~\citen{EWH26}, \S~49.2 in Ref.~\citen{TB65}). In this connection, we note that although the sum-function corresponding to a non-uniformly convergent series of continuous functions need not be discontinuous, such discontinuity is fundamentally ruled out in the domain where the latter series is uniformly convergent (\S~45 in Ref.~\citen{TB65}, \S~344 in Ref.~\citen{EWH07}). We point out that series corresponding to sequences of non-negative (or non-positive) terms have the special property that their corresponding sequence of partial sums are monotonic (\S~102 in Ref.~\citen{EWH26}). Consequently, these series either converge or diverge, but never oscillate (\S~7 in Ref.~\citen{TB65}). This aspect is of utmost significance in the context of our present considerations, for, in view of Eq.~(\ref{e107}), nowhere where $\mathrm{Im}[\t{\Sigma}_{\sigma}({\bm k};\varepsilon-i0^+)]$ is bounded, can the sum in Eq.~(\ref{e104}) diverge, a fact that is implicit in the convergence of the sequence $\{ \t{\Sigma}_{\sigma}^{[m]}({\bm k};z)\,\|\, m\}$ towards $\t{\Sigma}_{\sigma}({\bm k};z)$, discussed in Sec.~\ref{ss53s2}.

To summarise, we have demonstrated that the series in Eq.~(\ref{e104}) is \emph{uniformly} convergent for almost all ${\bm k}$ and $\varepsilon \in \mathds{R}$.

\subsubsection{Remarks}
\label{ss53s6}

The proposition that $\mathrm{Im}[\t{\Sigma}_{\sigma}({\bm k};\varepsilon-i 0^+)]$ and $\mathrm{Im}[\t{\Sigma}_{\sigma}^{(\nu)}({\bm k};\varepsilon-i 0^+)]$ are continuous for almost all ${\bm k}$ and $\varepsilon$, may be accepted as being self-evidently true. Otherwise, one should recall that continuity of these functions for almost all ${\bm k}$ and $\varepsilon$ is in fact prerequisite for these functions to correspond to functions in the space-time domain by Fourier transformation. To clarify this statement, let us first consider the relationship between the time-dependent function $f(t)$ and its Fourier transform $F(\varepsilon)$ \cite{Note6}. The correspondence between $f(t)$ and $F(\varepsilon)$, one being the Fourier transform of the other, is based on the fundamental requirement (\S~481 in Ref.~\citen{EWH26}) that both $f(t)$ and $F(\varepsilon)$ be functions of bounded variation (\S~243 in Ref.~\citen{EWH27}) over their respective domains of definition. Consequently, the set of points of discontinuity of these functions is enumerable (\S~243 in Ref.~\citen{EWH27}), that is that these functions must be continuous almost everywhere. The Fourier-transform pairs, such as $f(t)$ and $F(\varepsilon)$, being related through integrations with respect to $t$ and $\varepsilon$, it is interesting to note that the necessary and sufficient condition for a bounded function to be integrable (in the sense of Riemann) is that points of discontinuity of this function over the interval of integration form a set of measure zero (\S~333 in Ref.~\citen{EWH27}).

The same statement as above applies for the space-dependent function $f({\bm r})$ (or $f({\bm R}_j)$, if $f({\bm r})$ is defined on the lattice $\{ {\bm R}_j\}$) and its Fourier transform $F({\bm k})$; for $d>1$, additional conditions, concerning boundedness of the partial derivatives of these functions, are required to be fulfilled (\S~465 in Ref.~\citen{EWH26}), which we need not consider here.

\subsubsection{Second step}
\label{ss53s7}

Here we investigate the mode of convergence of the series in Eq.~(\ref{e105}). To this end, we employ the Kramers-Kr\"onig relationship in Eq.~(\ref{ec73}) in conjunction with the series in Eq.~(\ref{e104}). In this connection, we note that (\S~4.5 in Ref.~\citen{WW62})
\begin{equation}
\mathscr{P}\!\!\int_{-\infty}^{\infty} {\rm d}\varepsilon'\; (\dots)
\equiv \lim_{\epsilon\downarrow 0} \Big\{
\int_{-\infty}^{\varepsilon-\epsilon} {\rm d}\varepsilon'\; (\dots)
+ \int_{\varepsilon+\epsilon}^{\infty} {\rm d}\varepsilon'\; (\dots)
\Big\}. \label{e108}
\end{equation}
The specific form in which the quantity $\epsilon$ features in the boundaries of the integrals on the RHS of Eq.~(\ref{e108}) is in conformity with the definition of the principal-value integral according to Cauchy, which underlies the Kramers-Kr\"onig relationship in Eq.~(\ref{ec73}).

Using the series in Eq.~(\ref{e104}) one has
\begin{equation}
\int_{-\infty}^{\varepsilon-\epsilon} \frac{{\rm
d}\varepsilon'}{\pi}\; \frac{\mathrm{Im}[\t{\Sigma}_{\sigma}({\bm
k};\varepsilon'-i 0^+)]}{\varepsilon -\varepsilon'} = \sum_{\nu=2}^{\infty} \int_{-\infty}^{\varepsilon-\epsilon} \frac{{\rm
d}\varepsilon'}{\pi}\; \frac{\mathrm{Im}[\t{\Sigma}_{\sigma}^{(\nu)}({\bm k};\varepsilon'-i 0^+)]}{\varepsilon -\varepsilon'}, \label{e109}
\end{equation}
where the exchange of the orders of integration and summation is justified on account of the uniformity of convergence of the series in Eq.~(\ref{e104}) \emph{and} the continuity of integrand of the integral on the RHS of Eq.~(\ref{e109}) for all $\varepsilon' \le \varepsilon-\epsilon <\varepsilon$ when the $i0^+$ herein is replaced by $i\eta$, where $\eta$ is arbitrarily small but positive (\S~4.7 in Ref.~\citen{WW62} and \S\S~379-388 in Ref.~\citen{EWH07}, \S~45 (2) in Ref.~\citen{TB65}).\footnote{See Remarks in Sec.~\protect\ref{ss53s8} for the reason underlying the proposed substitution of $i\eta$ for $i0^+$. In essence, here we are stating that the limit $\eta\downarrow 0$ should be taken \emph{after} the evaluation of the pertinent integral with respect to $\varepsilon'$. Note that the numerical value of $0^+$ is \emph{exactly} equal to zero; the $+$ in $0^+$ solely indicates that $\eta$ in, e.g., $\t{\Sigma}_{\sigma}({\bm k};\varepsilon'-i \eta)$ has approached $0$ from above.} We point out that for the case at hand it is irrelevant that the lower bounds of the integrals in Eq.~(\ref{e109}) are equal to $-\infty$, since on replacing these integrals by $\int_{-E}^{\varepsilon-\epsilon} {\rm d}\varepsilon'\; (\dots)$, it can be demonstrated that the resulting integrals converge \emph{uniformly} for $E\to\infty$ (\S\S~4.42 and 4.44 in Ref.~\citen{WW62} and \S~378 in Ref.~\citen{EWH07}).

Following Eq.~(\ref{e107}), on account of $\varepsilon' \le\varepsilon -\epsilon < \varepsilon$ one observes that the summand of the sum on the RHS of Eq.~(\ref{e109}) is non-negative. Consequently, the established convergence of the series on the RHS of Eq.~(\ref{e109}) for almost all ${\bm k}$ and $\varepsilon$ implies that it is \emph{uniformity} convergent for almost all ${\bm k}$ and $\varepsilon$ (\S~347 in Ref.~\citen{EWH07}); this result can in fact be directly established, without recourse to any general theorem, by verifying that the series on the RHS of Eq.~(\ref{e109}) indeed satisfies the necessary and sufficient conditions (\S~3.31 in Ref.~\citen{WW62}) required of uniformly-convergent series.

Along the same line of reasoning as above, one can demonstrate that the series on the RHS of
\begin{equation}
\int_{\varepsilon+\epsilon}^{\infty} \frac{{\rm
d}\varepsilon'}{\pi}\; \frac{\mathrm{Im}[\t{\Sigma}_{\sigma}({\bm
k};\varepsilon'-i 0^+)]}{\varepsilon -\varepsilon'} = \sum_{\nu=2}^{\infty} \int_{\varepsilon+\epsilon}^{\infty} \frac{{\rm
d}\varepsilon'}{\pi}\; \frac{\mathrm{Im}[\t{\Sigma}_{\sigma}^{(\nu)}({\bm k};\varepsilon'-i 0^+)]}{\varepsilon -\varepsilon'} \label{e110}
\end{equation}
is \emph{uniformly} convergent for almost all ${\bm k}$ and $\varepsilon$. Key to this result is the fact that the integrand of the integral on the RHS of Eq.~(\ref{e110}) is non-positive for all $\varepsilon' \ge \varepsilon+\epsilon > \varepsilon$, rendering the series on the RHS a monotonic one.

Combining the series in Eqs.~(\ref{e109}) and (\ref{e110}), we thus arrive at the conclusion that indeed \emph{the series in Eq.~(\ref{e105}) is uniformly convergent for almost all ${\bm k}$ and $\varepsilon$}.

\subsubsection{Remarks}
\label{ss53s8}

Above we have arrived at the conclusion that the series on the RHSs of Eqs.~(\ref{e109}) and (\ref{e110}) are uniformly convergent for almost all ${\bm k}$ and $\varepsilon$ by relying on a general theorem concerning admissibility of the term-by-term integration of uniformly-convergent series of functions which are \emph{continuous} over the entire range of integration. In doing so, we have relied on the continuity of the pertinent functions when $i0^+$ in their respective arguments is replaced by $i\eta$, with $\eta$ arbitrarily small but positive. In this connection, it is relevant to recall that $\t{\Sigma}_{\sigma}({\bm k};\varepsilon'-i 0^+)$ and $\t{\Sigma}_{\sigma}^{(\nu)}({\bm k};\varepsilon'-i 0^+)$ are strictly continuous for \emph{almost} all $\varepsilon'$ but not necessarily for \emph{all} $\varepsilon'$. The question arises as to the relevance of relying on the property of continuity of functions whose arguments are different from those of the actual functions being integrated. In other words, on what ground is the prescription of taking the limit $\eta\downarrow 0$ subsequent to the evaluation of the pertinent integrals based?

The question posed above is immediately answered by appreciating that the Kramers-Kr\"onig relations are deduced through reliance on the use of the Cauchy residue theorem (\S~5.2 in Ref.~\citen{WW62}), requiring, amongst other things, that the function under consideration (here $\t{\Sigma}_{\sigma}({\bm k};z)$) be \emph{continuous} on the contour of integration (appendix \ref{sc});\footnote{See also in particular the last but one footnote on p.~85 of Ref.~\protect\citen{WW62}.} taking the limit $\eta\downarrow 0$ prior to evaluating the pertinent integrals with respect to $\varepsilon'$ would contravene use of the Cauchy theorem in the event that $\t{\Sigma}_{\sigma}({\bm k};\varepsilon'-i 0^+)$ turns out to be discontinuous at some $\varepsilon' \in \mathds{R}$. Since $\t{\Sigma}_{\sigma}({\bm k};\varepsilon'-i \eta)$ and $\t{\Sigma}_{\sigma}^{(\nu)}({\bm k};\varepsilon'-i \eta)$ are strictly continuous for $\eta>0$, $\forall\varepsilon' \in \mathds{R}$, one observes that our above prescription amounts to the very way in which the Kramers-Kr\"onig relations must be used.

\subsubsection{Third step}
\label{ss53s9}

In view of the above results concerning the series in Eqs.~(\ref{e104}) and (\ref{e105}), we have thus shown that the series in Eq.~(\ref{e103}) is a uniformly convergent series for almost all ${\bm k}$ and $\varepsilon \in \mathds{R}$. The uniformity of convergence of the series in Eq.~(\ref{e103}) for almost all $\varepsilon \in \mathds{R}$, allied with the analyticity of $\t{\Sigma}_{\sigma}^{(\nu)}({\bm k};z)$, $\forall\nu$, in the lower half of the $z$ plane, implies (\S~5.3 in Ref.~\citen{WW62}) that the series in Eq.~(\ref{e85}) is \emph{uniformly} convergent for all $z$ in the lower half of the complex plane, that is, in the region $\mathrm{Im}(z) <0$.

\subsubsection{Fourth and last step}
\label{ss53s10}

Since
\begin{equation}
\t{\Sigma}_{\sigma}({\bm k};z^*) = \t{\Sigma}_{\sigma}^*({\bm k};z)
\label{e111}
\end{equation}
and
\begin{equation} \t{\Sigma}_{\sigma}^{(\nu)}({\bm k};z^*) =
\t{\Sigma}_{\sigma}^{(\nu)*}({\bm k};z),\;\; \forall\nu > 1,
\label{e112}
\end{equation}
for all $z$ satisfying $\mathrm{Im}(z)\not=0$, on taking the complex
conjugates of the both sides of Eq.~(\ref{e85}) one deduces that this series is also (see the third step) \emph{uniformly} convergent in the upper half of the $z$ plane, that is, in the region $\mathrm{Im}(z)> 0$.

\subsubsection{Summary}
\label{ss53s11}

We have completed the proof of a main result which will play an essential role is answering the three questions, marked (i), (ii) and (iii), posed in Sec.~\ref{ss52s4}. Central to this proof has been the result in Eq.~(\ref{e107}) (or equivalently $\mathrm{Im}[\t{\Sigma}_{\sigma}({\bm k};\varepsilon+i 0^+)] \le 0$, $\forall\varepsilon\in \mathds{R}$, $\forall\nu$), which encodes a very significant analytic property of the perturbative contributions $\{\t{\Sigma}_{\sigma}^{(\nu)}({\bm k};z)\,\|\,\nu\}$ to $\t{\Sigma}_{\sigma}({\bm k};z)$.

We now proceed with investigating whether the above-mentioned three steps (i), (ii) and (iii) are allowable.

\subsubsection{Concerning step (i)}
\label{ss53s12}

Consider the \emph{convergent} series $f_1(z) + f_2(z)+\dots$, where $z$ may be real or complex. In general, validity of the relationship
\begin{equation}
\frac{\rm d}{{\rm d}z} \sum_{\nu=1}^{\infty} f_{\nu}(z) =
\sum_{\nu=1}^{\infty} \frac{\rm d}{{\rm d}z}\, f_{\nu}(z)\label{e113}
\end{equation}
for $z$ inside some region is dependent on the condition that the series on the \emph{RHS} converge \emph{uniformly} inside this region (\S~4.7 in Ref.~\citen{WW62}, \S~46 in Ref.~\citen{TB65}, \S~396 in Ref.~\citen{EWH07}). However, for the sequence $\{ f_{\nu}(z)\}$ consisting of functions analytic along a \emph{closed} contour and throughout its interior, the uniformity of convergence of the series on the \emph{LHS} of Eq.~(\ref{e113}) in the latter region of the complex $z$ plane, suffices for Eq.~(\ref{e113}) to be valid in this region (\S~5.3 in Ref.~\citen{WW62}). Since $\t{\Sigma}_{\sigma}^{(\nu)}({\bm k};z)$, $\nu=1,2\dots$, are analytic in the entire $z$ plane away from the real axis ($\t{\Sigma}_{\sigma}^{(1)}({\bm k};z)$ being independent of $z$, it is analytic everywhere), on account of the uniformity of convergence of the series in Eq.~(\ref{e85}) for all $z$ outside the real axis, it follows that the relationship
\begin{equation}
\frac{\partial}{\partial z} \sum_{\nu=1}^{\infty}
\t{\Sigma}_{\sigma}^{(\nu)}({\bm k};z) = \sum_{\nu=1}^{\infty}
\frac{\partial}{\partial z}\, \t{\Sigma}_{\sigma}^{(\nu)}({\bm k};z)
\label{e114}
\end{equation}
is valid for at least all $z$ away from the real axis. Thus, with the possible exception of $z=\mu$, the result in Eq.~(\ref{e114}) is valid for all $z \in \mathscr{C}(\mu)$.

Concerning the case $z=\mu$, since the functions on both sides of Eq.~(\ref{e114}) are analytic everywhere on the $z$ plane away from the real axis, their demonstrated equality in the region $\mathrm{Im}(z)\not=0$ implies that these functions are in fact identical everywhere on the $z$ plane (\S~4.1 in Ref.~\citen{ECT85}, \S~5.51 in Ref.~\citen{WW62}). Consequently, Eq.~(\ref{e114}) cannot fail at $z=\mu$. Evidently, depending on ${\bm k}$, the functions on both sides of Eq.~(\ref{e114}) may be unbounded at $z=\mu$, in which case the equality in Eq.~(\ref{e114}) at $z=\mu$ should be understood as signifying that both functions diverge identically as $z\to\mu$.

Hereby we have completed the proof that step (i) is indeed admissible.

\subsubsection{Concerning step (ii)}
\label{ss53s13}

Having demonstrated the general validity of the result in Eq.~(\ref{e114}), we can make progress by first demonstrating that the series on the RHS of Eq.~(\ref{e114}) is uniformly convergent for almost all ${\bm k}$ and $z$, similar to the series on the LHS of Eq.~(\ref{e114}). To this end and for simplicity of notation we revert to the general expression in Eq.~(\ref{e113}). In the following we shall assume that the series on the LHS of the Eq.~(\ref{e113}) is uniformly convergent for all $z$ inside a domain $D$ of the $z$ plane.

The uniformity of convergence of the series on the LHS of the Eq.~(\ref{e113}) for all $z$ inside $D$ implies that for an arbitrary positive number $\epsilon$, there exists an integer $n$ for which one has
\begin{equation}
\vert \mathcal{R}_{p}(z)\vert < \epsilon,\;\; \forall p > n,
\;\;\;\forall z\in D, \label{e115}
\end{equation}
where
\begin{equation}
\mathcal{R}_p(z) {:=} \sum_{\nu=p+1}^{\infty} f_{\nu}(z).
\label{e116}
\end{equation}
It will be convenient to express $\mathcal{R}_p(z)$ as
\begin{equation}
\mathcal{R}_p(z) = \epsilon\, \mathcal{Q}_p(z), \label{e117}
\end{equation}
where $\mathcal{Q}_p(z)$ is analytic inside $D$ and for which one
has $\vert \mathcal{Q}_p(z)\vert < 1$, $\forall p > n$, $\forall
z\in D$.

Since $\mathcal{Q}_{p}(z)$ is analytic inside $D$, it follows that $\mathcal{Q}_p'(z) \equiv {\rm d} \mathcal{Q}_p(z)/{\rm d}z$ is bounded inside $D$, that is $\vert \mathcal{Q}_p'(z)\vert <K$, where $K$ is some finite positive constant independent of $p$ and $z$; this result can be explicitly demonstrated with the aid of the Cauchy theorem (\S~5.22 in Ref.~\citen{WW62}). Consequently, for any positive number $\epsilon'$ there exists an integer $n'$ for which one has
\begin{equation}
\vert \mathcal{R}_{p}'(z) \vert < \epsilon',\;\; \forall p >
n',\;\;\;\forall z\in D, \label{e118}
\end{equation}
where $\mathcal{R}_p'(z) \equiv {\rm d}\mathcal{R}_p(z)/{\rm d}z$. An upper bound for $n'$ is obtained by replacing the $\epsilon$ on the RHS of the inequality in Eq.~(\ref{e115}) by $\epsilon/K$ and identifying $n'$ with the smallest $n$ for which the resulting inequality is satisfied. Hereby we have completed the proof of the uniformity of convergence of the series on the RHS of Eq.~(\ref{e113}), for all $z$ inside $D$.

We proceed by employing the following result which applies to \emph{uniformly} convergent series $\sum_{\nu=1}^{\infty} g_{\nu}(z)$ corresponding to \emph{continuous} functions $\{g_{\nu}(z)\}$ inside $D$ (\S~4.7 in Ref.~\citen{WW62}):
\begin{equation}
\int_{C} {\rm d}z\; \sum_{\nu=1}^{\infty} g_{\nu}(z) =
\sum_{\nu=1}^{\infty} \int_{C} {\rm d}z\; g_{\nu}(z), \label{e119}
\end{equation}
where $C$ is a \emph{path} (thus not necessarily a closed contour) inside $D$. In applying the result in Eq.~(\ref{e119}) we have in mind that
\begin{equation}
g_{\nu}(z) \equiv \t{G}_{\sigma}({\bm k};z) \frac{\partial}{\partial
z} \t{\Sigma}_{\sigma}^{(\nu)}({\bm k};z), \label{e120}
\end{equation}
and that $C \equiv\mathscr{C}(\mu)$. In this connection, the following remarks are in order:
\begin{itemize}
\item[(a)] The dependence of the $g_{\nu}(z)$ in Eq.~(\ref{e120}) on $\nu$ originating from $\t{\Sigma}_{\sigma}^{(\nu)}({\bm k};z)$, the uniformity of convergence of the series on the RHS of Eq.~(\ref{e114}) for almost all ${\bm k}$ and $z$ implies that $\sum_{\nu=1}^{\infty} g_{\nu}(z)$ is indeed uniformly convergent for almost all ${\bm k}$ and $z$.
\item[(b)] Since both $\t{G}_{\sigma}({\bm k};z)$ and $\t{\Sigma}_{\sigma}^{(\nu)}({\bm k};z)$ are analytic everywhere in the $z$ plane away from the real axis (appendix \ref{sc}), it follows that the $g_{\nu}(z)$ in Eq.~(\ref{e120}) is continuous at all $z$ outside the real axis.
\item[(c)] The possibility that the $g_{\nu}(z)$ in Eq.~(\ref{e120}) may not be continuous at $z=\mu \in \mathscr{C}(\mu)$ (depending on ${\bm k}$), can be shown to pose no problems.
\item[(d)] Although the extant proof of the result in Eq.~(\ref{e119}) (\S~4.7 in Ref.~\citen{WW62}) is specific to paths $C$ whose lengths $l$ are finite, it can be shown that for the $g_{\nu}(z)$ in Eq.~(\ref{e120}) the result in Eq.~(\ref{e119}) indeed applies when $C \equiv\mathscr{C}(\mu)$, this in spite of the fact that the length $l$ of $\mathscr{C}(\mu)$ is infinitely large.
\item[(e)] Since the $g_{\nu}(z)$ in Eq.~(\ref{e120}) is further a function of ${\bm k}$, which for $z\in D$ is bounded for all ${\bm k}$, the proof of convergence of the series on the RHS of Eq.~(\ref{e119}) (\S~4.7 in Ref.~\citen{WW62}) can be readily verified to imply uniformity of convergence of this series for all ${\bm k}$.
\end{itemize}

For establishing the validity of the statements (c) and (d), it is helpful to reproduce the proof of the result in Eq.~(\ref{e119}) (\S~4.7 in Ref.~\citen{WW62}), which we shall do now.

On account of the uniformity of convergence of the series $\sum_{\nu=1}^{\infty} g_{\nu}(z)$ for $z\in D$, one has the essential property that for an arbitrary positive number $\epsilon$ there exists an integer $n$ (independent of $z$) for which
\begin{equation}
\vert R_p(z)\vert <\epsilon,\;\;\; \forall p>n,\;\;\;\;\forall z\in D,
\label{e121}
\end{equation}
where
\begin{equation}
R_p(z) {:=} \sum_{\nu=p+1}^{\infty} g_{\nu}(z). \label{e122}
\end{equation}
It will prove advantageous to denote the smallest integer $n$ for which the inequality in Eq.~(\ref{e121}) applies by $n_{\epsilon}$. Since by assumption $g_{\nu}(z)$ is continuous for all $z\in D$, $\forall\nu\in \mathds{N}$, the uniformity of convergence of the series on the RHS of Eq.~(\ref{e122}) implies that $R_p(z)$ is \emph{continuous} for all $z\in D$ (\S~3.32 in Ref.~\citen{WW62}, \S~344 in Ref.~\citen{EWH07}).

With $p$ a \emph{finite} integer (not necessarily greater than $n_{\epsilon}$), one has the exact result
\begin{equation}
\int_{C} {\rm d}z\; \sum_{\nu=1}^{\infty} g_{\nu}(z) =
\sum_{\nu=1}^{p} \int_{C} {\rm d}z\; g_{\nu}(z) + \int_{C} {\rm
d}z\; R_p(z). \label{e123}
\end{equation}
On account of the continuity of $\vert R_{p}(z)\vert$ for $z\in C$ and of the result in Eq.~(\ref{e121}), one arrives at the inequality (\S~4.62 in Ref.~\citen{WW62})
\begin{equation}
\left| \int_{C} {\rm d}z\; R_p(z) \right| < \epsilon\, l
\;\;\;\mbox{\rm for}\;\;\; p > n_{\epsilon},\label{e124}
\end{equation}
where $l$ denotes the length of $C$. One observes that, for any finite $l$, the absolute value of the second integral on the RHS of Eq.~(\ref{e123}) can be made as small as desired by decreasing $\epsilon$ below a sufficiently small non-vanishing amount. Since for $\epsilon\downarrow 0$ one has $n_{\epsilon} \uparrow \infty$ (unless $g_{\nu}(z) \equiv 0$ for all $\nu> \mathcal{N}$, where $\mathcal{N}$ is some finite integer, in which case for $p\ge \mathcal{N}$ the second term on the RHS of Eq.~(\ref{e123}) disappears), on the basis of Eqs.~(\ref{e123}) and (\ref{e124}) one arrives at the result in Eq.~(\ref{e119}).

In our actual considerations, where $C \equiv \mathscr{C}(\mu)$, and thus $l=\infty$ (see item (d) above), the inequality in Eq.~(\ref{e124}) fails to be of direct use. We circumvent this problem by subdividing $\mathscr{C}(\mu)$ into a finite path $C'$ and a remaining infinite path $\mathscr{C}(\mu)\backslash C'$ located in the region of the $z$ plane corresponding to large values of $\vert z\vert$. Neglecting for the moment the fact that $R_p(z)$ need not be continuous at $z=\mu \in C'$, the integral of $R_p(z)$ along $C'$ can be dealt with by relying on the general result in Eq.~(\ref{e124}); from the details in the following it will become apparent that the possibility of $R_p(z)$ being discontinuous at $z=\mu \in C'$ is of no detrimental consequence (see item (c) above). As regards the integral of $R_p(z)$ over $\mathscr{C}(\mu)\backslash C'$, making use of an explicit asymptotic expression for $R_p(z)$ corresponding to large values of $\vert z\vert$ (appendix \ref{sc}), we directly demonstrate that also the absolute value of this integral is as small as desired for $p>n'$, where $n'$ is some finite integer.

We proceed by parameterising $\mathscr{C}(\mu)$ according to $z = \mu + i y$, where $y\uparrow_{-\infty}^{+\infty}$, so that
\begin{equation}
\int_{\mathscr{C}(\mu)} {\rm d}z\; R_{p}(z) =
i\int_{-\infty}^{\infty} {\rm d}y\; R_{p}(\mu+i y)
\equiv i \lim_{\substack{y_0\downarrow 0\\
y_1\to \infty}} \int_{y_0}^{y_1} {\rm d}y\; \big(R_{p}(\mu-i y) +
R_{p}(\mu+i y)\big). \label{e125}
\end{equation}
We note in passing that for the $R_p(z)$ at hand (based on the $g_{\nu}(z)$ in Eq.~(\ref{e120})) one has $R_p(\mu-iy) \equiv R_p^*(\mu+iy)$, for all $y>0$, so that the integral on the LHS of Eq.~(\ref{e125}) is purely imaginary for all ${\bm k}$ and $p$; consequently, the integral on the LHS of Eq.~(\ref{e125}) is real for all ${\bm k}$ and $p$.

To make progress, one needs to establish that the integral on the RHS of Eq.~(\ref{e125}) is bounded, for all relevant ${\bm k}$, as $y_0$ is made to approach $0$ and $y_1$ to approach $\infty$. That this is indeed the case follows directly from the existence of a similar integral (which we have considered in Sec.~\ref{ss41}; see also appendix \ref{sd}) concerning $R_p(z)\vert_{p=1}$ (recall that $R_0(z) \equiv R_1(z)$), for which, on account of Eq.~(\ref{e114}), one has
\begin{equation}
R_1(z) \equiv \t{G}_{\sigma}({\bm k};z) \frac{\partial}{\partial z}
\t{\Sigma}_{\sigma}({\bm k};z). \label{e126}
\end{equation}
We should emphasise that existence of the integral on the RHS of Eq.~(\ref{e125}), in the limits $y_0=0$ and $y_1=\infty$, for $p \ge p_0 > 1$, would not necessarily imply existence of this integral for $1\le p <p_0$.

In the light of the above observations, we express Eq.~(\ref{e125}) in the following equivalent form:
\begin{equation}
\int_{\mathscr{C}(\mu)} {\rm d}z\; R_p(z) = i \!\int_{0^+}^{y_1} {\rm d}y\; \big( R_p(\mu-i y) + R_p(\mu+i y)\big) + i \!\int_{y_1}^{\infty} {\rm d}y\; \big( R_p(\mu-i y) + R_p(\mu+i y)\big), \label{e127}
\end{equation}
where $y_1>0$ is a \emph{free} parameter; by $0^+$ in the lower bound of the first integral on the RHS we intend to emphasize the fact that the integrand of the corresponding integral is continuous over the entire range of integration. In what follows we shall assume that $y_1$ is \emph{finite}, however sufficiently large so that in determining the second integral on the RHS of Eq.~(\ref{e127}) we can replace the $g_{\nu}(\mu\pm i y)$, $\forall\nu >p$, in the defining expression for $R_p(\mu\pm i y)$ by its leading-order asymptotic term corresponding to $y\to\infty$.

On account of the general result in Eq.~(\ref{e124}), the continuity of $R_p(\mu- iy) + R_p(\mu +iy)$ for \emph{all} $y\in [0^+,y_1]$ implies that the magnitude of the first integral on the RHS of Eq.~(\ref{e127}) can be made as small as desired for $p> n_{\epsilon}$ by equating $\epsilon$ with a sufficiently small positive value. Consequently, it remains only to investigate the behaviour of the second integral on the RHS of Eq.~(\ref{e127}). To this end we point out that for the specific case of short-range two-body potentials and lattice models one has (appendix \ref{sc}, Sec.~\ref{ssc7})
\begin{equation}
\t{G}_{\sigma}({\bm k};z) \frac{\partial}{\partial
z} \t{\Sigma}_{\sigma}^{(\nu)}({\bm k};z) \sim -\hbar\, (\nu-1)\,
\frac{{\sf S}_{\sigma}^{(\nu)}({\bm k})}{z^{\nu+1}}\;\;\;\mbox{\rm for}\;\;\; \vert z\vert\to \infty\;\;\; (\nu\ge 2), \label{e128}
\end{equation}
where ${\sf S}_{\sigma}^{(\nu)}({\bm k})$ is a well-specified real-valued function; since the LHS of Eq.~(\ref{e128}) is identically vanishing for $\nu=1$, this asymptotic result can be declared valid also for $\nu=1$. We point out that although it is in general not permissible to differentiate asymptotic series (\S~8.31 in Ref.~\citen{WW62}), the large-$\vert z\vert$ asymptotic series expansion of $\t{\Sigma}_{\sigma}^{(\nu)}({\bm k};z)$ (and of $\t{\Sigma}_{\sigma}({\bm k};z)$) can be differentiated in the region $\mathrm{Im}(z) \not=0$; this follows on account of the considerations by Ritt \cite{JFR18}.

Since asymptotic series can be integrated term-by-term (\S~8.31 in Ref.~\citen{WW62}, \S~113 in Ref.~\citen{TB65}), one obtains that
\begin{eqnarray}
\int_{y_1}^{\infty} {\rm d}y\; \big( R_p(\mu-i y)
+ R_p(\mu+i y)\big) &\sim& \frac{2\hbar\, (-1)^{\nu_p+1} (2\nu_p-2)\,
{\sf S}_{\sigma}^{(2\nu_p-1)}({\bm k})}{(2\nu_{p} -1)\, y_1^{2\nu_p
-1}} \nonumber\\
&\sim& \frac{2 \hbar\, (2\nu_p-2)\,
\t{\Sigma}_{\sigma}^{(2\nu_p-1)}({\bm k};\pm i y_1)}{(2\nu_p-1)\,
y_1} \;\;\;\mbox{\rm for}\;\;\; y_1\to\infty, \nonumber\\
\label{e129}
\end{eqnarray}
where
\begin{equation}
\nu_p \equiv \left\{ \begin{array}{ll} \frac{1}{2} p +1, & p =
\mbox{\rm even},\\ \\
\frac{1}{2} (p+1) + 1, & p = \mbox{\rm odd}. \end{array} \right.
\label{e130}
\end{equation}
The first asymptotic expression in Eq.~(\ref{e129}) underlines the fact, pointed out earlier, that the integral in Eq.~(\ref{e129}) is real-valued. In this connection, note that since $2\nu_{\rm p}
-1$ is odd, one has
\begin{equation}
\t{\Sigma}_{\sigma}^{(2\nu_p-1)}({\bm k};+i y_1) \sim
\t{\Sigma}_{\sigma}^{(2\nu_p-1)}({\bm k};-i y_1)\;\;\;\mbox{\rm
for}\;\;\; y_1\to\infty, \label{e131}
\end{equation}
so that, on account of $\t{\Sigma}_{\sigma}^{(\nu)}({\bm k};z^*) \equiv \t{\Sigma}_{\sigma}^{(\nu) *}({\bm k};z)$ for $\mathrm{Im}(z) \not=0$, $\t{\Sigma}_{\sigma}^{(2\nu_p -1)}({\bm k};\pm iy_1)$ are real to leading order in $1/y_1$. Since $(2\nu_p-2)/(2\nu_p -1) \sim 1$ for $p\to\infty$, one thus arrives at
\begin{equation}
\int_{y_1}^{\infty} {\rm d}y\; \big( R_p(\mu-i y) +
R_p(\mu+i y)\big) \sim \frac{2 \hbar}{y_1}\,
\t{\Sigma}_{\sigma}^{(2\nu_p-1)}({\bm k};\pm i y_1)
\;\;\; \mbox{\rm for}\;\;\; y_1\to\infty,\; p\to\infty.
\label{e132}
\end{equation}

On account of the uniformity of convergence of the series in Eq.~(\ref{e85}) for all $z$ away from the real axis, it follows that for an arbitrary $\epsilon>0$ there exists an integer $n$ so that (\S~3.31 in Ref.~\citen{WW62}, \S\S~44 and 83 in Ref.~\citen{TB65}, \S~66 in Ref.~\citen{EWH26})
\begin{equation}
\vert \mathscr{R}_{n,p}(z) \vert < \epsilon,\;\;\; \forall p \in
\mathds{N},\;\; \mathrm{Im}(z) \not=0, \label{e133}
\end{equation}
where
\begin{equation}
\mathscr{R}_{n,p}(z) {:=} \sum_{\nu=n+1}^{n+p}
\t{\Sigma}_{\sigma}^{(\nu)}({\bm k};z). \label{e134}
\end{equation}
It follows that there exists a finite integer $n_{\epsilon}'$ corresponding to an arbitrary $\epsilon>0$ so that
\begin{equation}
\left| \t{\Sigma}_{\sigma}^{(2\nu_p-1)}({\bm k};\pm i y_1)\right| <
\frac{y_1}{2\hbar}\,\epsilon \;\;\;\mbox{\rm for}\;\;\; p>
n_{\epsilon}',\;\; y_1>0. \label{e135}
\end{equation}
From Eqs.~(\ref{e132}) and (\ref{e135}) one thus infers that
\begin{equation}
\left| \int_{y_1}^{\infty} {\rm d}y\; \big(R_p(\mu
- i y) + R_p(\mu + i y)\big)\right| < \epsilon \;\;\;\mbox{\rm for}\;\;\; p> n_{\epsilon}',\;\; y_1\to\infty. \label{e136}
\end{equation}

Denoting by $n_{\epsilon}''$ the smallest $p$ for which the magnitudes of the two integrals on the RHS of Eq.~(\ref{e127}) are less than $\epsilon/2$, in view of the above considerations and on the basis of the triangle inequality (item 3.2.5 in Ref.~\citen{AS72}) we arrive at the result
\begin{equation}
\left| \int_{\mathscr{C}(\mu)} {\rm d}z\; R_p(z)\right| <
\epsilon\;\;\;\mbox{\rm for}\;\;\; p > n_{\epsilon}'',
\label{e137}
\end{equation}
thus completing the proof of the statement that Eq.~(\ref{e119}) is valid for $C \equiv\mathscr{C}(\mu)$, with $g_{\nu}(z)$ as defined in Eq.~(\ref{e120}).

Finally, with reference to item (e) above, we remark that since the $g_{\nu}(z)$ in Eq.~(\ref{e120}) is bounded for all ${\bm k}$ when $z\in D$, the \emph{finite} integer $n$ in Eq.~(\ref{e121}) can be chosen sufficiently large so that for a given $\epsilon$ the inequality in Eq.~(\ref{e121}) applies not only for all $z\in D$, but also for \emph{all} relevant ${\bm k}$. Owing to this possibility, the series on the RHS of Eq.~(\ref{e119}) is not only convergent (\S~4.7 in Ref.~\citen{WW62}), but is \emph{uniformly} convergent for all ${\bm k}$. Note that without restricting $z$ to be located inside $D$, the $g_{\nu}(z)$ in Eq.~(\ref{e120}) is bounded for \emph{almost} all ${\bm k}$.

Hereby we have demonstrated that step (ii) is indeed admissible.

\subsubsection{Concerning step (iii)}
\label{ss53s14}

We now proceed with investigating whether the relationship
\begin{equation}
\sum_{\bm k} \sum_{\nu=2}^{\infty} f_{\sigma}^{(\nu)}({\bm k}) =
\sum_{\nu=2}^{\infty} \sum_{\bm k} f_{\sigma}^{(\nu)}({\bm k})
\label{e138}
\end{equation}
is valid, where
\begin{equation}
f_{\sigma}^{(\nu)}({\bm k}) \equiv \int_{\mathscr{C}(\mu)}
\frac{{\rm d}z}{2\pi i}\; \t{G}_{\sigma}({\bm k};z)
\frac{\partial}{\partial z} \t{\Sigma}_{\sigma}^{(\nu)}({\bm k};z).
\label{e139}
\end{equation}
In this notation, Eq.~(\ref{e87}) has the form
\begin{equation}
\sum_{\bm k} f_{\sigma}^{(\nu)}({\bm k}) = 0,\;\;\; \nu \in
\mathds{N}, \label{e140}
\end{equation}
so that, in view of our earlier observations concerning steps (i) and (ii), validity of Eq.~(\ref{e138}) is tantamount to validity of the Luttinger-Ward identity, Eq.~(\ref{e44}) (see Eqs.~(\ref{e43}) and (\ref{e85})).

Below we shall first consider finite systems and subsequently macroscopic systems. As will become evident, macroscopic systems can be dealt with in two principally distinct ways, corresponding to taking the thermodynamic limit \emph{before} and \emph{after} evaluating the sums on both sides of Eq.~(\ref{e138}). In the latter approach, the mathematical reasonings that establish validity of the result in Eq.~(\ref{e138}) are entirely identical to those specific to finite systems so that within the framework of this approach the proof of the expression in Eq.~(\ref{e138}) follows, on dividing both sides of this expression by an extensive quantity, immediately from the proof specific to finite systems. The former approach, of taking the thermodynamic limit prior to evaluating the sums on both sides of Eq.~(\ref{e138}), coincides with the conventional procedure, according to which $\sum_{\bm k}$ is at the outset replaced by $V \int {\rm d}^dk/(2\pi)^d$, where $V$ is the macroscopic volume of the system under consideration. We shall investigate both of these approaches and demonstrate that within these frameworks the expression in Eq.~(\ref{e138}) holds true in the thermodynamic limit.

\subsubsection{Some nomenclature and technical details}
\label{ss53s15}

The details in this section will facilitate the subsequent discussions.

Let $\{ a_{i,j} \,\|\, i,j\in \mathds{N}\}$ denote a sequence of numbers. The sum $\sum_{i,j} a_{i,j}$ is referred to as a \emph{double series}, and $\sum_i \sum_j a_{i,j}$ and $\sum_{j} \sum_{i} a_{i,j}$ as \emph{repeated series} (also known as \emph{iterated series}), with the former being more specifically called \emph{sum by rows} and the latter \emph{sum by columns} (\S\S~2.5, 2.51 in Ref.~\citen{WW62}, \S\S~29, 30 in Ref.~\citen{TB65}). For $m$ and $n$ \emph{finite} integers, the value of the partial sum $S_{m,n}$ is independent of the order in which summations are carried out, that is
\begin{equation}
S_{m,n} = \sum_{i=1}^{m} \sum_{j=1}^{n} a_{i,j} \equiv
\sum_{j=1}^{n} \sum_{i=1}^{m} a_{i,j}. \label{e141}
\end{equation}

The \emph{double series} $\sum_{i,j} a_{i,j}$ is said to converge to $S$ if for any $\epsilon> 0$ it is possible to find integers $m$ and $n$ so that (\S\S~2.5, 2.51 in Ref.~\citen{WW62}, \S\S~29, 30 in Ref.~\citen{TB65})
\begin{equation}
\vert S_{m+p,n+q} - S\vert< \epsilon,\;\;\; \forall p, q \in
\mathds{N}. \label{e142}
\end{equation}
Following Ref.~\citen{TB65}, we call $S$ the Pringsheim sum of the double series $\sum_{i,j} a_{i,j}$. It should be evident that the Pringsheim sum $S$ corresponds to the \emph{double limit} (\S~302 in Ref.~\citen{EWH27})
\begin{equation}
\lim_{m,n\to\infty} S_{m,n}\nonumber
\end{equation}
of the sequence $\{ S_{m,n} \}$ of partial sums, whereas the sums by rows and columns correspond to \emph{repeated limits} (\S\S~302-306 in Ref.~\citen{EWH27}) of the same sequence, that is
\begin{equation}
\lim_{m\to\infty} \lim_{n\to\infty} S_{m,n}\;\;\;\; \mbox{\rm
and}\;\;\;\; \lim_{n\to\infty} \lim_{m\to\infty} S_{m,n}, \nonumber
\end{equation}
respectively.

According to Stolz (\S~2.5 in Ref.~\citen{WW62}), the necessary and sufficient condition for the double series $\sum_{i,j} a_{i,j}$ to converge is that for an arbitrary $\epsilon> 0$ one can find integers $m_0$ and $n_0$ such that for all $m>m_0$ and $n>n_0$ one has
\begin{equation}
\vert S_{m+p,n+q} - S_{m,n}\vert< \epsilon,\;\;\;\forall p, q \in \mathds{Z}^*. \label{e143}
\end{equation}
The Pringsheim theorem (\S~2.51 in Ref.~\citen{WW62}, \S~29 in Ref.~\citen{TB65}, \S~336 in Ref.~\citen{EWH07}) states that:
\begin{itemize}
\item[{}]\emph{If $S$ exists and the sums by rows and columns exist, then each of these is equal to $S$}.
\end{itemize}

We point out that if the sum by rows and the sum by columns of a double series exist, it is not necessary that $S$ should exist, nor is it necessary that the latter two sums should be equal. An example may be clarifying. In considering the electrostatics of charged spheres in contact, one encounters the \emph{repeated series}
\begin{equation}
\sum_{i=1}^{\infty} \sum_{j=1}^{\infty} \frac{(-1)^{i+j} i
j}{(i+j)^2} \nonumber
\end{equation}
for which one obtains $\frac{1}{6} (\log2 -\frac{1}{4})$; in view the symmetry of the summand, the same value is obtained for
\begin{equation}
\sum_{j=1}^{\infty} \sum_{i=1}^{\infty} \frac{(-1)^{i+j} i
j}{(i+j)^2}. \nonumber
\end{equation}
However, the Pringsheim sum of the underlying double series does not exist, but oscillates between the limits of indeterminacy (Sec.~\ref{ss53s3}) $\frac{1}{6}(\log2-\frac{5}{8})$ and $\frac{1}{6} (\log2+\frac{1}{8})$ (\S~33 in Ref.~\citen{TB65}).

\subsubsection{Finite systems}
\label{ss53s16}

Since we deal with \emph{uniform} GSs, the finite systems that we consider here must be defined on \emph{finite} lattices without boundary, consisting of $\mathcal{N}_{\Sc l} <\infty$ lattice sites. On indexing the set of the relevant ${\bm k}$ points and denoting them by ${\bm k}_j$, $j=1,\dots,\mathcal{N}_{\Sc l}$, Eq.~(\ref{e138}) can be equivalently expressed as
\begin{equation}
\sum_{j=1}^{\mathcal{N}_{\Sc l}} \sum_{\nu=2}^{\infty} f_{\sigma}^{(\nu)}({\bm k}_j) = \sum_{\nu=2}^{\infty} \sum_{j=1}^{N_{\Sc l}} f_{\sigma}^{(\nu)}({\bm k}_j). \label{e144}
\end{equation}
Following Eq.~(\ref{e140}), the RHS of Eq.~(\ref{e144}) is vanishing. With reference to the Pringsheim theorem stated in Sec.~\ref{ss53s15}, the most significant aspect of the latter observation lies in its implication that the RHS of Eq.~(\ref{e144}) has a \emph{definite} value; that this definite value is equal to zero, although vitally relevant to the validity of the Luttinger-Ward identity, is wholly irrelevant to the validity of the equality in Eq.~(\ref{e144}).

In considering step (ii) in Sec.~\ref{ss53s13}, we showed that $\sum_{\nu=2}^{\infty} f_{\sigma}^{(\nu)}({\bm k})$ is a uniformly convergent series, $\forall {\bm k}$, of the bounded sequence $\{ f_{\sigma}^{(\nu)}({\bm k})\,\|\, \nu\}$, so that $\sum_{\nu=2}^{\infty} f_{\sigma}^{(\nu)}({\bm k})$ is bounded. Consequently, since $\mathcal{N}_{\Sc l}$ is finite, the double series $\sum_{j,\nu} f_{\sigma}^{(\nu)}({\bm k}_j)$ exists. This is established as follows: on account of the uniformity of convergence of $\sum_{\nu=2}^{\infty} f_{\sigma}^{(\nu)}({\bm k}_j)$ for all $j$, it follows that for an arbitrary $\epsilon> 0$ there exists an integer $n$ such that  (\S~3.31 in Ref.~\citen{WW62}, \S\S~44 and 83 in Ref.~\citen{TB65}, \S~66 in Ref.~\citen{EWH26})
\begin{equation}
\left| \sum_{\nu=n+1}^{n+q} f_{\sigma}^{(\nu)}({\bm k}_j) \right| <
\epsilon,\;\;\;\forall q\in \mathds{N}, \label{e145}
\end{equation}
implying, by the triangle inequality (item 3.2.6 in Ref.~\citen{AS72}), that for any $m\in \mathds{Z}^*$ and $p\in \mathds{N}$ for which $m+p \le \mathcal{N}_{\Sc l}$, one has
\begin{equation}
\left| \sum_{j=m+1}^{m+p} \sum_{\nu=n+1}^{n+q}
f_{\sigma}^{(\nu)}({\bm k}_j) \right| \le \sum_{j=m+1}^{m+p} \left|
\sum_{\nu=n+1}^{n+q} f_{\sigma}^{(\nu)}({\bm k}_j) \right| <
p\,\epsilon \le \mathcal{N}_{\Sc l}\, \epsilon. \label{e146}
\end{equation}
Since $\mathcal{N}_{\Sc l}$ is finite, by writing $\epsilon =\epsilon'/\mathcal{N}_{\Sc l}$ one observes that for an arbitrary value of $\epsilon'>0$ there exists an $n_0$ for which LHS of Eq.~(\ref{e146}) is less that $\epsilon'$ for all $n>n_0$ and $m>0$ ($m<\mathcal{N}_{\Sc l}$). It thus follows that the \emph{double series} $\sum_{j,\nu} f_{\sigma}^{(\nu)}({\bm k}_j)$ indeed exists. For clarity, the left-most quantity in Eq.~(\ref{e146}) stands for the $\vert S_{m+p,n+q} - S_{m,n}\vert$ introduced in Eq.~(\ref{e143}) in connection with the Stolz theorem.

Since $\sum_{\nu=2}^{\infty} f_{\sigma}^{(\nu)}({\bm k}_j)$ is bounded for all $j \in \{1,2,\dots, \mathcal{N}_{\Sc l}\}$, the repeated sum on the LHS of Eq.~(\ref{e144}) exists for any finite $\mathcal{N}_{\Sc l}$. As we pointed out above, following Eq.~(\ref{e144}), the repeated sum on the RHS of Eq.~(\ref{e144}) also exists. Consequently, by the Pringsheim theorem, quoted in Sec.~\ref{ss53s15}, the equality in Eq.~(\ref{e144}) follows.

On account of the above observation, and in view of Eq.~(\ref{e140}), it follows that \emph{the Luttinger-Ward identity is valid for uniform GSs of finite systems.} This result applies irrespective of the strength of correlation in the underlying uniform GSs.

\subsubsection{Remarks}
\label{ss53s17}

Recently Kokalj and Prelov\u{s}ek \cite{KP07} numerically examined\footnote{With the aid of exact diagonalisation, using Lanczos technique and twisted boundary conditions.} the Luttinger theorem for the Hubbard Hamiltonian on \emph{finite} lattices in one and two space dimensions. They found no indication suggestive of failure of the Luttinger theorem up to vary large values of the on-site interaction energy $U$ with respect to the nearest-neighbour hopping parameter $t$. The contrary observation by the authors concerning the $t$-$J$ model is not relevant to us, since a variety of aspects pertinent to the proof of the Luttinger theorem need not be valid for the $t$-$J$ Hamiltonian, which is defined in terms of constrained, as opposed to canonical, creation and annihilation operators. We remark that even in the strong-coupling regime of the Hubbard Hamiltonian, the corresponding $t$-$J$ Hamiltonian does not fully account for terms proportional to $t^2/U$, as it discards a three-site term \cite{ED94,EE96}.

For completeness, earlier, using three different criteria, Putikka, Luchini and Singh \cite{PLS98} reported violation of the Luttinger theorem for the metallic GSs of the $t$-$J$ Hamiltonian in two space dimensions; the authors determined the required ${\sf n}_{\sigma}({\bm k})$, on which these three criteria are based, with the aid of a twelfth-order high-temperature series expansion of ${\sf n}_{\sigma}({\bm k})$. Here ${\sf n}_{\sigma}({\bm k})$ stands for the ensemble average, at temperature $T$, of the partial-number-density operator in the ${\bm k}$ space. We remark that one of the criteria adopted by the authors, namely that $\mathcal{S}_{\Sc f;\sigma}$ consisted of the locus of the ${\bm k}$ points for which $\lim_{T\to 0}{\sf n}_{\sigma}({\bm k}) = \frac{1}{2}$ (see Fig.~1 in Ref.~\citen{PLS98}) is not justified \cite{BF03,BF04}, a fact also indicated in Ref.~\citen{PLS98}. Interestingly, $\lim_{T\to 0}{\sf n}_{\sigma}({\bm k}) \not= \frac{1}{2}$ for a ${\bm k}\in \mathcal{S}_{\Sc f;\sigma}$ directly contravenes $\lim_{T\to 0} {\rm d} {\sf n}_{\sigma}({\bm k})/{\rm d}T = 0$ as being the criterion for ${\bm k}\in \mathcal{S}_{\Sc f;\sigma}$ \cite{RDC95} (this is another of the three criteria considered in Ref.~\citen{PLS98}).

\subsubsection{Macroscopic systems}
\label{ss53s18}

As we have indicated in Sec.~\ref{ss53s14}, macroscopic systems can be dealt with in two principally distinct ways, corresponding to taking the thermodynamic limit \emph{before} and \emph{after} evaluating the sums on the RHS of Eq.~(\ref{e138}); since the sum $\sum_{\bm k}$ on the LHS of Eq.~(\ref{e138}) is the \emph{outer} sum, the order in which thermodynamic limit is taken is of no consequence to the LHS of Eq.~(\ref{e138}). Evidently, prior to taking the thermodynamic limit both sides of Eq.~(\ref{e138}) have to be divided by an extensive quantity, such as the volume of the system, so that in this limit both sides of the resulting expression are finite.

\subsubsection{Preliminaries}
\label{ss53s19}

According to the conventional approach, thermodynamic limit is taken prior to evaluating the sums in Eq.~(\ref{e138}). This is achieved by effecting the transition
\begin{equation}
\frac{1}{V}\sum_{\bm k} f({\bm k}) \rightharpoonup \int \frac{{\rm
d}^dk}{(2\pi)^d}\; f({\bm k}), \label{e147}
\end{equation}
where $V$ is the macroscopic volume of the system and $f({\bm k})$ stands for $\sum_{\nu=2}^{\infty} f_{\sigma}^{(\nu)}({\bm k})$ and $f_{\sigma}^{(\nu)}({\bm k})$, $\nu=2,3,\dots$~. The integral on the RHS of Eq.~(\ref{e147}) is over the space obtained on taking the continuum limit of the set of discrete ${\bm k}$ points over which the sum on the LHS is carried out.

The expression on the RHS of Eq.~(\ref{e147}) is the leading-order asymptotic term in the asymptotic series expansion of the function on the LHS of Eq.~(\ref{e147}) for $V\to\infty$; under some general conditions concerning the behaviour of $f({\bm k})$,\footnote{For instance, for any finite $V$, $f({\bm k})$ must be capable of being meaningfully extended to the continuum of ${\bm k}$ points. It should be noted that for any finite $V$, the set $\{ {\bm k}\}$ is \emph{not} dense in any closed subset of $\mathds{R}^d$.} and assuming that the extensions of the system under consideration are of the same order of magnitude in all $d$ spatial dimensions, one can readily demonstrate that the next-to-leading-order term in this asymptotic series expansion scales like $1/V^{1+1/d}$ followed by a term scaling like $1/V^{1+2/d}$, etc.\footnote{These observations are relevant for the considerations related to $d=\infty$.} This is achieved by expressing $\sum_{\bm k}$ in terms of a \emph{repeated sum} $\sum_{k_1}\dots \sum_{k_d}$ and subsequently replacing each sum by the associated Euler-Maclaurin summation formula (items 23.1.30 and 25.4.7 in Ref.~\citen{AS72}, \S~7.21 in Ref.~\citen{WW62}), giving rise to a \emph{repeated integral} of which the integral in Eq.~(\ref{e147}) is the associated \emph{multiple integral}. Here $k_j$, $j=1,\dots,d$, stands for the $j$th coordinate of ${\bm k}$ with respect to the primitive basis vectors $\{ {\bm b}_1,\dots,{\bm b}_d\}$ \cite{Note7}. In Sec.~\ref{ss53s21} we shall discuss the conditions under which, for $d>1$, \emph{multiple integrals} are equal to their corresponding \emph{repeated integrals}.

Dividing both sides of Eq.~(\ref{e138}) by $V/(2\pi)^d$ and taking the thermodynamic limit, following Eq.~(\ref{e147}) one obtains that
\begin{equation}
\int {\rm d}^dk\; \sum_{\nu=2}^{\infty} f_{\sigma}^{(\nu)}({\bm k})
= \sum_{\nu=2}^{\infty} \int {\rm d}^dk\; f_{\sigma}^{(\nu)}({\bm
k}), \label{e148}
\end{equation}
where, on account of Eq.~(\ref{e140}), the RHS is identically vanishing. It is to be noted that in contrast to both sides of Eq.~(\ref{e138}), those of Eq.~(\ref{e148}) are intensive quantities.

Following Eq.~(\ref{e144}), the above-mentioned two ways of taking the thermodynamic limit are embodied by the following two expressions:
\begin{equation}
\lim_{\mathcal{N}_{\Sc l}\to\infty}\frac{1}{\mathcal{N}_{\Sc l}} \sum_{j=1}^{\mathcal{N}_{\Sc l}}
\sum_{\nu=2}^{\infty} f_{\sigma}^{(\nu)}({\bm k}_j) =
\sum_{\nu=2}^{\infty} \lim_{\mathcal{N}_{\Sc l}\to\infty} \frac{1}{\mathcal{N}_{\Sc l}} \sum_{j=1}^{\mathcal{N}_{\Sc l}} f_{\sigma}^{(\nu)}({\bm k}_j), \label{e149}
\end{equation}
\begin{equation}
\lim_{\mathcal{N}_{\Sc l}\to\infty}\frac{1}{\mathcal{N}_{\Sc l}} \sum_{j=1}^{\mathcal{N}_{\Sc l}}
\sum_{\nu=2}^{\infty} f_{\sigma}^{(\nu)}({\bm k}_j) = \lim_{\mathcal{N}_{\Sc l}\to\infty} \sum_{\nu=2}^{\infty} \frac{1}{\mathcal{N}_{\Sc l}} \sum_{j=1}^{\mathcal{N}_{\Sc l}} f_{\sigma}^{(\nu)}({\bm k}_j). \label{e150}
\end{equation}
For the case at hand, the expression in Eq.~(\ref{e149}) is, up to a bounded dimensional scaling factor  (in contrast to $V$, $\mathcal{N}_{\Sc l}$ is dimensionless), equivalent to the conventional expression in Eq.~(\ref{e148}). This follows from the observation that the sum with respect to $j$ on the LHS of Eq.~(\ref{e149}) is the Riemann sum of $\{\sum_{\nu=2}^{\infty} f_{\sigma}^{(\nu)}({\bm k}_j)\,\|\, j\}$, and that on the RHS of Eq.~(\ref{e149}) is the Riemann sum of $\{ f_{\sigma}^{(\nu)}({\bm k}_j)\,\|\, j\}$, both to be identified with the Riemann, or Darboux, integrals of the latter functions as $\mathcal{N}_{\Sc l}\to\infty$ (\S\S~331, 338 and 369 in Ref.~\citen{EWH27}). We point out that since the functions $\{\sum_{\nu=2}^{\infty} f_{\sigma}^{(\nu)}({\bm k}_j)\}$ and $\{ f_{\sigma}^{(\nu)}({\bm k}_j)\}$ are Riemann integrable,\footnote{Since both functions are continuous over the entire relevant ${\bm k}$ space, they are Jordan measurable (\S~142 in Ref.~\protect\citen{EWH27}), and thus Riemann integrable.} one does not need to be concerned about the possibility of the limits in Eq.~(\ref{e149}) not existing.

The expression in Eq.~(\ref{e150}) is valid on account of the validity of the expression in Eq.~(\ref{e144}) by the following reasoning: Eq.~(\ref{e144}) applies for \emph{any} finite $\mathcal{N}_{\Sc l}$ (demonstrated in Sec.~\ref{ss53s16}) and the division by $\mathcal{N}_{\Sc l}$ of both sides of Eq.~(\ref{e144}) removes the only cause of the failure Eq.~(\ref{e144}) in the limit $\mathcal{N}_{\Sc l}=\infty$.

In the light of the above observations, below we shall investigate the validity of the expression in Eq.~(\ref{e149}). We shall further present some details which are relevant to the expression in Eq.~(\ref{e148}).

\subsubsection{Proof}
\label{ss53s20}

Here we demonstrate the validity of the expression in Eq.~(\ref{e149}). This is readily achieved by visiting the three elements of the proof of the expression in Eq.~(\ref{e144}) which corresponds to $\mathcal{N}_{\Sc l} <\infty$.

As we have indicated in the previous section, the limits on both sides of Eq.~(\ref{e149}) exist; these limits amount to the Riemann integrals of Riemann-integrable functions. In fact, the RHS of Eq.~(\ref{e149}) being vanishing on account of Eq.~(\ref{e140}), the existence of the limit on the RHS of Eq.~(\ref{e149}) is \emph{a priori} established. Hence, in view of the Pringsheim theorem, discussed in Sec.~\ref{ss53s15}, it remains only to show that the \emph{double series} $\sum_{j,\nu} f_{\sigma}^{(\nu)}({\bm k}_j)/\mathcal{N}_{\Sc l}$ exists for $\mathcal{N}_{\Sc l}\to \infty$. That this is indeed the case follows directly from the counterpart of the inequalities in Eq.~(\ref{e146}) specific to the present case, for which one has
\begin{equation}
\left| \sum_{j=m+1}^{m+p} \sum_{\nu=n+1}^{n+q}
\!\!\frac{f_{\sigma}^{(\nu)}({\bm k}_j)}{\mathcal{N}_{\Sc l}} \right| \le \!\!\sum_{j=m+1}^{m+p} \left| \sum_{\nu=n+1}^{n+q}
\!\!\frac{f_{\sigma}^{(\nu)}({\bm k}_j)}{\mathcal{N}_{\Sc l}} \right| <
\frac{p}{\mathcal{N}_{\Sc l}}\,\epsilon. \label{e151}
\end{equation}
Similar to Eq.~(\ref{e146}), here $m+p \le \mathcal{N}_{\Sc l}$ so that $p\,\epsilon$ is maximally equal to $\mathcal{N}_{\Sc l}\,\epsilon$. Consequently, $p\,\epsilon/\mathcal{N}_{\Sc l}$ is maximally equal to $\epsilon$, no matter how large $\mathcal{N}_{\Sc l}$ may be. Hence, by the Stolz theorem (\S~2.5 in Ref.~\citen{WW62}), the double sum $\sum_{j,\nu} f_{\sigma}^{(\nu)}({\bm k}_j)/\mathcal{N}_{\Sc l}$ exists in the limit $\mathcal{N}_{\Sc l}=\infty$. It follows that the expression in Eq.~(\ref{e149}) is indeed valid. In the light of the equivalence of the expression in Eq.~(\ref{e149}) with that in Eq.~(\ref{e148}), the validity of the latter expression is thus also hereby established.

With reference to Eq.~(\ref{e140}), we have thus completed the proof of the Luttinger-Ward identity for uniform GSs in the thermodynamic limit.

\subsubsection{Remarks}
\label{ss53s21}

In Sec.~\ref{ss53s19} we indicated that for $d>1$ the \emph{multiple integral} on the RHS of Eq.~(\ref{e147}) corresponds to the \emph{repeated integral} $\int {\rm d}k_1 \dots \int {\rm d}k_d$, which,  by employing the Euler-Maclaurin summation formula, is deduced as the leading-order term in the asymptotic series expansion of the \emph{repeated sum} $\sum_{k_1}\dots \sum_{k_d}$ for $V\to\infty$. It is therefore relevant that we consider the conditions under which a multiple integral can be inequivalent to one of its repeated forms. To this end, we first consider the case of $d=2$. Subsequently we briefly deal with the case of a general $d\ge 2$.

Defining
\begin{equation}
\t{f}(k_1,k_2) \equiv f({\bm k}), \label{e152}
\end{equation}
one has the following \emph{double integral} and \emph{repeated integrals}:
\begin{equation}
I\equiv \int {\rm d}^2k\; f({\bm k}),\;\;\; I_{i,j}\equiv \int {\rm
d}k_i \!\int {\rm d}k_j\; \t{f}(k_1,k_2), \label{e153}
\end{equation}
where $(i,j) = (1,2), (2,1)$, and where for simplicity we assume that the double integral $I$ is over the bounded square region $[a,b]\times [a,b]$ and that $k_i, k_j \in [a,b]$. Assuming the integrals in Eq.~(\ref{e153}) to be Riemann integrals (Ch. VI in Ref.~\citen{EWH27}), on describing $I$, $I_{1,2}$ and $I_{2,1}$ as limits of the pertinent Riemann sums, one immediately observes the resemblance of $I$ with the Pringsheim sum $S$, $I_{1,2}$ with the sum by rows and $I_{2,1}$ with the sum by columns of the double series $\sum_{i,j} a_{i,j}$ considered in Sec.~\ref{ss53s15}. Since for a relatively well-behaved function $f({\bm k})$ the elements of the sequence $\{\t{f}(k_1^{(i)},k_2^{(j)})\,\|\, i,j\in \mathds{N}\}$ are not entirely independent, one expects that the conditions required for the equality of $I$, $I_{1,2}$ and $I_{2,1}$ may not be as stringent as is the case for the equality of the Pringsheim sum $S$ and the sums by rows and by columns of the double series $\sum_{i,j} a_{i,j}$ corresponding to an in principle arbitrary sequence $\{a_{i,j}\,\|\, i,j \in \mathds{N}\}$.

Indeed, it has been shown by Du Bois-Reymond and Schoenflies (\S\S~362 and 363 in Ref.~\citen{EWH27}), that for $f({\bm k})$ a bounded function, existence of $I$ implies both existence \emph{and} equality of $I_{1,2}$ with $I_{2,1}$. The converse of this result, namely that existence of either $I_{1,2}$ or $I_{2,1}$, or existence of both and $I_{1,2} = I_{2,1}$, would imply existence of $I$, is not necessarily true (\S~365 in Ref.~\citen{EWH27}). Since however the functions $\sum_{\nu=2}^{\infty} f_{\sigma}^{(\nu)}({\bm k})$ and $f_{\sigma}^{(\nu)}({\bm k})$ encountered in our present considerations are bounded and continuous functions of ${\bm k}$ over the entire relevant ${\bm k}$ space, it follows that on identifying $f({\bm k})$ with one or the other of the latter two functions, we have that $I$ exists \emph{and} $I = I_{1,2} = I_{2,1}$.

Ettlinger \cite{E26} has demonstrated that in $d$ dimensions the necessary and sufficient condition for the existence of the Riemann integral $\int {\rm d}^dk\; f({\bm k})$ of the bounded function $f({\bm k})$ over the hypercube $[a,b]\times\dots\times [a,b]$ ($d$ times) is the continuity of $f({\bm k})$ for almost all ${\bm k}$ inside this hypercube (that is, $f({\bm k})$ may be discontinuous over a subset of measure zero of this hypercube). Ettlinger \cite{E26} has further shown that existence of this integral implies existence and equality of all the possible repeated integrals ($d!$ in number) associated with $\int {\rm d}^dk\; f({\bm k})$. For $f({\bm k})$ Lebesgue integrable over a bounded measurable set in $d$ dimensions, and $\int {\rm d}^dk\; f({\bm k})$ considered to be the Lebesgue integral of $f({\bm k})$ over the latter set, $\int {\rm d}^dk\; f({\bm k})$ is equal to all its associated repeated Lebesgue integrals (\S\S~427-429 in Ref.~\citen{EWH27}). Similarly as in the case of $d=2$, for $d>2$ and $f({\bm k})$ identified with one or the other of $\sum_{\nu=2}^{\infty} f_{\sigma}^{(\nu)}({\bm k})$ and $f_{\sigma}^{(\nu)}({\bm k})$, one has that $\int {\rm d}^dk\; f({\bm k})$ both exists and is equal to all its corresponding repeated integrals.

The same results as presented above apply when $a=-\infty$ and/or $b=\infty$; the only additional requirement to be demanded is that the underlying integrals exist for all $a$ and $b$ as $a\to -\infty$ and/or $b\to\infty$. For the relevant details, the reader may consult Ref.~\citen{EWH27}, in particular \S~354 herein.

Hereby we have completed the proof that step (iii) is indeed admissible.

\subsection{Summary}
\label{ss54}

Above we have identified two instances in the original proof of the Luttinger theorem where the weak-coupling many-body perturbation theory plays an \emph{essential} role. Both instances rely on the series expansion of the proper self-energy $\t{\Sigma}_{\sigma}({\bm k};z)$ in terms of connected skeleton diagrams, each expressed in terms of the exact single-particle Green functions $\{ \t{G}_{\sigma'}({\bm k};z)\,\|\, \sigma'\}$.

We have explicitly demonstrated that the above-mentioned series for $\t{\Sigma}_{\sigma}({\bm k};z)$ is not only convergent, but is \emph{uniformly} convergent for almost all ${\bm k}$ and $z$. On the basis of this result, and of the analyticity of $\t{\Sigma}_{\sigma}({\bm k};z)$ for all $z$ away from the real axis, we have subsequently demonstrated that the three major mathematical operations which are fundamental to the proof by Luttinger and Ward \cite{LW60} of the Luttinger-Ward identity are fully justified; above we have referred to these operations as constituting steps (i), (ii) and (iii). Thus we have arrived at the conclusion that the Luttinger-Ward identity is an exact result which applies to the uniform GSs of \emph{all} systems, irrespective of the strength of correlation in the underlying GSs and independent of whether these GSs are metallic or insulating. We have further explicitly demonstrated that the latter conclusion applies both to finite and macroscopic systems.

In the light of the above observations, and of those made in Sec.~\ref{ss43}, we conclude that the Luttinger theorem is unreservedly valid for \emph{all} uniform GSs corresponding to Hamiltonians defined specifically on lattices and in terms of short-range two-body interaction potentials.

\subsection{Comments on two existing confirmations of the Luttinger theorem}
\label{ss55}

Here we briefly discuss two papers, by Oshikawa \cite{MO00} and Praz
\emph{et al.} \cite{PFKT05}, which have direct bearing on the
Luttinger theorem under consideration.

\subsubsection{The first paper}
\label{ss55s1}

In Ref.~\citen{MO00} Oshikawa presented an argument based on topological considerations, purporting to demonstrate, in a non-perturbative way \cite{Note8}, the validity of the Luttinger theorem for \emph{Fermi-liquid} metallic states. For reasons that we shall present below, the observation by Oshikawa is in fact axiomatically true in the framework of the Landau Fermi-liquid theory. To conform with the notation used in the literature to be cited below, in what follows we shall suppress nearly all spin indices and assume that $\sigma$ is included in ${\bm k}$.

Within the framework of the phenomenological theory of Landau for normal Fermi liquids \cite{PN66,PN64}, one encounters two ``ideal'' distributions of \emph{non-interacting} particles, denoted by $n_{\bm k}^0$ and $n_{\bm k}$. Of these, $n_{\bm k}^0$ corresponds to the distribution of $N$ non-interacting particles that through an adiabatic switching-on of interaction evolves into the distribution of particles in the $N$-particle \emph{GS} of the interacting Hamiltonian $\wh{H}$, that is $\{{\sf n}_{\sigma}({\bm k})\,\|\,\sigma\}$ (see Eqs.~(\ref{e1}) and (\ref{e2})). With $\wh{H}_0$ denoting the non-interacting part of $\wh{H}$, in general $n_{\bm k}^0$ does \emph{not} correspond to the distribution of particles in the $N$-particle \emph{GS} of $\wh{H}_0$. This is in particular the case for anisotropic models for which interaction in general leads to deformation of the Fermi surface corresponding to the GS of $\wh{H}_0$, a possibility that undermines the process of adiabatic evolution of the $N$-particle GS of $\wh{H}_0$ into the $N$-particle GS of $\wh{H}$ (Ch.~5, \S~7 in Ref.~\citen{PN64}). We point out that even for an isotropic $\wh{H}$, the GS of $\wh{H}$ may be anisotropic, implying a Pomeranchuk instability \cite{IIP58,PN66} of the isotropic Fermi surface corresponding to the GS of $\wh{H}_0$ at some intermediate value of the coupling constant of interaction. Hence, \emph{within the framework of the Landau theory, $n_{\bm k}^0$ is equal to unity inside the Fermi sea corresponding to the $N$-particle GS of $\wh{H}$ (not to be confused with $\wh{H}_0$) and equal to zero outside.} The necessity for this choice is reinforced by the way in which within the framework of Landau's Fermi-liquid theory excited states are formally constructed (see later). With reference to our statements in Sec.~\ref{ss21}, we point out that the Fermi sea of a \emph{normal}, or \emph{conventional}, Fermi liquid metallic state is \emph{closed} so that its \emph{boundary}, the Fermi surface, is equal to its \emph{frontier}.

The second ``ideal'' distribution function, $n_{\bm k}$, corresponds to the distribution of $N'$ non-interacting particles that through an adiabatic switching-on of interaction evolves into the distribution of particles corresponding to one's desired $N'$-particle \emph{eigenstate} of $\wh{H}$; the latter eigenstate may be the GS or an excited state of $\wh{H}$. One may have $N'=N$, $N' = N\pm 1$, $N' = N\pm 2$, etc. It is owing to the fundamental theoretical problem of associating an arbitrary eigenstate of $\wh{H}$ with an eigenstate of $\wh{H}_0$ through a reversible adiabatic process \footnote{Reversibility is implicit by the identification of the time $t=-\infty$ with the time $t=+\infty$. } (what one also refers to as `the one-to-one correspondence' between the two types of eigenstates) that Landau's phenomenological theory can in principle only apply to properties determined by the \emph{low-lying} eigenstates of $\wh{H}$.\footnote{Physically, the elementary excitations corresponding to highly-excited states are too short-lived (as a consequence of high degree of degeneracy, or near degeneracy, of highly-exited states) to be describable by means of an \emph{adiabatic} evolution of a pure eigenstate of $\wh{H}_0$. }

In the framework of the Landau theory, the function of central significance is
\begin{equation}
\delta n_{\bm k} \equiv n_{\bm k} - n_{\bm k}^0, \label{e154}
\end{equation}
as it characterises the excited states in relation to the $N$-particle GS of $\wh{H}$; in contrast, the functions $n_{\bm k}^0$ and $n_{\bm k}$ are merely formal mathematical tools that solely serve to endow $\delta n_{\bm k}$ with the status of being a \emph{reliable} characteristic of the low-lying excited states of $\wh{H}$ (recall `the one-to-one correspondence' to which we referred above). For example, $\delta n_{\bm k} \not\equiv 0$ but $\sum_{\bm k} \delta n_{\bm k} =0$ corresponds to an $N$-particle excited state of $\wh{H}$. Further, in considering the $(N+1)$-particle eigenstate of $\wh{H}$ corresponding to wavevector ${\bm k}_0$, where ${\bm k}_0$ is \emph{outside} the underlying Fermi sea, one has $\delta n_{\bm k} = \delta_{{\bm k},{\bm k}_0}$. Similarly, in dealing with the $(N-1)$-particle eigenstate of $\wh{H}$ corresponding to wavevector ${\bm k}_0$, where ${\bm k}_0$ is \emph{inside} the underlying Fermi sea, one has $\delta n_{\bm k} = -\delta_{{\bm k},{\bm k}_0}$. The eigenstates of $\wh{H}$ corresponding to the latter two $\delta n_{\bm k}$ functions are said to contain respectively a \emph{quasielectron} and a \emph{quasihole} with wavevector ${\bm k}_0$.

We should emphasise that although $n_{\bm k}^0$ and $n_{\bm k}$ are often referred to as ``the distribution of quasiparticles,'' \cite{PN66,PN64} the relationships between $n_{\bm k}^0$, $n_{\bm k}$ and quasiparticles are indirect ones: far from being the distribution of $N$ ($N'$) `quasiparticles', or directly describing $N$ ($N'$) `quasiparticles', $n_{\bm k}^0$ ($n_{\bm k}$) merely refers to the \emph{original} distribution of $N$ ($N'$) particles, prior to turning on, adiabatically, the full interaction amongst these $N$ ($N'$) particles; to put it differently, $n_{\bm k}^0$ ($n_{\bm k}$) refers to $N$ ($N'$) `real', as opposed to `quasi', non-interacting particles as encountered in some $N$-particle ($N'$-particle) eigenstate of $\wh{H}_0$ (not to be confused with $\wh{H}$). The same statements apply to $\delta n_{\bm k}$, which is often referred to as the ``distribution of excited quasiparticles'' \cite{PN66,PN64}.

Above we indicated that the non-interacting Fermi sea corresponding to $n_{\bm k}^{0}$ coincides, by construction, with that of the $N$-particle GS of $\wh{H}$. One observes that if this were not the case, one would be confronted with the fundamental problem of constructing low-lying excited states of $\wh{H}$. In the example that we just presented, one would be at a loss to construct excited $(N+1)$-particle states containing a quasielectron and excited $(N-1)$-particle states containing a quasihole; one would not know whether the location of the wavevector ${\bm k}_0$ would be appropriate for constructing, through an adiabatic process, a quasielectron or a quasihole. This fundamental difficulty is of the same nature as that of constructing the $N$-particle GS of $\wh{H}$ by means of an adiabatic evolution of the $N$-particle GS of $\wh{H}_0$ in the cases where the Fermi seas corresponding to the two GSs do not coincide (Ch.~5, \S~7 in Ref.~\citen{PN64}).

We have thus clarified the fundamental significance, within the framework of the Landau Fermi-liquid theory, of the equivalence of the Fermi sea specific to the eigenstate of $\wh{H}_0$ corresponding to $n_{\bm k}^0$ with that of the $N$-particle GS of $\wh{H}$. Consequently, within this framework, the Luttinger theorem under consideration is valid by construction, since the number of ${\bm k}$ points for which $n_{\bm k}^0 =1$ is by design equal to $N$. We should underline the fact that metallic states can be characterised as being conventional Fermi liquids by some specific analytic properties of the corresponding $\t{\Sigma}_{\sigma}({\bm k};z)$ (see Sec.~\ref{ss43} as well as Sec.~\ref{ssc2s1}). In this connection, our last statement does \emph{not} imply that those metallic GSs that are Fermi liquids in this analytic sense, must necessarily conform with the Luttinger theorem at hand; there is no \emph{a priori} reason for this to be the case, even though the Luttinger theorem does apply to these GSs. Rather, our above statement has direct bearing on the \emph{phenomenological} theory of Landau in which idealised theoretical considerations merely provide a framework in which to rationalise the elements of the theory; within this framework, the stated conditions are more stringent than strictly necessary.

Finally, we point out that Oshikawa's proof of the Luttinger theorem for Fermi-liquid metallic states relies on the application of an exact result, obtained through topological considerations (i.e. Eq.~(3) in Ref.~\citen{MO00}), to the eigenstate of $\wh{H}_0$ corresponding to $n_{\bm k}^0$: the result in Eq.~(6) of Ref.~\citen{MO00} follows from this application. In doing so, Oshikawa \cite{MO00} identified $n_{\bm k}^0$ with the distribution of $N$ non-excited quasiparticles, in contravention of the fact, emphasised above, that $n_{\bm k}^0$ and $n_{\bm k}$ do not directly describe the physical quasiparticles. By identifying $n_{\bm k}^0$ with the distribution of the $N$-particle GS of the physical quasiparticles, whatever they may be, one in fact identifies the exact (i.e. physical) $\t{G}_{\sigma}({\bm k};z)$ with the $\t{G}_{\sigma;0}({\bm k};z)$ introduced in Eq.~(\ref{e89}), in which $\t{\varepsilon}_{{\bm k};\sigma}$ coincides with the energy dispersion of an isolated quasiparticle characterised by $({\bm k},\sigma)$.\footnote{For the reason stated above, the energy dispersion $\t{\varepsilon}_{{\bm k};\sigma}$ by construction reproduces the exact Fermi surface $\mathcal{S}_{\Sc f;\sigma}$ corresponding to the $N$-particle GS of $\wh{H}$.} The self-energy corresponding to $\t{G}_{\sigma;0}({\bm k};z)$, i.e. $\hbar^{-1} v_{\sigma}({\bm k})$, being independent of $z$, the Luttinger theorem as applied to $\t{G}_{\sigma;0}({\bm k};z)$ is automatically satisfied (Sec.~\ref{ss52s3}).

\subsubsection{The second paper}
\label{ss55s2}

Praz \emph{et al.} \cite{PFKT05,AP04} demonstrated, in a perturbative sense (see below), the validity of the Luttinger theorem for (Fermi-liquid) metallic states in two and three space dimensions. Explicitly, for $d=2$ and $3$, and under some restrictive conditions \cite{PFKT05,AP04}, the authors demonstrated two main results:
\begin{itemize}
\item[(1)] that the `volume' enclosed by the Fermi surface of a Fermi-liquid metallic state is an analytic function of the coupling constant $\lambda$ of the two-body interaction potential in a neighbourhood of $\lambda=0$, allowing for the possibility that the radius $\lambda_{\rm r}$ of this region may be, if non-vanishing, smaller than the physical value $\lambda_{\rm p}$ of the coupling constant of interaction, and,
\item[(2)] that the coefficient of $\lambda^n$ in the Taylor series expansion of this `volume' function at $\lambda=0$ is vanishing for all $n$ excluding $n=0$.
\end{itemize}
Thus although $\lambda_{\rm p}$ may lie outside the domain of validity of the above-mentioned Taylor series, the authors have shown that for $d=2$ and $3$, and $\lambda <\lambda_{\rm r}$, the `volume' enclosed by the Fermi surface of an interacting (Fermi-liquid) metallic state is strictly equal to its value corresponding to $\lambda=0$.

The work by Praz \emph{et al.} \cite{PFKT05,AP04} is remarkable in that by misreading \cite{Note9} the work by Luttinger and Ward \cite{LW60}, the authors have set out and solved problems that Luttinger and Ward \cite{LW60} have in fact circumnavigated. It is therefore not the approach by Praz \emph{et al.} \cite{PFKT05,AP04} that we intend to criticise here, but their imprecise description of the original proof of the Luttinger theorem at hand. For instance, the statements in Ref.~\citen{PFKT05} that ``From a mathematical point of view, the proof given by Luttinger is unsatisfactory. The manipulations of conditionally convergent integrals that lead to Luttinger's result have to be controlled rigorously.'' have no bearing, whatever, on the actual proof of the Luttinger theorem by Luttinger and Ward \cite{LW60}.

To clarify the above statement, we recall that Luttinger and Ward \cite{LW60} (or Luttinger \cite{JML60} for that matter) do \emph{not} employ an \emph{explicit} series expansion in powers of the interaction strength $\lambda$; the only expansion on which the original proof of the Luttinger theorem crucially depends is an \emph{implicit} series expansion for $\t{\Sigma}_{\sigma}({\bm k};z)$ in terms of skeleton self-energy diagrams, involving the \emph{interacting} single-particle Green functions $\{\t{G}_{\sigma'}({\bm k};z)\,\|\, \sigma'\}$. By employing this series expansion, Luttinger and Ward \cite{LW60} avoided two problems which Praz \emph{et al.} \cite{PFKT05,AP04} explicitly confronted and dealt with in their approach:

Firstly, following the exposition in Sec.~\ref{ss53s1}, it is evident that by relying on skeleton self-energy diagrams one circumvents all self-energy contributions whose `inner kernels' can be non-integrably singular in the thermodynamic limit (contrast this with the `conditionally convergent integrals' referred to by Praz \emph{et al.}\cite{PFKT05}).

Secondly, since in the considerations by Luttinger and Ward \cite{LW60} skeleton self-energy diagrams are expressed in terms of $\{ \t{G}_{\sigma'}({\bm k};z)\,\|\, \sigma'\}$, the perturbative contributions to $\t{\Sigma}_{\sigma}({\bm k};z)$ encountered in these considerations take full account of the \emph{exact} Fermi surface (should the underlying GS be metallic). Consequently, such remarks as \cite{AP04} ``The divergences as appearing in the Green functions at fixed $\lambda$ are not physical, but reflect the fact that the Fermi surface of a system of interacting fermions gets distorted. The diagrammatic expansion is then performed in the vicinity of the wrong surface of singularities.'' can have no bearing on the approach by Luttinger and Ward in Ref.~\citen{LW60}. In other words, the renormalization procedure that is central to the proof by Praz \emph{et al.} \cite{PFKT05,AP04}, is not required in the context of the original proof of the Luttinger theorem; the series expansion as employed by Luttinger and Ward \emph{is} around the `distorted' Fermi surface (for an exposition of the problems associated with the perturbation expansion around an `undistorted' Fermi surface, the reader may consult Ch.~5, \S~7 in Ref.~\citen{PN64}).

To summarise, the perturbative proof as provided by Praz \emph{et al.} \cite{PFKT05,AP04} concerning the Luttinger theorem in essence coincides with the original proof of this theorem by Luttinger and Ward \cite{LW60}. Further, the criticism by Praz \emph{et al.} \cite{PFKT05,AP04} of the proof by Luttinger and Ward is unjustified, as this criticism relates to an approach that is not Luttinger and Ward's, but is only perceived to be theirs.

\section{Examination of some observations regarding ``breakdown'' of the Luttinger theorem}
\label{s6}

In this section we examine a number of papers that report breakdown of the Luttinger theorem. We consider these papers under five different headings, marked as case I, case II, etc. In dealing with each of these cases, we focus on one primary publication and, where appropriate, make reference to other related publications. The cases that we explicitly consider in this section, expose a multiplicity of mechanisms that result in failure of the Luttinger theorem; as will become evident, all of these mechanisms contravene some basic principles underlying the proof of the Luttinger theorem. Case I is however somewhat exceptional, in that in the local (that is, zero-hopping) limit and the non-local but symmetric cases, the observed breakdowns are genuine from the perspective of the conventional wisdom, according to which for insulating GSs the chemical potential $\mu$ may be chosen to be located anywhere inside the finite gap $(\mu_N^-,\mu_N^+)$.

\subsection{Case I}
\label{ss61}

Here we consider a recent work by Rosch \cite{AR06} concerning breakdown of the Luttinger theorem for some Mott-insulating GSs. As mentioned above, for the local limit and the symmetric non-local cases, the observation by Rosch \cite{AR06} with regard to breakdown of the Luttinger theorem is entirely valid; the observed breakdown of the Luttinger theorem in the asymmetric non-local cases \cite{AR06} is however caused by a mechanism that we shall show to be external to the Luttinger theorem (Sections \ref{ss61s4} and \ref{ss61s8} --- \ref{ss61s13}).

Below we shall demonstrate that the breakdown of the Luttinger theorem as detected by Rosch \cite{AR06} signals a hitherto unacknowledged problem which, as we have indicated earlier (Sections \ref{s1} and \ref{ss41}), is not internal to the Luttinger theorem, but emerges from a \emph{false} zero-temperature limit that in principle can arise in the case of any interacting insulating GS on taking the zero-temperature limit for $\mu\not=\mu_{\infty}$ (cf. Eq.~(\ref{e25})). By explicitly considering the behaviour of the function $\b{N}_{\sigma}^{(2)}$, Eqs.~(\ref{e34}) and (\ref{e36}), we shall rigorously establish that the Luttinger-Ward identity, Eq.~(\ref{e44}), is obtained by equating $\mu$ with either $\mu_{\beta} \equiv\mu(\beta,N,V)$ or $\mu_{\infty}$ prior to effecting the zero-temperature limit $\beta\to\infty$.

The existence of the above-mentioned false limit is not specific to the Luttinger theorem; in principle, in determining limits of all multi-variable functions, one has to make distinction between the so-called \emph{multiple limit} and a variety of \emph{repeated limits} (\S\S~302-306 in Ref.~\citen{EWH27}); we have already encountered these limits in Sec.~\ref{ss53s15}, while dealing with \emph{multiple series} and \emph{repeated series}, as well as in Sec.~\ref{ss53s21}, in considering \emph{multiple integrals} and \emph{repeated integrals}.

As mentioned above, we shall further establish the reason underlying the apparent failure of the Luttinger theorem in the asymmetric non-local cases, for which one has $\mathcal{D}(-\omega) \not\equiv \mathcal{D}(\omega)$. Here $\mathcal{D}(\omega)$ denotes the density-of-states function corresponding to the energy dispersion $\tau_{\bm k}$ associated with particle hopping; in the local limit $\tau_{\bm k} \equiv 0$, $\forall {\bm k}$. Briefly, we shall rigorously demonstrate that irrespective of how minute the deviation of $\mathcal{D}(-\omega)$ from $\mathcal{D}(\omega)$ may be, such deviation transforms the insulating GS, as calculated to leading order in $\tau_{\bm k}$, into a metallic GS (assuming that one insists on the GS to remain half-filled). Technically, for $\mathcal{D}(-\omega) \not\equiv \mathcal{D}(\omega)$ the chemical potential $\mu_{\infty}$ corresponding to half-filling comes to be located near one the poles of $\t{G}_{\sigma}({\bm k};z)$, a neighbourhood along the real energy axis where, as conceded by Rosch \cite{AR06}, the regular perturbation expansion for $\t{G}_{\sigma}({\bm k};z)$, employed in Ref.~\citen{AR06}, has no validity.

To make contact with the notation in Ref.~\citen{AR06}, in this section we identify $\hbar$ with unity (so that $\varepsilon\equiv \omega$) and suppress the spin index $\sigma$; we do \emph{not} evaluate the trace with respect to $\sigma$, but assume that the mean values of the numbers of particles of both spin species are equal.\footnote{The practice of leaving out the trace with respect to $\sigma$ leads to some inconvenience in a context where in addition to an $N$-particle GS one deals with $N\pm 1$-particle GSs (appendix \protect\ref{sc}). Our present choice is therefore not ideal and amounts to a compromise.} As elsewhere in this paper, however, our notation in this section will conform with that of Luttinger and Ward's \cite{LW60} in considering $\mathscr{G}_{\sigma}$ and $\mathscr{S}_{\!\sigma}$ as explicit functions of $\zeta_m$, instead of $i\hbar\omega_m \equiv \zeta_m -\mu$ (see Sec.~\ref{ss30s1}). For clarity and later reference, we point out that what Rosch \cite{AR06} refers to as the zero-temperature limit of the chemical potential (corresponding to half-filling) in the `canonical ensemble' (denoted in Ref.~\citen{AR06} by $\mu_n$), coincides with the zero-temperature limit $\mu_{\infty}$ of $\mu(\beta,N,V)$ in the grand-canonical ensemble (Sec.~\ref{ss23}).

\subsubsection{Preliminaries (local limit)}
\label{ss61s1}

In Ref.~\citen{AR06} Rosch considered the insulating phase of a two-band Hubbard Hamiltonian at half filling. In the local limit, and for a specific choice of the various parameters of the model, the single-particle Green function $G_{\rm loc}(\omega)$ and the self-energy $\Sigma_{\rm loc}(\omega)$ have the forms \cite{AR06}
\begin{equation}
G_{\rm loc}(\omega) = \frac{1}{2} \Big( \frac{1}{\omega - \t{U}/2} +
\frac{1}{\omega+\t{U}/2} \Big), \label{e155}
\end{equation}
\begin{equation}
\Sigma_{\rm loc}(\omega) = \frac{\t{U}}{2} +
\frac{(\t{U}/2)^2}{\omega}, \label{e156}
\end{equation}
where
\begin{equation}
\t{U} = U + \frac{3}{2}\, J, \label{e157}
\end{equation}
in which $U$ is the on-site interaction energy and $J$ the spin-exchange-coupling energy.

Rosch obtained that \cite{AR06} (below, the last equality applies only for $\vert\mu\vert <\t{U}/2$):
\begin{equation}
\int_{\mu-i\infty}^{\mu+i\infty} \frac{{\rm d}z}{2\pi i}\;
\t{G}_{\rm loc}(z) \frac{\partial}{\partial z} \t{\Sigma}_{\rm
loc}(z) \equiv -\int_{\mu-i\infty}^{\mu+i\infty} \frac{{\rm d}z}{2\pi i}\; \t{\Sigma}_{\rm loc}(z) \frac{\partial}{\partial z} \t{G}_{\rm
loc}(z) = \frac{1}{2}\, \mathrm{sgn}(\mu), \label{e158}
\end{equation}
implying violation of the Luttinger-Ward identity, Eq.~(\ref{e44}), for $\mu\not=0$. As should be evident from the general considerations in Sections \ref{ss23} and \ref{ss41}, outwardly the validity of none of the steps leading to the Luttinger-Ward identity appears to be dependent on any other restriction on $\mu$ than $\mu\in (\mu_N^-, \mu_N^+)$. From this perspective, the deviation from zero of the RHS of Eq.~(\ref{e158})
for all $\mu \in (-\t{U}/2,\t{U}/2)$, except $\mu=0$, can indeed be viewed as the failure of the Luttinger-Ward identity. Since in the case at hand the violation of the Luttinger-Ward identity is accompanied by a concomitant violation of the Luttinger theorem (Sec.~\ref{ss43}),\cite{Note10} it follows that for $\mu\not=0$, however $\mu \in (-\t{U}/2,\t{U}/2)$, the Luttinger theorem breaks down in the local limit. In the next section we shall establish the reason underlying these failures.

\subsubsection{The problem of false zero-temperature limits}
\label{ss61s2}

Using the first expression on the RHS of Eq.~(\ref{e21}), for the Luttinger number per site and per spin species, $n_{\Sc l}$, one has
\begin{equation}
n_{\Sc l} = \frac{2}{\mathcal{N}_{\Sc l}} \sum_{\bm k} \Theta\big(\mu - (\t{U}/2)^2/\mu\big) \equiv  2\,\Theta\big(\mu - (\t{U}/2)^2/\mu\big),
\label{e159}
\end{equation}
where $\mathcal{N}_{\Sc l}$ denotes the number of lattice sites, not to be confused with the total Luttinger number $N_{\Sc l} \equiv \sum_{\sigma} N_{\Sc l;\sigma}$. The argument $\mu - (\t{U}/2)^2/\mu$ is
\begin{itemize}
\item[(1)] positive for $-\t{U}/2 <\mu <0$ and $\mu>\t{U}/2$,
\item[(2)] zero at $\mu=\pm\t{U}/2$,
\item[(3)] negative for $\mu<-\t{U}/2$, $0<\mu<\t{U}/2$, and
\item[(4)] \emph{undefined} for $\mu=0$.
\end{itemize}
If instead of the first expression on the RHS of Eq.~(\ref{e21}) we had used the second expression, we would be confronted with a situation where the argument of the $\Theta$ function would be equal to zero at $\mu=0$ but undefined at $\mu=\pm\t{U}/2$. Use of the first expression on the RHS of Eq.~(\ref{e21}) is appropriate, since it is the primary expression for $N_{\Sc l;\sigma}$; the second expression on the RHS of Eq.~(\ref{e21}) is a derived one (Sec.~(\ref{ss43}).

The expression in Eq.~(\ref{e159}) exposes the peculiarity that $\mu=0$, which is the only value of $\mu \in (\mu_{N}^-,\mu_{N}^+)$ for which the Luttinger-Ward identity, Eq.~(\ref{e158}), is satisfied, is the very value for which the argument of the pertinent $\Theta$ function is undefined. Evidently, were it not for the local nature of the self-energy $\Sigma_{\rm loc}(\omega)$ and of the corresponding Green function $G_{\rm loc}(\omega)$, we would not be confronted with such a dire condition where a subset of measure zero of $\mathds{R}$, consisting of the single real number $0$, is both significant for the validity of the Luttinger-Ward identity and over which $n_{\Sc l}$ is apparently ill-defined. These observations signify, firstly, the peculiarity of the local limit and, secondly and more importantly, the significance of the order in which various limits are taken. The \emph{repeated limit} that proves appropriate for the calculation of $n_{\Sc l}$ (or equivalently, of $\b{N}_{\sigma}^{(1)}$ for $\beta\to\infty$, Eqs.~(\ref{e34}), (\ref{e35}) and (\ref{e65})) is naturally related to the \emph{repeated limit} that is appropriate for the calculation of $\b{N}_{\sigma}^{(2)}$.

Noting that the first $\Theta$ function in Eq.~(\ref{e159}) is the limit of the function on the RHS of Eq.~(\ref{e53}), in which $\mu+i 0^+$ is the \emph{limit} of $z=\mu+i\eta$ for $\eta\downarrow 0$, for the $n_{\Sc l}$ in the case at hand we obtain that
\begin{equation}
n_{\Sc l} = -\frac{2}{\pi} \lim_{\eta\downarrow 0} \mathrm{Arctan}\Big( \frac{ (\mu^2 + \eta^2) + (\t{U}/2)^2}{(\mu^2+\eta^2) - (\t{U}/2)^2} \,\frac{\eta}{\mu}\Big). \label{e160}
\end{equation}
This expression clearly prescribes that the value of $n_{\Sc l}$ for in particular $\mu=0$ is to be obtained by taking the limit $\eta\downarrow 0$ \emph{after} having effected $\mu\to 0$; we draw attention to the fact that for an arbitrary small but non-vanishing value of $\eta$, $n_{\Sc l}$ is a continuous function of $\mu$ at $\mu=0$. This observation is of considerable significance in that it shows that if one substitutes $\mu_{\beta}\equiv \mu(\beta,N,V)$ for all the incidents of $\mu$ in Eq.~(\ref{e160}), and if $\mu_{\beta} \sim 0$ for $\beta\to\infty$, the correct value for $n_{\Sc l}$ corresponds to the \emph{repeated limit} according to which the limit $\beta\to\infty$ is effected \emph{prior} to effecting the limit $\eta\downarrow 0$ in Eq.~(\ref{e160}). Remarkably, following this procedure, one obtains the same value for $n_{\Sc l}$ irrespective of the specific way in which $\mu_{\beta}$ approaches zero for $\beta\to\infty$ (assuming that $\mu_{\beta}$ indeed approaches zero; see Sec.~\ref{ss61s3}); this is due to the fact that Eq.~(\ref{e53}), and thus Eq.~(\ref{e160}), is in terms of the Green function corresponding to the $N$-particle \emph{GS} of $\wh{H}$ (cf. Eq.~(\ref{e48})); owing to the requirement of dealing with the real and imaginary parts of the \emph{inverse} of $\mathscr{G}_{\sigma}({\bm k};z)$, explicit determination of the low-temperature asymptotic series expansion of the first term on the RHS of Eq.~(\ref{e48}) is disproportionately cumbersome, leading us to avoid such undertaking.

Since in the case at hand, the Luttinger theorem applies \emph{if and only if} the Luttinger-Ward identity applies (Sec.~\ref{ss43}), from the above observation it follows that in calculating $\b{N}_{\sigma}^{(2)}$ one must unmistakably obtain the exact result $\b{N}_{\sigma}^{(2)}=0$ in the zero-temperature limit by identifying the $\mu$ on which $\b{N}_{\sigma}^{(2)}$ explicitly depends with $\mu_{\infty}$; we shall show that $\mu_{\infty}$ is equal to zero. With the aid of a closed algebraic expression for $\b{N}_{\sigma}^{(2)}$, below we shall demonstrate that this is indeed the case. This algebraic expression will further shed light on the way in which the repeated limits
\begin{equation}
\lim_{\mu\to\mu_{\infty}} \lim_{\beta\to\infty}\;\;\;\; \mbox{\rm and}\;\;\;\;  \lim_{\beta\to\infty} \lim_{\mu\to\mu_{\infty}}
\nonumber
\end{equation}
differ.

In the case at hand and for sufficiently large $\beta$, $\b{N}_{\sigma}^{(2)}$ can be calculated to exponential accuracy without recourse to the expression in Eq.~(\ref{e41}). For the counterpart of Eq.~(\ref{e158}) one has
\begin{eqnarray}
\frac{1}{\beta} \sum_{m} \t{G}_{\rm loc}(\zeta_m)\,\frac{\partial \tilde{\Sigma}_{\rm loc}(\zeta_m)}{\partial \zeta_m} &=& \frac{1}{4} \Big[2\tanh\big(\frac{\beta\mu}{2}\big) - \tanh\big(\frac{\beta}{2} (\mu -\frac{\t{U}}{2})\big) -\tanh\big(\frac{\beta}{2} (\mu +\frac{\t{U}}{2})\big)\Big]\nonumber\\
&\equiv& \Phi(\beta\mu,\beta\t{U}). \label{e160a}
\end{eqnarray}
The equality in this expression holds for \emph{all} $\beta$. We point out that $\Phi(\beta\mu,\beta\t{U})$ differs from the local counterpart of $\b{\nu}_{\sigma}^{(2)}({\bm k})$, Eq.~(\ref{e36}), only in that $\Phi(\beta\mu,\beta\t{U})$ is defined in terms of the zero-temperature counterparts of $\mathscr{G}_{\sigma}({\bm k};\zeta_m)$ and $\mathscr{S}_{\!\sigma}({\bm k};\zeta_m)$ in the local limit. As will become evident in Sec.~\ref{ss61s3}, for $\mu\in (-\t{U}/2,\t{U}/2)$ and $\mu$ away from $\pm\t{U}/2$, $\Phi(\beta\mu,\beta\t{U})$ differs by an exponentially small correction from the $\b{\nu}_{\sigma}^{(2)}({\bm k})$ pertaining to the local limit, or indeed from $\b{N}_{\sigma}^{(2)}/\mathcal{N}_{\Sc l}$, for sufficiently large $\beta$. Thus by assuming that $\vert\mu\vert \ll \t{U}/2$, not only is $\Phi(\beta\mu,\beta\t{U})$ to exponential accuracy equal $\b{N}_{\sigma}^{(2)}/\mathcal{N}_{\Sc l}$, but also
\begin{equation}
\Phi(\beta\mu,\beta\t{U}) \sim \Psi(\beta\mu) \equiv \frac{1}{2} \tanh\big(\frac{\beta\mu}{2}\big)\;\;\; \mbox{\rm for}\;\;\; \beta\to\infty. \label{e160b}
\end{equation}
For $\beta\to\infty$, the deviation of $\Psi(\beta\mu)$ from $\Phi(\beta\mu,\beta\t{U})$ is comparable with that of $\Phi(\beta\mu,\beta\t{U})$ from $\b{N}_{\sigma}^{(2)}/\mathcal{N}_{\Sc l}$.

The results in Eqs.~(\ref{e160a}) and (\ref{e160b}) clearly expose the problematic nature of the zero-temperature limit: the scale on which $\beta$ can be considered large, is determined by the largest of $1/\vert\mu\vert$ and $1/\t{U}$, so that for $\mu\to 0$ no finite $\beta$ may be considered large. This problem disappears if in the framework of the Luttinger theorem concerning insulating GSs, one identifies $\mu \in (\mu_N^-,\mu_N^+)$ with $\mu_{\infty}$ or $\mu_{\beta} \equiv \mu(\beta,N,V)$:\footnote{Assuming that $\beta$ is sufficiently large so that $\mu_{\beta}\in (\mu_N^-,\mu_N^+)$.} we shall demonstrate that, for the case at hand $\mu_{\beta} = o(1/\beta)$ as $\beta\to\infty$  (Eq.~(\ref{e183}) below), so that indeed $\Psi(\beta\mu_{\beta}) \sim 0$ for $\beta\to\infty$.

We are now in a position to consider the validity or otherwise of the Luttinger theorem in the local limit. To do so we determine $\mu_{\beta}$ for large values of $\beta$.

\subsubsection{The Luttinger theorem (the local limit)}
\label{ss61s3}

For the single-particle spectral function $A({\bm k};\omega)$, Eq.~(\ref{ec40}), in the local limit, which we denote by $A_{\rm loc}(\omega)$, one has
\begin{equation}
A_{\rm loc}(\omega) = \frac{1}{2}\,\delta(\omega-\t{U}/2) + \frac{1}{2}\, \delta(\omega+\t{U}/2). \label{e161}
\end{equation}
Following Eqs.~(\ref{e1}) and (\ref{e2}), in the local limit for the number of particles per site \emph{per spin species} at zero temperature, $n$, one obtains that
\begin{equation}
n \equiv 2\!\int_{-\infty}^{\mu} {\rm d}\omega\; A_{\rm loc}(\omega)
= \Theta(\mu+\t{U}/2) + \Theta(\mu-\t{U}/2), \label{e162}
\end{equation}
where the pre-factor $2$ accounts for the trace over the two \emph{orbital indices} in the model under consideration. The expression in Eq.~(\ref{e162}) implies that in the half-filled case under consideration, corresponding to $n=1$ (recall that we do not take trace over spin indices), $\mu$ can be assigned any value inside the interval $(-\t{U}/2,\t{U}/2)$. From this perspective, it appears that in general the Luttinger-Ward identity should be violated in the local limit.\cite{AR06}

Following Eq.~(\ref{e31}), for the mean value of the number of particles corresponding to arbitrary $\beta$, $\b{n}_{\beta}$, one has
\begin{equation}
\b{n}_{\beta} \equiv 2\int_{-\infty}^{\infty} {\rm d}\omega\; \frac{\mathscr{A}_{\rm loc}(\omega)}{\mathrm{e}^{\beta (\omega-\mu)}+1}, \label{e163}
\end{equation}
where
\begin{equation}
\mathscr{A}_{\rm loc}(\omega) {:=} \frac{1}{\mathcal{N}_{\Sc l}} \sum_{\bm k} \mathscr{A}_{\sigma}({\bm k};\omega), \label{e163a}
\end{equation}
in which\footnote{We suppress $\sigma$ in $\mathscr{A}_{\rm loc}(\omega)$ for uniformity of presentation.} $\mathscr{A}_{\sigma}({\bm k};\omega)$ is the thermal single-particle spectral function, Eq.~(\ref{e32}); in the present case, $A_{\rm loc}(\omega)$ is the zero-temperature limit of $\mathscr{A}_{\sigma}({\bm k};\omega)$ (cf. Eq.~(\ref{e33})). We should emphasise that although it is tempting to assume that in the case at hand $\mathscr{A}_{\sigma}({\bm k};\omega)$ were local, for the reason that will become apparent below, this is generally not the case. In view of  $\frac{1}{\mathcal{N}_{\Sc l}} \sum_{\bm k} 1 = 1$, our following analysis applies irrespective of whether $\mathscr{A}_{\sigma}({\bm k};\omega)$ is local or otherwise.

In appendix \ref{sf} we present the first three leading-order terms in the low-temperature asymptotic series expansion of $\mathscr{G}_{\sigma}({\bm k};z)$, from which one obtains that (cf. Eqs.~(\ref{ef141}) and (\ref{ef142}))
\begin{equation}
\mathscr{A}_{\sigma}({\bm k};\varepsilon) \sim A_{N;\sigma}({\bm k};\varepsilon) + B_{N;\sigma}^+({\bm k};\varepsilon)\,\mathrm{e}^{-\beta (\mu_{N}^+ -\mu)} +B_{N;\sigma}^-({\bm k};\varepsilon)\,\mathrm{e}^{-\beta (\mu-\mu_{N}^-)}\;\,\mbox{\rm for}\;\, \beta\to\infty, \label{e164}
\end{equation}
where $\mu_{N}^{\pm}$ are defined in appendix \ref{sc}, and $A_{N;\sigma}({\bm k};\varepsilon)$ is the single-particle spectral function corresponding to the $N$-particle GS of the system under investigation (thus far we have denoted this function as $A_{\sigma}({\bm k};\varepsilon)$); the functions $B_{N;\sigma}^{\pm}({\bm k};\varepsilon)$ are defined according to
\begin{equation}
B_{N;\sigma}^+({\bm k};\varepsilon) {:=} A_{N+1;\sigma}({\bm k};\varepsilon) - A_{N;\sigma}({\bm k};\varepsilon), \label{e165}
\end{equation}
\begin{equation}
B_{N;\sigma}^-({\bm k};\varepsilon) {:=} A_{N-1;\sigma}({\bm k};\varepsilon) - A_{N;\sigma}({\bm k};\varepsilon). \label{e166}
\end{equation}
Evidently, $B_{N\pm 1;\sigma}^{\mp}({\bm k};\varepsilon) \equiv -B_{N;\sigma}^{\pm}({\bm k};\varepsilon)$. On general grounds, one can demonstrate that, with the underlying $N$-particle GS being an insulating state, the $N\pm 1$-particle GSs to which $A_{N\pm 1;\sigma}({\bm k};\varepsilon)$ correspond are metallic; the Fermi energies corresponding to these states are up to infinitesimal corrections equal to $\mu_{N}^{\pm}$ respectively (appendix \ref{sf}). This in turn implies that even though for an insulating $N$-particle GS $A_{N}({\bm k};\varepsilon)$ may be accurately described by a local function, in general the associated functions $A_{N\pm 1;\sigma}({\bm k};\varepsilon)$ cannot be accurately described by local functions. Therefore $B_{N;\sigma}^{\pm}({\bm k};\varepsilon)$ are in general non-trivial functions of ${\bm k}$, rendering $\mathscr{A}_{\sigma}({\bm k};\varepsilon)$ a non-trivial function of ${\bm k}$ for all $\beta <\infty$, even for $A_{N;\sigma}({\bm k};\varepsilon) \equiv A_{\rm loc}(\varepsilon)$ (cf. Eq.~(\ref{e164})).

In the local limit, for $\mu_{N}^-$ and $\mu_{N}^+$ one has
\begin{equation}
\mu_{N}^{\pm} = \pm \frac{1}{2} \t{U}, \label{e167}
\end{equation}
so that for $\mu_{N}^- < \mu <\mu_{N}^+$ and $\beta\to\infty$
\begin{equation}
\mathscr{A}_{\rm loc}(\omega) \sim A_{\rm loc}(\omega) + \mathrm{e}^{-\beta\t{U}/2}\,\big(B_{\rm loc}^+(\omega)\, \mathrm{e}^{\beta\mu}
+ B_{\rm loc}^-(\omega) \,\mathrm{e}^{-\beta\mu}\big), \label{e168}
\end{equation}
where for conciseness we have suppressed the subscript $N$ associated with the functions $A_{N;\rm loc}(\omega)$ and $B_{N;\rm loc}^{\pm}(\omega)$. The functions $B_{N;\rm loc}^{\pm}(\omega)$ are
the averages with respect to ${\bm k}$ of $B_{N;\sigma}^{\pm}({\bm k};\varepsilon)$ (cf. Eq.~(\ref{e163a})). Considering
\begin{equation}
\epsilon \equiv \mathrm{e}^{-\beta\t{U}/2} \label{e169}
\end{equation}
as the small parameter of expansion, the expression on the RHS of Eq.~(\ref{e168}) is exactly equal to $\mathscr{A}_{\rm loc}(\omega)$ to linear order in $\epsilon$.

From the expression in Eq.~(\ref{e163}) one obtains that
\begin{equation}
\b{n}_{\beta} = \frac{1}{\mathrm{e}^{\beta (\t{U}/2-\mu)}+1} + \frac{1}{\mathrm{e}^{\beta (-\t{U}/2-\mu)}+1} + \delta\b{n}_{\beta}, \label{e170}
\end{equation}
where $\delta\b{n}_{\beta}$ represents the contribution to $\b{n}_{\beta}$ arising from the temperature-dependent part of $\mathscr{A}_{\rm loc}(\omega)$. Since the temperature dependence of $\mathscr{A}_{\sigma}({\bm k};\varepsilon)$ is entirely due to particle-particle interaction, it follows that the property $\delta\b{n}_{\beta} \not\equiv 0$ is directly attributable to this interaction.

From the expression in Eq.~(\ref{e168}), making use of Eq.~(\ref{e163}), one has
\begin{equation}
\delta\b{n}_{\beta} \sim \mathrm{e}^{-\beta\t{U}/2} (\chi^+\, \mathrm{e}^{\beta\mu} + \chi^-\, \mathrm{e}^{-\beta\mu}),
\label{e171}
\end{equation}
in which $\chi^{\pm} \equiv \chi_{\sigma}^{\pm}$, where
\begin{equation}
\chi_{\sigma}^{\pm} {:=} \frac{2}{\mathcal{N}_{\Sc l}} \sum_{\bm k}\int_{-\infty}^{\mu_{\infty}} \frac{{\rm d}\varepsilon}{\hbar} \; B_{N;\sigma}^{\pm}({\bm k};\varepsilon). \label{e172}
\end{equation}

Combining Eqs.~(\ref{e170}) and (\ref{e171}), one arrives at
\begin{equation}
\b{n}_{\beta} \sim 1 + \mathrm{e}^{-\beta\t{U}/2} (\mathrm{e}^{\beta\mu} - \mathrm{e}^{-\beta\mu} + \chi^+\, \mathrm{e}^{\beta\mu} + \chi^-\, \mathrm{e}^{-\beta\mu})\;\;\; \mbox{\rm for}\;\;\; \beta\to\infty. \label{e173}
\end{equation}
Assuming that $\mu \in (-\t{U}/2,\t{U}/2)$ is far away from $\pm \t{U}/2$, the difference between the exact $\b{n}_{\beta}$ and the RHS of Eq.~(\ref{e173}) is of the form $o(\epsilon)$; in the cases where $\mathscr{A}_{\rm loc}(\omega)$ has a regular expansion to second order in $\epsilon$ (appendix \ref{sf}), the latter $o(\epsilon)$ coincides with $O(\epsilon^2)$ (for the $o-O$ notation see \S~5 in Ref.~\citen{EWH26}).

Since $A_{M;\sigma}({\bm k};\varepsilon) \ge 0$, $\forall {\bm k},\varepsilon$, on account of the exact sum rule (appendix \ref{sc})
\begin{equation}
\frac{1}{\hbar} \int_{-\infty}^{\infty} {\rm d}\varepsilon\; A_{M;\sigma}({\bm k};\varepsilon) =1,\;\; \forall M \in \mathds{N}, \label{e174}
\end{equation}
and of $\frac{1}{\mathcal{N}_{\Sc l}} \sum_{\bm k} 1 = 1$, one readily obtains that
\begin{equation}
\vert \chi_{\sigma}^{\pm} \vert <2. \label{e175}
\end{equation}
The value $2$ has its origin in the $2$ on the RHS of the defining expression in Eq.~(\ref{e172}). For the case at hand, where $N$ corresponds to half-filling, one similarly readily obtains that
\begin{equation}
\vert \chi_{\sigma}^{\pm} \vert <1\;\;\;\; \mbox{\rm (for $N$ corresponding to half-filling)}. \label{e176}
\end{equation}

Further, making use of the properties
\begin{equation}
\sum_{{\bm k},\sigma} \int_{-\infty}^{\mu_{\infty}} \frac{{\rm d}\varepsilon}{\hbar}\, A_{N;\sigma}({\bm k};\varepsilon) = N\;\;\; \mbox{\rm and}\;\;\; \sum_{{\bm k},\sigma} \int_{-\infty}^{\mu_{N;\sigma}^{\pm}} \frac{{\rm d}\varepsilon}{\hbar}\, A_{N\pm 1;\sigma}({\bm k};\varepsilon) = N\pm 1, \nonumber
\end{equation}
where $\mu_{N;\sigma}^{\pm}$ are defined in Eqs.~(\ref{ec5}) and (\ref{ec6}), following some elementary manipulations one obtains that
\begin{eqnarray}
\sum_{\sigma}\chi_{\sigma}^{\pm} &=& \pm \frac{2}{\mathcal{N}_{\Sc l}} -\frac{2}{\mathcal{N}_{\Sc l}}\sum_{{\bm k},\sigma} \int_{\mu_{\infty}}^{\mu_{N;\sigma}^{\pm}} \frac{{\rm d}\varepsilon}{\hbar}\; A_{N\pm 1;\sigma}({\bm k};\varepsilon) \nonumber\\
&\sim& -\frac{2}{\mathcal{N}_{\Sc l}}\sum_{{\bm k},\sigma} \int_{\mu_{\infty}}^{\mu_{N;\sigma}^{\pm}} \frac{{\rm d}\varepsilon}{\hbar}\; A_{N\pm 1;\sigma}({\bm k};\varepsilon)\;\;\;\mbox{\rm for}\;\;\; \mathcal{N}_{\Sc l}\to\infty. \label{e177}
\end{eqnarray}
Since $\mu_{N;\sigma}^- < \mu_{\infty} <\mu_{N;\sigma}^+$ and $A_{N\pm 1;\sigma}({\bm k};\varepsilon)\ge 0$, one observes that, for sufficiently large $\mathcal{N}_{\Sc l}$,
\begin{equation}
\chi_{\sigma}^{\pm} \lessgtr 0. \label{e178}
\end{equation}
For uncorrelated semiconductors or insulators, the magnitude of the RHS of Eq.~(\ref{e177}) is equal to $2/\mathcal{N}_{\Sc l}$ so that for these systems $\chi_{\sigma}^{\pm} =0$. This is in conformity with the fact that $\delta\b{n}_{\beta} \not\equiv 0$ is due to particle-particle interaction. Hence the magnitudes of $\chi_{\sigma}^{\pm}$ are quantitative measures indicative of the strength of correlation effects in the $N\pm 1$-particle GSs of the interacting systems whose $N$-particle GSs are insulating.

Making use of the expression in Eq.~(\ref{e173}), the chemical potential corresponding to $\b{n}_{\beta} = 1$, that is $\mu_{\beta} \equiv \mu(\beta,N,V)$, with $N$ corresponding to half-filling, is obtained by solving the following equation, which is exact to linear order in $\epsilon$:
\begin{equation}
\mathrm{e}^{\beta\mu} - \mathrm{e}^{-\beta\mu} + \chi^+\, \mathrm{e}^{\beta\mu} + \chi^-\, \mathrm{e}^{-\beta\mu} = 0 \iff (1 + \chi^+) \zeta - (1-\chi^-) \frac{1}{\zeta} = 0,\;\; \zeta \equiv \mathrm{e}^{\beta\mu}, \label{e179}
\end{equation}
where $\zeta$ is the fugacity. Since this equation corresponds to $\b{n}_{\beta} = 1$ (half-filling), the $\chi^{\pm}$ as encountered here are subject to the inequalities in Eq.~(\ref{e176}). We point out that the zeros on the RHSs of the equations in Eq.~(\ref{e179}) take the place of a function whose magnitude is of the  order of $\epsilon$; it can be readily verified that the approximation of this function by $0$ is fully justified, except for the cases where $(1-\chi^{-}) (1+\chi^{+})$ is comparable with or smaller than $\epsilon^2$. Since, however, $\vert\chi^{\pm}\vert$ are in general closer to zero than to unity, it follows that the latter possibility is rather improbable, if at all feasible.

The exact solution of Eq.~(\ref{e179}) is of the form $a/\beta$, where
\begin{equation}
a = \frac{1}{2} \ln(\frac{1-\chi^-}{1+\chi^+}), \label{e180}
\end{equation}
so that for the exact $\mu_{\beta}$ one has (for the $o-O$ notation see \S~5 in Ref.~\citen{EWH26})
\begin{equation}
\mu_{\beta} = \frac{a}{\beta} + o(1/\beta), \label{e181}
\end{equation}
Since, on account of Eq.~(\ref{e181}), the zero temperature limit of $\mu_{\beta}$ is equal to zero, in view of Eq.~(\ref{e167}) $\mu_{\infty}$ is located exactly at the middle point of the interval $[\mu_{N;\sigma}^-,\mu_{N;\sigma}^+]$ (here $[\mu_{N}^-, \mu_{N}^+]$). On the basis of this observation and of the particle-hole symmetry associated with the $N$ corresponding to half-filling, whereby\footnote{With reference to Eq.~(\protect\ref{e163a}) and the remarks following Eq.~(\protect\ref{e166}), $A_{N\pm 1;\rm loc}(\omega) $ is defined according to $\frac{1}{\mathcal{N}_{\Sc l}} \sum_{\bm k} A_{N\pm 1} ({\bm k};\omega)$.}
\begin{equation}
A_{N+1;\rm loc}(\omega) \equiv A_{N-1;\rm loc}(-\omega),\;\;\; \forall\omega \in \mathds{R}, \label{e181a}
\end{equation}
from Eq.~(\ref{e177}) one obtains that for the $\chi^{\pm}$ in Eq.~(\ref{e180}) one has\footnote{Starting from the defining expressions for $\chi_{\sigma}^{\pm}$, Eq.~(\protect\ref{e172}), one can rigorously prove that for $N$ corresponding to half-filling and $\mu_{\infty}=0$, the equality in Eq.~(\protect\ref{e182}) \emph{necessarily} leads to
$\int_{-\infty}^0 {\rm d}\omega\; \big(A_{N+1;\rm loc}(\omega) - A_{N-1;\rm loc}(-\omega)\big) =0$, which is trivially satisfied by the identity in Eq.~(\protect\ref{e181a}).}
\begin{equation}
\chi^{-} = -\chi^{+}. \label{e182}
\end{equation}
This equality implies that in the case at hand, and for sufficiently large $\beta$,
\begin{equation}
\mu_{\beta} = o(1/\beta).
\label{e183}
\end{equation}
That is, $\mu_{\beta}$ decays faster towards zero than $1/\beta$ for $\beta\to\infty$. Following this result, one observes that for $\mu=\mu_{\beta}$, the RHS of Eq.~(\ref{e160a}) approaches zero for $\beta\to\infty$ (cf. Eq.~(\ref{e160b})).

To calculate the leading-order term in the asymptotic series expansion corresponding to $\beta\to\infty$ of the $o(1/\beta)$ in Eq.~(\ref{e183}), one has to introduce the following two modifications: firstly, one has to redefine $\chi_{\sigma}^{\pm}$ according to (cf. Eq.~(\ref{e172}))
\begin{equation}
\chi_{\sigma}^{\pm} {:=} \frac{2}{\mathcal{N}_{\Sc l}} \sum_{\bm k} \int_{-\infty}^{\infty} \frac{{\rm d}\varepsilon}{\hbar} \frac{B_{N;\sigma}^{\pm}({\bm k};\varepsilon)}{\mathrm{e}^{\beta (\varepsilon-\mu)} +1}, \label{e184}
\end{equation}
and secondly, one has to rely on the extension of the expression in Eq.~(\ref{e173}) which is correct to quadratic order in $\epsilon$. As for the latter, one can readily verify that the contributions to $\b{n}_{\beta}$ which are quadratic in $\epsilon$, give rise to contributions to $\mu_{\beta}$ which are linear in $\epsilon$, that is they are exponentially small. Consequently, $\mu_{\beta}$ has to exponential accuracy the form $a/\beta$, where $a$ is defined according to the expression in Eq.~(\ref{e180}), with the $\chi^{\pm}$ herein determined on the basis of the expressions in Eq.~(\ref{e184}). Note that for the calculation of the leading-order corrections to $\chi_{\sigma}^{\pm}$ that follow the zero-temperature values $\lim_{\beta\to\infty} \chi_{\sigma}^{\pm}$, one can replace the $\mu$ on the RHS of Eq.~(\ref{e184}) by $\mu_{\infty}$.

For calculating the asymptotic series expansion of $\chi_{\sigma}^{\pm}$ for $\beta\to\infty$, one may employ the technique underlying the well-known Sommerfeld expansion (Appendix C in Ref.~\citen{AM76}). In this way one obtains series for $\chi_{\sigma}^{\pm}$ in terms of the asymptotic sequence $\{1, 1/\beta^2, 1/\beta^4, \dots\}$. Whether $\chi_{\sigma}^{+}$ and $\chi_{\sigma}^{-}$ admit of such regular asymptotic series, depends on the behaviours of $B_{N;\sigma}^{\pm}({\bm k};\varepsilon)$ as functions of $\varepsilon$. In fact, for the present case we do not rule out the possibility that, for sufficiently large values of $\beta$, $\chi_{\sigma}^{\pm}$ may be to exponential accuracy equal to $\lim_{\beta\to\infty}\chi_{\sigma}^{\pm}$.

We have thus established that in the local limit
\begin{equation}
\mu_{\infty} \equiv \lim_{\beta\to\infty} \mu_{\beta} =0. \label{ek48}
\end{equation}
Following Eq.~(\ref{e160a}), one observes that for $\mu=\mu_{\infty}$ the function $\Phi(\beta\mu,\beta \t{U})$ is identically vanishing for all $\beta$. This result is in conformity with the fact that, on account of $\mathrm{sgn}(0)=0$, the RHS of Eq.~(\ref{e158}) is vanishing for $\mu=\mu_{\infty}$. We remark that owing to Eq.~(\ref{e183}), one further has $\lim_{\beta\to\infty} \Phi(\beta\mu_{\beta},\beta \t{U}) = 0$. We thus conclude that for $\mu=\mu_{\infty}$ \emph{the Luttinger-Ward identity is indeed valid in the local limit. Consequently, the Luttinger theorem is unconditionally valid in the local limit}.

\subsubsection{Consequences of particle hopping}
\label{ss61s4}

Concerning the consequences of the hopping terms in the Hamiltonian under consideration (see Ref.~\citen{AR06}), to leading order in the hopping parameters $\{ t_{\ell,\ell'}\}$ the self-energy remains local \cite{AR06} and for\footnote{Here $\varepsilon_{\alpha}$ denotes orbital energy and $\alpha$ orbital index; the $2$ in Eq.~(\protect\ref{e162}), as well as Eq.~(\protect\ref{e163}), corresponds to a trace over this index.} $\varepsilon_{\alpha} = -\t{U}/2$ the single-particle Green function has the form
\begin{equation}
G({\bm k};\omega) = \frac{1}{\omega +\t{U}/2 - \tau_{\bm k} - \Sigma_{\rm loc}(\omega)}, \label{e185}
\end{equation}
where
\begin{equation}
\tau_{\bm k} {:=} \sum_{\ell=1}^{\mathcal{N}_{\Sc l}} t_{\ell,\ell'}\, \mathrm{e}^{-i {\bm k}\cdot ({\bm R}_{\ell}-{\bm R}_{\ell'})} \equiv \sum_{\ell=1}^{\mathcal{N}_{\Sc l}} t_{\ell,\ell'}\, \cos\big({\bm k}\cdot ({\bm R}_{\ell}-{\bm R}_{\ell'})\big), \label{e186}
\end{equation}
in which $\{ {\bm R}_{\ell} \,\|\, \ell =1,2,\dots, \mathcal{N}_{\Sc l}\}$ denotes the Bravais lattice on which the system under investigation is defined. We point out that
\begin{equation}
t_{\ell,\ell'} =\frac{1}{\mathcal{N}_{\Sc l}} \sum_{\bm k} \tau_{\bm k}\, \mathrm{e}^{i {\bm k}\cdot ({\bm R}_{\ell}-{\bm R}_{\ell'})}, \label{e187}
\end{equation}
where $\sum_{\bm k} \equiv \sum_{{\bm k}\in \mathrm{1BZ}}$ denotes summation over the $\mathcal{N}_{\Sc l}$ points of which the $\mathrm{1BZ}$ corresponding to $\{ {\bm R}_{\ell}\}$ consists. Consequently, since $t_{\ell,\ell} = 0$, $\forall\ell$, it follows that
\begin{equation}
\sum_{\bm k} \tau_{\bm k} = 0. \label{e188}
\end{equation}
As we shall see later, the simplicity to which this expression gives rise can be very deceptive, concealing a fundamental difference between symmetric cases, where $\tau_{\bm k}$ oscillates symmetrically around zero, and non-symmetric ones.

Denoting the two simple poles of the Green function in Eq.~(\ref{e185}) by $\omega_{\pm}({\bm k})$, one has
\begin{equation}
\omega_{\pm}({\bm k}) = \frac{ \tau_{\bm k}}{2}\pm \frac{1}{2} \Big(\tau_{\bm k}^2 +\t{U}^2\Big)^{1/2} \sim \frac{\tau_{\bm k}}{2} \pm \frac{\t{U}}{2}\;\;\;\mbox{\rm for}\;\;\; \frac{\vert\tau_{\bm k}\vert}{U} \to 0. \label{e189}
\end{equation}
One readily verifies that
\begin{eqnarray}
&&\hspace{0.3cm}\frac{1}{\mathcal{N}_{\Sc l}}\sum_{\bm k}\int_{\mu-i\infty}^{\mu+i\infty} \frac{{\rm d}z}{2\pi i}\; \t{G}({\bm k};z) \frac{\partial}{\partial z} \t{\Sigma}_{\rm loc}(z) \nonumber\\
&&\hspace{1.0cm}
= \frac{(\t{U}/2)^2}{\mathcal{N}_{\Sc l}}\sum_{\bm k} \frac{1}{\omega_+({\bm k})-\omega_-({\bm k})} \Big\{ \frac{\Theta(\omega_+({\bm k})-\mu) \,\Theta(\mu)}{\omega_+({\bm k})} +
\frac{\Theta(\mu-\omega_-({\bm k})) \,\Theta(-\mu)}{\omega_-({\bm k})} \Big\} \nonumber\\
&&\hspace{1.0cm}\sim \frac{1}{2 \mathcal{N}_{\Sc l}}\, \sum_{\bm k} \Big\{(1 - \frac{\tau_{\bm k}}{\t{U}})\, \Theta(\frac{\t{U}}{2}+\frac{\tau_{\bm k}}{2} -\mu)\,\Theta(\mu) -(1 + \frac{\tau_{\bm k}}{\t{U}})\, \Theta(\mu+\frac{\t{U}}{2}-\frac{\tau_{\bm k}}{2})\,\Theta(-\mu)\Big\}\nonumber\\
&&\hspace{10.3cm}\;\;\;\mbox{\rm for}\;\;\;\frac{\vert\tau_{\bm k}\vert}{U} \to 0,\; \forall {\bm k}  \nonumber\\
&&\hspace{1.0cm}\equiv \frac{1}{2}\, \mathrm{sgn}(\mu)\;\;\; \mbox{\rm for}\;\;\; -\frac{1}{2} (\t{U} - \max_{\bm k} \tau_{\bm k}) < \mu < \frac{1}{2} (\t{U} + \min_{\bm k} \tau_{\bm k}), \label{e190}
\end{eqnarray}
where in arriving at the last expression we have used $\sum_{\bm k} 1 = \mathcal{N}_{\Sc l}$ and the property in Eq.~(\ref{e188}). Note that Eq.~(\ref{e190}) reduces to that in Eq.~(\ref{e158}) for $\tau_{\bm k}\equiv 0$. The first expression on the RHS of Eq.~(\ref{e190}) can be explicitly shown to be vanishing for $\mu=0$, similar to the second expression, on account of $\mathrm{sgn}(0)=0$.

In analogy with Eq.~(\ref{e160a}), we obtain that
\begin{eqnarray}
&&\hspace{-0.0cm}\frac{1}{\mathcal{N}_{\Sc l}} \sum_{\bm k} \frac{1}{\beta} \sum_m \t{G}({\bm k};\zeta_m) \frac{\partial}{\partial\zeta_m} \t{\Sigma}_{\rm loc}(\zeta_m) \sim \Psi(\beta\mu)\nonumber\\
&&\hspace{0.6cm} -\frac{1}{4 \mathcal{N}_{\Sc l}}\sum_{\bm k} \Big\{\tanh\big(\frac{\beta}{2} (\mu -\frac{\tau_{\bm k}}{2} -\frac{\t{U}}{2})\big) +\tanh\big(\frac{\beta}{2} (\mu -\frac{\tau_{\bm k}}{2} +\frac{\t{U}}{2})\big) \Big\},\;\;\; \frac{\vert\tau_{\bm k}\vert}{\t{U}} \to 0,\;\forall {\bm k}, \nonumber\\
\label{e190a}
\end{eqnarray}
where $\Psi(\beta\mu)$ is defined in Eq.~(\ref{e160b}). For $\vert\mu\vert \ll \t{U}/2$ and sufficiently large $\beta$, one can to exponential accuracy approximate the RHS of Eq.~(\ref{e190a}) by $\Psi(\beta\mu)$. It follows that provided that $\mu_{\infty}=0$, also in the present case the apparent failure of the Luttinger-Ward identity is due to a false limit, associated with effecting $\beta\to\infty$ for $\mu\not=\mu_{\infty}$. As we shall see below, $\mu_{\infty}=0$ is only possible for the cases where the density of states function corresponding to $\tau_{\bm k}$, Eq.~(\ref{e196}) below, is symmetric with respect to $\omega=0$.

In view of (cf. Eq.~(\ref{e161}))
\begin{equation}
A({\bm k};\omega) = \frac{1-\tau_{\bm k}/(\tau_{\bm k}^2 +\t{U}^2)^{1/2}}{2}\, \delta(\omega-\omega_-({\bm k})) + \frac{1+\tau_{\bm k}/(\tau_{\bm k}^2+\t{U}^2)^{1/2}}{2}\, \delta(\omega-\omega_+({\bm k})), \label{e191}
\end{equation}
for the number of particles per site \emph{per spin species} (at zero temperature) one obtains that (cf. Eq.~(\ref{e162}))
\begin{eqnarray}
n &\equiv& \frac{2}{\mathcal{N}_{\Sc l}} \sum_{\bm k} \int_{-\infty}^{\mu} {\rm d}\omega\; A({\bm k};\omega)\nonumber\\
&=& \frac{1}{\mathcal{N}_{\Sc l}} \sum_{\bm k} \Big(\big(1-\frac{\tau_{\bm k}}{(\tau_{\bm k}^2+\t{U}^2)^{1/2}}\big)\, \Theta\big(\mu-\omega_-({\bm k})\big) + \big(1+\frac{\tau_{\bm k}}{(\tau_{\bm k}^2+/\t{U}^2)^{1/2}}\big)\, \Theta\big(\mu-\omega_+({\bm k})\big) \Big) \nonumber\\
&\sim& \frac{1}{\mathcal{N}_{\Sc l}} \sum_{\bm k} \Big(\big(1-\frac{\tau_{\bm k}}{\t{U}}\big)\, \Theta\big(\mu+\frac{\t{U}}{2}-\frac{\tau_{\bm k}}{2}\big) + \big(1+\frac{\tau_{\bm k}}{\t{U}}\big)\, \Theta\big(\mu-\frac{\t{U}}{2}-\frac{\tau_{\bm k}}{2}\big) \Big)\;\;\;\mbox{\rm for}\;\;\; \frac{\vert\tau_{\bm k}\vert}{\t{U}}\to 0. \nonumber\\
\label{e192}
\end{eqnarray}

Let
\begin{equation}
\tau_{\rm min} \equiv \min_{\bm k} \tau_{\bm k},\;\;\;\;\;\;
\tau_{\rm max} \equiv \max_{\bm k} \tau_{\bm k}, \label{e193}
\end{equation}
where, following Eq.~(\ref{e188}), one has
\begin{equation}
\tau_{\rm min} <0\;\;\;\;\;\mbox{\rm and}\;\;\;\;\; \tau_{\rm max} >0.
\label{e194}
\end{equation}
On the basis of Eq.~(\ref{e188}) and $\sum_{\bm k} 1 = \mathcal{N}_{\Sc l}$, from Eq.~(\ref{e192}) one obtains that up to and including the second order in $\tau_{\bm k}/\t{U}$ one has
\begin{equation}
n \equiv 1 \iff -\frac{\t{U}}{2} + \frac{\tau_{\rm max}}{2}
\lesssim \mu \lesssim \frac{\t{U}}{2} + \frac{\tau_{\rm min}}{2}. \label{e195}
\end{equation}
This apparent freedom in the choice of the $\mu$ corresponding to $n=1$, implying a non-vanishing contribution on the RHS of Eq.~(\ref{e190}) for the cases where $\mu\not=0$, has led Rosch \cite{AR06} to conclude that, similar to the local limit, the Luttinger theorem breaks down in the cases where effects of particle hopping are taken into account, albeit to leading order. Rosch \cite{AR06} has further concluded that unless $\tau_{\rm min} = -\tau_{\rm max}$, this violation is definite in the `canonical' ensemble (see our final remark in Sec.~\ref{ss61}).

It will be instructive to introduce a foretaste of what follows before entering into detailed calculations. To this end we define the normalised density-of-states function
\begin{equation}
\mathcal{D}(\omega) {:=} \frac{1}{\mathcal{N}_{\Sc l}} \sum_{\bm k} \delta(\omega-\tau_{\bm k}), \label{e196}
\end{equation}
for which one has
\begin{equation}
\int {\rm d}\omega\; \mathcal{D}(\omega) = 1,\;\;\;\;\;\; \int {\rm d}\omega\; \omega\, \mathcal{D}(\omega) = 0, \label{e197}
\end{equation}
where the second expression follows from that in Eq.~(\ref{e188}).

In terms of $\mathcal{D}(\omega)$, from the exact expression in Eq.~(\ref{e192}) one arrives at
\begin{equation}
n = 1 - \mathcal{C}\;\;\; \mbox{\rm for}\;\;\; \max_{\bm k} \omega_-({\bm k}) < \mu < \min_{\bm k} \omega_+({\bm k}), \label{e198}
\end{equation}
where
\begin{equation}
\mathcal{C} {:=} \int {\rm d}\omega\; \frac{\mathcal{D}(\omega)\, \omega}{(\omega^2+\t{U}^2)^{1/2}}. \label{e199}
\end{equation}

There are two classes of cases to be considered, corresponding to
\begin{equation}
\mathcal{D}(-\omega) \equiv \mathcal{D}(\omega),\;\forall\omega\;\;\;\mbox{\rm (symmetric class)}, \label{e200}
\end{equation}
\begin{equation}
\mathcal{D}(-\omega) \not\equiv \mathcal{D}(\omega)\;\;\;\;\,\mbox{\rm (asymmetric class)}. \label{e201}
\end{equation}
We note in passing that in the cases where only isotropic nearest-neighbour hoppings are operative (that is $t_{\ell,\ell'} \equiv t\not=0$ \emph{only if} ${\bm R}_{\ell}$ and ${\bm R}_{\ell'}$ are nearest neighbours), $\mathcal{D}(\omega) \equiv \mathcal{D}(-\omega)$ is assured for \emph{bipartite} lattices (\S~4.4 in Ref.~\citen{PF99}).

Evidently,
\begin{equation}
\mathcal{C} \equiv 0\;\;\;\mbox{\rm for symmetric cases}, \label{e202}
\end{equation}
so that, in view of Eq.~(\ref{e198}), $n=1$ is satisfied for $\mu$ inside a finite non-vanishing interval, suggesting a similarity between symmetric cases and the local limit discussed in Sec.~\ref{ss61s3}.

The most peculiar aspect of the expression in Eq.~(\ref{e198}) comes to the fore by realising that
\begin{equation}
\mathcal{C} \not\equiv 0\;\;\;\mbox{\rm for asymmetric cases}, \label{e203}
\end{equation}
which implies that in these cases \emph{no} $\mu$ inside the open interval indicated in Eq.~(\ref{e198}) can result in $n=1$. For clarity, the expression in Eq.~(\ref{e203}) does \emph{not} imply that $\mathcal{C}$ should be non-vanishing for all $\t{U}$; this expression allows for $\mathcal{C}$ to be vanishing for \emph{some} $\t{U}$.

The above observation, that in the cases where $\mathcal{D}(-\omega) \not\equiv \mathcal{D}(\omega)$, and excluding possibly some discrete values of $\t{U}$, the chemical potential \emph{cannot} be inside a finite interval of width approximately equal to $\t{U}$ centred around $0$, is at variance with the finding by Rosch \cite{AR06} that in the `canonical' ensemble (Sec.~\ref{ss61}) $\mu = (\tau_{\rm min}+\tau_{\max})/4$; this value of $\mu$ is of the order of $\tau_{\bm k}$, while for $\mu$ to be located outside the last-mentioned interval, it must take a value of the order of $\t{U}$. We shall return to this aspect in Sec.~\ref{ss61s8}. For now we mention that the origin of the latter incorrect result by Rosch lies in the deceptive nature of the second expression in Eq.~(\ref{e197}) which leads one to consider the symmetric and asymmetric cases as nearly identical for sufficiently small $\vert\tau_{\bm k}\vert/\t{U}$, for in asymmetric cases $\mathcal{C}$ is vanishing up to and including at least the second order in $\tau_{\bm k}/\t{U}$. As we shall see in more detail in Sec.~\ref{ss61s8}, \emph{no matter how small $\vert\mathcal{C}\vert$ may be, the consequences of a $\mathcal{C}\not=0$ cannot be captured by means of a finite-order perturbation series in powers of $\tau_{\bm k}/\t{U}$}.

Concerning arbitrary temperatures, making use of the finite-temperature result in Eq.~(\ref{e31}), one deduces that
\begin{eqnarray}
\b{n}_{\beta} &=& \int {\rm d}\omega\; \mathcal{D}(\omega)\, \Big(\frac{1-\omega/(\omega^2+\t{U}^2)^{1/2}}{\mathrm{e}^{\beta (\omega/2 -(\omega^2+\t{U}^2)^{1/2}/2-\mu)} +1} + \frac{1+\omega/(\omega^2+\t{U}^2)^{1/2}}{\mathrm{e}^{\beta (\omega/2 +(\omega^2+\t{U}^2)^{1/2}/2-\mu)} +1} \Big) + \delta\b{n}_{\beta} \nonumber\\
&\equiv& \mathcal{I}^{<} + \mathcal{I}^{>} + \delta\b{n}_{\beta}, \label{e204}
\end{eqnarray}
where, as earlier (cf. Eq.~(\ref{e170})), $\delta\b{n}_{\beta}$ represents the contribution to $\b{n}_{\beta}$ arising from the temperature-dependent part of $\mathscr{A}({\bm k};\omega)$. Assuming that $\mu$ is both inside $(\mu_N^-,\mu_N^+)$ and is sufficiently far away from both $\mu_{N}^-$ and $\mu_{N}^+$, to linear order in the small parameter $\epsilon$, Eq.~(\ref{e169}), one has
\begin{equation}
\delta\b{n}_{\beta} \sim \chi^+\, \mathrm{e}^{-\beta (\mu_N^+ -\mu)} + \chi^-\, \mathrm{e}^{-\beta (\mu -\mu_N^-)}\;\;\; \mbox{\rm as}\;\;\; \beta\to\infty, \label{e205}
\end{equation}
where $\chi^{\pm} \equiv \chi_{\sigma}^{\pm}$, in which $\chi_{\sigma}^{\pm}$ is defined in Eq.~(\ref{e172}).

Below we shall consider the cases corresponding to $\mathcal{D}(\omega) \equiv \mathcal{D}(-\omega)$, $\forall\omega$, and $\mathcal{D}(\omega) \not\equiv\mathcal{D}(-\omega)$. Before doing so, however, we present some details concerning $\mathcal{C}$.

\subsubsection{On the constant $\mathcal{C}$}
\label{ss61s5}

From the expression for $A({\bm k};\omega)$ in Eq.~(\ref{e191}) one observes that $\mathcal{C}$ amounts to the total spectral weight removed from (if $\mathcal{C}>0$) or added to (if $\mathcal{C}<0$) the single-particle states at $\omega_{-}({\bm k})$ for all ${\bm k}$ and $\alpha$ (as mentioned earlier, $\alpha \in \{1,2\}$ is an orbital index); this total spectral weight is necessarily, respectively, added to or removed from the single-particle states at $\omega_{+}({\bm k})$ for all ${\bm k}$ and $\alpha$. Thus with the exception of some discrete values of $\t{U}$ at which $\mathcal{C}$ may be vanishing (see below), hopping of particles in an asymmetric case gives rise to a net amount of inter-band spectral weight transfer; owing to the second expression in Eq.~(\ref{e197}), this transfer is at the lowest operative from the third order in $\tau_{\bm k}/\t{U}$.

To gain insight into the nature of $\mathcal{C}$, we assume that $\t{U} > \max(-\tau_{\rm min},\tau_{\rm max})$, Eq.~(\ref{e193}), so that $1/(\omega^2+\t{U}^2)^{1/2}$ can be expanded in the uniformly-convergent series in powers of $(\omega/\t{U})^2$ for all $\omega$ over the support of $\mathcal{D}(\omega)$. With
\begin{equation}
\mathcal{I}_j {:=} \int {\rm d}\omega\; \mathcal{D}(\omega)\,\omega^j,
\label{e206}
\end{equation}
the $j$th-order frequency moment of $\mathcal{D}(\omega)$,
from this series one obtains the uniformly-convergent series
\begin{equation}
\mathcal{C} = \frac{1}{\t{U}}\, \mathcal{I}_1 - \frac{1}{2\t{U}^3}\, \mathcal{I}_3 + \frac{3}{8\t{U}^5}\, \mathcal{I}_5 -\dots \label{e207}
\end{equation}
of which the first term is vanishing on account of the second expression in Eq.~(\ref{e197}); evidently, for a symmetric case all terms of this series are identically vanishing. For asymmetric cases, the expression in Eq.~(\ref{e207}) makes explicit that although $\mathcal{C}$ may be vanishing for some discrete values of $\t{U}$, it cannot be identically vanishing for all $\t{U}$.

In the cases where all $\mathcal{I}_{2j+1}$, $j=1,2,\dots$, have the same sign, the series on the RHS of Eq.~(\ref{e207}) is alternating for $\t{U}>0$. In such a case, on suppressing all terms beyond that proportional to $\mathcal{I}_{2m+1}$, the resulting $\mathcal{C}$ can have, by the Descartes sign rule \cite{AJS98}, $m-1$ \emph{finite} zeros along the \emph{positive} $\t{U}$-axis ($\t{U}=\infty$ is always a zero of $\mathcal{C}$); since the number of finite positive zeroes of the expanded $\mathcal{C}$ can be $m-1$, $m-3$, $\dots$ \cite{AJS98}, it follows that for $m$ an \emph{even} integer, $\mathcal{C}$ will have at least one zero at some finite positive $\t{U}$.

The distinctive consequences of $\mathcal{C}=0$ and $\mathcal{C}\not=0$, and in the latter case of $\mathcal{C}<0$ and $\mathcal{C}>0$ (Sec.~\ref{ss61s9}), shed light on an inherent inconsistency of the perturbation framework within which we investigate the validity or otherwise of the Luttinger theorem: although the framework takes account of the first-order contribution of $\tau_{\bm k}$ to $G^{-1}({\bm k};\omega)$, the constant $\mathcal{C}$, which at the lowest is of the third order in $\tau_{\bm k}$, turns out to exert the all-dominating influence; as we shall demonstrate below, irrespective of how small $\vert\mathcal{C}\vert \not=0$ may be, and provided that one insists on a half-filled GS, it transforms the underlying insulating GS into a metallic one.

An example should be clarifying. Consider the model function
\begin{equation}
\mathcal{D}(\omega) = (1-a)\, \delta(\omega+\omega_0) + a\, \delta(\omega-\omega_1),\;\;\; 0 < a < 1, \label{e208}
\end{equation}
where $\omega_0>0$ and
\begin{equation}
\omega_1 \equiv \frac{1-a}{a}\,\omega_0. \label{e209}
\end{equation}
The density-of-states function in Eq.~(\ref{e208}), which corresponds to the case of $\tau_{\bm k}$ taking solely two discrete values, $-\omega_0$ and $\omega_1$, for all ${\bm k}$, is reminiscent of that pertaining to a dilute concentration of acceptors and donors in an otherwise intrinsic semiconductor (Ch. 28 in Ref.~\citen{AM76}).

The function in Eq.~(\ref{e208}), which satisfies the two conditions in Eq.~(\ref{e197}), is asymmetric for $a\not=\frac{1}{2}$. For the corresponding $\mathcal{C}$ one has
\begin{equation}
\mathcal{C} = -(1-a)\,\omega_0 \Big(\frac{1}{(\omega_0^2 +\t{U}^2)^{1/2}} - \frac{1}{(\omega_1^2+\t{U}^2)^{1/2}}\Big). \label{e210}
\end{equation}
Evidently, $\omega_1 >\omega_0$ for $0<a<\frac{1}{2}$, and $\omega_1 <\omega_0$ for $\frac{1}{2} <a<1$. Consequently, $\mathcal{C}<0$ for all $\t{U}$ when $0<a<\frac{1}{2}$, and $\mathcal{C}>0$ for all $\t{U}$ when $\frac{1}{2}<a<1$; $\mathcal{C}\not=0$ when $a\not=\frac{1}{2}$, for all finite $\t{U}$. On the other hand, by expanding the $\mathcal{C}$ in Eq.~(\ref{e210}) to order $2m+1$ in $1/\t{U}$, one expects the resulting approximate $\mathcal{C}$ to be zero at some finite $\t{U}$ for even values of $m$. This follows from the fact that the series expansion of the $\mathcal{C}$ in Eq.~(\ref{e210}) in powers of $1/\t{U}$ is necessarily alternating; this is appreciated by realising that (i) the series expansion of $1/(1+x)^{1/2}$ in powers of $x$ is alternating, and that (ii) for $a\not=\frac{1}{2}$ one has $\omega_0\not=\omega_1$, so that one of the two functions on the RHS of Eq.~(\ref{e210}) dominates at all orders in $1/\t{U}$.

\begin{table}[t!]
\caption{The zero $\t{U}_{0}$ of the $\mathcal{C}$ in Eq.~(\protect\ref{e210}) as expanded to order $2m+1$ in $1/\t{U}$, where $m=2p$. The parameters used are $\omega_0 = 1$ and $a=\frac{4}{9}$, so that $\max(-\omega_0,\omega_1) = 1.25$. } \vspace{2pt}
\label{t2}
\begin{center}
\begin{tabular}{l|ccccccc} \hline\hline
$p$ & $1$ & $2$ & $3$ & $4$ & $5$ & $6$ & $7$ \\
\hline \vspace{-10pt} \\
$\t{U}_{0}$ & $1.3863$ & $1.3144$ & $1.2838$ & $1.2680$ & $1.2589$ & $1.2533$ & $1.2497$ \\
\hline
\end{tabular}
\end{center}
\end{table}

In Table \ref{t2} we present numerical results for the zero $\omega_{\rm c}$ of the $\mathcal{C}$ in Eq.~(\protect\ref{e210}) as expanded to order $2m+1$ in $1/\t{U}$, where $m=2p$, for $\omega_0=1$ and $a=\frac{4}{9}$. With $\max(-\omega_0,\omega_1) = 1.25$, one observes that only for an expansion beyond the order $(1/\t{U})^{25}$ will the expanded $\mathcal{C}$ be free from an artificial zero in the region where expansion of $\mathcal{C}$ in powers of $1/\t{U}$ is uniformly convergent. In the case at hand, for $0< \t{U} <\t{U}_{0}$ the sign of the expanded $\mathcal{C}$ is opposite to that of the exact $\mathcal{C}$. In this connection, we should emphasise that \emph{sign} of $\mathcal{C}$, no matter how small $\vert\mathcal{C}\vert \not=0$ may be, is of prime significance to the half-filled GS of the system under investigation (Sec.~\ref{ss61s9}).

Summarising, since not until the third order in $\tau_{\bm k}/\t{U}$ can $\mathcal{C}$ deviate from zero in asymmetric cases, we have thus the clearest evidence that \emph{first-order results are in principle incapable of establishing break-down of the Luttinger theorem in asymmetric cases}. We shall discuss higher-order results in Sec.~\ref{ss61s12}.

\subsubsection{Symmetric cases}
\label{ss61s6}

Let
\begin{equation}
{\sf S} \equiv [\tau_{\rm min},\tau_{\rm max}] \label{e211}
\end{equation}
denote the support of $\mathcal{D}(\omega)$. For $\mu$ satisfying
\begin{equation}
-\frac{1}{2} (\omega^2+\t{U}^2)^{1/2} +\frac{1}{2}\omega < \mu < \frac{1}{2} (\omega^2+\t{U}^2)^{1/2} +\frac{1}{2}\omega,\;\;\, \forall\omega\in {\sf S}, \label{e212}
\end{equation}
the task of calculating $\b{n}_{\beta}$ as a function of $\mu$ is technically the same irrespective of whether $\mathcal{D}(\omega)$ is symmetric or otherwise. The treatment in this section is therefore specific to symmetric cases only insofar as the $\mu$ corresponding to $\b{n}_{\beta} = 1$, that is $\mu_{\beta}$, in these cases satisfies the inequalities in Eq.~(\ref{e212}) (at least for sufficiently large $\beta$), a fact that we shall explicitly establish below; as should be evident (in view of Eq.~(\ref{e198})), and as we shall also explicitly establish in Sections \ref{ss61s9} and \ref{ss61s10}, this is not the case when the underlying $\mathcal{D}(\omega)$ is asymmetric.

Since for values of $\mu$ satisfying the inequalities in Eq.~(\ref{e212}) one has
\begin{equation}
0\le \mathrm{e}^{\beta (\pm\omega/2-(\omega^2+\t{U}^2)^{1/2}/2\mp\mu)} <1, \;\;\; \beta>0,\; \forall\omega \in {\sf S},
\label{e213}
\end{equation}
it follows that the integrands of $\mathcal{I}^{<}$ and $\mathcal{I}^{>}$, Eq.~(\ref{e204}), can be expanded in powers of
\begin{equation}
\mathrm{e}^{\beta (+\omega/2-(\omega^2+\t{U}^2)^{1/2}/2 -\mu)}
\;\;\mbox{\rm and}\;\; \mathrm{e}^{\beta (-\omega/2-(\omega^2+\t{U}^2)^{1/2}/2 +\mu)},
\label{e214}
\end{equation}
respectively. The resulting series being power series of functions whose magnitudes are less than unity for $\omega\in {\sf S}$, they are uniformly convergent for all $\omega\in {\sf S}$ (\S\S~2.6 and 3.7 in Ref.~\citen{WW62}). Since the functions in Eq.~(\ref{e214}) are in addition \emph{continuous} for all $\omega\in {\sf S}$, these series can be integrated term-by-term (\S~4.7 in Ref.~\citen{WW62}, \S~45 (2) in Ref.~\citen{TB65}).

For a $\mu$ satisfying the inequalities in Eq.~(\ref{e212}), one thus obtains that (see Eq.~(\ref{e204}))
\begin{equation}
\mathcal{I}^{<} = 1 -\mathcal{C} + \sum_{j=1}^{\infty} (-1)^j\, \mathcal{A}_+^{(j)}\, x^j\, \mathrm{e}^{-j\beta \t{U}/2}, \label{e215}
\end{equation}
\begin{equation}
\mathcal{I}^{>} = -\sum_{j=1}^{\infty} (-1)^j \mathcal{A}_{-}^{(j)} x^{-j}\, \mathrm{e}^{-j\beta\t{U}/2}, \label{e216}
\end{equation}
where
\begin{equation}
\mathcal{A}_{\pm}^{(j)} {:=} \int {\rm d}\omega\; \mathcal{D}(\omega)\,\Big(1 \mp \frac{\omega}{(\omega^2+\t{U}^2)^{1/2}}\Big)\, \mathrm{e}^{j\beta (\pm\omega-\phi(\omega))/2}, \label{e217}
\end{equation}
and
\begin{equation}
x\equiv \mathrm{e}^{-\beta\mu}, \label{e218}
\end{equation}
the inverse of the fugacity. The function $\phi(\omega)$ in Eq.~(\ref{e217}) is defined as
\begin{equation}
\phi(\omega) {:=} (\omega^2+\t{U}^2)^{1/2} - \t{U}. \label{e219}
\end{equation}

Since
\begin{equation}
\frac{\partial}{\partial\omega} (\varsigma\,\omega -\phi(\omega)) = 0,\; \varsigma \in \{+,-\} \iff \omega = \pm \frac{\t{U}}{\sqrt{3}}, \label{e220}
\end{equation}
it follows that for $-\tau_{\rm min}, \tau_{\rm max} \ll \t{U}$ the exponents $\pm\omega-\phi(\omega)$ in the defining expressions for $\mathcal{A}_{\pm}^{(j)}$ are strictly monotonic for $\omega\in {\sf S}$. This and the fact that $\pm\omega-\phi(\omega)$ are infinitely many times differentiable over ${\sf S}$, imply that for $\mathcal{D}(\omega)$ sufficiently many times differentiable in the \emph{interior} of ${\sf S}$, the low-temperature asymptotic series expansions of $\mathcal{A}_{\pm}^{(j)}$ can be obtained through a repeated process of applying integration by parts \cite{ETC65,BO99}. Consequently, since the global maximum of $\omega-\phi(\omega)$ ($-\omega-\phi(\omega)$) is located at $\omega=\tau_{\rm max}$ ($\omega=\tau_{\rm min}$), to exponential accuracy the low-temperature asymptotic series expansions of $\mathcal{A}_{\pm}^{(j)}$ are determined by the analytic behaviour of $\mathcal{D}(\omega)$ in the left (right) neighbourhood of $\omega=\tau_{\rm max}$ ($\omega=\tau_{\rm min}$). The exponential accuracy to which we just referred, is effective when $\beta (\tau_{\rm max}-\tau_{\rm min}) \gg 1$.

For lattice models, $\mathcal{D}(\omega)$ is in general (however not invariably) \emph{discontinuous} at $\omega=\tau_{\rm min}$ and $\omega=\tau_{\rm max}$ (see, e.g., Fig.~4.4 in Ref.~\citen{PF99}). In such cases, the lower and upper boundaries of the integral in Eqs.~(\ref{e217}) should be formally identified with respectively $\tau_{\rm min}^+\equiv \tau_{\rm min}+0^+$ (in dealing with $\mathcal{A}_{-}^{(j)}$) and $\tau_{\rm max}^- \equiv \tau_{\rm max}-0^+$ (in dealing with $\mathcal{A}_{+}^{(j)}$) prior to applying integration by parts. For simplicity, \emph{unless we indicate otherwise, below we consider the typical case where $\mathcal{D}(\omega)$ is \emph{discontinuous} at $\omega=\tau_{\rm min}$ and $\omega=\tau_{\rm max}$}. For the economy of notation we therefore introduce (cf. Eq.~(\ref{e193}))
\begin{equation}
\Omega_{-} {:=} -\tau_{\rm min} - 0^+ >0,\;\;\;\; \Omega_{+} {:=} \tau_{\rm max} - 0^+ >0. \label{e221}
\end{equation}
Other cases, where $\mathcal{D}(\omega)$ vanishes or (integrably) diverges for $\omega$ approaching the pertinent integration boundary (the lower boundary in the cases of $\mathcal{A}_{-}^{(j)}$ and the upper one in the cases of $\mathcal{A}_{+}^{(j)}$), can be equally straightforwardly dealt with.

Using integration by parts one obtains that
\begin{equation}
\mathcal{A}_{\pm}^{(j)} \sim \frac{\Gamma_{\pm}}{j\beta}\, \mathrm{e}^{j\beta (\Omega_{\pm} -\phi(\Omega_{\pm}))/2}\;\;\;\mbox{\rm for}\;\;\; \beta\to\infty, \label{e222}
\end{equation}
where
\begin{equation}
\Gamma_{\pm} \equiv \frac{2 \mathcal{D}(\pm\Omega_{\pm})}{1 -\phi'(\Omega_{\pm})}
\Big(1-\frac{\Omega_{\pm}}{(\Omega_{\pm}^2 +\t{U}^2)^{1/2}}\Big), \label{e223}
\end{equation}
in which $\phi'(\omega) \equiv {\rm d}\phi(\omega)/{\rm d}\omega$ so that, in view of $\phi(-\omega)\equiv \phi(\omega)$, one has $\phi'(-\omega) \equiv -\phi'(\omega)$. Note that
\begin{equation}
\phi'(\Omega_{\pm}) \sim \frac{\Omega_{\pm}}{\t{U}}, \label{e224} \end{equation}
which is relatively small with respect to unity. Consequently,
\begin{equation}
\Gamma_{\pm} \sim 2 \mathcal{D}(\pm\Omega_{\pm})\;\;\;\mbox{\rm for}\;\;\; \frac{\Omega_{\pm}}{\t{U}} \to 0. \label{e225}
\end{equation}

Using the standard result
\begin{equation}
\sum_{j=1}^{\infty} \frac{(-1)^j}{j}\, x^j = -\ln(1+x), \label{e226}
\end{equation}
from Eqs.~(\ref{e215}) and (\ref{e216}) and the asymptotic expressions in Eq.~(\ref{e222}) one obtains that
\begin{equation}
\mathcal{I}^{<} \sim 1 -\mathcal{C} - \Phi_{+}(x) \;\;\;\mbox{\rm for}\;\;\; \beta\to\infty, \label{e227}
\end{equation}
\begin{equation}
\mathcal{I}^{>} \sim \Phi_{-}(x)\;\;\; \mbox{\rm for}\;\;\; \beta\to\infty, \label{e228}
\end{equation}
where
\begin{equation}
\Phi_{\pm}(x) \equiv \frac{\Gamma_{\pm}}{\beta} \ln\Big(1 + x^{\pm 1}\, \mathrm{e}^{-\beta (\t{U}-\Omega_{\pm}+\phi(\Omega_{\pm}))/2}\Big).
\label{e229}
\end{equation}

Combining the results in Eqs.~(\ref{e227}) and (\ref{e228}), from Eq.~(\ref{e204}) one thus has
\begin{equation}
\b{n}_{\beta} \sim 1 -\mathcal{C} -\Phi_{+}(x) + \Phi_{-}(x) + \delta\b{n}_{\beta}\;\;\;\mbox{\rm for}\;\;\; \beta\to\infty. \label{e230}
\end{equation}
This result is applicable irrespective of whether $\mathcal{D}(\omega)$ is symmetric or not; the only condition for the validity of this result is that $\mu$ satisfy the inequalities in Eq.~(\ref{e212}).

For symmetric cases one has
\begin{equation}
\Omega_{-} = \Omega_{+} \equiv \Omega_0,\;\;\; \Gamma_{-} = \Gamma_{+} \equiv \Gamma_0, \label{e231}
\end{equation}
so that
\begin{equation}
\Phi_{+}(x) \equiv \Phi_{-}(1/x),\; \forall x. \label{e232}
\end{equation}
With $\mathcal{C}\equiv 0$ (a characteristic of any symmetric $\mathcal{D}(\omega)$), one thus obtains that
\begin{equation}
\b{n}_{\beta} = 1 \iff \frac{1 + \epsilon x}{1+\epsilon/x} \sim \mathrm{e}^{\beta \delta\b{n}_{\beta}/\Gamma_0}\;\;\mbox{\rm as}\;\; \beta\to\infty, \label{e233}
\end{equation}
where (cf. Eq.~(\ref{e169}))
\begin{equation}
\epsilon \equiv \mathrm{e}^{-\beta (\t{U} -\Omega_0 +\phi(\Omega_0))/2}. \label{e234}
\end{equation}
Since both $\Omega_0$ and $\phi(\Omega_0)$ are small in comparison with $\t{U}$, it follows that $\epsilon$ exponentially decays towards zero for $\beta\to\infty$. We point out that since $\Phi_{+}(x)$ is a monotonically increasing and $\Phi_{-}(x)$ a monotonically decreasing function of $x$ (both functions are positive for $x>0$), it follows that the RHS of Eq.~(\ref{e230}) is a monotonically decreasing function of $x$. Consequently, Eq.~(\ref{e233}) has a \emph{unique} solution for $x$, $\forall\beta$.

On the basis of the inequalities in Eq.~(\ref{e212}) one readily verifies that in general
\begin{equation}
\mu_{N}^{\pm} = \pm \frac{1}{2}\t{U} + \frac{1}{2} (\mp\Omega_{\mp} \pm \phi(\mp\Omega_{\mp})), \label{e235}
\end{equation}
so that in the symmetric case at hand
\begin{equation}
\mu_{N}^{+} = -\mu_{N}^{-} = \frac{1}{2} \t{U} -\frac{1}{2} (\Omega_0 - \phi(\Omega_0)). \label{e236}
\end{equation}
Consequently, from Eq.~(\ref{e205}) one obtains that (cf. Eq.~(\ref{e171}))
\begin{equation}
\frac{\delta\b{n}_{\beta}}{\epsilon} \sim \chi^+\,\mathrm{e}^{\beta\mu} +\chi^-\, \mathrm{e}^{-\beta\mu}\;\;\; \mbox{\rm as}\;\;\; \beta\to\infty. \label{e237}
\end{equation}
Thus, for $\mu$ sufficiently far away from both $-\t{U}/2$ and $\t{U}/2$, the function $\beta\delta\b{n}_{\beta}/\Gamma_0$ is small for $\beta\to\infty$ so that one can employ the Taylor expansion
\begin{equation}
\mathrm{e}^{\beta \delta\b{n}_{\beta}/\Gamma_0} = 1 + \frac{\beta\epsilon}{\Gamma_0} (\chi^+\, \mathrm{e}^{\beta\mu} + \chi^-\,\mathrm{e}^{-\beta\mu}) + \dots~. \label{e238}
\end{equation}

Employing the first two terms on the RHS of Eq.~(\ref{e238}), using the defining expression in Eq.~(\ref{e218}), for sufficiently large $\beta$ Eq.~(\ref{e233}) can be expressed as
\begin{equation}
\big(1-\frac{\beta\chi^-}{\Gamma_0}\big) x^3 - \epsilon\, \frac{\beta\chi^-}{\Gamma_0} x^2 - \big(1+\frac{\beta\chi^+}{\Gamma_0}\big) x - \epsilon\, \frac{\beta\chi^+}{\Gamma_0} \sim 0. \label{e239}
\end{equation}
We now decompose $x$ as
\begin{equation}
x = x_0 +\delta x, \label{e240}
\end{equation}
where $x_0$ denotes the solution of Eq.~(\ref{e239}) corresponding to $\epsilon=0$. For $\epsilon=0$ Eq.~(\ref{e239}) has two solutions, of which $x=0$ is not physical  for $\beta<\infty$ on account of Eq.~(\ref{e218}); for the other solution, which we identify with $x_0$, one has
\begin{equation}
x_0 = \Big(\frac{1 +\beta\chi^+/\Gamma_0}{1-\beta\chi^-/\Gamma_0}\Big)^{1/2}. \label{e241}
\end{equation}
On the basis of this expression and of Eq.~(\ref{e218}), to leading order in $\epsilon$ one has (cf. Eqs.~(\ref{e181}) and (\ref{e180}))
\begin{equation}
\mu_{\beta} \sim \frac{1}{2\beta} \ln\Big(\frac{1 -\beta\chi^-/\Gamma_0}{1+\beta\chi^+/\Gamma_0}\Big)\;\;\; \mbox{\rm for}\;\;\; \beta\to\infty. \label{e242}
\end{equation}
Thus
\begin{equation}
\lim_{\beta\to\infty} \mu_{\beta} = 0. \label{e243}
\end{equation}
On account of this result it follows that the $\chi^{\pm}$ as encountered in Eqs.~(\ref{e239}) and (\ref{e242}) satisfy the relationship in Eq.~(\ref{e182}). Consequently, to first order in $1/\beta$ the RHS of Eq.~(\ref{e242}) is vanishing. Comparing the expression in Eq.~(\ref{e242}) with that in Eq.~(\ref{e181}), with the $a$ herein as presented in Eq.~(\ref{e180}), one notes that, insofar as $\mu_{\beta}$ is concerned, incorporation of hopping terms has given rise to an \emph{explicit} magnification of $\chi^{\pm}$ by the dimensionless factor $\beta/\Gamma_0$.

To determine $\delta x$, we linearise Eq.~(\ref{e239}) with respect to $\delta x$, upon which we obtain that
\begin{equation}
\delta x \approx \frac{\epsilon\beta\chi^+/\Gamma_0}{3 x_0 (1-\beta\chi^-/\Gamma_0) - (1+\beta\chi^+/\Gamma_0) - 2\epsilon\beta\chi^-/\Gamma_0}, \label{e244}
\end{equation}
from which one observes that $\delta x = O(\epsilon)$; one can readily verify that for symmetric cases, where $\chi^- =-\chi^+ \equiv \chi_0$ (corresponding to $\mu_{\infty}=0$; cf. Eq.~(\ref{e182})), to leading order in $\epsilon$ the RHS of Eq.~(\ref{e244}) is equal to $\epsilon/2$ for $\beta\to\infty$. Thus, for $\beta\to\infty$, $\mu_{\beta}$ is to exponential accuracy (explicitly, to an error of the order of $\epsilon$) described by an expression similar to that on the RHS of Eq.~(\ref{e242}), corrected for the higher-order terms in the asymptotic series expansions of $\mathcal{A}_{\pm}^{(j)}$ corresponding to $\beta\to\infty$, neglected in Eq.~(\ref{e222}); as it stands, the numerator and denominator of the expression for $x_0$ in Eq.~(\ref{e241}) are correct up to errors of the order of $1/\beta$. As a consequence, to exponential accuracy the temperature dependence of $\mu_{\beta}$ is determined by the temperature dependence of, amongst other things,\footnote{Under the conditions for which the expressions in Eq.~(\protect\ref{e222}) apply, the asymptotic series expansion of $\mathcal{A}_{\varsigma}^{(j)}$, $\varsigma \in \{+,-\}$, for $\beta\to\infty$, consists of an exponential factor times a pre-exponential function which is expressible in terms of the asymptotic sequence $\{1/\beta, 1/\beta^2,\dots\}$. From this function we have only taken account of the leading-order asymptotic term. Consequently, while accounting for the sub-leading terms in the asymptotic series expansions of $\chi^{\pm}$ for $\beta\to\infty$, one has to equally account for the comparable terms in the last-mentioned asymptotic series expansion of the pre-exponential part of $\mathcal{A}_{\varsigma}^{(j)}$. For completeness, our above statement, that ``the numerator and denominator of the expression for $x_0$ in Eq.~(\protect\ref{e241}) are correct up to errors of the order of $1/\beta$'', is based on the same consideration as indicated here.} the $\chi^{\pm}$ as determined from the defining expression in Eq.~(\ref{e184}). We shall not go into any further details concerning the dependence of $\mu_{\beta}$ on $\beta$. For the purpose of the present considerations, it suffices to know that $\mu_{\beta} = o(1/\beta)$ for sufficiently large $\beta$ (cf. Eq.~(\ref{e183})).

In view of the above observations, it follows that for $\mu=\mu_{\beta}$ the RHS of Eq.~(\ref{e190a}) approaches zero as $\beta\to\infty$; for $\mu = \mu_{\infty}$, the RHS of Eq.~(\ref{e190a}) can be shown to be identically vanishing for all $\beta$. Thus, as in the local cases, we conclude that for $\mu =\mu_{\infty}$ \emph{the Luttinger-Ward identity applies in the symmetric non-local cases; consequently, the Luttinger theorem equally applies for these cases.} For the symmetric non-local cases we have thus shown that, similar to the local cases, Sec.~\ref{ss61s3}, the observed failure of the Luttinger theorem\cite{AR06} corresponds to a false zero-temperature limit arising from effecting $\beta\to\infty$ for a $\mu \in (\mu_{N}^-,\mu_{N}^+)$ which is different from either $\mu_{\infty}$ or $\mu_{\beta}$.

\subsubsection{Remarks}
\label{ss61s7}

For the $\mathcal{A}_{\pm}^{(j)}$ as defined in Eq.~(\ref{e217}) one has
\begin{equation}
\left. \mathcal{A}_{\pm}^{(j)}\right|_{\mathcal{D}(\omega)=\delta(\omega)} = 1,\;\;\forall\beta,\; \forall j. \label{e245}
\end{equation}
These results are distinctly different from those presented in Eq.~(\ref{e222}). As a matter of fact, on taking the limit $\mathcal{D}(\omega) \to\delta(\omega)$, the expression in Eq.~(\ref{e223}) becomes meaningless. It follows that the limits $\beta\to\infty$ and $\mathcal{D}(\omega)\to\delta(\omega)$ do not commute. These aspects are readily understood by recalling that the leading-order results on the RHS of Eq.~(\ref{e222}) are deduced under the assumption that variation of $\mathcal{D}(\omega)$ over the \emph{interior} of ${\sf S}$ is weak in comparison with that of the exponential function in the integrand of the expression on the RHS of Eq.~(\ref{e217}) for $\beta\to\infty$. This assumption is clearly invalid when $\mathcal{D}(\omega)$ is identified with $\delta(\omega)$.

\subsubsection{Asymmetric cases}
\label{ss61s8}

For the reason that we indicated in Sec.~\ref{ss61s4} (see also Sec.~\ref{ss61s6}), in the cases where $\mathcal{C}\not=0$ one of the inequalities in Eq.~(\ref{e212}) is violated for $\omega$ over a finite subset of ${\sf S}$ when $\mu=\mu_{\infty}$. Although in asymmetric cases $\mathcal{C}$ can be vanishing for some values of $\t{U}$, it cannot be identically vanishing for all $\t{U}$ (Sec.~\ref{ss61s5}). In the following we shall assume that the $\t{U}$ under consideration does not coincide with any of the possible zeroes of $\mathcal{C}$.

From the zero-temperature result presented in Eq.~(\ref{e198}) one observes that, for $\mathcal{C} <0$ and $\mathcal{C} >0$ the chemical potential $\mu_{\beta}$ corresponding to $\b{n}_{\beta} =1$  should be in the vicinity of respectively $-\t{U}/2$ and $+\t{U}/2$ as $\beta\to\infty$. Consequently, for this $\mu_{\beta}$ and $\beta\to\infty$, the RHS of the asymptotic expression
\begin{equation}
\b{n}_{\beta} \sim \left\{ \begin{array}{ll}
\mathcal{I}^{<} + \delta\b{n}_{\beta}, & \mathcal{C}<0,\\ \\
1-\mathcal{C} + \mathcal{I}^{>} + \delta\b{n}_{\beta}, & \mathcal{C}>0, \end{array} \right. \label{e246}
\end{equation}
is to exponential accuracy equal to the RHS of Eq.~(\ref{e204}). This observation simplifies calculation of $\mu_{\beta}$ for large values of $\beta$.

Since the GS of the system under consideration is metallic at half-filling, the zero-temperature value of the chemical potential corresponding to $n=1$ coincides with the zero-temperature \emph{limit} of the $\mu_{\beta}$ corresponding to $\b{n}_{\beta} =1$. Consequently, calculation of the low-temperature behaviour of the latter $\mu_{\beta}$ is not strictly necessary. In the following we shall therefore focus on the \emph{limit} of $\mu_{\beta}$ at $\beta=\infty$. For the purpose of
contrasting the symmetric and the asymmetric cases, below we shall also consider, however relatively briefly, the behaviour of the $\mu_{\beta}$ corresponding to $\b{n}_{\beta} =1$ at large values of $\beta$.

\subsubsection{The zero-temperature limit (asymmetric cases)}
\label{ss61s9}

One readily verifies that for $\mathcal{C}\not=0$ there exists an $\omega$, which we denote by $\omega_0$, satisfying
\begin{equation}
\omega_0 \in {\sf S}, \label{e247}
\end{equation}
for which
\begin{equation}
\lim_{\beta\to\infty} \mu_{\beta} = \mp\frac{1}{2} (\omega_0^2 + \t{U}^2)^{1/2} + \frac{1}{2} \omega_0,\;\; \mathcal{C} \lessgtr 0.
\label{e248}
\end{equation}
Using graphical representations of $\pm\frac{1}{2} (\omega^2+\t{U}^2)^{1/2} + \frac{1}{2}\omega$, one verifies that for sufficiently small $\vert\mathcal{C}\vert$ one must have
\begin{equation}
\omega_0 \gtrless 0,\;\;\; \mathcal{C} \lessgtr 0. \label{e249}
\end{equation}
Further,
\begin{equation}
\omega_0 \to \pm\Omega_{\pm} \;\;\;\mbox{\rm for}\;\;\; \left\{\begin{array}{l} \mathcal{C}\uparrow 0,\\ \\
\mathcal{C} \downarrow 0. \end{array} \right.
\label{e250}
\end{equation}
In connection with in particular Eq.~(\ref{e249}), it is instructive to note that
\begin{equation}
-\frac{1}{2} (\omega^2+\t{U}^2)^{1/2} + \frac{1}{2}\omega\;\;\;\mbox{\rm and}\;\; +\frac{1}{2} (\omega^2+\t{U}^2)^{1/2} + \frac{1}{2}\omega \nonumber
\end{equation}
are respectively negative and positive for \emph{all} $\omega$ and that the former function is equal to $-\t{U}/2$ and the latter equal to $+\t{U}/2$ at $\omega=0$; importantly, both are monotonically increasing functions of $\omega$.

Since $\lim_{\beta\to\infty} \delta\b{n}_{\beta} =0$, making use of Eq.~(\ref{e197}), from the low-temperature asymptotic expressions in Eq.~(\ref{e246}) one thus obtains the exact results
\begin{equation}
\lim_{\beta\to\infty} \b{n}_{\beta} = 1 -\mathcal{C} +\int_{\pm\Omega_{\pm}}^{\omega_0} {\rm d}\omega\; \mathcal{D}(\omega) \Big( 1 \mp\frac{\omega}{(\omega^2+\t{U}^2)^{1/2}}\Big), \;\; \mathcal{C} \lessgtr 0. \label{e251}
\end{equation}

To make progress, we assume that $\vert\mathcal{C}\vert$ is small (see later) so that, following Eq.~(\ref{e250}), $\omega_0$ is close to either $-\Omega_{-}$ or $+\Omega_{+}$. On account of this, and depending on $\mathcal{C} \lessgtr 0$, the integral on the RHS of Eq.~(\ref{e251}) can be Taylor expanded with respect to $\omega_0$ around $\pm\Omega_{\pm}$. We thus trivially obtain that
\begin{equation}
\lim_{\beta\to\infty} \b{n}_{\beta} = 1 -\mathcal{C} +\mathcal{D}(\pm\Omega_{\pm}) \Big(1 -\frac{\Omega_{\pm}}{(\Omega_{\pm}^2+\t{U}^2)^{1/2}}\Big)\, (\omega_0 \mp\Omega_{\pm}) + \dots,\;\; \mathcal{C} \lessgtr 0.
\label{e252}
\end{equation}
Truncating the series on the RHS of this expression to linear order in $(\omega_0\mp\Omega_{\pm})$, for the $\omega_0$ corresponding to $\lim_{\beta\to\infty} \b{n}_{\beta} = 1$ one deduces that
\begin{equation}
\omega_0 \approx \pm\Omega_{\pm} + \frac{\mathcal{C}}{\mathcal{D}(\pm\Omega_{\pm}) (1- \Omega_{\pm}/(\Omega_{\pm}^2+\t{U}^2)^{1/2})},\;\; \mathcal{C} \lessgtr 0. \label{e253}
\end{equation}
These expressions are in conformity with those in Eq.~(\ref{e250}). Since $\Omega_{\pm} \ll \t{U}$, for transparency below we shall employ
\begin{equation}
\omega_0 \approx \pm\Omega_{\pm} \mp \frac{\vert\mathcal{C}\vert}{\mathcal{D}(\pm\Omega_{\pm})},\;\; \mathcal{C} \lessgtr 0. \label{e254}
\end{equation}
Note that $\mathcal{C}$ is at the largest of the order of $(\Omega_{\pm}/\t{U})^3$ (see Eq.~(\ref{e207}) and recall that $\mathcal{I}_1=0$). From Eq.~(\ref{e254}) one infers that, in the context of the present considerations, smallness of $\vert\mathcal{C}\vert$ should be measured in relation to $\Omega_{\pm}\, \mathcal{D}(\pm\Omega_{\pm})$ for $\mathcal{C} \lessgtr 0$.

Substituting the result in Eq.~(\ref{e254}) into the expression on the RHS of Eq.~(\ref{e248}), one obtains that (cf. Eq.~(\ref{e235}))
\begin{eqnarray}
\lim_{\beta\to\infty} \mu_{\beta} &\approx& \mp\frac{1}{2}(\Omega_{\pm}^2+\t{U}^2)^{1/2} +\frac{1}{2} \big(\pm\Omega_{\pm} \mp\frac{\vert\mathcal{C}\vert}{\mathcal{D}(\pm\Omega_{\pm})}\big) \nonumber\\
&=& \left.\mu_{N}^{\mp}\right|_{\mathcal{C}=0} \mp\frac{\vert\mathcal{C}\vert}{2\mathcal{D}(\pm\Omega_{\pm})},\;\; \mathcal{C} \lessgtr 0, \label{e255}
\end{eqnarray}
which applies for $\vert\mathcal{C}\vert \to 0^+$. In view of the results in Eqs.~(\ref{e255}) and (\ref{e243}), one observes that on changing $\mathcal{C}$ from an infinitesimally small negative value to an infinitesimally small positive value, the $\mu_{\infty}$ corresponding to half-filling changes from a value approximately equal to $-\t{U}/2$ to $0$, at $\mathcal{C}=0$, to a value approximately equal to $+\t{U}/2$; in other words, for small values of $\vert\mathcal{C}\vert$, $\mu_{\infty}$ can to leading order in $\t{U}$ be expressed as $\frac{1}{2}\t{U}\, \mathrm{sgn}(\mathcal{C})$, with $\mathrm{sgn}(x)$ defined to be equal to zero at $x=0$.

We thus conclude that \emph{in asymmetric cases, and excluding the possible values of $\t{U}$ for which $\mathcal{C}=0$, the Luttinger-Ward identity and the Luttinger theorem are violated.} In Sec.~\ref{ss61s12} we shall demonstrate that these apparent failures are artifacts of the leading-order perturbative expression for $G({\bm k};\omega)$, Eq.~(\ref{e185}), employed here; these failures cannot therefore signify possible shortcomings in the Luttinger-Ward identity and the Luttinger theorem.

\subsubsection{Low-temperature limit (asymmetric cases)}
\label{ss61s10}

Consider a $\mu$ for which the equation (cf. Eq.~(\ref{e248}))
\begin{equation}
\mp\frac{1}{2} (\omega_0^2 + \t{U}^2)^{1/2} + \frac{1}{2}\omega_0 = \mu,\;\; \mathcal{C} \lessgtr 0, \label{e256}
\end{equation}
has a solution $\omega_0$ inside ${\sf S}$. Aside from this requirement, the $\mu$ in Eq.~(\ref{e256}) is entirely arbitrary; in our specific applications, we shall however ultimately identify $\mu$ with the $\mu_{\beta}$ corresponding to $\b{n}_{\beta} = 1$.

Following Eq.~(\ref{e246}), for sufficiently large $\beta$ one needs only to deal with either $\mathcal{I}^{<}$ or $\mathcal{I}^{>}$, depending on whether $\mathcal{C} <0$ or $\mathcal{C}>0$ respectively. On the basis of a similar approach as in the symmetric cases (Sec.~\ref{ss61s6}), one readily obtains that for $\beta\to\infty$
\begin{equation}
\mathcal{I}^{\lessgtr} \sim \int_{-\Omega_{-}}^{\omega_0} {\rm d}\omega\; \mathcal{D}(\omega) \Big(1\mp\frac{\omega}{(\omega^2+\t{U}^2)^{1/2}}\Big) +\frac{\Gamma^{\lessgtr}}{\beta} \ln\Big(\frac{1+ x^{-1} \mathrm{e}^{\pm\beta (\t{U}\mp\omega_0+\phi(\omega_0))/2}}{1+ x\, \mathrm{e}^{\mp\beta (\t{U}\mp\omega_0+\phi(\omega_0))/2}}\Big), \label{e257}
\end{equation}
where
\begin{equation}
\Gamma^{\lessgtr} \equiv \frac{2\mathcal{D}(\omega_0)}{1\mp\phi'(\omega_0)} \Big(1\mp \frac{\omega_0}{(\omega_0^2+\t{U}^2)^{1/2}}\Big). \label{e258}
\end{equation}
As before, $x$ is defined according to Eq.~(\ref{e218}). In arriving at the expressions in Eq.~(\ref{e257}), we have subdivided the integrals with respect to $\omega$, in the defining expressions for $\mathcal{I}^{\lessgtr}$, into sub-integrals over $[-\Omega_-,\omega_0]$ and $[\omega_0,+\Omega_+]$. Following this, we have employed geometric series expansions similar to those in Sec.~\ref{ss61s6}; for $\mathcal{I}^{\lessgtr}$ and in dealing with integrals over $[-\Omega_-,\omega_0]$ and $[\omega_0,+\Omega_+]$, we have employed series expansions in powers of (cf. Eq.~(\ref{e214}))
\begin{equation}
\mathrm{e}^{\mp\beta (\mp\omega/2 +(\omega^2+\t{U}^2)^{1/2}/2 \pm\mu)}\;\;\mbox{\rm and}\;\; \mathrm{e}^{\pm\beta (\mp\omega/2 +(\omega^2+\t{U}^2)^{1/2}/2 \pm\mu)} \label{e259}
\end{equation}
respectively. We point out that the expressions in Eq.~(\ref{e257}) apply for a somewhat different class of $\mathcal{D}(\omega)$ than we considered for symmetric cases; for instance, the results in Eq.~(\ref{e257}) apply even if $\mathcal{D}(\omega)$ may be vanishing, or integrably diverging, at one or both of $\tau_{\rm min}$ and $\tau_{\rm max}$.

Making use of Eqs.~(\ref{e246}) and (\ref{e257}), from $\b{n}_{\beta} =1$ we deduce the following asymptotic expressions corresponding to $\beta\to\infty$:
\begin{equation}
x \sim \mathrm{e}^{\pm \beta (\t{U}\mp\omega_0 +\phi(\omega_0))/2}\, \mathrm{e}^{-\beta Q^{\lessgtr}},\;\;\; \mathcal{C} \lessgtr 0, \label{e260}
\end{equation}
where
\begin{equation}
Q^{\lessgtr} \equiv \frac{1}{\Gamma^{\lessgtr}} \Big\{\mathcal{C} -\delta\b{n}_{\beta} - \int_{\pm\Omega_{\pm}}^{\omega_0} {\rm d}\omega\; \mathcal{D}(\omega) \Big(1\mp \frac{\omega}{(\omega^2+\t{U}^2)^{1/2}}\Big)\Big\}. \label{e261}
\end{equation}

With reference to Eq.~(\ref{e218}), for the $\mu_{\beta}$ corresponding to $\b{n}_{\beta} =1$ we thus obtain the following implicit expressions specific to $\beta\to\infty$:
\begin{equation}
\mu_{\beta} \sim \mp \frac{1}{2} (\t{U} \mp\omega_0+\phi(\omega_0)) + Q^{\lessgtr},\;\;\; \mathcal{C} \lessgtr 0. \label{e262}
\end{equation}
Comparing these expressions with those in Eq.~(\ref{e256}), with $\mu$ herein identified with $\mu_{\beta}$, one concludes that unless $\Gamma^{\lessgtr}$ is unbounded, $\omega_0$ must satisfy
\begin{equation}
\int_{\pm\Omega_{\pm}}^{\omega_0} {\rm d}\omega\; \mathcal{D}(\omega) \Big(1\mp \frac{\omega}{(\omega^2+\t{U}^2)^{1/2}}\Big) = \mathcal{C} -\delta\b{n}_{\beta},\;\; \mathcal{C} \lessgtr 0. \label{e263}
\end{equation}
For $\beta=\infty$, where $\delta\b{n}_{\beta}$ is vanishing, these equations identically coincide with those deduced from Eq.~(\ref{e251}) through the requirement $\lim_{\beta\to\infty}\b{n}_{\beta} = 1$. We observe that to the order in which the asymptotic expressions in Eqs.~(\ref{e257}) are exact, in the case at hand the temperature dependence of $\mu_{\beta}$ is entirely determined by that of $\delta\b{n}_{\beta}$; this dependence is mediated through the dependence on temperature of $\omega_0$, brought about through Eq.~(\ref{e263}). Note that since in the present case the GS of the system under investigation is metallic, the expression for $\delta\b{n}_{\beta}$ in Eq.~(\ref{e205}) does not apply; for the relevant technical details, we refer the reader to appendix \ref{sf}.

\subsubsection{Some approaches that lead to erroneous solutions}
\label{ss61s11}

Let us consider a case where $x = O(1)$. Since
\begin{equation}
\mathrm{e}^{-\beta (\t{U}-\Omega_{\pm}+\phi(\Omega_{\pm}))/2} \ll 1
\nonumber
\end{equation}
for sufficiently large $\t{U}$ and $\beta\to\infty$, in this case one may suffice to approximate the sums on the RHS of Eqs.~(\ref{e215}) and (\ref{e216}) by the terms corresponding to $j=1$. Taking into account the fact that $\mathcal{C}$ is vanishing to at least quadratic order in $\tau_{\bm k}/\t{U}$ and that $\delta\b{n}_{\beta}$ is exponentially diminishing for $\beta\to\infty$, the equation $\b{n}_{\beta} = 1$ may be approximated as (cf. Eq.~(\ref{e204}))
\begin{equation}
-\mathcal{A}_{+}^{(1)}\, x + \mathcal{A}_{-}^{(1)}\, x^{-1} = 0, \label{e264}
\end{equation}
from which one obtains that
\begin{equation}
x = \Big(\frac{\mathcal{A}_{-}^{(1)}}{\mathcal{A}_{+}^{(1)}}\Big)^{1/2}, \label{e264a}
\end{equation}
and consequently (see Eq.~(\ref{e218}))
\begin{equation}
\mu_{\beta} = \frac{1}{2\beta} \ln\Big(\frac{\mathcal{A}_{+}^{(1)}}{\mathcal{A}_{-}^{(1)}}\Big). \label{e264b}
\end{equation}

Making use of the asymptotic expressions in Eq.~(\ref{e222}), from Eq.~(\ref{e264b}) one deduces that
\begin{equation}
\mu_{\beta} \sim \frac{\Omega_{+} - \Omega_{-} -(\phi(\Omega_+)-\phi(\Omega_-))}{4} + \frac{1}{2\beta} \ln\Big(\frac{\Gamma_+}{\Gamma_-}\Big)\;\;\;\mbox{\rm for}\;\;\; \beta\to\infty. \label{e264c}
\end{equation}
Consequently (cf. Eq.~(\ref{e221}))
\begin{equation}
\lim_{\beta\to\infty} \mu_{\beta} = \frac{\Omega_{+} - \Omega_{-} -(\phi(\Omega_+)-\phi(\Omega_-))}{4} \approx \frac{\Omega_{+} - \Omega_{-}}{4} \equiv \frac{\tau_{\rm min} +\tau_{\rm max}}{4}. \label{e264d}
\end{equation}
This result identically coincides with the zero-temperature limit of the chemical potential in the `canonical' ensemble deduced in Ref.~\citen{AR06} and denoted by $\mu_{n}$ (see the final remark in Sec.~\ref{ss61}). It is on the basis of this result that Rosch \cite{AR06} arrived at the conclusion that the Luttinger theorem is valid in the `canonical' ensemble in symmetric cases where $\tau_{\rm min} = -\tau_{\rm max}$, so that $\lim_{\beta\to\infty} \mu_{\beta} = 0$. Evidently, the condition $\tau_{\rm min} = -\tau_{\rm max}$ (or $\Omega_- =\Omega_+$) is only necessary but not sufficient for $\mathcal{D}(\omega)\equiv \mathcal{D}(-\omega)$, and consequently $\mathcal{C} \equiv 0$; hence, even when $\tau_{\rm min} = -\tau_{\rm max}$, Eq.~(\ref{e264d}) need not apply (Sec.~\ref{ss61s9}). Note however that according to the expression in Eq.~(\ref{e264a}), for the cases where $\tau_{\rm min} = -\tau_{\rm max}$ one has $x=1$, in conformity with the assumption $x=O(1)$ underlying Eq.~(\ref{e264}).

Although for symmetric cases the expression in Eq.~(\ref{e264d}) yields the correct value for the zero-temperature limit of the $\mu_{\beta}$ corresponding to $\b{n}_{\beta} =1$, this expression is manifestly incorrect for asymmetric cases: with reference to the exact result in Eq.~(\ref{e248}) (see also Eq.~(\ref{e255})), no $\mu_{\infty}$ whose magnitude is of the order of $\tau_{\bm k}$ can lead to satisfaction of $\b{n}_{\infty} =1$ when $\vert\tau_{\bm k}\vert/\t{U} \ll 1$, $\forall {\bm k}$.

To clarify the reason underlying the above incorrect result, one needs only to substitute the expression on the RHS of Eq.~(\ref{e264c}) in the expression on the RHS of Eq.~(\ref{e218}). One immediately observes that, unless $\Omega_- = \Omega_+$ (more precisely, $\Omega_{+} -\phi(\Omega_+) = \Omega_{-} -\phi(\Omega_-)$), the resulting $x$ exponentially approaches $0$ or $\infty$, depending in whether $\Omega_{+} >\Omega_{-}$ or $\Omega_{+} <\Omega_{-}$ respectively. Both of these results are in violation of the assumption $x = O(1)$ underlying Eq.~(\ref{e264}). In this connection, note that for $x\ll 1$, $\mathcal{I}^{>}$ dominates (owing to $x^{-j}$, $j=1,2,\dots$) and for $x\gg 1$, $\mathcal{I}^{<}$ dominates (see Eqs.~(\ref{e215}) and (\ref{e216})); these aspects can be explicitly verified on the basis of the explicit expressions for $\Phi_{\pm}(x)$ in Eq.~(\ref{e229}). In contrast, for $x=O(1)$ both $\mathcal{I}^{<}$ and $\mathcal{I}^{>}$ are relevant.

Now we restore $\mathcal{C}$ and for transparency neglect $\delta\b{n}_{\beta}$. By the same reasoning as leading to Eq.~(\ref{e264}), one obtains that
\begin{equation}
-\mathcal{A}_{+}^{(1)}\, x + \mathcal{A}_{-}^{(1)}\, x^{-1} = \mathcal{C}\, \mathrm{e}^{\beta\t{U}/2}. \label{e265}
\end{equation}
This equation clearly underlines the significance of a non-vanishing $\mathcal{C}$, on account of the magnifying effect of the exponentially divergent factor $\mathrm{e}^{\beta\t{U}/2}$ for $\beta\to\infty$. In spite of this fact, one should note that in Eq.~(\ref{e265}) $\beta$ and $\mathcal{C}$ do not occur in the combination  $\beta \mathcal{C}$, which would be required for reproducing a result for $\mu_{\infty}$ similar to that in Eq.~(\ref{e255}). One easily verifies that the appearance of $\mathcal{C}$ in the expression for $\mu_{\beta}$ in general, and in that for $\mu_{\infty}$ in particular, is a direct consequence of \emph{infinite} sums, such as those on the RHSs of Eqs.~(\ref{e215}) and (\ref{e216}).

\subsubsection{The Luttinger-Ward identity and the Luttinger theorem revisited (asymmetric cases)}
\label{ss61s12}

The above considerations, demonstrating the significance of a non-vanishing $\mathcal{C}$, irrespective of its magnitude, raises the question whether perturbative corrections, not taken account of by the $\t{G}({\bm k};z)$ and $\t{\Sigma}_{\rm loc}(z)$ in Eq.~(\ref{e190}), can lead to restoration of the Luttinger-Ward identity, and thus the Luttinger theorem, for general asymmetric cases. In view of the detailed analysis that we have already presented in this paper concerning the general validity of the Luttinger theorem, the answer to this question is necessarily in the positive. Therefore in what follows we shall not attempt to construct an explicit proof of the Luttinger theorem as applied to the specific problem at hand. Instead, we shall show that corrections to the local self-energy $\Sigma_{\rm loc}(\omega)$ have dramatic consequences.

We proceed by introducing the following ansatz (cf. Eq.~(\ref{e156})):
\begin{equation}
\Sigma({\bm k};\omega) = \frac{\t{U}}{2} + \frac{(\t{U}/2)^2}{\omega-\zeta({\bm k};\omega)}, \label{e266}
\end{equation}
where $\zeta({\bm k};\omega)$ is some function whose various relevant properties will be established in the course of the considerations that follow. For completeness, although it would seem to be more encompassing to multiply the numerator of the second term on the RHS of Eq.~(\ref{e266}) with a function $\xi({\bm k};\omega)$, it can be explicitly shown that by necessity $\xi({\bm k};\omega)$ must approach $1$ for $\tau_{\bm k}/\t{U}\to 0$, $\forall {\bm k}$; both for this reason and for keeping the following algebra transparent, we have decided not to introduce $\xi({\bm k};\omega)\not\equiv 1$ in the ansatz in Eq.~(\ref{e266}).

In analogy with Eq.~(\ref{e185}), for the Green function associated with the self-energy in Eq.~(\ref{e266}) one has
\begin{equation}
G({\bm k};\omega) = \frac{1}{\omega -\tau_{\bm k} -(\t{U}/2)^2/(\omega-\zeta({\bm k};\omega))}. \label{e267}
\end{equation}
In the limit $\vert t_{\ell,\ell'}\vert \ll \vert J\vert \ll U$, $\forall\ell,\ell'$, Rosch \cite{AR06} obtained that
\begin{equation}
G({\bm k};\omega) \sim \frac{-12}{U^3 J}\,\gamma_{\bm k}\;\;\;\mbox{\rm as}\;\;\; \omega\to 0, \label{e268}
\end{equation}
where
\begin{equation}
\gamma_{\bm k} {:=} \sum_{\ell} (t_{\ell,\ell'})^3\, \cos\big({\bm k}\cdot ({\bm R}_{\ell}-{\bm R}_{\ell'})\big). \label{e269}
\end{equation}
On the basis of the result in Eq.~(\ref{e268}) and the expression in Eq.~(\ref{e267}), one readily obtains that
\begin{equation}
\zeta({\bm k};\omega) \sim \frac{3}{U J}\,\gamma_{\bm k}\;\;\; \mbox{\rm as}\;\;\; \omega\to 0. \label{e270}
\end{equation}
This result shows that the lower-order non-local correction to self-energy, giving rise to the result in Eq.~(\ref{e268}) for $\omega\to 0$, brings about a displacement of the simple pole of $\Sigma_{\rm loc}(\omega)$ at $\omega=0$ to
\begin{equation}
\omega \approx \frac{\zeta({\bm k};0)}{1-\zeta'({\bm k};0)}
\sim \zeta({\bm k};0) \equiv \frac{3}{U J}\,\gamma_{\bm k}\;\;\;\mbox{\rm for}\;\;\; \vert\zeta'({\bm k};0)\vert \ll 1, \label{e271}
\end{equation}
where $\zeta'({\bm k};\omega) \equiv \partial\zeta({\bm k};\omega)/\partial\omega$; we shall see in Sec.~\ref{ss61s13} that on general grounds $\zeta'({\bm k};0)$ should to leading order scale like $\tau_{\bm k}/\t{U}$ for sufficiently large $\t{U}$. According to Eq.~(\ref{e270}), the amount of shift in the pole of $\Sigma({\bm k};\omega)$, with respect to that of $\Sigma_{\rm loc}(\omega)$ at $\omega=0$, scales with the third power of $\{ t_{\ell,\ell'}\}$ for $\vert\zeta'({\bm k};0)\vert \ll 1$.

On account of the latter observation and of the result for $\mu_{\infty}$ in Eq.~(\ref{e264d}), which scales with the first power of $\{ t_{\ell,\ell'}\}$, Rosch \cite{AR06} concluded that higher-order non-local corrections to self-energy are not capable of rendering the Luttinger-Ward identity, and therefore the Luttinger theorem, valid in the `canonical' ensemble. The reasoning underlying this statement, which is implicit in the discussions in Ref.~\citen{AR06}, is as follows: the $\mathrm{sgn}(\mu)$ on the RHS of Eq.~(\ref{e190}) would have been $\mathrm{sgn}(\mu-\tau_0)$ if $\Sigma_{\rm loc}(\omega)$ had its simple pole at $\omega=\tau_0$, rather than $\omega=0$; if $\zeta({\bm k};0)$ had turned out to be equal to
\begin{equation}
\tau_0 {:=} \frac{\tau_{\rm min}+\tau_{\rm max}}{4},
\label{e272}
\end{equation}
in view of Eq.~(\ref{e264d}) one would have a restored Luttinger-Ward identity in the `canonical' ensemble\cite{AR06} (Sec.~\ref{ss61}), for
in such case $\mathrm{sgn}(\mu-\tau_0) = \mathrm{sgn}(0) \equiv 0$.

Assuming that $\zeta({\bm k};\omega) \equiv \tau_0$, $\forall {\bm k},\omega$, it can be shown that at half-filling $\vert \mu_{\infty} -\tau_0\vert = O(\t{U})$, irrespective of the value of $\tau_0\not=0$. This follows from the fact that for $\tau_0\not=0$ the energy dispersions $\omega_{\pm}({\bm k})$ as presented in Eq.~(\ref{e189}) change into
\begin{equation}
\omega_{\pm}({\bm k}) = \tau_0 + \frac{1}{2} (\tau_{\bm k} -\tau_0) \pm \frac{1}{2} \Big( (\tau_{\bm k} -\tau_0)^2 + \t{U}^2\Big)^{1/2}. \label{e273}
\end{equation}
Consequently, on effecting the transformations
\begin{equation}
\mathcal{D}(\omega) \rightharpoonup \mathcal{D\,}'(\omega) \equiv \mathcal{D}(\omega+\tau_0),\;\;\; \mu \rightharpoonup \mu' \equiv \mu-\tau_0, \label{e274}
\end{equation}
the present problem becomes mathematically identical to the one that we have extensively considered in Sections \ref{ss61s8} --- \ref{ss61s10}. Although $\mathcal{D\,}'(\omega)$ satisfies the first expression in Eq.~(\ref{e197}), instead of the second expression in Eq.~(\ref{e197})
it however satisfies
\begin{equation}
\int {\rm d}\omega\; \omega\,\mathcal{D\,}'(\omega) = -\tau_0. \label{e275}
\end{equation}
This implies that (cf. Eq.~(\ref{e199}))
\begin{equation}
\mathcal{C}' {:=}\int {\rm d}\omega\; \frac{\mathcal{D\,}'(\omega)\,\omega}{(\omega^2 + \t{U}^2)^{1/2}}
\label{e276}
\end{equation}
cannot be an identically-vanishing function of $\t{U}$. More generally, whatever the nature of $\mathcal{D}(\omega)$, whether symmetric or otherwise, $\mathcal{D\,}'(\omega)$ is an asymmetric function of $\omega$ for $\tau_0\not=0$. From Eq.~(\ref{e276}) one deduces that
\begin{equation}
\mathcal{C}' \sim -\frac{\tau_0}{\t{U}}\;\;\; \mbox{\rm for} \;\;\;\t{U} \to \infty, \label{e277}
\end{equation}
to be contrasted with the leading-order term in the asymptotic series expansion of $\mathcal{C}$ for $\t{U}\to\infty$, which at the slowest decays like $1/\t{U}^{3}$ for $\t{U}\to\infty$ (see Eq.~(\ref{e207}) and note that $\mathcal{I}_1 = 0$ on account of the second expression in Eq.~(\ref{e197})). Thus $\mathcal{C}'$ cannot be identically vanishing for all $\t{U}$ when $\tau_0\not=0$.

From the above observations it follows that irrespective of whether $\mathcal{D}(\omega)$ is symmetric or otherwise, for the zero-temperature limit of the chemical potential specific to half-filling and corresponding to the cases where $\tau_0\not=0$, one has (cf. Eq.~(\ref{e248}))
\begin{equation}
\lim_{\beta\to\infty} \mu_{\beta} = \tau_0 \mp \frac{1}{2} (\omega_0^2 +\t{U}^2)^{1/2} + \frac{1}{2} \omega_0,\;\;\; \mathcal{C}' \lessgtr 0, \label{e278}
\end{equation}
where $\omega_0$ is the solution of (cf. Eq.~(\ref{e263}))
\begin{equation}
\int_{\pm\Omega_{\pm}'}^{\omega_0} {\rm d}\omega\; \mathcal{D\,}'(\omega) \Big(1 \mp \frac{\omega}{(\omega^2+\t{U}^2)^{1/2}}\Big) = \mathcal{C}',\;\;\; \mathcal{C}' \lessgtr 0, \label{e279}
\end{equation}
in which
\begin{equation}
\Omega_{\pm}' \equiv \Omega_{\pm} \mp \tau_0. \label{e280}
\end{equation}
Following Eq.~(\ref{e255}), for sufficiently small $\vert\mathcal{C}'\vert$ one has
\begin{equation}
\lim_{\beta\to\infty} \mu_{\beta} \approx \left. \mu_{N}^{\mp}\right|_{\mathcal{C}'=0} \mp \frac{\vert\mathcal{C}'\vert}{2\mathcal{D\,}'(\pm\Omega_{\pm}')},\;\;\; \mathcal{C}' \lessgtr 0, \label{e281}
\end{equation}
where (cf. Eq.~(\ref{e235}))
\begin{equation}
\left. \mu_{N}^{\pm}\right|_{\mathcal{C}'=0} = \tau_0 \pm \frac{1}{2} \t{U} +\frac{1}{2} (\mp\Omega_{\mp}' \pm \phi(\mp\Omega_{\mp}')). \label{e282}
\end{equation}
Note that $\mathcal{D\,}'(\pm\Omega_{\pm}') \equiv \mathcal{D}(\pm\Omega_{\pm})$.

We have thus shown that for an asymmetric $\mathcal{D}(\omega)$, the Luttinger theorem \emph{cannot} be rendered valid for $\zeta({\bm k};\omega) \equiv \tau_0$, $\forall {\bm k},\omega$, whatever the value of $\tau_0\not=0$ may be. That this result cannot be held against the general validity of the Luttinger theorem, is demonstrated by the following considerations which show that $\zeta({\bm k};\omega) \equiv \tau_0$, $\forall {\bm k},\omega$, leads to contradiction for any $\tau_0\not=0$. Explicitly, the exact $\zeta({\bm k};\omega)$ is a non-trivial function of both ${\bm k}$ and $\omega$.

\subsubsection{Calculation of $\zeta({\bf k};\omega_{\pm})$}
\label{ss61s13}

We proceed by first calculating the `poles' of the Green function in Eq.~(\ref{e267}) which we denote by $\omega_{\pm}({\bm k})$ (at times also by $\omega_{\pm}$ for conciseness). Assuming that
\begin{equation}
\vert \tau_{\bm k} + \zeta({\bm k};\omega_{\pm})\vert \ll \t{U},
\label{e283}
\end{equation}
one readily obtains the following \emph{implicit} expressions for $\omega_{\pm}({\bm k})$ (cf. Eq.~(\ref{e189})):
\begin{equation}
\omega_{\pm}({\bm k}) \sim \frac{1}{2} \big(\tau_{\bm k} + \zeta({\bm k};\omega_{\pm})\big) \pm \frac{1}{2} \t{U}. \label{e284}
\end{equation}
The ${\bm k}$-dependent functions $\zeta({\bm k};\omega_{\pm})$ can be calculated on the basis of the fact that by construction
\begin{equation}
G^{-1}({\bm k};\omega_{\pm}({\bm k})) \equiv 0, \;\;\;\forall {\bm k},
\label{e285}
\end{equation}
so that from the Dyson equation (Eq.~(\ref{ec53})) one has
\begin{equation}
\Sigma({\bm k};\omega_{\pm}({\bm k})) = G_0^{-1}({\bm k};\omega_{\pm}({\bm k})),\;\;\; \forall {\bm k}. \label{e286}
\end{equation}
Here $G_0({\bm k};\omega)$ denotes the `unperturbed' Green function,  for which one has
\begin{equation}
G_{0}({\bm k};\omega) = \frac{1}{\omega -\tau_{\bm k} +\t{U}/2}. \label{e287}
\end{equation}

Using the implicit asymptotic expressions for $\omega_{\pm}({\bm k})$ in Eq.~(\ref{e284}), form Eq.~(\ref{e286}) one obtains that
\begin{equation}
\frac{\t{U}^2}{\tau_{\bm k}-\zeta({\bm k};\omega_{\pm}) \pm \t{U}} \sim -\big(\tau_{\bm k} - \zeta({\bm k};\omega_{\pm}) \mp \t{U}\big). \label{e288}
\end{equation}
One observes that on neglecting $\tau_{\bm k}$ and $\zeta({\bm k};\omega_{\pm})$, which is justified for $\vert\tau_{\bm k}\vert/\t{U} \ll 1$ and $\vert\zeta({\bm k};\omega_{\pm})\vert/\t{U} \ll 1$, the expressions in Eq.~(\ref{e288}) reduce to identities. The condition $\vert\tau_{\bm k}\vert/\t{U} \ll 1$ is satisfied by construction and we shall see below that $\zeta({\bm k};\omega_{\pm}) \sim \tau_{\bm k}$ so that $\vert\zeta({\bm k};\omega_{\pm})\vert/\t{U} \ll 1$ is also satisfied. These facts combine into an \emph{a posteriori} justification of our neglect of a second function, $\xi({\bm k};\omega) \not\equiv 1$, in the expression for $\Sigma({\bm k};\omega)$ in Eq.~(\ref{e266}).

Solving the equations in Eq.~(\ref{e288}), one obtains that
\begin{equation}
\zeta({\bm k};\omega_{\pm}) \sim \tau_{\bm k} \iff \omega_{\pm}({\bm k}) \sim \tau_{\bm k} \pm \frac{1}{2} \t{U}. \label{e289}
\end{equation}
These results are interesting for several reasons. For instance, in the light of the result in Eq.~(\ref{e270}), one observes that $\zeta({\bm k};\omega)$ is a non--trivial function of both ${\bm k}$ and $\omega$.
Further, by considering $\zeta({\bm k};0)$ as vanishingly small with respect to $\zeta({\bm k};\omega_{\pm})$, in view of $\omega_{\pm}({\bm k}) \approx \pm\t{U}/2$, it follows that in the cases where $\zeta({\bm k};\omega)$ is a smooth function of $\omega$ in the regions $\omega \gtrless 0$, $\partial \zeta({\bm k};\omega)/\partial\omega$ is approximately equal to $\pm 2\tau_{\bm k}/\t{U}$ for $\omega \gtrless 0$. This is naturally a very crude approximation, since $\zeta({\bm k};\omega)$ is likely to vary strongest for $\omega$ in the neighbourhoods of $\mu_{N}^{-}$ and $\mu_{N}^{+}$. Whether $\zeta({\bm k};\omega)$ is a smooth function of $\omega$ or not, the above observations establish the condition $\zeta({\bm k};\omega) \equiv \tau_0$, $\forall {\bm k},\omega$, as fundamentally incorrect.

The dependence of $\zeta({\bm k};\omega)$ on $\omega$ is particularly significant in connection with the Luttinger-Ward identity in which one encounters $\partial\t{\Sigma}_{\sigma}({\bm k};z)/\partial z$, leading to appearance of $1-\partial\zeta({\bm k};z)/\partial z$ in the expression for $\b{N}_{\sigma}^{(2)}$, Eqs.~(\ref{e43}) and (\ref{e44}). It should be noted that the unconditional validity of the Luttinger theorem in the local limit (Sec.~\ref{ss61s3}) is consistent with the viewpoint that $\partial\zeta({\bm k};\omega)/\partial\omega$ may scale like $\tau_{\bm k}/\t{U}$, in which case $\partial\zeta({\bm k};z)/\partial z \to 0$ for $\tau_{\bm k}/\t{U}\to 0$.

Rosch \cite{AR06} has emphasised that for $\omega$ close to poles of $G_{\rm loc}(\omega)$, such arguments as leading to the expression in Eq.~(\ref{e268}) are insufficient, and that in these regions perturbation series in powers of $\{ t_{\ell,\ell'}\}$ has to be summed to infinite order for obtaining the leading-order correction to $G_{\rm loc}(\omega)$. As we have shown, for asymmetric cases the $\mu_{\infty}$ corresponding to $\b{n}_{\infty} =1$ is located in an immediate neighbourhood of one of the poles of $G_{\rm loc}(\omega)$. It should be noted that since in asymmetric cases the $N$-particle GSs corresponding to half-filling are metallic, the observed breakdown of the Luttinger theorem is unrelated to the possibility of this breakdown arising from a false limit associated with evaluating the limit $\beta\to\infty$ for $\mu\not=\mu_{\infty}$; for metallic GSs, $\mu$ can deviate from $\mu_{\infty}$ by an amount of the order of $1/N$ (appendix \ref{sc}), which is vanishing in the thermodynamic limit.

\subsubsection{Summary and concluding remarks}
\label{ss61s14}

We have demonstrated that the Luttinger theorem is unconditionally valid in the local limit of the model considered by Rosch \cite{AR06}. The same statement unequivocally applies for the cases where the influence of hopping terms in the underlying Hamiltonian is taken into account to leading order, provided that the density-of-states function $\mathcal{D}(\omega)$ corresponding to the hopping energy dispersion function $\tau_{\bm k}$ satisfies $\mathcal{D}(-\omega) \equiv \mathcal{D}(\omega)$. Key to our findings has been the observation that,
for the insulating GSs considered in this section, effecting the zero-temperature limit, which constitutes a nontrivial step in the derivation of the Luttinger theorem, for $\mu\not=\mu_{\infty}$ or $\mu\not=\mu_{\beta}$, leads to false limits, one corresponding to $\mu<\mu_{\infty}$ and one to $\mu>\mu_{\infty}$.

We have further established that the Luttinger theorem breaks down in the cases where $\mathcal{D}(-\omega) \not\equiv \mathcal{D}(\omega)$ when $\Sigma({\bm k};\omega)$ is approximated by its local limit $\Sigma_{\rm loc}(\omega)$, and $G^{-1}({\bm k};\omega)$ by one corrected with respect to $G_{\rm loc}^{-1}(\omega)$ to leading order in $\tau_{\bm k}/\t{U}$.
By demonstrating that this apparent breakdown is mediated by the deviation from zero of a constant $\mathcal{C}$, Eq.~(\ref{e199}), irrespective of how small $\vert\mathcal{C}\vert$ may be,\footnote{Formally, $\mathcal{C}$ decays like $1/\t{U}^{2m+1}$ as $\t{U}\to\infty$, for $m$ some integer satisfying $1\le m<\infty$.} we have made explicit that it is \emph{a priori} unjustified to declare the Luttinger theorem as failing for asymmetric (in the sense indicated above) Mott-insulating GSs. In support of this statement, we have shown that the leading-order correction to the position of the pole of the local self-energy $\Sigma_{\rm loc}(\omega)$, at $\omega=0$, is dependent on both ${\bm k}$ and $\omega$, and that for $\omega$ in the neighbourhoods of the poles of $G_{\rm loc}(\omega)$ at $\omega=\pm\t{U}/2$, the magnitude of this correction behaves like $\tau_{\bm k}$ for sufficiently large $\t{U}$. Remarkably, we have shown that, for $\mathcal{C}\not=0$, the zero-temperature limit of the chemical potential $\mu_{\beta}$ corresponding to half-filling is located in the vicinity of one of the poles of $G_{\rm loc}(\omega)$; whether this chemical potential is close to $-\t{U}/2$ or $\t{U}/2$, turns out to be determined by the \emph{sign} and not by the magnitude of $\mathcal{C}\not=0$. In fact, for $\mathcal{C}\not=0$, independent of how small $\vert\mathcal{C}\vert$ may be, the GS of the system under investigation is a metallic one at half-filling.\footnote{This statement has bearing on the above-mentioned approximations for $G({\bm k};\omega)$ and $\Sigma({\bm k};\omega)$.}

Recently, Stanescu, Phillips and Choy \cite{SPC07} interpreted the
findings by Rosch \cite{AR06} as signifying breakdown of the many-body perturbation theory in the strong-coupling regime; they questioned the possibility that the self-energy in Eq.~(\ref{e156}) ``can be
generated from the noninteracting limit''. They subsequently put forward a ``modified Luttinger theorem'' that should be valid not only in
the Mott-insulating phase, but also in the phases associated with large values of the on-site repulsion energy $U$ in comparison with the nearest-neighbour hopping-integral parameter $t$.

Although Stanescu \emph{et al.} \cite{SPC07} conceded that in the particle-hole symmetric case, signified amongst others by ``$\mu=0$'', the Luttinger theorem is valid, it is unclear on which grounds the authors justified the specific choice $\mu=0$ as uniquely relevant to the Luttinger theorem as applied to the GS at hand. In this connection, one should recall that the expressions in Eqs.~(\ref{e158}) and (\ref{e190}) are indeed vanishing for $\mu=0$ and that the conclusion arrived at by Rosch in Ref.~\citen{AR06} concerning breakdown of the Luttinger theorem, even in symmetric cases, is not based on any denial of these basic facts, rather on the consideration that insofar as the Luttinger theorem is concerned, for insulating GSs the value of $\mu$
should not be bound by any other restriction than $\mu\in (\mu_{N}^-, \mu_{N}^+)$ (Sections \ref{s1}, \ref{ss23} and \ref{ss61s1}). It is only through the explicit expressions in Eqs.~(\ref{e160a}) and (\ref{e190a}) that it becomes evident that without $\mu=\mu_{\infty}$ (or $\mu=\mu_{\beta}$) the zero-temperature limiting process underlying the Luttinger-Ward identity can potentially result in spurious zero-temperature limits.

In conclusion, we wish to emphasise that the question whether a self-energy ``can be generated from the noninteracting limit'' is \emph{not} material to the proof of the Luttinger theorem: the role of ``the non-interacting limit'' in this proof is restricted to the requirement that in the limit of vanishing coupling-constant of interaction, self-energy must vanish; this requirement, summarised in Eq.~(\ref{e78}), merely serves to provide the first-order differential equation in Eq.~(\ref{e70}) with the initial condition in Eq.~(\ref{e69}), leading to the equality in Eq.~(\ref{e71}).\footnote{We draw the attention of the reader to the footnote related to Eq.~(\protect\ref{e78}).} Disposing of this only requirement would necessarily devoid the Luttinger theorem of its significance. As regards the use by Luttinger and Ward \cite{LW60} of the series $\sum_{\nu=1}^{\infty} \t{\Sigma}_{\sigma}^{(\nu)}({\bm k};z)$ for $\t{\Sigma}_{\sigma}({\bm k};z)$, the reader may in particular consult the considerations in Sec.~\ref{ss53s2}; a simple example that we shall present in Sec.~\ref{s7}, should further elucidate some of the abstract notions encountered in Sec.~\ref{ss53s2}.

\subsection{Case II}
\label{ss62}

For a model describing a spin-density-wave state on a two-dimensional square lattice, numerical results by Chubukov and collaborators \cite{CMS96,CM97} revealed violation of the Luttinger theorem, Eq.~(\ref{e18}). This finding has been confirmed analytically by Altshuler \emph{et al.} \cite{ACDFM98} who identified breakdown of the Luttinger-Ward identity, Eq.~(\ref{e44}), as the origin of the failure of the Luttinger theorem for the model under consideration.\cite{Note10} Altshuler \emph{et al.} \cite{ACDFM98} ascribed this failure to ``a hidden anomaly,'' ``similar to the chiral anomaly in quantum electrodynamics,'' whose influence would not be captured by the perturbative approach of Luttinger and Ward; only a ``proper regularization'' of the formalism would bring the impact of this anomaly, such as the breakdown of the Luttinger-Ward identity, to light.

Below we show that the breakdown of the Luttinger-Ward identity is in fact rooted in a finite cut-off energy, $\Omega_0$, in the underlying model which gives rise to a non-analyticity in the self-energy $\t{\Sigma}_{\sigma}({\bm k};z)$ away from the real axis of the complex $z$ plane, one that the exact self-energy cannot possess (appendix \ref{sc}). The consequences of a finite $\Omega_0$ is most profound for the cases where the coupling constant of interaction, $\Delta$, is comparable with or in excess of $\Omega_0$.\footnote{Unlike $\lambda$ elsewhere in this paper, which is dimensionless, $\Delta$ has the dimension of energy.} Our numerical results, which, where comparison has been possible, have proved fully to conform with those reported in Refs.~\citen{CMS96,CM97}, establish that indeed for $\Delta/\Omega_0 \to 0$ the Luttinger-Ward identity is fully restored. The latter possibility has been overlooked in the investigations by Altshuler \emph{et al.} \cite{ACDFM98} through neglect of a contribution to self-energy on account of its purported insignificance \cite{CMS96} in the strong-coupling regime.

The diagnosis that Altshuler \emph{et al.} \cite{ACDFM98} presented for the failure of the Luttinger-Ward identity in the case at hand amounts to an undue oversimplification of the proof by Luttinger and Ward \cite{LW60} of the Luttinger-Ward identity. As our detailed examination of this proof must have made abundantly evident, the only perturbation series employed by Luttinger and Ward is that for $\t{\Sigma}_{\sigma}({\bm k};z)$ represented by means of skeleton diagrams and evaluated in terms of the \emph{interacting} single-particle Green functions $\{ \t{G}_{\sigma'}({\bm k};z)\,\|\, \sigma'\}$ and the bare two-body interaction potential. In contrast, the explicit expansion of $\sum_{\sigma} \b{N}_{\sigma}^{(2)}$ (denoted by $I_2$ in Ref.~\citen{ACDFM98}) in powers of $\Delta$ (see Eq.~(12) and the subsequent discussions in Ref.~\citen{ACDFM98}) unduly suggests that the proof by Luttinger and Ward must have relied on the use of a perturbation series for $\t{\Sigma}_{\sigma}({\bm k};z)$ in terms of the \emph{bare} Green functions $\{ \t{G}_{\sigma';0}({\bm k};z)\,\|\, \sigma'\}$.\footnote{As we pointed out in Sec.~\protect\ref{ss52s3}, this series, Eq.~(\protect\ref{e92}), need not converge.}

The ``regularization'' to which Altshuler \emph{et al.} \cite{ACDFM98} referred is implicit in the original proof by Luttinger and Ward \cite{LW60} of the Luttinger-Ward identity. What this proof does not account for, however, are the peculiarities of self-energies and Green functions of models, such as the model by Chubukov and collaborators, which are artificial and specific to these models.

\begin{table}[t!]
\caption{Symbols adopted for the same quantities in various relevant publications. Key: CMS (Chubukov, \emph{et al.} \protect\cite{CMS96}), CM (Chubukov and Morr \protect\cite{CM97}), ACDFM (Altshuler, \emph{et al.} \protect\cite{ACDFM98}). The symbol $\omega_0$, adopted in Ref.~\protect\citen{CM97}, should not be confused with the Matsubara frequency $\omega_m$ at $m=0$. With reference to $g_{\rm e}^{(3)} \equiv g_{\rm cr}^{(2)}/2$, we point out that ``$g_{\rm cr}^{(2)} \approx 0.82$'' in the caption of Fig.~15 in Ref.~\protect\citen{CM97} should have been $g_{\rm cr}^{(2)}/2 \approx 0.82$; this can be readily verified on the basis of the unnumbered expression for $g_{\rm cr}^{(2)}$ following Eq.~(35) in Ref.~\protect\citen{CM97} (see also the captions to Figs.~1 and 2 in Ref.~\protect\citen{CMS96}). }
\label{t3}
\begin{center}
\begin{tabular}{lccl} \hline\hline
{\footnotesize CMS} & {\footnotesize CM} & {\footnotesize ACDFM} & {\footnotesize Present} \\
\hline \vspace{-10pt} \\
$\;\;\;C$ & $\omega_0$ & --- & $\;\;\;\Omega_0$ \\
$\;\;\;\delta$ & $\Delta_0$ & --- & $\;\;\;\Delta_0$ \\
$\;\;\;g_{\rm e}$ & $g/2$ & $\Delta$ & $\;\;\;\Delta$ \\
$\;\;\;g_{\rm e}^{(3)}$ & $g_{\rm cr}^{(2)}/2$ & --- &
$\;\;\;\Delta_{\rm c}$ \\
\hline
\end{tabular}
\end{center}
\end{table}

In what follows, where no confusion can arise we suppress the spin indices of functions; for instance, below $\mathscr{S}({\bm k};\zeta_m)$ will denote what we have thus far denoted by $\mathscr{S}_{\!\sigma}({\bm k};\zeta_m)$. To facilitate inspection, in Table \ref{t3} we present some relevant symbols used in Refs.~\citen{CMS96,CM97,ACDFM98} as well as in the following considerations.

\subsubsection{Preliminaries}
\label{ss62s1}

For the full self-energy of the model by Chubukov and collaborators \cite{CMS96,CM97} one has (for a detailed derivation of this expression see appendix A in Ref.~\citen{CM97})
\begin{equation}
\mathscr{S}({\bm k};\zeta_m) = \left\{\begin{array}{ll}
\mathscr{S}_{>}({\bm k};\zeta_m), & (\varepsilon_{\bm k}-\mu)^2 -
(\zeta_m-\mu)^2 \ge \Omega_0^2,\\ \\
\mathscr{S}_{<}({\bm k};\zeta_m), & (\varepsilon_{\bm k}-\mu)^2 -
(\zeta_m-\mu)^2 \le \Omega_0^2, \end{array} \right. \label{e290}
\end{equation}
where
\begin{equation}
\mathscr{S}_{\gtrless}({\bm k};\zeta_m) = \frac{1}{\hbar}\,
\frac{\Delta^2}{\zeta_m - \varepsilon_{{\bm k}+{\bm Q}}}\,
\phi_{{\bm k}+{\bm Q}}^{\gtrless}(\zeta_m), \label{e291}
\end{equation}
in which
\begin{equation}
\phi_{\bm k}^{\gtrless}(\zeta) \equiv \left\{ \begin{array}{ll}
1, & \\ \\
\frac{\sqrt{(\varepsilon_{\bm
k}-\mu)^2-(\zeta-\mu)^2 + \Delta_0^2} - \Delta_0}{\sqrt{\Omega_0^2 +
\Delta_0^2} -\Delta_0}. &  \end{array} \right.
\label{e292}
\end{equation}
The quantity $\Delta$ is the coupling constant of interaction, $\Delta_0 \equiv \hbar\, c_{\rm sw}/\xi$, where $c_{\rm sw}$ is the spin-wave velocity and $\xi$ spin-correlation length; $\Omega_0$ is a cut-off energy, Table \ref{t3}, and $\zeta_m \equiv i\hbar\omega_m + \mu$, Eq.~(\ref{e27}). The energy dispersion $\varepsilon_{\bm k}$ is of the form \cite{CM97}
\begin{equation}
\varepsilon_{\bm k} = -2t \big(\cos(k_x) + \cos(k_y)\big) -4 t'\,
\cos(k_x) \cos(k_y), \label{e293}
\end{equation}
which is specific to a two-dimensional square lattice with lattice constant $a$ which is taken as the unit of length so that $(k_x,k_y)$, the coordinates of ${\bm k}$ with respect to the principal axes in the reciprocal space, are in units of the inverse of $a$; thus $k_x, k_y \in [-\pi,\pi)$. In the same units and with respect to the same Cartesian basis,
\begin{equation}
{\bm Q} = (\pi,\pi). \label{e294}
\end{equation}

We shall denote the single-particle Green function corresponding to $\mathscr{S}_{\gtrless}({\bm k};\zeta_m)$, through the Dyson equation, by $\mathscr{G}_{\gtrless}({\bm k};\zeta_m)$, where in particular $\mathscr{G}_{>}({\bm k};\zeta_m)$ has the simple form
\begin{equation}
\mathscr{G}_{>}({\bm k};\zeta_m) \equiv \hbar\,\frac{\zeta_m
-\varepsilon_{{\bm k}+{\bm Q}}}{(\zeta_m - E_{\bm k}^-) (\zeta_m -
E_{\bm k}^+)},\;\;\; \forall m, \label{e295}
\end{equation}
in which
\begin{equation}
E_{\bm k}^{\pm} \equiv \frac{1}{2} (\varepsilon_{\bm
k}+\varepsilon_{{\bm k}+{\bm Q}}) \pm \Big( \Delta^2 +
\big(\frac{1}{2} [\varepsilon_{\bm k}-\varepsilon_{{\bm k}+{\bm
Q}}]\big)^2\Big)^{1/2}. \label{e296}
\end{equation}
In Sec.~\ref{ss62s2} it will become evident how the two branches of $\mathscr{G}({\bm k};\zeta)$, that is $\mathscr{G}_{>}({\bm k};\zeta)$ and $\mathscr{G}_{<}({\bm k};\zeta)$, contribute to $\b{N}_{\sigma}^{(1)}$ and $\b{N}_{\sigma}^{(2)}$ (see Eqs.~(\ref{e34}), (\ref{e35}) and (\ref{e36})).

We point out that whereas $\phi_{\bm k}^{>}(\zeta)$ is analytic in the entire $\zeta$ plane, $\phi_{\bm k}^{<}(\zeta)$ is a multivalued function of $\zeta$ on account of its two distinct branch points at $\zeta=\varepsilon_{\pm}({\bm k})$ along the real axis, where
\begin{equation}
\varepsilon_{\pm}({\bm k}) \equiv \mu \pm \big((\varepsilon_{\bm
k}-\mu)^2 +\Delta_0^2\big)^{1/2}. \label{e297}
\end{equation}
It can be readily verified that
\begin{equation}
\phi_{\bm k}^{<}(\varepsilon+i 0^+) \not\equiv
\phi^{<}_{\bm k}(\varepsilon-i 0^+) \;\;\; \mbox{\rm for}\;\;\; \varepsilon\in \big[-\infty,\varepsilon_-({\bm k})\big)\cup \big(\varepsilon_+({\bm k}),\infty\big]. \label{e298}
\end{equation}
These details clearly illustrate the distinctive natures of $\mathscr{S}_{>}({\bm k};\zeta_m)$ and $\mathscr{S}_{<}({\bm k};\zeta_m)$. In fact, since only for $y_{\bm k}$, satisfying the equation
\begin{equation}
(\varepsilon_{\bm k}-\mu)^2 + y_{\bm k}^2 = \Omega_0^2, \label{e299}
\end{equation}
one has
\begin{equation}
\phi_{\bm k}^{<}(\mu+iy_{\bm k}) = 1,\;\;\; \forall {\bm k},
\label{e300}
\end{equation}
it follows that unless $\zeta = \mu+iy_{\bm k}$,
\begin{equation}
\mathscr{S}_{<}({\bm k};\zeta) \not= \mathscr{S}_{>}({\bm k};\zeta).
\label{e301}
\end{equation}
Consequently, unless $\zeta = \mu+iy_{\bm k}$,
\begin{equation}
\mathscr{G}_{<}({\bm k};\zeta) \not= \mathscr{G}_{>}({\bm k};\zeta).
\label{e302}
\end{equation}
From Eq.~(\ref{e298}) one further deduces that on changing $\eta$ from $0^+$ to $0^-$, $\mathscr{G}_{<}({\bm k};\varepsilon + i\eta)$ undergoes a branch-cut discontinuity for
\begin{equation}
\varepsilon\in \big[-\infty,\varepsilon_{-}({\bm k}+{\bm Q})\big)\cup
\big(\varepsilon_{+}({\bm k}+{\bm Q}),\infty\big]. \label{e303}
\end{equation}

With reference to our earlier remark, we point out that in the considerations by Altshuler \emph{et al.} \cite{ACDFM98}, $\mathscr{G}({\bm k};\zeta)$ has been identified with $\mathscr{G}_{>}({\bm k};\zeta)$ for \emph{all} ${\bm k}$ and $\zeta$; this has been argued \cite{CMS96} to be justified in the regime of large $\Delta$ where $\vert\mu\vert$ would be equally large. This statement is based on the consideration that for sufficiently large $\vert\mu\vert$ no $m$ can exist for which $(\varepsilon_{\bm k} -\mu)^2 + \hbar^2 \omega_m^2 \le \Omega_0^2$ would apply (recall that $\zeta_m = i\hbar\omega_m + \mu$). With the smallest $\hbar\vert\omega_m\vert$ being equal to $\pi/\beta$, one observes that for sufficiently large $\vert\mu\vert$ indeed the condition $(\varepsilon_{\bm k} -\mu)^2 + \hbar^2 \omega_m^2 \le \Omega_0^2$ cannot be fulfilled for any $m$. Two aspects are to be taken into account, however. Firstly, so long as $\mu$ is located inside the interval covered by $\varepsilon_{\bm k}$ (that is, inside $[-4(t+t'), 4(t-t')]$ for $t >t'$), there exists a non-vanishing region in the ${\bm k}$ space where $(\varepsilon_{\bm k} -\mu)^2$ can be as small as desired, whereby, for sufficiently large $\beta$, $(\varepsilon_{\bm k} -\mu)^2 +\hbar^2\omega_m^2 \le \Omega_0^2$ is satisfied for \emph{some} $m$, no matter how small $\Omega_0>0$ may be. Secondly, and importantly, even for a large $\vert\mu\vert$, the condition $(\varepsilon_{\bm k} -\mu)^2 + \hbar^2 \omega_m^2 \le \Omega_0^2$ can be met for some $m$, provided that the cut-off energy $\Omega_0$ is sufficiently large. In other words, strong correlation (say, $\Delta \gtrsim 4t$) cannot be an unequivocal justification for identifying $\mathscr{G}({\bm k};\zeta)$ with $\mathscr{G}_{>}({\bm k};\zeta)$ for all ${\bm k}$ and $\zeta$.

\subsubsection{Details}
\label{ss62s2}

We proceed by determining $\b{\nu}_{\sigma}^{(1)}({\bm k})$, hereafter to be denoted by $\b{\nu}^{(1)}({\bm k})$ (see Eqs.~(\ref{e35}) and (\ref{e37})). The function $\b{\nu}_{\sigma}^{(2)}({\bm k})$, hereafter $\b{\nu}^{(2)}({\bm k})$, is readily determined by following the same strategy (see Eqs.~(\ref{e36}) and (\ref{e43})). For a correct determination of these functions, corresponding to the model under consideration, it is crucial that $\mathscr{S}({\bm k};\zeta)$ ($\mathscr{G}({\bm k};\zeta)$) be correctly identified with the appropriate branches $\mathscr{S}_{>}({\bm k};\zeta)$ and $\mathscr{S}_{<}({\bm k};\zeta)$ ($\mathscr{G}_{>}({\bm k};\zeta)$ and $\mathscr{G}_{<}({\bm k};\zeta)$) when $\zeta$ is not equal to $\zeta_m$ for some $m$; for $\zeta=\zeta_m$ the defining expression in Eq.~(\ref{e290}) is unambiguous.

To determine $\b{\nu}^{(1)}({\bm k})$ we subdivide the closed contour $\Gamma_j^{-}$, $j=1,2$ (see Sections \ref{ss41} and \ref{ss41s1}), into the closed contour $\mathfrak{g}_j^-$ and the remaining part $\mathfrak{G}_j^-$, where $\mathfrak{g}_1^-$ ($\mathfrak{g}_2^-$) encloses the segment of the line $\mu+iy$ corresponding to $y \in (y_0,y_1)$ ($y\in (-y_1,-y_0)$), where
\begin{equation}
y_0 \in (0,\hbar\omega_0)\;\;\; \mbox{\rm and}\;\;\; y_1\in
(\hbar\omega_{m_0},\hbar\omega_{m_0+1}). \label{e304}
\end{equation}
Here $m_0 > 0$ denotes the value of $m$ for which one has
\begin{equation}
(\varepsilon_{{\bm k}+{\bm Q}}-\mu)^2 + (\hbar\omega_{m_0})^2 <\Omega_0^2 \;\;\;\;\mbox{\rm and}\;\;\;\; (\varepsilon_{{\bm k}+{\bm Q}}-\mu)^2 + (\hbar\omega_{m_0+1})^2 >\Omega_0^2. \label{e305}
\end{equation}
Evidently, $m_0$ is a function of ${\bm k}$ and $\mu$ (as well as $\beta$, of which also $y_0$ and $y_1$ are functions). It is possible that for given values of ${\bm k}$ and $\mu$ no positive $m_0$ may satisfy the inequalities in Eq.~(\ref{e305}). In such a case, we formally identify $y_1$ with $y_0$ whereby $\mathfrak{G}_j^-$ identically coincides with $\Gamma_j^-$, $j=1,2$. As in the case of $\Gamma^{-}$, for which one has $\Gamma^{-} = \Gamma_1^{-} \oplus \Gamma_2^{-}$ (Sections \ref{ss41} and \ref{ss41s1}), below $\mathfrak{g}^- = \mathfrak{g}_1^{-} \oplus \mathfrak{g}_2^{-}$.

The above subdivision of the contour $\Gamma_j^-$ into $\mathfrak{g}_j^-$ and $\mathfrak{G}_j^-$, $j=1,2$, is useful in that in replacing the relevant sums with respect to Matsubara frequencies by their corresponding Mellin-Barnes-type contour integrals (\S\S~14.5 and 16.4 in Ref.~\citen{WW62}), one can \emph{unequivocally} identify the analytic continuation of $\mathscr{G}({\bm k};\zeta_m)$ onto $\mathfrak{g}_j^-$ with $\mathscr{G}_{<}({\bm k};\zeta)$, $j=1,2$, and the analytic continuation of $\mathscr{G}({\bm k};\zeta_m)$ onto $\mathfrak{G}_j^-$ with $\mathscr{G}_{>}({\bm k};\zeta)$, $j=1,2$ (see Eq.~(\ref{e295})). For completeness, the contour integrals over $\mathfrak{g}_j^-$ and $\mathfrak{G}_j^-$, $j=1,2$, have the following associations with sums with respect to Matsubara frequencies:
\begin{eqnarray}
&&\hspace{-0.5cm} \int_{\mathfrak{g}_1^-} \frac{{\rm d}\zeta}{2\pi
i} \; \frac{\mathrm{e}^{\zeta\, 0^+}}{\mathrm{e}^{\beta (\zeta-\mu)}
+ 1}\;(\dots) = \frac{1}{\beta}\sum_{m=0 (1)}^{m_0}
\mathrm{e}^{\zeta_m 0^+} (\dots),
\nonumber\\
&&\hspace{-0.5cm} \int_{\mathfrak{g}_2^-} \frac{{\rm d}\zeta}{2\pi
i} \; \frac{\mathrm{e}^{\zeta\, 0^+}}{\mathrm{e}^{\beta (\zeta-\mu)}
+ 1}\;(\dots) = \frac{1}{\beta} \sum_{m=-m_0}^{-1 (0)}
\mathrm{e}^{\zeta_m 0^+} (\dots),
\nonumber\\
&&\hspace{-0.5cm} \int_{\mathfrak{G}_1^-} \frac{{\rm d}\zeta}{2\pi
i} \; \frac{\mathrm{e}^{\zeta\, 0^+}}{\mathrm{e}^{\beta (\zeta-\mu)}
+ 1}\;(\dots) = \frac{1}{\beta}\!\sum_{m=m_0+1}^{\infty}\!\!
\mathrm{e}^{\zeta_m 0^+} (\dots),
\nonumber\\
&&\hspace{-0.5cm} \int_{\mathfrak{G}_2^-} \frac{{\rm d}\zeta}{2\pi
i}\; \frac{\mathrm{e}^{\zeta\, 0^+}}{\mathrm{e}^{\beta (\zeta-\mu)}
+ 1}\;(\dots) = \frac{1}{\beta} \sum_{m=-\infty}^{-m_0-1}
\mathrm{e}^{\zeta_m 0^+} (\dots). \nonumber\\
\label{e306}
\end{eqnarray}

Since for a finite $\Omega_0$, $y_1$ is finite, it follows that the contours $\mathfrak{g}_j^-$ and $\mathfrak{G}_j^-$ cannot be deformed in the way that $\Gamma_j^-$ can, $j=1,2$. This aspect is reflected in the following expression deduced from that in Eq.~(\ref{e35}) (cf. Eq.~(\ref{e37})):
\begin{equation}
\b{\nu}^{(1)}({\bm k}) = \int_{\Gamma^-} \frac{{\rm
d}\zeta}{2\pi i}\; \frac{\mathrm{e}^{\zeta\, 0^+}}{\mathrm{e}^{\beta
(\zeta-\mu)} +1}\; \frac{\partial}{\partial\zeta}\,
\ln\big(-\beta\hbar\, \mathscr{G}_{>}^{-1}({\bm k};\zeta)\big)
+\int_{\mathfrak{g}^+} \frac{{\rm d}\zeta}{2\pi i}\; \frac{\mathscr{K}({\bm k};\zeta)}{\mathrm{e}^{\beta (\zeta-\mu)} +1}, \label{e307}
\end{equation}
where
\begin{equation}
\mathscr{K}({\bm k};\zeta) \equiv
\!\frac{\partial}{\partial\zeta}\Big\{\! \ln\big(\!-\beta\hbar\,
\mathscr{G}_{>}^{-1}({\bm k};\zeta)\big) - \ln\big(\!-\beta\hbar\,
\mathscr{G}_{<}^{-1}({\bm k};\zeta)\big)\!\Big\}. \label{e308}
\end{equation}
The second integral on the RHS of Eq.~(\ref{e307}), over the contour $\mathfrak{g}^+ \equiv \mathfrak{g}_1^+ \oplus \mathfrak{g}_2^+$, is missing in the analysis presented in Ref.~\citen{ACDFM98}. That $\mathscr{K}({\bm k};\zeta)$ cannot be identically vanishing is evident from Eq.~(\ref{e302}).

It can be readily verified that
\begin{eqnarray}
&&\hspace{-0.0cm}\lim_{\beta\to\infty} \int_{\Gamma^-} \frac{{\rm d}\zeta}{2\pi
i}\; \frac{\mathrm{e}^{\zeta 0^+}}{\mathrm{e}^{\beta (\zeta-\mu)}
+1}\; \frac{\partial}{\partial\zeta}\, \ln\big(-\beta\hbar\,
\mathscr{G}_{>}^{-1}({\bm k};\zeta)\big)\nonumber\\
&&\hspace{0.0cm} = -\frac{1}{\pi} \mathrm{Arctan}\Big(\frac{\mathrm{Im}\big[\t{G}_{>}^{-1}({\bm
k};\mu+i 0^+)\big]}{\mathrm{Re}\big[\t{G}_{>}^{-1}({\bm k};\mu+i
0^+)\big]}\Big) \equiv \Theta(\mu - E_{\bm k}^-) + \Theta(\mu -
E_{\bm k}^+) - \Theta(\mu-\varepsilon_{{\bm k}+{\bm Q}}),\nonumber\\
\label{e309}
\end{eqnarray}
where $\mathrm{Arctan}(y/x)$, $x, y\in \mathds{R}$, is defined in Eq.~(\ref{e49}). Further, $\t{G}_{\gtrless}({\bm k};z)$ denote the zero-temperature limits of $\mathscr{G}_{\gtrless}({\bm k};z)$ (cf. Eq.~(\ref{e29})).

To evaluate the contour integral over $\mathfrak{g}^+$ on the RHS of Eq.~(\ref{e307}), it is convenient to conceive of $\mathfrak{g}_1^+$ ($\mathfrak{g}_2^+$) as a closed \emph{circle} of radius $\rho$ centred at $\mu + i\rho$ ($\mu - i\rho$), where
\begin{equation}
\rho = \left\{ \begin{array}{ll} 0,
& \vert\varepsilon_{{\bm k}+{\bm Q}}-\mu\vert > \Omega_0,\\ \\
\displaystyle\frac{1}{2} \big(\Omega_0^2-(\varepsilon_{{\bm k}+{\bm
Q}}-\mu)^2\big)^{1/2}, &\vert\varepsilon_{{\bm k}+{\bm Q}}-\mu\vert
< \Omega_0.
\end{array} \right. \label{e310}
\end{equation}
Evidently, $0\le \rho \le \frac{1}{2}\Omega_0$. Thus,
\begin{equation}
\int_{\mathfrak{g}_{j}^+} \frac{{\rm d}\zeta}{2\pi
i}\; \frac{\mathscr{K}({\bm k};\zeta)}{\mathrm{e}^{\beta
(\zeta-\mu)} +1}=\rho\!\int_0^{2\pi} \frac{{\rm
d}\varphi}{2\pi}\,\left.\frac{\mathrm{e}^{i\varphi}\,
\mathscr{K}({\bm k};\zeta)}{\mathrm{e}^{\beta (\zeta-\mu)}
+1}\right|_{\zeta=\mu+\rho (s i+\,\mathrm{e}^{i\varphi})}\!\!,
\label{e311}
\end{equation}
where
\begin{equation}
s = \left\{ \begin{array}{ll} +1, & j=1, \\ \\
-1, & j=2. \end{array} \right. \label{e312}
\end{equation}
In both cases, corresponding to $j=1,2$, the integration interval $[0,2\pi]$ in Eq.~(\ref{e311}) is reduced to $[\pi/2,3\pi/2]$ as $\beta\to\infty$. On the basis of this and of the observation that $\mathscr{K}({\bm k};\zeta^*) = \mathscr{K}^*({\bm k};\zeta)$ for $\mathrm{Im}(\zeta)\not=0$, one deduces that
\begin{eqnarray}
&&\hspace{-0.0cm}\lim_{\beta\to\infty} \int_{\mathfrak{g}^+} \frac{{\rm d}\zeta}{2\pi i}\; \frac{\mathscr{K}({\bm k};\zeta)}{\mathrm{e}^{\beta (\zeta-\mu)} +1} = -\rho\,
\lim_{\beta\to\infty} \mathrm{Re}\Big[\int_{0}^{\pi/2}\frac{{\rm d}\varphi}{\pi}\; \mathrm{e}^{i\varphi}\, \big\{\mathscr{K}({\bm
k};\mu+i\rho-\rho\,\mathrm{e}^{i\varphi}+i 0^+) \nonumber\\
&&\hspace{8.9cm}
+ \mathscr{K}({\bm k};\mu-i\rho-\rho\,\mathrm{e}^{i\varphi}-i 0^+) \big\}\Big].\nonumber\\ \label{e313}
\end{eqnarray}
Making use of the defining expression in Eq.~(\ref{e308}) and of $\partial/\partial\zeta = (i/\rho)\,\mathrm{e}^{-i\varphi} \partial/\partial\varphi$, the expression in Eq.~(\ref{e313}) can be brought into the following closed form:
\begin{eqnarray}
&&\hspace{-1.1cm}\lim_{\beta\to\infty} \int_{\mathfrak{g}^+} \frac{{\rm
d}\zeta}{2\pi i}\; \frac{\mathscr{K}({\bm k};\zeta)}{\mathrm{e}^{\beta (\zeta-\mu)} +1}\nonumber\\
&&\hspace{-0.6cm} = \frac{1}{\pi} \lim_{\beta\to\infty}\mathrm{Im}\Big[
\big\{\ln\big(\!-\beta\hbar\, \mathscr{G}_{>}^{-1}({\bm k};\mu +i
0^+)\big) - \ln\big(\!-\beta\hbar\, \mathscr{G}_{<}^{-1}({\bm
k};\mu+i 0^+)\big)\big\}\nonumber\\
&&\hspace{1.35cm} + \big\{\ln\big(\!-\beta\hbar\,
\mathscr{G}_{>}^{-1}({\bm k};\mu-2i\rho)\big) -
\ln\big(\!-\beta\hbar\, \mathscr{G}_{<}^{-1}({\bm
k};\mu-2i\rho)\big)\big\}\Big] \nonumber\\
&&\hspace{-0.6cm} = \frac{1}{\pi}
\Big\{\mathrm{Arctan}\Big(\frac{\mathrm{Im}\big[\t{G}_{>}^{-1}({\bm
k};\mu+i 0^+)\big]}{\mathrm{Re}\big[\t{G}_{>}^{-1}({\bm k};\mu+i
0^+)\big]}\Big)
-\mathrm{Arctan}\Big(\frac{\mathrm{Im}\big[\t{G}_{<}^{-1}({\bm
k};\mu+i 0^+)\big]}{\mathrm{Re}\big[\t{G}_{<}^{-1}({\bm k};\mu+i
0^+)\big]}\Big)\nonumber\\
&&\hspace{0.05cm}
+\mathrm{Arctan}\Big(\frac{\mathrm{Im}\big[\t{G}_{>}^{-1}({\bm
k};\mu-2i\rho)\big]}{\mathrm{Re}\big[\t{G}_{>}^{-1}({\bm
k};\mu-2i\rho)\big]}\Big)
-\mathrm{Arctan}\Big(\frac{\mathrm{Im}\big[\t{G}_{<}^{-1}({\bm
k};\mu-2i\rho)\big]}{\mathrm{Re}\big[\t{G}_{<}^{-1}({\bm
k};\mu-2i\rho)\big]}\Big) \Big\}.\label{e314}
\end{eqnarray}
With reference to Eq.~(\ref{e309}), one observes that the contribution of the first integral on the RHS Eq.~(\ref{e307}) is identically canceled by the first of the four terms, presented on the RHS of Eq.~(\ref{e314}), contributing to the second integral on the RHS of Eq.~(\ref{e307}). Consequently one has
\begin{eqnarray}
&&\hspace{0.0cm}\lim_{\beta\to\infty} \b{\nu}^{(1)}({\bm k})
= \frac{-1}{\pi} \Big\{
\mathrm{Arctan}\Big(\frac{\mathrm{Im}\big[\t{G}_{<}^{-1}({\bm
k};\mu+i 0^+)\big]}{\mathrm{Re}\big[\t{G}_{<}^{-1}({\bm k};\mu+i
0^+)\big]}\Big)
-\mathrm{Arctan}\Big(\frac{\mathrm{Im}\big[\t{G}_{>}^{-1}({\bm
k};\mu-2i\rho)\big]}{\mathrm{Re}\big[\t{G}_{>}^{-1}({\bm
k};\mu-2i\rho)\big]}\!\Big) \nonumber\\
&&\hspace{3.05cm}
+\mathrm{Arctan}\Big(\frac{\mathrm{Im}\big[\t{G}_{<}^{-1}({\bm
k};\mu-2i\rho)\big]}{\mathrm{Re}\big[\t{G}_{<}^{-1}({\bm
k};\mu-2i\rho)\big]}\Big) \Big\}.\label{e315}
\end{eqnarray}

Following the same approach as above for obtaining the closed expression for $\lim_{\beta\to\infty}\b{\nu}^{(1)}({\bm k})$, we arrive at
\begin{eqnarray}
&&\hspace{-1.2cm}\lim_{\beta\to\infty}\b{\nu}^{(2)}({\bm
k}) = -\lim_{\beta\to\infty} \b{\nu}^{(1)}({\bm k})\nonumber\\
&&\hspace{-0.5cm} + \frac{1}{E_{\bm k}^+ - E_{\bm k}^-}
\Big\{ (E_{\bm k}^+ - \varepsilon_{{\bm k}+{\bm Q}})\, \Theta(\mu -
E_{\bm k}^+) - (E_{\bm k}^- - \varepsilon_{{\bm k}+{\bm Q}})\,
\Theta(\mu - E_{\bm k}^-)\Big\} \nonumber\\
&&\hspace{-0.5cm} + \frac{\rho}{\hbar}\, \mathrm{Re}\Big[
\int_0^{\pi/2} \frac{{\rm d}\varphi}{\pi}\;
\mathrm{e}^{i\varphi}\,\Big\{ \t{G}_{<}({\bm k};\mu +i\rho -\rho\,
\mathrm{e}^{i\varphi}) - \t{G}_{>}({\bm k};\mu +i\rho
-\rho\, \mathrm{e}^{i\varphi}) \nonumber\\
&&\hspace{3.0cm} + \t{G}_{<}({\bm k};\mu -i\rho -\rho\,
\mathrm{e}^{i\varphi}) - \t{G}_{>}({\bm k};\mu -i\rho -\rho\,
\mathrm{e}^{i\varphi})\Big\}\Big], \label{e316}
\end{eqnarray}
where the $\lim_{\beta\to\infty} \b{\nu}^{(1)}({\bm k})$ on the RHS stands for the expression on the RHS of Eq.~(\ref{e307}) (or, equivalently, Eq.~(\ref{e315})). We have made no attempt to obtain a closed expression for the last integral on the RHS of Eq.~(\ref{e316}) and for the purpose of the following considerations have determined it numerically.

One notes that for (see Eq.~(\ref{e310}))
\begin{equation}
\vert \varepsilon_{{\bm k}+{\bm Q}} -\mu\vert <\Omega_0, \label{e317}
\end{equation}
the last contribution on the RHS of Eq.~(\ref{e316}) is non-vanishing by virtue of the distinction between the two branches of the Green function, $\t{G}_{<}({\bm k};z)$ and $\t{G}_{>}({\bm k};z)$, on the contours
\begin{equation}
z= \mu \pm i\rho -\rho\,\mathrm{e}^{i\varphi},\;\;\; \varphi\in
[0,\frac{\pi}{2}]. \label{e318}
\end{equation}
Consequently, for sufficiently large $\Omega_0$, the last integral on the RHS of Eq.~(\ref{e316}) contributes to $\lim_{\beta\to\infty}\b{\nu}^{(2)}({\bm k})$ for \emph{all} ${\bm k}$.

Finally, from Eq.~(\ref{e46}) it follows that the last two terms on the RHS of Eq.~(\ref{e316}) amount to the GS momentum distribution function ${\sf n}_{\sigma}({\bm k})$.

\subsubsection{Quantitative results}
\label{ss62s3}

\begin{figure}[t!]
\begin{center}
\includegraphics[width=3.2in]{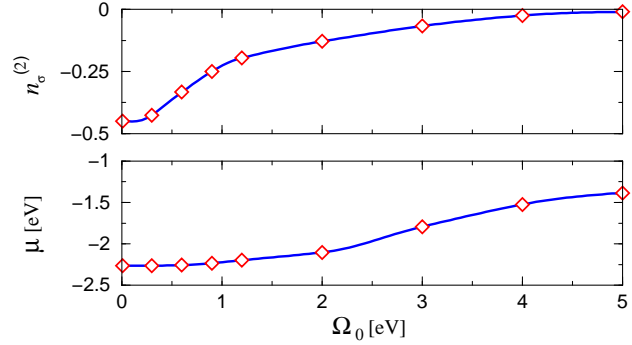}
\caption{\label{f1} The self-consistently calculated values
(\textcolor{red}{${\bm\diamond}$}) for $n_{\sigma}^{(2)}$, Eq.~(\protect\ref{e319}), and the chemical potential $\mu$. The
solid lines connecting the calculated data guide the eyes. The
parameters used in the calculations are: $t=1$, $t'=-0.45$,
$\Delta_0 = 0.1$, $\Delta = 2$ (all energies are in eV) and
$n_{\sigma} = 0.45$, $\forall\sigma$, where $n_{\sigma}$ denotes the
number of spin-$\sigma$ particles per site. The data displayed here
correspond to $\beta=\infty$. Whereas for $\Omega_0 = 0.01$,
$n_{\sigma}^{(2)} = -0.450 \approx -n_{\sigma}$, for $\Omega_0 = 5$,
$n_{\sigma}^{(2)} \approx -0.010$, confirming our statement that for
the model under consideration the Luttinger-Ward identity is
restored as $\Delta/\Omega_0 \to 0$. We note that for $t=1$ and
$t'=-0.45$, $\min_{\bm k} \varepsilon_{\bm k}\equiv -4 (t+t') =
-2.2$ and $\max_{\bm k} \varepsilon_{\bm k}\equiv 4 (t-t') = 5.8$.
Thus, for $\Omega_0 = 0.01, 0.3, 0.6, 0.9$ the corresponding $\mu$
($-2.26, -2.26, -2.26, -2.23$, respectively) lie outside $[\min_{\bm
k}\varepsilon_{\bm k}, \max_{\bm k}\varepsilon_{\bm k}]$. }
\end{center}
\end{figure}

In the upper panel of Fig.~\ref{f1} we present the self-consistently calculated quantity (see Eqs.~(\ref{e34}), (\ref{e44}), (\ref{e315}) and (\ref{e316}))
\begin{equation}
n_{\sigma}^{(j)} {:=} \lim_{\beta\to\infty} \frac{\b{N}_{\sigma}^{(j)}}{\mathcal{N}_{\Sc l}} \equiv \frac{1}{\mathcal{N}_{\Sc l}} \sum_{{\bm k}\in \mathrm{1BZ}} \lim_{\beta\to\infty}\b{\nu}_{\sigma}^{(j)}({\bm k}) \label{e319}
\end{equation}
for $j=2$, where $\mathcal{N}_{\Sc l}$ is the number of lattice points.\footnote{Not to be confused with $N_{\Sc l} \equiv \sum_{\sigma} N_{{\Sc l};\sigma}$, the total Luttinger number.} The $\mathrm{1BZ}$ corresponds to a two-dimensional square lattice of lattice constant $a$, which we consider as the unit of length, so that $\mathrm{1BZ}
= [-\pi,\pi) \times [-\pi,\pi)$. On transforming $\sum_{{\bm k}\in \mathrm{1BZ}}$ into an integral over the $\mathrm{1BZ}$, we have determined this integral by employing an adaptive Monte-Carlo method \protect\cite{Note11}, using maximally $N_{\Sc m\Sc c} = 60,000$, $100,000$ and $120,000$ Monte-Carlo sampling points for respectively small, intermediate and large values of $\Omega_0$, ensuring relative accuracies of better than $1$ part in $10^3$ in $n_{\sigma}^{(j)}$, $j=1,2$.

\begin{figure}[t!]
\begin{center}
\includegraphics[width=3.2in]{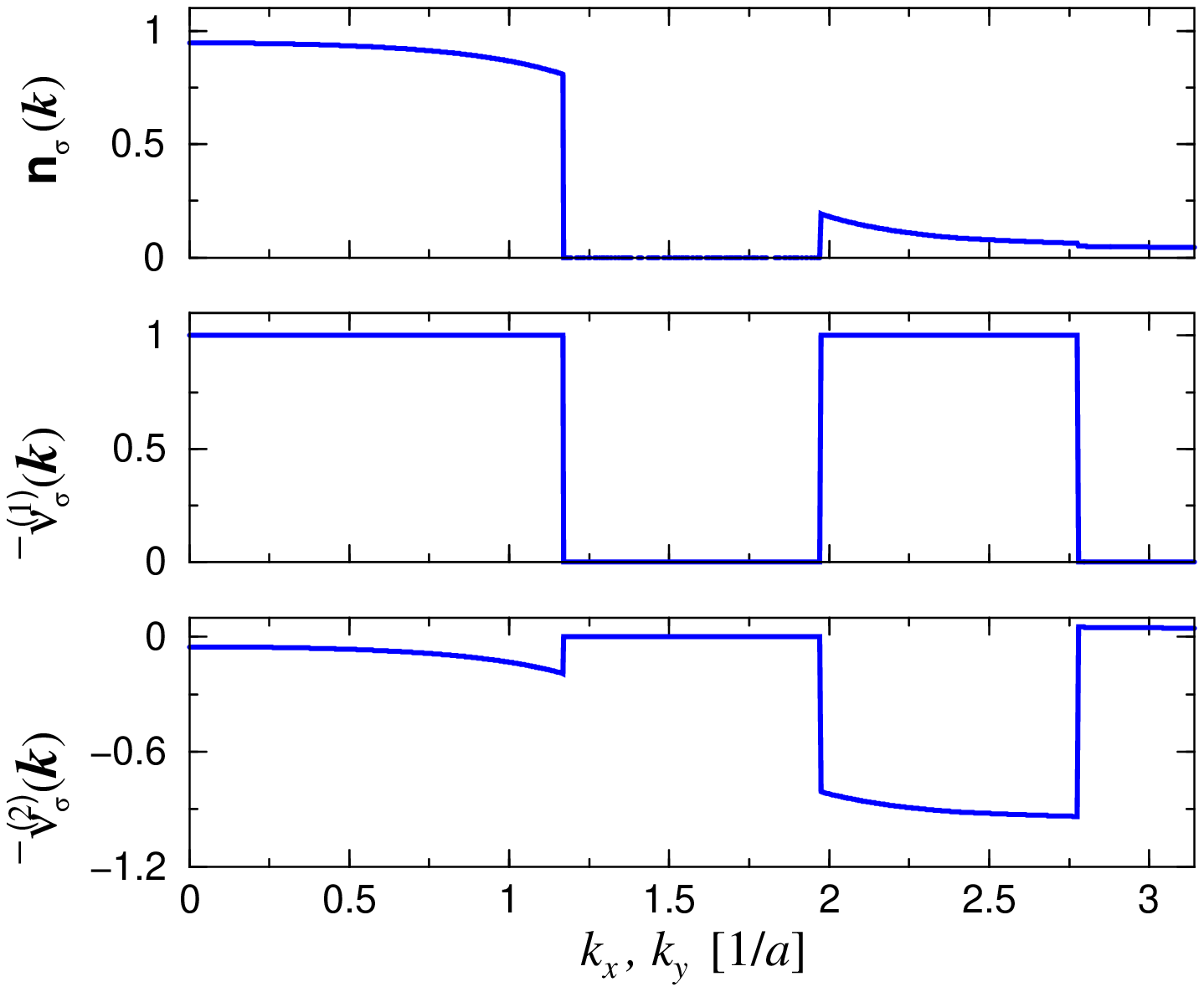}
\caption{\label{f2} The functions ${\sf n}_{\sigma}({\bm k})$,
$\b{\nu}_{\sigma}^{(1)}({\bm k})$ and $\b{\nu}_{\sigma}^{(2)}({\bm
k})$ for ${\bm k}$ along the diagonal of the square $\mathrm{1BZ}$,
$k_x = k_y \in [0,\pi)$. The displayed data correspond to $t=1$,
$t'=-0.45$, $\Delta_0=0.1$, $\Delta=2$, $\Omega_0=0.3$, $n_{\sigma}
= 0.45$, $\beta=\infty$, and have been obtained from the expressions
in Eqs.~(\protect\ref{e315}), (\protect\ref{e316}) and
(\protect\ref{e46}). Maximally $N_{\Sc m\Sc c} = 100,000$
Monte-Carlo samplings of the $\mathrm{1BZ}$ have been used
\protect\cite{Note11}. The following results correspond to the data
displayed here: $\mu=-2.2641$, $n_{\sigma}^{(1)}=0.8764$,
$n_{\sigma}^{(2)}=-0.4265$, $n_{\sigma}^{(1)} + n_{\sigma}^{(2)}=
0.4499$. The deviation of the latter number from $n_{\sigma} = 0.45$
is due to the finite precision with which we have calculated $\mu$;
with $n_{\sigma}^{(1)}$ and $n_{\sigma}^{(2)}$ containing stochastic
noise, arising from a finite $N_{\Sc m\Sc c}$, the accuracy with
which $\mu$ can be calculated is limited by the magnitude of noise
superimposed on $n_{\sigma}^{(1)} + n_{\sigma}^{(2)}$.}
\end{center}
\end{figure}

\begin{figure}[t!]
\begin{center}
\includegraphics[width=3.2in]{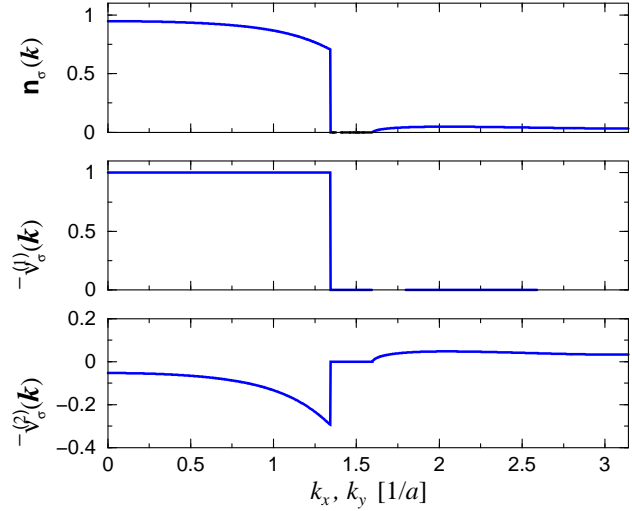}
\caption{\label{f3} Similar to Fig.~\protect\ref{f2} except that
$\Omega_0=2$~eV. The following results correspond to the data
displayed here: $\mu=-2.1024$, $n_{\sigma}^{(1)}=0.5784$,
$n_{\sigma}^{(2)}= -0.1286$, $n_{\sigma}^{(1)} + n_{\sigma}^{(2)}=
0.4498$. }
\end{center}
\end{figure}

All the numerical results presented in this section correspond to $\beta=\infty$. The parameters $t$, $t'$, $\Delta_0$ and $n_{\sigma}$ to which the data in Fig.~\ref{f1} (as well as those in Figs.~\ref{f2} and \ref{f3}) correspond, coincide with those adopted in Refs.~\citen{CMS96,CM97}. We have chosen the coupling constant of interaction $\Delta$ relatively large, equal to $2$~eV, in order to achieve pronounced effects in the behaviours of $\b{\nu}_{\sigma}^{(1)}({\bm k})$, $\b{\nu}_{\sigma}^{(2)}({\bm k})$ and ${\sf n}_{\sigma}({\bm k})$, presented in Figs.~\ref{f2} and \ref{f3}, which for relatively small values of $\Delta$ (say, for $\Delta=0.3$ or smaller) would prove difficult to reproduce in print. Extensive experiments have shown that our numerical results are in full accord with those presented in Refs.~\citen{CMS96,CM97}; more explicitly, we reproduce the data in Figs.~1 and 2 in Ref.~\citen{CMS96} as well as those in Figs.~14, 15 and 16 in Ref.~\citen{CM97}.

The data in Fig.~\ref{f1} concerning $n_{\sigma}^{(2)}$ clearly demonstrate violation of the Luttinger-Ward identity over the range of $\Omega_0$ shown. One observes however that for decreasing values of $\Delta/\Omega_0$, the quantity $\vert n_{\sigma}^{(2)}\vert$ monotonically decreases; already for $\Delta/\Omega_0 = 0.2$ one has $\vert n_{\sigma}^{(2)}\vert \approx 0.010$, a forty-five (forty-three) fold suppression with respect to the value corresponding to $\Omega_0 =0.01$ ($\Omega_0=0.3$). We conclude that for the model under consideration the violation of the Luttinger theorem, brought about by the breakdown of the Luttinger-Ward identity, is indeed a consequence of the cut-off energy $\Omega_0$ being small with respect to the coupling-constant of interaction. In this connection, we should emphasise that this observation is not specific to $\Delta=2$, but applies to all values of $\Delta$ that we have examined numerically.

The data presented in Figs.~\ref{f2} and \ref{f3} illustrate the behaviours of ${\sf n}_{\sigma}({\bm k})$, $\b{\nu}_{\sigma}^{(1)}({\bm k})$ and $\b{\nu}_{\sigma}^{(2)}({\bm k})$ as functions of ${\bm k}$. By comparing the data in Figs.~\ref{f2} and \ref{f3} one gains insight into the way in which these functions, corresponding to the same value of $\Delta$, are affected by an increase in the value of the cut-off energy $\Omega_0$. We should emphasize that in order for the Luttinger-Ward identity to be satisfied, it is \emph{not} necessary that $\b{\nu}_{\sigma}^{(2)}({\bm k}) \equiv 0$, $\forall {\bm k}$; it is only necessary that $n_{\sigma}^{(2)} = 0$ (see Eq.~(\ref{e19})). Nonetheless, it appears that, for the model under consideration, decrease in $\vert n_{\sigma}^{(2)}\vert$ is accompanied by an overall suppression of $\b{\nu}_{\sigma}^{(2)}({\bm k})$.

The role of a finite cut-off energy $\Omega_0$ in the violation of the Luttinger-Ward identity can be understood by concentrating on one essential element in the proof of the Luttinger-Ward identity (see Sec.~\ref{ss52s2}), namely that the validity of the Luttinger-Ward identity crucially depends on the possibility to apply integration by parts to the contour integral over $\mathscr{C}(\mu)$ as encountered in the expression on the LHS of Eq.~(\ref{e88}). As we have seen, by applying integration by parts various contributions corresponding to $\lim_{\beta\to\infty}\partial Y_{\sigma}^{\prime (\nu)}/\partial\mu$, $\nu=2,3,\dots$, pairwise cancel, leading to the result presented in Eq.~(\ref{e87}). If, however, the contributions to $\t{\Sigma}_{\sigma}^{(\nu)}({\bm k};z)$ turn out not to be continuously differentiable (\S~351 in Ref.~\citen{EWH27}; see also Ref.~\citen{Note4}) for all $z\in \mathscr{C}(\mu)\backslash\mu$, it becomes necessary that application of integration by parts be preceded by a subdivision of $\mathscr{C}(\mu)\backslash\mu$ into sub-contours over each of which $\t{\Sigma}_{\sigma}^{(\nu)}({\bm k};z)$ is continuously differentiable. Application of integration by parts over each finite sub-contour gives rise to non-vanishing boundary contributions; evidently, $\mathscr{C}(\mu)$ being closed at the point of infinity, no such boundary contributions arise when $\t{\Sigma}_{\sigma}^{(\nu)}({\bm k};z)$ is continuously differentiable for all $z\in \mathscr{C}(\mu)\backslash\mu$ (see Sec.~\ref{ss52s2}).

The breakdown of the Luttinger-Ward identity as we have encountered in dealing with the model by Chubukov and collaborators \cite{CMS96,CM97}, signifies that for this model the above-mentioned non-vanishing boundary contributions do not add up to zero. The fact that for $\Delta/\Omega_0 \to 0$ the Luttinger-Ward identity is recovered (see Fig.~\ref{f1}), is in conformity with the fact that for sufficiently large $\vert z\vert$, both $\t{G}_{\sigma}({\bm k};z)$ and $\partial\t{\Sigma}_{\sigma}({\bm k};z)/\partial z$ are decreasing functions of $z$ for increasing $\vert z\vert$, leading to a steady marginalization of the above-mentioned boundary contributions for increasing values of $\Omega_0$. In this connection, we remark that, according to our numerical results, $n_{\sigma}^{(2)}$ approaches zero \emph{continuously} for $\Delta/\Omega_0 \to 0$, supporting the viewpoint that indeed the boundary contributions responsible for the violation of the Luttinger-Ward identity are steadily marginalized for $\Delta/\Omega_0 \to 0$.

\subsubsection{Summary}
\label{ss62s4}

We have explicitly demonstrated that the breakdown of the Luttinger theorem as observed for the metallic GSs of the model proposed by Chubukov and collaborators \cite{CMS96,CM97}, reflects an artificial aspect of this model, for which the underlying $\t{G}({\bm k};z)$ and $\t{\Sigma}({\bm k};z)$ are non-analytic in the region $\mathrm{Im}(z)\not=0$, rather than a failure of the Luttinger theorem; since the exact $\t{G}({\bm k};z)$ and $\t{\Sigma}({\bm k};z)$ are fully analytic in the region $\mathrm{Im}(z)\not=0$ (appendix \ref{sc}), it is evident that the above-mentioned non-analyticity cannot have been anticipated and thus accounted for in the proof of the Luttinger theorem. We have further established the mechanism by which this non-analyticity undermines one of the fundamental conditions under which the Luttinger-Ward identity has been deduced, the condition being that the integrand of the integral on the RHS of Eq.~(\ref{e43}) be continuously differentiable for \emph{all} $z\in \mathscr{C}(\mu)\backslash\mu$ (see Sec.~\ref{ss52s2}).

\subsection{Case III}
\label{ss63}

Schmalian \emph{et al.} \cite{SLGB96a} reported breakdown of the Luttinger theorem for the uniform metallic GSs of the single-band Hubbard Hamiltonian, on a two-dimensional square lattice, within the framework of the self-consistent fluctuation-exchange (FLEX) approximation \cite{BS89}. More explicitly, Schmalian \emph{et al.} \cite{SLGB96a} found that $n - n_{\Sc l}$ deviates from zero in the hole-doped region, away from half-filling and for doping concentrations $x \equiv 1 - n$ less than $0.2$, with $n - n_{\Sc l}>0$ decreasing towards zero for $x$ increasing towards $0.2$. Here $n$ denotes the total site density, with $n=1$ the site density at half-filling, and $n_{\Sc l}$ the total Luttinger site density.

Although the calculations by Schmalian \emph{et al.} \cite{SLGB96a} correspond to finite temperatures, variation of $n - n_{\Sc l}$ as function of temperature suggests that the observed value for $n - n_{\Sc l}$ at the lowest temperature considered (that is $T=63$~K) is not attributable to $T$ being non-vanishing (see Sec.~\ref{ss52s1} and appendix \ref{sf}). For clarity, $T = 63$~K is to be compared with $4t_0/k_{\Sc b}$, the total width, expressed in temperature, of the non-interacting energy dispersion $\varepsilon_{\bm k}$ underlying the calculations by Schmalian \emph{et al.} \cite{SLGB96a}, where $t_0$ is the nearest-neighbour hopping parameter which in Ref.~\citen{SLGB96a} is set equal to $0.25$~eV, for which $4t_0/k_{\Sc b} \approx 1.2 \times 10^4$~K.

Schmalian \emph{et al.} \cite{SLGB96a} traced back the result $n \not= n_{\Sc l}$ to $\mathrm{Im}[\Sigma_{\sigma}({\bm k};\mu)] \not\equiv 0$. In other words, the breakdown of the Luttinger theorem as observed by Schmalian \emph{et al.} \cite{SLGB96a} largely originates from the second term on the RHS of Eq.~(\ref{e57}) and \emph{not} from the breakdown of the Luttinger-Ward identity in Eq.~(\ref{e44});\cite{Note10} the deviation from zero of $\sum_{\sigma}\b{N}_{\sigma}^{(2)}$ per lattice site for the temperatures considered by the authors is by one order of magnitude smaller than $n-n_{\Sc l}$ \cite{SLGB96a}: the former quantity, denoted by $I(T)$ in Ref.~\citen{SLGB96a}, is approximately equal to $10^{-3}$, to be compared with $n-n_{\Sc l} \approx 10^{-2}$ at low doping concentrations \cite{SLGB96a}. It is noteworthy that the breakdown of the Luttinger theorem as reported by Chubukov \emph{et al.} \cite{CMS96,CM97} (see Sec.~\ref{ss62}) has been traced back \cite{ACDFM98} to the breakdown of the Luttinger-Ward identity.

The result $\mathrm{Im}[\Sigma_{\sigma}({\bm k};\mu)] \not\equiv 0$ is in violation of Eq.~(\ref{e8}) whose exactness we rigorously demonstrate in appendix \ref{sc} (see also Sec.~\ref{ss21s2}); it is worth mentioning that, insofar as metallic GSs are concerned, this demonstration does not rely on any Fermi-liquid assumption and is applicable to both isotropic and anisotropic GSs. In fact, the numerical data concerning $\mathrm{Im}[\t{\Sigma}_{\sigma}({\bm k};\varepsilon+ i 0^+)]$ presented in the papers by Schmalian \emph{et al.} \cite{SLGB96a,SLGB96b} and Langer \emph{et al.} \cite{LSGB95,LSGB96}, show, without exception, that this function violates Eq.~(\ref{e17}), which in Sec.~\ref{ss43} we have explicitly shown to be valid for a widely distinct variety of metallic states. The reader may consult, for instance, Fig.~1 in Ref.~\citen{SLGB96a} from which one observes that for $x=0.16, 0.12$ and $T=63$~K, $\mathrm{Im}[\Sigma_{\sigma}({\bm k};\mu)]$ is considerable for ${\bm k}$ in the vicinity of the region in the $\mathrm{1BZ}$ which supposedly coincides with the underlying Fermi surface $\mathcal{S}_{\Sc f;\sigma}$, this on account of the inset of Fig.~2 in Ref.~\citen{SLGB96a}.

In the light of the above observations we conclude that the GSs which Schmalian \emph{et al.} \cite{SLGB96a,SLGB96b} and Langer \emph{et al.} \cite{LSGB95,LSGB96} qualified as `non-Fermi-liquid' metallic states are in fact \emph{pathological} rather than metallic states of any type (see Sec.~\ref{ss21s2}).

In this section we establish that\footnote{Since the lowest $T$ in the calculations under consideration is non-vanishing, it would be more accurate to refer to the relatively \emph{large} values of $\mathrm{Im}[\mathscr{S}_{\!\sigma}({\bm k};\mu+i 0^+)]$ than to $\mathrm{Im}[\Sigma_{\sigma}({\bm k};\mu)] \not\equiv 0$.}  $\mathrm{Im}[\Sigma_{\sigma}({\bm k};\mu)] \not\equiv 0$ as observed by Schmalian \emph{et al.} \cite{SLGB96a,SLGB96b}, and Langer \emph{et al.} \cite{LSGB95,LSGB96}, is almost entirely, if not entirely, attributable to an essential contribution to $\t{\Sigma}_{\sigma}({\bm k};z)$ that the authors have inadvertently and incorrectly not taken into account. This contribution which within the exact framework is vanishing, is non-vanishing in practical calculations where integrals with respect to $\varepsilon$ are carried out over a finite interval, say $[-E,E]$, instead of $[-\infty,\infty]$, leading to violation of causality. We demonstrate that failure to take appropriate account of the non-causal contribution to the `retarded' self-energy $\t{\Sigma}_{\sigma}({\bm k};\varepsilon+i 0^+)$ results in a significant contribution to $\mathrm{Im}[\t{\Sigma}_{\sigma}({\bm k};\varepsilon+i 0^+)]$ which is operative over the entire range of $\varepsilon$. We explicitly calculate the asymptotic series expansion of this contribution for $E\to\infty$ in terms of the coefficients of the asymptotic series expansion of $\t{\Sigma}_{\sigma}({\bm k};z)$ corresponding to large values of $\vert z\vert$ (appendix \ref{sc}). Our analysis reveals that the leading-order contribution to the `missing' part of the self-energy under consideration, which is correct to order $1/E$, nearly perfectly accounts for the deviation from zero of the $\mathrm{Im}[\Sigma_{\sigma}({\bm k};\mu)]$ as calculated by Schmalian \emph{et al.} \cite{SLGB96a,SLGB96b}, and Langer \emph{et al.} \cite{LSGB95,LSGB96}. We thus conclude that \emph{the breakdown of the Luttinger theorem as reported by Schmalian \emph{et al.} \cite{SLGB96a} amounts to a computational artifact.}

\subsubsection{Preliminaries}
\label{ss63s1}

The numerical method employed by Schmalian \emph{et al.} \cite{SLGB96a} consists of performing the self-consistent calculations at the complex energy $z=\varepsilon + i\gamma$, where $\gamma$ is a finite positive constant, equated in the actual calculations \cite{SLGB96b} with $\pi k_{\Sc b} T/2$ (that is one half of the Matsubara energy $\hbar\omega_m$ at $m=0$). On completing the self-consistent calculation of $\mathscr{S}_{\!\sigma}({\bm k};\varepsilon+i \gamma)$, the authors determined the self-consistent retarded self-energy $\mathscr{S}_{\!\sigma}({\bm k};\varepsilon +i 0^+)$ by employing an extrapolation scheme, effecting $\gamma \to 0^+$, based on the Pad\'e approximation of the self-consistent $\mathscr{S}_{\!\sigma}({\bm k};\varepsilon+i\gamma)$ \cite{SLGB96b}. The scheme developed and employed by the authors of Refs.~\citen{SLGB96b,SLGB96a} thus bypasses the conventional approach \cite{VS77} of calculating $\mathscr{S}_{\!\sigma}({\bm k};\varepsilon+i 0^+)$ through performing, numerically, analytic continuation towards the real energy axis of $\{\mathscr{S}_{\!\sigma}({\bm k};\zeta_m)\,\|\, m=0, \pm 1,\dots\}$ (see Eq.~(\ref{e27})). In this way, the authors achieved a higher resolution concerning the behaviour of $\mathscr{S}_{\!\sigma}({\bm k};\varepsilon +i 0^+)$ as function of $\varepsilon$ than is achievable by means of the conventional analytic-continuation approach.

For the self-consistent calculation of $\mathscr{S}_{\!\sigma}({\bm k};\varepsilon+i \gamma)$, Schmalian \emph{et al.} \cite{SLGB96a} relied on the use of a combination of Fourier and one-sided Laplace transformations, regarding the time and energy dependence of various functions, which are carried out by approximating continuous integrations over unbounded intervals by means of finite discrete sums which are efficiently evaluated with the aid of the fast-Fourier-transformation (FFT) technique \cite{SLGB96b} (Ch.~12 in Ref.~\citen{PTVF}). In fact, by employing a $(2+1)$-dimensional FFT technique, the authors performed the underlying self-consistent calculations partly on the real-space lattice and the time domain.

For our analysis it suffices to consider $\t{\Sigma}_{\sigma}({\bm k};\varepsilon + i \gamma)$, the zero-temperature limit of $\mathscr{S}_{\!\sigma}({\bm k};\varepsilon+ i\gamma)$, Eq.~(\ref{e30}). Since our main interest concerns the behaviour of $\t{\Sigma}_{\sigma}({\bm k};\varepsilon + i \gamma)$ as function of $\varepsilon$, we explicitly deal with (see Eqs.~(\ref{ec70}) and (\ref{ec71}))
\begin{equation}
\t{\Sigma}_{\sigma}^{\times}({\bm k};z) {:=}
\t{\Sigma}_{\sigma}({\bm k};z) -\Sigma_{\sigma}^{\Sc h\Sc f}({\bm
k}). \label{e320}
\end{equation}
We thus define
\begin{equation}
\mathfrak{S}_{\sigma}^{(\gamma)}({\bm k}; t) {:=} \mathrm{e}^{\gamma
t/\hbar} \int_{-\infty}^{\infty} \frac{{\rm
d}\varepsilon}{2\pi\hbar}\; \mathrm{e}^{-i\varepsilon t/\hbar}\,
\t{\Sigma}_{\sigma}^{\times}({\bm k};\varepsilon + i \gamma),
\label{e321}
\end{equation}
which is the Fourier transform of $\t{\Sigma}_{\sigma}^{\times}({\bm k};\varepsilon + i \gamma)$ into the time domain.

For clarity, since $\Sigma_{\sigma}^{\Sc h\Sc f}({\bm k})$ is independent of ${\bm k}$ and $\sigma$ in the uniform and non-magnetic GSs of Hamiltonians in which the two-body interaction potential is contact-type (such as is the case for the conventional Hubbard Hamiltonian), in dealing with such GSs, it is common practice to absorb the underlying $\Sigma_{\sigma}^{\Sc h\Sc f}({\bm k})$ in the chemical potential; this is the practice adopted in Refs.~\citen{SLGB96a,SLGB96b,LSGB95,LSGB96}. Consequently, the function $\t{\Sigma}_{\sigma}^{\times}({\bm k};\varepsilon + i \gamma)$ dealt with here is the counterpart of the function $\t{\Sigma}_{\sigma}({\bm k};\varepsilon + i \gamma)$ (more precisely $\Sigma_{\bm k}(\omega+i\gamma)$) encountered in Refs.~\citen{SLGB96a,SLGB96b,LSGB95,LSGB96}.

Since $\t{\Sigma}_{\sigma}^{\times}({\bm k};z + i \gamma)$ is analytic in the region $\mathrm{Im}(z) > -\gamma$ of the $z$ plane (appendix \ref{sc}), it follows that
\begin{equation}
\mathfrak{S}_{\sigma}^{(\gamma)}({\bm k};t) \equiv 0,\;\;\; \forall
t< 0, \label{e322}
\end{equation}
as befits a retarded function. The general expression
\begin{equation}
\t{\Sigma}_{\sigma}^{\times}({\bm k};\varepsilon + i \gamma) =
\int_{-\infty}^{\infty} {\rm d}t\; \mathrm{e}^{i (\varepsilon +
i\gamma) t/\hbar}\, \mathfrak{S}_{\sigma}^{(\gamma)}({\bm k};t),
\label{e323}
\end{equation}
which is the inverse of the transformation in Eq.~(\ref{e321}), thus reduces to
\begin{equation}
\t{\Sigma}_{\sigma}^{\times}({\bm k};\varepsilon + i \gamma) =
\int_0^{\infty} {\rm d}t\; \mathrm{e}^{i (\varepsilon + i\gamma)
t/\hbar}\, \mathfrak{S}_{\sigma}^{(\gamma)}({\bm k};t), \label{e324}
\end{equation}
according to which $\t{\Sigma}_{\sigma}^{\times}({\bm k};\varepsilon + i \gamma)$ is the one-sided Laplace transform of $\mathfrak{S}_{\sigma}^{(\gamma)}({\bm k};t)$, with $p = (\gamma -i\varepsilon)/\hbar$ the pertinent complex variable in the Laplace domain (Ch. 29 in Ref.~\citen{AS72}).

The technical aspects concerning the way in which Schmalian \emph{et al.} \cite{SLGB96a} calculated $\t{\Sigma}_{\sigma}^{\times}({\bm k};\varepsilon + i \gamma)$ can be found in Ref.~\citen{SLGB96b}. From the perspective of our present considerations, it is only relevant to note that, in the actual calculations the expression in Eq.~(\ref{e324}), and \emph{not} that in Eq.~(\ref{e323}), was used as the link between $\t{\Sigma}_{\sigma}^{\times}({\bm k};\varepsilon + i \gamma)$ and $\mathfrak{S}_{\sigma}^{(\gamma)}({\bm k};t)$. This aspect is significant in that in the calculations where integrations with respect to $\varepsilon$ over $[-\infty,\infty]$ are carried out over $[-E,E]$, with $E<\infty$, Eq.~(\ref{e322}) is \emph{not} satisfied, thus rendering the representation in Eq.~(\ref{e324}) \emph{incomplete}. We shall explicitly demonstrate this statement in Sec.~\ref{ss63s2}. Although not as relevant, we also point out that in the calculations under consideration \cite{SLGB96a,SLGB96b} the integral on the RHS of Eq.~(\ref{e324}) was reduced to one over $[0,\tau_0]$ \cite{Note12}; in Sec.~\ref{ss63s4} we shall indicate how a finite $\tau_0$ affects the results and how this can be easily corrected for in the cases where $\tau_0$ is sufficiently large.

Below we explicitly demonstrate that failure of the Luttinger theorem as reported by Schmalian \emph{et al.} \cite{SLGB96a} is a direct consequence of the use of a finite value for $E$ (better, from the practical perspective, a relatively small value for $E$) on the one hand, and employing the representation in Eq.~(\ref{e324}), instead of that in Eq.~(\ref{e323}), on the other. Anticipating future self-consistent calculations of self-energy on the basis of a modified version of the computational method put forward by Schmalian \emph{et al.} \cite{SLGB96b}, in the following we present more details than are strictly necessary for the purpose of our present analysis.

\subsubsection{Violation of causality}
\label{ss63s2}

We demonstrate that approximating the integral on the RHS of Eq.~(\ref{e321}) by one over the finite interval $[-E,E]$, $E<\infty$, gives rise to violation of Eq.~(\ref{e322}). To this end, and with reference to Eq.~(\ref{e321}), we introduce the function
\begin{equation}
\mathfrak{S}_{\sigma;E}^{(\gamma)}({\bm k}; t) {:=}
\mathrm{e}^{\gamma t/\hbar} \int_{-E}^{E} \frac{{\rm
d}\varepsilon}{2\pi\hbar}\; \mathrm{e}^{-i\varepsilon t/\hbar}\,
\t{\Sigma}_{\sigma}^{\times}({\bm k};\varepsilon + i \gamma),
\label{e325}
\end{equation}
for which one has
\begin{equation}
\mathfrak{S}_{\sigma}^{(\gamma)}({\bm k}; t) =
\left.\mathfrak{S}_{\sigma;E}^{(\gamma)}({\bm k};
t)\right|_{E=\infty}. \label{e326}
\end{equation}
For the reasons that will become apparent below, here we strictly distinguish between $E=\infty$ and $E\to\infty$. Briefly, the limits $t\to 0$ and $E\to\infty$ do not commute.

Let now $\Gamma_{1;E}^+$ be the counter-clockwise oriented semicircle of radius $E$ in the upper half of the $z$ plane (Sec.~\ref{ss41s1}), centred at the origin. Owing to the analyticity of $\t{\Sigma}_{\sigma}^{\times}({\bm k};z)$ in the upper-half of the $z$ plane (appendix \ref{sc}), by the Cauchy theorem one can express $\mathfrak{S}_{\sigma;E}^{(\gamma)}({\bm k}; t)$ as
\begin{equation}
\mathfrak{S}_{\sigma;E}^{(\gamma)}({\bm k}; t) = -\mathrm{e}^{\gamma
t/\hbar} \int_{\Gamma_{1;E}^+} \frac{{\rm d}z}{2\pi\hbar}\;
\mathrm{e}^{-i z t/\hbar}\, \t{\Sigma}_{\sigma}^{\times}({\bm
k};z+i\gamma). \label{e327}
\end{equation}
For $t\le 0$, from this expression one can obtain the asymptotic series for $\mathfrak{S}_{\sigma;E}^{(\gamma)}({\bm k}; t)$ corresponding to $E\to\infty$ by employing the following asymptotic series (appendix \ref{sc}):
\begin{equation}
\t{\Sigma}_{\sigma}^{\times}({\bm k};z+i\gamma)
\sim \frac{\Sigma_{\sigma;\infty_1}({\bm k})}{z} +
\frac{\Sigma_{\sigma;\infty_2}({\bm k})
-i\gamma\,\Sigma_{\sigma;\infty_1}({\bm k})}{z^2}+\dots\;\;\mbox{\rm for}\;\;\; \vert z\vert\to\infty. \label{e328}
\end{equation}
On the basis of this expression, from Eq.~(\ref{e327}) for $t\le 0$ and $E\to\infty$ one obtains that
\begin{equation}
\mathfrak{S}_{\sigma;E}^{(\gamma)}({\bm k}; t) \sim
\frac{1}{2\pi i\hbar}\, \mathrm{e}^{\gamma t/\hbar} \Big\{
\Sigma_{\sigma;\infty_1}({\bm k})\, \mathcal{I}_{0}\big(\frac{E t}{\hbar}\big) + \frac{\Sigma_{\sigma;\infty_2}({\bm k})
-i\gamma\,\Sigma_{\sigma;\infty_1}({\bm k})}{E}\,
\mathcal{I}_{1}\big(\frac{E t}{\hbar}\big) + \dots
\Big\},\label{e329}
\end{equation}
where
\begin{equation}
\mathcal{I}_{j}(x) {:=} \int_0^{\pi} {\rm d}\varphi\;\mathrm{e}^{-i
x (\cos\varphi + i\sin\varphi)}\, \mathrm{e}^{-i j\varphi},\;\;
j \in \mathds{Z}^*. \label{e330}
\end{equation}
It can be readily verified that
\begin{equation}
\mathcal{I}_{0}(x) = 2\, \mathrm{Si}(x) + \pi,\;\;\;\forall x,
\label{e331}
\end{equation}
\begin{equation}
\mathcal{I}_{1}(x) = -i \big(2 x\,\mathrm{Si}(x) + 2 \cos(x) +\pi x
\big),\;\;\;\forall x, \label{e332}
\end{equation}
where $\mathrm{Si}(x)$ is the sine-integral function (item 5.2.1 in Ref.~\citen{AS72}). One observes that $\mathcal{I}_0(x)$ is real and $\mathcal{I}_1(x)$ is purely imaginary for $x\in \mathds{R}$. More generally, for $x\in \mathds{R}$, $\mathcal{I}_j(x)$ can be shown to be real for even values of $j$ and purely imaginary for odd values of $j$. This even-odd effect will prove to be of considerable physical consequence.

Using the asymptotic series for $\mathrm{Si}(x)$ (items 5.2.8, 5.2.34 and 5.2.35 in Ref.~\citen{AS72}) one obtains the following leading-order expressions:
\begin{equation}
\mathcal{I}_0\big(\frac{E t}{\hbar}\big) \sim \left\{
\begin{array}{ll} \displaystyle\frac{-2 \cos(E t/\hbar)}{E t/\hbar},
& E t/\hbar \to-\infty,\\ \\
\pi, & t=0, \end{array} \right.
\label{e333}
\end{equation}
\begin{equation}
\frac{1}{E}\, \mathcal{I}_1\big(\frac{E t}{\hbar}\big) \sim \left\{
\begin{array}{ll} \displaystyle\frac{2i}{E}\, \frac{\sin(E t/\hbar)}{E t/\hbar},
& E t/\hbar \to-\infty,\\ \\
\displaystyle -\frac{2 i}{E}, & t=0. \end{array} \right.
\label{e334}
\end{equation}
From the expression in Eq.~(\ref{e329}) and the above asymptotic results, one infers the way in which the exact result in Eq.~(\ref{e322}) is obtained for $E=\infty$; it is evident that for any $E<\infty$, $\mathfrak{S}_{\sigma;E}^{(\gamma)}({\bm k};t)\not\equiv 0$ for $t<0$, contradicting the exact result in Eq.~(\ref{e322}). Note in passing the evident fact that the exact results in Eqs.~(\ref{e333}) and (\ref{e334}) corresponding to $t=0$ \emph{cannot} be recovered from the leading-order asymptotic results corresponding to $t<0$ and $E\to\infty$, that is $E t/\hbar\to -\infty$. This phenomenon is principally the same as we encountered in Sec.~\ref{ss61} in dealing with $\mu\to \mu_{\infty}$ and $\beta\to\infty$; in both cases one observes a manifestation the same mathematical principle, that different \emph{repeated limits} of multi-variable function are in general inequivalent (\S\S~302-306 in Ref.~\citen{EWH27}).

On the basis of the results in Eqs.~(\ref{e329}), (\ref{e333}) and (\ref{e334}) one further deduces that
\begin{equation}
\mathfrak{S}_{\sigma;E}^{(\gamma)}({\bm k};0) \sim
\frac{\Sigma_{\sigma;\infty_1}({\bm k})}{2 i\hbar}
-\frac{\Sigma_{\sigma;\infty_2}({\bm k})
-i\gamma\,\Sigma_{\sigma;\infty_1}({\bm k})}{\pi\hbar\, E} +
\dots\;\;\mbox{\rm for}\;\;\; E\to\infty, \label{e335}
\end{equation}
exposing the way in which $\mathfrak{S}_{\sigma;E}^{(\gamma)}({\bm k};0)$ converges towards the exact $\mathfrak{S}_{\sigma}^{(\gamma)}({\bm k};0)$ for increasing values of $E$. In appendix \ref{sc} we demonstrate that for the interacting GSs of the conventional Hubbard Hamiltonian (amongst others), $\Sigma_{\sigma;\infty_1}({\bm k}) >0$, $\forall {\bm k}$ (Eqs.~(\ref{ec88}) and (\ref{ec90})). Consequently, in view of Eq.~(\ref{e322}), $\mathfrak{S}_{\sigma}^{(\gamma)}({\bm k};t)$ is \emph{strictly discontinuous} at $t=0$ for these GSs. In contrast, $\mathfrak{S}_{\sigma;E}^{(\gamma)}({\bm k};t)$ is continuous at $t=0$ for any $E<\infty$.

The discontinuity of $\mathfrak{S}_{\sigma}^{(\gamma)}({\bm k};t)$ at $t=0$ implies that $\lim_{t\to 0}$ and the integration with respect to $\varepsilon$ on the RHS of Eq.~(\ref{e321}) do not commute: on effecting
\begin{equation}
\lim_{t\to 0} \int_{-\infty}^{\infty} {\rm d}\varepsilon\; (\dots)
\rightharpoonup \int_{-\infty}^{\infty} {\rm d}\varepsilon\;
\lim_{t\to 0}\, (\dots),\label{e336}
\end{equation}
one obtains $\mathfrak{S}_{\sigma}^{(\gamma)}({\bm k};0)$, irrespective of whether $t\uparrow 0$ or $t\downarrow 0$ \cite{Note13}. This aspect has its root partly, but very essentially, in the underlying interval of integration being unbounded. For other part, since $\t{\Sigma}_{\sigma}^{\times}({\bm k};\varepsilon+i\gamma)$ decays like $1/\varepsilon$ for $\vert\varepsilon\vert\to\infty$ (see Eq.~(\ref{e328})), this function cannot be absolutely integrable (\S~4.43 in Ref.~\citen{WW62}) over $[-\infty,\infty]$. Consequently, there exists no non-negative function $g(\varepsilon)$ satisfying both $\vert \t{\Sigma}_{\sigma}({\bm k};\varepsilon+i\gamma)\vert \le g(\varepsilon)$ \emph{and} $\int_{-\infty}^{\infty} {\rm d}\varepsilon\; g(\varepsilon) <\infty$, whereby, on account of the Lebesgue dominated-convergence theorem (theorem 8.6 in Ref.~\citen{HS91}, \S~399 in Ref.~\citen{EWH27}), the substitution in Eq.~(\ref{e336}) would be allowable, in consequence of which $\mathfrak{S}_{\sigma}^{(\gamma)}({\bm k};0^-)$, $\mathfrak{S}_{\sigma}^{(\gamma)}({\bm k};0)$ and $\mathfrak{S}_{\sigma}^{(\gamma)}({\bm k};0^+)$ would be necessarily identical. The expression in Eq.~(\ref{e335}) implies that
\begin{equation}
\mathfrak{S}_{\sigma}^{(\gamma)}({\bm k};0) = -\frac{i}{2\hbar}\,
\Sigma_{\sigma;\infty_1}({\bm k}), \label{e337}
\end{equation}
and later (see the discussions following Eq.~(\ref{e353}) below) we shall deduce that
\begin{equation}
\mathfrak{S}_{\sigma}^{(\gamma)}({\bm k};0^+) =  2\,
\mathfrak{S}_{\sigma}^{(\gamma)}({\bm k};0) \equiv -\frac{i}{\hbar}\,
\Sigma_{\sigma;\infty_1}({\bm k}). \label{e338}
\end{equation}
These results are to be compared with that in Eq.~(\ref{e322}), according to which one in particular has $\mathfrak{S}_{\sigma}^{(\gamma)}({\bm k};0^-) \equiv 0$, $\forall {\bm k}$. These observations should be taken account of in any reliable implementation of the basic method described in Ref.~\citen{SLGB96b}.

Introducing $\t{\Sigma}_{\sigma;E}^{\times}({\bm k};\varepsilon+i\gamma)$ as the function corresponding to $\mathfrak{S}_{\sigma;E}^{(\gamma)}({\bm k};t)$ (cf. Eqs.~(\ref{e321}) and (\ref{e324})), one has the exact correspondence (cf. Eq.~(\ref{e323}))
\begin{equation}
\t{\Sigma}_{\sigma;E}^{\times}({\bm k};\varepsilon+i\gamma) =
\int_{-\infty}^{\infty} {\rm d}t\; \mathrm{e}^{i
(\varepsilon+i\gamma)t/\hbar}\,
\mathfrak{S}_{\sigma;E}^{(\gamma)}({\bm k};t), \label{e339}
\end{equation}
from which one obtains the identity
\begin{equation}
\t{\Sigma}_{\sigma;E}^{\times}({\bm k};\varepsilon+i\gamma) =
\t{\Sigma}_{\sigma}^{\times}({\bm k};\varepsilon+i\gamma) +\delta
\t{\Sigma}_{\sigma;E}^{\times}({\bm k};\varepsilon+i\gamma),
\label{e340}
\end{equation}
where
\begin{eqnarray}
&&\hspace{-0.4cm}\delta \t{\Sigma}_{\sigma;E}^{\times}({\bm
k};\varepsilon+i\gamma) =
\delta_1\t{\Sigma}_{\sigma;E}^{\times}({\bm k};\varepsilon+i\gamma)
+ \delta_2\t{\Sigma}_{\sigma;E}^{\times}({\bm
k};\varepsilon+i\gamma) \nonumber\\
&&\hspace{0.3cm} \equiv \int_{0}^{\infty} {\rm d}t\; \mathrm{e}^{i
(\varepsilon+i\gamma)t/\hbar}\, \big[
\mathfrak{S}_{\sigma;E}^{(\gamma)}({\bm k};t)-
\mathfrak{S}_{\sigma}^{(\gamma)}({\bm k};t)\big] +\int_{-\infty}^0 {\rm d}t\; \mathrm{e}^{i (\varepsilon+i\gamma)t/\hbar}\,
\mathfrak{S}_{\sigma;E}^{(\gamma)}({\bm k};t). \nonumber\\
\label{e341}
\end{eqnarray}
With reference to the last integral, recall that $\mathfrak{S}_{\sigma}^{(\gamma)}({\bm k};t)\equiv 0$ for $t< 0$, Eq.~(\ref{e322}).

We now denote the self-energy $\t{\Sigma}_{\sigma;E}^{\times}({\bm k};\varepsilon+i\gamma)$ as calculated by Schmalian \emph{et al.} \cite{SLGB96a,SLGB96b} and Langer \emph{et al.} \cite{LSGB95,LSGB96} by $\t{\Sigma}_{\sigma;E}^{\times\prime}({\bm k};\varepsilon+i\gamma)$. Assuming that the FLEX approximation \cite{BS89}, adopted by Schmalian \emph{et al.} and Langer \emph{et al.}, is exact, disregarding errors that are necessarily part of any numerically-calculated quantity (i.e., round-off errors, aliasing errors, etc.), and neglecting for the time being the fact that in the calculations under consideration integrals with respect to $t$ have been carried out over the finite interval $[0,\tau_0]$ \cite{Note12}, rather than $[0,\infty]$, one readily verifies that (cf. Eq.~(\ref{e340}))
\begin{equation}
\t{\Sigma}_{\sigma;E}^{\times\prime}({\bm k};\varepsilon+i\gamma) =
\t{\Sigma}_{\sigma}^{\times}({\bm k};\varepsilon+i\gamma) +\delta_1
\t{\Sigma}_{\sigma;E}^{\times}({\bm k};\varepsilon+i\gamma).
\label{e342}
\end{equation}
That is, Schmalian \emph{et al.} \cite{SLGB96a,SLGB96b} and Langer \emph{et al.} \cite{LSGB95,LSGB96} have neglected $\delta_2\t{\Sigma}_{\sigma;E}^{\times}({\bm k};\varepsilon+i\gamma)$ altogether. Were it not for the fact that $\delta_2\t{\Sigma}_{\sigma;E}^{\times}({\bm k};\varepsilon+i\gamma)$ almost perfectly cancels the imaginary part of $\delta_1\t{\Sigma}_{\sigma;E}^{\times}({\bm k};\varepsilon+i\gamma)$, $\t{\Sigma}_{\sigma;E}^{\times\prime}({\bm k};\varepsilon+i\gamma)$ would have possibly been, in view of Eqs.~(\ref{e340}) and (\ref{e341}), closer to the exact $\t{\Sigma}_{\sigma}^{\times}({\bm k};\varepsilon+i\gamma)$ than $\t{\Sigma}_{\sigma;E}^{\times}({\bm k};\varepsilon+i\gamma)$.

Below we demonstrate that for $\varepsilon\in \mathds{R}$ and $\gamma$ finite, but satisfying $\gamma\ll E$, $\delta_2\t{\Sigma}_{\sigma;E}^{\times}({\bm k};\varepsilon+i\gamma)$ is to leading order in $1/E$ equal to the complex conjugate of $\delta_1\t{\Sigma}_{\sigma;E}^{\times}({\bm k};\varepsilon+i\gamma)$. The analysis leading to this result further reveals that
\begin{equation}
\delta_1\t{\Sigma}_{\sigma;E}^{\times}({\bm k};\varepsilon+i 0^+) =
\delta_2\t{\Sigma}_{\sigma;E}^{\times *}({\bm k};\varepsilon+i
0^+),\;\;\varepsilon\in \mathds{R}, \label{e343}
\end{equation}
to \emph{all} orders of $1/E$ as $E\to\infty$.\footnote{In view of the analyses in appendix \protect\ref{sc}, these two functions can differ by an exponentially small function for all $\varepsilon$; description of such function is outside the domain of a Poincar\'e-type asymptotic series expansion (\S~8.2 in Ref.~\protect\citen{WW62}) in terms of the asymptotic sequence $\{1,1/E,1/E^2,\dots\}$.} In practice, where integrations with respect to $t$ are carried out over finite intervals, such as $[-\tau_0,0]$ and $[0,\tau_0]$, the result in Eq.~(\ref{e343}) will be valid up an error of the order of $\gamma_0/E$, where $\gamma_0=\hbar/\tau_0$.

\subsubsection{Equality of $\delta_2\t{\Sigma}_{\sigma;E}^{\times}({\bf k};\varepsilon+i\gamma)$ with $\delta_1\t{\Sigma}_{\sigma;E}^{\times *}({\bf k};\varepsilon+i\gamma)$ to leading order in $1/E$}
\label{ss63s3}

Consider the complex-valued function $f(t)$ and its Fourier transform
\begin{equation}
F(\varepsilon) = \int_{-\infty}^{\infty} {\rm d}t\;
\mathrm{e}^{i\varepsilon t/\hbar}\, f(t) \label{e344}
\end{equation}
which we express as
\begin{equation}
F(\varepsilon) \equiv F_{-}(\varepsilon) + F_{+}(\varepsilon),
\label{e345}
\end{equation}
where (cf. Eq.~(\ref{e341}))
\begin{equation}
F_{-}(\varepsilon) {:=} \int_{-\infty}^{0} {\rm d}t\;
\mathrm{e}^{i\varepsilon t/\hbar}\, f(t),\;\;\; F_{+}(\varepsilon)
{:=} \int_0^{\infty} {\rm d}t\; \mathrm{e}^{i\varepsilon t/\hbar}\,
f(t). \label{e346}
\end{equation}
It can be readily verified that for $f(t)$ a \emph{purely imaginary} and \emph{odd} function of $t$, or a \emph{purely real} and \emph{even} function of $t$, one has
\begin{equation}
F_{+}(\varepsilon) = F_{-}^*(\varepsilon),\;\;\; \varepsilon\in
\mathds{R}. \label{e347}
\end{equation}
In other words, in such cases, $F(\varepsilon)$ is real for real values of $\varepsilon$.

It will prove useful in the following to bear in mind that for $f(t)$ a \emph{purely imaginary} function of $t$, $\mathrm{Re}[F_{-}(\varepsilon)]$ ($\mathrm{Im}[F_{-}(\varepsilon)]$) is an odd (even) function of $\varepsilon$ for real values of $\varepsilon$, and that for $f(t)$ a \emph{purely real} function of $t$, $\mathrm{Re}[F_{-}(\varepsilon)]$ ($\mathrm{Im}[F_{-}(\varepsilon)]$) is an even (odd) function of $\varepsilon$, for real values of $\varepsilon$.

With the above facts in mind, it is evident that if (cf. Eq.~(\ref{e341}))
\begin{equation}
f(t) {:=} \mathrm{e}^{-\gamma t/\hbar}
\big[\mathfrak{S}_{\sigma;E}^{(\gamma)}({\bm k};t)-
\mathfrak{S}_{\sigma}^{(\gamma)}({\bm k};t)\big] \label{e348}
\end{equation}
is a \emph{purely imaginary} and \emph{odd} function of $t$, or a
\emph{purely real} and \emph{even} function of $t$, then
\begin{equation}
\delta_1\t{\Sigma}_{\sigma;E}^{\times}({\bm k};\varepsilon+i\gamma)
= \delta_2\t{\Sigma}_{\sigma;E}^{\times *}({\bm
k};\varepsilon+i\gamma)\;\;\; \mbox{\rm for}\;\;\; \varepsilon\in
\mathds{R}. \label{e349}
\end{equation}
For clarity and latter reference, since $\mathfrak{S}_{\sigma}^{(\gamma)}({\bm k};t)\equiv 0$ for $t<0$, one has (cf. Eq.~(\ref{e341}))
\begin{equation}
\delta_1 \t{\Sigma}_{\sigma;E}^{\times}({\bm k};\varepsilon+i\gamma) \equiv \int_0^{\infty} {\rm d}t\; \mathrm{e}^{i\varepsilon t/\hbar}\, f(t),\;\;\; \delta_2 \t{\Sigma}_{\sigma;E}^{\times}({\bm k};\varepsilon+i\gamma) \equiv \int_{-\infty}^0 {\rm d}t\; \mathrm{e}^{i\varepsilon t/\hbar}\, f(t), \label{e350}
\end{equation}
from which one observes that $\delta_1 \t{\Sigma}_{\sigma;E}^{\times}({\bm k};\varepsilon+i\gamma)$ ($\delta_2 \t{\Sigma}_{\sigma;E}^{\times}({\bm k};\varepsilon+i\gamma)$) is to be compared with the $F_+(\varepsilon)$ ($F_-(\varepsilon)$) in Eq.~(\ref{e346}) (see Sec.~\ref{ss63s4}).

Clearly, the $f(t)$ in Eq.~(\ref{e348}) cannot in general be either purely imaginary and odd or purely real and even, however, as we shall demonstrate below, to leading order in $1/E$ this function is a purely imaginary and odd function of $t$, irrespective of the value of the real quantity $\gamma$. For $\gamma=\pm 0^+$, however, one can explicitly demonstrate that the contribution of order $1/E^j$ to $f(t)$ is a purely imaginary and odd function of $t$ for $j$ odd, and purely real and even function of $t$ for $j$ even. This aspect is related to the fact, remarked earlier, that $\mathcal{I}_j(x)$, $x\in \mathds{R}$, Eq.~(\ref{e330}), is real for even values of $j$, and purely imaginary for odd values of $j$.

From the defining expressions in Eqs.~(\ref{e321}) and (\ref{e325}) one has
\begin{eqnarray}
f(t) &=& -\big[\int_{-\infty}^{-E} +
\int_E^{\infty}\big]\, \frac{{\rm d}\varepsilon}{2\pi\hbar}\;
\mathrm{e}^{-i\varepsilon t/\hbar}\,
\t{\Sigma}_{\sigma}^{\times}({\bm k};\varepsilon+i\gamma)
\nonumber\\
&\sim&  \frac{\Sigma_{\sigma;\infty_1}({\bm
k})}{2\pi i\hbar}\,  \mathcal{J}_0\big(\frac{E t}{\hbar}\big)
+\frac{\Sigma_{\sigma;\infty_2}({\bm k}) - i\gamma\, \Sigma_{\sigma;\infty_1}({\bm k})}{2\pi i\hbar}\,
\frac{1}{E}\, \mathcal{J}_1\big(\frac{E t}{\hbar}\big) + \dots,\;\; \mbox{\rm as}\;\; E\to\infty,\nonumber\\
\label{e351}
\end{eqnarray}
where
\begin{equation}
\mathcal{J}_0(x) {:=} 2 x\, \mathrm{Si}(x) - \pi\,
\mathrm{sgn}(x),\;\;\;\forall x, \label{e352}
\end{equation}
\begin{equation}
\mathcal{J}_1(x) {:=} -i\big( 2 x\, \mathrm{Si}(x) + 2\cos(x) -\pi
x\,\mathrm{sgn}(x)\big),\;\;\;\forall x. \label{e353}
\end{equation}
Since $\Sigma_{\sigma;\infty_1}({\bm k}) \in \mathds{R}$ (appendix \ref{sc}), from Eq.~(\ref{e351}) one observes that to leading order in $1/E$ the function $f(t)$ is purely imaginary. Since $\mathrm{Si}(-x) =-\mathrm{Si}(x)$ (item 5.2.19 in Ref.~\citen{AS72}), from Eq.~(\ref{e351}) one further infers that $f(t)$ is to leading order in $1/E$  an \emph{odd} functions of $t$. Since $\Sigma_{\sigma;\infty_2}({\bm k}) \in \mathds{R}$ (appendix \ref{sc}), for $\gamma= 0^{\pm}$ the next-to-leading-order asymptotic contribution to $f(t)$ is evidently real; this contribution is clearly an even function of $t$, irrespective of the value of $\gamma$. It is on the basis of the alternating change, from purely imaginary and odd to purely real and even, of the terms in the asymptotic series expansion for $f(t)$ corresponding to $E\to\infty$ and $\gamma= 0^{\pm}$, that Eq.~(\ref{e343}) can be shown to be valid to all orders in $1/E$. Making use of Eq.~(\ref{e351}), in conjunction with the results in Eqs.~(\ref{e322}) and (\ref{e337}), one readily arrives at the result in Eq.~(\ref{e338}).

In deducing the expression in Eq.~(\ref{e351}) we have used the asymptotic series in Eq.~(\ref{e328}). Since $\mathfrak{S}_{\sigma}^{(\gamma)}({\bm k};t)\equiv 0$ for $t<0$, it should not come as a surprise that for $t<0$ the series on the RHS of Eq.~(\ref{e351}) is equivalent with that on the RHS of Eq.~(\ref{e329}), which concerns $\mathfrak{S}_{\sigma;E}^{(\gamma)}({\bm k};t)$ for $t\le 0$. In this connection, note that
\begin{equation}
\mathcal{J}_j(x) \equiv \mathcal{I}_j(x)\;\;\; \mbox{\rm for}\;\;\;
x<0,\; \forall j \in \mathds{Z}^*, \label{e354}
\end{equation}
and that similar to $\mathcal{I}_j(x)$, for $x\in\mathds{R}$, $\mathcal{J}_j(x)$ is real for even values of $j$ and purely imaginary for odd values of $j$.

We should point out that appearance of $E$ in the argument of the transcendental functions $\mathrm{Si}(x)$ and $\cos(x)$ on the RHS of Eq.~(\ref{e351}) implies that the asymptotic series for $f(t)$ corresponding to $E\to\infty$ is not of the Poincar\'e type (\S~8.2 in Ref.~\citen{WW62}). Nonetheless, since $\mathcal{J}_j(x)$, $j \in \mathds{Z}^*$, are \emph{bounded} functions of $x$ for all real values of $x$, the non-Poincar\'e-type nature of the latter series will not be of any consequence to the main conclusions of our present analysis. Numerical results, to be presented in Sec.~\ref{ss63s4}, will corroborate this statement.

\subsubsection{Quantitative results}
\label{ss63s4}

Defining
\begin{equation}
C_{E;0}(\zeta) {:=} \frac{{\sf e}_0}{2\pi i\hbar}\int_{-\infty}^0
{\rm d}t\; \mathrm{e}^{i \zeta t/\hbar}\, \mathcal{I}_0\Big(\frac{E
t}{\hbar}\Big), \;\;\;\mathrm{Im}(\zeta)<0,\label{e355}
\end{equation}
and
\begin{equation}
C_{E;1}(\zeta) {:=} \frac{{\sf e}_0^2}{2\pi i\hbar} \int_{-\infty}^0
{\rm d}t\; \mathrm{e}^{i \zeta t/\hbar}\,
\frac{1}{E}\,\mathcal{I}_1\Big(\frac{E
t}{\hbar}\Big),\;\;\;\mathrm{Im}(\zeta)<0, \label{e356}
\end{equation}
in view of Eqs.~(\ref{e350}), (\ref{e351}) and (\ref{e354}), one observes that for $E\to\infty$
\begin{equation}
\delta_2\t{\Sigma}_{\sigma;E}^{\times}({\bm k};\varepsilon+i\gamma) \sim \frac{\Sigma_{\sigma;\infty_1}({\bm k})}{{\sf e}_0}\, C_{E;0}(\varepsilon+i 0^+) + \frac{\Sigma_{\sigma;\infty_2}({\bm k}) -i\gamma\Sigma_{\sigma;\infty_1}({\bm k})}{{\sf e}_0^2}\, C_{E;1}(\varepsilon+i 0^+), \label{e357}
\end{equation}
where $C_{E;0}(\varepsilon+i 0^+)$ and $C_{E;1}(\varepsilon+i 0^+)$ stand for the analytic continuations (\S~5.5 in Ref.~\citen{WW62}) of the functions in respectively Eqs.~(\ref{e355}) and (\ref{e356}) into the region corresponding to $\mathrm{Im}(\zeta)>0$ (see later); the integrals on the RHSs of Eqs.~(\ref{e355}) and (\ref{e356}) do not exist for $\mathrm{Im}(\zeta)> 0$. Above, the quantity ${\sf e}_0$ has the dimension of energy so that $C_{E;0}(\zeta)$ and $C_{E;1}(\zeta)$ are dimensionless; ${\sf e}_0$ may be identified with the total width of the non-interacting energy dispersion $\varepsilon_{\bm k}$.

The apparent disparity between the $\varepsilon+i\gamma$ on the LHS of Eq.~(\ref{e357}) and the $\varepsilon+i 0^+$ on the RHS originates from the fact that we have not explicitly incorporated $\mathrm{e}^{\gamma t/\hbar}$ in the definitions of $C_{E;0}(\zeta)$ and $C_{E;1}(\zeta)$; consequently, in expressing $\delta_2\t{\Sigma}_{\sigma;E}^{\times}({\bm k};z)$ at $z=\varepsilon+i\gamma$ in terms of $C_{E;0}(\zeta)$ and $C_{E;1}(\zeta)$, we have relied on the equivalence
\begin{equation}
z=\varepsilon+i\gamma \,\Longleftrightarrow\,\zeta = \varepsilon + i 0^+. \label{e358}
\end{equation}

Using the explicit expressions for $\mathcal{I}_0(x)$ and $\mathcal{I}_1(x)$ in Eqs.~(\ref{e331}) and (\ref{e332}), we obtain that
\begin{equation}
C_{E;0}(\zeta) = \frac{{\sf e}_0}{2\pi i}\,
\frac{\displaystyle\ln\Big(\frac{\zeta+E}{\zeta-E}\Big) - \pi
i}{\zeta},\;\;\;\mathrm{Im}(\zeta) <0, \label{e359}
\end{equation}
\begin{equation}
C_{E;1}(\zeta) = \frac{{\sf e}_0^2}{2\pi i}\,
\frac{\displaystyle\ln\Big(\frac{\zeta+E}{\zeta-E}\Big) - \frac{2
\zeta}{E} - \pi i}{\zeta^2},\;\;\;\mathrm{Im}(\zeta) <0,\label{e360}
\end{equation}
where $\ln(z)$ is the principal branch of the logarithm function (\S~4.1 in Ref.~\citen{AS72}).

The need to calculate $C_{E;0}(\zeta)$ and $C_{E;1}(\zeta)$ at $\zeta=\varepsilon+i 0^+$, where $\varepsilon\in \mathds{R}$, requires that we determine the analytic continuations of these functions into the upper half of the $\zeta$ plane. To this end, we note that for $x\in \mathds{R}$, $\mathcal{I}_0(x)$ is real and $\mathcal{I}_1(x)$ is purely imaginary. Consequently, from the defining expressions in Eqs.~(\ref{e355}) and (\ref{e356}) one deduces that
\begin{equation}
C_{E;0}(\zeta^*) \equiv -C_{E;0}^*(-\zeta), \label{e361}
\end{equation}
\begin{equation}
C_{E;1}(\zeta^*) \equiv C_{E;1}^*(-\zeta). \label{e362}
\end{equation}
Hence, for the sought-after analytic continuations one has
\begin{equation}
C_{E;0}(\zeta) = -C_{E;0}^*(-\zeta),\;\;\; \mathrm{Im}(\zeta)>0,
\label{e363}
\end{equation}
\begin{equation}
C_{E;1}(\zeta) = C_{E;1}^*(-\zeta),\;\;\; \mathrm{Im}(\zeta)>0,
\label{e364}
\end{equation}
where the functions $C_{E;0}(-\zeta)$ and $C_{E;1}(-\zeta)$ on the RHSs of Eqs.~(\ref{e363}) and (\ref{e364}) are determined from the expressions in Eqs.~(\ref{e359}) and (\ref{e360}) respectively.

Since in practice the integration with respect to $t$ in Eq.~(\ref{e324}) is carried out over a finite interval, say $[0,\tau_0]$, where $\tau_0>0$ (this is in particular the case in the calculations relying on the FFT technique \cite{Note12}), it is reasonable that the integrals in Eqs.~(\ref{e355}) and (\ref{e356}) be over $[-\tau_0,0]$ instead of $[-\infty,0]$. Denoting the corresponding functions by $C_{E;0}^{(\tau_0)}(\zeta)$ and $C_{E;1}^{(\tau_0)}(\zeta)$, respectively, one can readily verify that for $\varepsilon\in \mathds{R}$, and sufficiently large $\tau_0$, one has
\begin{equation}
C_{E;j}^{(\tau_0)}(\varepsilon) \approx C_{E;j}(\varepsilon+i
\gamma_0),\;\; \gamma_0 \equiv \frac{\hbar}{\tau_0},\;\;\; \mbox{\rm where}\;\;\; j=0,1. \label{e365}
\end{equation}
The constant $\gamma_0$ is not to be confused with $\gamma$. Numerical tests confirm that the approximate expressions in Eq.~(\ref{e365}) are very accurate. The results in Eq.~(\ref{e365}) in conjunction with the analytic expressions in Eqs.~(\ref{e359}) and (\ref{e360}) can be employed to eliminate, in an approximate way, the consequences of employing a finite value for $\tau_0$ in practical calculations.

We now define the functions (cf. Eqs.~(\ref{e355}) and (\ref{e356}))
\begin{equation}
C_{E;0}'(\zeta) {:=} \frac{{\sf e}_0}{2\pi i\hbar}\int_0^{\infty}
{\rm d}t\; \mathrm{e}^{i \zeta t/\hbar}\, \mathcal{J}_0\Big(\frac{E
t}{\hbar}\Big), \;\;\;\mathrm{Im}(\zeta)>0,\label{e366}
\end{equation}
\begin{equation}
C_{E;1}'(\zeta) {:=} \frac{{\sf e}_0^2}{2\pi i\hbar} \int_0^{\infty}
{\rm d}t\; \mathrm{e}^{i \zeta t/\hbar}\,
\frac{1}{E}\,\mathcal{J}_1\Big(\frac{E
t}{\hbar}\Big),\;\;\;\mathrm{Im}(\zeta)>0. \label{e367}
\end{equation}
In the light of Eqs.~(\ref{e350}) and (\ref{e351}), one observes that for $E\to\infty$
\begin{equation}
\delta_1\t{\Sigma}_{\sigma;E}^{\times}({\bm k};\varepsilon+i\gamma) \sim \frac{\Sigma_{\sigma;\infty_1}({\bm k})}{{\sf e}_0}\, C_{E;0}'(\varepsilon+i 0^+) + \frac{\Sigma_{\sigma;\infty_2}({\bm k}) -i\gamma\Sigma_{\sigma;\infty_1}({\bm k})}{{\sf e}_0^2}\, C_{E;1}'(\varepsilon+i 0^+), \label{e368}
\end{equation}
where, since $\mathrm{Im}(\varepsilon+i 0^+)>0$ for $\varepsilon\in \mathds{R}$, $C_{E;0}'(\varepsilon+i 0^+)$ and $C_{E;1}'(\varepsilon+i 0^+)$ can be directly determined from the defining expressions in respectively Eqs.~(\ref{e366}) and (\ref{e367}). Although the integrals in the latter expressions do not exist for $\mathrm{Im}(\zeta)<0$, nonetheless $C_{E;0}'(\zeta)$ and $C_{E;1}'(\zeta)$ can be determined for this region of the complex $\zeta$ plane by means of analytic continuation (\S~5.5 in Ref.~\citen{WW62}). In this light, where in the following we encounter $C_{E;0}'(\zeta)$ and $C_{E;1}'(\zeta)$ for a $\zeta$ in the lower half of the $\zeta$ plane, we consider $C_{E;0}'(\zeta)$ and $C_{E;1}'(\zeta)$ as denoting the analytic continuations of the functions defined in respectively Eqs.~(\ref{e366}) and (\ref{e367}) into this half-plane.

Following Eq.~(\ref{e354}), one readily verifies that
\begin{equation}
C_{E;0}'(\zeta) = -C_{E;0}(-\zeta), \label{e369}
\end{equation}
\begin{equation}
C_{E;1}'(\zeta) = C_{E;1}(-\zeta). \label{e370}
\end{equation}
In view of Eqs.~(\ref{e363}) and (\ref{e364}) one thus has
\begin{equation}
C_{E;j}'(\zeta) \equiv C_{E;j}^*(\zeta)\;\;\;\mbox{\rm for}
\;\;\;\mathrm{Im}(\zeta)>0,\; j=0,1. \label{e371}
\end{equation}
This result is the generalised version of that in Eq.~(\ref{e347})
which is specific to $\varepsilon\in\mathds{R}$.

\begin{figure}[t!]
\begin{center}
\includegraphics[width=3.2in]{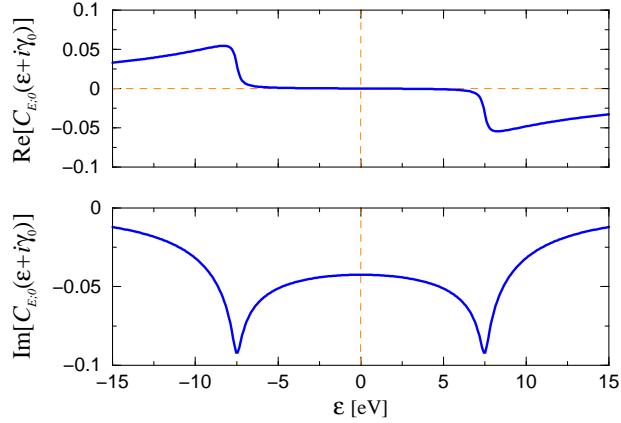}
\caption{\label{f4} The real and imaginary parts of $C_{E;0}(\zeta)$, Eq.~(\ref{e355}), for $\zeta=\varepsilon+i\gamma_0$ (see Eq.~(\protect\ref{e365})), as functions of $\varepsilon$. The parameters to which the functions correspond are $E=7.5$~eV, ${\sf e}_0 = 1.0$~eV and $\gamma_0= 0.2$~eV. The broken lines are guides to the eye. Note the odd-even behaviours.}
\end{center}
\end{figure}

\begin{figure}[t!]
\begin{center}
\includegraphics[width=3.2in]{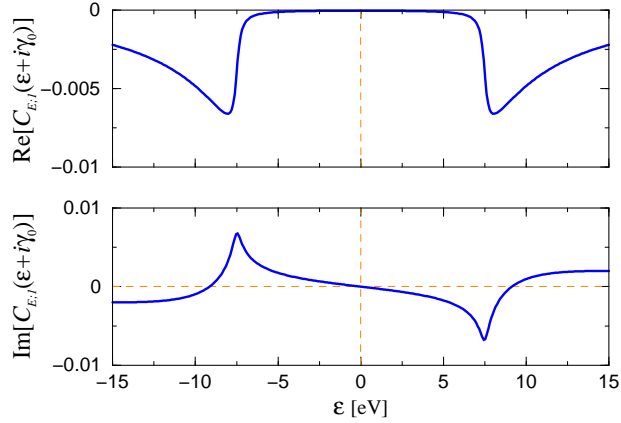}
\caption{\label{f5} The real and imaginary parts of $C_{E;1}(\zeta)$, Eq.~(\ref{e356}), for $\zeta=\varepsilon+i\gamma_0$, as functions of $\varepsilon$. The relevant parameters are the same as those presented in the caption of Fig.~\protect\ref{f4}. The broken lines are guides to the eye. Note the even-odd behaviours.}
\end{center}
\end{figure}

In Figs.~\ref{f4} and \ref{f5} we present the real and imaginary parts of $C_{E;0}(\varepsilon+i\gamma_0)$ and $C_{E;1}(\varepsilon+i\gamma_0)$ as functions of $\varepsilon$. These results correspond to the choice ${\sf e}_0 = 4 t_0$ and $E=30 t_0$, where $t_0= 0.25$~eV, conform the value adopted for the nearest-neighbour hopping integral $t_0$ in Refs.~\citen{SLGB96a,SLGB96b}. The value $\gamma_0 = 0.2$~eV, which is far greater than the value specific to the calculations in Refs.~\citen{SLGB96a,SLGB96b} (explicitly, $\hbar/\tau_0 = 2\times 30 t_0/4096 \approx 3.7\times 10^{-3}$~eV \cite{Note12}), is only responsible for the relatively smooth features of the data in Figs.~\ref{f4} and \ref{f5}. One can verify that for a sufficiently large value of $E$, $C_{E;0}(\varepsilon+i\gamma_0)$ and $C_{E;1}(\varepsilon+i\gamma_0)$ only weakly depend on the value of $\gamma_0$ for $\varepsilon$ in the vicinity of $\varepsilon =0$; this is owing to the fact that, for $\vert\varepsilon+i\gamma_0\vert \ll E$, the dependence of these functions on $\gamma_0$ is to leading order of the form $(\varepsilon+i\gamma_0)/E$. One can similarly show that, in contrast, $C_{E;0}(\varepsilon+i\gamma_0)$ and $C_{E;1}(\varepsilon+i\gamma_0)$ sensitively depend on the value of $\gamma_0$ in the immediate neighbourhoods of $\varepsilon = \pm E$. In view of Eq.~(\ref{e371}), the same observations apply to $C_{E;0}'(\varepsilon+i\gamma_0)$ and $C_{E;1}'(\varepsilon+i\gamma_0)$.

It is interesting to note that the range of variation of $\vert C_{E;1}(\varepsilon+i\gamma_0)\vert$ is suppressed by nearly one order of magnitude with respect to that of $\vert C_{E;0}(\varepsilon+i\gamma_0)\vert$ over a wide range of $\varepsilon$. This amount of suppression is consistent with the value of ${\sf e}_0/E$ which in the present calculations is approximately equal to $1.3\times 10^{-1}$. Thus, in spite of the fact that the series in Eq.~(\ref{e351}) is not a Poincar\'e-type asymptotic series, the bounded nature of the functions $\mathcal{J}_j(x)$, $j \in \mathds{Z}^*$, for real values of $x$, ensures that the series obtained from Eq.~(\ref{e351}) for $\delta_1\t{\Sigma}_{\sigma;E}^{\times}({\bm k};\varepsilon+i\gamma)$ and $\delta_2\t{\Sigma}_{\sigma;E}^{\times}({\bm k};\varepsilon+i\gamma)$ are indeed appropriate asymptotic series corresponding to $E\to\infty$.

It is remarkable that the real parts of $C_{E;0}(\varepsilon+i\gamma_0)$ and $C_{E;1}(\varepsilon+i\gamma_0)$ are negligibly small over a range as wide as approximately $[-E,E]$. If fact, for a wide range of $\varepsilon$ centred at $\varepsilon=0$ the absolute values of $\mathrm{Re}[C_{E;0}(\varepsilon+i\gamma_0)]$ and $\mathrm{Re}[C_{E;1}(\varepsilon+i\gamma_0)]$ are further diminished by decreasing the value of $\gamma_0$. In view of Eq.~(\ref{e371}), the same applies for $\mathrm{Re}[C_{E;0}'(\varepsilon+i\gamma_0)]$ and $\mathrm{Re}[C_{E;1}'(\varepsilon+i\gamma_0)]$. It follows that for sufficiently large values of $E$, and for $\varepsilon$ not too far removed from $\varepsilon=0$, $\mathrm{Re}[\t{\Sigma}_{\sigma;E}^{\times}({\bm k};\varepsilon+i\gamma)]$ is to a very good approximation equal to $\mathrm{Re}[\t{\Sigma}_{\sigma}^{\times}({\bm k};\varepsilon+i\gamma)]$ (see Eqs.~(\ref{e340}) and (\ref{e341})); similarly for $\mathrm{Re}[\t{\Sigma}_{\sigma;E}^{\times\prime}({\bm k};\varepsilon+i\gamma)]$ (see Eq.~(\ref{e342})).

Following Eq.~(\ref{e371}), however, whereas the above statement similarly applies to $\mathrm{Im}[\t{\Sigma}_{\sigma;E}^{\times}({\bm k};\varepsilon+i\gamma)]$, the imaginary part of the self-energy as calculated by Schmalian \emph{et al.} \cite{SLGB96a,SLGB96b} and Langer \emph{et al.} \cite{LSGB95,LSGB96}, that is $\mathrm{Im}[\t{\Sigma}_{\sigma;E}^{\times\prime}({\bm k};\varepsilon+i\gamma)]$, is considerably in error. This follows in particular from the fact that $\Sigma_{\sigma;\infty_1}({\bm k}) > 0$, $\forall {\bm k}$ (appendix \ref{sc}). We should emphasise that on account the self-consistency effects, it is not ruled out that $\mathrm{Re}[\t{\Sigma}_{\sigma;E}^{\times\prime}({\bm k};\varepsilon+i\gamma)]$ may be farther removed from $\mathrm{Re}[\t{\Sigma}_{\sigma}^{\times}({\bm k};\varepsilon+i\gamma)]$ than our present analysis suggests.

\subsubsection{A quantitative analysis}
\label{ss63s5}

For the purpose of a quantitative analysis, we introduce the following empirical ansatz concerning the $\Sigma_{\sigma;\infty_1}({\bm k})$ corresponding to the conventional Hubbard Hamiltonian in two space dimensions:
\begin{equation}
\Sigma_{\sigma;\infty_1}({\bm k}) \approx -5\,U\,
\mathrm{Re}[\Sigma_{\sigma}^{\times}({\bm k};\mu+U/2)], \label{e372}
\end{equation}
where $U>0$ is the on-site interaction energy in the underlying Hubbard Hamiltonian, and $\mu$ the short-hand for $\mu_{\infty} \equiv \varepsilon_{\Sc f}$. This ansatz clearly breaks down in the cases where $\mathrm{Re}[\Sigma_{\sigma}^{\times}({\bm k};\mu+U/2)]$ turns out to be positive, since it is required that $\Sigma_{\sigma;\infty_1}({\bm k})>0$, $\forall {\bm k}$ (appendix \ref{sc}). The ansatz in Eq.~(\ref{e372}) encapsulates the following considerations which can be verified to apply reasonably well to the data presented in Ref.~\citen{ZSS95} (these data are presented with $\mu_{\infty}$ chosen as the origin of the $\varepsilon$ axis):
\begin{itemize}
\item[(1)] $\mathrm{Re}[\Sigma_{\sigma}^{\times}({\bm k};\varepsilon)]$ passes through zero at $\varepsilon \approx \mu+U$;
\item[(2)] a well-behaved $\mathrm{Re}[\Sigma_{\sigma}^{\times}({\bm k};\varepsilon)]$ should therefore take its maximum magnitude at $\varepsilon\approx \mu+U/2$,
\item[(3)] $\mathrm{Re}[\Sigma_{\sigma}^{\times}({\bm k};\varepsilon)]$ should start to behave like $\Sigma_{\sigma;\infty_1}({\bm k})/(\varepsilon-\mu)$ for $\varepsilon \gtrsim \mu+3U/2$, and
\item[(4)] $\mathrm{Re}[\Sigma_{\sigma}^{\times}({\bm k};\mu+3U/2)] = -\alpha\,\mathrm{Re}[\Sigma_{\sigma}^{\times}({\bm k};\mu+U/2)]$ where $\alpha\approx 3-4$.
\end{itemize}
We point out that in considering the behaviour of $\t{\Sigma}_{\sigma}({\bm k};z)$ for $\vert z\vert\to\infty$, one may employ the asymptotic sequence $\{1, 1/(z-\varepsilon_0), 1/(z-\varepsilon_0)^2,\dots\}$, where $\varepsilon_0\in \mathds{R}$ need not be equal to zero. One trivially verifies that the coefficient of $1/(z-\varepsilon_0)^j$ is independent of $\varepsilon_0$ for $j=0,1$; for all finite $\varepsilon_0$, the coefficients corresponding to $j=0$ and $1$ are equal to $\Sigma_{\sigma;\infty_0}({\bm k})$ and $\Sigma_{\sigma;\infty_1}({\bm k})$ respectively (appendix \ref{sc}).

From the data in panel (c) of Fig.~1 in Ref.~\citen{LSGB96} we deduce for $\hbar\mathrm{Re}[\Sigma_{\sigma}^{\times}({\bm k};\mu+U/2)]$ a value between $-0.1$~eV and $-0.15$~eV; here $U= 4 t_0$ and $t_0 = 0.25$~eV. Although the data in Fig.~1 of Ref.~\citen{LSGB96} correspond to $\varepsilon\in [\mu-0.3,\mu+0.3]$~eV, so that $\mu+U/2 = 0.5$~eV lies outside the frame of panel (c), since $\mathrm{Re}[\Sigma_{\sigma}^{\times}({\bm k};\varepsilon)]$ should be approximately stationary in the vicinity of $\varepsilon \approx \mu+U/2$ (see Fig.~1 in Ref.~\citen{ZSS95}), we do not expect $\mathrm{Re}[\Sigma_{\sigma}^{\times}({\bm k};\mu+U/2)]$ to deviate significantly from, say, $\mathrm{Re}[\Sigma_{\sigma}^{\times}({\bm k};\mu+3U/10)]$.

From the above results and Eq.~(\ref{e372}) we thus conclude that for $U=4 t_0 = 1$~eV and ${\sf e}_0= 1$~eV one should have
\begin{equation}
\frac{\hbar\Sigma_{\sigma;\infty_1}({\bm k})}{{\sf e}_0} \approx 0.5
- 0.75~\mathrm{eV} = O(1)~\mathrm{eV}. \label{e92fx}
\end{equation}
On the basis of this result and of $\mathrm{Im}[C_{E;0}(\varepsilon+i\gamma_0)] \approx -0.05$ for a wide range of $\varepsilon$ centred around $\varepsilon =0$ (see Fig.~\ref{f4}), which is specific to in particular\footnote{No mention can be found in Ref.~\protect\citen{LSGB96} concerning the value used for $E$.} $E=7.5$, we expect that
\begin{equation}
\hbar\,\mathrm{Im}[\delta_1\t{\Sigma}_{\sigma}^{\times}({\bm
k};\varepsilon+i 0^+)] \approx - \hbar\,
\mathrm{Im}[\delta_2\t{\Sigma}_{\sigma}^{\times}({\bm
k};\varepsilon+i 0^+)]
\approx 25 - 38~\mathrm{meV}. \label{e373}
\end{equation}
According to the data in panel (a) of Fig.~1 in Ref.~\citen{LSGB96}, one has $-\hbar\mathrm{Im}[\t{\Sigma}_{\sigma;E}^{\times\prime}({\bm k};\mu)] \approx 37 - 50$~meV, in relatively good agreement with the result in Eq.~(\ref{e373}) (in Ref.~\citen{LSGB96}, $\mu=0$).

By identifying $\hbar\Sigma_{\sigma;\infty_1}({\bm k})/{\sf e}_0$ with $1$~eV for ${\bm k} = (7\pi/8,0)$ (the coordinates being with respect to the basis vectors along the principal axes of the square lattice and in units of the inverse lattice constant), on account of the above considerations we expect that $-\hbar\mathrm{Im}[\t{\Sigma}_{\sigma}^{\times\prime}({\bm k};\mu)] \approx 50$~meV, which is in a very good quantitative agreement with the value presented for $-\hbar\mathrm{Im}[\t{\Sigma}_{\sigma}^{\times\prime}({\bm k};\mu)]$ in Ref.~\citen{SLGB96a} (see Fig.~2 in Ref.~\citen{SLGB96a}); the fact that $-\hbar\mathrm{Im}[\t{\Sigma}_{\sigma}^{\times\prime}({\bm k};\mu)] \approx 130$~meV for ${\bm k} = (\pi,\pi/8)$ \cite{SLGB96a}, implies that $\Sigma_{\sigma;\infty_1}({\bm k})$ must be a relatively strongly varying function of ${\bm k}$ in the neighbourhood of the ``Fermi surface'' of the GS under consideration.

\subsubsection{Summary}
\label{ss63s6}

We have shown that the erroneous property (Sec.~\ref{ss21s2}) $\mathrm{Im}[\Sigma_{\sigma}({\bm k};\mu)]\not\equiv 0$ obtained by Schmalian \emph{et al.} \cite{SLGB96a,SLGB96b}, and Langer \emph{et al.}, \cite{LSGB95,LSGB96} and established and reported by Schmalian \emph{et al.} \cite{SLGB96a} as the cause of the breakdown of the Luttinger theorem, is an artifact of employing a one-sided Fourier representation (or a one-sided Laplace transformation) for what in an ideal calculation would be the \emph{retarded} self-energy. This function is however not retarded in such calculations as those by the latter authors, where integrations with respect to $\varepsilon$ are carried out over $[-E,E']$, $0<E, E' <\infty$. Consequently, the above-mentioned one-sided representation is incomplete; it neglects the advanced part of the `retarded' self-energy. In this section we elucidated the mechanism through which the latter incompleteness gives rise to $\mathrm{Im}[\Sigma_{\sigma}({\bm k};\mu)]\not\equiv 0$. By considering the above-mentioned finite interval to be $[-E,E]$, we have deduced the asymptotic series expansion, corresponding to $E\to\infty$, for the missing part of the self-energy as calculated by Schmalian \emph{et al.} \cite{SLGB96a,SLGB96b}, and Langer \emph{et al.} \cite{LSGB95,LSGB96}. A quantitative analysis of the leading-order term in the latter series revealed that this missing part almost entirely accounts for the deviation from zero of the $\mathrm{Im}[\Sigma_{\sigma}({\bm k};\mu)]$ calculated by the latter authors. We have therefore shown that the breakdown of the Luttinger theorem as observed by Schmalian \emph{et al.} \cite{SLGB96a} amounts to a manifestation of the above-mentioned artificial aspect of their numerical results and that this theorem proves valid on removing this defect.

Our considerations in this section have further shown the way in which the shortcomings associated with a finite value of $E$ can be relatively easily corrected for. This is achieved through employing Eq.~(\ref{e342}), according to which the exact self-energy $\t{\Sigma}_{\sigma}^{\times}({\bm k};\varepsilon+i\gamma)$ (i.e. the one corresponding to $E=\infty$) is obtained through subtracting $\delta_1\t{\Sigma}_{\sigma;E}^{\times}({\bm k};\varepsilon+i\gamma)$ from the self-energy as calculated by means of the method employed by Schmalian \emph{et al.} \cite{SLGB96a,SLGB96b}, and Langer \emph{et al.} \cite{LSGB95,LSGB96}, that is $\t{\Sigma}_{\sigma;E}^{\times\prime}({\bm k};\varepsilon+i\gamma)$. For sufficiently large $E$, the function $\delta_1\t{\Sigma}_{\sigma;E}^{\times}({\bm k};\varepsilon+i\gamma)$ is accurately described in terms of its asymptotic series expansion corresponding to $E\to\infty$, the leading two terms of which we have presented in Eq.~(\ref{e368}); the leading-order term in this series is fully determined in terms of $\Sigma_{\sigma;\infty_1}({\bm k})$ and the next-to-leading-order term by $\Sigma_{\sigma;\infty_1}({\bm k})$ (assuming a finite $\gamma$, as opposed to $\gamma=0^{\pm}$) and $\Sigma_{\sigma;\infty_2}({\bm k})$. Both of these functions can be readily deduced from Roth's \cite{LMR69} two-pole approximation for the single-particle Green function, to be discussed in Sec.~\ref{ss65} (see also appendix \ref{sc}). The data presented in Figs.~\ref{f4} and \ref{f5} show that for a reasonably large value of $E$, already the first term on the RHS of Eq.~(\ref{e368}) can amount to a very accurate approximation of $\delta_1\t{\Sigma}_{\sigma;E}^{\times}({\bm k};\varepsilon+i\gamma)$ (for the relationships between $C_{E;j}'(\zeta)$ and $C_{E;j}(\zeta)$, $j=0,1$, see Eqs.~(\ref{e369}) and (\ref{e370})). On the basis of the same approach and following Eqs.~(\ref{e340}), (\ref{e341}) and (\ref{e342}), one can also calculate the $\t{\Sigma}_{\sigma;E}^{\times}({\bm k};\varepsilon+i\gamma)$ (not to be confused with $\t{\Sigma}_{\sigma;E}^{\times\prime}({\bm k};\varepsilon+i\gamma)$) corresponding to large values of $E$. By doing so, one can obtain the results that Schmalian \emph{et al.} \cite{SLGB96a,SLGB96b}, and Langer \emph{et al.} \cite{LSGB95,LSGB96} would have obtained if they had employed a two-sided Laplace transform instead of a one-sided one.

In closing this section, we remark that using the method of `dynamical cluster approximation', Maier, Pruschke and Jarrell \cite{MPJ02} also observed violation of the Luttinger theorem for the metallic GSs of the
single-band Hubbard Hamiltonian on a square lattice corresponding to low concentration of holes, away from half-filling. The authors identified an occurrence of this violation as coinciding with emergence of a non-Fermi-liquid metallic state; for relatively large doping concentrations, the authors observed both Fermi-liquid metallic states and satisfaction of the Luttinger theorem. By considering the function $\mathrm{Im}[\Sigma_{\sigma}({\bm k};\varepsilon)]$ (more precisely $\mathrm{Im}[\Sigma({\bm k};\omega)]$) as displayed in Figs.~2 and 3 of Ref.~\citen{MPJ02} (panel (d), in both figures), one observes that for a low doping concentration ($\delta =0.05$) the behaviour of this function markedly deviates from that expected from the exact results in Eqs.~(\ref{e8}) and (\ref{e17}) (in the pertinent calculations, $\mu=0$), in contrast to the case corresponding to a relatively high doping concentration ($\delta=0.2$). For the reasons indicated in Sec.~\ref{ss21s2}, we conclude that the states for which the Luttinger theorem was found to be violated, and which Maier, Pruschke and Jarrell \cite{MPJ02} considered as `non-Fermi-liquid' metallic states, are pathological states, rather than metallic states of any kind.

\subsection{Case IV}
\label{ss64}

Using the single-particle spectral function $A_{\sigma}({\bm k};\varepsilon)$ (appendix \ref{sc}) corresponding to a single-band Hubbard Hamiltonian defined on a two-dimensional square lattice and calculated with the aid of the quantum-Monte-Carlo technique, Gr\"ober, Eder and Hanke \cite{GEH00} arrived at the conclusion that in the regime of low hole-doping concentrations, away from half-filling, the Luttinger theorem breaks down. Below we demonstrate that this result is a direct consequence of misapplying the Luttinger theorem.

\subsubsection{Preliminaries}
\label{ss64s1}

The calculations by Gr\"ober, Eder and Hanke \cite{GEH00} are carried out on a finite lattice, consisting of $\mathcal{N}_{\Sc l}$ lattice points\footnote{$\mathcal{N}_{\Sc l}$ is to be distinguished from $N_{\Sc l} \equiv \sum_{\sigma} N_{\Sc l;\sigma}$, the total Luttinger number.} so that the number of the relevant ${\bm k}$ points is equal to $\mathcal{N}_{\Sc l}$. We denote this set of points by $\{ {\bm k}_j\,\|\, j=1,2,\dots,\mathcal{N}_{\Sc l}\}$. The authors of Ref.~\citen{GEH00} calculated the Luttinger number per lattice site $n_{\Sc l;\sigma} \equiv N_{\Sc l;\sigma}/\mathcal{N}_{\Sc l}$, denoted in Ref.~\citen{GEH00} by $V_{\Sc f}$, through employing the expression
\begin{equation}
n_{\Sc l;\sigma} = \frac{1}{\mathcal{N}_{\Sc l}}
\sum_{j=1}^{\mathcal{N}_{\Sc l}} w_{\sigma}({\bm k}_j), \label{e374}
\end{equation}
where $w_{\sigma}({\bm k}_j)$ is assigned one of the three values $1$, $\frac{1}{2}$ and $0$ according to the following prescription \cite{GEH00}:
\begin{itemize}
\item[(a)]
$w_{\sigma}({\bm k}_j)=1$, if $A_{\sigma}({\bm k}_j;\varepsilon)$
has a peak at $\varepsilon$ less than the chemical potential $\mu$;
\item[(b)]
$w_{\sigma}({\bm k}_j)=\frac{1}{2}$, if the peak in $A_{\sigma}({\bm
k}_j;\varepsilon)$ is located at $\varepsilon=\mu$;
\item[(c)] $w_{\sigma}({\bm k}_j) =0$, otherwise.
\end{itemize}

Using the above prescription, Gr\"ober, Eder and Hanke \cite{GEH00} obtained that for $n$, the total number of particles per lattice site, increasing from approximately $n=0.8$ towards $n=1$ (i.e. half-filling), $n_{\Sc l;\sigma}$ increasingly deviated from $n_{\sigma} \equiv n/2$ (see Fig.~12 in Ref.~\citen{GEH00}), reaching the value $1$ (i.e. twice the expected value), or very nearly $1$, for $n/2=0.97/2\equiv 0.485$.

\subsubsection{Observations}
\label{ss64s2}

For interacting GSs, the expression in Eq.~(\ref{e374}) is fundamentally different from that in Eq.~(\ref{e21}) (or Eq.~(\ref{e7}), since the GSs under consideration are metallic); only for non-interacting GSs are the two expressions strictly equivalent.

The above statement is immediately appreciated by the fact (more about this in Sec.~\ref{ss64s3}) that for an interacting GS a peak in $A_{\sigma}({\bm k};\varepsilon)$ at ${\bm k}={\bm k}_0$ and $\varepsilon=\varepsilon_0$, with $\varepsilon_0 <\mu$, is \emph{not} necessarily indicative of the wave-vector ${\bm k}_0$ being inside the underlying Fermi sea (or Luttinger sea, if the GS under consideration is non-metallic). Clearly, however, this association of ${\bm k}_0$ with an interior point of the underlying Fermi sea is a perfectly valid one for non-interacting GSs, or more generally, mean-field GSs for which one has $A_{\sigma}({\bm k};\varepsilon) \equiv A_{\sigma;0}({\bm k};\varepsilon)$, where (cf. Eq.~(\ref{e89}))
\begin{equation}
A_{\sigma;0}({\bm k};\varepsilon) = \hbar\,
\delta(\varepsilon-\t{\varepsilon}_{{\bm k};\sigma}). \label{e375}
\end{equation}
The degree to which the above-mentioned association breaks down is thus a measure of the strength of correlation in the underlying GS. Therefore, the observations by Gr\"ober, Eder and Hanke \cite{GEH00}, considered by the authors as signifying violation of the Luttinger theorem, are clear indications of the fact that the strength of correlation in the GSs considered by the authors is an increasing function of $n$ for $n$ increasing towards $1$.

It is significant that in the Luttinger sums on the RHSs of Eqs.~(\ref{e21}) and (\ref{e7}), the energy $\varepsilon$ pertaining to $G_{\sigma}({\bm k};\varepsilon)$ is not a parameter to be varied, but one to be fixed at $\mu$. Consequently, insofar as $n_{\Sc l;\sigma}$ is concerned, it is in principle immaterial whether at a specific ${\bm k}$ the exact spectral function $A_{\sigma}({\bm k};\varepsilon)$ may be peaked at some $\varepsilon$, say $\varepsilon_0$, which may or may not be less than $\mu$. This is evidenced by the fact that even though the expression in Eq.~(\ref{e20}) is, for interacting GSs, a manifestly incorrect expression regarding $\t{G}_{\sigma}({\bm k};z)$, nonetheless it is a perfectly valid expression to be used in conjunction with the Luttinger sums on the RHSs of Eqs.~(\ref{e21}) and (\ref{e7}). In this connection, one should note that for the spectral function associated with the Green function in Eq.~(\ref{e20}), Eq.~(\ref{ec40}), one has
\begin{equation}
\left. A_{\sigma}({\bm
k};\varepsilon)\right|_{\mathrm{Eq.}~(\protect\ref{e20})} = \hbar\,
\delta\big(\varepsilon - \varepsilon_{\bm k} -
\hbar\Sigma_{\sigma}({\bm k};\varepsilon_{\Sc f})\big), \label{e376}
\end{equation}
which is characteristic of the single-particle spectral function within the framework of a mean-field theory where self-energy is a static quantity (cf. Eq.~(\ref{e375}) and see Sec.~\ref{ss52s3}).

\subsubsection{A quantitative analysis}
\label{ss64s3}

We now proceed with a quantitative analysis of the observation in Ref.~\citen{GEH00} that for $n=2\, n_{\sigma}$ close to $1$, $n_{\Sc l;\sigma}$ (i.e. $V_{\Sc f}$ in Ref.~\citen{GEH00}) is almost twice as large as $n_{\sigma}$. To this end, we introduce the following model single-particle spectral function:\cite{BF03,BF04a}
\begin{equation}
\mathcal{A}_{\sigma}({\bm k};\varepsilon) = \hbar\,{\sf
n}_{\sigma}({\bm k})\,\delta(\varepsilon-\varepsilon_{{\bm
k};\sigma}^{<}) + \hbar\,\big(1-{\sf n}_{\sigma}({\bm k})\big)\,
\delta(\varepsilon-\varepsilon_{{\bm k};\sigma}^{>}), \label{e377}
\end{equation}
where
\begin{equation}
\varepsilon_{{\bm k};\sigma}^{<} = \int_{-\infty}^{\mu} {\rm
d}\varepsilon\; \varepsilon\, P_{\sigma}^{<}({\bm
k};\varepsilon),\;\;\; \varepsilon_{{\bm k};\sigma}^{>} =
\int_{\mu}^{\infty} {\rm d}\varepsilon\; \varepsilon\,
P_{\sigma}^{>}({\bm k};\varepsilon), \label{e378}
\end{equation}
in which
\begin{equation}
P_{\sigma}^{<}({\bm k};\varepsilon) = \frac{1}{\hbar}
\frac{A_{\sigma}({\bm k};\varepsilon)}{{\sf n}_{\sigma}({\bm
k})},\;\;\; P_{\sigma}^{>}({\bm k};\varepsilon) = \frac{1}{\hbar}
\frac{A_{\sigma}({\bm k};\varepsilon)}{1-{\sf n}_{\sigma}({\bm k})}.
\label{e379}
\end{equation}
One has $P_{\sigma}^{\lessgtr}({\bm k};\varepsilon) \ge 0$, $\forall
\varepsilon$. Both here and below, $\mu$ denotes $\mu_{\infty}$, Eq.~(\ref{e25}).

On account of (Eq.~(\ref{ec39}))
\begin{equation}
\frac{1}{\hbar} \int_{-\infty}^{\infty} {\rm d}\varepsilon\;
A_{\sigma}({\bm k};\varepsilon) = 1,\;\;\;\forall {\bm k},
\label{e380}
\end{equation}
and (Eq.~(\ref{ec36}))
\begin{equation}
{\sf n}_{\sigma}({\bm k}) = \frac{1}{\hbar} \int_{-\infty}^{\mu}
{\rm d}\varepsilon\; A_{\sigma}({\bm k};\varepsilon), \label{e381}
\end{equation}
one has
\begin{equation}
\int_{-\infty}^{\mu} {\rm d}\varepsilon\; P_{\sigma}^{<}({\bm
k};\varepsilon) = 1,\;\;\; \int_{\mu}^{\infty} {\rm d}\varepsilon\;
P_{\sigma}^{>}({\bm k};\varepsilon) = 1. \label{e382}
\end{equation}
Thus $P_{\sigma}^{<}({\bm k};\varepsilon)$ and $P_{\sigma}^{>}({\bm k};\varepsilon)$ are appropriately normalised distribution functions for $\varepsilon\in [-\infty,\mu]$ and $\varepsilon\in [\mu,\infty]$ respectively. According to the expressions in Eq.~(\ref{e378}), $\varepsilon_{{\bm k};\sigma}^{<}$ and $\varepsilon_{{\bm k};\sigma}^{>}$ are thus the mean values of $\varepsilon$ distributed according to respectively $P_{\sigma}^{<}({\bm k};\varepsilon)$ and $P_{\sigma}^{>}({\bm k};\varepsilon)$, reflecting, for a specific ${\bm k}$, some global properties of these distribution functions which in turn are determined by $A_{\sigma}({\bm k};\varepsilon)$.

Following the above considerations, it is reasonable to identify $\varepsilon_{{\bm k};\sigma}^{<}$ ($\varepsilon_{{\bm k};\sigma}^{>}$) as the energy in the interval $[-\infty,\mu]$ ($[\mu,\infty]$) at or in the neighbourhood of which $A_{\sigma}({\bm k};\varepsilon)$ is peaked \cite{BF04a}. In this connection, it is relevant to enumerate the following facts. Firstly, $\mathcal{A}_{\sigma}({\bm k};\varepsilon)$ trivially satisfies the exact sum rule in Eq.~(\ref{e380}). Secondly,\cite{BF03}
\begin{equation}
\varepsilon_{{\bm k};\sigma}^{<} < \mu <\varepsilon_{{\bm
k};\sigma}^{>},\;\;\; \forall {\bm k}. \label{e383}
\end{equation}
Thirdly, on account of the inequalities in Eq.~(\ref{e383}), substitution of the $A_{\sigma}({\bm k};\varepsilon)$ on the RHS of Eq.~(\ref{e381}) by the model spectral function $\mathcal{A}_{\sigma}({\bm k};\varepsilon)$, Eq.~(\ref{e377}), yields the \emph{exact} ${\sf n}_{\sigma}({\bm k})$. And lastly, replacing the exact $A_{\sigma}({\bm k};\varepsilon)$ by $\mathcal{A}_{\sigma}({\bm k};\varepsilon)$ in the expression for the GS energy,
\begin{equation}
E_{N;0} = \frac{1}{2\hbar} \sum_{{\bm k},\sigma}
\int_{-\infty}^{\mu} {\rm d}\varepsilon\; (\varepsilon_{\bm k} +
\varepsilon)\, A_{\sigma}({\bm k};\varepsilon), \label{e384}
\end{equation}
yields the \emph{exact} energy of the $N$-particle GS of the system under consideration \cite{BF03}.

From Eq.~(\ref{e377}) one observes that in the cases of weakly correlated GSs, where ${\sf n}_{\sigma}({\bm k})$ takes values close to $1$ ($0$) for ${\bm k}$ inside (outside) the underlying Fermi sea, indeed the expression in Eq.~(\ref{e374}), with $w_{\sigma}({\bm k})$ as prescribed subsequent to Eq.~(\ref{e374}), should yield a value close to $n_{\sigma}$: with ${\sf n}_{\sigma}({\bm k}) \approx 0$ for ${\bm k}$ outside Fermi sea, according to Eq.~(\ref{e377}) the peak in $A_{\sigma}({\bm k};\varepsilon)$ at or in the close neighbourhood of $\varepsilon_{{\bm k};\sigma}^{<} <\mu$ should not be `visible', eliminating thus the possibility of incorrectly marking this ${\bm k}$ as a constituent point of the pertinent Fermi sea.

On the other hand, for strongly-correlated GSs, where ${\sf n}_{\sigma}({\bm k})$ is of the same order of magnitude for almost all ${\bm k}$ (here, for almost all $\{ {\bm k}_j\}$), the peak in $A_{\sigma}({\bm k};\varepsilon)$ associated with $\varepsilon_{{\bm k};\sigma}^{<}$ should be `visible' for all ${\bm k}$, including the points outside the relevant Fermi sea,\footnote{Note that $\varepsilon_{{\bm k};\sigma}^{<} <\mu$, $\forall {\bm k}$ (see Eq.~(\protect\ref{e383})). In this connection, we should emphasise that both $\varepsilon_{{\bm k};\sigma}^{<}$ and $\varepsilon_{{\bm k};\sigma}^{>}$ are defined for \emph{all} ${\bm k} \in \mathrm{1BZ}$. Thus, unless ${\sf n}_{\sigma}({\bm k})$ is sufficiently small for ${\bm k}$ outside (inside) the Fermi or Luttinger sea, $A_{\sigma}({\bm k};\varepsilon)$ should possess, in this region of the ${\bm k}$ space, a prominent peak at or in a close neighbourhood of $\varepsilon=\varepsilon_{{\bm k};\sigma}^{<} <\mu$ ($\varepsilon=\varepsilon_{{\bm k};\sigma}^{>} >\mu$).} incorrectly suggesting, to the observer who tends to view the peaks in $A_{\sigma}({\bm k};\varepsilon)$ as though this function were comparable with $A_{\sigma;0}({\bm k};\varepsilon)$, that the Fermi (or Luttinger) sea consisted of the entire ${\bm k}$ space. This clarifies the observation by Gr\"ober, Eder and Hanke \cite{GEH00} that for $n_{\sigma} = n/2$ approaching $1/2$ from below, the $n_{\Sc l;\sigma}$ calculated according to the expression in Eq.~(\ref{e374}) approaches $1$.

\subsubsection{Summary and remarks}
\label{ss64s4}

We have shown that the apparent breakdown of the Luttinger theorem, as reported by Gr\"ober, Eder and Hanke \cite{GEH00}, is a direct consequence of the invalidity of the expression for the Luttinger number $n_{\Sc l;\sigma}$, Eq.~(\ref{e374}), employed by the authors. On general grounds we demonstrated that Eq.~(\ref{e374}) is strictly valid for GSs whose underlying momentum distribution function ${\sf n}_{\sigma}({\bm k})$ takes the value $1$ ($0$) for ${\bm k}$ inside (outside) the underlying Fermi sea. Therefore, the increased deviation of the calculated $n_{\Sc l;\sigma}$ from $n_{\sigma}$ for $n_{\sigma} \to \frac{1}{2}$ is indicative of the fact that for $n_{\sigma} \to \frac{1}{2}$ the underlying GSs are increasingly more correlated, whereby the corresponding ${\sf n}_{\sigma}({\bm k})$ progressively takes comparable values for all ${\bm k}\in \mathrm{1BZ}$. This trend in the behaviour of ${\sf n}_{\sigma}({\bm k})$ as a function of the strength of correlation in the underlying GS is unambiguously apparent from the exact results for the ${\sf n}_{\sigma}({\bm k})$ pertaining to the Gutzwiller wave function in one space dimension \cite{MV87} (see Fig.~1 in Ref.~\citen{MV87} which corresponds to half-filling and in which $g=1$ corresponds to zero correlation and $g=0$ to infinite correlation). The considerations in Sec.~\ref{ss65} will shed some additional light on the observations made in this section.

\subsection{Case V}
\label{ss65}

Using the Roth two-pole approximation for the single-particle Green function \cite{LMR69}, Beenen and Edwards \cite{BE95} observed breakdown of the Luttinger theorem for the nearly half-filled metallic GSs of the single-band Hubbard Hamiltonian in two space dimensions. Here we clarify the reason for this apparent breakdown; we show that this breakdown is directly related to the approximative nature of the two-pole scheme of Roth and signifies no inherent shortcoming of the Luttinger theorem. We further deduce a closed expression for $\b{N}_{\sigma}^{(2)}$, Eq.~(\ref{e34}), which explicitly demonstrates the break-down of the Luttinger-Ward identity, Eq.~(\ref{e44}), within the framework of Roth's two-pole approximation programme. This expression, which reduces to a simple algebraic formula for $N_{\sigma} -N_{\Sc l;\sigma}$ in the strong-coupling regime, reproduces the numerical results for $N_{\sigma} -N_{\Sc l;\sigma}$ as calculated by Beenen and Edwards \cite{BE95} and further sheds light on the observations by Gr\"ober, Eder and Hanke \cite{GEH00} discussed in Sec.~\ref{ss64}.

\subsubsection{Preliminaries}
\label{ss65s1}

The single-particle Green function employed by Beenen and Edwards \cite{BE95} has the form
\begin{equation}
\t{G}_{\sigma}({\bm k};z) = \frac{\hbar\,\alpha_{\sigma;1}({\bm
k})}{z- \varepsilon_{\sigma;1}({\bm k})} + \frac{\hbar\,
\alpha_{\sigma;2}({\bm k})}{z- \varepsilon_{\sigma;2}({\bm k})},
\label{e385}
\end{equation}
where the four functions $\alpha_{\sigma;j}({\bm k})$, $\varepsilon_{\sigma;j}({\bm k})$, $j=1,2$, satisfy such relationships that in principle ensure equality of the first four moments integrals $G_{\sigma;\infty_j}({\bm k})$, $j=1,2,3,4$, Eq.~(\ref{ec67}), corresponding to the Green function in Eq.~(\ref{e385}) with those of the exact Green function. For the explicit expressions pertinent to $\alpha_{\sigma;j}({\bm k})$, $\varepsilon_{\sigma;j}({\bm k})$, $j=1,2$, the reader is referred to Refs.~\citen{BE95,LMR69}. For later reference we mention that
\begin{equation}
\alpha_{\sigma;1}({\bm k}) + \alpha_{\sigma;2}({\bm k}) = 1,
\label{e386}
\end{equation}
which ensures the $G_{\sigma;\infty_1}({\bm k})$ corresponding to the Green function in Eq.~(\ref{e385}) to coincide with that of the exact Green function.

It will be instructive to write the expression in Eq.~(\ref{e385}) in the following equivalent form (for conciseness, we suppress the arguments of $\alpha_{\sigma;j}({\bm k})$, $\varepsilon_{\sigma;j}({\bm k})$, $j=1,2$):
\begin{equation}
\t{G}_{\sigma}({\bm k};z) = \hbar\, \frac{z -
(\alpha_{\sigma;1} \varepsilon_{\sigma;2} + \alpha_{\sigma;2}
\varepsilon_{\sigma;1})}{z^2 -
(\varepsilon_{\sigma;1}+\varepsilon_{\sigma;2}) z +
\varepsilon_{\sigma;1} \varepsilon_{\sigma;2}}
\equiv \frac{\hbar}{z}\, \frac{ 1 -
\big(\alpha_{\sigma;1}\varepsilon_{\sigma;2} + \alpha_{\sigma;2}
\varepsilon_{\sigma;1}\big)/z}{1 - \big(\varepsilon_{\sigma;1} +
\varepsilon_{\sigma;2}\big)/z + \varepsilon_{\sigma;1}
\varepsilon_{\sigma;2}/z^2}, \label{e387}
\end{equation}
where we have partly used the result in Eq.~(\ref{e386}). The last expression in Eq.~(\ref{e387}) makes explicit that $z/\hbar$ times the Green function under consideration is simply the $(1,2)$ Pad\'e approximant (\S~19.7 in Ref.~\citen{BD02}) of $z/\hbar$ times the exact single-particle Green function, constructed from a fourth-order asymptotic series expansion of the exact $\t{G}_{\sigma}({\bm k};z)$ corresponding to $\vert z\vert \to\infty$ in terms of the asymptotic sequence $\{1/z, 1/z^2, \dots\}$, Eq.~(\ref{ec64}). It is immediately evident that the expression in Eq.~(\ref{e387}) reproduces the exact leading-order result $\t{G}_{\sigma}({\bm k};z) \sim \hbar/z$ corresponding to $\vert z\vert\to\infty$ (cf. Eqs.~(\ref{ec64}) and (\ref{ec66})). The expression in Eq.~(\ref{e385}), and thus that in Eq.~(\ref{e387}), will in principle fail correctly to reproduce the contributions of the form $G_{\sigma;\infty_j}({\bm k})/z^j$ for $j>4$ in the last-mentioned asymptotic series expansion.

Since the model under consideration is defined on a Bravais lattice \emph{and} the two-body interaction potential is short-range, the coefficient functions $\{ G_{\sigma;\infty_j}({\bm k})\,\|\, j\in \mathds{N}\}$ and $\{ \Sigma_{\sigma;\infty_j}({\bm k})\,\|\, j \in \mathds{Z}^*\}$ exist and the following statements apply (appendix \ref{sc}, in particular Sec.~\ref{ssc7}):
\begin{itemize}
\item[(1)] For $m\ge 2$, $G_{\sigma;\infty_m}({\bm k})$ is determined by the set of functions $\Sigma_{\sigma;\infty_j}({\bm k})$, $j=0,1,\dots, m-2$, where $\Sigma_{\sigma;\infty_0}({\bm k})$ coincides with the exact Hartree-Fock contribution to $\t{\Sigma}_{\sigma}({\bm k};z)$.
\item[(2)] $\Sigma_{\sigma;\infty_j}({\bm k})$, $j\ge 1$, is fully determined by the contributions arising from $\t{\Sigma}_{\sigma}^{(\nu)}({\bm k};z)$, $\nu =2,\dots,j+1$, where $\t{\Sigma}_{\sigma}^{(\nu)}({\bm k};z)$ denotes the total contribution of all $\nu$th-order skeleton self-energy diagrams in terms of the exact Green functions $\{\t{G}_{\sigma'}({\bm k};z)\}$ and the bare interaction potential.
\end{itemize}
From (1) and (2) it follows that by calculating the self-energy $\t{\Sigma}_{\sigma}({\bm k};z)$ in terms of skeleton diagrams and the exact Green functions $\{\t{G}_{\sigma'}({\bm k};z)\}$ to third order in the bare interaction potential and subsequently determining $\t{G}_{\sigma}({\bm k};z)$ from this self-energy through the Dyson equation, Eq.~(\ref{ec53}), this Green function, similar to that in Eq.~(\ref{e385}), reproduces the exact $G_{\sigma;\infty_j}({\bm k})$ for $j=1,2,3,4$. Conversely, since the Green function in Eq.~(\ref{e385}) in principle does not reproduce the exact $G_{\sigma;\infty_j}({\bm k})$ for any $j>4$, it follows that its corresponding self-energy, obtained through the Dyson equation, cannot correctly describe $\t{\Sigma}_{\sigma}^{(\nu)}({\bm k};z)$ for $\nu>3$.

On employing the Dyson equation, for the self-energy corresponding to the $\t{G}_{\sigma}({\bm k};z)$ in Eq.~(\ref{e387}) one obtains that
\begin{equation}
\hbar\,\t{\Sigma}_{\sigma}({\bm k};z) = z -\varepsilon_{\bm k} -
\frac{z^2 - (\varepsilon_{\sigma;1} + \varepsilon_{\sigma;2}) z +
\varepsilon_{\sigma;1} \varepsilon_{\sigma;2}}{z -
(\alpha_{\sigma;1} \varepsilon_{\sigma;2} + \alpha_{\sigma;2}
\varepsilon_{\sigma;1})}. \label{e388}
\end{equation}
With reference to our above observations, the first three terms in the asymptotic series expansion of the self-energy in Eq.~(\ref{e388}) corresponding to $\vert z\vert\to\infty$ coincide with the first three terms in the same series pertaining to the exact $\t{\Sigma}_{\sigma}({\bm k};z)$, the coefficients of the three terms, decaying like $1/z^j$, being $\Sigma_{\sigma;\infty_j}({\bm k})$, $j=0,1,2$.

With reference to Eq.~(\ref{e85}), we now express a general self-energy, whether exact or approximate, as follows:
\begin{equation}
\t{\Sigma}_{\sigma}({\bm k};z) = \sum_{\nu=1}^{\infty} \t{S}_{\sigma}^{(\nu)}({\bm k};z), \label{e389}
\end{equation}
where $\t{S}_{\sigma}^{(\nu)}({\bm k};z)$ originates from $\t{\Sigma}_{\sigma}^{(\nu)}({\bm k};z)$, $\nu\ge 1$. For a self-energy constructed by a prescription similar to that on which the self-energy in Eq.~(\ref{e388}) is based, the set $\{\t{S}_{\sigma}^{(\nu)}({\bm k};z)\,\|\, \nu=2,\dots,j+1\}$ is constrained by $j$ conditions, namely that the coefficients of $1/z$, \dots, $1/z^j$ in the large-$\vert z\vert$ asymptotic series expansion of the self-energy on the LHS of Eq.~(\ref{e389}) coincide with those of the exact self-energy; see statement (2) presented above, and note that $\t{\Sigma}_{\sigma}^{(\nu)}({\bm k};z)$, $\nu\ge 2$, contributes to $\Sigma_{\sigma;\infty_j}({\bm k};z)$ for \emph{all} $j\in \mathds{N}$. It should therefore be evident that there are infinitely many different self-energies that similar to the $\t{\Sigma}_{\sigma}({\bm k};z)$ in Eq.~(\ref{e388}) reproduce the exact $\Sigma_{\sigma;\infty_j}({\bm k})$ for $j=0,1,2$; of these, only one self-energy is the exact $\t{\Sigma}_{\sigma}({\bm k};z)$.

In view of Eq.~(\ref{e85}), the above observations imply that, aside from
\begin{equation}
\t{S}_{\sigma}^{(1)}({\bm k};z) = \t{\Sigma}_{\sigma}^{(1)}({\bm k};z) \equiv \Sigma_{\sigma}^{\Sc h\Sc f}({\bm k}), \label{e390}
\end{equation}
one can \emph{at best} have
\begin{equation}
\t{S}_{\sigma}^{(\nu)}({\bm k};z) \approx \t{\Sigma}_{\sigma}^{(\nu)}({\bm k};z)\;\;\;\mbox{\rm for}\;\;\; \nu=2,3, \label{e391}
\end{equation}
and that such approximate equality can in general not apply for any $\nu>3$. To emphasise the approximate equality in Eq.~(\ref{e391}), we point out that the exact $\t{\Sigma}_{\sigma}^{(\nu)}({\bm k};z)$ has a complicated analytic structure for $\nu\ge 2$ (for instance it possesses branch-cuts along the real energy axis), whereas the full self-energy in Eq.~(\ref{e388}) is a relatively simple function of $z$; this would not have been the case if instead of the approximate equality in Eq.~(\ref{e391}) one had an exact equality, for $\nu=2,3$.

\subsubsection{Analysis}
\label{ss65s2}

We are now in a position to clarify the reason underlying the observed breakdown of the Luttinger theorem for the Green function in Eq.~(\ref{e385}). To this end we consider the Luttinger-Ward identity whose violation in the present case\cite{Note10} leads to the breakdown of the Luttinger theorem, and vice versa (Sec.~\ref{ss43}). As we have pointed out earlier, the Luttinger-Ward identity, Eq.~(\ref{e44}), follows from the identity in Eq.~(\ref{e87}) for all $\nu\ge 1$.

If the Green function in Eq.~(\ref{e385}) were exact \emph{and} if the approximate equalities in Eq.~(\ref{e391}) were exact equalities, then the equation in Eq.~(\ref{e87}) would have been identically satisfied for $\nu=1,2,3$; for an exact $\t{G}_{\sigma}({\bm k};z)$, the equation in Eq.~(\ref{e87}) would have failed to be satisfied for $\nu >3$. As it stands, however, neither the Green function in Eq.~(\ref{e385}) is exact, nor can the $\t{S}_{\sigma}^{(\nu)}({\bm k};z)$ corresponding to the self-energy in Eq.~(\ref{e388}) be identical with the exact $\t{\Sigma}_{\sigma}^{(\nu)}({\bm k};z)$ for $\nu=2,3$. It follows that for the case under investigation there is no \emph{a priori} reason, whatever, why the Luttinger-Ward identity would be satisfied. Since in the case at hand breakdown of the Luttinger-Ward identity necessarily leads to breakdown of the Luttinger theorem, a similar statement applies as regards the Luttinger theorem.

It is important to draw attention to a result to which we have referred in Sec.~\ref{ss53s2}, namely that the Luttinger-Ward identity applies for $\t{\Sigma}_{\sigma}^{[m]}({\bm k};z)$, defined as the \emph{solution} of the equation in Eq.~(\ref{e101}), and the corresponding Green function $\t{G}_{\sigma}^{[m]}({\bm k};z)$, which is deduced from $\t{\Sigma}_{\sigma}^{[m]}({\bm k};z)$ through the Dyson equation. Since $\t{\Sigma}_{\sigma}^{[m]}({\bm k};z)$ and $\t{G}_{\sigma}^{[m]}({\bm k};z)$ do not coincide with the exact $\t{\Sigma}_{\sigma}({\bm k};z)$ and $\t{G}_{\sigma}({\bm k};z)$, respectively, for any $m<\infty$, it follows that for $m<\infty$ in principle $G_{\sigma;\infty_j}^{[m]}({\bm k})$, $j>1$, and $\Sigma_{\sigma;\infty_j}^{[m]}({\bm k})$ need not coincide with respectively the exact $G_{\sigma;\infty_j}({\bm k})$, $j>1$, and $\Sigma_{\sigma;\infty_j}({\bm k})$ for any $j$. In other words, the essential property of the Green function in Eq.~(\ref{e385}), described following Eq.~(\ref{e385}), does \emph{not} constitute a sufficient condition for the Luttinger-Ward identity to hold for this function and its corresponding self-energy. This aspect is reinforced by comparing the results in Eqs.~(\ref{e93}) and (\ref{e94}) with those in respectively Eqs.~(\ref{e97}) and (\ref{e98}); one observes that an apparently `inferior' Green function satisfies an identity reminiscent of the Luttinger-Ward identity, a task failed by its `superior' counterpart.

\subsubsection{The Luttinger-Ward identity revisited}
\label{ss65s3}

From the expressions in Eqs.~(\ref{e385}) and (\ref{e388}) one readily obtains that
\begin{eqnarray}
&&\hspace{-1.2cm} \sum_{\bm k} \int_{\mathscr{C}(\mu)} \frac{{\rm d}z}{2\pi i}\; \t{G}_{\sigma}({\bm k};z) \frac{\partial}{\partial z}
\t{\Sigma}_{\sigma}({\bm k};z) \nonumber\\
&&\hspace{-0.7cm} = N_{\sigma} - \sum_{\bm k} \Big\{
\Theta(\mu-\varepsilon_{\sigma;1}) +
\Theta(\mu-\varepsilon_{\sigma;2})
-\Theta\big(\mu-(\alpha_{\sigma;1}\varepsilon_{\sigma;2} +
\alpha_{\sigma;2}\varepsilon_{\sigma;1})\big)\Big\}, \label{e392}
\end{eqnarray}
where (see Eqs.~(\ref{e1}) and (\ref{e2}))
\begin{equation}
N_{\sigma} = \sum_{\bm k} \alpha_{\sigma;1}\,
\Theta(\mu-\varepsilon_{\sigma;1}) + \sum_{\bm k}
\alpha_{\sigma;2}\, \Theta(\mu-\varepsilon_{\sigma;2}). \label{e393}
\end{equation}
It is not difficult to convince oneself that the expression on the RHS of Eq.~(\ref{e392}) is in general non-vanishing, in violation of the Luttinger-Ward identity, Eq.~(\ref{e44}).

To gain insight, we consider the strong-coupling region in which one has\footnote{See Eq.~(3.12) in Ref.~\citen{BE95} and, with reference to Eq.~(2.16) in Ref.~\protect\citen{BE95}, note that the $n$ in the latter reference is identical to the $n_{\sigma}$ in our considerations.}
\begin{equation}
\alpha_{\sigma;1} \sim 1-n_{\sigma},\;\;\;\;\; \alpha_{\sigma;2}
\sim n_{\sigma}. \label{e394}
\end{equation}
Further, in this region the energy band $\varepsilon_{\sigma;1}$ is (partially) occupied and $\varepsilon_{\sigma;2}$ is empty (see Fig.~1 in Ref.~\citen{BE95}). Consequently, close to half-filling, corresponding to $n_{\sigma} \approx 1/2$, one has $\alpha_{\sigma;1} \approx \alpha_{\sigma;2} \approx 1/2$. Thus with $\varepsilon_{\sigma;2}$ sufficiently large, in the strong-coupling regime the term corresponding to $\Theta\big(\mu-(\alpha_{\sigma;1}\varepsilon_{\sigma;2} + \alpha_{\sigma;2}\varepsilon_{\sigma;1})\big)$ on the RHS of Eq.~(\ref{e392}) does not contribute. In this regime one thus has
\begin{equation}
N_{\sigma} \sim (1-n_{\sigma}) \sum_{\bm k}
\Theta(\mu-\varepsilon_{\sigma;1}), \label{e395}
\end{equation}
\begin{equation}
\sum_{\bm k} \int_{\mathscr{C}(\mu)} \frac{{\rm d}z}{2\pi i}\;
\t{G}_{\sigma}({\bm k};z) \frac{\partial}{\partial z}
\t{\Sigma}_{\sigma}({\bm k};z) \sim \frac{-n_{\sigma}}{1-n_{\sigma}}\,
N_{\sigma}. \label{e396}
\end{equation}
This result is in conformity with the observation by Beenen and Edwards \cite{BE95} concerning the breakdown of the Luttinger theorem in the case at hand. The fact that the RHS of Eq.~(\ref{e396}) is \emph{negative}, is in full agreement with the observations in Ref.~\citen{BE95} that the area of the occupied region of the $\mathrm{1BZ}$ is \emph{in excess of} the area corresponding to $N_{\sigma}$ points, that is the area of the non-interacting Fermi sea (see Fig.~3 in Ref.~\citen{BE95}). In this connection, note that, according to Eq.~(\ref{e34}), at $T=0$ one has\footnote{We suppress the bars on $\b{N}_{\sigma}$, etc., on account of the remarks concerning $T=0$ in Sec.~\protect\ref{s3}.} $N_{\sigma} = N_{\sigma}^{(1)} + N_{\sigma}^{(2)}$, where $N_{\sigma}^{(1)}$ is the number of ${\bm k}$ points inside the Fermi sea, the Luttinger number $N_{\Sc l;\sigma}$, and $N_{\sigma}^{(2)}$ the contribution arising from the Luttinger-Ward integral; following Eq.~(\ref{e396}) for the case at hand one has
\begin{equation}
N_{\sigma}^{(1)} \equiv N_{\Sc l;\sigma} \sim N_{\sigma} + \frac{n_{\sigma}}{1-n_{\sigma}}\,
N_{\sigma}. \label{e397}
\end{equation}
Note that for $n_{\sigma} \to \frac{1}{2}$ one has $N_{\Sc l;\sigma} \to 2 N_{\sigma}$. These results should be compared with the observations made by Gr\"ober, Eder and Hanke \cite{GEH00} discussed in Sec.~\ref{ss64}.

\subsubsection{Summary and remarks}
\label{ss65s4}

We have shown that the apparent breakdown of the Luttinger theorem as observed by Beenen and Edwards \cite{BE95} is a direct consequence the underlying Green function (the two-pole approximation due to Roth) not possessing the essential properties that are directly relevant to the validity of the Luttinger-Ward identity. These properties are encoded by the functional form of the exact self-energy $\t{\Sigma}_{\sigma}({\bm k};z)$ in its dependence on the exact Green functions $\{\t{G}_{\sigma'}({\bm k};z)\}$; diagrammatically, $\t{\Sigma}_{\sigma}({\bm k};z)$ consists of contributions arising from skeleton diagrams expressed in terms of $\{\t{G}_{\sigma'}({\bm k};z)\}$ and the bare two-body potential. We have shown that such functional relationship cannot be appropriately represented within a formalism that \emph{solely} reproduces a proper subset of all terms in the asymptotic series of $\t{G}_{\sigma}({\bm k};z)$ corresponding to $\vert z\vert \to\infty$; the two-pole approximation of Roth for $\t{G}_{\sigma}({\bm k};z)$ is capable of reproducing the four leading terms in this asymptotic series.

Since the coefficients $\{G_{\sigma;\infty_j}({\bm k};z)\,\|\, j \in \mathds{N}\}$ of the asymptotic series expansion of $\t{G}_{\sigma}({\bm k};z)$ corresponding to $\vert z\vert\to\infty$ coincide with the frequency moments of the single-particle spectral function $A_{\sigma}({\bm k};\varepsilon)$, Eq.~(\ref{ec67}), it follows that for a ${\bm k}$, say ${\bm k}={\bm k}_0$, at which $A_{\sigma}({\bm k};\varepsilon)$ has a single prominent peak at $\varepsilon =\varepsilon_0$, an approximation such as that by Roth for $\t{G}_{\sigma}({\bm k};z)$ should be in general accurate in determining $\varepsilon_0$. On the other hand, as we have shown in our examination of the work by Gr\"ober, Eder and Hanke \cite{GEH00} in Sec.~\ref{ss64}, observation of a prominent peak in $A_{\sigma}({\bm k}_0;\varepsilon)$ at $\varepsilon=\varepsilon_0 <\mu$ ($\varepsilon_0 >\mu$) does not imply ${\bm k}_0$ to be inside (outside) the underlying Fermi sea (or Luttinger sea), establishing that for in particular strongly-correlated metallic GSs a mere knowledge of the peak positions of $A_{\sigma}({\bm k};\varepsilon)$ in the $({\bm k},\varepsilon)$ space does not suffice to establish validity or failure of the Luttinger theorem. \emph{This statement applies irrespective of whether the underlying $A_{\sigma}({\bm k};\varepsilon)$ is based on theoretical calculations or experimental observations.}

In the light of the above considerations, one observes that the overestimation of the size of the Fermi sea by the two-pole approximation of Roth (see Eq.~(\ref{e397})) and a similar result deduced by Gr\"ober, Eder and Hanke \cite{GEH00}, on the basis of tracing the peak positions of $A_{\sigma}({\bm k};\varepsilon)$ in the $({\bm k},\varepsilon)$ space, are very closely related. We recall that the $A_{\sigma}({\bm k};\varepsilon)$ employed by Gr\"ober, Eder and Hanke \cite{GEH00} is calculated with the aid of the quantum Monte-Carlo technique on a finite lattice and that Beenen and Edwards \cite{BE95} emphasised very close agreement between the dispersion of the peak positions in the $A_{\sigma}({\bm k};\varepsilon)$ corresponding to the two-pole approximation of $\t{G}_{\sigma}({\bm k};z)$ and that deduced by Bulut \emph{et al.} \cite{BSW94} on the basis of the peak positions of the $A_{\sigma}({\bm k};\varepsilon)$ calculated by means of quantum Monte-Carlo simulations in conjunction with the maximum-entropy analytic continuation technique.

Lastly, using a self-consistent projection-operator method \cite{KF04}, Kakehashi and Fulde \cite{KF05} also observed violation of the Luttinger theorem in their calculations corresponding to a Hubbard Hamiltonian on a two-dimensional square lattice. Owing to the complexity of the analytic expressions underlying these calculations (which are strongly numerical in character), it is not possible to specify in a short space such as this the specific reason for this observation. However, the projection-operator method being in essence a moments-expansion formalism (which is closely akin to the continued-fraction expansion of positive functions), we believe that the apparent breakdown of the Luttinger theorem as observed by Kakehashi and Fulde \cite{KF05} is due to exactly the same reason as described in this section in regard to the calculations by Beenen and Edwards \cite{BE95}.

\section{Summary and concluding remarks}
\label{s7}

As regards metallic GSs,\footnote{Luttinger and Ward explicitly considered \emph{metallic} GSs in Ref.~\citen{LW60}.} we have shown that the Luttinger theorem is unequivocally valid under the conditions specified by Luttinger and Ward \cite{LW60}. These conditions consist of $(a)$ uniformity of the underlying $N$-particle GSs and $(b)$ existence of the contributions $\{ \t{\Sigma}_{\sigma}^{(\nu)}({\bm k};z)\,\|\, \nu\in \mathds{N}\}$ of skeleton self-energy diagrams to $\t{\Sigma}_{\sigma}({\bm k};z)$, expressed in terms of the \emph{interacting} single-particle Green functions $\{\t{G}_{\sigma'}({\bm k};z)\,\|\, \sigma'\}$ and the bare two-body interaction potential. Condition $(b)$ is guaranteed to hold when $(b1)$ the two-body interaction potential is short-range, and $(b2)$ either the underlying ${\bm k}$ space is bounded or the problem involves a natural ultraviolet cut-off. Condition $(b1)$ circumvents the possibility of the underlying wave-vector integrals being infrared divergent, and condition $(b2)$ the possibility of these being ultraviolet divergent. Since the $N$-particle GSs of interest are uniform, barring the uniform GS of the electron-gas model, these GSs must necessarily correspond to models defined on discrete lattices. The inverse of the shortest distance between lattice points in any of these lattices providing a finite ultraviolet cut-off,\footnote{For Bravais lattices, the relevant ${\bm k}$ space consists of the (bounded) $\mathrm{1BZ}$ of the corresponding reciprocal space. The ultraviolet cut-off for these lattices is therefore sharp.} it follows that, leaving aside the uniform electron-gas model, the existence of $\t{\Sigma}_{\sigma}^{(\nu)}({\bm k};z)$ for any finite value of $\nu$ is guaranteed by the above-mentioned conditions $(a)$ and $(b1)$. In this connection, as has been indicated by Luttinger \cite{JML61} (see Sec.~\ref{ss53s1}), on account of a specific characteristic of \emph{skeleton} self-energy diagrams,\footnote{That is, that they do not contain self-energy sub-diagrams capable of being removed through cutting two lines representing Green functions.} the integrals in terms of which the contributions to $\t{\Sigma}_{\sigma}^{(\nu)}({\bm k};z)$ are expressed, are well-defined; explicitly, the integrands of these integrals are free from ``repeated denominators'' \cite{JML61}.

As regards insulating GSs, the (generalised) Luttinger theorem is equally unconditionally valid under the conditions specified above, so long as the chemical potential $\mu$ is identified with the zero-temperature limit of the chemical potential $\mu_{\beta} \equiv \mu(\beta,N,V)$ satisfying the equation of state corresponding to the interacting grand-canonical ensemble whose mean value of particles $\b{N}$ is equal to $N$. The possibility of breakdown of the Luttinger theorem for a $\mu$, $\mu \in (\mu_{N}^-,\mu_{N}^+)$, deviating from $\mu_{\infty}$, was first detected by Rosch \cite{AR06}. In this paper we have shown that this breakdown is not related to a possible shortcoming in the Luttinger theorem, but signals a false limit arising from evaluating the zero-temperature limit without identifying $\mu$ with $\mu_{\infty}$, or $\mu_{\beta}$. It is a well-known mathematical fact that not all \emph{repeated} limits of a multi-variable function need be equal (\S\S~302-306 in Ref.~\citen{EWH27}). We have presented both physical (Sections \ref{s1} and \ref{ss23}) and mathematical (Sections \ref{ss61s2} and \ref{ss61s4}) reasons establishing the appropriate repeated limit as consisting of $\mu\to \mu_{\infty}$ followed by $\beta\to\infty$; for the specific cases that we have explicitly considered in Sections \ref{ss61s2} and \ref{ss61s4}, the simple limit $\beta\to\infty$ subsequent to identifying $\mu$ with $\mu_{\beta}$ has proved equally appropriate. By adopting the repeated limit $\lim_{\beta\to\infty} \lim_{\mu\to\mu_{\infty}}$ in the context of the Luttinger theorem, one treats insulating GSs on exactly the same footing as metallic GSs.

We have in some detail considered the problem regarding convergence of the series $\sum_{\nu=1}^{\infty} \t{\Sigma}_{\sigma}^{(\nu)}({\bm k};z)$ and shown that this series is not only convergent for all $({\bm k},z,\sigma)$ for which $\t{\Sigma}_{\sigma}({\bm k};z)$ is bounded, but that it is \emph{uniformly} convergent for almost all $({\bm k},z,\sigma)$. Two aspects are central to these properties. Firstly, that the elements of the sequence $\{ \t{\Sigma}_{\sigma}^{(\nu)}({\bm k};z)\}$ are not arbitrary, but are functionals of the exact $\{\t{\Sigma}_{\sigma'}({\bm k};z)\,\|\, \sigma'\}$. In consequence of this, the partial sums of these elements cannot diverge for those $({\bm k},z,\sigma)$ at which $\t{\Sigma}_{\sigma}({\bm k};z)$ is bounded; at worst, the sequence $\sum_{\nu=1}^{\ell} \t{\Sigma}_{\sigma}^{(\nu)}({\bm k};z)$ can \emph{oscillate} (Sec.~\ref{ss53s2}) between two \emph{finite} limits (Sec.~\ref{ss53s3}) for increasing values of $\ell$. Secondly, we have ruled out the latter possibility by the fact that $\mathrm{Im}[\t{\Sigma}_{\sigma}^{(\nu)}({\bm k};\varepsilon -i0^+)]
\equiv -\mathrm{Im}[\t{\Sigma}_{\sigma}^{(\nu)}({\bm k};\varepsilon -i0^+)] \ge 0$, $\forall\varepsilon \in \mathds{R}$ and $\forall\nu$; violation of the latter inequality is tantamount to an instability of the underlying GS (Sec.~\ref{ss53s4}). On its own, the latter inequality implies that the sequence $\sum_{\nu=1}^{\ell} \mathrm{Im}[\t{\Sigma}_{\sigma}^{(\nu)}({\bm k};\varepsilon -i0^+)]$, $\forall\varepsilon \in \mathds{R}$, can either converge or diverge as $\ell\to\infty$, but cannot oscillate. Divergence of this sequence is ruled out at those $({\bm k},\varepsilon,\sigma)$ for which $\t{\Sigma}_{\sigma}({\bm k};\varepsilon \pm i0^+)$ is bounded. To appreciate this aspect, one should recall the requirement of \emph{self-consistency}, whereby the elements of the sequence $\{ \t{\Sigma}_{\sigma}^{(\nu)}({\bm k};z) \,\|\, \nu\}$ are functionals of the same functions as resulting from $\sum_{\nu=1}^{\infty} \t{\Sigma}_{\sigma}^{(\nu)}({\bm k};z)$, $\forall\sigma$.

A very simple example should illustrate the principles discussed above. Consider the sequence $\{1,x,x^2,\dots\}$ and the sequence $\{ S_{\ell}(x)\}$ of the partial sums $S_{\ell} = \sum_{\nu=0}^{\ell} x^{\nu}$, $\ell \in \mathds{Z}^*$, for which one has
\begin{equation}
S_{\ell}(x) = \frac{x^{\ell+1}-1}{x-1}. \label{e398}
\end{equation}
As is evident, for $\vert x\vert<1$ one has $\lim_{\ell\to\infty} S_{\ell}(x) = 1/(1-x)$. For $\vert x\vert>1$, the sequence $S_{\ell}(x)$ has no limit for $\ell\to\infty$. For transparency, let us consider $x\in \mathds{R}$. In this case, $S_{\ell}(x)$ diverges \emph{monotonically} for $\ell\to\infty$ when $x>1$, and it diverges in an \emph{oscillating} fashion when $x<-1$; for $x<-1$ and $\ell\to\infty$, $S_{\ell}(x)$ approaches $+\infty$ for \emph{even} values of $\ell$, and $-\infty$ for \emph{odd} values of $\ell$.

With reference to the aspect of self-consistency referred to above, let us now consider the equation
\begin{equation}
x = S_{\ell}(x), \label{e399}
\end{equation}
of which we seek a real solution. One readily verifies that for \emph{even} values of $\ell$ (excluding $\ell=0$), Eq.~(\ref{e399}) has no real solution, however for \emph{odd} values of $\ell$ this equation has a real solution, $X_{\ell}$, strictly less than $-1$, that is outside the region of convergence of $S_{\ell}(x)$ for $\ell\to\infty$; one trivially verifies that\footnote{Note that the equation $x = 1/(1-x)$ has no real solution, even through $\{S_{\ell}(x)\}$ is conditionally convergent (\S~2.32 in Ref.~\protect\citen{WW62}) at $x=-1$. Note further that $1/(1-x)$ is equal to $1/2$ at $x=-1$.}
\begin{equation}
X_{2\ell+1} \sim -1 - \frac{\xi}{2\ell}\;\;\;\mbox{\rm for}\;\;\;\ell \to\infty\;\;\; (\ell\in \mathds{N}), \label{e400}
\end{equation}
where $\xi \approx 1$; the exact value of $\xi$ is to be obtained through solving a transcendental equation of which $\xi=1$ is the leading-order solution, $\xi=\sqrt{5}-1$ the next-to-leading-order solution, etc. This trivial example clearly shows that even though $S_{\ell}(x)$ has no limit for $\vert x\vert >1$ as $\ell\to\infty$, nonetheless the `self-consistency condition' in Eq.~(\ref{e399}) yields a finite limit, $X_{2\ell+1}$, for all $\ell\in \mathds{N}$, in the very region where $S_{2\ell+1}(x)$ has no limit for $\ell\to\infty$. In this light, the absurdity of the conception of $\sum_{\nu=1}^{\ell} \t{\Sigma}_{\sigma}^{(\nu)}({\bm k};z)$ diverging for $\ell\to\infty$, at a given $({\bm k},z,\sigma)$ for which $\t{\Sigma}_{\sigma}({\bm k};z)$ is bounded, should be apparent.

On the basis of the uniformity of convergence for almost all $({\bm k},z)$ of the sequence $\sum_{\nu=1}^{\ell} \t{\Sigma}_{\sigma}^{(\nu)}({\bm k};z)$ as $\ell\to\infty$, and on account of the analyticity of all relevant functions everywhere on the complex $z$ plane away from the real axis of this plane, we have shown that the proof by Luttinger and Ward \cite{LW60} of the Luttinger-Ward identity is mathematically fully justified. This, together with the specific behaviours for $z\to\mu$ of the $\mathrm{Re}[\t{\Sigma}_{\sigma}({\bm k};z)]$ and $\mathrm{Im}[\t{\Sigma}_{\sigma}({\bm k};z)]$ pertaining to stable GSs (Sec.~\ref{ss43}),  have led us to the above-mentioned conclusions concerning the validity of the Luttinger theorem for metallic and insulating GSs.

In view of the above observations, it should not come as a surprise that none of the reported failures of the Luttinger theorem that we have explicitly considered in this paper (Sec.~\ref{s6}) has implied a shortcoming in this theorem. For each and all of these we have been able to identify the reason or reasons underlying the apparent failure of the Luttinger theorem and all have proved to be unrelated to this theorem. The observation by Rosch \cite{AR06} concerning breakdown of the Luttinger theorem in the case of a Mott-insulating GS (Sec.~\ref{ss61}), is somewhat exceptional in that from the perspective of the conventional wisdom that this theorem would apply for \emph{all} $\mu \in (\mu_{N}^-,\mu_{N}^+)$, it indicates a genuine failure of the Luttinger theorem. We have shown that this failure is not related to an inherent shortcoming of the Luttinger theorem, but rather signals the fact that the zero-temperature limiting process underlying the Luttinger theorem may produce a spurious zero-temperature limit in the cases where this process is carried out prior to identifying $\mu$ with $\mu_{\infty}$, or $\mu_{\beta}$.

Finally, we point out that some of the observations that we have made in
Sections \ref{ss64} and \ref{ss65} of this paper, are of experimental relevance. Similarly, the details presented in appendix \ref{sf} can be used in analyzing the temperature dependence of the experimentally determined single-particle spectral functions at low temperatures; it is our understanding that it is common practice to ascribe the full  temperature dependence of the measured single-particle spectral functions (i.e. the photoemission part of these) as arising from the temperature dependence of the Fermi function $1/(\mathrm{e}^{\beta (\varepsilon-\mu)}+1)$ (see, e.g., Refs.~\citen{RDC95,DHS03}), a procedure that is strictly valid only for systems of non-interacting particles (see, for instance, Eqs.~(5.53)-(5.55) in Ref.~\citen{NO98}).

\section*{Dedication}
\label{dedic}

I dedicate this work to the memories of Joaquin Mazdak Luttinger (1923-1997) and John Clive Ward (1924-2004).

%
\begin{appendix}

\section{Lists of acronyms and some symbols employed in this paper}
\label{sa}

\begin{wraptable}{c}{\halftext}
\caption{Acronyms } \vspace{2pt}
\label{ta1}
\begin{center}
\begin{tabular}{lcl}
\hline\hline
GS  & {} & Ground state \\
LHS & {} & Left-hand side\\
RHS & {} & Right-hand side\\
1BZ & {} & The first Brillouin zone\\
\hline
\end{tabular}
\end{center}
\end{wraptable}

\vspace{0.6cm}

\begin{wraptable}{c}{\halftext}
\caption{Some symbols} \vspace{-10pt}
\label{ta2}
\begin{center}
\begin{tabular}{lcl}
\hline\hline
$\mathds{C}\;$  & {} & Set of complex numbers\\
$\mathds{N}\;$  & {} & Set of positive integers; the same as $\mathds{Z}^+$\\
$\mathds{R}\;$  & {} & Set of real numbers \\
$\mathds{Z}\;$  & {} & Set of all integers, negative, zero and positive\\
$\mathds{Z}^*\;$  & {} & Set of non-negative integers\\
$\varnothing\;$ & {} & Empty set\\
$o$, $O\;$ & {} & Order symbols (see \S~5 in Ref.~\protect\citen{EWH26})\\
\hline
\end{tabular}
\end{center}
\end{wraptable}

\section{Some theoretical details}
\label{sc}

Here we present some theoretical details to which we frequently refer throughout this paper. The bulk of the results in this appendix apply to \emph{uniform} GSs of both finite and macroscopic systems; by necessity, these finite systems must be defined on finite lattices without boundary. Since, however, for finite systems the supports of $\mathrm{Im}[\t{G}_{\sigma}({\bm k};\varepsilon\pm i0^+)]$ and $\mathrm{Im}[\t{\Sigma}_{\sigma}({\bm k};\varepsilon\pm i0^+)]$ along the real $\varepsilon$ axis are not necessarily dense in any closed subset of $\mathds{R}$, our analyses concerning in particular the asymptotic behaviours of these two functions for $\vert\varepsilon\vert\to\infty$ have in principle only bearing on macroscopic systems. For the same reason, when referring to branch-cut discontinuities of $\t{G}_{\sigma}({\bm k};z)$ and $\t{\Sigma}_{\sigma}({\bm k};z)$ along the real $\varepsilon$ axis, we have in principle exclusively macroscopic systems in mind.

\subsection{Definitions}
\label{ssc1}

The central expression in this appendix is the Lehmann
representation \cite{JML61,FW03,AGD75,BF03} for $\t{G}_{\sigma}({\bm k};z)$:
\begin{eqnarray}
\t{G}_{\sigma}({\bm k};z) &=& \hbar \sum_{s} \frac{\vert{\sf
f}_{s;\sigma}^-({\bm k})\vert^2}{z - \varepsilon_{s;\sigma}^-} +
\hbar \sum_{s} \frac{\vert{\sf f}_{s;\sigma}^+({\bm k})\vert^2}{z -
\varepsilon_{s;\sigma}^+} \nonumber \\
&\equiv& \t{G}_{\sigma}^{-}({\bm k};z) + \t{G}_{\sigma}^{+}({\bm
k};z), \label{ec1}
\end{eqnarray}
where $z \in \mathds{C}$, and
\begin{equation}
{\sf f}_{s;\sigma}^{\mp}({\bm k}) {:=} \left\{ \begin{array}{l}
\langle \Psi_{N_{\sigma}-1,N_{\b\sigma};s}\vert \h{c}_{{\bm
k};\sigma}
\vert\Psi_{N;0}\rangle,\\ \\
\langle \Psi_{N;0}\vert \h{c}_{{\bm k};\sigma}
\vert\Psi_{N_{\sigma}+1,N_{\b\sigma};s}\rangle, \end{array} \right.
\label{ec2}
\end{equation}
\begin{equation}
\varepsilon_{s;\sigma}^{\mp} {:=} \left\{ \begin{array}{l} E_{N;0} -
E_{N_{\sigma}-1,N_{\b\sigma};s},\\ \\
E_{N_{\sigma}+1,N_{\b\sigma};s} - E_{N;0}. \end{array} \right.
\label{ec3}
\end{equation}
Above $N = N_{\sigma} + N_{\b\sigma}$ denotes the total number of
particles\footnote{If $\sigma =\uparrow$, then $\b\sigma =\downarrow$, and \emph{vice versa}.} and $\vert\Psi_{M_{\sigma},M_{\b\sigma};s}\rangle$ an
$M$-particle normalised eigenstate of $\wh{H}$, corresponding to eigenvalue $E_{M_{\sigma},M_{\b\sigma};s}$. The
state $\vert\Psi_{M_{\sigma},M_{\b\sigma};s}\rangle$ is fully
specified in terms of $\{ M_{\sigma},M_{\b\sigma}\}$, where
$M_{\sigma}+M_{\b\sigma} = M$, and the set of quantum numbers
denoted by the compound index $s$. In the present context, where we
consider uniform GSs, $s$ may be expressed as
$({\bm\kappa},\varpi)$, where ${\bm\kappa}$ is wavevector, located
in the same space as ${\bm k}$, and $\varpi$, which is also a
compound index, is the so-called `parameter of degeneracy'
\cite{KP58}. Further, $\vert\Psi_{N;0}\rangle$ denotes the
normalised $N$-particle GS of $\wh{H}$ and $E_{N;0}$ the
corresponding eigenenergy.

By the assumed stability of the underlying GS (see the following paragraph), one has
\begin{equation}
\mu_{N;\sigma}^- <\mu_{N;\sigma}^+,\;\; \forall\sigma, \label{ec4}
\end{equation}
where
\begin{equation}
\mu_{N;\sigma}^- \equiv E_{N;0}-E_{N_{\sigma}-1,N_{\b\sigma};0},
\label{ec5}
\end{equation}
\begin{equation}
\mu_{N;\sigma}^{+} \equiv E_{N_{\sigma}+1,N_{\b\sigma};0}-E_{N;0}.
\label{ec6}
\end{equation}
In defining $\mu_{N;\sigma}^-$ we have assumed that $N_{\sigma} \ge 1$ so that $N_{\sigma}-1$ is not negative. The eigenenergies $E_{N_{\sigma}\pm 1,N_{\b\sigma};0}$ are the lowest energies $E_{M_{\sigma},M_{\b\sigma};s}$ corresponding to $M_{\sigma} =N_{\sigma}\pm 1$ and $M_{\b\sigma}=N_{\b\sigma}$; they are thus upper bounds to $E_{N\pm 1;0}$, the energies of the $(N\pm 1)$-particle GSs of $\wh{H}$: whereas $E_{N\pm 1;0}$ is the minimum of all $E_{N\pm 1;s}$, with $M_{\sigma}$ and $M_{\b\sigma}$ constrained only by $M_{\sigma}+M_{\b\sigma}=N\pm 1$, $E_{N_{\sigma}\pm 1,N_{\b\sigma};0}$ is the minimum of all $E_{N\pm 1;s}$ corresponding to $M_{\b\sigma} =N_{\b\sigma}$, whereby $M_{\sigma} = N\pm 1 - N_{\b\sigma}$,
where $N_{\b\sigma}$ is specific to the $N$-particle \emph{GS} of $\wh{H}$. Note however that by definition $E_{N_{\sigma},N_{\b\sigma};0} = E_{N;0}$ for $N = N_{\sigma} +N_{\b\sigma}$, to be contrasted with $E_{N\pm 1;0} \le E_{N_{\sigma}\pm 1,N_{\b\sigma};0}$ and $E_{N\pm 1;0} \le E_{N_{\sigma},N_{\b\sigma}\pm 1;0}$.

On account of a sum rule concerning the Lehmann amplitudes $\{ {\sf f}_{s;\sigma}^{\mp}({\bm k})\}$ that we shall present later in this appendix (see Eq.~(\ref{ec36}); cf. Eqs.~(\ref{ec34})), and of Eq.~(\ref{e1}), one deduces that at zero temperature the chemical potential $\mu$ corresponding to $N$ particles is bound to satisfy the following inequalities:
\begin{equation}
\mu_{N;\sigma}^- < \mu < \mu_{N;\sigma}^{+},\;\;\forall\sigma. \label{ec7}
\end{equation}
With reference to the considerations in Sections \ref{ss23} and \ref{ss41}, the above $\mu$ need not be equal to the zero-temperature limit of $\mu(\beta,N,V)$ in the case of insulating $N$-particle GSs. Later in this appendix we shall examine the range of variation of
\begin{equation}
\Delta_{\sigma} {:=} \mu_{N;\sigma}^+ - \mu_{N;\sigma}^-,
\;\;\forall\sigma, \label{ec8}
\end{equation}
both for metallic and insulating $N$-particle GSs. As we shall see, the strictness of the inequality $\mu_{N;\sigma}^- < \mu_{N;\sigma}^+$ for even metallic GSs and $N<\infty$ is of some mathematical significance. For instance, owing to this strict inequality, those in Eq.~(\ref{ec7}) are feasible, whereby even for metallic GSs one has $\mathrm{Im}[\Sigma_{\sigma}({\bm k};\mu)] \equiv 0$, $\forall {\bm k}$ (see also Sec.~\ref{ss21s2}).

Concerning the inequality in Eq.~(\ref{ec4}), we note that
\begin{equation}
\mu_{N;\sigma}^- \le \mu_{N;\sigma}^+ \Longleftrightarrow  E_{N;0}
\le \frac{1}{2} \big( E_{N_{\sigma}-1,N_{\b\sigma};0} +
E_{N_{\sigma}+1,N_{\b\sigma};0}\big), \label{ec9}
\end{equation}
where the right-most Jensen inequality \cite{J06} is a manifestation of the \emph{convexity} of the GS energy as a function of number of particles,\footnote{Although $N \in \mathds{Z}^*$, here we rely on the fact that in the \emph{algebraic expression} for $E_{N;0}$ the domain of variation of $N$ can be extended to $\mathds{R}$. The same applies for $N_{\sigma}$ and $N_{\b\sigma}$ in $E_{N_{\sigma},N_{\b\sigma};0}$. As we shall see below, owing to the extensive natures of $E_{N;0}$ and $E_{N_{\sigma},N_{\b\sigma};0}$, for macroscopic systems we shall not need to rely on these observations.} \footnote{The \emph{convexity} of $E_{N;0}$ as a function of $N$ implies that $E_{N;0} \le (E_{N-1;0}+E_{N+1;0})/2$. Owing to $E_{N\pm 1;0} \le E_{N_{\sigma}\pm 1,N_{\b\sigma};0}$ (see above), the right-most inequality in Eq.~(\ref{ec9}) is seen to be a less stringent condition imposed on $E_{N;0}$ for it to be a convex function of $N$.} which is required for the thermodynamic stability of the system against \emph{implosion} \cite{WT90}. One readily verifies that, unless the GS energy is a strictly linear function of number of particles (corresponding to an infinitely compressible state), the strict equality $\mu_{N;\sigma}^- =\mu_{N;\sigma}^+$ implies that the GS energy must be, as function of number of particles $M$, \emph{concave} in some neighbourhood of $M=N$ inside the interval $(N-1,N+1)$. This together with the assumption concerning stability of the $N$-particle GS $\vert\Psi_{N;0}\rangle$ against implosion, account for the strict inequality $\mu_{N;\sigma}^- <\mu_{N;\sigma}^+$ in Eq.~(\ref{ec4}). It follows that the GS energy $E_{M;0}$ is a \emph{strictly convex} function of $M$ for in particular $M\in (N-1,N+1)$. We note in passing that for non-interacting free fermions in $d$ dimensions, the energy density (i.e. energy per unit `volume') is equal to $A_d\, n^{1+2/d}$, where $n\equiv N/V$ and $A_d$ a well-specified positive constant (see Ref.~\citen{Note14}), with the inequality $1+2/d >1$ for $d<\infty$ manifesting the exclusion principle.

Following the inequalities in Eq.~(\ref{ec7}), one can define
\begin{equation}
{\sf f}_{s;\sigma}({\bm k}) {:=} \left\{ \begin{array}{ll} {\sf
f}_{s;\sigma}^{-}({\bm k}), & \varepsilon_{s;\sigma} < \mu, \\ \\
{\sf f}_{s;\sigma}^{+}({\bm k}), & \varepsilon_{s;\sigma} > \mu,
\end{array} \right. \label{ec10}
\end{equation}
where $\varepsilon_{s;\sigma}$ is specified according to the
self-referencing expression
\begin{equation}
\varepsilon_{s;\sigma} {:=} \left\{ \begin{array}{ll}
\varepsilon_{s;\sigma}^{-}, & \varepsilon_{s;\sigma} < \mu, \\ \\
\varepsilon_{s;\sigma}^{+}, & \varepsilon_{s;\sigma} > \mu.
\end{array} \right. \label{ec11}
\end{equation}
One can thus express Eq.~(\ref{ec1}) more concisely as
\begin{equation}
\t{G}_{\sigma}({\bm k};z) = \hbar \sum_{s} \frac{\vert{\sf
f}_{s;\sigma}({\bm k})\vert^2}{z - \varepsilon_{s;\sigma}}.
\label{ec12}
\end{equation}
The less compact expression in Eq.~(\ref{ec1}) has the merit that it reveals the two physically distinctive contributions, $\t{G}_{\sigma}^-({\bm k};z)$ and $\t{G}_{\sigma}^+({\bm k};z)$, of which $\t{G}_{\sigma}({\bm k};z)$ consists.

\subsubsection{Remarks}
\label{ssc1s1}

Since $E_{N_{\sigma},N_{\b\sigma};0}\equiv E_{N;0}$ is an extensive quantity (Ch. 2, \S~4 in Ref.~\citen{FW03}, Ch. 3, \S~1.6 in Ref.~\citen{CL00}), for macroscopic systems one has\footnote{Strictly, $\mathcal{E}_0(n_{\uparrow},n_{\downarrow})$ is the leading-order term in the asymptotic series expansion of $E_{N_{\uparrow},N_{\downarrow};0}/V$ for $V\to\infty$. Consequently, for the following analyses in terms of $\mathcal{E}_0(n_{\uparrow},n_{\downarrow})$ to be exact, it is required that the term subsequent to $\mathcal{E}_0(n_{\uparrow},n_{\downarrow})$ in the last-mentioned asymptotic series scale like $1/V^{\alpha}$, with $\alpha\ge 1$. Should this term scale like $\ln(V)/V^{\alpha}$, one must have $\alpha>1$.}
\begin{equation}
E_{N_{\uparrow},N_{\downarrow};0} \equiv V\, \mathcal{E}_0(n_{\uparrow},n_{\downarrow}),\;\;\; n_{\sigma} \equiv \frac{N_{\sigma}}{V},\; \forall\sigma, \label{ec13}
\end{equation}
where $V$ stands for the macroscopic volume of the system.\footnote{For lattice models, one replaces $V$ by the number of lattice cites $\mathcal{N}_{\Sc l}$ and subsequently deals with $E_{N_{\sigma},N_{\b{\sigma}};0} \equiv \mathcal{N}_{\Sc l}\, \mathcal{E}_0(n_{\uparrow},n_{\downarrow})$, where $n_{\sigma}\equiv N_{\sigma}/\mathcal{N}_{\Sc l}$.} Denoting the \emph{left} derivatives by $\partial/\partial_-$ and the \emph{right} derivatives by $\partial/\partial_+$, to corrections of order $1/V$ one has
\begin{equation}
\mu_{N;\sigma}^{\mp} = \frac{\partial}{\partial_{\mp} n_{\sigma}}\, \mathcal{E}_0(n_{\uparrow},n_{\downarrow}),\;\;\sigma \in \{\uparrow,\downarrow\}, \label{ec14}
\end{equation}
implying that
\begin{equation}
\mu_{N;\sigma}^- < \mu<\mu_{N;\sigma}^+ \iff \frac{\partial}{\partial_{-} n_{\sigma}}\, \mathcal{E}_0(n_{\uparrow},n_{\downarrow}) < \mu< \frac{\partial}{\partial_{+} n_{\sigma}}\, \mathcal{E}_0(n_{\uparrow},n_{\downarrow}),\;\forall\sigma. \label{ec15}
\end{equation}

A macroscopic $N$-particle GS is formally a \emph{metallic state} if up to an infinitesimally small correction of the order of $1/N$ (see later) the following applies:
\begin{equation}
\mu_{N;\sigma}^{-} = \mu_{N;\sigma'}^{+} =\mu,\;\;\;\;\mbox{\rm where}\;\;\; \sigma, \sigma' \in \{\uparrow,\downarrow\}. \label{ec16}
\end{equation}
As we shall see later, $\sigma$ and $\sigma'$ need not be equal. Following Eq.~(\ref{ec14}), for the case at hand one has, up to infinitesimal corrections,
\begin{equation}
\frac{\partial}{\partial_{-} n_{\sigma}}\, \mathcal{E}_0(n_{\uparrow},n_{\downarrow}) = \frac{\partial}{\partial_{+} n_{\sigma'}}\, \mathcal{E}_0(n_{\uparrow},n_{\downarrow}) = \mu. \label{ec17}
\end{equation}
For the specific $\sigma$ and $\sigma'$ in Eq.~(\ref{ec16}), by the above-mentioned convexity condition one has
\begin{equation}
\mu_{N;\sigma'}^{-} \le \mu_{N;\sigma}^{-} = \mu= \mu_{N;\sigma'}^{+} \le \mu_{N;\sigma}^{+}. \label{ec18}
\end{equation}
The inequalities in Eq.~(\ref{ec18}) have their roots in the fact that $E_{N_{\sigma}\pm 1,N_{\b\sigma};0}$, or $E_{N_{\sigma},N_{\b\sigma}\pm 1;0}$, are in general upper bounds to $E_{N\pm 1;0}$ (see above).

With
\begin{equation}
\mu_{N}^{\mp} \equiv \mp (E_{N\mp 1;0} - E_{N;0}), \label{ec19}
\end{equation}
\emph{it may or may not be true that}
\begin{equation}
\mu_{N}^- = \max(\mu_{N;\sigma}^-, \mu_{N;\b\sigma}^-),\;\;\;
\mu_{N}^+ = \min(\mu_{N;\sigma}^+, \mu_{N;\b\sigma}^+), \label{ec20}
\end{equation}
irrespective of whether the underlying $N$-particle GS is metallic or insulating. It is however always true that
\begin{equation}
\max(\mu_{N;\sigma}^-, \mu_{N;\b\sigma}^-) \le \mu_{N}^- < \mu_{N}^+ \le \min(\mu_{N;\sigma}^+, \mu_{N;\b\sigma}^+). \label{ec21}
\end{equation}
The condition $\mu_{N}^- <\mu_{N}^+$ follows from the strict convexity of $E_{N;0}$ as a function of $N$; for $N$-particle metallic GSs $\mu_{N}^{+} -\mu_{N}^{-}$ is infinitesimally small, of the order of $1/N$ (see later). We note that the convexity condition together with the extensive nature of the GS energy suffice to demonstrate the so-called \emph{subadditivity} condition \cite{WT90} which safeguards the system against \emph{explosion}.

Following the above statements, it is relevant to examine the magnitude of $\Delta_{\sigma}$, Eq.~(\ref{ec8}), for the cases of $N$-particle metallic GSs. In this connection, note that we have defined $N$-particle \emph{metallic} GSs as those for which the quantity (cf. Eq.~(\ref{ec16}))
\begin{equation}
\Delta_{\sigma',\sigma} {:=} \mu_{N;\sigma'}^+ - \mu_{N;\sigma}^-,\;\;\; \mbox{\rm for \emph{some}}\;\;\sigma, \sigma' \in \{\uparrow,\downarrow\}, \label{ec22}
\end{equation}
is microscopically small, scaling like $1/N$ for $N\to\infty$; according to this definition, the quantity $\Delta_{\sigma}$, Eq.~(\ref{ec8}), may in principle be finite for $N$-particle metallic GSs. The characterisation of metallic states in terms of $\Delta_{\sigma',\sigma}$ is somewhat different from the prevalent characterisation in the literature (see for instance Ref.~\citen{FW03}, p.~75, and Ref.~\citen{AGD75}, p.~56). We shall return to this aspect below.

On account of the extensive nature of $E_{N;0}$, one has (cf. Eq.~(\ref{ec13}) and consult, e.g., appendix B in Ref.~\citen{AM76}, \S~7.6 in Ref.~\citen{KH87} and \S~3.6 in Ref.~\citen{CL00})
\begin{equation}
E_{N;0} \equiv V\, \t{\mathcal{E}}_0(n),\;\;\;\mbox{\rm where}\;\;\; n \equiv n_{\uparrow} + n_{\downarrow}, \label{ec23}
\end{equation}
so that, to an error of the order of $1/V$, from Eq.~(\ref{ec19}) one obtains that
\begin{equation}
\mu_{N}^{\mp} \equiv \frac{\rm d}{{\rm d}_{\mp} n} \t{\mathcal{E}}_0(n). \label{ec24}
\end{equation}
For $N$-particle metallic GSs one has
\begin{equation}
\Delta {:=} \mu_{N}^+ - \mu_{N}^- = O(\frac{1}{N}), \label{ec25}
\end{equation}
which follows from the fact that $E_{N;0}$ is extensive and that for stable GSs the isothermal compressibility is \emph{strictly positive}. For illustration, for free fermions in $d$ dimensions (see above) one has $\Delta = (\frac{2d +4}{d^2} A_d\, n^{2/d})/N + O(1/N^3)$ (for $A_d$ see Ref.~\citen{Note14}). In view of the inequalities in Eq.~(\ref{ec21}), it is evident that the result in Eq.~(\ref{ec25}) is not capable of constraining the magnitude of $\Delta_{\sigma}$, Eq.~(\ref{ec8}).

Above we have indicated that characterisation of metallic states as being those for which $\Delta_{\sigma',\sigma} = O(1/N)$, differs from the predominant definition in the literature. One can readily verify that according to the latter definition, metallic states are those for which $\Delta_{\sigma} = O(1/N)$. It is evident that the two definitions coincide in the cases where $\mu_{N;\sigma}^{\pm} = \mu_{N;\b\sigma}^{\pm}$. A necessary condition (also a sufficient one, provided that ${\sf f}_{0;\sigma}^{\pm}({\bm k})\equiv {\sf f}_{0;\b\sigma}^{\pm}({\bm k})\not\equiv 0$, $\forall {\bm k}$) for this be the case, is that for these GSs $\t{G}_{\sigma}({\bm k};z) \equiv \t{G}_{\b\sigma}({\bm k};z)$, a property that in general only applies for time-reversal-invariant isotopic GSs (pp. 75 and 76 in Ref.~\citen{FW03} and p.~60 in Ref.~\citen{PN64}).\footnote{In Ref.~\protect\citen{AGD75}, p.~52, only time-reversal symmetry has been emphasised (``absence of ferromagnetism and of an external magnetic field''); this is owing to the fact that in Ref.~\protect\citen{AGD75} assumption with regard to symmetry under spatial (proper) continuous rotations, i.e. isotropy, is implicit. In Ref.~\protect\citen{FW03}, pp.~75 and 76, emphasis has been laid on isotropy \emph{and} the property of invariance under spatial reflections. The latter is implied by the time-reversal symmetry of the underlying GS.} The crucial property of isotropy is absent in the cases of lattice models.

The question arises whether the distinction between the two definitions is relevant. The answer to this question is in the affirmative: a GS for which $\Delta_{\sigma,\b\sigma} = O(1/N)$ \emph{is} experimentally metallic, even though both $\Delta_{\sigma}$ and $\Delta_{\b\sigma}$ may turn out to be finite. This possibility has been overlooked in the extant theoretical literature by the fact that in these the quantity $\Delta$, Eq.~(\ref{ec25}), is inadvertently identified with $\Delta_{\sigma}$, Eq.~(\ref{ec8}): since for metallic $N$-particle GSs one indeed has $\Delta = O(1/N)$ (see Eq.~(\ref{ec25})), the last-mention inappropriate identification of $\Delta$ with $\Delta_{\sigma}$ has necessarily implied that $\Delta_{\sigma} = O(1/N)$, $\forall\sigma$, in contradiction with the inequalities in Eq.~(\ref{ec21}), according to which $\Delta = O(1/N)$ \emph{cannot} imply $\Delta_{\sigma} = O(1/N)$. In contrast, as we shall explicitly demonstrate below, $\Delta = O(1/N)$ necessarily implies $\Delta_{\sigma',\sigma} = O(1/N)$ for \emph{some} $\sigma,\sigma' \in \{\uparrow,\downarrow\}$. Consequently, knowledge of $A_{\sigma}({\bm k};\varepsilon)$ for either $\sigma=\uparrow$ or $\sigma=\downarrow$ is in principle \emph{not} sufficient for establishing whether the underlying GS is metallic; in contrast, knowledge of $\sum_{\sigma} A_{\sigma}({\bm k};\varepsilon)$ suffices, this owing to the fact that $A_{\sigma}({\bm k};\varepsilon)\ge 0$, $\forall {\bm k}, \varepsilon, \sigma$, so that contributions of $A_{\uparrow}({\bm k};\varepsilon)$ and $A_{\downarrow}({\bm k};\varepsilon)$ in the latter sum cannot cancel.

In order to establish that for metallic GSs $\Delta_{\sigma',\sigma} = O(1/N)$ for \emph{some} $\sigma$ and $\sigma'$, we proceed as follows. With reference to our earlier statements following Eq.~(\ref{ec6}), there exist \emph{integers} $m_{\pm} \in \mathds{Z}$ and $m_{\pm}' \in \mathds{Z}$ for which one has
\begin{equation}
E_{N\pm 1;0} = E_{N_{\uparrow}+m_{\pm}, N_{\downarrow}+m_{\pm}';0},\;\;\; m_{\pm} + m_{\pm}' = \pm 1. \label{ec26}
\end{equation}
These expressions are exact so long as the true $(N\pm 1)$-particle GSs $\vert\Psi_{N\pm 1;0}\rangle$ of $\wh{H}$ are simultaneous eigenstates of $\wh{N}_{\sigma}$, $\forall\sigma$. In this connection, we remark that in this paper $\wh{H}$ is assumed to satisfy $[\wh{H},\wh{N}_{\sigma}]_- = 0$, $\forall\sigma$. Evidently, $m_{\pm}$ and $m_{\pm}'$ are bound by the requirements that $N_{\uparrow}+m_{\pm} \ge 0$ and $N_{\downarrow} + m_{\pm}' \ge 0$.

For macroscopic systems, from Eqs.~(\ref{ec13}), (\ref{ec23}) and (\ref{ec26}) one obtains that (to errors of order $1/V$)
\begin{equation}
\mu_{N}^{\mp} = \mp \big(m_{\mp}\, \mu_{N;\uparrow}^{\varsigma(m_{\mp})} + m_{\mp}'\, \mu_{N;\downarrow}^{\varsigma(m_{\mp}')}\big),\;\;\ m_{\mp} + m_{\mp}' = \mp 1, \label{ec27}
\end{equation}
where $\varsigma(x)$,
\begin{equation}
\varsigma(x) = \pm \;\;\;\mbox{\rm for}\;\;\; x\gtrless 0, \label{ec28}
\end{equation}
is the signature of $x$. In arriving at the expression in Eq.~(\ref{ec27}) we have assumed that $m_{\pm}/N = o(1)$ (for the $o-O$ notation see \S~5 in Ref.~\citen{EWH26}). We have further assumed, through our reliance on the expression in Eq.~(\ref{ec13}), that the function $\mathcal{E}_0(n_{\uparrow},n_{\downarrow})$, which is appropriate for the neighbourhoods of the $n_{\uparrow}$ and $n_{\downarrow}$ corresponding to the $N$-particle GS of $\wh{H}$, is \emph{the same} function appropriate for the neighbourhoods of the $n_{\uparrow}$ and $n_{\downarrow}$ corresponding to the $(N\pm 1)$-particle GSs of $\wh{H}$. The two assumptions are not entirely independent.

Since the expressions in Eq.~(\ref{ec27}) are exact up to, but not including, terms of the form $O(1/V)$, or $O(1/N)$, in view of Eq.~(\ref{ec25}) the $\Delta$ as calculated in terms of the
expressions for $\mu_{N}^{\mp}$ in Eq.~(\ref{ec27}) must be exactly equal to zero. For $N$-particle metallic GSs one thus arrives at the equation
\begin{equation}
m_+ \, \mu_{N;\uparrow}^{\varsigma(m_+)} + (1-m_+)\, \mu_{N;\downarrow}^{\varsigma(1-m_+)} + m_-\, \mu_{N;\uparrow}^{\varsigma(m_-)} - (1+m_-)\, \mu_{N;\downarrow}^{-\varsigma(1+m_-)} = 0, \label{ec29}
\end{equation}
where $m_+ \in \mathds{Z}$ and $m_- \in \mathds{Z}$ can in principle take arbitrary values (however conform the conditions $m_{\pm}/N = o(1)$). The expression in Eq.~(\ref{ec29}) is specific to metallic states only by virtue of the zero on its RHS, which follows from Eq.~(\ref{ec25}).

From Eq.~(\ref{ec29}) one immediately obtains the following results (cf. Eq.~(\ref{ec18}))
\begin{eqnarray}
\mbox{\rm (a)}\;\;\; m_+ =+1,\; m_-=-1 &\Longrightarrow& \mu_{N;\uparrow}^+ = \mu_{N;\uparrow}^-,\nonumber\\
\mbox{\rm (b)}\;\;\; m_+ =0 \phantom{+},\; m_-=-1 &\Longrightarrow& \mu_{N;\downarrow}^+ = \mu_{N;\uparrow}^-,\nonumber\\
\mbox{\rm (c)}\;\;\; m_+ =+1,\; m_-=0\phantom{+} &\Longrightarrow&  \mu_{N;\uparrow}^+ = \mu_{N;\downarrow}^-,\nonumber\\
\mbox{\rm (d)}\;\;\; m_+ =0\phantom{+},\; m_-=0\phantom{+} &\Longrightarrow& \mu_{N;\downarrow}^+ = \mu_{N;\downarrow}^-. \label{ec30}
\end{eqnarray}
One observes that for the combination of the pairs $(m_+,m_-)$ considered above, only the equality, to an error of order $1/N$, between two `partial chemical potentials' is fixed; the relationships between these with the other two partial chemical potentials is undetermined.

Following some rearrangement of terms, from Eq.~(\ref{ec29}) one further obtains that
\begin{eqnarray}
&&\hspace{-0.2cm} m_+\ge +2,\, m_- \le -2 \nonumber\\
&&\hspace{0.8cm}\Longrightarrow (m_+ -1) (\mu_{N;\uparrow}^+ -\mu_{N;\downarrow}^-) + (-1-m_-) (\mu_{N;\downarrow}^+ -\mu_{N;\uparrow}^-) + (\mu_{N;\uparrow}^+ -\mu_{N;\uparrow}^-) =0, \nonumber\\
&&\hspace{-0.2cm} m_+\le -2,\, m_- \le -2 \nonumber\\
&&\hspace{0.8cm}\Longrightarrow -(m_+ + m_-) (\mu_{N;\downarrow}^+ -\mu_{N;\uparrow}^-) =0, \nonumber\\
&&\hspace{-0.2cm} m_+\ge +2,\, m_- \ge +2 \nonumber\\
&&\hspace{0.8cm}\Longrightarrow (m_+ + m_-) (\mu_{N;\uparrow}^+ -\mu_{N;\downarrow}^-) =0, \nonumber\\
&&\hspace{-0.2cm} m_+\le -2,\, m_- \ge +2 \nonumber\\
&&\hspace{0.8cm}\Longrightarrow -m_+ (\mu_{N;\downarrow}^+ -\mu_{N;\uparrow}^-) + m_- (\mu_{N;\uparrow}^+ -\mu_{N;\downarrow}^-) + (\mu_{N;\downarrow}^+ -\mu_{N;\downarrow}^-) =0.
\label{ec31}
\end{eqnarray}
From the inequalities in Eq.~(\ref{ec18}) it is evident that the sums in Eq.~(\ref{ec31}) all consist of non-negative terms. One thus trivially arrives at the following results:
\begin{eqnarray}
\mbox{\rm (e)}\;\;\; m_+\ge +2,\, m_- \le -2 &\Longrightarrow& \mu_{N;\uparrow}^+ =\mu_{N;\uparrow}^- = \mu_{N;\downarrow}^+ = \mu_{N;\downarrow}^-, \nonumber\\
\mbox{\rm (f)}\;\;\; m_+\le -2,\, m_- \le -2 &\Longrightarrow& \mu_{N;\downarrow}^+ = \mu_{N;\uparrow}^-,  \nonumber\\
\mbox{\rm (g)}\;\;\; m_+\ge +2,\, m_- \ge +2 &\Longrightarrow& \mu_{N;\uparrow}^+ = \mu_{N;\downarrow}^-, \nonumber\\
\mbox{\rm (h)}\;\;\; m_+\le -2,\, m_- \ge +2 &\Longrightarrow& \mu_{N;\uparrow}^+ =\mu_{N;\uparrow}^- = \mu_{N;\downarrow}^+ = \mu_{N;\downarrow}^-.
\label{ec32}
\end{eqnarray}
Evidently, only for the cases (e) and (h) are all four `partial chemical potentials' $\mu_{N;\sigma}^{\mp}$, $\forall\sigma$, equal (up to errors of the order of $1/N$). We have thus demonstrated the general validity of the result in Eq.~(\ref{ec18}) for $N$-particle metallic GSs. To determine the appropriate values for $m_+$ and $m_-$, one needs to perform explicit calculations.

By the same reasoning as above, for \emph{insulating} $N$-particle GSs one has
\begin{equation}
\mu_{N;\sigma'}^{-} \le \mu_{N;\sigma}^{-} \le \mu_{N}^- <\mu < \mu_{N}^+ \le \mu_{N;\sigma'}^{+} \le \mu_{N;\sigma}^{+},\;\;\;\;\mbox{\rm where}\;\;\; \sigma, \sigma' \in \{\uparrow,\downarrow\}, \label{ec33}
\end{equation}
where $\Delta\equiv \mu_{N}^+ -\mu_{N}^-$ is finite. For insulating $N_0$-particle GSs, where the quantity $\Delta_{\sigma}$, Eq.~(\ref{ec8}), is \emph{finite} for $N=N_0 \equiv N_{0;\uparrow} + N_{0;\downarrow}$, the inequalities in Eq.~(\ref{ec33}) (see Eq.~(\ref{ec15})) imply that the function $\mathcal{E}_0(n_{\uparrow},n_{\downarrow})$ has cusps at $n_{\sigma} = n_{0;\sigma} \equiv N_{0;\sigma}/V$ for both $\sigma = \uparrow$ and $\sigma=\downarrow$. Similarly for $\t{\mathcal{E}}_0(n)$, which has a cusp at $n= n_0 \equiv N_0/V$.

\subsection{General results}
\label{ssc2}

Since $\varepsilon_{s;\sigma} \in \mathds{R}$ (see Eqs.~(\ref{ec3}) and (\ref{ec11})), from Eq.~(\ref{ec12}) one observes that $\t{G}_{\sigma}({\bm k};z)$ is analytic everywhere in the complex $z$ plane outside the real axis \cite{JML61}. Evidently, $\t{G}_{\sigma}^{-}({\bm k};z)$ and $\t{G}_{\sigma}^{+}({\bm k};z)$, Eq.~(\ref{ec1}), are in addition analytic along the real intervals $(\mu_{N;\sigma}^-,\infty]$ and $[-\infty,\mu_{N;\sigma}^+)$ respectively.

Owing to the analyticity of $\t{G}_{\sigma}({\bm k};z)$ outside the real axis of the $z$ plane, $\t{G}_{\sigma}({\bm k};\varepsilon \pm i\eta)$ are \emph{continuous} for $\eta>0$ and \emph{all} real values of $\varepsilon$. In this connection, divergence of $\t{G}_{\sigma}({\bm k};\varepsilon\pm i 0^+)$ at specific values $\varepsilon\in \mathds{R}$ reflects the fact that in $\t{G}_{\sigma}({\bm k};\varepsilon\pm i 0^+)$ the limit $\eta\downarrow 0$ has been effected;\footnote{The numerical value of $0^+$ is \emph{exactly} equal to zero; the $+$ in $0^+$ solely indicates that $\eta$ has approached $0$ from above, signified by means of the notation $\eta\downarrow 0$.} for any $\eta>0$, no matter how small $\eta$ may be, the functions $\t{G}_{\sigma}({\bm k};\varepsilon \pm i\eta)$ are bounded for all $\varepsilon \in \mathds{R}$.

For $\mu$ satisfying the inequalities in Eq.~(\ref{ec7}), one obtains that
\begin{equation}
\frac{1}{\hbar} \int_{\mathscr{C}_+(\mu)} \frac{{\rm d}z}{2\pi i}\;
\t{G}_{\sigma}({\bm k};z) =\sum_{s} \vert {\sf f}_{s;\sigma}^-({\bm
k})\vert^2, \label{ec34}
\end{equation}
where $\mathscr{C}_+(\mu)$ is defined in the text following Eq.~(\ref{e2}). Making use of the closure relation
\begin{equation}
\sum_{s} \vert\Psi_{M_{\sigma},M_{\b\sigma};s}\rangle
\langle\Psi_{M_{\sigma},M_{\b\sigma};s}\vert =
\h{I}_{M_{\sigma},M_{\b\sigma}}, \label{ec35}
\end{equation}
where $\h{I}_{M_{\sigma},M_{\b\sigma}}$ denotes the unit operator in the Hilbert space of the $(M_{\sigma}+M_{\b\sigma})$-particle eigenstates of $\wh{H}$, one readily deduces that
\begin{equation}
\sum_{s} \vert {\sf f}_{s;\sigma}^-({\bm k})\vert^2 =
\langle\Psi_{N;0}\vert \h{c}_{{\bm k};\sigma}^{\dag} \h{c}_{{\bm
k};\sigma} \vert\Psi_{N;0}\rangle \equiv {\sf n}_{\sigma}({\bm k}).
\label{ec36}
\end{equation}
Hereby is the validity of Eq.~(\ref{e2}) established. Note that since for $\wh{N}_{\sigma}$ one has
\begin{equation}
\wh{N}_{\sigma} = \sum_{\bm k} \h{c}_{{\bm k};\sigma}^{\dag}
\h{c}_{{\bm k};\sigma}, \label{ec37}
\end{equation}
Eq.~(\ref{e1}) directly follows from Eqs.~(\ref{ec34}) and (\ref{ec36}).

Similarly as above, one obtains that
\begin{equation}
\sum_{s} \vert {\sf f}_{s;\sigma}^+({\bm k})\vert^2 =
\langle\Psi_{N;0}\vert \h{c}_{{\bm k};\sigma} \h{c}_{{\bm
k};\sigma}^{\dag} \vert\Psi_{N;0}\rangle \equiv 1 - {\sf
n}_{\sigma}({\bm k}), \label{ec38}
\end{equation}
where for the last equality we have used the canonical anti-commutation relation $[\h{c}_{{\bm k};\sigma}, \h{c}_{{\bm
k};\sigma}^{\dag}]_{+} = 1$ and
$\langle\Psi_{N;0}\vert\Psi_{N;0}\rangle =1$. Combining
Eqs.~(\ref{ec36}) and (\ref{ec38}) one has (cf.
Eq.~(\ref{ec10}))
\begin{equation}
\sum_s \vert {\sf f}_{s;\sigma}({\bm k})\vert^2 = 1,\;\;\; \forall
{\bm k},\sigma. \label{ec39}
\end{equation}

With the single-particle spectral function $A_{\sigma}({\bm k};\varepsilon)$ defined according to
\begin{eqnarray}
A_{\sigma}({\bm k};\varepsilon) {:=} \pm\frac{1}{\pi}\,
\mathrm{Im}[\t{G}_{\sigma}({\bm k};\varepsilon\mp i 0^+)],
\label{ec40}
\end{eqnarray}
from the Lehmann representation in Eq.~(\ref{ec12}) one obtains that
\begin{equation}
A_{\sigma}({\bm k};\varepsilon) = \hbar \sum_{s} \vert {\sf
f}_{s;\sigma}({\bm k})\vert^2\,
\delta(\varepsilon-\varepsilon_{s;\sigma}) \ge 0. \label{ec41}
\end{equation}
Consequently,
\begin{equation}
A_{\sigma}({\bm k};\varepsilon) \equiv 0\;\;\;\mbox{\rm for}\;\;\;
\mu_{N;\sigma}^- < \varepsilon < \mu_{N;\sigma}^+,\;\;\forall\sigma. \label{ec42}
\end{equation}
From Eq.~(\ref{ec40}) it follows that those closed regions of the real $\varepsilon$ axis for which $A_{\sigma}({\bm k};\varepsilon) \not=0$ signify the branch cuts of $\t{G}_{\sigma}({\bm k};z)$ along the real $\varepsilon$ axis. Across these branch cuts the imaginary part of $\t{G}_{\sigma}({\bm k};z)$ undergoes a discontinuity; since $\t{G}({\bm k};z^*) \equiv \t{G}_{\sigma}^*({\bm k};z)$ for all $z$ satisfying $\mathrm{Im}(z)\not=0$ (cf. Eq.~(\ref{ec12})), it follows that, aside from those points for which the $G_{\sigma}({\bm k};\varepsilon)$ is unbounded, the real part of $\t{G}_{\sigma}({\bm k};\varepsilon +i\eta)$ is continuous at $\eta=0$ along the entire real $\varepsilon$ axis.\footnote{By extending the notion of continuity to unbounded functions (\S~219 in Ref.~\protect\citen{EWH27}), one can show that $\mathrm{Re}[\t{G}_{\sigma}({\bm k};\varepsilon + i\eta)]$, $\varepsilon\in \mathds{R}$, is continuous at $\eta=0$ for \emph{all} $\varepsilon\in\mathds{R}$. }

Since (cf. Eq.~(\ref{e5}))
\begin{equation}
A_{\sigma}({\bm k};\varepsilon) = \frac{\hbar}{\pi}
\frac{\hbar\left|\mathrm{Im}[\Sigma_{\sigma}({\bm
k};\varepsilon)]\right|}{\big(\varepsilon -\varepsilon_{\bm k}
-\hbar \mathrm{Re}[\Sigma_{\sigma}({\bm k};\varepsilon)]\big)^2\!
+\! \big(\hbar\mathrm{Im}[\Sigma_{\sigma}({\bm
k};\varepsilon)]\big)^2}, \label{ec43}
\end{equation}
where $\varepsilon_{\bm k}$ is the non-interacting single-particle
energy dispersion, from Eq.~(\ref{ec42}) it follows that
\begin{equation}
\mathrm{Im}[\Sigma_{\sigma}({\bm k};\varepsilon)] \equiv
0\;\;\;\mbox{\rm for}\;\;\; \mu_{N;\sigma}^- < \varepsilon <
\mu_{N;\sigma}^+,\;\;\forall {\bm k}. \label{ec44}
\end{equation}
Although this result may appear to be only relevant to insulating GSs, for which $\Delta_{\sigma}$, Eq.~(\ref{ec8}), has a finite non-vanishing value, it is in fact equally relevant to metallic GSs: irrespective of whether the underlying uniform GS is metallic or insulating, for $\mu$ satisfying the inequalities in Eq.~(\ref{ec7}) one has (cf. Eq.~(\ref{e8}))
\begin{equation}
\mathrm{Im}[\Sigma_{\sigma}({\bm k};\mu)] \equiv 0,\;\;\;
\forall{\bm k}. \label{ec45}
\end{equation}
This result implies that for both types of GSs one has
\begin{equation}
\mathrm{Im}[G_{\sigma}({\bm k};\mu)] \equiv 0,\;\;\; \forall {\bm
k}. \label{ec46}
\end{equation}
Conversely, Eq.~(\ref{ec46}) implies Eq.~(\ref{ec45}). In Sec.~\ref{ss21s2} we considered the result in Eq.~(\ref{ec45}), having in mind in particular metallic $N$-particle GSs for which $\Delta$, Eq.~(\ref{ec25}), is microscopically small. We now present a demonstration of the result in Eq.~(\ref{ec46}) for these states, focusing on the crucial role played by the \emph{strict} inequality $\mu_{N;\sigma}^- <\mu_{N;\sigma}^+$ whose significance may not be apparent in the case of metallic GSs.

We start from the following exact spectral representation (see
Eqs.~(\ref{ec12}) and (\ref{ec41}))
\begin{equation}
\t{G}_{\sigma}({\bm k};z) = \int_{-\infty}^{\infty} {\rm
d}\varepsilon'\; \frac{A_{\sigma}({\bm
k};\varepsilon')}{z-\varepsilon'}, \label{ec47}
\end{equation}
which on account of Eq.~(\ref{ec42}) can be equivalently written as
\begin{equation}
\t{G}_{\sigma}({\bm k};z) = \int_{-\infty}^{\mu_{N;\sigma}^-} {\rm
d}\varepsilon'\; \frac{A_{\sigma}({\bm
k};\varepsilon')}{z-\varepsilon'} + \int_{\mu_{N;\sigma}^+}^{\infty}
{\rm d}\varepsilon'\; \frac{A_{\sigma}({\bm
k};\varepsilon')}{z-\varepsilon'} \equiv \t{G}_{\sigma}^-({\bm k};z) + \t{G}_{\sigma}^+({\bm k};z). \label{ec48}
\end{equation}
Following Eq.~(\ref{ec7}), it is evident that $\mu = (\mu_{N;\sigma}^- + \mu_{N;\sigma}^+)/2$ qualifies as a chemical potential specific to the $N$-particle GS under consideration. On equating $\mu$ with $(\mu_{N;\sigma}^- + \mu_{N;\sigma}^+)/2$, for $z=\mu$ and metallic GSs, Eq.~(\ref{ec48}) can be expressed as
\begin{equation}
\t{G}_{\sigma}({\bm k};\mu) = \mathscr{P}\!\!\int_{-\infty}^{\infty}
{\rm d}\varepsilon'\; \frac{A_{\sigma}({\bm k};\varepsilon')}{\mu -
\varepsilon'}, \;\;\; \forall {\bm k},\label{ec49}
\end{equation}
where we have used the definition for the Cauchy principal value of integrals (\S~4.5 in Ref.~\citen{WW62}), presented in Eq.~(\ref{e108}), in which we have identified $\epsilon$ with $\Delta_{\sigma}/2$, Eq.~(\ref{ec8}); this quantity is indeed positive for any $N<\infty$ and is infinitesimally small as $N\to\infty$. Evidently, Eq.~(\ref{ec49}) applies equally to insulating GSs, in which case the principal-value integration can be replaced by an ordinary integration with no consequence, whether detrimental or otherwise.

The result in Eq.~(\ref{ec49}) is equivalent to that in Eq.~(\ref{ec46}). For clarity, on employing $1/(x\mp i 0^+) = \mathscr{P}(1/x) \pm i\pi\delta(x)$, $x\in \mathds{R}$, for $z=\varepsilon\mp i 0^+$, with $\varepsilon \in \mathds{R}$, from Eq.~(\ref{ec47}) one obtains two results. Of these, the first is identical to the defining expression for $A_{\sigma}({\bm k};\varepsilon)$, Eq.~(\ref{ec40}), and the second result is
\begin{equation}
\mathrm{Re}[G_{\sigma}({\bm k};\varepsilon)] =
\mathscr{P}\!\!\int_{-\infty}^{\infty} {\rm d}\varepsilon'\;
\frac{A_{\sigma}({\bm k};\varepsilon')}{\varepsilon -
\varepsilon'},\;\;\; \varepsilon \in \mathds{R}, \label{ec50}
\end{equation}
where on the basis of the continuity of
$\mathrm{Re}[\t{G}_{\sigma}({\bm k};\varepsilon \mp i \eta)]$ at
$\eta=0$ (discussed above) we have replaced
$\mathrm{Re}[\t{G}_{\sigma}({\bm k};\varepsilon \mp i 0^+)]$ by the
more compact, but equivalent, notation $\mathrm{Re}[G_{\sigma}({\bm
k};\varepsilon)]$. It is observed that for $\varepsilon=\mu$ the RHS
of Eq.~(\ref{ec50}) is identical to the RHS of Eq.~(\ref{ec49}) so that indeed
\begin{equation}
\t{G}({\bm k};\mu) \equiv \mathrm{Re}[G_{\sigma}{\bm
k};\mu)], \;\;\;\forall {\bm k}. \label{ec51}
\end{equation}

We point out that our above specific choice for $\mu$, namely $\mu = (\mu_{N;\sigma}^- + \mu_{N;\sigma}^+)/2$, required in order for the principal-value integral on the RHS of Eq.~(\ref{ec49}) to be \emph{in the sense of Cauchy} (\S~4.5 in Ref.~\citen{WW62}), is fully in conformity with the expression $1/(x\mp i 0^+) = \mathscr{P}(1/x) \pm i\pi \delta(x)$, $x\in \mathds{R}$, in which $\mathscr{P}(1/x)$ explicitly refers to the Cauchy principal value of the integrals involving $1/x$. This follows from the fact that the \emph{distribution} $\delta(x)$ is, as commonly understood, the limit of a sequence of functions which are symmetric with respect to $x=0$; this symmetry is reflected in the symmetry of the limiting process according to which the Cauchy principal value of integrals is defined.

The question arises whether in the case under consideration a choice for $\mu$ different form $\mu = (\mu_{N;\sigma}^- + \mu_{N;\sigma}^+)/2$, but satisfying the inequalities in Eq.~(\ref{ec7}), can alter the conclusion in Eq.~(\ref{ec51}). The answer to this question is in the negative, this on account of the property of analyticity of $\t{G}_{\sigma}({\bm k};z)$ for all $z$ away from the real axis, whereby the limit of $\t{G}_{\sigma}({\bm k};z)$, with $z = \mu + \eta\, \mathrm{e}^{\pm i \vartheta}$, for $\eta\downarrow 0$ is the same so long as $0 <\vartheta<\pi$ (see a similar and related discussion in Sec.~\ref{ss21s2}). Note that the exclusion of $\vartheta=0,\pi$ is consistent with the definition of the `physical' $G_{\sigma}({\bm k};\varepsilon)$, $\varepsilon\in \mathds{R}$, in terms of $\t{G}_{\sigma}({\bm k};z)$, $z\in \mathds{C}$ (see Eq.~(\ref{e3})).

We should point out that the interval $(\mu_{N;\sigma}^-, \mu_{N;\sigma}^+)$ is special in that $\mu_{N;\sigma}^-$ and $\mu_{N;\sigma}^+$ are \emph{end-point singularities} of respectively $\t{G}_{\sigma}^-({\bm k};z)$ and $\t{G}_{\sigma}^+({\bm k};z)$, Eq.~(\ref{ec48}); although the behaviours of these functions for respectively $z\to \mu_{N;\sigma}^-$ and $z\to \mu_{N;\sigma}^+$ are dependent on the value of $N$, nonetheless $\mu_{N;\sigma}^-$ and $\mu_{N;\sigma}^+$ remain end-point singularities for all $N$ as $N\to\infty$. It is owing to this fact that $\t{G}_{\sigma}({\bm k};z)$ is real-valued at $z=\mu$, no matter how large $N$ may be.

Owing to $\varepsilon_{s;\sigma}\in \mathds{R}$ (see
Eqs.~(\ref{ec3}) and (\ref{ec11})), the Lehmann representation in
Eq.~(\ref{ec12}) can be equivalently expressed as
\begin{equation}
\t{G}_{\sigma}({\bm k};z) = \hbar \sum_s (z^*
-\varepsilon_{s;\sigma})\, \left|\frac{{\sf f}_{s;\sigma}({\bm
k})}{z -\varepsilon_{s;\sigma}}\right|^2, \label{ec52}
\end{equation}
from which one observes that $\t{G}_{\sigma}({\bm k};z)$
\emph{cannot} be equal to zero for any $z$ satisfying
$\mathrm{Im}(z) \not=0$; in this connection, note that although ${\sf f}_{s;\sigma}({\bm k})$ may be vanishing at ${\bm k}={\bm k}_0$ for some $s$, on account of the sum rule in Eq.~(\ref{ec39}) it cannot be vanishing for all $s$. Consequently, $\t{G}_{\sigma}^{-1}({\bm k};z)$ cannot be singular away from the real axis. The same statement applies to $\t{G}_{\sigma;0}^{-1}({\bm k};z)$, a fact which is apparent from the explicit expression $\t{G}_{\sigma;0}^{-1}({\bm k};z) = \frac{1}{\hbar} (z -\varepsilon_{\bm k})$, where $\varepsilon_{\bm k} \in \mathds{R}$. One evidently has
\begin{equation}
\mathrm{Im}[\t{G}_{\sigma}({\bm k};z))] \lessgtr 0\;\;\; \mbox{\rm for}\;\;\; \mathrm{Im}(z) \gtrless 0. \label{ec52a}
\end{equation}

From the above observations and the Dyson equation
\begin{equation}
\t{\Sigma}_{\sigma}({\bm k};z) = \t{G}_{\sigma;0}^{-1}({\bm k};z) -
\t{G}_{\sigma}^{-1}({\bm k};z) \label{ec53}
\end{equation}
it thus follows that $\t{\Sigma}_{\sigma}({\bm k};z)$ is analytic everywhere in the complex $z$ plane outside the real axis \cite{JML61}. Consequently, $\t{\Sigma}_{\sigma}({\bm k};\varepsilon \pm i\eta)$ is \emph{continuous} for real values of $\varepsilon$ when $\eta>0$; this is reminiscent of the observation made above concerning the continuity of $\t{G}_{\sigma}({\bm k};\varepsilon\pm i\eta)$ for $\eta>0$, even if $\vert\t{G}_{\sigma}({\bm k};\varepsilon\pm i\eta)\vert$ may increase unboundedly for $\eta$ approaching zero. This property plays a vital role in our proof of the \emph{uniformity} of convergence of the series in
Eq.~(\ref{e85}) for almost all ${\bm k}$ and all $z$ over the entire $z$ plane outside the real axis of this plane.

From the Dyson equation and the expression in Eq.~(\ref{ec52}), one further obtains that
\begin{equation}
\mathrm{Im}[\t{\Sigma}_{\sigma}({\bm k};z)] =
\frac{1}{\hbar} \mathrm{Im}(z) +
\frac{\mathrm{Im}[\t{G}_{\sigma}({\bm k};z)]}{\vert
\t{G}_{\sigma}({\bm k};z)\vert^2} \equiv -\frac{1}{\hbar} \mathrm{Im}(z)\,
\Upsilon_{\!\sigma}({\bm k};z), \label{ec54}
\end{equation}
where
\begin{equation}
\Upsilon_{\!\sigma}({\bm k};z) {:=} \frac{\displaystyle\sum_s \left|
\frac{{\sf f}_{s;\sigma}({\bm
k})}{z-\varepsilon_{s;\sigma}}\right|^2} {\left| \displaystyle\sum_s
\frac{\vert {\sf f}_{s;\sigma}({\bm
k})\vert^2}{z-\varepsilon_{s;\sigma}}\right|^2} -1. \label{ec55}
\end{equation}
Note that $\Upsilon_{\!\sigma}({\bm k};z)\in \mathds{R}$, $\forall z$, and that $\Upsilon_{\!\sigma}({\bm k};z)\equiv 0$ for non-interacting GSs. Making use of Eq.~(\ref{ec39}) and the Cauchy-Schwarz-H\"older inequality (item 3.2.8 in Ref.~\citen{AS72} for $p=q=2$) one infers that
\begin{equation}
\sum_s \left| \frac{{\sf f}_{s;\sigma}({\bm
k})}{z-\varepsilon_{s;\sigma}}\right|^2 \equiv \sum_s \left|
\frac{{\sf f}_{s;\sigma}({\bm
k})}{z-\varepsilon_{s;\sigma}}\right|^2 \, \sum_s \vert {\sf
f}_{s;\sigma}({\bm k})\vert^2 \ge \left( \sum_s \frac{\vert {\sf
f}_{s;\sigma}({\bm k})\vert^2}{\vert z-\varepsilon_{s;\sigma}\vert}
\right)^2. \label{ec56}
\end{equation}
On the other hand,
\begin{equation}
\left| \sum_s \frac{\vert {\sf f}_{s;\sigma}({\bm
k})\vert^2}{z-\varepsilon_{s;\sigma}}\right|^2 \le  \left( \sum_s
\frac{\vert {\sf f}_{s;\sigma}({\bm k})\vert^2}{\vert
z-\varepsilon_{s;\sigma}\vert} \right)^2. \label{ec57}
\end{equation}
From the inequalities in Eqs.~(\ref{ec56}) and (\ref{ec57}) one deduces that
\begin{equation}
\Upsilon_{\!\sigma}({\bm k};z) \ge 0,\;\;\; \forall {\bm k}, z.
\label{ec58}
\end{equation}
This inequality together with the expression in Eq.~(\ref{ec54}) lead to the conclusion that, similar to $\mathrm{Im}[\t{G}_{\sigma}({\bm k};z)]$, Eq.~(\ref{ec52a}), \emph{the sign of $\mathrm{Im}[\t{\Sigma}_{\sigma}({\bm k};z)]$ is always opposite to that of $\mathrm{Im}(z)$} (Eq.~(\ref{e14}). This observation is consistent with the fact that for stable GSs $\mathrm{Im}[\t{\Sigma}_{\sigma}({\bm k};\varepsilon-i 0^+)] \ge 0$, or, equivalently, $\mathrm{Im}[\t{\Sigma}_{\sigma}({\bm k};\varepsilon+i 0^+)] \le 0$, $\forall\varepsilon \in \mathds{R}$ (cf. Eq.~(\ref{e16})).

From Eq.~(\ref{ec54}) it further follows that at any real energy $\varepsilon$ for which $\mathrm{Im}[\t{\Sigma}_{\sigma}({\bm k};\varepsilon \mp i 0^+)]$ is non-vanishing, the corresponding $\Upsilon_{\!\sigma}({\bm k};\varepsilon \mp i\eta)$ must be diverging like $1/\eta^{\gamma}$, $\gamma\ge 1$, as $\eta> 0$ is made to approach zero, with the condition $\gamma=1$ corresponding to a bounded and non-vanishing $\mathrm{Im}[\t{\Sigma}_{\sigma}({\bm k};\varepsilon \mp i 0^+)]$. We note that although $\Upsilon_{\!\sigma}({\bm k};\varepsilon \mp i\eta)$ may in principle to leading order diverge like $\ln(\eta)/\eta^{\gamma}$ for $\eta\downarrow 0$, such behaviour is ruled out if $\mathrm{Im}[\t{\Sigma}_{\sigma}({\bm k};\varepsilon \mp i 0^+)]$ is to be non-vanishing (ruling out $\gamma<1$) and finite for the ${\bm k}$ and $\varepsilon$ under consideration.

\subsubsection{Remarks}
\label{ssc2s1}

It is instructive to apply the above observations to a well-characterised state, namely the conventional Fermi-liquid metallic states. To this end, we introduce
\begin{equation}
\Upsilon_{\sigma}({\bm k};\mu + i\eta) \sim u_{\sigma}({\bm k})\,
\eta^{-\gamma_{\sigma}({\bm k})},\;\;\forall {\bm k}, \label{ec59}
\end{equation}
as representing the leading-order asymptotic series expansion of $\Upsilon_{\sigma}({\bm k};\mu + i\eta)$ for $\eta\downarrow 0$.
Following Eq.~(\ref{ec58}), one must have $u_{\sigma}({\bm k})\le 0$, $\forall {\bm k}$. However, since by the Cauchy-Riemann differential equation for analytic functions (\S~5.12 in Ref.~\citen{WW62}), $\mathrm{Im}[\t{\Sigma}_{\sigma}({\bm k};\mu+i\eta)] \not\equiv 0$ for $\eta$ increased from zero towards some non-vanishing value (unless one deals with a mean-field Hamiltonian, for which $\t{\Sigma}_{\sigma}({\bm k};z)\equiv 0$, $\forall {\bm k}, z$), one in fact has $u_{\sigma}({\bm k})< 0$, $\forall {\bm k}$.

For the conventional Fermi-liquid states, and sufficiently small $\vert\varepsilon-\mu\vert$, one has\cite{Note15}
\begin{equation}
\Sigma_{\sigma}({\bm k};\varepsilon) \sim \Sigma_{\sigma}({\bm k};\mu) + \beta_{\sigma}({\bm k}) (\varepsilon -\mu) \mp i \alpha_{\sigma}({\bm k}) (\varepsilon-\mu)^2 \;\;\; \mbox{\rm for}\;\;\; \varepsilon \gtrless \mu.
\label{ec60}
\end{equation}
where $\Sigma_{\sigma}({\bm k};\mu), \beta_{\sigma}({\bm k}), \alpha_{\sigma}({\bm k}) \in \mathds{R}$, $\forall {\bm k}$. Stability of the underlying GS requires that $\alpha_{\sigma}({\bm k})\ge 0$, $\forall {\bm k}$. Using the Kramers-Kr\"onig relationship (Sec.~\ref{ssc5}) one readily obtains that
\begin{equation}
\beta_{\sigma}({\bm k}) = \frac{1}{\pi} \int_{0}^{\infty} \frac{{\rm d}\varepsilon}{\varepsilon^2}\; \Big(\mathrm{Im}[\Sigma_{\sigma}({\bm k};\mu+\varepsilon)] - \mathrm{Im}[\Sigma_{\sigma}({\bm k};\mu-\varepsilon)]\Big), \label{ec61}
\end{equation}
from which and from the inequalities in Eq.~(\ref{e16}) one deduces that, unless the interaction strength is set equal to zero, $\beta_{\sigma}({\bm k}) >0$ (strictly positive) for all ${\bm k}$. Note that, following the expression in Eq.~(\ref{ec60}), the integrand of the integral on the RHS of Eq.~(\ref{ec61}) is bounded at $\varepsilon=0$.
Although this would not have been the case if $\vert\mathrm{Im}[\Sigma_{\sigma}({\bm k};\varepsilon)]\vert$ would have diminished like $\vert\varepsilon-\mu\vert^{1+\epsilon}$ for $\varepsilon\to\mu$, with $\epsilon<1$, nonetheless
$\beta_{\sigma}({\bm k})$ would be bounded for all $\epsilon>0$.

Since $\Sigma_{\sigma}({\bm k};\mu)\in \mathds{R}$, $\forall {\bm k}$, from the expression in Eq.~(\ref{ec60}) one readily deduces that
\begin{equation}
\mathrm{Im}[\t{\Sigma}_{\sigma}({\bm k};\mu + i\eta)] \sim
\beta_{\sigma}({\bm k}) \eta + \alpha_{\sigma}({\bm k}) \eta^2 \Leftrightarrow \Upsilon_{\sigma}({\bm k};\mu+i\eta) \sim -\hbar\beta_{\sigma}({\bm k}) -\hbar \alpha_{\sigma}({\bm k}) \eta\;\;\mbox{\rm for}\;\, \eta\downarrow 0. \label{ec62}
\end{equation}
Comparison of this expression with that in Eq.~(\ref{ec59}), one arrives at the following results, applicable to conventional Fermi-liquid metallic states:
\begin{equation}
u_{\sigma}({\bm k}) \equiv -\hbar\beta_{\sigma}({\bm k})<0, \;\;\; \gamma_{\sigma}({\bm k}) \equiv 0,\;\;\; \forall {\bm k}. \label{ec63}
\end{equation}
Thus, in the case at hand indeed $u_{\sigma}({\bm k})<0$, and $\gamma_{\sigma}({\bm k})<1$, $\forall {\bm k}$, the latter in conformity with the property $\mathrm{Im}[\Sigma_{\sigma}({\bm k};\mu)]\equiv 0$, $\forall {\bm k}$.

With reference to the details in appendix \ref{sd} concerning the Luttinger-liquid metallic states of the one-dimensional Luttinger model for spin-less fermions, we note in passing that what we have in this section denoted by $\gamma_{\sigma}({\bm k})$ is equal to $2\gamma_0(k)$ at the underlying Fermi points and to $\gamma_0(k)$ elsewhere, where, depending on whether one considers the left-moving or the right-moving fermions, $k$ is measured with respect to $-k_{\Sc f}$ and $+k_{\Sc f}$ respectively. The dependence of $\gamma_0(k)$ on $k$ is fully determined by the dependencies on $k$ of the interaction potentials $g_2(k)$ and $g_4(k)$.

\subsection{Asymptotic series of $\t{G}_{\sigma}({\bf k};z)$ for
$\vert z\vert\to\infty$}
\label{ssc3}

From the Lehmann representation in Eq.~(\ref{ec12}) one obtains the
formal asymptotic series \cite{BF02}
\begin{equation}
\t{G}_{\sigma}({\bm k};z) \sim
\frac{G_{\sigma;\infty_1}({\bm k})}{z}+
\frac{G_{\sigma;\infty_2}({\bm k})}{z^2}+\dots \;\;\; \mbox{\rm
as}\;\;\; \vert z\vert\to \infty, \label{ec64}
\end{equation}
where
\begin{equation}
G_{\sigma;\infty_j}({\bm k}) \equiv \hbar\sum_s
(\varepsilon_{s;\sigma})^{j-1}\, \vert {\sf f}_{s;\sigma}({\bm
k})\vert^2,\;\;\; j\ge 1. \label{ec65}
\end{equation}
Evidently, when $G_{\sigma;\infty_j}({\bm k})$ exits (see later), it is positive for odd values of $j$. From the sum rule in Eq.~(\ref{ec39}), one immediately deduces that
\begin{equation}
G_{\sigma;\infty_1}({\bm k}) = \hbar. \label{ec66}
\end{equation}
The leading-order asymptotic expression $\t{G}_{\sigma}({\bm k};z)
\sim \hbar/z$ for $z = \varepsilon \pm i 0^+$ and
$\vert\varepsilon\vert\to\infty$ can be found in various textbooks
(see, e.g., Eq.~($7.21'$) in Ref.~\citen{AGD75} and Eq.~(7.72) in Ref.~\citen{FW03}).

It can be shown that \cite{HL67,BF02}
\begin{equation}
G_{\sigma;\infty_j}({\bm k}) = \hbar\,
\langle\Psi_{N;0}\vert [\wh{L}^{j-1} \h{c}_{{\bm k};\sigma},
\h{c}_{{\bm k};\sigma}^{\dag} ]_{+}\vert\Psi_{N;0}\rangle
\equiv \int_{-\infty}^{\infty} {\rm d}\varepsilon\;
\varepsilon^{j-1}\, A_{\sigma}({\bm k};\varepsilon),\;\; j\ge 1,
\label{ec67}
\end{equation}
where $\wh{L}$ is the Liouville super-operator corresponding to the interacting Hamiltonian $\wh{H}$, defined according to
\begin{equation}
\wh{L}\, \h{c}_{{\bm k};\sigma} {:=} [\h{c}_{{\bm k};\sigma},\wh{H}]_{-}. \label{ec68}
\end{equation}
Above $[{\;},{\;}]_{-}$ denotes a commutator and $[{\;},{\;}]_{+}$ an anti-commutator. The first expression in Eq.~(\ref{ec67}), which can be directly deduced \cite{BF02} from the Mori-Zwanzig representation of $\t{G}_{\sigma}({\bm k};z)$ \cite{PF91}, or in fact from the equation-of-motion of $\t{G}_{\sigma}({\bm k};z)$ in the time domain \cite{BF02}, provides one with a recipe according to which to calculate $G_{\sigma;\infty_j}({\bm k})$ without prior knowledge of $A_{\sigma}({\bm k};\varepsilon)$, and the second expression, which is deduced from the exact spectral representation in Eq.~(\ref{ec47}), reveals the significance of $G_{\sigma;\infty_j}({\bm k})$ as the $(j-1)$th energy moment of the single-particle spectral function.

The second expression in Eq.~(\ref{ec67}) further gives insight into the maximum value of $j$ for which $G_{\sigma;\infty_j}({\bm k})$ can be a bounded function. Evidently, for $G_{\sigma;\infty_j}({\bm k})$ to be finite for an arbitrary finite value of $j$, it is required that $A_{\sigma}({\bm k};\varepsilon)$, as function of $\varepsilon$, possess a bounded support (which may or may not be of measure zero) or otherwise decay at least exponentially for $\vert\varepsilon\vert\to\infty$. In view of Eq.~(\ref{ec43}), and since the denominator of the expression on the RHS of Eq.~(\ref{ec43}) scales to leading order like $\varepsilon^2$ as $\vert\varepsilon\vert\to\infty$, one observes that in order for $G_{\sigma;\infty_j}({\bm k})$ to be bounded for an arbitrary finite value of $j$, it is required that $\mathrm{Im}[\Sigma_{\sigma}({\bm k};\varepsilon)]$ either possess a bounded support or decay at least exponentially for
$\vert\varepsilon\vert\to\infty$.

From the first expression in Eq.~(\ref{ec67}) it is evident that $G_{\sigma;\infty_j}({\bm k})$ must be bounded for an arbitrary finite value of $j$ in the cases of models for which the wave-vector integrals, in terms of which $G_{\sigma;\infty_j}({\bm k})$ is expressed, are bounded. The sufficient conditions for this are exactly the same as those that are required in order for $\t{\Sigma}_{\sigma}^{(\nu)}({\bm k};z)$ to be well-defined for an arbitrary finite value of $\nu$; this aspect will be illuminated in Sec.~\ref{ssc7}. Explicitly, $G_{\sigma;\infty_j}({\bm k})$ is well-defined, for an arbitrary finite value of $j$, for systems defined on lattices and in which the two-body interaction potential is short range. It follows that \emph{the exact $A_{\sigma}({\bm k};\varepsilon)$ pertaining to the uniform GSs of these systems, assumed to be macroscopic (see the introductory section of the present appendix), must decay at least exponentially for $\vert\varepsilon\vert\to\infty$.} With reference to Eq.~(\ref{ec43}), this conclusion also applies to the corresponding $\mathrm{Im}[\Sigma_{\sigma}({\bm k};\varepsilon)]$, with $\varepsilon\in \mathds{R}$. In Sections \ref{ssc5} and \ref{ssc6} we shall explicitly consider the behaviour of $\mathrm{Im}[\Sigma_{\sigma}({\bm k};\varepsilon)]$ for $\varepsilon\in \mathds{R}$ and $\vert\varepsilon\vert\to\infty$.

For illustration, for a continuum model in $d=3$ concerning particles whose non-interacting energy dispersion $\varepsilon_{\bm k}$ is proportional to $\|{\bm k}\|^2$ and interact through the long-range Coulomb potential, $G_{\sigma;\infty_j}({\bm k})$ is bounded for $j=1, 2, 3$, and unbounded for $j\ge 4$ \cite{BF02}.

As is evident from Eq.~(\ref{ec65}), $G_{\sigma;\infty_j}({\bm k})$ is real for all $j \in \mathds{N}$ so that truncating the asymptotic series in Eq.~(\ref{ec64}) at some \emph{finite} order (less than the order for which the corresponding $G_{\sigma;\infty_j}({\bm k})$ is unbounded), one obtains an asymptotic expression for $\t{G}_{\sigma}({\bm k};z)$ whose imaginary part is infinitesimally small for all $z$ (excluding $z=0$) in the infinitesimal neighbourhood of the real energy axis. This aspect is noteworthy, since, in view of Eqs.~(\ref{ec40}) and (\ref{ec41}), $\mathrm{Im}[\t{G}_{\sigma}({\bm k};\varepsilon\mp i 0^+)]\not\equiv 0$ for large $\vert\varepsilon\vert$ in the cases where $\{\varepsilon_{s;\sigma}\,\|\, s\}$ is an unbounded continuum set.

To underline that the above-mentioned aspect of a finite-order asymptotic expansion for $\t{G}_{\sigma}({\bm k};z)$ corresponding to $\vert z\vert\to\infty$ does not signify a contradiction, consider $\sum_{j=1}^{m} G_{\sigma;\infty_j}({\bm k})/z^j$, where $m$ is an integer less than the smallest value of $j$ for which $G_{\sigma;\infty_j}({\bm k})$ is unbounded (see the one but the previous paragraph). In order for this finite-order series to be an asymptotic series expansion of $\t{G}_{\sigma}({\bm k};z)$ in the region $\vert z\vert\to\infty$, it is required, following Poincar\'e (\S~8.2 in Ref.~\citen{WW62}), that
\begin{equation}
\left| z^m \Big(\t{G}_{\sigma}({\bm k};z) - \sum_{j=1}^m
\frac{G_{\sigma;\infty_j}({\bm k})}{z^j}\Big) \right| = o(1) \;\;\;
\mbox{\rm for}\;\;\; \vert z\vert\to\infty, \label{ec69}
\end{equation}
where $o(1)$ is a real-valued function of $z$ which approaches zero for $\vert z\vert \to\infty$ (\S~2.11 in Ref.~\citen{WW62}). One readily verifies that for $G_{\sigma;\infty_j}({\bm k})$, $j=1,\dots, m$, determined according to the expression in Eq.~(\ref{ec65}), indeed Eq.~(\ref{ec69}) is satisfied, with $o(1) = O(1/\vert z\vert)$ for $\vert z\vert\to\infty$ in the event that $G_{\sigma;\infty_{m+1}}({\bm k})$ is bounded (for the $o-O$ notation see \S~5 in Ref.~\citen{EWH26}).

Being a \emph{single} constraint concerning the behaviour of the \emph{complex-valued} function $\t{G}_{\sigma}({\bm k};z)$ for $\vert z\vert\to\infty$, Eq.~(\ref{ec69}) cannot be expected to restrict the behaviours of $\mathrm{Re}[\t{G}_{\sigma}({\bm k};z)]$ and $\mathrm{Im}[\t{G}_{\sigma}({\bm k};z)]$ for $\vert z\vert\to\infty$ as tightly as the \emph{two} constraints associated with the $m$th-order asymptotic series expansions specific to the \emph{real-valued} functions $\mathrm{Re}[\t{G}_{\sigma}({\bm k};z)]$ and $\mathrm{Im}[\t{G}_{\sigma}({\bm k};z)]$. This fact is clearly manifested by the real value of $\sum_{j=1}^{m} G_{\sigma;\infty_j}({\bm k})/z^j$ for $z\in \mathds{R}$. We shall return to this subject matter later in Sections \ref{ssc5} and \ref{ssc6}, where we explicitly investigate the asymptotic behaviours of $\mathrm{Im}[\t{\Sigma}_{\sigma}({\bm k};\varepsilon \mp i 0^+)]$ and $\mathrm{Re}[\t{\Sigma}_{\sigma}({\bm k};\varepsilon \mp i 0^+)]$ for $\varepsilon\in \mathds{R}$ and $\vert \varepsilon\vert\to\infty$.

We remark that the asymptotic series under consideration being in terms of the asymptotic sequence $\{1/z, 1/z^2,\dots \}$, it is fundamentally incapable of describing exponentially decaying contributions to, for instance, $\mathrm{Im}[\t{G}_{\sigma}({\bm k};\varepsilon \mp i 0^+)]$ and $\mathrm{Im}[\t{\Sigma}_{\sigma}({\bm k};\varepsilon \mp i 0^+)]$ when these series are truncated at some finite order; this is evident from the Poincar\'e definition, presented in Eq.~(\ref{ec69}), which remains satisfied on adding an exponentially decaying contribution (for $\vert z\vert\to\infty$) to $\sum_{j=1}^m G_{\sigma;\infty_j}({\bm k})/z^j$ (\S~8.32 in Ref.~\citen{WW62}). Such contributions are dealt with by so-called hyperasymptotics \cite{BH90}. A relatively informal review of super- and hyper-asymptotics can be found in Ref.~\citen{JPB99}.

\subsection{Asymptotic series of $\t{\Sigma}_{\sigma}({\bf k};z)$ for $\vert z\vert\to\infty$}
\label{ssc4}

From the Dyson equation and the asymptotic expression in Eq.~(\ref{ec64}) one readily obtains the formal asymptotic series \cite{BF02}
\begin{equation}
\t{\Sigma}_{\sigma}({\bm k};z) \sim \Sigma_{\sigma;\infty_0}({\bm
k}) + \frac{\Sigma_{\sigma;\infty_1}({\bm k})}{z} + \dots
\;\;\;\mbox{\rm as}\;\;\; \vert z\vert\to\infty, \label{ec70}
\end{equation}
where \cite{BF99,BF02}
\begin{equation}
\Sigma_{\sigma;\infty_0}({\bm k}) =
\left.\t{\Sigma}_{\sigma}^{(\nu)}({\bm k};z)\right|_{\nu=1} \equiv
\Sigma_{\sigma}^{\Sc h\Sc f}({\bm k}), \label{ec71}
\end{equation}
the exact Hartree-Fock self-energy, and
\begin{equation}
\Sigma_{\sigma;\infty_1}({\bm k}) = \frac{1}{\hbar^3} \big( \hbar
G_{\sigma;\infty_3}({\bm k}) - G_{\sigma;\infty_2}^2({\bm k}) \big).
\label{ec72}
\end{equation}
We note that $\Sigma_{\sigma;\infty_1}({\bm k})$ is bounded for the continuum model in $d=3$ referred to above \cite{BF02}. For this model, $\Sigma_{\sigma;\infty_2}({\bm k})$ is unbounded however; on performing partial infinite summations over relevant unbounded contributions arising from $\Sigma_{\sigma;\infty_j}({\bm k})/z^j$ for $j\ge 2$, one obtains that in this case $\Sigma_{\sigma;\infty_1}({\bm k})/z$ is followed by terms scaling like $1/z^{3/2}$, $\ln(z)/z^2$ and $1/z^2$ \cite{BF02}. The contribution scaling like $1/z^{3/2}$ has its origin in an interplay between the unbounded nature of $\varepsilon_{\bm k} \propto \|{\bm k}\|^2$ for $\|{\bm k}\| \to \infty$ and the divergence of the Coulomb potential $v_{\Sc c}({\bm r}-{\bm r}')$ for $\|{\bm r}-{\bm r}'\|\to 0$; the contribution scaling like $\ln(z)/z^2$ has its root in the long range of the Coulomb potential for $\|{\bm r}-{\bm r}'\|\to \infty$ \cite{BF02}.

It can be shown \cite{BF02} that $\Sigma_{\sigma;\infty_j}({\bm k})$, $j\ge 1$, is fully expressed in terms of $G_{\sigma;\infty_2}({\bm k})$, $\dots$, $G_{\sigma;\infty_{j+2}}({\bm k})$ (cf. Eq.~(\ref{ec72})). Further, similar to $G_{\sigma;\infty_j}({\bm k})$, $\Sigma_{\sigma;\infty_j}({\bm k})$ can be shown to be real valued for all $j$ \cite{BF02}; the expressions in Eqs.~(\ref{ec71}) and (\ref{ec72}) are seen to testify to this fact. Consequently, retaining a finite number of the leading-order terms on the RHS of Eq.~(\ref{ec70}) (assuming that these are bounded), the imaginary part of the resulting asymptotic expression is infinitesimally small for $z$ in an infinitesimal neighbourhood of the real axis. This aspect is clarified by the same reasoning as presented above for the $m$th-order asymptotic series expansion of $\t{G}_{\sigma}({\bm k};z)$ in the light of the defining expression in Eq.~(\ref{ec69}).

\subsection{Asymptotic series of $\mathrm{Im}[\t{\Sigma}_{\sigma}({\bf k};\varepsilon \mp i 0^+)]$ and $\mathrm{Re}[\t{\Sigma}_{\sigma}({\bf k};\varepsilon \mp i 0^+)]$ for $\varepsilon \in \mathds{R}$ and $\vert \varepsilon\vert\to\infty$}
\label{ssc5}

Since $\t{\Sigma}_{\sigma}({\bm k};z)$ is analytic over the entire complex $z$ plane away from the real axis (Sec.~\ref{ssc2}), and since, following Eqs.~(\ref{ec70}) and (\ref{ec71}), $\t{\Sigma}_{\sigma}({\bm k};z) - \Sigma_{\sigma}^{\Sc h\Sc f}({\bm k})$ approaches zero for $\vert z\vert\to\infty$, on using the Cauchy residue theorem one obtains the following pair of Kramers-Kr\"onig relationships
\begin{equation}
\mathrm{Re}[{\Sigma}_{\sigma}({\bm k};\varepsilon)] =
\Sigma_{\sigma}^{\Sc h\Sc f}({\bm k}) +
\mathscr{P}\!\!\int_{-\infty}^{\infty} \frac{{\rm
d}\varepsilon'}{\pi}\; \frac{\mathrm{Im}[\t{\Sigma}_{\sigma}({\bm
k};\varepsilon'-i 0^+)]}{\varepsilon-\varepsilon'}, \label{ec73}
\end{equation}
\begin{equation}
\mathrm{Im}[\t{\Sigma}_{\sigma}({\bm k};\varepsilon-i 0^+)] =
-\mathscr{P}\!\!\int_{-\infty}^{\infty} \frac{{\rm
d}\varepsilon'}{\pi}\; \frac{\mathrm{Re}[{\Sigma}_{\sigma}({\bm
k};\varepsilon')] -\Sigma_{\sigma}^{\Sc h\Sc f}({\bm
k})}{\varepsilon-\varepsilon'}. \label{ec74}
\end{equation}
Recall that (see Eq.~(\ref{e5}))
\begin{equation}
\mathrm{Re}[\t{\Sigma}_{\sigma}({\bm k};\varepsilon\mp i 0^+)] \equiv \mathrm{Re}[\Sigma_{\sigma}({\bm k};\varepsilon)],\;\; \mathrm{Im}[\t{\Sigma}_{\sigma}({\bm k};\varepsilon+ i 0^+)] \equiv -\mathrm{Im}[\t{\Sigma}_{\sigma}({\bm k};\varepsilon-i 0^+)],\;\, \varepsilon\in\mathds{R}. \label{ec75}
\end{equation}
We point out that were it not for the continuity of $\t{\Sigma}_{\sigma}({\bm k};\varepsilon\mp i\eta)$ for $\varepsilon\in \mathds{R}$ and $\eta>0$, use of the Cauchy's residue theorem, leading to the results in Eqs.~(\ref{ec73}) and (\ref{ec74}), would not have been justified; the proof of this theorem crucially depends on the \emph{continuity} of the integrand \emph{on} the closed contour of integration (\S~5.2 in Ref.~\citen{WW62}; see in particular the last but one footnote on p.~85 of Ref.~\citen{WW62}).

\subsubsection{Concerning
$\mathrm{Im}[{\Sigma}_{\sigma}({\bf k};\varepsilon-i 0^+)]$}
\label{ssc5s1}

Following some algebra, from Eq.~(\ref{ec74}) one obtains the following formal asymptotic expression \cite{BF02}
\begin{equation}
\mathrm{Im}[\t{\Sigma}_{\sigma}({\bm k};\varepsilon-i 0^+)] \sim
\frac{\Xi_{\sigma;\infty_1}({\bm k})}{\varepsilon} +
\frac{\Xi_{\sigma;\infty_2}({\bm k})}{\varepsilon^2} +\dots
\;\;\;\mbox{\rm for}\;\;\; \vert\varepsilon\vert\to\infty,
\label{ec76}
\end{equation}
where
\begin{equation}
\Xi_{\sigma;\infty_1}({\bm k}) \equiv  \int_0^{\infty} \frac{{\rm
d}\varepsilon'}{\pi}\; \Big( 2 \Sigma_{\sigma;\infty_0}({\bm k})
-\mathrm{Re}[{\Sigma}_{\sigma}({\bm k};\varepsilon')
+{\Sigma}_{\sigma}({\bm k};-\varepsilon')]\Big), \label{ec77}
\end{equation}
and
\begin{equation}
\Xi_{\sigma;\infty_2}({\bm k}) \equiv \int_0^{\infty} \frac{{\rm
d}\varepsilon'}{\pi}\; \Big( 2 \Sigma_{\sigma;\infty_1}({\bm k})
-\varepsilon' \,\mathrm{Re}[{\Sigma}_{\sigma}({\bm k};\varepsilon')
-{\Sigma}_{\sigma}({\bm k};-\varepsilon')] \Big). \label{ec78}
\end{equation}
Assuming that $\Xi_{\sigma;\infty_1}({\bm k})$ exists (see later), the requirement $\mathrm{Im}[\t{\Sigma}_{\sigma}({\bm k};\varepsilon-i 0^+)] \ge 0$, $\forall {\bm k}$, $\forall\varepsilon \in \mathds{R}$, Eq.~(\ref{e14}), leads to the exact sum rule
\begin{equation}
\Xi_{\sigma;\infty_1}({\bm k}) = 0,\;\;\; \forall {\bm k}, \sigma.
\label{ec79}
\end{equation}
Assuming that $\Xi_{\sigma;\infty_2}({\bm k})$ exists (see later), the result in Eq.~(\ref{ec79}) in combination with the requirement $\mathrm{Im}[\t{\Sigma}_{\sigma}({\bm k};\varepsilon-i 0^+)] \ge 0$, $\forall {\bm k}$, $\forall\varepsilon \in \mathds{R}$, lead to the condition that
\begin{equation}
\Xi_{\sigma;\infty_2}({\bm k}) \ge 0,\;\;\; \forall {\bm k}, \sigma.
\label{ec80}
\end{equation}

The formal asymptotic series in Eq.~(\ref{ec76}) is valid up to term of maximal order $j$ in $1/\varepsilon$ if $\Xi_{\sigma;\infty_{j+1}}({\bm k})$ is unbounded. The explicit dependence of $\Xi_{\sigma;\infty_j}({\bm k})$, $j=1,2$, on $\Sigma_{\sigma;\infty_{j-1}}({\bm k})$ implies that breakdown of the formal series in Eq.~(\ref{ec70}) at the $(j-1)$th order in $1/z$ is \emph{sufficient} for the series in Eq.~(\ref{ec76}) to fail at the $j$th order in $1/\varepsilon$. As regards the existence of $\Xi_{\sigma;\infty_j}({\bm k})$, $j=1,2$, it is required not only that $\Sigma_{\sigma;\infty_{j-1}}({\bm k})$ exist, but also, in the case of $j=1$, that (for the $o-O$ notation see \S~5 in Ref.~\citen{EWH26})
\begin{equation}
2\Sigma_{\sigma;\infty_0}({\bm k})
-\mathrm{Re}[{\Sigma}_{\sigma}({\bm k};\varepsilon)
+{\Sigma}_{\sigma}({\bm k};-\varepsilon)] =
o(1/\varepsilon) \;\;\;\mbox{\rm as}\;\;\; \varepsilon\to\infty,
\label{ec81}
\end{equation}
and, in the case of $j=2$, that
\begin{equation}
2\Sigma_{\sigma;\infty_1}({\bm k}) -\varepsilon
\,\mathrm{Re}[{\Sigma}_{\sigma}({\bm k};\varepsilon)
-{\Sigma}_{\sigma}({\bm k};-\varepsilon)] =
o(1/\varepsilon) \;\;\; \mbox{\rm as}\;\;\; \varepsilon\to\infty.
\label{ec82}
\end{equation}
Evidently, Eq.~(\ref{ec81}) is satisfied if $\Sigma_{\sigma;\infty_1}({\bm k})$ exists and Eq.~(\ref{ec82}) is satisfied if $\Sigma_{\sigma;\infty_2}({\bm k})$ exists. Note that if $\Sigma_{\sigma;\infty_2}({\bm k})$ exists, for the $o(1/\varepsilon)$ on the RHS of Eq.~(\ref{ec81}) one has $o(1/\varepsilon) = O(1/\varepsilon^2)$.

It appears therefore that the existence of the Poincar\'e-type series in Eq.~(\ref{ec76}) to order $j$ in $1/\varepsilon$ is dependent on the existence of the Poincar\'e-type series in Eq.~(\ref{ec70}) for $\t{\Sigma}_{\sigma}({\bm k};z)$ to order $j$ in $1/z$, and \emph{vice versa}. In view of Eq.~(\ref{ec71}), the existence of $\Sigma_{\sigma;\infty_0}({\bm k})$ is in no doubt. Since, as we have indicated earlier, $\Sigma_{\sigma;\infty_j}({\bm k})$, $j\ge 1$, is fully determined in terms of $G_{\sigma;\infty_2}({\bm k})$, $\dots$, $G_{\sigma;\infty_{j+2}}({\bm k})$ \cite{BF02}, we thus conclude that $\Xi_{\sigma;\infty_1}({\bm k})$ exists if $G_{\sigma;\infty_3}({\bm k})$ is bounded, and that $\Xi_{\sigma;\infty_2}({\bm k})$ exists if $G_{\sigma;\infty_4}({\bm k})$ is bounded. For the continuum model in $d=3$, to which we have referred twice before in this appendix, $G_{\sigma;\infty_j}({\bm k})$ exists for $j=1,2,3$, however it does not exist for $j=4$ \cite{BF02}. Consequently, for this model $\Xi_{\sigma;\infty_1}({\bm k})$ exists, and thus on account of Eq.~(\ref{ec79}) is vanishing, however $\Xi_{\sigma;\infty_2}({\bm k})$ does not exist. Existence of $\Xi_{\sigma;\infty_j}({\bm k})$ and non-existence of $\Xi_{\sigma;\infty_{j+1}}({\bm k})$ signal the fact that in the large-$\vert\varepsilon\vert$ asymptotic series for $\mathrm{Im}[\t{\Sigma}_{\sigma}({\bm k};\varepsilon-i0^+)]$ the term $\Xi_{\sigma;\infty_j}({\bm k})/\varepsilon^j$ is immediately followed by a term more dominant than one proportional $1/\varepsilon^{j+1}$ for $\vert\varepsilon\vert\to\infty$ \cite{BF02}.

\subsubsection{Technicalities}
\label{ssc5s2}

Here we present an outline of the algebraic manipulations that lead to the expressions presented in Eqs.~(\ref{ec76}), (\ref{ec77}) and (\ref{ec78}). Evidently, these results originate from the expression in Eq.~(\ref{ec74}), which, on subdividing the interval $[-\infty,\infty]$ of integration with respect to $\varepsilon'$ into $[-\infty,0]$ and $[0,\infty]$ and subsequently applying the variable transformation $\varepsilon' \rightharpoonup 1/\varepsilon'$ in both of the resulting integrals, can be expressed in the following equivalent form:
\begin{eqnarray}
&&\hspace{-0.2cm}\mathrm{Im}[\t{\Sigma}_{\sigma}({\bm k};\varepsilon-i 0^+)]=\frac{-1}{\pi\varepsilon} \mathscr{P}\!\!\int_{0}^{\infty}
\frac{{\rm d}\varepsilon'}{{\varepsilon'}^2}\; \nonumber\\
&&\hspace{0.2cm} \times\frac{\displaystyle(1
+ \frac{1}{\varepsilon\varepsilon'})
\mathrm{Re}[{\Sigma}_{\sigma}({\bm k};\frac{1}{\varepsilon'})] + (1
- \frac{1}{\varepsilon\varepsilon'})
\mathrm{Re}[{\Sigma}_{\sigma}({\bm k};\frac{-1}{\varepsilon'})] - 2
\Sigma_{\sigma;\infty_0}({\bm k}) - \frac{2}{\varepsilon}
\Sigma_{\sigma;\infty_1}({\bm k})}{\displaystyle 1 -
\frac{1}{\varepsilon^2 {\varepsilon'}^2}}. \nonumber\\
\label{ec83}
\end{eqnarray}
In arriving at this result, we have inserted the last constant (that is, independent of $\varepsilon'$) in the numerator of the integrand on account of the fact that
\begin{equation}
\lim_{E\to\infty}\mathscr{P}\!\!\int_{-E}^{E} \frac{{\rm
d}\varepsilon'}{\varepsilon-\varepsilon'} = \lim_{E\to\infty}
\ln\left|\frac{E+\varepsilon}{E-\varepsilon}\right| = 0,\;\;\forall
\varepsilon\in \mathds{R}. \label{ec84}
\end{equation}

The two terms explicitly displayed on the RHS of Eq.~(\ref{ec76}) arise from the exact expression on the RHS of Eq.~(\ref{ec83}) on suppressing the term $1/[\varepsilon^2 {\varepsilon'}^2]$ in the denominator of this expression. On the condition that Eq.~(\ref{ec81}) is satisfied, it can be shown (see next paragraph) that the deviation of $\Xi_{\sigma;\infty_1}({\bm k})/\varepsilon$ from $\mathrm{Im}[\t{\Sigma}_{\sigma}({\bm k};\varepsilon-i 0^+)]$ is vanishing with respect to $1/\varepsilon$ as $\vert\varepsilon\vert\to\infty$; if in addition Eq.~(\ref{ec82}) is satisfied, it can be similarly demonstrated that the deviation of $\Xi_{\sigma;\infty_1}({\bm k})/\varepsilon + \Xi_{\sigma;\infty_2}({\bm k})/\varepsilon^2$ from $\mathrm{Im}[\t{\Sigma}_{\sigma}({\bm k};\varepsilon-i 0^+)]$ is vanishing with respect to $1/\varepsilon^2$ as $\vert\varepsilon\vert\to\infty$.

The denominator of the expression on the RHS of Eq.~(\ref{ec83}) deviates considerably from unity for $\varepsilon'\in (0,u/\vert\varepsilon\vert)$, where $u\not=1$ is a finite positive constant for which no definite value needs to be specified. On account of Eqs.~(\ref{ec81}) and (\ref{ec82}) one should thus consider the behaviour of the integral
\begin{equation}
I {:=} \frac{-1}{\pi\varepsilon}
\mathscr{P}\!\!\int_{0}^{u/\vert\varepsilon\vert} \frac{{\rm
d}\varepsilon'}{{\varepsilon'}^2}\; \frac{o(\varepsilon')}{1 -
1/[\varepsilon^2 {\varepsilon'}^2]} \label{ec85}
\end{equation}
for $\vert\varepsilon\vert\to\infty$. It can be readily verified that $I = o(1/\varepsilon)$ for $\vert\varepsilon\vert\to\infty$. For instance, on replacing the $o(\varepsilon')$ in the numerator of the integrand in Eq.~(\ref{ec85}) by $O({\varepsilon'}^{1+\gamma})$, where $\gamma> 0$, one deduces that $I = O(1/\varepsilon^{1+\gamma})$. With reference to our remarks following Eq.~(\ref{ec82}), hereby is, under the conditions specified, the validity of the asymptotic series in Eq.~(\ref{ec76}) established.

\subsubsection{Concerning
$\mathrm{Re}[{\Sigma}_{\sigma}({\bf k};\varepsilon)]$}
\label{ssc5s3}

From Eq.~(\ref{ec73}) one obtains the formal asymptotic series
\begin{equation}
\mathrm{Re}[{\Sigma}_{\sigma}({\bm k};\varepsilon)] \sim
\Sigma_{\sigma}^{\Sc h\Sc f}({\bm k})+
\frac{\Pi_{\sigma;\infty_1}({\bm k})}{\varepsilon} + \dots\;\;\;
\mbox{\rm as}\;\;\; \vert\varepsilon\vert \to \infty, \label{ec86}
\end{equation}
where
\begin{equation}
\Pi_{\sigma;\infty_j}({\bm k}) \equiv \int_{-\infty}^{\infty}
\frac{{\rm d}\varepsilon}{\pi}\; \varepsilon^{j-1}\,
\mathrm{Im}[\t{\Sigma}_{\sigma}({\bm k};\varepsilon'-i 0^+)],\;\;\;
j\ge 1. \label{ec87}
\end{equation}
Whether $\Pi_{\sigma;\infty_j}({\bm k})$ is bounded is dependent on whether $\mathrm{Im}[\t{\Sigma}_{\sigma}({\bm k};\varepsilon-i 0^+)]$ decays sufficiently steeply for $\vert\varepsilon\vert\to\infty$. From Eq.~(\ref{ec76}) and Eq.~(\ref{ec79}), which depends on the existence of $\Xi_{\sigma;\infty_1}({\bm k})$, one concludes that existence of $\Xi_{\sigma;\infty_1}({\bm k})$ is \emph{sufficient} for the existence of $\Pi_{\sigma;\infty_1}({\bm k})$. Excluding the possibility of the leading-order asymptotic term in the series expansion of $\mathrm{Im}[\t{\Sigma}_{\sigma}({\bm k};\varepsilon-i 0^+)]$ for $\vert\varepsilon\vert\to\infty$ behaving like, for instance, $\sin(\varepsilon/\varepsilon_0)/\varepsilon$ (which possibility signals non-existence of $\Xi_{\sigma;\infty_1}({\bm k})$), one observes that existence of $\Xi_{\sigma;\infty_1}({\bm k})$ is also \emph{necessary} for the existence of $\Pi_{\sigma;\infty_1}({\bm k})$.

Assuming that $\Pi_{\sigma;\infty_1}({\bm k})$ exists, on account of the fact that for an interacting system $\mathrm{Im}[\t{\Sigma}_{\sigma}({\bm k};\varepsilon-i 0^+)] \not\equiv 0$ and, moreover, $\mathrm{Im}[\t{\Sigma}_{\sigma}({\bm k};\varepsilon-i 0^+)] \ge 0$, $\forall\varepsilon\in \mathds{R}$, it follows that
\begin{equation}
\Pi_{\sigma;\infty_1}({\bm k}) > 0,\;\;\; \forall {\bm k}, \sigma,
\label{ec88}
\end{equation}
which is to be compared with the result in Eq.~(\ref{ec79}).

\subsection{Asymptotic series of $\mathrm{Im}[\t{\Sigma}_{\sigma}({\bf k};\varepsilon \mp i 0^+)]$ and $\mathrm{Re}[\t{\Sigma}_{\sigma}({\bf k};\varepsilon \mp i 0^+)]$ revisited}
\label{ssc6}

Here we deal specifically with the cases where $G_{\sigma;\infty_j}({\bm k})$ is bounded for all finite values of $j$, so that $\Sigma_{\sigma;\infty_j}({\bm k})$ is also bounded for all finite values of $j$ (Sec.~\ref{ssc4}). As we have indicated earlier, this is the case for the uniform GSs of all systems defined on lattices (for instance, Bravais lattices, for which the corresponding $\mathrm{1BZ}$s are bounded), and the Fourier transform of the bare two-body interaction potential is bounded everywhere in the relevant ${\bm k}$ space. With reference to the second expression on the RHS of Eq.~(\ref{ec67}), we have earlier indicated that boundedness of $G_{\sigma;\infty_j}({\bm k})$ for all finite values of $j$ is indicative of $A_{\sigma}({\bm k};\varepsilon)$ possessing a bounded support or of $A_{\sigma}({\bm k};\varepsilon)$ decaying at least exponentially for $\vert\varepsilon\vert\to\infty$.

In the light of Eq.~(\ref{ec43}), boundedness of $G_{\sigma;\infty_j}({\bm k})$ for all finite values of $j$ is thus indicative of $\mathrm{Im}[\Sigma_{\sigma}({\bm k};\varepsilon)]$, and thus equivalently $\mathrm{Im}[\t{\Sigma}_{\sigma}({\bm k};\varepsilon \mp i 0^+)]$, possessing either a bounded support or decaying at least exponentially for $\vert\varepsilon\vert\to\infty$. With reference to Eq.~(\ref{ec76}), in such cases one must have (cf. Eq.~(\ref{ec79}))
\begin{equation}
\Xi_{\sigma;\infty_j}({\bm k}) = 0,\;\;\; \forall {\bm k},\sigma,
\label{ec89}
\end{equation}
for all finite values of $j$. Consequently,
\begin{equation}
\Pi_{\sigma;\infty_j}({\bm k}) \equiv \Sigma_{\sigma;\infty_j}({\bm
k}),\label{ec90}
\end{equation}
for all finite values of $j \ge 1$.

The result in Eq.~(\ref{ec90}) is remarkable in that it relates the moments integrals of $\mathrm{Im}[\t{\Sigma}_{\sigma}({\bm k};\varepsilon-i 0^+)]$ with those of the single-particle spectral function $A_{\sigma}({\bm k};\varepsilon)$. For instance, following Eqs.~(\ref{ec90}) and (\ref{ec72}) one must have
\begin{equation}
\Pi_{\sigma;\infty_1}({\bm k}) = \frac{1}{\hbar^3} \big( \hbar
G_{\sigma;\infty_3}({\bm k}) - G_{\sigma;\infty_2}^2({\bm
k})\big),\;\;\; \forall {\bm k}. \label{ec91}
\end{equation}
Making use of the expression in Eq.~(\ref{ec43}) for $A_{\sigma}({\bm k};\varepsilon)$ in calculating $G_{\sigma;\infty_j}({\bm k})$ according to the second expression in Eq.~(\ref{ec67}), one observes that Eq.~(\ref{ec91}) amounts to a self-consistency condition for $\mathrm{Im}[\t{\Sigma}_{\sigma}({\bm k};\varepsilon-i 0^+)]$; this function determines not only $\Pi_{\sigma;\infty_1}({\bm k})$ according to Eq.~(\ref{ec87}), but also $\mathrm{Re}[\Sigma_{\sigma}({\bm k};\varepsilon)]$, which is encountered in the above-mentioned expression for $A_{\sigma}({\bm k};\varepsilon)$, through the Kramers-Kr\"onig relation, Eq.~(\ref{ec73}). The condition in Eq.~(\ref{ec89}), in particular for $j=1,2$ (see Eqs.~(\ref{ec77}) and (\ref{ec78})), is seen further to restrict the behaviour of $\mathrm{Re}[\Sigma_{\sigma}({\bm k};\varepsilon)]$ as a function of $\varepsilon$ for any given ${\bm k}$. We note that Eq.~(\ref{ec91}) reduces to the identity $0=0$ in the limit of vanishing particle-particle interaction.

\subsubsection{Remarks}
\label{ssc6s1}

Some analysis of the asymptotic expression in Eq.~(\ref{ec86}) is in order. Making use of the identity (\S~62 in Ref.~\citen{TB65})
\begin{equation}
\frac{1}{1-x} = \sum_{j=0}^{n-1} x^j + \frac{x^{n}}{1-x},\;\;\; n\ge
1, \label{ec92}
\end{equation}
from Eq.~(\ref{ec73}) one obtains the \emph{exact} result
\begin{equation}
\mathrm{Re}[{\Sigma}_{\sigma}({\bm k};\varepsilon)] =
\Sigma_{\sigma}^{\Sc h\Sc f}({\bm k})+ \sum_{j=1}^{n}
\frac{\Pi_{\sigma;\infty_j}({\bm k})}{\varepsilon^j} +
R_{\sigma}^{(n)}({\bm k};\varepsilon), \label{ec93}
\end{equation}
where
\begin{equation}
R_{\sigma}^{(n)}({\bm k};\varepsilon) \equiv
\frac{1}{\varepsilon^{n+1}} \mathscr{P}\!\!\int_{-\infty}^{\infty}
\frac{{\rm d}\varepsilon'}{\pi}\; \frac{{\varepsilon'}^{n}\,
\mathrm{Im}[\t{\Sigma}_{\sigma}({\bm k};\varepsilon'-i 0^+)]}{1-
\varepsilon'/\varepsilon}. \label{ec94}
\end{equation}
It follows that so long as
\begin{equation}
\mathscr{P}\!\!\int_{-\infty}^{\infty} \frac{{\rm
d}\varepsilon'}{\pi}\; \frac{{\varepsilon'}^{n}\,
\mathrm{Im}[\t{\Sigma}_{\sigma}({\bm k};\varepsilon'-i 0^+)]}{1-
\varepsilon'/\varepsilon} = o(\varepsilon)\;\;\;\mbox{\rm for}\;\;\;
\vert\varepsilon\vert\to \infty, \label{ec95}
\end{equation}
on suppressing the $R_{\sigma}^{(n)}({\bm k};\varepsilon)$ on the RHS of Eq.~(\ref{ec93}) one obtains the $(n+1)$th-order Poincar\'e asymptotic series for $\mathrm{Re}[{\Sigma}_{\sigma}({\bm k};\varepsilon)]$ corresponding to $\vert\varepsilon\vert\to\infty$ in terms of the asymptotic sequence $\{1, 1/\varepsilon, 1/\varepsilon^2,\dots\}$. This is indeed the case for the function $\mathrm{Im}[\t{\Sigma}_{\sigma}({\bm k};\varepsilon'-i 0^+)]$ considered in Sec.~\ref{ssc6}.

To gain some insight concerning the function to be expected for the $o(\varepsilon)$ on the RHS of Eq.~(\ref{ec95}), it is instructive to replace the $\mathrm{Im}[\t{\Sigma}_{\sigma}({\bm k};\varepsilon'-i 0^+)]$ on the LHS of Eq.~(\ref{ec95}) by $f(\varepsilon)$ and express the resulting integral as follows
\begin{equation}
\mathscr{P}\!\!\int_{-\infty}^{\infty} \frac{{\rm
d}\varepsilon'}{\pi}\; \frac{{\varepsilon'}^{n}\, f(\varepsilon)}
{1-\varepsilon'/\varepsilon} = (-1)^n
\mathscr{P}\!\!\int_{0}^{\infty} \frac{{\rm d}\varepsilon'}{\pi}\;
\frac{{\varepsilon'}^{n}\, f(-\varepsilon')}
{1+\varepsilon'/\varepsilon}+\mathscr{P}\!\!\int_{0}^{\infty} \frac{{\rm
d}\varepsilon'}{\pi}\; \frac{{\varepsilon'}^{n}\, f(\varepsilon')}
{1-\varepsilon'/\varepsilon}. \label{ec96}
\end{equation}
For $f(\varepsilon) \equiv \mathrm{e}^{-\vert\varepsilon\vert}$ one has
\begin{equation}
\mathscr{P}\!\!\int_{0}^{\infty} \frac{{\rm d}\varepsilon'}{\pi}\;
\frac{{\varepsilon'}^{n}\, f(-\varepsilon')}
{1+\varepsilon'/\varepsilon} = \frac{n!}{\pi} \,\left\{
\begin{array}{ll}  \Lambda_n(\varepsilon), & \varepsilon> 0, \\ \\
\ol{\Lambda}_n(\varepsilon), &
\varepsilon< 0,\\
\end{array} \right. \label{ec97}
\end{equation}
\begin{equation}
\mathscr{P}\!\!\int_{0}^{\infty} \frac{{\rm d}\varepsilon'}{\pi}\;
\frac{{\varepsilon'}^{n}\, f(\varepsilon')}
{1-\varepsilon'/\varepsilon} = \frac{n!}{\pi} \,\left\{
\begin{array}{ll}  \ol{\Lambda}_n(-\varepsilon), &\varepsilon> 0, \\ \\
\Lambda_n(-\varepsilon), &\varepsilon< 0,\\
\end{array} \right. \label{ec98}
\end{equation}
where $\Lambda_s(z)$ and $\ol{\Lambda}_s(-z)$ are two of the four so-called `basic converging factors' introduced and analyzed by Dingle \cite{RBD58}; $\Lambda_s(z)$ is in fact $z/\Gamma(s+1)$ times the Stieltjes transform of $\varepsilon^s\, \mathrm{e}^{-\varepsilon}$ \cite{WGCB90}, and for real values of $s$ and $\varepsilon> 0$, $\ol{\Lambda}_s(-\varepsilon) = \mathrm{Re}[\Lambda_s(-\varepsilon)]$.

For $\vert z\vert\gg 1$ and $\vert z\vert\gg \vert s\vert$ Dingle \cite{RBD58} obtained that (Eqs.~(47) and (48) in Ref.~\citen{RBD58})
\begin{equation}
\Lambda_s(z) \sim 1 - \frac{s+1}{z} + \frac{(s+1) (s+2)}{z^2} -\dots,
\label{ec99}
\end{equation}
and for either $\vert z\vert\gg 1$ or $\vert s\vert\gg 1$, or both, that
\begin{equation}
\Lambda_s(z) \sim \frac{z}{s+z} \Big\{ 1 - \frac{z}{(s+z)^2} -
\frac{z (s-2z)}{(s+z)^4} - \dots \Big\}. \label{ec100}
\end{equation}
For real and positive values of $z$, the expansions corresponding to $\ol{\Lambda}_s(-z)$ are obtained from those in Eqs.~(\ref{ec99}) and (\ref{ec100}) by reversing herein the sign of $z$ \cite{RBD58}. For details concerning the cases where $z$ is near $s$, the reader is referred to Ref.~\citen{RBD58}, where it has been shown that, for instance, to order $1/s^3$ one has the exact result (Eq.~(56) in
Ref.~\citen{RBD58})
\begin{equation}
\Lambda_s(s) = \frac{1}{2} - \frac{1}{8 s} + \frac{1}{32 s^2} +
\frac{1}{128 s^3}. \label{ec101}
\end{equation}

If $\mathrm{Im}[\t{\Sigma}_{\sigma}({\bm k};\varepsilon-i 0^+)]$ were identical with $\mathrm{e}^{-\vert\varepsilon\vert}$, then the above details would suffice to construct both the superasymptotic and hyperasymptotic series \cite{BH90,JPB99} for $\mathrm{Re}[\Sigma_{\sigma}({\bm k};\varepsilon)]$ corresponding to large values of $\vert\varepsilon\vert$.

\subsection{Asymptotic series of $\t{\Sigma}_{\sigma}^{(\nu)}({\bf k};z)$ for $\vert z\vert\to\infty$}
\label{ssc7}

\begin{table}[t!]
\caption{The function $\t{\Sigma}_{\sigma}^{(\nu)}({\bm k};z)$, $\nu\ge 2$, contributes to \emph{all} $\Sigma_{\sigma;\infty_j}({\bm k})$ with $j=\nu-1, \nu, \nu+1, \dots$. In other words,  $\t{\Sigma}_{\sigma}^{(\nu)}({\bm k};z)$ contributes to \emph{all} functions in the second row directly beneath it and to its right, but \emph{never} to a function in this row to its left. The function $\t{\Sigma}_{\sigma}^{(\nu)}({\bm k};z)$ cannot contribute to $\Sigma_{\sigma;\infty_j}({\bm k})$, with $j\le\nu-2$, owing to the
following three facts: firstly, the \emph{explicit} dependence of $\t{\Sigma}_{\sigma}^{(\nu)}({\bm k};z)$ on $\lambda$ is of the form $\lambda^{\nu}$, where $\lambda$ denotes the coupling constant of interaction; secondly (Sec.~\protect\ref{ssc4}), $\Sigma_{\sigma;\infty_j}({\bm k})$, $j\ge 1$, is fully determined in terms of \protect\cite{BF02} $G_{\sigma;\infty_2}({\bm k})$, $\dots$, $G_{\sigma;\infty_{j+2}}({\bm k})$ (cf. Eq.~(\protect\ref{ec72})), and, thirdly, $G_{\sigma;\infty_k}({\bm k})$, $k\ge 2$, consists of a superposition of terms proportional to \protect\cite{BF02} $1$, $\lambda$, \dots, $\lambda^{k-1}$ (see the first equality in Eq.~(\protect\ref{ec67})).\vspace{5pt}}
\label{ta3}
\begin{center}
\begin{tabular}{llll}
\hline\hline
\vspace{+3pt}
$\t{\Sigma}_{\sigma}^{(2)}({\bm k};z)$ & $\t{\Sigma}_{\sigma}^{(3)}({\bm k};z)$ & $\t{\Sigma}_{\sigma}^{(4)}({\bm k};z)$ & \dots \\
$\Sigma_{\sigma;\infty_1}({\bm k})$  & $\Sigma_{\sigma;\infty_2}({\bm k})$ & $\Sigma_{\sigma;\infty_3}({\bm k})$ & \dots \\
\hline
\end{tabular}
\end{center}
\end{table}

It can be shown that \cite{BF02}, $\Sigma_{\sigma;\infty_j}({\bm k})$, $j\ge 1$, is fully determined by contributions arising from $\t{\Sigma}_{\sigma}^{(\nu)}({\bm k};z)$ for $\nu =2,\dots, j+1$ \cite{BF02}, where, as elsewhere in this paper, $\t{\Sigma}_{\sigma}^{(\nu)}({\bm k};z)$ denotes the total contributions of the $\nu$th-order skeleton self-energy diagrams expressed in terms of the \emph{bare} two-body interaction potential and the \emph{exact} single-particle Green functions $\{\t{G}_{\sigma'}({\bm k};z)\,\|\, \sigma'\}$. Consequently, the leading-order term in the asymptotic series expansion of $\t{\Sigma}_{\sigma}^{(\nu)}({\bm k};z)$ for $\vert z\vert\to\infty$ \emph{cannot} decay slower than $1/\vert z\vert^{\nu-1}$, $\nu\ge 1$, for if it did, $\t{\Sigma}_{\sigma}^{(\nu)}({\bm k};z)$ would contribute to $\Sigma_{\sigma;\infty_{\nu-2}}({\bm k})$ for $\nu\ge 2$ (Table \ref{ta3}). We thus arrive at the following leading-order asymptotic expression
\begin{equation}
\t{\Sigma}_{\sigma}^{(\nu)}({\bm k};z) \sim \frac{{\sf
S}_{\sigma}^{(\nu)}({\bm k})}{z^{\nu-1}}\;\;\;\mbox{\rm for}\;\;\;
\vert z\vert\to\infty\;\;\; (\nu\ge 2), \label{ec102}
\end{equation}
where ${\sf S}_{\sigma}^{(\nu)}({\bm k})$ is a well-defined real-valued function whose further specification is not necessary for the considerations of this paper. As regards $\t{\Sigma}_{\sigma}^{(1)}({\bm k};z)$, this function is independent of $z$ (see Eq.~(\ref{ec71})); one can thus declare Eq.~(\ref{ec102}) valid for $\nu\ge 1$ by identifying ${\sf S}_{\sigma}^{(1)}({\bm k})$ with $\Sigma_{\sigma}^{\Sc h\Sc f}({\bm k})$.

By expressing the asymptotic expression in Eq.~(\ref{ec102}) more completely as
\begin{equation}
\t{\Sigma}_{\sigma}^{(\nu)}({\bm k};z) \sim \frac{{\sf
S}_{\sigma;\infty_{\nu-1}}^{(\nu)}({\bm k})}{z^{\nu-1}} + \frac{{\sf
S}_{\sigma;\infty_{\nu}}^{(\nu)}({\bm k})}{z^{\nu}} + \frac{{\sf
S}_{\sigma;\infty_{\nu+1}}^{(\nu)}({\bm k})}{z^{\nu+1}} +\dots\;\;\;\mbox{\rm for}\;\;\;
\vert z\vert\to\infty, \label{ec103}
\end{equation}
one readily obtains that
\begin{equation}
\Sigma_{\sigma;\infty_j}({\bm k}) = \sum_{\nu=2}^{j+1} {\sf
S}_{\sigma;\infty_j}^{(\nu)}({\bm k}),\;\; j\in \mathds{N}. \label{ec104}
\end{equation}
This expression makes explicit the above-mentioned fact that $\Sigma_{\sigma;\infty_j}({\bm k})$, $j\ge 1$, is fully determined by contributions arising from $\t{\Sigma}_{\sigma}^{(\nu)}({\bm k};z)$ for $\nu =2,\dots, j+1$. Since existence of $\t{\Sigma}_{\sigma}^{(\nu)}({\bm k};z)$ is a \emph{necessary} condition for existence of ${\sf
S}_{\sigma;\infty_j}^{(\nu)}({\bm k})$, $j=\nu-1, \nu,\dots$ (this condition is not \emph{a priori} sufficient), Eq.~(\ref{ec104}) reveals that existence $\t{\Sigma}_{\sigma}^{(\nu)}({\bm k};z)$ for an arbitrary finite $\nu$ is also \emph{necessary} for existence $\Sigma_{\sigma;\infty_j}({\bm k})$ for an arbitrary finite $j$. In this connection, note that for any finite $j$, the sum on the RHS of Eq.~(\ref{ec104}) amounts to a finite series. As we have indicated in Sec.~\ref{ssc3}, the same conditions that are sufficient for $\t{\Sigma}_{\sigma}^{(\nu)}({\bm k};z)$ to exist for an arbitrary finite $\nu$, are also sufficient for $G_{\sigma;\infty_j}({\bm k})$, and therefore $\Sigma_{\sigma;\infty_j}({\bm k})$, to exist for an arbitrary finite $j$ \cite{BF02}.

In the most general case, $\t{\Sigma}_{\sigma}^{(\nu)}({\bm k};z)$ is likely to be unbounded for $\nu\ge \nu_{\star} = j_{\star}$, where $j_{\star}$ is the smallest $j$ for which $\Sigma_{\sigma;\infty_j}({\bm k})$ is unbounded (see the details following Eq.~(\ref{ec72})); the actual value of $j_{\star}$ depends, amongst other things, on the nature of the two-body interaction potential \emph{and} the dimensionality $d$ of space. In such a case, the very expression in Eq.~(\ref{ec102}) is meaningless for $\nu\ge \nu_{\star}$. Technically, and with the exception of the leading-order terms $\Sigma_{\sigma;\infty_j}({\bm k})/z^j$, $j=0,1,\dots,j_{\star}-1$, each term in the asymptotic series expansion of $\t{\Sigma}_{\sigma}({\bm k};z)$ corresponding to $\vert z\vert\to\infty$ arises from contributions of infinite number of terms, each of which (whether bounded or unbounded) corresponds to some $\t{\Sigma}_{\sigma}^{(\nu)}({\bm k};z)$, $\nu\ge 2$.

\section{The thermal Green function at low temperatures}
\label{sf}

In this appendix we establish the validity of the result in Eq.~(\ref{e29}). We further determine the subsequent two leading-order terms in the low-temperature asymptotic series expansion of $\mathscr{G}_{\sigma}({\bm k};z)$ for the cases where the underlying $N$-particle GSs are insulating.\footnote{As will become evident, these two terms constitute the second leading-order contribution to $\mathscr{G}_{\sigma}({\bm k};z)$ for $\beta\to\infty$ when $\mu = (\mu_{N}^- + \mu_{N}^+)/2$.} These terms are essential for calculating the value of the chemical potential $\mu(\beta,N,V)$ in the zero-temperature \emph{limit} (see Sec.~\ref{ss61}); $\mu(\beta,N,V)$ being the solution of Eq.~(\ref{e23}), one needs to calculate $\b{N}$ for law temperatures as a function of $\mu$, for which one may employ the expression in Eq.~(\ref{e31}), thus necessitating calculation of $\mathscr{G}_{\sigma}({\bm k};z)$ at low temperatures. As for the cases where the underlying $N$-particle GSs are metallic, since in these cases the zero-temperature value of the chemical potential corresponding to $N$ particles is unique, the zero-temperature \emph{limit} of $\mu(\beta,N,V)$ necessarily coincides with this value and therefore need not be explicitly determined (see Sec.~\ref{ss23}). Both for this and for the reason of keeping away from lengthy expressions, we shall not present the explicit expression for the next-to-leading term in the low-temperature asymptotic series expansion of $\mathscr{G}_{\sigma}({\bm k};z)$ for the cases where the underlying $N$-particle GSs are metallic. From the details that we present in this appendix one will be able to infer that, under some well-specified conditions, this term diminishes like $1/\beta$ for finite systems,\footnote{There is a fundamental difficulty in ascertaining that a finite system is metallic. Consequently, the above-mentioned decay of the form $1/\beta$ is applicable so long as $\beta <\beta_{\rm c}$, where $1/\beta_{\rm c}$ is of the order of $\mu_{N}^+ -\mu_{N}^-$; for $\beta> \beta_{\rm c}$, the last-mentioned power-law decay crosses over into an exponential decay.} and more rapidly than $1/\beta$ for macroscopic systems, as $\beta\to\infty$.

As will become evident, calculation of the asymptotic series expansion of the $\mathscr{G}_{\sigma}({\bm k};z)$ corresponding to finite systems is relatively more laborious than that of the $\mathscr{G}_{\sigma}({\bm k};z)$ corresponding to macroscopic systems. Explicitly, with $V$ denoting the volume occupied by the system under investigation, calculations are considerably simplified on first taking the limit $V\to\infty$ and subsequently considering to increase $\beta$ towards $\infty$; this, as we shall see, is due to the fact that effecting the thermodynamic limit for a finite value of $\beta$ leads to suppression to zero of a multiplicity of contributions to $\mathscr{G}_{\sigma}({\bm k};z)$ that all decay according to the same power of $1/\beta$ for $\beta\to\infty$. In this connection, it is relevant to recall that the well-known problem of the zero-temperature Brueckner-Goldstone perturbation series, first identified by Kohn and Luttinger \cite{KL60}, arises from taking the limit $\beta\to \infty$ first and the limit $V\to \infty$ afterwards; this problem disappears on taking the latter limits in the opposite order.

We note in passing that the above-mentioned discrepancy between the results corresponding to $\lim_{\beta\to\infty} \lim_{V\to\infty}$ and $\lim_{V\to\infty} \lim_{\beta\to\infty}$ is not specific to many-body problems or to $\mathscr{G}_{\sigma}({\bm k};z)$; this discrepancy amounts to a manifestation of the mathematical fact, encountered in this paper for a number of times, that in general various \emph{repeated} limits of multi-variable functions need not be equal (\S\S~302-306 in Ref.~\citen{EWH27}). As regards asymptotic series expansions, for illustration one may consider the Bessel function of the first kind $J_{\nu}(z)$ \cite{AS72}. Prior to calculating a finite-order asymptotic series expansion for $J_{\nu}(z)$ corresponding to $\vert z\vert\to\infty$, one must specify the relationship between $z$ and $\nu$, that is whether $\vert z\vert/\vert\nu\vert \ll 1$, $\vert z\vert/\vert\nu\vert \approx 1$ or $\vert z\vert/\vert\nu\vert \gg 1$ \cite{AS72,GNW22}.

\subsection{Preliminaries}
\label{ssf1}

Let
\begin{equation}
\wh{K} {:=} \wh{H} - \mu \wh{N}, \label{ef1}
\end{equation}
where $\wh{H}$ is the interacting Hamiltonian and $\wh{N}$ the number operator. Let further $\{ \vert\nu\rangle \}$ denote the complete set of eigenstates of $\wh{K}$, spanning the Fock space of the system under consideration. Since $\wh{H}$ and $\wh{N}$ commute, we assume that each element of $\{ \vert\nu\rangle \}$ is a simultaneous eigenstate of $\wh{H}$ and $\wh{N}$, so that in
\begin{equation}
\wh{K} \vert\nu\rangle = K_{\nu} \vert\nu\rangle \label{ef2}
\end{equation}
one has
\begin{equation}
K_{\nu} = E_{\nu} - \mu N_{\nu}, \label{ef3}
\end{equation}
in which $E_{\nu}$ and $N_{\nu}$ are eigenvalues of $\wh{H}$ and $\wh{N}$ respectively.

It will prove convenient (amongst other things, for making contact with some relevant details discussed in appendix \ref{sc}) to make the following identification:
\begin{equation}
\nu = (N;s), \label{ef4}
\end{equation}
in which $s$, a compound index (appendix \ref{sc}), is defined as representing all indices of which $\nu$ consists excluding $N$. Consequently, in this appendix we shall interchangeably employ the following notations:
\begin{equation}
\vert\nu\rangle \equiv \vert\Psi_{N;s}\rangle,\;\;\;
E_{\nu} \equiv E_{N;s} \equiv E_s(N).
\label{ef5}
\end{equation}
Similarly, $\vert\nu'\rangle \equiv \vert\Psi_{N';s'}\rangle$, etc. As in appendix \ref{sc}, in this appendix we shall signify the $s$ specific to GSs by $s=0$. This indicates our implicit assumption in this appendix that the GS of $\wh{H}$ is up to a trivial phase factor unique.

\subsection{The leading-order asymptotic term}
\label{ssf2}

Expressing the thermal single-particle spectral function $\mathscr{A}_{\sigma}({\bm k};\varepsilon)$, Eq.~(\ref{e32}), as
\begin{equation}
\mathscr{A}_{\sigma}({\bm k};\varepsilon) \equiv \mathscr{A}_{\sigma}^+({\bm k};\varepsilon) + \mathscr{A}_{\sigma}^-({\bm k};\varepsilon), \label{ef6}
\end{equation}
one has the following spectral representations (see Eq.~(5.45) in Ref.~\citen{NO98}, Eq.~(31.34) in Ref.~\citen{FW03}) \cite{Note16}
\begin{equation}
\mathscr{A}_{\sigma}^+({\bm k};\varepsilon) = \frac{\hbar}{\mathcal{Z}_{\Sc g}} \sum_{\nu,\nu'} \left|\langle\nu\vert \h{c}_{{\bm k};\sigma}\vert \nu'\rangle\right|^2 \mathrm{e}^{-\beta K_{\nu}}\, \delta(\varepsilon + K_{\nu} - K_{\nu'} + \mu), \label{ef7}
\end{equation}
\begin{equation}
\mathscr{A}_{\sigma}^-({\bm k};\varepsilon) = \frac{\hbar}{\mathcal{Z}_{\Sc g}} \sum_{\nu,\nu'} \left|\langle\nu\vert \h{c}_{{\bm k};\sigma}\vert \nu'\rangle\right|^2 \mathrm{e}^{-\beta K_{\nu'}}\, \delta(\varepsilon + K_{\nu} - K_{\nu'} + \mu), \label{ef8}
\end{equation}
where
\begin{equation}
\mathcal{Z}_{\Sc g} {:=} \mathrm{Tr}\big(\mathrm{e}^{-\beta \wh{K}}\big)
\equiv \sum_{\nu} \mathrm{e}^{-\beta K_{\nu}} \label{ef9}
\end{equation}
is the grand partition function. Owing to $\langle\nu\vert\h{c}_{{\bm k};\sigma}\vert\nu'\rangle$, in the above spectral representations one has
\begin{equation}
N_{\nu'} = N_{\nu} + 1, \label{ef10}
\end{equation}
so that, in the same representations (thus not outside these where Eq.~(\ref{ef10}) does not need to apply),
\begin{equation}
K_{\nu}-K_{\nu'} + \mu = E_{\nu} - E_{\nu'} \equiv E_{N;s} - E_{N+1;s'}. \label{ef11}
\end{equation}
It follows that the dependence of $\mathscr{A}_{\sigma}({\bm k};\varepsilon)$ (and thus of $\mathscr{G}_{\sigma}({\bm k};z)$) on $\mu$ is solely through the dependence of $\mathcal{Z}_{\Sc g}$ and $\mathrm{e}^{-\beta K_{\nu}}$ on $\mu$ .

We proceed by first introducing a fundamental result; later in this appendix we shall present the basic principles on which this result rests. One has (for $\mu_{N}^{\mp}$ see appendix \ref{sc})
\begin{equation}
\frac{1}{\mathcal{Z}_{\Sc g}} \sum_{\nu} f(\nu)\, \mathrm{e}^{-\beta K_{\nu}} \sim \b{f}(N,0)\;\;\;\mbox{\rm for}\;\;\;
\mu \in (\mu_{N}^-,\mu_{N}^+)\;\;\;\mbox{\rm as}\;\;\;
\beta\to\infty, \label{ef12}
\end{equation}
where (cf. Eq.~(\ref{ef4}))
\begin{equation}
f(\nu) \equiv \b{f}(N,s) \label{ef13}
\end{equation}
is a function, formally supposed to be `continuous' at $\nu = (N,0)$.
The expression presented in Eq.~(\ref{ef12}) amounts to a generalisation of a result deduced by the Laplace method concerning Laplace integrals (Ch. 5 in Ref.~\citen{ETC65}, \S~6.4 in Ref.~\citen{BO99}).

Following Eq.~(\ref{ef12}) one obtains that
\begin{equation}
\lim_{\beta\to\infty} \mathscr{A}_{\sigma}^{\pm}({\bm k};\varepsilon)
= \hbar \sum_{s} \left|{\sf f}_{s;\sigma}^{\pm}({\bm k})\right|^2 \delta(\varepsilon - \varepsilon_{s;\sigma}^{\pm}), \label{ef14}
\end{equation}
where ${\sf f}_{s;\sigma}^{\pm}({\bm k})$ and $\varepsilon_{s;\sigma}^{\pm}$ are defined in Eqs.~(\ref{ec2}) and (\ref{ec3}) respectively. In arriving at the expression in Eq.~(\ref{ef14}), we have used (Eq.~(\ref{ef10}))
\begin{equation}
\langle\Psi_{N_0;0}\vert \h{c}_{{\bm k};\sigma}\vert\Psi_{N';s'}\rangle
= \langle\Psi_{N_0;0}\vert \h{c}_{{\bm k};\sigma}\vert\Psi_{N_0+1;s'}\rangle\, \delta_{N_0+1,N'}, \label{ef15}
\end{equation}
\begin{equation}
\langle\Psi_{N;s}\vert \h{c}_{{\bm k};\sigma}\vert\Psi_{N_0;0}\rangle
= \langle\Psi_{N_0-1;s}\vert \h{c}_{{\bm k};\sigma}\vert\Psi_{N_0;0}\rangle\, \delta_{N,N_0-1}. \label{ef16}
\end{equation}
With reference to Eq.~(\ref{ec41}), hereby is the formal proof of the result in Eq.~(\ref{e33}) complete. Similarly for the results in Eqs.~(\ref{e29}) and (\ref{e30}).

For completeness, we point out that (cf. Eq.~(\ref{ec47}))
\begin{equation}
\mathscr{G}_{\sigma}({\bm k};z) = \int_{-\infty}^{\infty} {\rm d}\varepsilon'\; \frac{\mathscr{A}_{\sigma}({\bm k};\varepsilon')}{z-\varepsilon'}, \label{ef17}
\end{equation}
and further that (cf. Eq.~(\ref{ec39}))
\begin{equation}
\frac{1}{\hbar}\int_{-\infty}^{\infty} {\rm d}\varepsilon\; \mathscr{A}_{\sigma}({\bm k};\varepsilon) = \frac{[\h{c}_{{\bm k};\sigma},\h{c}_{{\bm k};\sigma}^{\dag}]_+}{\mathcal{Z}_{\Sc g}} \sum_{\nu} \mathrm{e}^{-\beta K_{\nu}} \equiv 1,\;\; \forall {\bm k},\sigma, \label{ef18}
\end{equation}
which applies for \emph{all} $\beta$. The second equality in Eq.~(\ref{ef18}) follows from the fact that $\h{c}_{{\bm k};\sigma}$ and $\h{c}_{{\bm k};\sigma}^{\dag}$ are canonical fermion operators.\footnote{For some consequences of employing non-canonical fermion operators, such as encountered in the $t$-$J$ Hamiltonian, see Appendix A in Ref.~\protect\citen{BF03}.}

\subsection{Technicalities}
\label{ssf3}

For the purpose of establishing the fundamental result in Eq.~(\ref{ef12}), here we discuss some general technical details.

\subsubsection{Some basic results}
\label{ssf3s1}

First we express the LHS of Eq.~(\ref{ef12}) as
\begin{equation}
\frac{1}{\mathcal{Z}_{\Sc g}} \sum_{\nu} f(\nu)\, \mathrm{e}^{-\beta K_{\nu}} = \frac{1}{\mathcal{Z}_{\Sc g}} \sum_{N=0}^{\infty} \mathrm{e}^{\beta \mu N} \sum_{s} \b{f}(N,s)\, \mathrm{e}^{-\beta E_s(N)}. \label{ef19}
\end{equation}
We shall disregard the question whether or not the multiple sum $\sum_{\nu}$ and the repeated sums $\sum_{N} \sum_s$ and $\sum_s \sum_{N}$ are invariably equivalent (Sec.~\ref{ss53s15}). For the $\b{f}(N,s)$ associated with $\mathscr{G}_{\sigma}({\bm k};z)$ one has
\begin{equation}
\b{f}(N,s) = \hbar \sum_{s'} \Big\{ \frac{\left|\langle\Psi_{N-1;s'}\vert \h{c}_{{\bm k};\sigma}\vert\Psi_{N;s}\rangle \right|^2}{z - (E_{N,s} - E_{N-1,s'})} + \frac{\left|\langle\Psi_{N;s}\vert \h{c}_{{\bm k};\sigma}\vert\Psi_{N+1;s'}\rangle \right|^2}{z - (E_{N+1,s'}-E_{N,s})} \Big\}. \label{ef20}
\end{equation}
Evidently (cf. Eq.~(\ref{ec1})),
\begin{equation}
\b{f}(N,0) \equiv \t{G}_{\sigma}^-({\bm k};z) + \t{G}_{\sigma}^+({\bm k};z) \equiv \t{G}_{\sigma}({\bm k};z),
\label{ef21}
\end{equation}
the zero-temperature Green function corresponding to the $N$-particle GS of the system under investigation. Combining this result with the leading-order asymptotic result in Eq.~(\ref{ef12}), one arrives at the result in Eq.~(\ref{e29}). For our later considerations it is relevant to appreciate that the $\b{f}(N,s)$ in Eq.~(\ref{ef20}) does not scale with the size of the system.

In Sec.~\ref{ss23} we indicated that $\t{G}_{\sigma}({\bm k};z)$ being the zero-temperature \emph{limit} of $\mathscr{G}_{\sigma}({\bm k};z)$, it implicitly depends on the value of the thermodynamic variable $\mu$. Evidently, the result in Eq.~(\ref{ef12}), with the $\b{f}(N,0)$ herein equal to the $\b{f}(N,0)$ in Eq.~(\ref{ef21}), would not have been obtained if $\mu \not\in (\mu_{N}^-,\mu_{N}^+)$. Given the fact that for metallic $N$-particle GSs the width of the latter interval is microscopically small, of the order of $1/N$ (appendix \ref{sc}), it follows that for these GSs $\t{G}_{\sigma}({\bm k};z)$ indeed implicitly depends on $\mu$. Such dependence is however not apparent as regards insulating $N$-particle GSs for which $\mu_{N}^+ -\mu_N^-$ is non-vanishing and finite, whereby irrespective of the value of $\mu$, assumed however to satisfy $\mu \in (\mu_{N}^-,\mu_{N}^+)$, one arrives at the same function $\b{f}(N,0)$; from this perspective, for these GSs $\t{G}_{\sigma}({\bm k};z)$ exhibits no implicit dependence on $\mu$. This aspect has its root in the fact that the expression in Eq.~(\ref{ef12}) amounts to a strict equality only for $\beta=\infty$; for any $\beta<\infty$, no matter how large $\beta$ may be, this expression differs from an exact equality. The expressions that we shall deduce in Sec.~\ref{ssf5} for the sub-leading terms in the asymptotic series expansion of $\mathscr{G}_{\sigma}({\bm k};z)$ corresponding to $\beta\to\infty$, make the dependence of $\mathscr{G}_{\sigma}({\bm k};z)$ on $\mu$ in the zero-temperature \emph{limit} apparent (see Eq.~(\ref{ef140}) below). Although these sub-leading terms are exponentially vanishing for $\beta\to\infty$, the considerations in Sec.~\ref{ss61} reveal that they nonetheless have considerable influence on the behaviour of $\mu_{\beta} \equiv \mu(\beta,N,V)$ as a function of $\beta$ for $\beta\to\infty$.

Let now $\mathfrak{S}(N)$ denote the set of all $s$ characterising all $N$-particle eigenstates $\vert\Psi_{N;s}\rangle$ of $\wh{H}$. One readily recognises that for an interacting system the measure of the set $\mathfrak{S}(N)$ is unknown (see later however). Nonetheless, $\mathfrak{S}(N)$ being a measurable set, by means of the measure function $\mathfrak{m}$, one can express the sum with respect to $s$ on the RHS of Eq.~(\ref{ef19}) as a Lebesgue integral over $\mathfrak{S}(N)$ (Ch.~VII in Ref.~\citen{EWH27}, Ch. 6, \S~7 in Ref.~\citen{HS91}). On doing so, one can establish that the conventional Laplace method concerning Laplace integrals (Ch. 5 in Ref.~\citen{ETC65}, \S~6.4 in Ref.~\citen{BO99}), which are traditionally Riemann integrals, applies and, assuming that $\b{f}(N,s)$ is continuous at $s=0$,\footnote{The measure function $\mathfrak{m}$ induces a norm inside $\mathfrak{S}(N)$, with the aid of which the notion of continuity can be defined.} one thus arrives at the leading-order result
\begin{equation}
\sum_{s} \b{f}(N,s)\, \mathrm{e}^{-\beta E_s(N)} \sim \b{f}(N,0)  \sum_{s} \mathrm{e}^{-\beta E_s(N)} \equiv \b{f}(N,0)\, \mathrm{e}^{- \beta F(N)}\;\;\;\mbox{\rm as}\;\;\; \beta\to\infty, \label{ef22}
\end{equation}
where
\begin{equation}
F(N) \equiv F(\beta,N,V) \label{ef23}
\end{equation}
is the Helmholtz free energy \cite{KH87} in the canonical ensemble of $N$ particles confined to volume $V$. The result in Eq.~(\ref{ef22}) is specific to the cases where the $N$-particle GSs are unique; in the cases where there are a countable number of degenerate GSs, the result in Eq.~(\ref{ef22}) is easily generalised by replacing the $\b{f}(N,0)$ on the RHS by a superposition of the $\b{f}(N,0_j)$ corresponding to all GSs of $\wh{H}$.

As we shall see in Sec.~\ref{ssf3s2}, for in particular macroscopic systems, a specific measure of $\mathfrak{S}(N)$ determines the entropy $S(N,E)$ in the $N$-particle micro-canonical ensemble corresponding to energy $E$. With $E(N)$ denoting the free energy of the $N$-particle canonical ensemble of the system under consideration,
\begin{equation}
S(N) \equiv S(N,E(N)) \label{ef24}
\end{equation}
amounts to the entropy in this ensemble. Since $S(N)$ is extensive, it follows that $S(N,E)$ is also extensive for $E$ in the vicinity of $E(N)$. For reasons that will become evident later in this appendix, this observation signifies the fact that even though $\beta$ may be large on the absolute scale, so long as $E(N)\equiv E(N,\beta)$ is not sufficiently close to the $N$-particle GS energy $E_0(N)$, the result in Eq.~(\ref{ef22}) may not be sufficiently accurate; we emphasise however that Eq.~(\ref{ef22}) is exact for $\beta$ in the limit $\beta=\infty$.

Below we shall demonstrate that in the cases where
\begin{equation}
\b{f}(N,s) \equiv \varphi(N,E_s(N)), \label{ef25}
\end{equation}
in which $\varphi(N,E)$ is a well-defined function which is continuous at $E=E(N)$, the low-temperature asymptotic expression
\begin{equation}
\sum_{s} \b{f}(N,s)\, \mathrm{e}^{-\beta E_s(N)}\sim \b{f}(N,s(N))\, \mathrm{e}^{- \beta F(N)}\;\;\mbox{\rm as}\;\; \beta\to\infty, \label{ef26}
\end{equation}
is superior to that in Eq.~(\ref{ef22}). Here
\begin{equation}
s(N) \equiv s(N,\beta) \label{ef27}
\end{equation}
is the index $s$ corresponding to the $N$-particle eigenstate $\vert\Psi_{N;s}\rangle$ of $\wh{H}$ for which, at a given $\beta$, one has
\begin{equation}
\left. E_s(N)\right|_{s=s(N)} = E(N); \label{ef28}
\end{equation}
as will become evident, for our considerations it is immaterial whether the equality in Eq.~(\ref{ef28}) is exactly satisfied or is merely satisfied to leading order in $1/\beta$ for $\beta\to\infty$. Should the ansatz in Eq.~(\ref{ef25}) be exact, our analysis will further reveal that
\begin{equation}
\lim_{N\to\infty} \frac{1}{\b{f}(N,s(N))} \frac{\sum_{s} \b{f}(N,s)\, \mathrm{e}^{-\beta E_s(N)}}{\sum_{s} \mathrm{e}^{-\beta E_s(N)}} = 1\;\;\;\mbox{\rm for}\;\;\;\beta>0, \label{ef29}
\end{equation}
where, physically, $N\to\infty$ corresponds to a finite value of $n \equiv N/V$. We have expressed the result in Eq.~(\ref{ef29}) in a different form than that in Eq.~(\ref{ef26}) so as to avoid the meaningless identity $\infty=\infty$ in the cases where $\b{f}(N,s(N))$ diverges for $N\to\infty$.

On identifying $\b{f}(N,s)$ with $1$, from Eq.~(\ref{ef22}) one obtains that
\begin{equation}
\mathcal{Z}_{\Sc g} = \sum_{N=0}^{\infty} \mathrm{e}^{-\beta (F(N) -\mu N)},\;\;\; \forall\beta, \label{ef30}
\end{equation}
where the equality sign, rather than $\sim$, is a consequence of $1$ not depending on $s$. Consequently, for $\beta\to\infty$ one has
\begin{equation}
\frac{1}{\mathcal{Z}_{\Sc g}} \sum_{\nu} f(\nu)\, \mathrm{e}^{-\beta K_{\nu}} \sim \frac{\sum_{N=0}^{\infty} \b{f}(N,0)\, \mathrm{e}^{-\beta (F(N) -\mu N)}}{\sum_{N=0}^{\infty} \mathrm{e}^{-\beta (F(N) -\mu N)}}. \label{ef31}
\end{equation}
Similarly, from Eq.~(\ref{ef26}) one obtains that
\begin{equation}
\frac{1}{\mathcal{Z}_{\Sc g}} \sum_{\nu} f(\nu)\, \mathrm{e}^{-\beta K_{\nu}} \sim \frac{\sum_{N=0}^{\infty} \b{f}(N,s(N))\, \mathrm{e}^{-\beta (F(N) -\mu N)}}{\sum_{N=0}^{\infty} \mathrm{e}^{-\beta (F(N) -\mu N)}}. \label{ef32}
\end{equation}
For the reason indicated above, the expression in Eq.~(\ref{ef32}) is more accurate than that in Eq.~(\ref{ef31}). As in the case of Eq.~(\ref{ef26}), which in the thermodynamic limit reduces to the equality in Eq.~(\ref{ef29}) for all $\beta>0$, we shall demonstrate that the asymptotic expression in Eq.~(\ref{ef32}) turns into an equality, valid for all $\beta>0$, in the same limit.

\subsubsection{Details}
\label{ssf3s2}

To make progress, we need to rely on some knowledge concerning the dependence of $\b{f}(N,s)$ on $s$. Therefore, in what follows we shall employ the ansatz in Eq.~(\ref{ef25}), where $\varphi(N,x)$ is a function which we assume to be continuous at $x=E(N)$, the internal energy of the $N$-particle canonical ensemble of the system under investigation (see later). Evidently, in calculating $\mathcal{Z}_{\Sc g}$ one deals with the specific case where $\varphi(N,x)\equiv 1$, $\forall N, x$.

For the particular case of $\mathscr{G}_{\sigma}({\bm k};z)$, the underlying function $\b{f}(N,s)$ is that presented in Eq.~(\ref{ef20}) whose dependence on $s$ is indeed in part through its explicit dependence on $E_s(N)$. Nonetheless, for the ansatz in Eq.~(\ref{ef25}) to amount to an exact statement in regard to the function $\b{f}(N,s)$ in Eq.~(\ref{ef20}), one will have to demonstrate that the dependence on $s$ of the matrix elements in Eq.~(\ref{ef20}), involving $\vert\Psi_{N;s}\rangle$, is mediated through $E_s(N)$. Although this possibility seems too remote, in anticipation of what follows we point out that insofar as our specific considerations in this appendix are concerned, this possibility needs to apply only for those $N$-particle eigenstates $\vert\Psi_{N;s}\rangle$ of $\wh{H}$ whose energies $E_s(N)$ are located in a small neighbourhood of $E(N)$, which to leading-order in $1/\beta$ is equal to the GS energy $E_0(N)$. In other words, from the relatively narrow perspective of our present considerations, it is wholly immaterial if the ansatz in Eq.~(\ref{ef25}) turns out to be inadequate for those $s$ whose corresponding $E_s(N)$ are outside an arbitrary small (but non-vanishing) closed interval surrounding $E(N)$. The situation at hand may be likened with what one encounters in the phenomenological theory of Landau (see, e.g., Ref.~\citen{PN66}; see also Sec.~\ref{ss55s1}), where the low-lying exited states are described in terms of the distribution $\delta n_{\bm k}$, Eq.~(\ref{e154}), of quasi-particles, rather than in terms of the deviation of the compound index $s$ from the compound index $0$ corresponding to the underlying GS; in the present case, the deviation of $E_s(N)$ from $E_0(N)$, as opposed to the deviation of $s$ from $0$, plays a similar role as the quasi-particle distribution function $\delta n_{\bm k}$ in the Landau theory.

In view of the assumption in Eq.~(\ref{ef25}), we introduce the many-body density-of-states function
\begin{equation}
\omega(N,E) {:=} \sum_s  \delta\big(E-E_s(N)\big). \label{ef33}
\end{equation}
With the entropy in the micro-canonical ensemble defined according to \cite{KH87,CL00}
\begin{equation}
S(N,E) \equiv k_{\Sc b} \ln\big(\omega(N,E)\, \Delta E\big), \label{ef34}
\end{equation}
where $\Delta E >0$ (strictly positive, however supposed to satisfy $\vert \Delta E/E\vert \ll 1$) is the amount of uncertainty in the energy $E$ of the micro-canonical ensemble under consideration, one has
\begin{equation}
\sum_{s} \b{f}(N,s)\, \mathrm{e}^{-\beta E_s(N)} = \frac{1}{\Delta E}\int {\rm d}E\; \varphi(N,E)\, \mathrm{e}^{-\beta \psi(N,E)}, \label{ef35}
\end{equation}
where
\begin{equation}
\psi(N,E) {:=} E - \frac{1}{k_{\Sc b} \beta}\, S(N,E). \label{ef36}
\end{equation}

For the following considerations it will be crucial that we can consider $\omega(N,E)$, and therefore $S(N,E)$, as a function which is at least twice differentiable with respect to $E$ in a finite neighbourhood of the above-mentioned internal energy $E(N)$. This requirement implies that $\{ E_s(N)\}$ must consist of a dense subset at least in a finite neighborhood of $E(N)$. This condition cannot be ascertained to be valid in general; it particular, it is highly unlikely to be valid in the cases of isolated small quantum systems. Within the theoretical framework pertaining to canonical ensembles (\S~7.1 in Ref.~\citen{KH87}), where each member of a canonical ensemble of $N$ particles stands in thermal contact with a heat bath, of temperature $T$ and consisting of $\mathcal{N}$ particles, where $\mathcal{N}\gg N$ (ideally, $\mathcal{N}=\infty$), one may reasonably expect that in particular interaction of the particles in the system with those in the heat bath is capable of causing $\{ E_s(N) \}$ to consist of a dense subset in at least a finite neighbourhood of $E(N)$.

The problem just described, signals one of the main theoretical difficulties concerning treatment of finite quantum systems in any framework, such as the present one, where one takes no explicit account of the full range of the physical implications of a heat bath and merely considers temperature as a given parameter. To bypass the mathematical problems arising from this shortcoming, one may replace the $\delta$ function in Eq.~(\ref{ef33}), which is a distribution function, by a conventional function of which the $\delta$ function is the limit; one could for instance consider the function $(\eta/\pi)/(x^2 +\eta^2)$ whose limit for $\eta\downarrow 0$ is $\delta(x)$ (see Ref.~\citen{Note17}). By doing so, the limit corresponding to the realisation of $\delta(x)$, such as $\eta\downarrow 0$ in $(\eta/\pi)/(x^2 +\eta^2)$, should be effected as the last step of the calculations. One may also argue that the uncertainty $\Delta E$ in the energy of the micro-canonical ensemble (see Eq.~(\ref{ef34})) should imply that $\eta$ should retain a non-vanishing value. \emph{In the following we shall assume that $\omega(N,E)$ is a sufficiently smooth function of $E$ in a finite neighbourhood of $E(N)$, bearing in mind the above-mentioned limitation of this assumption in the cases where $N$ is finite.} In Sec.~\ref{ssf3s3} we shall briefly deal with the cases where the underlying $N$-particle GSs are insulating; for sufficiently low temperatures, in these cases $E(N)$ is up to an exponentially small correction equal to $E_0(N)$, the energy of the $N$-particle GS of $\wh{H}$, which is separated from the energy of the lowest-lying $N$-particle excited states of $\wh{H}$ by a finite amount.

The integral in Eq.~(\ref{ef35}) is of the Laplace type and the asymptotic series expansion of this integral for $\beta\to\infty$ can be determined by means of the Laplace method \cite{ETC65,BO99}. An aspect that is fundamental to the statistical mechanics of macroscopic systems is that for these systems both $E(N)$ and $S(N)$, Eq.~(\ref{ef24}), are macroscopic quantities (both quantities are extensive), so that for $E$ in the neighbourhood of $E(N)$ the function $\psi(N,E)$ is macroscopically large. It follows that for such systems the exponent of the exponential function on the RHS of Eq.~(\ref{ef35}) is large for \emph{any} $\beta>0$. Note that since $S(N)$ is an extensive quantity, from the expressions in Eqs.~(\ref{ef33}) and (\ref{ef34}) it follows that $\omega(N,E)$ diverges like some power of $\mathrm{e}^{N}$ for $E$ in the vicinity of $E(N)$ as $N\to\infty$.

Denoting the location of the global minimum of $\psi(N,E)$ along the $E$ axis by $E(N)$, and assuming that $\psi(N,E)$ is at least once continuously differentiable at $E=E(N)$, the value of $E(N)$ is obtained from \cite{KH87}
\begin{equation}
\left.\frac{{\rm d}\psi(N,E)}{{\rm d}E}\right|_{E=E(N)} = 0 \iff \left.\frac{{\rm d} S(N,E)}{{\rm d}E}\right|_{E=E(N)} = \frac{1}{T}, \label{ef37}
\end{equation}
which implies that $E(N)$ is the internal energy of the canonical ensemble of $N$ particles. Assuming that $\psi(N,E)$ is at last twice continuously differentiable at $E=E(N)$, expanding $\psi(N,E)$ around $E=E(N)$ to second order in $E-E(N)$, making use of the fact that
\begin{equation}
\frac{\partial^2 S(N,E)}{\partial E^2} = \frac{-1}{C_{\Sc v} T^2}, \label{ef38}
\end{equation}
where
\begin{equation}
C_{\Sc v} \equiv \left. T \frac{\partial S}{\partial T}\right|_{V,N} \label{ef39}
\end{equation}
is the heat capacity at constant volume, for $E$ in the vicinity of $E(N)$ one can write
\begin{equation}
\psi(N,E) \sim \psi(N,E(N)) + \frac{1}{2 C_{\Sc v} T} (E-E(N))^2. \label{ef40}
\end{equation}
For independent fermions in the thermodynamic limit, $C_{\Sc v} \propto T$ (Eq.~(5.59) in Ref.~\citen{FW03}), so that the pre-factor of the $(E-E(N))^2$ in Eq.~(\ref{ef40}) diverges like $1/T^2$ for $T\to 0$; this scaling applies also for conventional Fermi liquids.

With reference to Eqs.~(\ref{ef24}) and (\ref{ef36}), one has
\begin{equation}
\psi(N,E(N)) = E(N) - T S(N), \label{ef41}
\end{equation}
so that by applying the Laplace method \cite{ETC65,BO99}, to leading order one obtains that
\begin{eqnarray}
&&\hspace{-0.0cm}\sum_{s} \varphi(N,E_s(N))\, \mathrm{e}^{-\beta E_s} \sim \frac{\varphi(N,E(N))}{\Delta E}\, \mathrm{e}^{-\beta (E(N) - T S(N))} \! \int_{-\infty}^{\infty}\! {\rm d}E\; \mathrm{e}^{-(E-E(N))^2/(2 k_{\Sc b} C_{\Sc v} T^2)} \nonumber\\
&&\hspace{1.5cm}
= N \frac{(2\pi C_{\Sc v} T/N)^{1/2} \varphi(N,E(N))}{ (N \beta)^{1/2}\, \Delta E} \,\mathrm{e}^{-\beta (E(N) - T S(N))}\;\;\;\mbox{\rm as}\;\;\; N\beta\to\infty. \label{ef42}
\end{eqnarray}
We note that $C_{\Sc v}$ is an extensive quantity so that the quantity $(2\pi C_{\Sc v} T/N)^{1/2}$ is to leading order independent of $N$ for $N\to\infty$.

With $\varphi(N,x) \equiv 1$, $\forall N, x$, the expression on the LHS of Eq.~(\ref{ef42}) is equal to $\mathrm{e}^{-\beta F(N)}$ (see Eq.~(\ref{ef22})). Thus, for $N\beta\to\infty$ one has \cite{KH87}
\begin{equation}
F(N) \sim E(N) - T S(N) -\frac{1}{2\beta} \ln\Big(\frac{2\pi C_{\Sc v}/k_{\Sc b}}{(\beta\, \Delta E)^2}\Big). \label{ef43}
\end{equation}
Since $E(N)$ is the internal energy of the $N$-particle canonical ensemble under consideration, on account of the thermodynamic relationship
\begin{equation}
F = E - T S, \label{ef44}
\end{equation}
for the entropy $S_{\Sc c}(N,T)$ of the $N$-particle canonical ensemble under consideration one has
\begin{equation}
S_{\Sc c}(N,T) \sim S(N,E(N)) + \frac{k_{\Sc b}}{2} \ln\Big(\frac{2\pi C_{\Sc v}/k_{\Sc b}}{(\beta\, \Delta E)^2}\Big)\;\;\;\mbox{\rm for}\;\;\; N\beta\to\infty. \label{ef45}
\end{equation}
Since $C_{\Sc v}$ is extensive, one observes that, for a finite $\Delta E$, to leading order $S_{\Sc c}(N,T)$ deviates from the entropy of the corresponding micro-canonical ensemble of $N$ particles at energy $E(N)$ (cf. Eq.~(\ref{ef24})) by a term scaling like $\ln(N)$ \cite{KH87}.

The above considerations have prepared the ground for calculating the next-to-leading term in the low-temperature asymptotic series expansion of the expression on the LHS of Eq.~(\ref{ef42}). The following analysis will be most transparent by introducing
\begin{equation}
\t{\beta} \equiv N \beta,\;\;\;\t{\varphi}(\varepsilon) \equiv \varphi(N,N \varepsilon),\;\;\; \t{\psi}(\varepsilon) \equiv \frac{1}{N}\, \psi(N,N\varepsilon).  \label{ef46}
\end{equation}
Using these definitions, one obtains that
\begin{equation}
\int_{E_1}^{E_2} {\rm d}E\; \varphi(N,E)\, \mathrm{e}^{-\beta \psi(N,E)} \equiv N \int_{E_1/N}^{E_2/N} {\rm d}\varepsilon\; \t{\varphi}(\varepsilon)\, \mathrm{e}^{-\t{\beta} \t{\psi}(\varepsilon)}, \label{ef47}
\end{equation}
where we have displayed
\begin{equation}
E_1 \equiv \inf_{s} E_s(N)\;\;\;\mbox{\rm and}\;\;\; E_2 \equiv\sup_{s} E_s(N) \nonumber
\end{equation}
so as to make explicit all consequences of the transformation $E= N\varepsilon$ underlying the expression in Eq.~(\ref{ef47}). The pre-factor $N$ on the RHS of this expression is exactly the same multiplicative $N$ that one encounters on the RHS of Eq.~(\ref{ef42}). Note that $E_1 = E_0(N)$, the energy of the $N$-particle GS of $\wh{H}$.

So long as $E(N)$ is the only energy at which $\psi(N,E)$ acquires its absolute minimum and so long as $\psi(N,E)$ is infinitely many times differentiable with respect to $E$ at $E(N)$, and moreover $E_1< E(N)<E_2$ (strict inequalities), the asymptotic series expansion of the integral on the RHS of Eq.~(\ref{ef47}) corresponding to $\t{\beta}\to\infty$ is to exponential accuracy independent of the precise values of $E_1$ and $E_2$ and therefore the lower and upper boundaries of this integral may be identified with respectively $-\infty$ and $+\infty$. This is also the origin of the integral over $[-\infty,\infty]$ in Eq.~(\ref{ef42}). Hence below we shall focus on the following integral:
\begin{equation}
\mathcal{I} {:=} \int_{-\infty}^{\infty} {\rm d}\varepsilon\; \t{\varphi}(\varepsilon)\, \mathrm{e}^{-\t{\beta} \t{\psi}(\varepsilon)}. \label{ef48}
\end{equation}

For calculating the first two leading terms in the asymptotic series expansion of $\mathcal{I}$ corresponding to $\t{\beta}\to\infty$, one expands $\t{\psi}(\varepsilon)$ to fourth order in $(\varepsilon-\varepsilon_0)$, where
\begin{equation}
\varepsilon_0 \equiv \frac{E(N)}{N}, \label{ef49}
\end{equation}
for which one has $\t{\psi}'(\varepsilon_0) = 0$ (cf. Eq.~(\ref{ef37})), and subsequently employs the expansion
\begin{eqnarray}
&&\hspace{-1.2cm}\t{\varphi}(\varepsilon)\, \mathrm{e}^{-\t{\beta} (\frac{1}{3!} \t{\psi}'''(\varepsilon_0) (\varepsilon-\varepsilon_0)^3 + \frac{1}{4!} \t{\psi}''''(\varepsilon_0) (\varepsilon-\varepsilon_0)^4)}
\nonumber\\
&&\hspace{-0.7cm} \sim \big(\t{\varphi}(\varepsilon_0) + \t{\varphi}'(\varepsilon_0) (\varepsilon-\varepsilon_0) +\frac{1}{2} \t{\varphi}''(\varepsilon_0) (\varepsilon-\varepsilon_0)^2 \big)\, \nonumber\\
&&\hspace{-0.3cm} \times \big(1 - \frac{\t{\beta} \t{\psi}'''(\varepsilon_0)}{6} (\varepsilon-\varepsilon_0)^3 - \frac{\t{\beta} \t{\psi}''''(\varepsilon_0)}{24} (\varepsilon-\varepsilon_0)^4 + \frac{(\t{\beta} \t{\psi}'''(\varepsilon_0))^2}{72} (\varepsilon-\varepsilon_0)^6\big). \label{ef50}
\end{eqnarray}
Of the latter expression one needs to retain the terms up to and including the sixth order in $(\varepsilon-\varepsilon_0)$. Further, of this sixth-order polynomial expression only terms proportional to even powers of $(\varepsilon-\varepsilon_0)$ contribute to $\mathcal{I}$; had we defined $\mathcal{I}$ in terms of an integral over $(E_1/N, E_2/N)$, contributions of these terms would be identically vanishing for $E_1=-E_2$ and exponentially small for $E_1 \not=-E_2$.

Making use of the variable transformation
\begin{equation}
x = \t{\beta}^{1/2} (\varepsilon-\varepsilon_0), \label{ef51}
\end{equation}
and the standard result
\begin{equation}
\int_{-\infty}^{\infty} {\rm d}x\; x^{2\alpha}\, \mathrm{e}^{-\frac{1}{2} x^2} =  2^{\alpha -1/2} \big(1 + \mathrm{e}^{2\pi\alpha i}\big)\, \Gamma\big(\alpha + \frac{1}{2}\big),\;\; \mathrm{Re}(\alpha)> -\frac{1}{2},
\label{ef52}
\end{equation}
one readily obtains a closed expression of the form
\begin{equation}
\mathcal{I} \sim \frac{\mathrm{e}^{-\t{\beta} \t{\psi}(\varepsilon_0)}}{\t{\beta}^{1/2}} \Big(A_0+ \frac{A_1}{\t{\beta}} \Big) \;\;\; \mbox{\rm for}\;\;\; \t{\beta}\to\infty. \label{ef53}
\end{equation}
The explicit expression for $A_0$ can be read off from the expression on the RHS of Eq.~(\ref{ef42}) and that for $A_1$ can be easily determined; it can also be reconstructed from a general expression presented in Eq.~(6.4.34) of Ref.~\citen{BO99}.

The result presented in Eq.~(\ref{ef53}) can be easily generalised. One readily verifies that this expansion contains the exponential factor $\mathrm{e}^{-\t{\beta} \t{\psi}(\varepsilon_0)}$ multiplied by a pre-exponential function consisting of terms decaying like $1/\t{\beta}^{1/2}$, $1/\t{\beta}^{3/2}$, $1/\t{\beta}^{5/2}$, $\dots$ for $\beta\to\infty$. The extent to which this regularity obtains is dependent on the maximum number of times that $\t{\varphi}(\varepsilon)$ and $\t{\psi}(\varepsilon)$ can be differentiated with respect to $\varepsilon$ at $\varepsilon=\varepsilon_0$; for instance, validity of the expression presented in Eq.~(\ref{ef53}) is dependent on the existence of the first four derivatives of $\t{\psi}(\varepsilon)$ and the first two derivatives of $\t{\varphi}(\varepsilon)$ at $\varepsilon=\varepsilon_0$.

In order to calculate all terms in the asymptotic series expansion of the above-mentioned pre-exponential function, correct up to and including the term decaying like $1/\t{\beta}^{(2m+1)/2}$, $m=0,1,\dots$ (assuming that $\t{\varphi}(\varepsilon)$ and $\t{\psi}(\varepsilon)$ are sufficiently many times differentiable at $\varepsilon=\varepsilon_0$), one has to replace $\t{\psi}(\varepsilon)$ by its expansion to order $2m+2$ in $(\varepsilon-\varepsilon_0)$, followed by expanding the corresponding
\begin{equation}
\mathrm{e}^{-\beta(\t{\psi}(\varepsilon) - [\t{\psi}(\varepsilon_0) +\frac{1}{2} \t{\psi}''(\varepsilon_0) (\varepsilon-\varepsilon_0)^2])}
\nonumber
\end{equation}
through employing $\mathrm{e}^{x} \sim \sum_{j=0}^{2m} x^j/j!$, and subsequently retaining all terms up to and including order $6m$ in $(\varepsilon-\varepsilon_0)$; on multiplying this with a series expansion to order $2m$ in $(\varepsilon-\varepsilon_0)$ of $\t{\varphi}(\varepsilon)$, retaining in the resulting expression the terms of up to and including order $6m$ in $(\varepsilon-\varepsilon_0)$, one obtains the sought-after asymptotic series expansion of $\mathcal{I}$ corresponding to $\t{\beta}\to\infty$ in terms of integrals of the form presented in Eq.~(\ref{ef52}). As earlier, the terms in the latter series corresponding to odd powers of $(\varepsilon -\varepsilon_0)$ do not contribute to $\mathcal{I}$. For clarity, the relationships between $(2m+1)/2$ and the above-mentioned integers $2m+2$, $6m$, $2m$ and $6m$ can be easily uncovered by means of power counting. The possibility of appearance of `anomalous' powers, or logarithmic corrections, in the asymptotic series expansion of the pre-exponential part of $\mathcal{I}$, for a predetermined value of $m$, arises from $\t{\varphi}(\varepsilon)$ and/or $\t{\psi}(\varepsilon)$ not being sufficiently many times differentiable at $\varepsilon =\varepsilon_0$.

Following the above considerations, and provided that $\t{\varphi}(\varepsilon)$ and $\t{\psi}(\varepsilon)$ are sufficiently many times differentiable at $\varepsilon=\varepsilon_0$, for $N\beta\to\infty$ one arrives at
\begin{equation}
\sum_{s} \varphi(N,E_s(N))\, \mathrm{e}^{-\beta E_s} \sim N\, \frac{1}{(N\beta)^{1/2}\, \Delta E} \Big(A_0 + \frac{A_1}{(N\beta)} + \frac{A_2}{(N\beta)^2} +\dots\Big)\, \mathrm{e}^{-\beta (E(N) - T S(N))}, \label{ef54}
\end{equation}
where, as may be gleaned from the expression in Eq.~(\ref{ef42}), the coefficients $A_0$, $A_1$, $\dots$, are functions of both $T$ and $N$; since $\t{\psi}(\varepsilon)$ is intensive, the scaling with $N$ of each of these coefficients, for $N\to\infty$, is fully determined by the scaling with $N$ of $\t{\varphi}(\varepsilon_0) \equiv \varphi(N,N \varepsilon_0)$ for $N\to\infty$; thus if $\varphi(N,N \varepsilon_0)$ is, for instance, intensive (extensive), the coefficients $\{A_j\}$ will be similarly intensive (extensive). Consequently, for the following considerations it proves convenient to introduce the intensive quantities $\{ a_j\}$, defined according to
\begin{equation}
A_j = \varphi(N,E(N))\, a_j,\;\;\; j=0,1,\dots~. \label{ef55}
\end{equation}

We denote the coefficients of the asymptotic series specific to $\varphi(N,x)\equiv 1$, $\forall N, x$, by $b_0$, $b_1$, $\dots$~. These coefficients are, similar to $A_0$, $A_1$, $\dots$, functions of both $T$ and $N$, however they are intensive owing to the underlying $\varphi(N,x)$ being a finite constant. From the general expression in Eq.~(\ref{ef54}) one thus obtains that for $N\beta\to\infty$
\begin{equation}
\frac{\sum_s \varphi(N,E_s(N))\, \mathrm{e}^{-\beta E_s}}{\sum_s \mathrm{e}^{-\beta E_s}} \sim \varphi(N,E(N))\, \frac{a_0 + a_1/(N\beta) +a_2/(N\beta)^2+\dots}{b_0 + b_1/(N\beta)+b_2/(N\beta)^2+\dots}. \label{ef56}
\end{equation}
Since both $\{ a_j\}$ and $\{ b_j\}$ are intensive, one observes that the decay of the terms of both series on the RHS of Eq.~(\ref{ef56}) is governed by the increasing powers of $1/(N\beta)$; thus for sufficiently large $N\beta$, irrespective of whether for instance $\beta$ is large, the rational function on the RHS of Eq.~(\ref{ef56}) is to an error of the order $1/(N\beta)$ determined by $a_0/b_0$, which one can verify to be equal to $1$.

The expression in Eq.~(\ref{ef56}) demonstrates that for $N\beta\to\infty$ the behaviour of the function on the LHS of Eq.~(\ref{ef56}) is up to possibly an exponentially small correction fully determined by pre-exponential functions of the form presented on the RHS of Eq.~(\ref{ef54}). We remark that in contrast to macroscopic systems, for finite systems (leaving aside the fundamental mathematical problems that these systems can in principle pose) the leading-order contribution of the RHS of Eq.~(\ref{ef56}), corresponding to
\begin{equation}
\frac{a_0 + a_1/(N\beta)+a_2/(N\beta)^2+\dots}{b_0 + b_1/(N\beta)+b_2/(N\beta)^2+\dots} \sim \frac{a_0}{b_0} =1, \nonumber
\end{equation}
may not accurately describe the function on the LHS for insufficiently large values of $\beta$.

From the above considerations we arrive at the expression
\begin{equation}
\frac{1}{\varphi(N,E(N))}\frac{\sum_s \varphi(N,E_s(N))\, \mathrm{e}^{-\beta E_s}}{\sum_s \mathrm{e}^{-\beta E_s}} = 1 + O\big(\frac{1}{N\beta}\big)\;\;\; \mbox{\rm for}\;\;\; N\beta\to\infty, \label{ef57}
\end{equation}
where $O(\frac{1}{N\beta}) \equiv \frac{(a_1 - a_0 b_1/b_0)/b_0}{N\beta}$, from which one deduces the exact result
\begin{equation}
\lim_{N\to\infty} \frac{1}{\varphi(N,E(N))} \frac{\sum_s \varphi(N,E_s(N))\, \mathrm{e}^{-\beta E_s}}{\sum_s \mathrm{e}^{-\beta E_s}} = 1,\;\;\; \forall\beta>0,
\label{ef58}
\end{equation}
where physically the limit $N\to\infty$ corresponds to a finite constant value of $n \equiv N/V$. The result in Eq.~(\ref{ef57}) can be equivalently expressed as
\begin{equation}
\frac{\sum_s \varphi(N,E_s(N))\, \mathrm{e}^{-\beta E_s}}{\sum_s \mathrm{e}^{-\beta E_s}} = \varphi(N,E(N)) + O\big(\frac{\varphi(N,E(N))}{N\beta}\big)\;\;\; \mbox{\rm for}\;\;\; N\beta\to\infty, \label{ef59}
\end{equation}
however it should be borne in mind that in the cases where $\varphi(N,E(N))$ is not intensive, this expression turns into the meaningless `identity' $\infty = \infty$ as $N\to\infty$ .

A corollary of Eq.~(\ref{ef57}), corresponding to $\varphi(N,x)\equiv x$, $\forall N, x$, is the exact result
\begin{equation}
\frac{\langle\wh{H}\rangle_N}{E(N)}
= 1 + O\big(\frac{1}{N\beta}\big)\;\;\;\mbox{\rm for}\;\;\; N\beta \to
\infty, \label{ef60}
\end{equation}
where
\begin{equation}
\langle \wh{H}\rangle_N \equiv \frac{\sum_{s} E_s(N)\, \mathrm{e}^{-\beta E_{s}(N)}}{\sum_{s} \mathrm{e}^{-\beta E_{s}(N)}}. \label{ef61}
\end{equation}
According to Eq.~(\ref{ef60}), for sufficiently large $N\beta$, the internal energy $E(N)$ corresponding to a canonical ensemble of $N$ particles, is up to a relative error of the order of $1/(N\beta)$ equal to the ensemble average of the Hamiltonian; this error vanishes in the thermodynamic limit for all $\beta>0$, however, since $E(N)$ is extensive, the difference between the two energies is to leading order independent of $N$ and decays like $1/\beta$ for $\beta\to\infty$. Since further (combine Eqs.~(3.2.14) and (3.2.23) in Ref.~\citen{CL00})
\begin{equation}
\langle (\wh{H}- \langle\wh{H}\rangle_N)^2\rangle_N \equiv \langle \wh{H}^2 \rangle_N - \langle \wh{H}\rangle_N^2 = k_{\Sc b} C_{\Sc v}\, T^2, \label{ef62}
\end{equation}
it follows that
\begin{equation}
\lim_{\beta\to\infty} E(N) = E_0(N), \label{ef63}
\end{equation}
the GS energy.

As the approach of $E(N)$ towards $E_0(N)$ is continuous for $T\to 0$, the assumption with regard to continuity of $\varphi(N,x)$, for $x$ in the interior of a connected interval containing $E(N)$ and $E_0(N)$, implies that for $\beta\to\infty$ Eq.~(\ref{ef58}) reduces to the leading-order asymptotic expression
\begin{equation}
\lim_{N\to\infty}\frac{1}{\varphi(N,E_0(N))} \frac{\sum_s \varphi(N,E_s(N))\, \mathrm{e}^{-\beta E_s}}{\sum_s \mathrm{e}^{-\beta E_s}} \sim 1. \label{ef64}
\end{equation}
It should be evident that the deviation of the LHS of Eq.~(\ref{ef64}) from unity is \emph{entirely} due to the deviation from $E(N)$ of the $E_0(N)$ encountered on the LHS of Eq.~(\ref{ef64}) (more about this later); following Eq.~(\ref{ef63}), this deviation is continuously reduced to zero for $\beta\to\infty$.

In order to determine the amount of the deviation of the LHS of Eq.~(\ref{ef64}) from unity, we proceed by introducing the thermodynamic relationships \cite{CL00}
\begin{equation}
\left. \frac{\partial F}{\partial T}\right|_{V,N} = -S,\;\;\;
\left. \frac{\partial^2 F}{\partial T^2}\right|_{V,N} = -\frac{C_{\Sc v}}{T}. \label{ef65}
\end{equation}
For conventional metals one has
\begin{equation}
C_{\Sc v}(T) \sim \Gamma\, T \;\;\; \mbox{\rm as} \;\;\; T\to 0, \label{ef66}
\end{equation}
where $\gamma = \Gamma/V$ is the well-known Sommerfeld constant of the specific heat $c_{\Sc v} \equiv C_{\Sc v}/V$ (p.~49 in Ref.~\citen{AM76}). In contrast, $C_{\Sc v}(T)$ vanishes exponentially for $T\to 0$ in the cases of insulators. It follows that for these two classes of states, $c_{\Sc v}/T$ takes a finite value at $T=0$, which for insulators is equal to zero. On the other hand, there are strongly-correlated metallic states for which $c_{\Sc v}/T$ diverges either logarithmically \cite{BGC01} or as a power-law \cite{CKZ05} (for a general review see Ref.~\citen{GRS01}). Since $S(T=0) = 0$ (the third law of thermodynamics \cite{KH87}), from Eq.~(\ref{ef65}) one observes that, in view of Eq.~(\ref{ef63}), $F(N)$ can be most generally expressed as (for the $o-O$ notation see \S~5 in Ref.~\citen{EWH26})
\begin{equation}
F(N) = E_0(N) + N\, o(1/\beta)\;\;\;\mbox{\rm for}\;\;\; \beta\to\infty, \label{ef67}
\end{equation}
where through pre-multiplying $o(1/\beta)$ by $N$ we have made explicit the fact that $F(N)-E_0(N)$ is an extensive quantity. For conventional metals, $o(1/\beta)$ is to leading-order proportional to $1/\beta^2$, and for insulators $o(1/\beta)$ is less dominant than any finite power of $1/\beta$ as $\beta\to\infty$. On the other hand, there are strongly-correlated metals for which $o(1/\beta)$ is to leading order proportional to either $\ln(\beta)/\beta^2$ or $1/\beta^{1+\alpha}$, where $0 <\alpha<1$ (see above). Note that the function $o(1/\beta)$ equally accounts for such functional form as $\ln(\beta)/\beta^{1+\alpha}$, where $\alpha>0$. For the cases where the underlying $N$-particle GSs are insulating, the $o(1/\beta)$ on the RHS of Eq.~(\ref{ef67}) is an exponentially decaying function of $\beta$ for $\beta\to\infty$ (see Sec.~\ref{ssf3s3}).

Since in general $S(T) = N\, o(1)$ for $T\to 0$,\footnote{For free fermions and Fermi-liquid metallic states, $S(T)$ is to leading order equal to $C_{\Sc v}(T)$. See, e.g., Eqs.~(5.58) and (5.59) in Ref.~\protect\citen{FW03}.} it follows that one in general can write
\begin{equation}
T S = N\, o(1/\beta)\;\;\; \mbox{\rm for}\;\;\; \beta\to\infty. \label{ef68}
\end{equation}
Similar to the case of $F(N) - E_0(N)$, the function $o(1/\beta)$ in the expression on the RHS of Eq.~(\ref{ef68}) is to leading order proportional to $1/\beta^2$ for conventional metals and decays exponentially for insulating GSs as $\beta\to\infty$.

In the light of the results in Eqs.~(\ref{ef67}) and (\ref{ef68}), from the thermodynamic relationship in Eq.~(\ref{ef44}) one obtains that
\begin{equation}
E(N) = E_0(N) + N\, o(1/\beta)\;\;\;\mbox{\rm for}\;\;\; \beta\to\infty. \label{ef69}
\end{equation}
This result implies that, for $\varphi(N,x)$ a continuous and differentiable function of $x$ at $x=E_0(N)$, one has
\begin{equation}
\varphi(N,E(N)) \sim \varphi(N,E_0(N)) + N\varphi'(N,E_0(N))\, o(1/\beta)\;\;\; \mbox{\rm as}\;\, \beta\to \infty, \label{ef70}
\end{equation}
where $\varphi'(N,x) \equiv \partial \varphi(N,x)/\partial x$. Evidently, for $\varphi'(N,E_0(N)) = 0$ and $\varphi(N,x)$ a twice continuously differentiable function of $x$ at $x=E_0(N)$, the deviation of $\varphi(N,E(N))$ from $\varphi(N,E_0(N))$ is of the second order in $o(1/\beta)$.

Assuming that $\varphi'(N,E_0(N)) \not=0$, the result in Eq.~(\ref{ef70}) implies that the LHS of Eq.~(\ref{ef64}) is equal to unity up to a correction of the order of $o(1/\beta)$ as $\beta\to \infty$ (cf. Eq.~(\ref{ef58})); for conventional metals, for instance, the deviation of the LHS of Eq.~(\ref{ef64}) from unity is to leading order proportional to $1/\beta^2$. We can thus express the result in Eq.~(\ref{ef64}) more explicitly as
\begin{equation}
\lim_{N\to\infty}\frac{1}{\varphi(N,E_0(N))} \frac{\sum_s \varphi(N,E_s(N))\, \mathrm{e}^{-\beta E_s}}{\sum_s \mathrm{e}^{-\beta E_s}} = 1 + o(1/\beta), \label{ef71}
\end{equation}
for sufficiently large $\beta$. Note that the $o(1/\beta)$ on the RHS of this expression is \emph{not} the same function $o(1/\beta)$ as encountered on the RHS of Eq.~(\ref{ef70}); to leading order in $1/\beta$, it is equal to the function $o(1/\beta)$ on the RHS of Eq.~(\ref{ef70}) times the intensive quantity $N\varphi'(N,E_0(N))/\varphi(N,E_0(N))$. A comparison of Eq.~(\ref{ef71}) with Eq.~(\ref{ef58}) reveals the impact of replacing in the argument of $\varphi(N,E(N))$ the internal energy $E(N)$ at $\beta$ with its zero-temperature limit $E_0(N)$, the GS energy.

On the basis of the result in Eq.~(\ref{ef71}), in particular on the basis of one's knowledge concerning the explicit form of the function $o(1/\beta)$, one can determine the number of exact terms that one can calculate in the low-temperature asymptotic series expansion of $\mathcal{Z}_{\Sc g}^{-1} \sum_{\nu} f(\nu)\, \mathrm{e}^{-\beta K_{\nu}}$ in terms of the leading-order expression in Eq.~(\ref{ef31}).

\subsubsection{Remarks}
\label{ssf3s3}

In Sec.~\ref{ssf5} of this appendix we shall need to rely on some knowledge concerning the behaviour of the function $o(1/\beta)$ on the RHS of Eq.~(\ref{ef67}) for the cases where the underlying $N$-particle GSs are insulating. To establish this behaviour, in analogy with $E_0(N)$ we denote the energy of the lowest-lying $N$-particle excited state(s) of $\wh{H}$ by $E_1(N)$, where the subscript $1$ has \emph{nearly} the same status as the subscript $0$ in $E_0(N)$; whereas in our considerations the $N$-particle GS of $\wh{H}$ is non-degenerate, this is in general not the case for the $N$-particle excited states of $\wh{H}$.

The considerations in this section will shed light on a mathematical problem associated with the process of directly evaluating $F(N)$. We shall fully establish to root cause of this problem and indicate the remedy for its resolution. As an aside, it is a well-known mathematical fact that in general it is not permissible to differentiate asymptotic series (\S~8.31 in Ref.~\citen{WW62}). Since development of an asymptotic series of a function of a function, of the form $f(g(x))$, based on an asymptotic series expansion of the inner function, $g(x)$, involves application of differentiation, in general such development must always be carried out with extreme care.

The value of $E_1(N) - E_0(N)$ is in general different from  $\mu_{N}^+ -\mu_{N}^-$ (appendix \ref{sc}); that $E_1(N) - E_0(N)$ may be slightly less than $\mu_{N}^+ -\mu_{N}^-$ can be appreciated by recalling the physical picture of Mott excitons in insulators (pp. 626-628 in Ref.~\citen{AM76}); in this picture, the amount by which $E_1(N) - E_0(N)$ is less than $\mu_{N}^+ -\mu_{N}^-$ accounts for the binding energy of the wave packets of electrons and holes participating in the formation of one such exciton. Theoretically, $E_1(N) - E_0(N)$ is the gap energy as observed in the spectral function corresponding to the density (or density-density) correlation function, which is deducible from the two-particle Green function (Ch. 5, \S~4 in Ref.~\citen{NO98}, Ch.~3, \S~2 in Ref.~\citen{PN64}, Ch.~9, \S~32 in Ref.~\citen{FW03}). Within the framework of the random-phase approximation for the density correlation function, $E_1(N) - E_0(N)$ is exactly equal to $\mu_{N}^+ -\mu_{N}^-$.

Following the equivalence relationship in Eq.~(\ref{ef22}), for $F(N)$ one has the exact expression (cf. Eq.~(\ref{ef67}))
\begin{equation}
F(N) = E_0(N) -\frac{1}{\beta} \ln\big( 1 + \Xi_{\beta}(N)\big),\;\;\; \Xi_{\beta}(N) {:=} \sum_{s\not=0} \mathrm{e}^{-\beta (E_s(N) - E_0(N))}. \label{ef72}
\end{equation}
Since the argument of the logarithm function is greater than unity, it follows that $F(N) < E_0(N)$ (strict inequality) for \emph{all} $\beta<\infty$ (cf. Eqs.~(\ref{ef44}) and (\ref{ef63})). Making use of the many-body density-of-states function $\omega(N,E)$, Eq.~(\ref{ef33}), and assuming that the internal energy $E(N)$ (cf.~(\ref{ef37})) is less than $E_1(N)$, through the application of integration by parts \cite{ETC65,BO99} we obtain that
\begin{eqnarray}
&&\hspace{-0.5cm}\Xi_{\beta}(N) \equiv \int_{E_1(N)}^{\infty} {\rm d}E\; \omega(N,E)\,  \mathrm{e}^{-\beta (E - E_0(N))} \nonumber\\
&&\hspace{0.0cm}\sim \frac{1}{\beta}\, \omega(N,E_1^+(N))\, \mathrm{e}^{-\beta (E_1(N) - E_0(N))}\;\;\;\mbox{\rm for}\;\;\; \beta\to\infty\;\; (E(N) <E_1(N)), \label{ef73}
\end{eqnarray}
where we have assumed that $\omega(N,E)$ is continuous in a right neighbourhood of $E_1(N)$ and that the integral of $\partial\omega(N,E)/\partial E$ over $(E_1(N), E_1(N)+\Delta)$ exists, where $\Delta$ is a finite positive constant. For clarity, in general the integrand of $\Xi_{\beta}(N)$ can be expressed as $\mathrm{e}^{-\beta (\psi(N,E) - E_0(N))}$ (see Eq.~(\ref{ef36})), so that the asymptotic series expansion of $\Xi_{\beta}(N)$ for $\beta\to\infty$ is determined by the behaviour of $\psi(N,E)$ in the neighbourhood of the absolute minimum of this function \emph{inside} $(E_1(N), E_1(N)+\Delta)$. For $E(N) <E_1(N)$, this absolute minimum is located at $E_1(N)$, which is not a stationary point of $\psi(N,E)$. It follows that, with $\beta_{\rm c}$ denoting the largest $\beta$ at which $E(N) = E_1(N)$, one expects a qualitative change in the behaviour of $\Xi_{\beta}(N)$ as $\beta$ is increased from below $\beta_{\rm c}$ to above $\beta_{\rm c}$. For $\beta$ large, however $\beta <\beta_{\rm c}$, the behaviour of $\Xi_{\beta}(N)$ is to be determined according to the same procedure as leading to Eq.~(\ref{ef42}).

It should be evident that Eq.~(\ref{ef73}) becomes meaningless if $E_1(N)$ is isolated\footnote{In the present context, `isolation' is a relative notion, defined in terms of the quantity $\Delta E$, the energy uncertainty in the micro-canonical ensemble.} and not part of a continuum of the eigenenergies $\{ E_s(N) \}$; should $E_1(N)$ be isolated, the leading-order asymptotic behaviour of $\Xi_{\beta}(N)$ for $\beta\to\infty$ is of the same form as that on the RHS of Eq.~(\ref{ef73}), with the $\omega(N,E_1^+(N))$ herein replaced by $g_1$, the degeneracy factor of the $N$-particle eigenstates of $\wh{H}$ corresponding to $E_1(N)$.

Assuming that $\beta$ is sufficiently large so that the expression in Eq.~(\ref{ef73}) is valid, we now consider two distinct cases. The first case corresponds to $\Xi_{\beta}(N) \gg 1$, for which one trivially obtains that
\begin{equation}
F(N) \sim E_1(N) - \frac{1}{\beta} \ln\big(\frac{\omega(N,E_1^+(N))}{\beta}\big). \label{ef73a}
\end{equation}
Since $\Xi_{\beta}(N)$ is exponentially decaying for $\beta\to\infty$, the condition $\Xi_{\beta}(N) \gg 1$ can only be relevant if $\omega(N,E_1^+(N)) \to\infty$ for $N\to\infty$, that is if $\omega(N,E_1^+(N))$ is not intensive. In fact, owing to the exact inequality $F(N) < E_0(N)$ (see above), the expression in Eq.~(\ref{ef73a}) can apply only if $\omega(N,E_1^+(N)) > \beta\, \mathrm{e}^{\beta (E_1(N) - E_0(N))}$. Since the RHS of this inequality is a monotonically increasing function of $\beta$, it follows that Eq.~(\ref{ef73a}) fails to be valid for $\beta$ greater than some finite value, say $\beta_{\star}$. It is conceivable that $\beta_{\star} < \beta_{\rm c}$, in which case Eq.~(\ref{ef73a}), having been deduced from Eq.~(\ref{ef73}), has no validity for any range of $\beta$.

The second case corresponds to $\Xi_{\beta}(N) \ll 1$. On account of $\ln(1+x) = x + O(x^2)$ as $x\to 0$, for this case one has (cf. Eq.~(\ref{ef33}))
\begin{equation}
F(N) \sim E_0(N) -\frac{1}{\beta^2}\, \omega(N,E_1^+(N))\,\mathrm{e}^{-\beta (E_1(N) - E_0(N))}\;\;\; \mbox{\rm as}\;\;\; \beta\to\infty. \label{ef74}
\end{equation}
Since by assumption $\omega(N,E_1^+(N)) >0$, one observes that according to this expression one indeed has $F(N) < E_0(N)$ for all $\beta<\infty$. Although the expression in Eq.~(\ref{ef74}) seems reasonable at first glance, it suffers from a so-called `size-inconsistency' problem. This statement is clarified as follows.

If the second term on the RHS of Eq.~(\ref{ef74}) is to have relevance for large values of $N$, then $\omega(N,E_1^+(N)))$ is to scale like $N$ for $N\to\infty$, that is, it has to be extensive. The prospect of this being the case, renders the expression in Eq.~(\ref{ef74}) as mathematically not fully satisfactory: the $\beta$ above which this expression may be viewed as a valid asymptotic expression for $F(N)$, is an increasing function of $N$. In this connection, recall that Eq.~(\ref{ef74}) is deduced under the condition that $\Xi_{\beta}(N) \ll 1$. Denoting the value of $\beta$ above which Eq.~(\ref{ef74}) applies by $\beta_{N} \equiv 1/(k_{\Sc b} T_N)$, with $\omega(N,E_1^+(N)) \sim a N$ for $N\to \infty$, to logarithmic accuracy one has
\begin{equation}
\beta_N \propto \frac{\ln(N)}{E_1(N)-E_0(N))} \iff T_{N} \propto \frac{E_1(N)-E_0(N)}{k_{\Sc b} \ln(N)}. \label{ef75}
\end{equation}
Consequently, Eq.~(\ref{ef74}) cannot apply for any finite $\beta$ on taking the thermodynamic limit. We should emphasise that this is a purely mathematical statement and that physically Eq.~(\ref{ef74}) cannot be \emph{a priori} ruled out. To clarify, consider a sample of the size $1$ cm$^3$, consisting of $N \approx 10^{23}$ electrons, and $E_1(N) - E_0(N) \approx 10^{-2}$~eV. With $1~\mathrm{eV}/k_{\Sc b} \approx 1.2 \times 10^{4}$~K, one obtains that $T_N \propto 2.1$~K. Increase of the size of the sample to $1$ m$^3$ only leads to $T_N \propto 1.7$~K.

To establish the reason underlying the `size-inconsistent' nature of the expression in Eq.~(\ref{ef74}), we introduce the Hamiltonian
\begin{equation}
\wh{H}_{\lambda} {:=} \lambda \wh{H},\;\;\; \lambda >0. \label{ef76}
\end{equation}
The $N$-particle eigenstates of $\wh{H}_{\lambda}$ are exactly those of $\wh{H}$ and the corresponding eigenvalues are equal to $\lambda$ times those of $\wh{H}$. Denoting the Helmholtz free energy corresponding to $\wh{H}_{\lambda}$ by $F_{\lambda}(N)$, following the approach as employed in determining the grand potential $\Omega(\beta,N,V)$ in Ref.~\citen{FW03} (see Ch.~7, \S~23 herein; see also Sec.~\ref{ss51s3}), we readily obtain that
\begin{eqnarray}
\hspace{-0.5cm}
\frac{\partial F_{\lambda}(N)}{\partial\lambda} &=&
\frac{1+\sum_{s\not=0} E_{s}(N) \, \mathrm{e}^{-\lambda\beta (E_{s}(N)-E_{0}(N))}}{1+\sum_{s\not=0} \mathrm{e}^{-\lambda\beta (E_{s}(N)-E_{0}(N))}} \nonumber\\
&\equiv& \frac{E_0(N) + \int_{E_1(N)}^{\infty} {\rm d}E\; \omega(N,E)\, E\, \mathrm{e}^{-\lambda\beta (E-E_0(N))}}{1 + \int_{E_1(N)}^{\infty} {\rm d}E\; \omega(N,E)\, \mathrm{e}^{-\lambda\beta (E-E_0(N))}} \nonumber\\
&\sim& \frac{E_0(N) + \frac{\omega(N,E_1^+(N))\, E_1(N)}{\lambda\beta}\,  \mathrm{e}^{-\lambda\beta (E_1(N) - E_0(N))}}{1 + \frac{\omega(N,E_1^+(N))}{\lambda\beta}\, \mathrm{e}^{-\lambda\beta (E_1(N)-E_0(N))}}\;\;\; \mbox{\rm for}\;\;\; \beta\to\infty,
\label{ef77}
\end{eqnarray}
where in arriving at the last asymptotic expression we have assumed that $\beta > \beta_{\rm c}$, so that $E(N) < E_1(N)$.

For $\omega(N,E_1^+(N))$ intensive, from Eq.~(\ref{ef77}) one immediately obtains that
\begin{equation}
\frac{\partial F_{\lambda}(N)}{\partial\lambda} \sim E_0(N) + \frac{\omega(N,E_1^+(N))\, E_1(N)}{\lambda\beta}\,  \mathrm{e}^{-\lambda\beta (E_1(N) - E_0(N))}\;\;\; \mbox{\rm as}\;\;\; \beta\to\infty, \label{ef78}
\end{equation}
which is an explicitly `size-consistent' expression, this owing to the fact that $E_1(N)$ is extensive. For an extensive  $\omega(N,E_1^+(N))$, on taking the thermodynamic limit for a finite value of $\lambda\beta$, from Eq.~(\ref{ef77}) one obtains that  $\lim_{N\to\infty} \partial F_{\lambda}(N)/\partial\lambda$ is to leading order in $1/(\lambda\beta)$ equal to $E_1(N)$, which is reasonable. We should emphasise however that for any case where the degeneracy factor of the underlying $N$-particle GS is equal to unity (or, more generally, intensive), the prospect of an extensive $\omega(N,E_1^+(N))$ raises the problem of the GS being thermodynamically irrelevant for all $\beta \lesssim \beta_{\rm c}$. In fact, it is not difficult to show that as $N\to\infty$, the condition $E(N) < E_1(N)$ cannot be fulfilled for any finite value of $\beta$.

For comparison we now consider the $\partial F_{\lambda}(N)/\partial\lambda$ corresponding to the $F(N)$ in Eq.~(\ref{ef74}). To this end, let $\omega_{\lambda}(N,E)$ denote the many-body density-of-states function, Eq.~(\ref{ef33}), corresponding to $\wh{H}_{\lambda}$; in this notation, $\omega(N,E) \equiv \omega_1(N,E)$. It can be readily verified that
\begin{equation}
\omega_{\lambda}(N,\lambda E) = \frac{1}{\lambda} \omega(N,E)\;\;\; \mbox{\rm for}\;\;\; \lambda>0. \label{ef79}
\end{equation}
From Eq.~(\ref{ef74}) one thus immediately obtains that
\begin{equation}
\left. F_{\lambda}(N)\right|_{\mathrm{ Eq.~(\protect\ref{ef74})}} \sim \lambda E_0(N) - \frac{\omega(N,E_1^+(N))}{\lambda \beta^2}\, \mathrm{e}^{-\lambda\beta (E_1(N)-E_0(N))},
\label{ef80}
\end{equation}
so that
\begin{equation}
\left.\frac{\partial F_{\lambda}(N)}{\partial\lambda}\right|_{\mathrm{ Eq.~(\protect\ref{ef80})}} \sim E_0(N) + \frac{\omega(N,E_1^+(N))\, (E_1(N) - E_0(N))}{\lambda\beta}\, \mathrm{e}^{-\lambda\beta (E_1(N)-E_0(N))}, \label{ef81}
\end{equation}
where for transparency we have suppressed a term smaller by $O((\lambda\beta)^{-1})$ than the last term shown. The expression in Eq.~(\ref{ef81}) should be compared with that in Eq.~(\ref{ef78}). Since $E_1(N)-E_0(N)$ is intensive, one observes a `size-inconsistency' also in the expression in Eq.~(\ref{ef81}).

It is not difficult to verify that the two terms on the RHS of Eq.~(\ref{ef81}) that are proportional to $E_0(N)$ are in fact the first two terms of the geometric series
\begin{equation}
\frac{E_0(N)}{1 + x} = E_0(N) \big(1 - x + x^2 - \dots \big), \label{ef82}
\end{equation}
where
\begin{equation}
x \equiv \frac{\omega(N,E_1^+(N))}{\lambda\beta}\, \mathrm{e}^{-\lambda\beta (E_1(N)-E_0(N))}, \label{ef83}
\end{equation}
and that the term on the RHS of Eq.~(\ref{ef81}) that is proportional to $E_1(N)$, is the leading term of the geometric series
\begin{equation}
\frac{x E_1(N)}{1+x} = x E_1(N) \big(1 - x + x^2 -\dots\big). \label{ef84}
\end{equation}
Although the sum of the rational functions in Eqs.~(\ref{ef82}) and (\ref{ef84}) yields the last expression on the RHS of Eq.~(\ref{ef77}), Eq.~(\ref{ef81}) fails to be `size consistent'. This is owing to the fact that in determining Eq.~(\ref{ef81}) no account has been taken of the actuality that the numerator of the last expression on the RHS of Eq.~(\ref{ef77}) \emph{cannot} be separated into its component parts on account of $x$, Eq.~(\ref{ef83}), being the expansion parameter.

Above we have focused on $\partial F_{\lambda}(N)/\partial\lambda$ instead of $F_{\lambda}(N)$, or $F(N)\equiv F_1(N)$. This we have done by the fact that calculation of a closed expression for the `size-consistent' leading-order term of the pre-exponential part of $F(N)-E_0(N)$ is relatively involved. Technically, this calculation requires determination of the cumulants \cite{RK62} $\{ \kappa_n\}$ of $\ln(1+\Xi_{\beta}(N))$ and evaluation of the cumulant series $\sum_{j=n}^{\infty} \kappa_n/\beta^n$. We note that $\kappa_n$ is expressible in terms of $\partial^j\omega(N,E)/\partial E^j$, $j=0,1,\dots,n$, evaluated at $E=E_1^+(N)$, and $\mathrm{e}^{-j\beta (E_1(N)-E_0(N))}$, $j=1,2,\dots,n+1$; consequently, applicability of the conventional cumulant expansion in the present context is conditional on $\omega(N,E)$ being infinitely many times right-differentiable with respect to $E$ at $E=E_1(N)$. In practice one bypasses use of such cumbersome approach, through integration of $\partial F_{\lambda}(N)/\partial\lambda$ with respect to $\lambda$. The considerations by Luttinger and Ward \cite{LW60}, involving $\partial\Omega(\lambda)/\partial\lambda$ and $\partial Y(\lambda)/\partial\lambda$, extensively discussed in Sec.~\ref{s5}, serve the same purpose. We note that although integration of the asymptotic expression in Eq.~(\ref{ef78}) is straightforward, full determination of $F_{\lambda}(N)$ requires knowledge of $F_{\lambda}(N)$ at $\lambda=\lambda_0$, where $\lambda_0$ is some arbitrary positive constant, which we do not possess. In practice, one considers such $\wh{H}_{\lambda}$ as $\wh{H}_0 + \lambda (\wh{H}-\wh{H}_0)$, where $\wh{H}_0$ is chosen such that $F_{\lambda}(N)$ can be explicitly calculated for $\lambda=\lambda_0\equiv 0$.

Although above we have not presented the explicit expression for the leading-order asymptotic term for $F(N)-E_0(N)$ corresponding to $\beta\to\infty$, we have made explicit that this term is proportional to $\mathrm{e}^{-\beta (E_1(N)-E_0(N))}$. This form is reminiscent of the form that one obtains within the framework of the microscopic BCS theory for the specific heat $C_{\Sc s}/V$ of an isotropic (or $s$-wave) superconductor (Ch. 13, \S~51 in Ref.~\citen{FW03}), where the superconducting gap energy $\Delta_0$ takes the place of the energy difference $E_1(N)-E_0(N)$ in the considerations of this section. In this connection, it is relevant to note that the expressions for both $C_{\Sc s}$ and $\Omega_{\Sc s}$ are `size consistent' and that the coefficients of $\mathrm{e}^{-\beta\Delta_0}$ in the expressions for $C_{\Sc s}$ and $\Omega_{\Sc s}$ are extensive. This is not surprising, as these expressions have their roots in an integration with respect to the coupling constant of the pairing interaction in the BCS Hamiltonian (Ch. 13, \S~51 in Ref.~\citen{FW03}).

\subsection{Metallic ground states; two leading-order asymptotic terms}
\label{ssf4}

For the considerations in this section we rely on the expression in Eq.~(\ref{ef32}). Since here and in the subsequent section $N$ will be a summation variable, $N \in \mathds{Z}^*$, in the following we denote the number of particles in the GS of the system under investigation by $N_0$.

With the above convention in mind, we proceed by introducing the following expansion:
\begin{eqnarray}
&&\hspace{-1.0cm} F(N) - \mu N = (F(N_0) - \mu N_0) + (F'(N_0) - \mu)\, (N-N_0) \nonumber\\
&&\hspace{1.5cm} +\frac{1}{2} F''(N_0)\, (N-N_0)^2 + \frac{1}{6} F'''(N_0)\, (N-N_0)^3 + \dots, \nonumber\\
\label{ef85}
\end{eqnarray}
where
\begin{equation}
F'(M) \equiv \left.\frac{{\rm d}F(M)}{{\rm d}M}\right|_{\beta,V},\;\; F''(M) \equiv \left.\frac{{\rm d}^2F(M)}{{\rm d}M^2}\right|_{\beta,V}, \;\; \dots~.
\label{ef86}
\end{equation}
By the thermodynamic relationship (Eq.~(4.6) in Ref.~\citen{FW03})
\begin{equation}
\mu(\beta,N,V) = \left.\frac{\partial F(\beta,N,V)}{\partial N}\right|_{\beta, V},
\label{ef87}
\end{equation}
it follows that, for the $\mu$ in Eq.~(\ref{ef85}) identified with $\mu(\beta,N_0,V)$, the expression for $F(N) - \mu N$ transforms into a parabolic one for $N$ in a neighbourhood of $N_0$. One could also consider $\mu$ as given and regard $N_0 \equiv N_0(\mu)$ as the solution of the equation
\begin{equation}
F'(N_0) = \mu. \label{ef88}
\end{equation}
In such case, $F(N)-\mu N$ is similarly a parabolic function of $N$ in a neighbourhood of $N=N_0$. It should be evident that the solution $N_0(\mu)$ of Eq.~(\ref{ef88}) corresponds to the mean value of the particles in the grand canonical ensemble and therefore need not be an integer; for $N_0$ macroscopically large, deviation of $N_0(\mu)$ from an integer is irrelevant.

Some general remarks concerning the expression in Eq.~(\ref{ef85}) are in order. Since $F$ is an extensive quantity, one can write\footnote{A similar remark as the one concerning Eq.~(\protect\ref{ec13}) applies here. }
\begin{equation}
F(\beta,N,V) = V\, \phi(\beta,n),\;\;\; n \equiv \frac{N}{V}. \label{ef89}
\end{equation}
On the basis of this expression one can explicitly demonstrate (\S~7.6 in Ref.~\citen{KH87}) that
\begin{equation}
\frac{\partial P}{\partial V} \le 0 \iff F''(N) \ge 0, \label{ef90}
\end{equation}
where
\begin{equation}
P \equiv \left. -\frac{\partial F}{\partial V}\right|_{\beta,N} \label{ef91}
\end{equation}
is pressure. Consequently, for systems which at temperature $T$ are thermodynamically stable, one has $F''(N_0)>0$. Thus, for these systems
$F(N)-\mu N$ is minimal at $T$ and $N=N_0$, with $N_0$ satisfying Eq.~(\ref{ef88}). From the expression in Eq.~(\ref{ef89}) one further infers that $F'(N_0)$ is intensive (conform Eq.~(\ref{ef88}), in which $\mu$ is intensive) and that $F''(N_0)$ scales like $1/V \equiv n/N_0$ for $V\to\infty$; in general, the $m$th derivative with respect to $N$ of $F(N)$ scales like $1/V^{m-1}$. The same applies for $E_0'(N_0)$, $E_0''(N_0)$, $\dots$, respectively.

Now we consider the general expression
\begin{equation}
\mathcal{J} {:=} \sum_{N=0}^{\infty} g(N) \,\mathrm{e}^{-\beta (F(N)-\mu N)}, \label{ef92}
\end{equation}
where $g(N)$ stands for $\b{f}(N,E(N))$ (unity) in dealing with the numerator (denominator) of the expression on the RHS of Eq.~(\ref{ef32}). The expansion in Eq.~(\ref{ef85}) reveals that in both cases $\mathcal{J}$ is proportional to $\mathrm{e}^{-\beta (F(N_0) -\mu N_0)}$, so that it is the behaviour of the pre-exponential part of $\mathcal{J}$ that ultimately determines the function on the RHS of Eq.~(\ref{ef32}).

Before determining the asymptotic series expansion of $\mathcal{J}$ for $\beta\to\infty$, we express $-\beta (F(N)-\mu N)$ as follows:
\begin{eqnarray}
&&\hspace{0.0cm} -\beta (F(N) -\mu N) = -\beta (F(N_0) -\mu N_0) \nonumber\\
&&\hspace{3.33cm} -\t{\beta} \Big\{ (h N -1)^2 + c_3 (h N -1)^3 + c_4 (h N -1)^4 + \dots\Big\} \nonumber\\
&&\hspace{2.94cm}
\equiv -\beta (F(N_0) -\mu N_0) - \t{\beta}\, \xi(h N -1), \label{ef93}
\end{eqnarray}
where
\begin{equation}
\t{\beta} \equiv \frac{1}{2} F''(N_0) N_0^2\,\beta,\;\;\; h \equiv \frac{1}{N_0}, \label{ef94}
\end{equation}
\begin{equation}
c_3 \equiv \frac{N_0 F'''(N_0)}{3 F''(N_0)},\;\; c_4 \equiv \frac{N_0^2 F''''(N_0)}{12 F''(N_0)}, \dots~. \label{ef95}
\end{equation}
The last expression in Eq.~(\ref{ef93}) defines the function $\xi(x)$ as
\begin{equation}
\xi(x) {:=} \sum_{j=2}^{\infty} c_j\, x^j, \label{ef96}
\end{equation}
in which $c_2=1$. Evidently, the absolute minimum of $\xi(x)$ is located at $x=0$. On account of Eq.~(\ref{ef89}) one readily verifies that $\t{\beta}$ is extensive (explicitly, for a constant density $n\equiv N_0/V$ it scales like $N_0$) and that the constants $c_2$, $c_3$, $c_4$, $\dots$, are all intensive.

Making use of the Euler-Maclaurin summation formula (item 23.1.30 in Ref.~\citen{AS72}), for $\beta>0$ to exponential accuracy one obtains that
\begin{equation}
\mathcal{J} = \mathrm{e}^{-\beta (F(N_0)-\mu N_0)}\, N_0\! \int_{-1}^{\infty} {\rm d}x\; \b{g}(x+1)\, \mathrm{e}^{-\t{\beta} \xi(x)}, \label{ef97}
\end{equation}
where
\begin{equation}
\b{g}(x) {:=} g(x/h) \equiv g(N_0 x). \label{ef98}
\end{equation}
The `exponential accuracy' to which we have just referred, concerns the \emph{ratio} of the exact $\mathcal{J}$ in Eq.~(\ref{ef92}) and the $\mathcal{J}$ in Eq.~(\ref{ef97}), which deviates from unity by an amount that to leading order decays like some power of $\mathrm{e}^{-N_0}$. Consequently, for all practical purposes the expression in Eq.~(\ref{ef97}) can be considered as the exact representation of the $\mathcal{J}$ in Eq.~(\ref{ef92}). In fact, to a similar exponential accuracy the lower bound of the integral on the RHS of Eq.~(\ref{ef97}) can be replaced by $-\infty$. One thus has
\begin{equation}
\mathcal{J} = \mathrm{e}^{-\beta (F(N_0)-\mu N_0)}\, N_0\! \int_{-\infty}^{\infty} {\rm d}x\; \b{g}(x+1)\, \mathrm{e}^{-\t{\beta} \xi(x)}. \label{ef99}
\end{equation}

Making use of the Laplace method (Ch. 5 in Ref.~\citen{ETC65}, \S~6.4 in Ref.~\citen{BO99}), extensively described in Sec.~\ref{ssf3s2}, one readily obtains the following asymptotic expression:\footnote{As regards the leading term, corresponding to $d_0$, the reader may consult problem 9.1, p.~308, in Ref.~\protect\citen{FW03}.}
\begin{equation}
\mathcal{J} \sim \frac{N_0\, \mathrm{e}^{-\beta (F(N_0)-\mu N_0)}}{\t{\beta}^{1/2}} \Big( d_0 + \frac{d_1}{\t{\beta}} +\dots \Big)\;\;\mbox{\rm as}\;\; \t{\beta}\to\infty, \label{ef100}
\end{equation}
where $\{ d_j\}$ are intensive (extensive) depending on whether $\b{g}(x) \equiv g(N_0 x)$ is intensive (extensive). For in particular $d_0$ one has
\begin{equation}
\frac{d_0}{\t{\beta}^{1/2}} = \b{g}(1) \int_{-\infty}^{\infty} {\rm d}x\; \mathrm{e}^{-\t{\beta} x^2} \iff d_0 = \sqrt{\pi}\, g(N_0). \label{ef101}
\end{equation}
We remark that as in the case of $\mathcal{I}$ considered in Sec.~\ref{ssf3s2}, the regularity of the asymptotic series expansion of the pre-exponential part of $\mathcal{J}$, expressed in terms of the asymptotic sequence $\{ 1/\t{\beta}^{1/2}, 1/\t{\beta}^{3/2}, \dots\}$ and displayed in Eq.~(\ref{ef100}), is dependent on the functions $\xi(x)$ and $\b{g}(x+1)$ being sufficiently many times differentiable at $x=0$.

Making use of the general expression in Eq.~(\ref{ef100}) and the result in Eq.~(\ref{ef101}) one arrives at
\begin{equation}
\frac{\sum_{N=0}^{\infty} \b{f}(N,s(N))\, \mathrm{e}^{-\beta (F(N) -\mu N)}}{\sum_{N=0}^{\infty} \mathrm{e}^{-\beta (F(N) -\mu N)}} = \b{f}(N_0,s(N_0)) + O(\frac{1}{\t{\beta}}), \label{ef102}
\end{equation}
where we have assumed that $s(N)$ is `continuous' at $N=N_0$. Since $\t{\beta}$ scales like $N_0$, it follows that the ratio of the LHS of Eq.~(\ref{ef102}) to $\b{f}(N_0,s(N_0))$ is equal to $1$ in the thermodynamic limit for \emph{all} $\beta>0$; for the $\b{f}(N,s)$ in Eq.~(\ref{ef20}), which is intensive, the LHS of Eq.~(\ref{ef102}) is exactly equal to $\b{f}(N_0,s(N_0))$ in the thermodynamic limit for all $\beta>0$. With reference to Eq.~(\ref{ef32}), hereby we have demonstrated that
\begin{equation}
\frac{1}{\mathcal{Z}_{\Sc g}}\sum_{\nu} f(\nu)\, \mathrm{e}^{-\beta K_{\nu}} \sim \b{f}(N_0,s(N_0))\;\;\; \mbox{\rm for}\;\;\; \beta\to\infty. \label{ef103}
\end{equation}
Since $s(N_0) \to 0$ for $\beta\to\infty$ (recall that here $s(N_0)$ and $0$ are compound indices), the expression in Eq.~(\ref{ef103}) reduces to that in Eq.~(\ref{ef12}) for $\beta\to\infty$.

Now we demonstrate that in the cases where the ansatz in Eq.~(\ref{ef25}) is exact, the expression in Eq.~(\ref{ef103}) becomes exact, valid for \emph{all} $\beta>0$, in the thermodynamic limit; for transparency, we assume that $\b{f}(N,s(N)) \equiv \varphi(N,E_s(N))$ is bounded for $N\to\infty$. To this end, we first express Eq.~(\ref{ef57}) as
\begin{equation}
\sum_s \varphi(N,E_s(N))\, \mathrm{e}^{-\beta E_s(N)} = \varphi(N,E(N))\, \mathrm{e}^{-\beta F(N)}
+ O(\frac{1}{N\beta})\, \mathrm{e}^{-\beta F(N)}, \label{ef104}
\end{equation}
which applies for sufficiently large $N\beta$. Multiplying both sides of this expression with $\mathrm{e}^{\beta \mu N}$ and applying $\sum_N$ to both sides of the resulting expression, one obtains that (cf. Eq.~(\ref{ef30}))
\begin{equation}
\sum_{N=0}^{\infty} \sum_s \varphi(N,E_s(N))\, \mathrm{e}^{-\beta (E_s(N)-\mu N)} = \sum_{N=0}^{\infty} \varphi(N,E(N))\, \mathrm{e}^{-\beta (F(N)-\mu N)}
+ O(\frac{1}{\beta N_0})\, \mathcal{Z}_{\Sc g},
\label{ef105}
\end{equation}
where we have used the fact that the essential contribution of the sum with respect to $N$, in particular as applied to the last term on the RHS of Eq.~(\ref{ef104}), originates from the neighbourhood of $N=N_0$, the width of this neighbourhood being of the order of $\sqrt{N_0}$. Dividing both sides of Eq.~(\ref{ef105}) by $\mathcal{Z}_{\Sc g}$, making use of Eq.~(\ref{ef102}), one obtains that for $\mu = \mu(\beta,N_0,V)$ and sufficiently large $\beta$ one has
\begin{equation}
\frac{1}{\mathcal{Z}_{\Sc g}}\sum_{N=0}^{\infty} f(\nu)\, \mathrm{e}^{-\beta K_{\nu}} = \b{f}(N_0,s(N_0)) + O(\frac{1}{\beta N_0}). \label{ef106}
\end{equation}
This result demonstrates that, in the thermodynamic limit the LHS of Eq.~(\ref{ef106}) is exactly equal to $\b{f}(N_0,s(N_0))$ for \emph{all} $\beta>0$. We should emphasize that this statement need not be valid for the $\b{f}(N,s)$ specific to $\mathscr{G}_{\sigma}({\bm k};z)$, Eq.~(\ref{ef20}); it only exactly applies for those $\b{f}(N,s)$ for which the ansatz in Eq.~(\ref{ef25}) is valid for the set of $s$ whose corresponding $E_s(N_0)$ are inside some neighbourhood of $E(N_0)$. For the same reason as presented earlier, replacing by $\b{f}(N_0,0)$ the $\b{f}(N_0,s(N_0))$ on the RHS of Eq.~(\ref{ef106}) leads to an error of the form $o(1/\beta)$ for sufficiently large $\beta$ (cf. Eq.~(\ref{ef71})).

\subsection{Insulating ground states; three leading-order asymptotic terms}
\label{ssf5}

Insulating $N_0$-particle GSs are characterised by a finite $\Delta \equiv \mu_{N_0}^+ -\mu_{N_0}^-$, Eq.~(\ref{ec19}). It follows that the expansion in Eq.~(\ref{ef85}) fails as $\beta\to\infty$ in the cases where the underlying $N_0$-particle GSs are insulating (see Eq.~(\ref{ec33}) and the subsequent text). For these cases, one needs to rely on \emph{two} separate expansions, suitable for the regions $N <N_0$ and $N >N_0$; for the region $N <N_0$ ($N >N_0$) one needs to rely on an expansion which only involves left (right) derivatives with respect to $N$ of $F(N)$, evaluated at $N=N_0-1$ ($N=N_0+1$).

Thus we employ the following expansions instead of that in Eq.~(\ref{ef85}):
\begin{eqnarray}
&&\hspace{-1.0cm}F(N) = F(N_0\pm 1) + F'(N_0\pm 1)\, \big(N-(N_0\pm 1)\big)\nonumber\\
&&\hspace{0.2cm} + \frac{1}{2} F''(N_0\pm 1)\, \big(N - (N_0\pm 1)\big)^2 + \dots \;\;\; \mbox{\rm for}\;\;\; N \gtrless N_0. \label{ef107}
\end{eqnarray}
Since the $N_0$-particle GSs of the systems under consideration are insulating, for sufficiently large $\beta$ to exponential accuracy we can replace $F(N_0\pm 1)$ and the derivatives of this function by respectively $E_0(N_0\pm 1)$ and the associated derivatives (see Sec.~\ref{ssf3s3})).\footnote{As will become evident, the distinction between $F(N)$ and $E_0(N)$ becomes relevant for $\mu$ too close to one of $\mu_{N}^{\pm}$; one can readily verify that the present substitutions will be of no consequence to the following results so long as $\max(\mu-\mu_{N}^-,\mu_{N}^+ -\mu) < \min(\mu-\mu_{N}^-,\mu_{N}^+ -\mu) + E_1(N)-E_0(N)$, where $E_1(N) - E_0(N) \approx \mu_{N}^+ -\mu_{N}^-$ (Sec.~\protect\ref{ssf3s3}). We should emphasise that the present substitutions are by no means essential; we could have maintained to deal with $F(N)$ only at the expense of introducing $\mu_{N}^{\pm}(\beta)$ whose zero-temperature limits are $\mu_{N}^{\pm}$.} Thus we employ
\begin{eqnarray}
&&\hspace{-0.5cm} F(N) -\mu N = \big(E_0(N_0\pm 1) - \mu (N_0\pm 1)\big) + \big(E_0'(N_0\pm 1) -\mu\big)\, \big(N - (N_0\pm 1)\big) \nonumber\\
&&\hspace{1.8cm} +\frac{1}{2} E_0''(N_0\pm 1)\, \big(N - (N_0\pm 1)\big)^2 + \dots\;\;\mbox{\rm for}\;\; N \gtrless N_0. \nonumber\\ \label{ef108}
\end{eqnarray}

Since (Eq.~(\ref{ec19}))
\begin{equation}
E_0(N_0\pm 1) = E_0(N_0) \pm \mu_{N_0}^{\pm}, \label{ef109}
\end{equation}
one can write
\begin{equation}
E_0(N_0\pm 1) - \mu (N_0\pm 1) \equiv \big(E_0(N_0) -\mu N_0\big) \pm (\mu_{N_0}^{\pm} -\mu). \label{ef110}
\end{equation}
Following Eq.~(\ref{ef109}) one further has (cf. Eqs.~(\ref{ec23}), (\ref{ec24}) and (\ref{ef88}))
\begin{equation}
E_0'(N_0\pm 1) = \mu_{N_0}^{\pm}. \label{ef111}
\end{equation}
Consequently, for $\mu_{N_0}^- < \mu< \mu_{N_0}^+$ the coefficients of $\big(N - (N_0 \pm 1)\big)$ in the expressions on the RHS of Eq.~(\ref{ef108}) are non-vanishing; the coefficient corresponding to $N >N_0$ is positive and that corresponding to $N <N_0$ negative. This is one of the major mathematical consequences associated with the non-differentiability of $F(N)$ at $N=N_0$ (at least for sufficiently low temperatures) for the cases where the $N_0$-particle GSs are insulating.

Following the above considerations and using the short-hand notation\footnote{Since $F(N)$ is to exponential accuracy equal to $E_0(N)$ for sufficiently large $\beta$ (see Sec.~\protect\ref{ssf3s3}), the distinction between $\b{f}(N,s(N))$ and $\b{f}(N,0)$ is negligible. For the definition of $s(N)$ see Eq.~(\protect\ref{ef28}) and note that, for the present case $F(N)$ is to exponential accuracy equal to $E(N)$ (see Eq.~(\ref{ef44})).} $g(N) \equiv \b{f}(N,0)$ we thus write
\begin{eqnarray}
&&\hspace{-0.0cm}\sum_{N=0}^{\infty} g(N)\, \mathrm{e}^{-\beta (F(N)-\mu N)} =  \mathrm{e}^{-\beta (E_0(N_0) - \mu N_0)} \Big\{ g(N_0)  \nonumber\\
&&\hspace{0.4cm} + \mathrm{e}^{-\beta (\mu_{N_0}^- -\mu)} \sum_{N=0}^{N_0-1} g(N)\, \mathrm{e}^{-\t{\beta}^- \zeta^-(h^- N -1)} + \mathrm{e}^{+\beta (\mu_{N_0}^+ -\mu)} \!\!\sum_{N=N_0+1}^{\infty} g(N)\, \mathrm{e}^{-\t{\beta}^+ \zeta^+(h^+ N -1)}\Big\}, \nonumber\\ \label{ef112}
\end{eqnarray}
where
\begin{equation}
h^{\pm} {:=} \frac{1}{N_0\pm 1},\;\; \t{\beta}^{\pm} {:=} \frac{\beta}{h^{\pm}} \equiv (N_0\pm 1)\,\beta, \label{ef113}
\end{equation}
\begin{equation}
\zeta^{\pm}(x) {:=} \sum_{j=1}^{\infty} e_j^{\pm}\, x^j, \label{ef114}
\end{equation}
in which
\begin{equation}
e_1^{\pm} \equiv \mu_{N_0}^{\pm} -\mu,\;\;\; e_2^{\pm} \equiv \frac{E_0''(N_0\pm 1)}{2 h^{\pm}},\;\;\; \dots~. \label{ef115}
\end{equation}
Since $E_0(N)$ is extensive, it follows that the coefficients $\{ e_j^{\pm} \}$ are intensive. Further, defining
\begin{equation}
\t{\zeta}(x) {:=} \left\{  \begin{array}{ll} \zeta^-(x), & x<0,\\ \\
\zeta^+(x), & x>0,\\ \end{array} \right. \label{ef116}
\end{equation}
one observes that $\t{\zeta}(x)$ has a \emph{cusp} at $x=0$; this is owing to the fact that $e_1^- <0$ and $e_1^+ >0$. This cusp is characteristic of insulating states. Note that since $\zeta^{\pm}(0) =0$, $\t{\zeta}(x)$ is continuous, and vanishing, at $x=0$. Note further that $\t{\zeta}(x)$ takes its absolute minimum value at $x=0$ so that $\t{\zeta}(x) >0$, for all $x$ away from $x=0$.

Below we determine the leading-order terms in the asymptotic series expansions of the sums in Eq.~(\ref{ef112}) for $\t{\beta}^{\pm}\to\infty$; evidently, in the thermodynamic limit $\t{\beta}^{\pm}$ is infinitely large for \emph{all} $\beta>0$. To this end we employ the Euler-Maclaurin summation formula (item 23.1.30 in Ref.~\citen{AS72}); as will become apparent, in the case at hand use of this formula does not fully dispose of discrete summations, nonetheless for sufficiently large values of $N_0$ the discrete sums to be evaluated can be readily expressed in closed forms. Below we consider the following generic functions:
\begin{equation}
S^- {:=} \sum_{N=0}^{N_0-1} \t{g}(h N-1)\, \mathrm{e}^{-\t{\beta} \t{\zeta}(h N -1)} , \label{ef117}
\end{equation}
and
\begin{equation}
S^+ {:=} \sum_{N=N_0+1}^{\infty} \t{g}(h N-1)\, \mathrm{e}^{-\t{\beta} \t{\zeta}(h N -1)}, \label{ef118}
\end{equation}
where
\begin{equation}
\t{g}(x) \equiv g(\frac{x+1}{h}). \label{ef119}
\end{equation}
In the expression for $S^-$ ($S^+$) $\t{\beta}$ and $h$ stand for $\t{\beta}^{-}$ and $h^{-}$ ($\t{\beta}^{+}$ and $h^{+}$) respectively; similarly, depending on whether one considers the $\t{g}(x)$ in the expression for $S^-$ or $S^+$, the $h$ in Eq.~(\ref{ef119}) is to be identified with $h^-$ and $h^+$ respectively. It will further be convenient to introduce
\begin{equation}
\phi(x) \equiv \t{g}(x)\, \mathrm{e}^{-\t{\beta} \t{\zeta}(x)}, \label{ef120}
\end{equation}
where depending on whether one deals with $S^-$ or $S^+$, $\t{\beta}$ stands for $\t{\beta}^-$ and $\t{\beta}^+$ respectively.

\subsubsection{The sum $S^-$}
\label{ssf5s1}

Using the Euler-Maclaurin summation formula (item 23.1.30 in Ref.~\citen{AS72}) one has
\begin{eqnarray}
S^- = \frac{1}{h} \int_{-1}^{0} {\rm d}x\; \phi(x) &+& \frac{1}{2}\big(\phi(0)+\phi(-1)\big) + \sum_{k=1}^{n-1} \frac{h^{2k-1} B_{2 k}}{(2k)!}\, \big( \phi^{(2k-1)}(0) - \phi^{(2k-1)}(-1)\big)
\nonumber\\
&+& \frac{h^{2n} B_{2 n}}{(2n)!} \sum_{k=0}^{N_0-2} \phi^{(2n)}\big((k +\theta) h -1\big), \label{ef121}
\end{eqnarray}
where $\{B_{m}\}$ are Bernoulli numbers ($B_0=1$, $B_1 =-\frac{1}{2}$, $B_2=\frac{1}{6}$, etc. \cite{AS72})\footnote{The ``Bernoullian'' numbers in Ref.~\protect\citen{WW62} (Ch. VII) differ from those in Ref.~\protect\citen{AS72} (Ch. 23); in the former one has $B_1 =\frac{1}{6}$, $B_2 = \frac{1}{30}$, etc. This should be taken into account when using the Euler-Maclaurin summation formula given in \S~7.21 of Ref.~\citen{WW62}.} and
\begin{equation}
\phi^{(m)}(x) \equiv \frac{{\rm d}^m\phi(x)}{{\rm d}x^m}.
\label{ef122}
\end{equation}
The integer $n\ge 2$ is arbitrary and $\theta \in (0,1)$. In Eq.~(\ref{ef121}) $\phi^{(m)}(0)$ stands for the $m$th  \emph{left} derivative of $\phi(x)$ at $x=0$.

Since $\phi(-1)$ and $\phi^{(m)}(-1)$, $m\ge 1$, are exponentially small for any $\beta>0$, one can identify these quantities with zero. This is not the case however as regards $\phi(0)$ and $\phi^{(m)}(0)$, $m\ge 1$, so that the sums on the RHS of Eq.~(\ref{ef121}) involving $\phi^{(2k-1)}(0)$ and $\phi^{(n)}(0)$ cannot be neglected. Further, since $\theta \in (0,1)$, $(k+\theta) h -1 \not=0$, the last sum on the RHS of Eq.~(\ref{ef121}) is an exponentially decaying function of $n$ for $\beta\to\infty$. We entirely dispose of the contribution of this sum by identifying $n$ with $\infty$; in exchange for this, the second sum on the RHS of Eq.~(\ref{ef121}) turns into an infinite series. Evaluation of this sum is facilitated by considering the fact that $h^{2k-1}$ decays to leading order like $1/N_0^{2k-1}$ for $N_0\to\infty$; on the basis of this observation, and since we ultimately wish to consider macroscopic systems, we only retain the leading-order term in the asymptotic series expansion of $\phi^{2k-1}(0)$ for $N_0\to\infty$. For $\t{\beta}\to\infty$ one has
\begin{equation}
\phi^{(m)}(x) \sim \big(-\t{\beta} \t{\zeta}'(x)\big)^m\, \phi(x)\;\;\;\mbox{\rm as}\;\;\; x\uparrow 0. \label{ef123}
\end{equation}
With reference to Eqs.~(\ref{ef113}), (\ref{ef114}) and (\ref{ef115}) one thus has
\begin{equation}
\phi^{(m)}(x) \sim (-\t{\beta} e_1)^m\, \t{g}(0) \equiv \frac{1}{h^m} (-\beta e_1)^m\, \t{g}(0)\;\;\mbox{\rm as}\;\; x\uparrow 0, \label{ef124}
\end{equation}
where clearly $e_1$ stands for $e_1^-$ and $h$ for $h^-\equiv N_0-1$.

Following the above considerations
\begin{eqnarray}
&&\hspace{-1.5cm}\sum_{k=1}^{\infty} \frac{h^{2k-1} B_{2k}}{(2k)!} \phi^{(2k-1)}(0) \sim \t{g}(0) \left.\sum_{k=1}^{\infty} \frac{B_{2k}}{(2k)!}\,
x^{2k-1}\right|_{x= -\beta e_1}\nonumber\\
&&\hspace{1.5cm} =\left. \t{g}(0) \Big(\frac{1}{2} \coth\big(\frac{x}{2}\big) -\frac{1}{x}\Big)\right|_{x= -\beta e_1}\;\mbox{\rm for}\;\;\; N_0\to\infty. \label{ef125}
\end{eqnarray}
Noting that
\begin{equation}
\frac{1}{2} \coth\big(\frac{x}{2}\big) -\frac{1}{x} \sim \pm\frac{1}{2} -\frac{1}{x} \pm \mathrm{e}^{\mp x}\;\;\; \mbox{\rm for}\;\;\; x\to \pm\infty, \label{ef126}
\end{equation}
for a macroscopically large $N_0$ and $\mu>\mu_{N_0}^-$, one has
\begin{equation}
\sum_{k=1}^{\infty} \frac{h^{2k-1} B_{2k}}{(2k)!} \phi^{(2k-1)}(0) = g(N_0-1) \Big\{\frac{1}{2} - \frac{1/(\mu-\mu_{N_0}^-)}{\beta} + O(\mathrm{e}^{-\beta (\mu-\mu_{N_0}^-)}) \Big\}\;\, \mbox{\rm for}\;\, \beta\to\infty. \label{ef127}
\end{equation}
With $\phi(0) = \t{g}(0) \equiv g(N_0-1)$, one thus arrives at
\begin{equation}
S^- \sim \frac{1}{h}\int_{-1}^0 {\rm d}x\; \phi(x) + g(N_0-1) \Big\{1 - \frac{1/(\mu-\mu_{N_0}^-)}{\beta} \Big\}\;\;\; \mbox{\rm for}\;\;\; \beta\to\infty. \label{ef128}
\end{equation}
As should be evident, the difference between the exact $S^-$ and the RHS of Eq.~(\ref{ef128}) is of the order of $g(N_0-1)$ times $\mathrm{e}^{-\beta (\mu-\mu_{N_0}^-)}$.

Mathematically, the assumption with regard to $\t{\zeta}'(x)\equiv {\rm d}\t{\zeta}(x)/{\rm d}x \not=0$ for $x\in [-1,0]$ is sufficient to obtain the asymptotic series expansion of the integral on the RHS of Eq.~(\ref{ef128}) through repeated application of integration by parts \cite{ETC65,BO99}. Explicitly, on applying integration by parts one obtains that
\begin{equation}
\int_{-1}^0 {\rm d}z\; \t{g}(x)\, \mathrm{e}^{-\t{\beta} \t{\zeta}(x)} = \left. \frac{-1}{\t{\beta}} \frac{\t{g}(x)}{\t{\zeta}'(x)}\,\mathrm{e}^{-\t{\beta} \t{\zeta}(x)} \right|_{x=-1}^0 + \frac{1}{\t{\beta}} \int_{-1}^0 {\rm d}x\; \Big\{ \frac{\partial}{\partial x} \frac{\t{g}(x)}{\t{\zeta}'(x)}\Big\}\, \mathrm{e}^{-\t{\beta} \t{\zeta}(x)}, \label{ef129}
\end{equation}
from which one observes that in order for the use of integration by parts to be justified, it is required that
\begin{itemize}
\item[(i)] $\t{\zeta}'(x) \not=0$ for $x=-1$ and $x=0$, and that
\item[(ii)] the function enclosed by the curly braces on the RHS of Eq.~(\ref{ef129}) be integrable over the interval $[-1,0]$.
\end{itemize}
When these two conditions are met, one can readily demonstrate that for $\t{\beta}\to\infty$ the second term on the RHS of Eq.~(\ref{ef129}) is asymptotically less relevant than the first term. In this connection, we note that on approximating $\t{\zeta}(x)$ for $x\in [-1,0]$ (that is $\zeta^-(x)$) by $e_1^- x + e_2^- x^2$, for the $x_0$ at which $\t{\zeta}'(x)=0$ one thus obtains that
\begin{equation}
x_0 \approx -\frac{e_1^-}{2 e_2^-} >0, \label{ef130}
\end{equation}
where the positivity of $-e_1^-/e_2^-$ follows from the fact that $e_1^- \equiv \mu_{N_0}^- -\mu <0$ and that $E_0''(N_0-1)>0$, the latter on account of the strict convexity of the GS energy (appendix \ref{sc}).

Provided that $\t{g}(x)$ and $\t{\zeta}(x)$ are sufficiently many times differentiable for $x\in [-1,0]$, one can in a similar fashion as above construct an asymptotic series expansion for the integral on the LHS of Eq.~(\ref{ef129}) in terms of the asymptotic sequence $\{ 1/\t{\beta}, 1/\t{\beta}^2,\dots \}$. From the perspective of our present applications, it is relevant that $\int_{-1}^0 {\rm d}x\; \phi(x)$ is pre-multiplied by $1/h$; with $\t{\beta} \equiv \beta/h$, and assuming that the second term in the large-$\t{\beta}$ asymptotic series expansion of the integral on the LHS of Eq.~(\ref{ef129}) decays like $1/\t{\beta}^2$, one obtains that
\begin{equation}
\frac{1}{h} \int_{-1}^0 {\rm d}x\; \phi(x) = g(N_0-1) \frac{1/(\mu-\mu_{N_0}^-)}{\beta} + \frac{1}{\beta}\, O(\frac{1}{\t{\beta}}). \label{ef131}
\end{equation}
This implies that, for macroscopic systems the first term on the RHS of Eq.~(\ref{ef131}) is exactly equal to the expression on the LHS of this equation for \emph{all} $\beta>0$. The same applies for the cases where for $\t{\beta}\to\infty$ the integral on the RHS of Eq.~(\ref{ef129}) to leading order decays like $1/\t{\beta}^{\alpha}$ or $\ln(\t{\beta})/\t{\beta}^{\alpha}$, with $\alpha>0$.

We note that the expression in Eq.~(\ref{ef131}) is based on the assumption that $g(N)$ is intensive. If $g(N)$ were extensive, i.e. scaling to leading order like $N$ for $N\to\infty$, then the second term on the RHS of Eq.~(\ref{ef131}) would be of the form $O(1/\beta^2)$. In such a case, for macroscopic systems and $\beta<\infty$ the exact result for the integral on the LHS of Eq.~(\ref{ef131}) is derivable from the first \emph{two} terms in the asymptotic series expansion of this integral.

Combining the expressions in Eqs.~(\ref{ef128}) and (\ref{ef131}), for macroscopic systems one obtains that
\begin{equation}
S^- \sim g(N_0-1)\;\;\; \mbox{\rm as}\;\;\; \beta\to\infty. \label{ef132}
\end{equation}
As indicated above, the deviation from unity of the ratio of $S^-$ to $g(N_0-1)$ is of the order of $\mathrm{e}^{-\beta (\mu-\mu_{N_0}^-)}$.

\subsubsection{The sum $S^+$}
\label{ssf5s2}

Using the Euler-Maclaurin summation formula (item 23.1.30 in Ref.~\citen{AS72}), on account of $\phi(\infty) =0$, $\phi^{(m)}(\infty)=0$, $m\ge 1$, and $\phi(0)= \t{g}(1) \equiv g(N_0+1)$, for $n=\infty$ one has (cf. Eq.~(\ref{ef121}))
\begin{equation}
S^+ = \frac{1}{h} \int_{0}^{\infty} {\rm d}x\; \phi(x) + \frac{1}{2} \t{g}(0) -\sum_{k=1}^{\infty} \frac{h^{2k-1} B_{2 k}}{(2k)!}\, \phi^{(2k-1)}(0). \label{ef133}
\end{equation}
Since (cf. Eq.~(\ref{ef124})),
\begin{equation}
\phi^{(m)}(x) \sim (-\t{\beta} e_1)^m\, \t{g}(0) \equiv \frac{1}{h^m} (-\beta e_1)^m\, \t{g}(0)\;\;\;\mbox{\rm as}\;\;\; x\downarrow 0, \label{ef134}
\end{equation}
where $e_1$ stands for $e_1^+$ and $h$ for $h^+$, for $N_0\to\infty$ one deduces that (cf. Eq.~(\ref{ef127}))
\begin{equation}
-\!\sum_{k=1}^{\infty} \frac{h^{2k-1} B_{2k}}{(2k)!} \phi^{(2k-1)}(0) = g(N_0+1) \Big\{\frac{1}{2} - \frac{1/(\mu_{N_0}^+ -\mu)}{\beta} + O(\mathrm{e}^{-\beta (\mu_{N_0}^+ -\mu)})\Big\}\;\, \mbox{\rm for}\;\, \beta\to\infty. \label{ef135}
\end{equation}
Since $\mu<\mu_{N_0}^+$, the last term enclosed by curly braces is exponentially small for $\beta\to\infty$.

To determine the integral on the RHS of Eq.~(\ref{ef133}) we proceed in exactly the same way as in the case of the integral on the RHS of Eq.~(\ref{ef121}) (or Eq.~(\ref{ef128})). To this end, we first note that on approximating $\t{\zeta}(x)$ for $x\downarrow 0$ by $e_1^+ x + e_2^+ x^2$, for the $x$ at which $\t{\zeta}'(x) = 0$, one obtains that
\begin{equation}
x_0 \approx -\frac{e_1^+}{2 e_2^+} <0, \label{ef136}
\end{equation}
where the negativity of $-e_1^+/e_2^+$ follows from the fact that $e_1^+ \equiv \mu_{N_0}^+ -\mu>0$ and that $E_0''(N_0+1)>0$, which follows from the strict convexity of the GS energy (appendix \ref{sc}). Applying integration by parts, one has (cf. Eq.~(\ref{ef129}))
\begin{equation}
\int_{0}^{\infty} {\rm d}z\; \t{g}(x)\, \mathrm{e}^{-\t{\beta} \t{\zeta}(x)} = g(N_0+1) \frac{1/(\mu_N^+ -\mu)}{\t{\beta}} + \frac{1}{\t{\beta}} \int_{0}^{\infty} {\rm d}x\; \Big\{ \frac{\partial}{\partial x} \frac{\t{g}(x)}{\t{\zeta}'(x)}\Big\}\, \mathrm{e}^{-\t{\beta} \t{\zeta}(x)}. \label{ef137}
\end{equation}
By the same reasoning as leading to the expression in Eq.~(\ref{ef131}), one thus arrives at
\begin{equation}
\frac{1}{h} \int_0^{\infty} {\rm d}x\; \phi(x) = g(N_0+1)  \frac{1/(\mu_N^+ -\mu)}{\beta} + \frac{1}{\beta} O(\frac{1}{\t{\beta}}). \label{ef138}
\end{equation}
For $g(N)$ an intensive function, the integral on the LHS of Eq.~(\ref{ef138}) is exactly equal to the first term on the RHS of Eq.~(\ref{ef138}) in the thermodynamic limit.

Combining the results in Eqs.~(\ref{ef135}) and (\ref{ef138}), from Eq.~(\ref{ef133}) one obtains that for $N_0\to\infty$ to exponential accuracy one has (cf. Eq.~(\ref{ef132}))
\begin{equation}
S^+ \sim g(N_0+1)\;\;\; \mbox{\rm for}\;\;\; \beta\to\infty. \label{ef139}
\end{equation}

\subsubsection{Analysis}
\label{ssf5s3}

On the basis of the expressions in Eqs.~(\ref{ef31}) (or (\ref{ef32})), (\ref{ef112}), (\ref{ef132}) and  (\ref{ef139}), one obtains that
\begin{eqnarray}
&&\hspace{0.0cm} \frac{1}{\mathcal{Z}_{\Sc g}} \sum_{\nu} f(\nu))\, \mathrm{e}^{-\beta K_{\nu}} \sim \frac{\b{f}(N_0,0) + \b{f}(N_0-1,0)\, \mathrm{e}^{-\beta(\mu-\mu_{N_0}^-)} +\b{f}(N_0+1,0)\, \mathrm{e}^{-\beta(\mu_{N_0}^+ -\mu)}}{1 + \mathrm{e}^{-\beta(\mu-\mu_{N_0}^-)} + \mathrm{e}^{-\beta(\mu_{N_0}^+ -\mu)}}\nonumber\\
&&\hspace{10.0cm} \mbox{\rm for}\;\;\; \beta\to\infty. \label{ef140}
\end{eqnarray}
Three cases can arise:
\begin{itemize}
\item[(i)] $\mu-\mu_{N_0}^- = \mu_{N_0}^+ -\mu \iff \mu = \frac{1}{2} (\mu_{N_0}^+ +\mu_{N_0}^-)$,
\item[(ii)] $\mu-\mu_{N_0}^- <\mu_{N_0}^+ -\mu$, and
\item[(iii)] $\mu-\mu_{N_0}^- >\mu_{N_0}^+ -\mu$.
\end{itemize}
From Eq.~(\ref{ef140}), up to and including the second leading term corresponding to $\beta\to\infty$, one obtains that
\begin{eqnarray}
&&\hspace{-0.5cm} \frac{1}{\mathcal{Z}_{\Sc g}} \sum_{\nu} f(\nu)\, \mathrm{e}^{-\beta K_{\nu}} \sim \b{f}(N_0,0) \nonumber\\
&&\hspace{1.5cm}
+\big(\b{f}(N_0-1,0)+\b{f}(N_0+1,0)-2\b{f}(N_0,0)\big)\, \mathrm{e}^{-\frac{1}{2}\beta (\mu_{N_0}^+ -\mu_{N_0}^-)}, \nonumber\\
&&\hspace{6.5cm} \mbox{\rm when}\;\;\; \mu = \frac{1}{2}(\mu_{N_0}^+ +\mu_{N_0}^-), \label{ef141}
\end{eqnarray}
\begin{eqnarray}
&&\hspace{0.0cm} \frac{1}{\mathcal{Z}_{\Sc g}} \sum_{\nu} f(\nu)\, \mathrm{e}^{-\beta K_{\nu}} \sim \b{f}(N_0,0) + \big( \b{f}(N_0\pm 1,0) - \b{f}(N_0,0)\big)\,
\mathrm{e}^{-\beta \vert \mu -\mu_{N_0}^{\pm}\vert},\nonumber\\
&&\hspace{6.8cm} \mbox{\rm when}\;\;\; \mu \gtrless \frac{1}{2} (\mu_{N_0}^+ +\mu_{N_0}^-). \label{ef142}
\end{eqnarray}
Thus $\mathcal{Z}_{\Sc g}^{-1} \sum_{\nu} f(\nu)\, \mathrm{e}^{-\beta K_{\nu}}$ decays fastest towards $\b{f}(N_0,0)$ for $\beta\to\infty$
when $\mu$ is located exactly at the mid-gap point.

\subsubsection{Identifying $\frac{1}{\mathcal{Z}_{\Sc g}} \sum_{\nu} f(\nu)\, \mathrm{e}^{-\beta K_{\nu}}$ with $\mathscr{G}_{\sigma}({\bf k};z)$}
\label{ssf5s4}

On considering $\mathcal{Z}_{\Sc g}^{-1} \sum_{\nu} f(\nu)\, \mathrm{e}^{-\beta K_{\nu}}$ to represent $\mathscr{G}_{\sigma}({\bm k};z)$, the function $\b{f}(N_0,0)$ in the above pertinent expressions is to be identified with the $\t{G}_{\sigma}({\bm k};z)$ corresponding to the $N_0$-particle GS $\vert\Psi_{N_0;0}\rangle$ of $\wh{H}$ (see Eq.~(\ref{ef21})). For clarity, we write
\begin{equation}
\b{f}(N_0,0) = \t{G}_{N_0;\sigma}({\bm k};z). \label{ef143}
\end{equation}
Similarly,
\begin{equation}
\b{f}(N_0\pm 1,0) = \t{G}_{N_0\pm 1;\sigma}({\bm k};z). \label{ef144}
\end{equation}
Since the $N_0$-particle GSs under consideration are insulating, the $N_0\pm 1$-particle GSs to which $\t{G}_{N_0\pm 1;\sigma}({\bm k};z)$ correspond are \emph{metallic}. This is readily established by employing the Lehmann representations of these functions. In the same way that one from this representation for $\t{G}_{N;\sigma}({\bm k};z)\equiv \t{G}_{\sigma}({\bm k};z)$ deduces the inequalities in Eq.~(\ref{ec7}) concerning the range of variation of the chemical potential $\mu$ corresponding to $N$-particle GSs, for $N_0+1$-particle GSs one obtains that
\begin{equation}
E_{N_0+1;0} - E_{N_0;0} <\mu < E_{N_0+2;0} -E_{N_0+1;0}, \label{ef145}
\end{equation}
and for $N_0-1$-particle GSs that
\begin{equation}
E_{N_0-1;0} - E_{N_0-2;0} <\mu < E_{N_0;0} - E_{N_0-1;0}. \label{ef146}
\end{equation}
With reference to Eq.~(\ref{ef109}), the energy difference in the left-most part of Eq.~(\ref{ef145}) is equal to $\mu_{N_0}^+$ and that in the right-most part of Eq.~(\ref{ef146}) is equal to $\mu_{N_0}^-$. Since above we have assumed $F(N)$ to be a smooth function of $N$ for $N<N_0$ and $N>N_0$ (Sec.~\ref{ssf5}), it follows that $E_{N_0+2;0} -E_{N_0+1;0}$ is to an error of the order of $1/N_0$ equal to $\mu_{N_0}^+$ and, similarly, $E_{N_0-1;0}-E_{N_0-2;0}$ is to an error of the order of $1/N_0$ equal to $\mu_{N_0}^-$. The strict convexity of $F(N)$ as a function of $N$ guarantees that the signs of the last-mentioned errors are such that the strict inequalities in Eqs.~(\ref{ef145}) and (\ref{ef146}) apply for any $N_0<\infty$. We have thus shown that the $N_0\pm 1$-particle states under consideration are indeed metallic.

In spite of the above fact, only for \emph{interacting} Hamiltonians the three Green functions $\t{G}_{N_0;\sigma}({\bm k};z)$, $\t{G}_{N_0\pm 1;\sigma}({\bm k};z)$ are dissimilar to one another. This follows from a `shake-up' effect which is absent in the GSs of non-interacting Hamiltonians. For these Hamiltonians the corresponding Green functions do not explicitly depend on $N$; for mean-field Hamiltonians, these Green functions implicitly depend of $N$ only for one specific value of $N$, say $N_0$, so that they remain unchanged upon changing $N$ from $N_0$ to $N_0\pm 1$. Further, since zero-temperature Green functions do not explicitly depend on $\mu$, the differences between the chemicals corresponding to $N_0$- and $N_0\pm 1$-particle GSs, discussed in the previous paragraph, have no impact on the behaviours of these Green functions. Explicitly, consider a mean-field Hamiltonian described by the single-particle energy dispersion $\t{\varepsilon}_{\bm k;\sigma}$ (Sec.~\ref{ss52s3}). Suppose that $\t{\varepsilon}_{\bm k;\sigma}$ is chosen such that the $N_0$-particle GS of this mean-field Hamiltonian is insulating. Although, by the arguments presented above, the $N_0\pm 1$-particle GSs of this Hamiltonian are metallic, since $\t{\varepsilon}_{\bm k;\sigma}$ is not recalculated for $N=N_0+1$ and $N=N_0-1$, it trivially follows that $\t{G}_{N_0;\sigma;0}({\bm k};z) \equiv \t{G}_{N_0+1;\sigma;0}({\bm k};z) \equiv \t{G}_{N_0-1;\sigma;0}({\bm k};z)$. In fact, even if $\t{\varepsilon}_{\bm k;\sigma}$ were to be recalculated for $N=N_0\pm 1$, for macroscopic systems the consequent changes in $\t{\varepsilon}_{\bm k;\sigma}$ would be of the order of $1/N_0$.

On the basis of the above observations, it follows that the second terms on the RHSs of Eqs.~(\ref{ef141}) and (\ref{ef142}) are identically vanishing when the expressions in these equations are applied to $\mathscr{G}_{\sigma;0}({\bm k};z)$. This is in conformity with the fact that the temperature dependence of $\mathscr{G}_{\sigma}({\bm k};z)$
is entirely due to particle-particle interaction.

As for \emph{interacting} Hamiltonians, consider the GSs of the Hubbard Hamiltonian for fermions involving nearest-neighbour hopping terms on a two-dimensional square lattice. The quantum Monte-Carlo calculations by Hirsch \cite{JEH85} on this system show that for $N_0$ corresponding to half-filling, and for at least $0< U/\vert t\vert <10$ (where $U$ is the on-site interaction energy and $t$ the nearest-neighbour hopping integral), $\vert\Psi_{N_0;0}\rangle$ is an anti-ferromagnetic \emph{insulating} state and $\vert\Psi_{N_0\pm 1;0}\rangle$ are paramagnetic \emph{metallic} states. For $N_0$ finite, however large, one may recall that for sufficiently large on-site energy $U$,  $\vert\Psi_{N_0+1;0}\rangle$ is a Nagaoka state \cite{YN65,YN66} (see also Ch.~8 in Ref.~\citen{PF99}). It follows that contrary to their non-interacting counterparts, $\t{G}_{N_0;\sigma}({\bm k};z)$,
$\t{G}_{N_0+1;\sigma}({\bm k};z)$ and $\t{G}_{N_0-1;\sigma}({\bm k};z)$
are distinctively different functions of ${\bm k}$, $z$ and $\sigma$. This, in turn, implies that the leading-order asymptotic contribution to $\mathscr{G}_{\sigma}({\bm k};z) - \t{G}_{N_0;\sigma}({\bm k};z)$, presented in Eqs.~(\ref{ef141} and (\ref{ef142}), is a non-trivial function of ${\bm k}$, $z$ and $\sigma$ for $\mu \in (\mu_{N_0}^-, \mu_{N_0}^+)$ and $\beta<\infty$.

\section{Some aspects concerning the metallic ground state of the one-dimensional Luttinger model for spin-less fermions}
\label{sd}

Here we make use of the expressions for the zero-temperature single-particle Green function and self-energy corresponding to the metallic GS of the one-dimensional Luttinger model for spin-less fermions \cite{JML63,ML65} deduced in Ref.~\citen{BF99} (see appendix D herein) from the explicit expression for the single-particle spectral function due to Voit \cite{JV93}. To make contact with the notation employed in Refs.~\citen{JV93} and \citen{BF99}, in the following we set $\hbar$ equal to unity, so that $\varepsilon \equiv \omega$, and identify $\mu$ with zero. We confine our consideration to the right branch of the single-particle spectrum, conventionally marked by $r=+$; thus, for instance, $G_{+}(k;\omega)$ denotes the single-particle Green function corresponding to the right-moving particles. Equating $k_{\Sc f}$ with zero, in the following $k \to 0$ implies approach of $k$ towards the right-most Fermi point.

\subsection{Preliminaries}
\label{ssa1}

With the density operator $\h{\rho}_r(k)$ defined as
\begin{equation}
\h{\rho}_r(k) \equiv \sum_{p} {:\!\h{c}_{r,k+p}^{\dag}
\h{c}_{r,p}\!:},\;\;\; r\in \{+,-\}, \label{ed1}
\end{equation}
where `${:\cdots:}$' denotes normal ordering \cite{JV94}, and
\begin{equation}
\wh{N} \equiv \wh{N}_+ + \wh{N}_-,\;\;\;\; \wh{J} \equiv \wh{N}_+ -
\wh{N}_-, \label{ed2}
\end{equation}
for the Hamiltonian of the system under consideration one has (see pp. 1002 and 1003 in Ref.~\citen{JV94})
\begin{eqnarray}
&&\hspace{-0.75cm}\wh{H} = \frac{\pi v_{\Sc f}}{L} \sum_{r, p}
{:\!\h{\rho}_r(p) \h{\rho}_r(-p)\!:} + \frac{\pi v_{\Sc f}}{2L}
(\wh{N}^2 + \wh{J}^{\;2}) \nonumber\\
&&\hspace{-0.0cm} + \frac{1}{L} \sum_p \Big( g_2(p)\, \h{\rho}_+(p)
\h{\rho}_-(p) + \frac{g_4(p)}{2} \sum_r {:\!\h{\rho}_r(p)
\h{\rho}_r(-p)\!:}\Big),\nonumber\\ \label{ed3}
\end{eqnarray}
where $L$ is the macroscopic length of the system and $g_2(p)$ and $g_4(p)$ are the forward-scattering interaction potentials. For this model the renormalised particle velocity $v(k)$ has the form
\cite{JV94}
\begin{equation}
v(k) = \sqrt{\Big( v_{\Sc f} + \frac{g_4(k)}{2\pi}\Big)^2 +
\Big(\frac{g_2(k)}{2\pi}\Big)^2}. \label{ed4}
\end{equation}

Below we shall encounter the parameter $\gamma_0(k)$ for which one has \cite{JV93}
\begin{equation}
\gamma_0(k) \equiv \frac{1}{4}\big( K(k) + 1/K(k) -2\big),
\label{ed5}
\end{equation}
where \cite{JV94}
\begin{equation}
K(k) \equiv \sqrt{\frac{2\pi v_{\Sc f} + g_4(k) - g_2(k)}{2\pi
v_{\Sc f} + g_4(k) + g_2(k)}}. \label{ed6}
\end{equation}
For `repulsive' (`attractive') interaction potential $g_2(k)>0$ ($<0$) one has $K(k) < 1$ ($>1$). A further parameter, denoted by $\mathcal{A}_{\alpha}(k)$, that we shall encounter below, is defined according to \cite{JV93,BF99}
\begin{equation}
\mathcal{A}_{\alpha}(k) \equiv
\frac{1}{\Gamma^2(\gamma_0(k))}\,\Big(\frac{\Lambda}{2
v(k)}\Big)^{\alpha}, \label{ed7}
\end{equation}
in which $\Lambda>0$ is the value for the cut-off wave-number determining $g_2(k)$ and $g_4(k)$ according to \cite{JV93}
\begin{equation}
g_i(k) = g_i \,\mathrm{e}^{-\Lambda \vert k\vert},\;\; i=2, 4.
\label{ed8}
\end{equation}
On assuming $\vert k\vert \ll \Lambda$, the functions $\mathcal{A}_{\alpha}(k)$, $\gamma_0(k)$ and $v(k)$ can be identified with $\mathcal{A}_{\alpha}(0) \equiv \mathcal{A}_{\alpha}$, $\gamma_0(0) \equiv \gamma_0$ and $v(0) \equiv v$ respectively.

As in Ref.~\citen{BF99}, here we distinguish between two cases, corresponding to $k=0$ and $k\not=0$.

\subsubsection{The case of $k=0$}
\label{ssa1s1}

In this case one has \cite{BF99}
\begin{equation}
G_{+}(0;\omega) \sim \pi \mathcal{A}_{\gamma_0+1}
\big\{ \cot(\pi\gamma_0) - i\big\}\, \mathrm{sgn}(\omega)
\vert\omega\vert^{2\gamma_0 -1}, \;\;\; \omega\to 0\;\;\; (0<\gamma_0<\frac{1}{2}).
\label{ed9}
\end{equation}
Noting that $G_{0;+}^{-1}(0;\omega)=\omega$, from this expression and the Dyson equation, for the real and imaginary parts of $\Sigma_{+}(0;\omega)$ one obtains that \cite{BF99}
\begin{equation}
\mathrm{Re}[\Sigma_+(0;\omega)] \sim \frac{-\cot(\pi\gamma_0)\,
\mathrm{sgn}(\omega)}{\pi
\mathcal{A}_{\gamma_0+1}\big(\cot^2(\pi\gamma_0)+ 1\big)}\,
\vert\omega\vert^{1-2\gamma_0},\;\;\omega\to 0, \label{ed10}
\end{equation}
\begin{equation}
\mathrm{Im}[\Sigma_+(0;\omega)] \sim
\frac{-\mathrm{sgn}(\omega)}{\pi
\mathcal{A}_{\gamma_0+1}\big(\cot^2(\pi\gamma_0)+ 1\big)}\,
\vert\omega\vert^{1-2\gamma_0},\;\;\omega\to 0. \label{ed11}
\end{equation}
We point out that $\cot(\pi\gamma_0) > 0$ for $0 < \gamma_0 < 1/2$.

\subsubsection{The case of $k\not=0$}
\label{ssa1s2}

In this case one has \cite{BF99}
\begin{equation}
G_{+}(k;\omega) \sim \pi \gamma(\gamma_0,\Lambda
k)\, \mathcal{A}_{\gamma_0} \big\{ \cot(\pi\gamma_0) - i\big\}\,
\mathrm{sgn}(\omega) \vert\omega\vert^{\gamma_0 -1}, \;\;\; \omega\to 0\;\;\; (0<\gamma_0< 1), \label{ed12}
\end{equation}
where $\gamma(a,z)$ is the incomplete Gamma function (item 6.5.2 in Ref.~\citen{AS72}) for which one has $\gamma(a,\infty) \equiv \Gamma(a)$ (item 6.1.1 in Ref.~\citen{AS72}); thus, for sufficiently large $\Lambda k$, $\gamma(\gamma_0,\Lambda k) \sim \Gamma(\gamma_0)$. Noting that $G_{0;+}^{-1}(k;\omega)=\omega - v k$, from the expression in Eq.~(\ref{ed12}) and the Dyson equation, for the real and imaginary parts of $\Sigma_{+}(k;\omega)$ one obtains that \cite{BF99}
\begin{equation}
\mathrm{Re}[\Sigma_+(k;\omega)] \sim
\frac{-\cot(\pi\gamma_0)\,
\mathrm{sgn}(\omega)}{\pi\gamma(\gamma_0,\Lambda k)
\mathcal{A}_{\gamma_0}\big(\cot^2(\pi\gamma_0)+ 1\big)}\,
\vert\omega\vert^{1-\gamma_0} - v k, \;\;\;\omega\to 0, \label{ed13}
\end{equation}
\begin{equation}
\mathrm{Im}[\Sigma_+(k;\omega)] \sim
\frac{-\mathrm{sgn}(\omega)}{\pi\gamma(\gamma_0,\Lambda k)
\mathcal{A}_{\gamma_0}\big(\cot^2(\pi\gamma_0)+ 1\big)}\,
\vert\omega\vert^{1-\gamma_0},\;\;\; \omega\to 0, \label{ed14}
\end{equation}
where $\cot(\pi\gamma_0)$ takes both positive and negative values
for $\gamma_0$ varying inside $(0,1)$.

\subsection{Concerning $\lim_{\beta\to\infty}\b{\nu}_{\sigma}^{(1)}({\bf k})$}
\label{ssa2}

Combining the results in Eqs.~(\ref{ed10}), (\ref{ed11}), (\ref{ed13}) and (\ref{ed14}), we arrive at
\begin{equation}
\frac{\mathrm{Im}[\Sigma_+(k;\omega)]}{\mathrm{Re}[\Sigma_+(k;\omega)]
- \Sigma_+(k;0)} \sim \tan(\pi\gamma_0),\;\;\; \omega\to 0,
\label{ed15}
\end{equation}
which applies both for $k=0$ and $k\not=0$. This result has been reproduced (albeit in a somewhat different notation) in Eq.~(\ref{e64}). We point out that as $g_2 \to 0$, $K\to 1$ so that $\gamma_0 \to 0$ (see Eqs.~(\ref{ed5}) and (\ref{ed6})). Consequently, in the limit $g_2 = 0$ the RHS of Eq.~(\ref{ed15}) is vanishing; this result is to be contrasted with that in Eq.~(\ref{e63}) which is specific to Fermi- and marginal-Fermi liquids and which applies irrespective of the strength of the particle-particle interaction potential.

\subsection{Concerning $\lim_{\beta\to\infty} \b{\nu}_{\sigma}^{(2)}({\bf k})$}
\label{ssa3}

In Sec.~\ref{ss43} we indicated that on account of Eq.~(\ref{e64}), $\lim_{\beta\to\infty}\b{\nu}_{\sigma}^{(1)}({\bm k})$ does not satisfy Eq.~(\ref{e65}) for the ${\bm k}$ points in the infinitesimal neighbourhoods of the Fermi points of the metallic state of the one-dimensional Luttinger model. Below we investigate the behaviour of $\lim_{\beta\to\infty} \b{\nu}_{\sigma}^{(2)}({\bm k})$ for in particular this GS and show that this function is well-defined for all ${\bm k}$.

Since both $\t{G}_{\sigma}({\bm k};z)$ and $\t{\Sigma}_{\sigma}({\bm k};z)$ are analytic in the complex $z$ plane away from the real axis and $\t{G}_{\sigma}({\bm k};z) \partial\t{\Sigma}_{\sigma}({\bm k};z)/\partial z$ decays sufficiently rapidly towards zero for $\vert z\vert\to\infty$ (appendix \ref{sc}), it follows that in our investigation we only need to concentrate on the behaviour of the integrand of the contour integral in Eq.~(\ref{e43}) in the neighbourhood of $z=\mu$; if $\lim_{\beta\to\infty} \b{\nu}_{\sigma}^{(2)}({\bm k})$ is to be unbounded, this must arise as a consequence of $\t{G}_{\sigma}({\bm k};z)\partial\t{\Sigma}_{\sigma}({\bm k};z)/\partial z$ diverging sufficiently strongly for $z\to\mu$.

For our following considerations it is convenient to parameterise $\mathscr{C}(\mu)$ according to $z = \mu+i y$, where $y\uparrow_{-\infty}^{+\infty}$; making use of $\t{G}_{\sigma}({\bm k};z^*) = \t{G}_{\sigma}^*({\bm k};z)$ and $\t{\Sigma}_{\sigma}({\bm k};z^*) = \t{\Sigma}_{\sigma}^{(\nu) *}({\bm k};z)$ for $\mathrm{Im}(z)\not=0$, we thus deduce that
\begin{equation}
\lim_{\beta\to\infty} \b{\nu}_{\sigma}^{(2)}({\bm k}) =
\frac{1}{\pi}\lim_{y_0\downarrow 0} \int_{y_0}^{\infty} {\rm d}y\;
\mathrm{Re}[\t{G}_{\sigma}({\bm k};\mu+iy)\,
\frac{\partial}{\partial\,iy}\, \t{\Sigma}_{\sigma}({\bm k};\mu+i
y)]. \label{ed16}
\end{equation}
For analyzing the behaviour of the integrand of the integral on the RHS of this equation we first note that since $y \ge y_0 > 0$, $\t{G}_{\sigma}({\bm k};\mu+iy)$ and $\t{\Sigma}_{\sigma}({\bm k};\mu+iy)$ correspond to analytic continuations of respectively ${G}_{\sigma}({\bm k};\varepsilon)$ and ${\Sigma}_{\sigma}({\bm k};\varepsilon)$ from $\varepsilon > \mu$ to the upper half of the complex $z$ plane (cf. Eqs.~(\ref{e3}) and (\ref{e5})). With reference to the conventions $\hbar=1$ and $\mu=0$ adopted in this appendix, we need therefore analytically to continue $G_+(k;\omega)$ and $\Sigma_+(k;\omega)$ from the region $\omega > 0$ to the upper half of the $z$ plane. We have (below $y\downarrow 0$):

\noindent
--- The case of $k=0$

\begin{equation}
\t{G}_+(0;iy) \sim \pi\,\mathcal{A}_{\gamma_0+1} \big\{
\cot(\pi\gamma_0) - i\big\}\, (i y)^{2\gamma_0-1}, \label{ed17}
\end{equation}
\begin{equation}
\t{\Sigma}_+(0;iy) \sim \frac{-
\big(\cot(\pi\gamma_0)+i\big)}{\pi\mathcal{A}_{\gamma_0+1}
\big(\cot^2(\pi\gamma_0) +1\big)}\, (iy)^{1-2\gamma_0}. \label{ed18}
\end{equation}
Consequently
\begin{equation}
\t{G}_+(0;iy)\, \frac{\partial}{\partial\,iy}\, \t{\Sigma}_+(0;iy)
\sim \frac{-(1-2\gamma_0)}{iy}, \label{ed19}
\end{equation}
which is \emph{purely imaginary} for $y\in \mathds{R}$.

\noindent
--- The case of $k\not=0$

\begin{equation}
\t{G}_+(k;iy) \sim \pi\gamma(\gamma_0,\Lambda\,k)\,
\mathcal{A}_{\gamma_0} \big\{ \cot(\pi\gamma_0)-i\big\}\,
(iy)^{\gamma_0-1}, \label{ed20}
\end{equation}
\begin{equation}
\t{\Sigma}_+(k;iy) \sim \frac{-\big(\cot(\pi\gamma_0)+i\big)}
{\pi\gamma(\gamma_0,\Lambda\,k)\mathcal{A}_{\gamma_0}
\big(\cot^2(\pi\gamma_0) +1\big)}\, (iy)^{1-\gamma_0}. \label{ed21}
\end{equation}
Consequently
\begin{equation}
\t{G}_+(k;iy)\, \frac{\partial}{\partial\,iy}\, \t{\Sigma}_+(k;iy)
\sim \frac{-(1-\gamma_0)}{iy}, \label{ed22}
\end{equation}
which is also \emph{purely imaginary} for $y\in \mathds{R}$.

We observe that for both $k=0$ and $k\not=0$ the leading-order term in the asymptotic series expansion of the integrand of the integral on the RHS of Eq.~(\ref{ed16}) is, if divergent, less divergent than $1/y$. Consequently, the limit on the RHS of Eq.~(\ref{ed16}) corresponding to $y_0 =0$ exists and thus $\lim_{\beta\to\infty} \b{\nu}_{\sigma}^{(2)}({\bm k})$ is indeed bounded. $\hfill\Box$

\end{appendix}

\end{document}